\documentclass[preprint,authoryear,11pt]{elsarticle}
\usepackage{deluxetable}
\usepackage{amssymb}
\newcommand{\kms}{\,{\rm km}\;{\rm s}^{-1}}
\newcommand{\cms}{\,{\rm cm}\;{\rm s}^{-1}}
\newcommand{\hubunits}{\,\kms\;{\rm Mpc}^{-1}}
\newcommand{\hmpc}{\,h^{-1}\;{\rm Mpc}}
\newcommand{\invhmpc}{\,h\;{\rm Mpc}^{-1}}

\newcommand{\hgpc}{\,h^{-1}\;{\rm Gpc}}
\newcommand{\msun}{\,M_\odot}
\newcommand{\hmsun}{h^{-1}\,M_\odot}
\newcommand{\K}{\,{\rm K}}

\newcommand{\Mpc}{\,{\rm Mpc}}

\newcommand{\lya}{Ly$\alpha$}
\newcommand{\mdeg}{\,{\rm deg}}
\newcommand{\micron}{\,{\mu}{\rm m}}
\newcommand{\arcsec}{^{\prime\prime}}
\newcommand\kmax{k_{\rm max}}

\newcommand{\rosat}{{\it ROSAT}}
\newcommand{\erosita}{{\it eROSITA}}
\newcommand{\chandra}{{\it Chandra}}
\newcommand{\xmm}{{\it XMM-Newton}}
\newcommand{\euclid}{{\it Euclid}}
\newcommand{\wfirst}{{\it WFIRST}}
\newcommand{\jwst}{{\it JWST}}
\newcommand{\hst}{{\it HST}}
\newcommand{\planck}{{\it Planck}}
\newcommand{\wmap}{{\it WMAP}}
\newcommand{\hipparcos}{{\it Hipparcos}}
\newcommand{\gaia}{{\it Gaia}}
\newcommand{\lisa}{{\it LISA}}
\newcommand{\spitzer}{{\it Spitzer}}
\newcommand{\wise}{{\it WISE}}
\newcommand{\jdem}{{\it JDEM}}

\newcommand{\lcdm}{{$\Lambda$}CDM}

\newcommand{\om}{{\Omega_{m}}}
\newcommand{\Om}{{\Omega_{m}}}
\newcommand{\op}{{\Omega_{\phi}}}

\newcommand{\ok}{{\Omega_{k}}}
\newcommand{\ob}{{\Omega_{b}}}

\newcommand{\ol}{{\Omega_\Lambda}}

\newcommand{\omo}{{\Omega_{m}}}
\newcommand{\opo}{{\Omega_{\phi}}}
\newcommand{\orado}{{\Omega_{r}}}
\newcommand{\oko}{{\Omega_{k}}}
\newcommand{\obo}{{\Omega_{b}}}
\newcommand{\oco}{{\Omega_{c}}}

\newcommand{\phidot}{\dot{\phi}}
\newcommand{\uphi}{u_\phi}
\newcommand{\uphio}{u_{\phi,0}}
\newcommand{\pphi}{p_\phi}

\newcommand{\zmax}{z_{\rm max}}
\newcommand{\lmax}{l_{\rm max}}
\newcommand{\fsky}{f_{\rm sky}}
\newcommand{\Ggr}{G_{\rm GR}}

\newcommand{\vx}{{\bf x}}
\newcommand{\vr}{{\bf r}}
\newcommand{\vk}{{\bf k}}
\newcommand{\vv}{{\bf v}}

\newcommand{\dvk}{\tilde{\delta}(\vk)}
\newcommand{\dvkprime}{\tilde{\delta}(\vk')}

\newcommand{\sigEight}{\sigma_8}
\newcommand{\sigEightz}{\sigma_8(z)}
\newcommand{\sigElev}{\sigma_{11,{\rm abs}}}
\newcommand{\sigElevz}{\sigma_{11,{\rm abs}}(z)}

\newcommand{\avg}[1]{\left\langle #1 \right\rangle}
\newcommand{\Var}{\mbox{Var}}

\newcommand{\slnx}{{[\mbox{Var}(\ln X | M,z)]^{1/2}}}

\newcommand{\dNN}{\Delta N/N}

\newcommand{\arcmin}{{\rm arcmin}}

\newcommand{\la}{\lesssim}
\newcommand{\ga}{\gtrsim}

\journal{Physics Reports}

\topmargin = -\topskip
\advance\topmargin by -\headsep
\begin{document}

\begin{frontmatter}

%% Title, authors and addresses

%% use the tnoteref command within \title for footnotes;
%% use the tnotetext command for the associated footnote;
%% use the fnref command within \author or \address for footnotes;
%% use the fntext command for the associated footnote;
%% use the corref command within \author for corresponding author footnotes;
%% use the cortext command for the associated footnote;
%% use the ead command for the email address,
%% and the form \ead[url] for the home page:
%%
%% \title{Title\tnoteref{label1}}
%% \tnotetext[label1]{}
%% \author{Name\corref{cor1}\fnref{label2}}
%% \ead{email address}
%% \ead[url]{home page}
%% \fntext[label2]{}
%% \cortext[cor1]{}
%% \address{Address\fnref{label3}}
%% \fntext[label3]{}

\title{Observational Probes of Cosmic Acceleration\tnoteref{$^*$}}
\tnotetext[$^*$]{To Appear in Physics Reports.}

%% use optional labels to link authors explicitly to addresses:
%% \author[label1,label2]{<author name>}
%% \address[label1]{<address>}
%% \address[label2]{<address>}

\author[1,2]{David H. Weinberg}
\author[2]{Michael J. Mortonson}
\author[3,4]{Daniel J. Eisenstein}
\author[5]{Christopher Hirata}
\author[6]{Adam G. Riess}
\author[7]{Eduardo Rozo}
\address[1]{Department of Astronomy, Ohio State University, Columbus,
OH}
\address[2]{Center for Cosmology and Astro-Particle Physics,
         Ohio State University, Columbus, OH}
\address[3]{Steward Observatory, University of Arizona, Tucson, AZ}
\address[4]{Harvard College Observatory, Cambridge, MA}
\address[5]{California Institute of Technology, Pasadena, CA}
\address[6]{Department of Physics and Astronomy, Johns Hopkins University,
            Baltimore, MD}
\address[7]{Kavli Institute for Cosmological Physics, University of
            Chicago, Chicago, IL}

\begin{abstract}
The accelerating expansion of the universe is the most surprising cosmological
discovery in many decades, implying that the universe is dominated by some form
of ``dark energy'' with exotic physical properties, or that Einstein's theory
of gravity breaks down on cosmological scales. The profound implications of 
cosmic acceleration have inspired ambitious efforts to understand its origin,
with experiments that aim to measure the history of expansion and growth of 
structure with percent-level precision or higher. We review in detail the four
most well established methods for making such measurements: Type Ia supernovae,
baryon acoustic oscillations (BAO), weak gravitational lensing, and the 
abundance of galaxy clusters. We pay particular attention to the systematic
uncertainties in these techniques and to strategies for controlling them at the
level needed to exploit ``Stage IV'' dark energy facilities such as BigBOSS,
LSST, \euclid, and \wfirst. We briefly review a number of other approaches 
including redshift-space distortions, the Alcock-Paczynski effect, and direct
measurements of the Hubble constant $H_0$. We present extensive forecasts for
constraints on the dark energy equation of state and parameterized deviations
from General Relativity, achievable with Stage III and Stage IV experimental
programs that incorporate supernovae, BAO, weak lensing, and cosmic microwave
background data. We also show the level of precision required for clusters or
other methods to provide constraints competitive with those of these fiducial
programs. We emphasize the value of a balanced program that employs several of
the most powerful methods in combination, both to cross-check systematic 
uncertainties and to take advantage of complementary information. Surveys to 
probe cosmic acceleration produce data sets that support a wide range of
scientific investigations, and they continue the longstanding astronomical 
tradition of mapping the universe in ever greater detail over ever larger 
scales.
\end{abstract}

\begin{keyword}
%% keywords here, in the form: keyword \sep keyword

%% MSC codes here, in the form: \MSC code \sep code
%% or \MSC[2008] code \sep code (2000 is the default)

\end{keyword}

\end{frontmatter}

% \linenumbers

\tableofcontents

\vfill\eject

%% main text
%\onehalfspacing
\section{Introduction}
\label{sec:intro}

Gravity pulls.  Newton's {\it Principia} generalized
this longstanding fact of human experience into a universal attractive force,
providing compelling explanations of an extraordinary range of
terrestrial and celestial phenomena.
Newtonian attraction weakens with distance, but it never vanishes,
and it never changes sign.
Einstein's theory of General Relativity (GR) reproduces Newtonian
gravity in the limit of weak spacetime curvature and
low velocities.  For a homogeneous universe filled with matter
or radiation, GR predicts that the cosmic expansion will slow down
over time, in accord with Newtonian intuition.
In the late 1990s, however, two independent studies of distant
supernovae found that the expansion of the universe has
accelerated over the last five billion years
\citep{riess98,perlmutter99}, a remarkable discovery
that is now buttressed by multiple lines of independent evidence.
On the scale of the cosmos, gravity repels.

Cosmic acceleration is the most profound puzzle in contemporary physics.
Even the least exotic explanations demand the existence of a pervasive
new component of the universe with unusual physical properties that
lead to repulsive gravity.  Alternatively, acceleration could be a sign
that GR itself breaks down on cosmological scales.
Cosmic acceleration may be the crucial empirical clue that leads
to understanding the interaction between gravity and the quantum vacuum,
or reveals the existence of extra spatial dimensions, or sheds light on the
nature of quantum gravity itself.

Because of these profound implications, cosmic acceleration has
inspired a wide range of ambitious experimental efforts, which
aim to measure the expansion history and growth of structure
in the cosmos with percent-level precision or better.
In this article, we review the observational methods that
underlie these efforts, with particular attention to techniques
that are likely to see major advances over the next decade.
We will emphasize the value of a balanced program that pursues
several of these methods in combination, both to cross-check
systematic uncertainties and to take advantage of complementary information.

The remainder of this introduction briefly recaps the history
of cosmic acceleration
and current theories for its origin,
then sets this article in the context of future experimental
efforts and other reviews of the field.  Section~\ref{sec:observables} 
describes
the basic observables that can be used to probe cosmic acceleration,
relates them to the underlying equations that govern the
expansion history and the growth of structure, and introduces
some of the parameters commonly used to define ``generic''
cosmic acceleration models.
It concludes with an overview of the leading methods for measuring
these observables.
In \S\S\ref{sec:sn}--\ref{sec:cl}
we go through the four most well developed
methods in detail: Type Ia supernovae, baryon acoustic oscillations (BAO),
weak gravitational lensing, and clusters of galaxies.
Section~\ref{sec:alternatives} 
summarizes several other potential probes, whose
prospects are currently more difficult to assess
but in some cases appear quite promising.
Informed by the discussions in these sections, \S\ref{sec:forecast} presents
our principal new results: forecasts of the constraints on
cosmic acceleration models that could be achieved by
combining results from these methods, based on ambitious but
feasible experiments like the ones endorsed by the Astro2010
Decadal Survey report, {\it New Worlds, New Horizons in Astronomy 
and Astrophysics}.
We summarize the implications of our analyses in \S\ref{sec:conclusions} .

\subsection{History}
\label{sec:history}

Just two years after the completion of General Relativity,
\cite{einstein17}
introduced the first modern cosmological model.
With little observational guidance, Einstein assumed (correctly)
that the universe is homogeneous on large scales, and he proposed
a matter-filled space with finite, positively curved, 3-sphere
geometry.  He also assumed (incorrectly) that the universe is static.
Finding these two assumptions to be incompatible, Einstein modified
the GR field equation to include the infamous ``cosmological term,''
now usually known as the ``cosmological constant'' and denoted $\Lambda$.
In effect, he added a new component whose repulsive gravity
could balance the attractive gravity of the matter
(though he did not describe his modification in these terms).
In the 1920s, \citet{friedmann22,friedmann24} and \cite{lemaitre27}
introduced GR-based cosmological
models with an expanding or contracting universe, some of them
including a cosmological constant, others not.
In 1929, Hubble discovered direct evidence for the expansion of
the universe \citep{hubble29}, thus removing the original motivation for
the $\Lambda$ term.\footnote{Several recent papers have addressed the
contributions of Lema\^itre, Friedmann, and Slipher 
to this discovery; the story is
interestingly tangled (see, e.g.,
\citealt{block11,vandenbergh11,livio11,belenkiy12,peacock13}).}
In 1965, the discovery and interpretation
of the cosmic microwave background (CMB; \citealt{penzias65,dicke65})
provided the pivotal evidence for a hot big bang origin of the cosmos.

{}From the 1930s through the 1980s, a cosmological constant seemed
unnecessary to explaining cosmological observations.
The ``cosmological constant problem'' as it was defined in
the 1980s was a theoretical one: why was the gravitational impact
of the quantum vacuum vanishingly small compared to the
``naturally'' expected value (see \S\ref{sec:theory})?
In the late 1980s and early 1990s, however, a variety of
indirect evidence began to accumulate in favor of a cosmological
constant.  Studies of large scale galaxy clustering, interpreted
in the framework of cold dark matter (CDM) models with inflationary initial
conditions, implied a low matter density parameter
$\Omega_m = \rho_m/\rho_{\rm crit} \approx 0.15-0.4$
(e.g., \citealt{maddox90,efstathiou90}), in agreement
with direct dynamical estimates that assumed galaxies to be
fair tracers of the mass distribution.
Reconciling this result with the standard inflationary cosmology
prediction of a spatially flat universe \citep{guth80}
required a new energy component with density parameter $1-\Omega_m$.
Open-universe inflation models were also considered, but explaining
the observed homogeneity of the CMB \citep{smoot92} in such
models required speculative appeals to quantum gravity effects
(e.g., \citealt{bucher95}) rather than the semi-classical explanation
of traditional inflation.

By the mid-1990s, many cosmological
simulation studies included both open-CDM models and $\Lambda$-CDM
models, along with $\Omega_m=1$ models incorporating tilted inflationary
spectra, non-standard radiation components, or massive neutrino
components (e.g., \citealt{ostriker96,cole97,gross98,jenkins98}).  
Once normalized to the observed
level of CMB anisotropies, the large-scale structure
predictions of open and flat-$\Lambda$ models differed at the
tens-of-percent level, with flat models generally yielding a more
natural fit to the observations (e.g., \citealt{cole97}).
Conflict between high values of the Hubble constant and the
ages of globular clusters also favored a cosmological constant
(e.g., \citealt{pierce94,freedman94,chaboyer96}),
though the frequency of gravitational lenses pointed in the 
opposite direction \citep{kochanek96}.
Thus, the combination of CMB data, large-scale structure data,
age of the universe,
and inflationary theory led many cosmologists to consider models
with a cosmological constant, and some to declare it as the
preferred solution (e.g., \citealt{efstathiou90,krauss95,ostriker95}).

Enter the supernovae.
In the mid-1990s, two teams set out to measure the cosmic deceleration
rate, and thereby determine the matter density parameter $\Omega_m$,
by discovering and monitoring high-redshift, Type Ia supernovae.
The recognition that the peak luminosity of supernovae was tightly
correlated with the shape of the light curve
\citep{phillips93,riess96} played a critical role in this strategy,
reducing the intrinsic distance error per supernova to $\sim 10\%$.
While the first analysis of a small sample indicated deceleration
\citep{perlmutter97}, by 1998 the two teams had converged on a remarkable
result: when compared to local Type Ia supernovae \citep{hamuy96}, 
the supernovae at
$z\approx 0.5$ were fainter than expected in a matter-dominated
universe with $\Omega_m\approx 0.2$ by about 0.2 mag, or 20\%
\citep{riess98,perlmutter99}.
Even an empty, freely expanding universe was inconsistent with
the observations.
Both teams interpreted their measurements as evidence for an accelerating
universe with a cosmological constant, consistent with a flat universe
($\Omega_{\rm tot}=1$) having $\Omega_\Lambda \approx 0.7$.

Why was the supernova evidence for cosmic acceleration accepted so quickly
by the community at large?  First, the internal checks carried out by
the two teams, and the agreement of their conclusions despite independent
observations and many differences of methodology, seemed to rule out
many forms of observational systematics, even relatively subtle effects
of photometric calibration or selection bias.
Second, the ground had been well prepared by the CMB and large scale
structure data, which already provided substantial indirect evidence
for a cosmological constant.  This confluence of arguments favored
the cosmological interpretation of the results over astrophysical
explanations such as evolution of the supernova population or grey dust
extinction that increased towards higher redshifts.  Third, the supernova
results were followed within a year by the results of balloon-borne CMB
experiments that mapped the first acoustic peak and measured its
angular location, providing strong evidence for spatial flatness
(\citealt{debernardis00,hanany00}; see \citealt{netterfield97} for
earlier ground-based measurements hinting at the same result).
On its own, the acoustic peak only implied $\Omega_{\rm tot} \approx 1$,
not a non-zero $\Omega_\Lambda$, but it dovetailed perfectly with the
estimates of $\Omega_m$ and $\Omega_\Lambda$ from large scale structure
and supernovae.  Furthermore, the acoustic peak measurement
implied that the alternative to $\Lambda$ was not an open universe
but a strongly decelerating, $\Omega_m=1$
universe that disagreed with the supernova data by 0.5 magnitudes,
a level much harder to explain with observational or astrophysical effects.
Finally, the combination of spatial flatness and improving
measurements of the Hubble constant
(e.g., $H_0 = 71 \pm 6 \hubunits$; \citealt{mould00})
provided an entirely independent
argument for an energetically dominant accelerating component:
a matter-dominated universe with $\Omega_{\rm tot} =1$
would have age
$t_0 = (2/3)H_0^{-1} \approx 9.5$ Gyr,
too young to accommodate the 12-14 Gyr ages estimated for globular
clusters (e.g., \citealt{chaboyer98}).

A decade later, the web of observational evidence for cosmic acceleration
is intricate and robust.  A wide range of observations --- including
larger and better calibrated supernova samples over a broader redshift
span, high-precision CMB data down to small angular scales,
the baryon acoustic scale in galaxy clustering, weak lensing measurements
of dark matter clustering, the abundance of massive clusters in
X-ray and optical surveys, the level of structure in the \lya\ forest,
and precise measurements of $H_0$ --- are all consistent with an inflationary
cold dark matter model with a cosmological constant, commonly abbreviated
as $\Lambda$CDM.\footnote{Many of
the relevant observational
references will appear in subsequent sections on specific
topics.}  Explaining all of these data simultaneously {\it requires} an
accelerating universe.  Completely eliminating any one class of constraints
(e.g., supernovae, or CMB, or $H_0$) would not change this conclusion,
nor would doubling the estimated systematic errors on all of them.
The question is no longer {\it whether} the universe is accelerating,
but {\it why}.

\subsection{Theories of Cosmic Acceleration}
\label{sec:theory}

A cosmological constant is the mathematically simplest solution to
the cosmic acceleration puzzle.  While Einstein introduced his
cosmological term as a modification to the curvature side of the
field equation, it is now more common to interpret $\Lambda$ as a
new energy component, constant in space and time.  For an ideal
fluid with energy density $u$ and pressure $p$, the effective
gravitational source term in GR is $(u+3p)/c^2$, reducing to the
usual mass density $\rho = u/c^2$ if the fluid is non-relativistic.
For a component whose energy density remains constant as the universe
expands, the first law of thermodynamics implies that when a comoving
volume element in the universe expands by a (physical) amount $dV$,
the corresponding change in energy is related to the pressure via
$-p dV = dU = udV$.  Thus, $p=-u$, making the gravitational
source term $(u+3p)/c^2 = -2u/c^2$.
A form of energy that is constant in space and time
must have a repulsive gravitational effect.

According to quantum field theory, ``empty'' space is filled with
a sea of virtual particles.  It would be reasonable to interpret
the cosmological constant as the gravitational signature of this
quantum vacuum energy, much as the Lamb shift is a signature of
its electromagnetic effects.\footnote{This interpretation of the
cosmological constant in the context of quantum field theory,
originally due to Wolfgang Pauli,
was revived in the late 1960s
by \citet{zeldovich68}.  For further discussion of the history
see \citet{peebles03}.  For a detailed discussion in the context
of contemporary particle theory, see the review by
\cite{martin12}.}
%and \cite{caldwell10}.}
The problem is one of magnitude.
Since virtual particles of any allowable mass can come into existence
for short periods of time, the ``natural'' value for the quantum
vacuum density is one Planck Mass per cubic Planck Length.
This density is about 120 orders of magnitude larger than the
cosmological constant suggested by observations: it would drive accelerated
expansion with a timescale of $t_{\rm Planck} \approx 10^{-43}\sec$
instead of $t_{\rm Hubble} \approx 10^{18}\sec$.
Since the only ``natural'' number close to $10^{-120}$ is zero,
it was generally assumed (prior to 1990) that a correct calculation
of the quantum vacuum energy would eventually show it to be zero, or at least
suppressed by an extremely large exponential factor 
(see review by \citealt{weinberg89}).
But the discovery of cosmic acceleration raises the possibility
that the quantum vacuum really does act as a cosmological constant,
and that its energy scale is $10^{-3}\,$eV rather than $10^{28}\,$eV for
reasons that we do not yet understand.  To date, there are no
compelling theoretical arguments that explain either why the
fundamental quantum vacuum energy might have this magnitude {\it or}
why it might be zero.

The other basic puzzle concerning a cosmological constant is:
Why now?  The ratio of a constant vacuum energy density to the matter
density scales as $a^3(t)$, so it has changed by a factor of
$\sim 10^{27}$ since big bang nucleosynthesis and by a factor
$\sim 10^{42}$ since the electroweak symmetry breaking epoch,
which seems (based on our current understanding of physics) like
the last opportunity for a major rebalancing of matter and energy
components.  It therefore seems remarkably coincidental for the vacuum
energy density and the matter energy density to have the same order
of magnitude today.  In the late 1970s, Robert Dicke used a similar
line of reasoning to argue for a spatially flat universe
(see \citealt{dicke79}),
an argument that provided much of the initial motivation
for inflationary theory \citep{guth80}.  However, while the universe
appears to be impressively close to spatial flatness, the existence
of two energy components with different $a(t)$ scalings means that
Dicke's ``coincidence problem'' is still with us.

One possible solution to the coincidence problem is anthropic: if the
vacuum energy assumes widely different values in different regions of
the universe, then conscious observers will find themselves in
regions of the universe where the vacuum energy is low enough to
allow structure formation \citep{efstathiou95,martel98}.  This type of
explanation finds a natural home in ``multiverse'' models of eternal
inflation, where different histories of spontaneous symmetry
breaking lead to different values of physical constants in each
non-inflating ``bubble'' \citep{linde87}, and it has gained new
prominence in the context of string theory, which predicts a
``landscape'' of vacua arising from different compactifications of
spatial dimensions \citep{susskind03}.  One can attempt to derive an
expectation value of the observed cosmological constant from
such arguments (e.g., \citealt{martel98}), but the results are
sensitive to the choice of parameters that are allowed to vary
\citep{tegmark98} and to the choice of measure on parameter space,
so it is hard to take such ``predictions'' beyond a qualitative level.
A variant on these ideas is that the effective value (and perhaps
even the sign) of the cosmological constant varies in time, and that
structure will form and observers arise during periods when its
magnitude is anomalously low compared to its natural
(presumably Planck-level) energy scale \citep{brandenberger02,griest02}.

A straightforward alternative to a cosmological constant is a field
with negative pressure (and thus repulsive gravitational effect) whose
energy density changes with time \citep{ratra88,frieman95,ferreira97}.  
In particular,
a canonical scalar field $\phi$ with potential $V(\phi)$ has
energy density and pressure
\begin{eqnarray}
\label{eqn:phi}
u_\phi &= {1 \over 2}{1 \over \hbar c^3} \phidot^2 + V(\phi),  \cr
p_\phi  &= {1 \over 2}{1\over \hbar c^3}\phidot^2 - V(\phi),
\end{eqnarray}
so if the kinetic term is subdominant, then $p_\phi\approx -u_\phi$.
A slowly rolling scalar field of this sort is analogous to the
inflaton field hypothesized to drive inflation, but at an energy
scale many, many orders of magnitude lower.  In general, a
scalar field has an equation-of-state parameter
\begin{equation}
\label{eqn:wdef}
w \equiv {p \over u}
\end{equation}
that is greater than $-1$ and varies in time, while a true cosmological
constant has $w=-1$ at all times.
Some forms of $V(\phi)$ allow attractor or ``tracker'' solutions
in which the late-time evolution of $\phi$ is insensitive to
the initial conditions \citep{ratra88,steinhardt99}, and a 
subset of these 
allow $u_\phi$ to track the matter energy density at early times,
ameliorating the coincidence problem \citep{skordis02}.
Some choices give a nearly constant $w$ that is different from $-1$,
while others have $w\approx -1$ as an asymptotic state at either early
or late times, referred to respectively as ``thawing'' or ``freezing''
solutions \citep{caldwell05}.

Scalar field models in which the energy density is dominated by $V(\phi)$
are popularly known as ``quintessence'' \citep{zlatev99}.
A number of variations have been proposed in which the energy density
of the field is dominated by kinetic, spin, or oscillatory
degrees of freedom (e.g., \citealt{armendariz01,boyle02}).  
Other models introduce
non-canonical kinetic terms or couple the field to dark matter.
Models differ in the evolution of
$u_\phi(a)$ and $w(a)$, and some have other distinctive features
such as large scale energy density fluctuations that can
affect CMB anisotropies.
Of course, none of these models addresses the original
``cosmological constant problem'' of why
the true vacuum energy is unobservably small. 

The alternative to introducing a new energy component is to modify
General Relativity itself on cosmological scales, for example by
replacing the Ricci scalar $R$ in the gravitational action with
some higher order function $f(R)$ 
(e.g., \citealt{capozziello02,carroll04}),
or by allowing gravity to
``leak'' into an extra dimension in a way that reduces its
attractive effect at large scales \citep{dvali00}.
GR modifications can alter the relation between the expansion
history and the growth of matter clustering, and, as discussed
in subsequent sections, searching for mismatches between observational
probes of expansion and observational probes of structure growth
is one generic approach to seeking signatures of modified gravity.
To be consistent with tight constraints from solar system tests,
modifications of gravity must generally be ``shielded'' on small scales,
by mechanisms such as the ``chameleon'' effect, the ``symmetron''
mechanism, or ``Vainshtein screening'' 
(see the review by \citealt{jain10}).
These mechanisms can have
the effect of introducing intermediate scale forces.
%\tbd{Indicate range?}  
GR modifications can also alter the relation
between non-relativistic matter clustering and gravitational
lensing, which in standard GR is controlled by two different
potentials that are equal to each other for fluids without
anisotropic stress.

The distinction between a new energy component and a modification of
gravity may be ambiguous.  The most obvious ambiguous case is the
cosmological constant itself, which can be placed on either the
``curvature'' side or the ``stress-energy'' side of the
Einstein field equation.
More generally, many theories with $f(R)$ modifications of
the gravitational action can be written in a mathematically
equivalent form of GR plus a scalar field with specified
properties \citep{chiba03,kunz07}.
Relative to expectations for a cosmological constant
or a simple scalar field model, models in which dark matter
decays into dark energy can produce a mismatch between
the histories of expansion and structure growth while maintaining GR
(e.g., \citealt{jain08,wei08}).
Thus, even perfect measurements of all relevant
observables may not uniquely locate the explanation of
cosmic acceleration
in the gravitational or stress-energy sector.

There is a minority opinion
(see \citealt{buchert11} for a recent review article)
that the phenomena interpreted as evidence for dark energy
could instead arise from the backreaction of small scale inhomogeneities
on the large scale cosmic expansion.  This line of argument contends
that the expansion rate of a universe with small scale inhomogeneity
can differ significantly from that of a homogeneous universe with the 
same average density.  In our view, the papers on this theme present
an incorrect interpretation of correct underlying equations, and
we do not see these ``inhomogeneous averaging'' effects as a 
viable alternative to dark energy.  \cite{baumann12} present a
detailed counter-argument, treating inhomogeneous matter as a fluid
with an effective viscosity and pressure and demonstrating that
the backreaction on the evolution of background expansion and
large scale perturbations is extremely small.
(See \citealt{green11a} for an alternative form of this
counter-argument and \citealt{peebles10} for a less formal
but still persuasive version.) In a somewhat
related vein, the suggestion that acceleration could arise from
superhorizon fluctuations generated by inflation
\citep{barausse05,kolb05} is ruled out by a complete
perturbation analysis \citep{hirata05}.

While the term ``dark energy'' seems to presuppose a stress-energy
explanation, in practice it has become a generic term for referring to the
cosmic acceleration phenomenon.  In particular, the phrase
``dark energy experiments'' has come to mean observational studies
aimed at measuring acceleration and uncovering its cause, regardless
of whether that cause is a new energy field or a
modification of gravity.  We will generally adopt this common
usage of ``dark energy'' in this review, though where the distinction
matters we will try to use ``cosmic acceleration''
as our generic term.  It is important to keep in mind that we
presently have strong observational evidence for accelerated
cosmic expansion but no compelling evidence that the cause of this
acceleration is really a new energy component.

The magnitude and coincidence problems are challenges for any
explanation of cosmic acceleration, whether a cosmological constant,
a scalar field, or a modification of GR.
The coincidence problem seems like an important clue for identifying
a correct solution, and some models at least reduce its severity
by coupling the matter and dark energy densities in some way.
Multiverse models with anthropic selection arguably offer a solution
to the coincidence problem, because if the probability distribution
of vacuum energy densities rises swiftly towards high values,
then structure may generically form at a time when the matter
and vacuum energy density values are similar, in that small
subset of universes where structure forms at all.  But sometimes
a coincidence is just a coincidence.
Essentially all current theories of cosmic acceleration have
one or more adjustable parameters whose value is tuned to give
the observed level of acceleration, and none of them yield this
level as a ``natural'' expectation unless they have built it in
ahead of time.  These theories are designed to explain acceleration
itself rather than emerging from independent theoretical
considerations or experimental constraints.
Conversely, a theory that provided a compelling account of
the observed magnitude of acceleration --- like GR's successful
explanation of the precession of Mercury --- would quickly jump
to the top of the list of cosmic acceleration models.

\subsection{Looking Forward}
\label{sec:forward}

The deep mystery and fundamental implications of cosmic acceleration
have inspired numerous ambitious observational efforts to measure
its history and, it is hoped, reveal its origin.
The report of the Dark Energy Task Force
(DETF; \citealt{albrecht06}) played a critical role in
systematizing the field, by categorizing experimental approaches
and providing a quantitative framework to compare their capabilities.
The DETF categorized then-ongoing experiments as ``Stage II''
(following the ``Stage I'' discovery experiments) and the
next generation as ``Stage III.''  It looked forward to
a generation of more capable (and more expensive) ``Stage IV''
efforts that might begin observations around the second half
of the coming decade.  The DETF focused on the same four methods
that will be the primary focus of this review: Type Ia supernovae,
baryon acoustic oscillations (BAO), weak gravitational lensing,
and clusters of galaxies.

Six years on, the main ``Stage II'' experiments have completed their
observations though not necessarily their final analyses.
Prominent examples include the supernova and weak lensing
programs of the CFHT Legacy Survey (CFHTLS; 
\citealt{Conley11,semboloni06,heymans_cfht}),
the ESSENCE supernova survey \citep{WoodVasey07},
BAO measurements from the Sloan Digital Sky Survey
(SDSS; \citealt{eisenstein05,percival10,padmanabhan12}), and the
SDSS-II supernova survey \citep{frieman08a}.
These have been complemented by extensive multi-wavelength
studies of local and high-redshift supernovae such as the
Carnegie Supernova Project \citep{hamuy06,freedman09}, by systematic searches
for $z>1$ supernovae with {\it Hubble Space Telescope} 
\citep{riess07,suzuki12},
by dark energy constraints from the evolution of X-ray or
optically selected clusters \citep{henry09,mantz09,vikhlinin09,rozo10},
by improved measurements of the Hubble constant 
\citep{riess09,riess11,freedman12},
and by CMB data from the \wmap\ satellite \citep{bennett03,larson11} and
from ground-based experiments that probe smaller angular 
scales.\footnote{We follow the convention in the astronomical literature
of italicizing the names and acronyms of space missions but not
of ground-based facilities.  For reference, note that the many
acronyms that appear in the article are all defined in Appendix A,
the glossary of acronyms and facilities.}
Most data remain consistent with a spatially flat universe and a
cosmological constant with $\Omega_\Lambda = 1-\Omega_m \approx 0.75$,
with an uncertainty in the equation-of-state parameter $w$ that
is roughly $\pm 0.1$ at the $1-2\sigma$ level.  Substantial
further improvement will in many cases require reduction in systematic
errors as well as increased statistical power from larger data sets.

The clearest examples of ``Stage III'' experiments, now in the
late construction or early operations phase, are the Dark Energy Survey
(DES), Pan-STARRS\footnote{Pan-STARRS, the Panoramic Survey Telescope
and Rapid Response System, is the acronym of the facility rather than
the project, but cosmological surveys are among its major goals.
Pan-STARRS eventually hopes to use four coordinated telescopes,
but the surveys currently underway (and now nearing completion)
use the first of these telescopes,
referred to as PS1.}, the
Baryon Oscillation Spectroscopic Survey (BOSS) of SDSS-III,
and the Hobby-Eberly Telescope Dark Energy Experiment 
(HETDEX).\footnote{The acronym and facilities glossary gives references
to web sites and/or publications that describe these and other
experiments.}
All four projects are being carried out by international, multi-institutional
collaborations.  Pan-STARRS and DES will both carry out large
area, multi-band imaging surveys that go a factor of ten or more
deeper (in flux) than the SDSS imaging survey \citep{abazajian09},
using, respectively, a 1.4-Gigapixel camera on the 1.8-m PS1 telescope
on Haleakala in Hawaii and a 0.5-Gigapixel camera on the 4-m Blanco
telescope on Cerro Tololo in Chile.  These imaging surveys will
be used to measure structure growth via weak lensing, to identify
galaxy clusters and calibrate their masses via weak lensing,
and to measure BAO in galaxy angular clustering using photometric
redshifts.  Each project also plans to carry out monitoring surveys
over smaller areas to discover and measure thousands of Type Ia
supernovae.  Fully exploiting BAO
requires spectroscopic redshifts, and BOSS will carry out a
nearly cosmic-variance limited survey (over $10^4$ deg$^2$) out
to $z\approx 0.7$ using a 1000-fiber spectrograph to measure
redshifts of 1.5 million luminous galaxies, and a pioneering quasar
survey that will measure BAO at $z\approx 2.5$ by using the \lya\
forest along 150,000 quasar sightlines to trace the underlying
matter distribution.
HETDEX plans a BAO survey of $10^6$ \lya-emitting galaxies at $z\approx 3$.

There are many other ambitious observational efforts that do not
fit so neatly into the definition of a ``Stage III dark energy
experiment'' but will nonetheless play an important role in
``Stage III'' constraints.
A predecessor to BOSS, the WiggleZ project
on the Anglo-Australian 3.9-m telescope, recently completed a
spectroscopic survey of 240,000 emission line galaxies out to
$z = 1.0$ \citep{blake11a}.  
The Hyper Suprime-Cam (HSC) facility on the Subaru telescope will
have wide-area imaging capabilities comparable to DES and Pan-STARRS,
and it is likely to devote substantial fractions of its time
to weak lensing surveys.
Other examples include intensive spectroscopic and photometric
monitoring of supernova samples aimed at calibration and understanding
of systematics, new {\it HST} searches for $z>1$ supernovae,
further improvements in $H_0$ determination,
deeper X-ray and weak lensing studies of samples of tens or hundreds
of galaxy clusters, and new cluster searches via the
Sunyaev-Zel'dovich (\citeyear{sunyaev70}) effect using the
South Pole Telescope (SPT), the Atacama Cosmology Telescope (ACT),
or the \planck\ satellite.
In addition, Stage III analyses will draw on primary CMB constraints
from \planck.

The Astro2010 report identifies cosmic acceleration
as one of the most pressing questions in contemporary astrophysics,
and its highest priority recommendations for new ground-based and
space-based facilities both have cosmic acceleration as a primary science
theme.\footnote{We will use the term ``Astro2010 report'' to refer 
collectively to {\it New Worlds, New Horizons} and to the
panel reports that supported it.  In particular, detailed 
discussion of these science themes and related facilities can 
be found in the individual reports of the
Cosmology and Fundamental Physics (CFP) Science Frontiers Panel
and the Electromagnetic Observations from Space
(EOS), Optical and Infrared Astronomy from the Ground 
(OIR), and Radio, Millimeter, 
and Sub-Millimeter Astronomy from the Ground
(RMS) Program Prioritization Panels.  Information on all
of these reports can be found at 
{\tt http://sites.nationalacademies.org/bpa/BPA\_049810}.}
On the ground, the Large Synoptic Survey Telescope (LSST),
a wide-field 8.4-m optical telescope equipped with a 3.2-Gigapixel camera,
would enable deep
weak lensing and optical cluster surveys over much of the sky, synoptic
surveys that would detect and measure tens of thousands of supernovae,
and photometric-redshift BAO surveys extending to $z\approx 3.5$.
BigBOSS, highlighted as an initiative that could be
supported by the proposed ``mid-scale innovation program,'' would
use a highly multiplexed fiber spectrograph on the
NOAO 4-m telescopes to carry out spectroscopic surveys of $\sim 10^7$
galaxies to $z\approx 1.6$ and \lya\ forest BAO measurements at $z>2.2$.
Another potential ground-based method for large volume BAO surveys
is radio ``intensity mapping,'' which seeks to trace the large scale
distribution of neutral hydrogen without resolving the scale of
individual galaxies.  In the longer run, the Square Kilometer Array
(SKA) could enable a
BAO survey of $\sim 10^9$ HI-selected galaxies and weak lensing
measurements of $\sim 10^{10}$ star-forming galaxies
using radio continuum shapes.

Space observations offer two critical advantages for cosmic acceleration
studies: stable high resolution imaging over large areas, and vastly
higher sensitivity at near-IR wavelengths.
(For cluster studies, space observations are also the only
route to X-ray measurements.)
These advantages inspired the Supernova Acceleration Probe
({\it SNAP}), initially designed with a concentration on
supernova measurements at $0.1 < z < 1.7$, and later
expanded to include a wide area weak lensing survey as a major
component.  Following the National Research Council's Quarks to Cosmos
report \citep{qtoc03}, NASA and the U.S. Department of Energy embarked
on plans for a Joint Dark Energy Mission ({\it JDEM}), which has
considered a variety of mission architectures for space-based supernova,
weak lensing, and BAO surveys.  The Astro2010 report endorsed as its
highest priority space mission a Wide Field Infrared Survey Telescope
({\it WFIRST}),
which would carry out imaging and dispersive prism spectroscopy
in the near-IR to support all of these methods, and, in addition,
a planetary microlensing program, a Galactic plane survey,
and a guest observer program.
The recently completed report of the \wfirst\ Science Definition Team
\citep{green12} presents detailed designs and operational concepts,
with a primary design reference mission that includes three years of
dark energy programs (out of a five year mission) on an unobstructed
1.3-meter telescope with a $0.375\mdeg^2$ near-IR focal plane
(150 million $0.18^{\prime\prime}$ pixels).
The recent transfer of two 2.4-meter diameter telescopes from the U.S.
National Reconnaissance Office (NRO) to NASA opens the door for a 
potential implementation of \wfirst\ on a larger platform; this possibility
is now a subject of active, detailed study
(see \citealt{dressler12} for an initial assessment).
\wfirst\ faces significant funding hurdles, 
despite its top billing in Astro2010,
but a launch in the early 2020s still appears likely.
%An updated technical and scientific plan for \wfirst\ appears in
%the preliminary report from the \wfirst\ Science Definition Team
%\citep{green11}.  The mission still faces substantial funding
%hurdles, so despite its top billing in Astro2010, its future
%is uncertain.
%SNAP in turn inspired competing proposals such as
%DESTINY and JEDI\footnote{See the Acronym Glossary appendix for
%acronyms not spelled out in the main text.} using alternative
%technical approaches, DUNE, focused on weak lensing rather than
%supernovae, and ADEPT, a space-based BAO experiment using
%near-IR slitless spectroscopy for a $\sim 10^8$-galaxy redshift survey.
On the European side, ESA recently
selected the \euclid\footnote{Not an acronym.} satellite
as a medium-class mission for its Cosmic Vision 2015-2025 program, with 
launch planned for 2020.
\euclid\ plans to carry out optical and near-IR imaging and
near-IR slitless spectroscopy over roughly 14,000$\mdeg^2$, for weak
lensing and BAO measurements.  
In its current design \citep{laureijs11}, \euclid\ utilizes a
1.2-meter telescope, a $0.56\mdeg^2$ optical focal
plane (604 million $0.10^{\prime\prime}$ pixels), and a near-IR
focal plane with similar area but larger pixels
(67 million $0.30^{\prime\prime}$ pixels).
Well ahead of either \euclid\ or \wfirst,
the European X-ray telescope \erosita\ (on the Russian
{\it Spectrum Roentgen Gamma} satellite) is expected
to produce an all-sky catalog of $\sim 10^5$ X-ray selected clusters,
with X-ray temperature measurements and resolved profiles for the
brighter clusters \citep{merloni12}.\footnote{More detailed description 
of \euclid\ and \wfirst\ can be found in \S\ref{sec:wl_prospects},
and of \erosita\ in \S\ref{sec:cl_space}.}

The completion of the Astro2010 Decadal Survey and the 
\euclid\ selection by ESA make this an opportune time to review the
techniques and prospects for probing cosmic acceleration with
ambitious observational programs.  Our goal is, in some sense, an
update of the DETF report \citep{albrecht06}, incorporating the many
developments in the field over the last few years and (the difference
between a report and a review) emphasizing explanation rather than
recommendation.  We aim to complement other reviews of the field
that differ in focus or in level of detail.  To mention just a
selection of these, we note that
\cite{frieman08b} and \cite{blanchard10} provide excellent overviews
of the field, covering theory, current observations, and future experiments,
while \cite{astier12} cover the observational approaches concisely;
\cite{peebles03} and \cite{copeland06}
are especially good on history of the subject
and on theoretical aspects of scalar field models;
\cite{jain10} review the observational and (especially) 
theoretical aspects of modified gravity models in much greater
depth than we cover here;
\cite{carroll04a} and \cite{linder03a,linder07} provide
accessible and informative introductions at the less forbidding
length of conference proceedings; \cite{linder10} provides a
review aimed at a general scientific audience; and the
conference proceedings by \cite{peebles10} nicely situates
the cosmic acceleration problem in the broader context of
contemporary cosmology.
The distinctive features of the present review are our in-depth discussion
of individual observational methods and our new quantitative forecasts
for how combinations of these methods can constrain parameters
of cosmic acceleration theories.

To the extent that we have a consistent underlying theme, it is the
importance of pursuing a balanced observational program.
We do not believe that all methods or all implementations of
methods are equal; some approaches have evident systematic
limitations that will prevent them reaching the sub-percent accuracy
level that is needed to make major contributions to the field over the
next decade, while others would require prohibitively expensive
investments to achieve the needed statistical precision.
However, for a given level of community investment, we think
there is more to be gained by doing a good job on the three or
four most promising methods than by doing a perfect job on one
at the expense of the others.  A balanced approach offers crucial
cross-checks against systematic errors, takes advantage of
complementary information contained in different observables
or complementary strengths in different redshift ranges, and
holds the best chance of uncovering ``surprises'' that do not
fit into the conventional categories of theoretical models.
This philosophy will emerge most clearly in \S\ref{sec:forecast},
where we present our quantitative forecasts.
For understandable reasons, most articles and proposals
(including some we have written ourselves) start from current
knowledge and show the impact of adding a particular new
experiment.  We will instead start from a ``fiducial program''
that assumes ambitious but achievable advances in several different
methods at once, then consider the impact of strengthening,
weakening, or omitting its individual elements.

We expect that different readers will want to approach this
lengthy article in different ways.  For a reader who is new
to the field and wants to learn it well, it makes sense to
start at the beginning and read to the end.
A reader interested in a specific method can skim 
\S\ref{sec:observables} 
to get a sense of our notation, then jump to the section that
describes that method (Type Ia supernovae in \S\ref{sec:sn},
BAO in \S\ref{sec:bao}, weak lensing in \S\ref{sec:wl}, and
clusters in \S\ref{sec:cl}).  We think that these sections will
provide useful insights even to experts in the field.
Section \ref{sec:alternatives} provides a brief overview of
emerging methods that could play an important role in future studies.
Readers interested mainly in the ways that different methods
contribute to constraints on cosmic acceleration models and
the quantitative forecasts for Stage III and Stage IV 
programs can jump directly to \S\ref{sec:forecast}.
Finally, \S\ref{sec:conclusions} provides a summary of our
findings and their implications for experimental programs,
and some readers may choose to start from the end
(we recommend including \S\S\ref{sec:forecast_aggregate}
and \ref{sec:forecast_multiprobe} as well as \S\ref{sec:conclusions}), then
work backwards to the supporting details.

\vfill\eject
\section{Observables, Parameterizations, and Methods}
\label{sec:observables}

The two top-level questions about cosmic acceleration are:
\begin{enumerate}
\item Does acceleration arise from a breakdown of GR on cosmological
scales or from a new energy component that exerts repulsive gravity
within GR?
\item If acceleration is caused by a new energy component,
is its energy density constant in space and time?
\end{enumerate}
As already discussed in \S\ref{sec:theory}, the distinction
between ``modified gravity'' and ``new energy component''
solutions may not be unambiguous.  However, the cosmological constant
hypothesis makes specific, testable predictions, and the combination
of GR with relatively simple scalar field models predicts
testable consistency relations between expansion and
structure growth.

The answer to these questions, or a major step towards an answer,
could come from a surprising direction: a theoretical breakthrough,
a revealing discovery in accelerator experiments,
a time-variation of a fundamental ``constant,'' or an experimental
failure of GR on terrestrial or solar system scales
(see \S\ref{sec:gravity} for brief discussion).
However, ``wait for a breakthrough'' is an unsatisfying recipe
for scientific progress, and there is one clear path forward:
measure the history of expansion and the growth of structure with
increasing precision over an increasing range of redshift and
lengthscale.

\subsection{Basic Equations}
\label{sec:basics}

In GR, the expansion of a homogeneous and isotropic universe is
governed by the Friedmann equation, which can be written in the form
\begin{equation}
\label{eqn:friedmann}
{H^2(z) \over H_0^2}= \omo(1+z)^3 + \orado(1+z)^4 + \oko(1+z)^2 +
                    \opo{\uphi(z) \over \uphi(z=0)} ~,
\end{equation}
where $(1+z)\equiv a^{-1}$ is the cosmological redshift
and $a(t)$ is the expansion factor relating physical separations
to comoving separations.
The Hubble parameter is $H(z) \equiv \dot{a}/a$, and 
$\omo$, $\orado$, and $\opo$ are the {\it present day}
energy densities of matter, radiation, and a generic form of
dark energy $\phi$.\footnote{We will refer to values of these parameters
at $z \neq 0$ as $\Omega_m(z)$, $\Omega_\phi(z)$, etc.  
For other quantities (e.g., $H_0$), we use
subscripts 0 to denote values at $z=0$.  When we 
{\it assume} a cosmological constant, we will replace $\Omega_\phi$
by $\Omega_\Lambda$.} 
These are expressed as ratios to the critical energy
density required for flat space geometry
\begin{equation}
\label{eqn:rhocrit}
\Omega_{x} = {u_{x} \over \rho_{{\rm crit}} c^2}~, \qquad
  \rho_{{\rm crit}} = {3H_0^2 \over 8\pi G}~.
\end{equation}
At higher redshifts,
\begin{equation}
\label{eqn:omegaz}
\Omega_m(z) \equiv {\rho_m(z) \over \rho_{\rm crit}(z)} =
  \omo (1+z)^3 {H_0^2 \over H^2(z)},
\end{equation}
where the second equality follows from the scaling
$\rho_m(z) = \rho_{m,0} \times (1+z)^3$ and from the
definition of $\rho_{\rm crit}(z)$.  In the
formulation~(\ref{eqn:friedmann}),
the impact of curvature on expansion is expressed like that of a
``dynamical''
component with scaled energy density
\begin{equation}
\label{eqn:omegak}
\Omega_k \equiv 1 - \Omega_m - \Omega_r - \Omega_\phi,
\end{equation}
with $\Omega_k=0$ for a spatially flat universe.
In a standard cold dark matter scenario, the matter
density is the sum of the densities of CDM, baryons,
and non-relativistic neutrinos, 
$\Omega_m = \Omega_c+\Omega_b+\Omega_\nu$.
In detail, one must beware that the neutrino energy density
does not scale as $(1+z)^3$ at higher redshifts, when they
are mildly relativistic, and that the clustering of neutrinos
on small scales is suppressed by their residual thermal velocities.

There are some routes to direct measurement of $H(z)$, most
notably via BAO (see \S\ref{sec:bao}).
For the most part, however, observations constrain $H(z)$ indirectly by
measuring the distance-redshift relation or the history of structure growth.

%{\bf Changed next paragraph to Chris's suggested notation; Chris
%please check.}

\cite{hogg99} provides a compact and pedagogical summary of cosmological
distance measures.  The comoving line-of-sight distance to an object at
redshift $z$ is
\begin{equation}
\label{eqn:dcomove}
D_C(z) = {c \over H_0} \int_0^z dz' {H_0 \over H(z')}~.
\end{equation}
Defining a dimensional (length$^{-2}$) curvature parameter
\begin{equation}
\label{eqn:Kdef}
K = - \Omega_k (c/H_0)^{-2} 
\end{equation}
allows us to write 
the comoving angular diameter distance,\footnote{Note that \cite{hogg99} 
refers to this quantity as the comoving transverse distance and uses $D_A$
to denote the quantity relating {\it physical} size to angular
size.} 
relating an object's comoving size $l$ to its angular size 
$\theta = l/D_A$, as
\begin{equation}
\label{eqn:adist}
D_A(z) = K^{-1/2} \sin\left(K^{1/2} D_C\right) ~,
\end{equation}
which applies for either sign of $\Omega_k$.\footnote{Recall
that $\sin(ix) = i\sinh(x)$.}
Noting that observations imply $|\oko| \ll 1$, we can Taylor expand
equation~(\ref{eqn:adist}) to write
\begin{equation}
\label{eqn:adistapprox}
D_A(z) \approx D_C \left[1 + {1\over 6} \oko \left({D_C \over c/H_0}\right)^2
                   \right]~,
\end{equation}
which also yields the correct result $D_A = D_C$ for $\ok=0$.
Note that positive space curvature ($\Omega_{\rm tot} > 1$, $K>0$)
corresponds to negative $\oko$,
hence a smaller $D_A$ and larger angular size than in a flat universe.
If $\uphi(z) > u_{\phi,0}$ then the Hubble parameter at $z>0$ is
higher compared to a cosmological constant model with the
same matter density and curvature
(eq.~\ref{eqn:friedmann}), and distances to redshifts $z>0$
are lower (eq.~\ref{eqn:adist}).

The luminosity distance relating an object's bolometric flux $f_{\rm bol}$
to its bolometric luminosity $L_{\rm bol}$ is
\begin{equation}
\label{eqn:ldist}
D_L = \sqrt{L_{\rm bol}/4\pi f_{\rm bol}} = D_A \times (1+z)~.
\end{equation}
The relation between luminosity and angular diameter distance is
independent of cosmology, so the two measures contain the same
information about
$H(z)$ and $\oko$.  For this reason, we will sometimes use $D(z)$
to stand in generically for either of these transverse distance measures.
Some methods (e.g., counts of galaxy clusters) effectively probe
the comoving volume element that relates solid angle and redshift
intervals to comoving volume $V_C$.  We will denote this quantity
\begin{equation}
\label{eqn:dvc}
dV_C(z) \equiv c H^{-1}(z) D_A^2(z) d\Omega\,dz.
\end{equation}

On large scales, the gravitational evolution of fluctuations
in pressureless
dark matter follows linear perturbation theory, according to which
\begin{equation}
\label{eqn:deltaxt}
\delta(\vx,t) ~\equiv~ {\rho_m(\vx,t)-\bar{\rho}_m(t) \over \bar{\rho}_m(t)}
              ~=~ \delta(\vx,t_i) \times {G(t) \over G(t_i)} ~,
\end{equation}
where $t_i$ is an arbitrarily chosen initial time,
the linear growth function $G(t)$ obeys the differential equation
\begin{equation}
\label{eqn:lingrowth}
\ddot \Ggr + 2H(z) \dot \Ggr - {3\over 2}\omo H_0^2 (1+z)^3 \Ggr =0 ~,
\end{equation}
and the GR subscript denotes the fact that this equation applies
in standard GR.\footnote{This equation applies on scales much
smaller than the horizon.  On scales close to the horizon one
must pay careful attention to gauge definitions.  \cite{yoo09b}
and \cite{yoo09c}
provide a unified and comprehensive discussion of the multiple
GR effects that influence observable large scale structure
on scales approaching the horizon.}
The solution to this equation can only be written in integral form
for specific forms of $H(z)$, and thus for specific dark energy
models specifying $\uphi(z)$.  However, to a very good approximation
the logarithmic growth rate of linear perturbations in GR is
\begin{equation}
\label{eqn:dlng.dlna}
f_{\rm GR}(z) \equiv {d\ln \Ggr \over d\ln a} \approx \left[\om(z)\right]^\gamma
 ~,
\end{equation}
where $\gamma \approx 0.55-0.6$ depends only weakly on cosmological
parameters \citep{peebles80,lightman90}.
Integrating this equation yields
\begin{equation}
\label{eqn:lin_approx}
{\Ggr(z) \over \Ggr(z=0)} \approx \exp
                \left[-\int_0^z {dz' \over 1+z'} [\om(z')]^{\gamma}\right] ~,
\end{equation}
where $\om(z)$ is given by equation~(\ref{eqn:omegaz}).
\cite{linder05} shows that equation~(\ref{eqn:lin_approx}) is accurate
to better than 0.5\% for a wide variety of dark energy models if one adopts
\begin{equation}
\label{eqn:gamma}
\gamma = 0.55 + 0.05[1+w(z=1)]
\end{equation}
(see also \citealt{wang98,weinberg05,amendola05}).
While the full solution of equation~(\ref{eqn:lingrowth}) should
be used for high accuracy calculations, equation~(\ref{eqn:lin_approx})
is useful for intuition and for approximate calculations.
Note in particular that if $\uphi(z) > \uphio$ then, relative to
a cosmological constant model, $\om(z) \propto H^{-2}(z)$ is lower
(eq.~\ref{eqn:omegaz}), so $\Ggr(z)/\Ggr(z=0)$ is higher --- i.e., there has
been {\it less} growth of structure between redshift $z$ and
the present day because matter has been a smaller fraction of
the total density over that time.
It is often useful to refer the growth factor not to its $z=0$ value
but to the value at some high redshift when, in typical models,
dark energy is dynamically negligible and $\om(z) \approx 1$.
We will frequently use $z=9$ as a reference epoch, in which case
equation~(\ref{eqn:lin_approx}) becomes
\begin{equation}
\label{eqn:lin_approx9}
{\Ggr(z) \over \Ggr(z=9)} \approx \exp
                \left[\int_z^9 {dz' \over 1+z'} [\om(z')]^{\gamma}\right] ~.
\end{equation}
In the limit $\om(z)\rightarrow 1$, $\Ggr(z) \propto (1+z)^{-1}$, i.e.,
the amplitude of linear fluctuations is proportional to $a(t)$.

\subsection{Model Parameterizations}
\label{sec:parameterizations}

The properties of dark energy influence the observables ---
$H(z)$, $D(z)$, and $G(z)$ --- through the history of
$\uphi(z)/\uphio$ in the Friedmann equation~(\ref{eqn:friedmann}).
This history is usually framed in terms of the value and evolution
of the equation-of-state parameter $w(z) = \pphi(z)/\uphi(z)$.
Provided that the field $\phi$ is not transferring energy
directly to or from other components (e.g., by decaying into
dark matter), applying the first law of thermodynamics $dU=-p\,dV$ to a 
comoving volume implies
\begin{eqnarray}\label{eqn:dlogu}
&&d(\uphi a^3) = -\pphi d(a^3) \\
\Longrightarrow~ &&a^3 d\uphi + 3\uphi a^2 da = -3w(z) \uphi a^2 da \\
\Longrightarrow~ &&d\ln\uphi = -3[1+w(z)] d\ln a = 3[1+w(z)] d\ln (1+z)~,
\end{eqnarray}
where the last equality uses the definition $a=(1+z)^{-1}$.
Integrating both sides implies
\begin{equation}
\label{eqn:uphi}
{\uphi(z) \over \uphi(z=0)} = \exp \left[ 3\int_0^z [1+w(z')]
  {dz' \over 1+z'} \right].
\end{equation}
For a constant $w$ independent of $z$ we find
\begin{equation}
\label{eqn:uphi_vs_w}
{\uphi(z) \over \uphi(z=0)} = (1+z)^{3(1+w)},
\end{equation}
which yields the familiar results $u \propto (1+z)^3$ for
pressureless matter and $u \propto (1+z)^4$ for radiation ($w=+\frac{1}{3}$),
and which shows once again that a cosmological constant
$\uphi(z)={\rm constant}$ corresponds to $w=-1$.

The first obvious way to parameterize $w(z)$ is with a Taylor expansion
$w(z) = w_0 + w'z + ...$, but this expansion becomes ill-behaved at high $z$.
A more useful two-parameter model
\citep{chevallier01,linder03b} is
\begin{equation}
\label{eqn:w0wa}
w(a) = w_0 + w_a (1-a),
\end{equation}
in which the value of $w$ evolves linearly with scale factor
from $w_0+w_a$ at small $a$
(high $z$) to $w_0$ at $z=0$.  Observations usually provide the best
constraint on $w$ at some intermediate redshift, not at $z=0$, so
statistical errors on $w_0$ and $w_a$ are highly correlated.  This problem
can be circumvented by recasting equation~(\ref{eqn:w0wa}) into the
equivalent form
\begin{equation}
\label{eqn:wpwa}
w(a) = w_p + w_a (a_p-a)
\end{equation}
and choosing the ``pivot'' expansion factor $a_p$ so that the observational
errors on $w_p$ and $w_a$ are uncorrelated (or at least weakly so).
The value of the pivot redshift
depends on what data sets are being considered,
but in practice it is usually close to 
$z_p \equiv a_p^{-1} -1 \approx 0.4-0.5$ (see Table~\ref{tbl:forecasts1}).
The best-fit $w_p$ is, approximately, the parameter
of the constant-$w$ model that would best reproduce the data.
A cosmological constant would be statistically ruled out either if $w_p$
were inconsistent with $-1$ or if $w_a$ were inconsistent with zero.
In practice, error bars on $w_a$ are generally much larger than error
bars on $w_p$, by a factor of $5-10$.  
More generally, it is much more difficult to detect time dependence
of $w$ than to show $w \neq -1$, typically requiring
sub-percent measurements of observables even if $w$ changes by
order unity in an interval $\Delta z < 1$ at low redshift \citep{kujat02}.
The DETF proposed a figure of merit (FoM) for dark energy programs
based on the expected error ellipse in the $w_0-w_a$ plane
(similar to the approach described by
\citeauthor{huterer01} [\citeyear{huterer01}]).
We will frequently refer to this DETF figure of merit, adopting
the definition
\begin{equation}
\label{eqn:fom}
{\rm FoM} = \left[\sigma(w_p)\sigma(w_a)\right]^{-1} ~,
\end{equation}
and we will refer to dark energy models defined by 
equations~(\ref{eqn:w0wa}) or~(\ref{eqn:wpwa}) as 
``$w_0-w_a$ models.''

An alternative parameterization
approach is to approximate $w(z)$ as a stepwise-constant
function defined by its values in a number of discrete bins, perhaps
with priors or constraints on the allowed values
(e.g., $-1 \leq w(z) \leq 1$).  For a given set of observations, this
function can then be decomposed into orthogonal
principal components (PCs), with
the first PC being the one that is best constrained by the data,
the second PC the next best constrained, and so forth
\citep{huterer03}.  Variants
of this approach have been widely adopted in recent investigations
(e.g., \citealt{albrecht07,sarkar08c,mortonson09a}), 
including the report of the \jdem\ Figure-of-Merit
Science Working Group \citep{albrecht09}.  The PCA approach has the
advantage of allowing quite general $w(z)$ histories to be represented,
though in practice only a few PCs can be constrained well; 
\cite{linder05b} and
\cite{deputter08} have argued that the $w_0-w_a$ parameterization has
equal power for practical purposes.  We will use both characterizations
for our forecasts in \S\ref{sec:forecast}.
For scalar field models, one can attempt to reconstruct the
potential $V(\phi)$ instead of $w(z)$ 
\citep{starobinsky98,huterer99,nakamura99}, an
approach that we discuss briefly at the end of \S\ref{sec:results_wzbin}.
\cite{gott11} emphasize that slowly rolling scalar field models
generically predict $1+w \approx (1+w_0)H_0^2/H^2(z)$ for
$|1+w| \ll 1$, making the space of $w(z)$ models, to leading order,
one-dimensional, rather than the two-dimensional
parameterization of $w_0-w_a$.  As a complement to parameterized
models, one can attempt to construct non-parametric ``null tests''
for a cosmological constant or scalar field models
\citep{sahni08}.

If $w \neq -1$, then the dark energy density should display spatial
inhomogeneities, but for simple scalar field models these 
inhomogeneities are strongly suppressed on scales below the horizon.
More complicated models that have a sound speed
($c_s^2 = \delta p/\delta\rho$) much smaller than $c$ allow
fluctuations to grow on sub-horizon scales
(e.g., \citealt{hu98,erickson02,weller03,dedeo03,bean04}).
\cite{deputter10} provide a clear discussion of the background physics
and observable consequences of dark energy inhomogeneities.
In general these inhomogeneities are very difficult to detect, because their
growth is significant only when $w$ is far from $-1$ and
$c_s \ll c$, and because the fluctuations in dark energy
density are much smaller than those in dark matter.
We will mostly ignore dark energy inhomogeneities in this
article, though we return to the subject briefly in \S\ref{sec:isw}.

Our equations so far have assumed that GR is correct.  The alternative
to dark energy is to modify GR in a way that produces accelerated
expansion.  One of the best-studied examples is DGP gravity
\citep{dvali00}, which posits a five-dimensional gravitational field
equation that leads to a Friedmann equation
\begin{equation}
\label{eqn:dgp}
H^2(z) = {8\pi G \over 3} \rho(z) \pm {c H \over r_c}
\end{equation}
for a spatially flat,
homogeneous universe confined to a $(3+1)$-dimensional brane.
Above the ``crossover scale'' $r_c$, which relates the five-dimensional and
four-dimensional gravitational constants, the gravitational force law
scales as $r^{-3}$ instead of the usual $r^{-2}$.
Choosing the positive sign for the second term in equation~(\ref{eqn:dgp})
and setting $r_c \sim c/H_0$ leads to an initially
decelerating universe that transitions to accelerating, and ultimately
exponential, expansion.
Other modifications to the gravitational action that replace the
curvature scalar $R$ by some function $f(R)$ will modify the
Friedmann equation in different ways, some of which can produce
late-time acceleration 
(e.g., \citealt{capozziello02,carroll04}).
Alternatively, one can simply
postulate a modified Friedmann equation without specifying a complete
gravitational theory, e.g., by replacing $\rho$ on the right hand side of
$H^2 \propto \rho$ with a parameterized function
$H^2 \propto g(\rho)$ \citep{freese02,freese05}.
Of course, there is no guarantee that such a function can in
fact be derived from a self-consistent gravitational theory.

Using equations~(\ref{eqn:friedmann}) and~(\ref{eqn:uphi}), 
one can express a modified
Friedmann equation in terms of an effective time-dependent
dark energy equation of state. In this review, we will use $w(z)$
to parameterize the expansion histories of both dark energy and modified
gravity theories. Given $w(z)$, $H(z)$ and $D(z)$ generally follow from the
same set of equations for both types of theories,
so observations that only probe the geometry of the universe are
incapable of distinguishing between the two possible explanations of
cosmic acceleration.
In addition to changing the Friedmann equation, however, a modified gravity
model may alter the equation~(\ref{eqn:lingrowth}) that relates the
growth of structure to the expansion history $H(z)$.
Therefore, one general approach to testing modified gravity
explanations is to search for inconsistency between observables that
probe $H(z)$ or $D(z)$ and observables that also probe the growth
function $G(z)$.  Some methods effectively measure $G(z)/G(z=0)$,
others measure $G(z)$ relative to an amplitude anchored in the CMB,
and others measure the logarithmic growth index $\gamma$ of
equation~(\ref{eqn:dlng.dlna}). For ``generic'' parameters that describe 
departures from GR-predicted
growth, we will use a parameter $G_9$ that characterizes an overall
multiplicative offset of the growth factor and a parameter
$\Delta\gamma$ that characterizes a change in the fluctuation growth
rate.  We define these parameters in \S\ref{sec:dependences} below,
following our review of CMB anisotropy and large scale structure.
These parameters serve as useful diagnostics for deviations from GR,
but they do not provide a complete description of the
effects of modified gravity theories.
In particular, it is also possible (see \S\ref{sec:gravity}) that modified
gravity will cause $G(z)$ to be scale-dependent, or that it will alter the
relation between gravitational lensing
and non-relativistic mass tracers,
or that it will reveal its presence through a high-precision test
on solar system or terrestrial scales.

The above considerations lead to the following general strategy for
probing the physics of cosmic acceleration: use observations to
constrain the functions $H(z)$, $D(z)$, and $G(z)$, and use these
constraints in turn to constrain the history of $w(z)$ for dark
energy models and to test for inconsistencies that could point to
a modified gravity explanation.  For pure $H(z)$ and $D(z)$ measurements,
the ``nuisance parameters'' in such a strategy are the values of
$\omo$ and $\oko$, in addition to parameters related directly to
the observational method itself
(e.g., the absolute luminosity of supernovae).
Assuming a standard radiation content, the value of
$\opo = 1-\omo-\orado-\oko$ is fixed once $\omo$ and $\oko$ are
known.  The effects of $\omo$ and $\oko$ are separable both from
their different redshift dependence in the Friedmann
equation~(\ref{eqn:friedmann}) and from the influence of $\oko$ on
transverse distances (eq.~\ref{eqn:adist}) via space curvature.

\subsection{CMB Anisotropies and Large Scale Structure}
\label{sec:cmb_lss}

CMB anisotropies have little direct constraining power on dark energy,
but they play a critical role in cosmic acceleration studies because
they often provide the strongest constraints on nuisance parameters
such as $\omo$, $\oko$, and the high-redshift normalization of
matter fluctuations.  In particular, the amplitudes of the acoustic peaks
in the CMB angular power spectrum
depend sensitively (and differently) on the matter and baryon densities,
and the locations of the peaks depend sensitively on spatial
curvature.  Using CMB constraints necessarily brings in additional
nuisance parameters such as the spectral index $n_s$ and curvature
$d n_s/d\ln k$ of the scalar fluctuation spectrum, the amplitude and
slope of the tensor (gravitational wave) fluctuation spectrum, the
post-recombination electron-scattering 
optical depth $\tau$, and the Hubble constant
\begin{equation}
\label{eqn:hdef}
h \equiv H_0 / (100\hubunits).
\end{equation}
However, some of these parameters are themselves relevant to cosmic
acceleration studies, and current CMB measurements yield tight constraints
even after marginalizing over many parameters
(e.g., \citealt{komatsu11}).  The strength
of these constraints depends significantly on the adopted parameter space ---
for example, current CMB data provide tight constraints on $h$ if one
assumes a flat universe with a cosmological constant, but these constraints
are much weaker if $\oko$ and $w$ are free parameters.

CMB data are usually incorporated into dark energy constraints, or
forecasts, by adding priors on parameters that are then marginalized
over in the analysis.  We will adopt this strategy in \S\ref{sec:forecast},
using the level of precision forecast for the \planck\ satellite.
However, it is worth noting some rules of thumb.
For practical purposes, \planck\ data will give near-perfect determinations
of $\omo h^2$ and $\obo h^2$ from the heights of the acoustic peaks,
where the $h^2$ dependence arises because it is the physical density that
affects the acoustic features, not the density relative to the
critical density.
``Near-perfect'' means that marginalizing over the expected uncertainties
in $\omo h^2$ and $\obo h^2$ adds little to the 
error bars on dark energy parameters even from ambitious ``Stage IV''
experiments, relative to assuming that they are known 
perfectly.\footnote{However, the effects of \planck-level CMB
uncertainties are not completely negligible.  For the fiducial
Stage IV program discussed in \S\ref{sec:forecast}, fixing $\om h^2$
and $\ob h^2$ instead of marginalizing increases the DETF FoM
from 664 to 876.}
\planck\ data will also give near-perfect determinations of the sound horizon
at recombination $r_s(z_*)$, 
which determines the physical scale of the
acoustic peaks in the CMB and the scale of BAO in large scale structure
(see \S\ref{sec:bao_method}).
Since the angular scale of the acoustic peaks is precisely measured,
\planck\ data should also yield a near-perfect determination of the angular
diameter distance to the redshift of recombination, 
$D_* \equiv D_A(z_*)$, where $z_* \approx 1091$.
Finally, the amplitude of CMB anisotropies gives a near-perfect determination
(after marginalizing over the optical depth $\tau$, which is constrained
by polarization data) of the amplitude of matter fluctuations at 
$z_*$,
and thus throughout the era in which dark energy (or deviation from GR)
is negligible.  
As emphasized by \citeauthor{hu05} (\citeyear{hu05}; an
excellent source for more detailed discussion of CMB anisotropies in
the context of dark energy constraints), these determinations all
depend on the assumptions of a standard thermal and recombination history,
but the CMB data themselves allow tests of these assumptions at the
required level of accuracy.
CMB data also allow tests of cosmic acceleration models via
the integrated Sachs-Wolfe (ISW) effect, which we discuss
briefly in \S\ref{sec:isw}.

If primordial matter fluctuations are Gaussian, as predicted by
simple inflation models and supported by most observational investigations
to date, then their statistical properties are fully specified by
the power spectrum $P(k)$ or its Fourier transform, the two-point
correlation function $\xi(r)$.  Defining the Fourier transform of the
density contrast\footnote{A variety of Fourier conventions float
around the cosmology literature.  Here we adopt the same Fourier
conventions and definitions as \cite{dodelson03}.}
\begin{equation}
\label{eqn:deltak}
\dvk = \int d^3r  e^{-i\vk\cdot\vr} \delta(\vr), \qquad
\delta(\vr) = (2\pi)^{-3} \int d^3k e^{i\vk\cdot\vr} \dvk,
\end{equation}
the power spectrum is defined by
\begin{equation}
\label{eqn:pkdef}
\langle \dvk \dvkprime \rangle = (2\pi)^3 P(k) \delta_D^3 (\vk - \vk'),
\end{equation}
where $\delta_D^3$ is a 3-d Dirac-delta function
and isotropy guarantees that $P(\vk)$ is a function of
$k=|\vk|$ alone.
The power spectrum has units of volume, and it is often more intuitive
to discuss the dimensionless quantity
\begin{equation}
\label{eqn:Delta}
\Delta^2(k) \equiv (2\pi)^{-3} \times 4\pi k^3 P(k) = {d\sigma^2 \over
d\ln k},
\end{equation}
which is the contribution to the variance
$\sigma^2 \equiv \langle \delta^2 \rangle$ of the density contrast
per logarithmic interval of $k$.  The variance 
of the density field smoothed with a
window $W_R(r)$ of scale $R$ is
\begin{equation}
\label{eqn:sigma2}
\sigma^2(R) = \int_0^\infty {dk \over k} \Delta^2(k) \widetilde{W}_R^2(k),
\end{equation}
where the Fourier transform of a top-hat window,
$W_R(r) = (4\pi R^3/3)^{-1}\Theta(1-r/R)$, is
\begin{equation}
\label{eqn:tophattransform}
\widetilde{W}_R(k) = {3 \over k^3 R^3} \left[\sin(kR)-kR\cos(kR)\right],
\end{equation}
and the Fourier transform of a Gaussian window,
$W_R(r) = (2\pi)^{-3/2} R^{-3} e^{-r^2/2R^2}$, is
\begin{equation}
\label{eqn:gaussiantransform}
\widetilde{W}_R(k) = e^{-k^2 R^2/2}.
\end{equation}
The correlation function is
\begin{equation}
\label{eqn:xipk}
\xi(r) \equiv \langle \delta(\vx)\delta(\vx+\vr)\rangle =
\int_0^\infty {dk\over k} \Delta^2(k) {\sin (kr) \over kr}.
\end{equation}

In linear perturbation theory, the power spectrum amplitude is proportional
to $G^2(z)$, and we will
take $P_{\rm lin}(k)$ to refer to the $z=0$ normalization
when the redshift is not otherwise specified:
\begin{equation}
\label{eqn:plinevol}
P_{\rm lin}(k,z) = {G^2(z) \over G^2(z=0)} P_{\rm lin}(k).
\end{equation}
We discuss the normalization of $G(z)$ and $P_{\rm lin}(k)$
more precisely in \S\ref{sec:dependences} below.
The evolution of $P(k)$ remains close to linear theory for scales
$k \ll k_{\rm nl}$, where
\begin{equation}
\label{eqn:knl}
\int_0^{k_{\rm nl}} {dk \over k} \Delta^2(k) = 1.
\end{equation}
For realistic power spectra, non-linear evolution on small scales
does not feed back to alter the linear evolution on large scales
\citep{peebles80,shandarin90,little91}.
However, the shape of the power spectrum does change on scales
approaching $k_{\rm nl}$, in ways that can be calculated using
N-body simulations \citep{heitmann10} or several variants of
cosmological perturbation theory (\citealt{carlson09} and references
therein).  Non-linear evolution is a significant effect for weak
lensing predictions and for the evolution of BAO, as we discuss
in the corresponding sections below.

While there are many ways of characterizing the matter distribution in
the non-linear regime, the two measures that matter the most for our
purposes are the mass function and clustering
bias of dark matter halos.  There
are several different algorithms for identifying halos in N-body
simulations, all of them designed to pick out collapsed, gravitationally
bound dark matter structures in approximate virial equilibrium.
It is convenient to express the halo mass function in the form
\begin{equation}
\label{eqn:massfunction}
{dn \over d\ln M} = f(\sigma) \bar{\rho}_m 
  \left|{d\ln \sigma \over dM}\right|,
\end{equation}
where $\sigma^2$ is the variance of the linear density field smoothed
with a top-hat filter of mass scale $M = {4\over 3}\pi R^3 \bar{\rho}_m$
(eqs.~\ref{eqn:sigma2} and~\ref{eqn:tophattransform}).
To a first approximation, the function $f(\sigma)$ is universal,
and the effects of power spectrum shape, redshift (and thus power
spectrum amplitude), and background cosmological model
(e.g., $\Omega_m$ and $\Omega_\Lambda$) enter only through
determining $|d\ln\sigma/dM|$ and $\bar{\rho}_m$.
The state-of-the-art numerical investigation is that of \cite{tinker08},
who fit a large number of N-body simulation results with the functional
form
\begin{equation}
\label{eqn:fsigma}
f(\sigma) = A \left[\left({\sigma \over b}\right)^{-a} + 1 \right]
          e^{-c/\sigma^2} ,
\end{equation}
finding best-fit values $A=0.186$, $a=1.47$, $b=2.57$, $c=1.19$
for $z=0$ halos, defined to be spherical regions centered on density
peaks enclosing a mean interior overdensity of 200 times the
cosmic mean density $\bar{\rho}_m$.  
(Different halo mass definitions lead to different coefficients.)
A similar functional form
was justified on analytic grounds by \cite{sheth99}, following a
chain of argument that ultimately traces back to \cite{press74} and
\cite{bond91}.
Discussions of the halo population frequently refer to the
characteristic mass scale $M^*$, defined by
\begin{equation}
\label{eqn:mstardef}
\sigma(M^*) = \delta_c = 1.686,
\end{equation}
which sets the location of the exponential cutoff in the
Press-Schechter mass function.
Here $\delta_c$ is the linear theory overdensity at which a
spherically symmetric perturbation would collapse.\footnote{See
\cite{gunn72}, but note that their argument must be corrected to
growing mode initial conditions, as is done in standard textbook
treatments.  The value $\delta_c=1.686$ is derived for $\om=1$,
but the cosmology dependence is weak.}

In detail, \cite{tinker08} find that $f(\sigma)$ depends on redshift
at the 10-20\% level, probably because of the dependence of halo
mass profiles on $\Omega_m(z)$.  At overdensities of $\sim 200$,
the baryon fraction in group and cluster mass halos ($M>10^{13}M_\odot$)
is expected to be close to the cosmic mean ratio $\Omega_b/\Omega_m$,
but gas pressure, dissipation, and feedback from star
formation and AGN can alter this fraction and change baryon density
profiles relative to dark matter profiles.  We discuss these issues
further in \S\ref{sec:cl}.

Massive halos are more strongly clustered than the underlying matter
distribution because they form near high peaks of the initial density
field, which arise more frequently in regions where the background
density is high \citep{kaiser84,bardeen86}.  On large scales, the
correlation function of halos of mass $M$ is a scale-independent multiple
of the matter correlation function $\xi_{hh}(r) = b_h^2(M)\xi_{mm}(r)$.
The halo-mass cross-correlation in this regime is
$\xi_{hm}(r) = b_h(M)\xi_{mm}(r)$, and similar scalings
($b_h^2$ and $b_h$) hold for the halo power spectrum and halo-mass
cross spectrum at low $k$.
Analytic arguments suggest a bias factor \citep{cole89,mo96}
\begin{equation}
\label{eqn:simplebias}
b_h(M) = 1 + {[\delta_c/\sigma(M)]^2-1 \over \delta_c}.
\end{equation}
There have been numerous refinements to this formula based on
analytic models and numerical calibrations.
The state-of-the-art numerical study is that of \cite{tinker10}.

Galaxies reside in dark matter halos, and they, too, are biased
tracers of the underlying matter distribution.  Here one must allow for
the fact that different kinds of galaxies reside in different mass
halos and that massive halos host multiple galaxies.
More massive or more luminous galaxies are more strongly clustered
because they reside in more massive halos that have higher $b_h(M)$.
At low redshift, the large scale bias factor is
$b_g \leq 1$ for galaxies below the characteristic cutoff $L^*$
of the \cite{schechter76} luminosity function, but it rises
sharply at higher luminosities \citep{norberg01,zehavi05,zehavi11}.

For a galaxy sample defined by a threshold $L_{\rm min}$
in optical or near-IR
luminosity (or stellar mass), theoretical models and empirical
studies (too numerous to list comprehensively, but our summary here
is especially influenced by \citealt{kravtsov04,conroy06,zehavi11})
suggest the following approximate model.
The minimum host halo mass is the one for which the comoving
space density $n(M_{\rm min})$ of halos above $M_{\rm min}$
matches the space density $n(L_{\rm min})$ of galaxies above the luminosity
threshold.  Each halo above $M_{\rm min}$ hosts one central
galaxy, and in addition each such halo hosts a mean number
of satellite galaxies
$\langle N_{\rm sat}\rangle = (M-M_{\rm min})/15M_{\rm min}$,
with the actual number of satellites drawn from a Poisson
distribution with this mean.\footnote{To make the model more
accurate, one should adjust $M_{\rm min}$ iteratively so that
the {\it total} space density of galaxies, central+satellite,
matches the observed $n(L_{\rm min})$, but this is usually a modest
correction because the typical fraction of galaxies that are satellites
is $5-20\%$.}  The large scale galaxy bias factor $b_g$ is
the average bias factor $b_h(M)$ of halos above $M_{\rm min}$,
with the average weighted by the product of the halo space
density and the average number of galaxies per halo.  In addition
to increasing $b_g$ by giving more weight to high mass halos,
satellite galaxies contribute to clustering on small scales,
where pairs or groups of galaxies reside in a single halo
\citep{seljak00,scoccimarro01,berlind02}.  In detail, at high
luminosities one must allow for scatter between galaxy luminosity
and halo mass, which reduces the bias below that of the sharp threshold
model described above.  Furthermore, selecting galaxies by color
or spectral type alters the relative fractions of central and satellite
galaxies; redder, more passive galaxies are more strongly clustered
because a larger fraction of them are satellites, and the
reverse holds for bluer galaxies with active star formation.
Thus each class of galaxies has its own halo occupation distribution
(HOD), which describes the probability $P(N|M)$ of finding $N$
galaxies in a halo of mass $M$ and specifies any relative bias
of galaxies and dark matter within halos.

On large scales, where $b_g^2\Delta_{\rm lin}^2(k,z) \ll 1$,
the galaxy power
spectrum should have the same shape as the linear matter power
spectrum, $P_{gg}(k,z) = b_g^2 P_{\rm lin}(k,z).$  However, scale-dependence
of bias at the 10-20\% level can persist to quite low $k$, especially
for luminous, highly biased galaxy populations,
and the effective ``shot noise'' contribution to $P_{gg}(k)$ can
differ from the naive $1/\bar{n}_g$ term expected for
Poisson statistics \citep{yoo09}.
Combinations of CMB power spectrum measurements with galaxy power
spectrum measurements can yield tighter cosmological parameter
constraints than either one in isolation (e.g., \citealt{cole05,reid10}).
In particular, this combination provides greater leverage on
the Hubble constant $h$, since CMB-constrained models predict
galaxy clustering in Mpc while galaxy redshift surveys measure
distances in $\hmpc$ (or, equivalently, in $\kms$).

Another complicating factor in galaxy clustering measurements is
redshift-space distortion (\citealt{kaiser87}; see \citealt{hamilton98}
for a comprehensive review), which arises because galaxy redshifts
measure a combination of distance and peculiar velocity rather
than true distance.  On small scales, velocity dispersions in
collapsed objects stretch structures along the line of sight.
On large scales, coherent inflow to overdense regions compresses
them in the line-of-sight direction, and coherent outflow from
underdense regions stretches them along the line of sight.
In linear perturbation theory, the divergence of the
peculiar velocity field is related to the density contrast field by
\begin{equation}
\label{eqn:divv}
{\vec\nabla}\cdot {\vv}(\vx,z) =
  -(1+z)^{-1} H(z) {d\ln G \over d\ln a} \delta(\vx,z)
\approx -(1+z)^{-1} H(z) [\Omega_m(z)]^\gamma \delta(\vx,z)~,
\end{equation}
with $\gamma$ defined by equation~(\ref{eqn:dlng.dlna}).
The galaxy redshift-space power spectrum in linear theory is
anisotropic, depending on the angle $\theta$ between the wavevector $\vk$
and the observer's line of sight as
\begin{equation}
\label{eqn:pkmu}
P_g(k,\mu) = b_g^2 (1+\beta\mu^2)^2 P(k) 
           = \left[b_g+\mu^2 f(z)\right]^2 P(k)~,
\end{equation}
where $P(k)$ is the real-space
matter power spectrum, $\mu \equiv \cos\theta$, 
and $f(z) \approx [\om(z)]^\gamma$ is the logarithmic
growth rate (eq.~\ref{eqn:dlng.dlna}).
The strength of the anisotropy depends on the ratio
$\beta \equiv f(z)/b_g$; because linear bias amplifies
galaxy clustering isotropically, more strongly biased galaxies
exhibit weaker redshift-space distortion.
A variety of non-linear effects, most notably the small scale
dispersion and its correlation with large scale density, mean that
equation~(\ref{eqn:pkmu}) is rarely an adequate approximation in
practice, even on quite large scales \citep{cole94,scoccimarro04}.  
In the galaxy correlation
function, one can remove the effects of redshift-space distortion
straightforwardly by projection, counting galaxy pairs as a function of
projected separation rather than 3-d redshift-space separation.
For the power spectrum, one can correct for redshift-space distortion,
but the analysis is more model-dependent
(see, e.g., \citealt{tegmark04}).  However, redshift-space distortion
can be an asset as well as a nuisance, since it provides a route
to measuring $d\ln G/d\ln a$.  We will discuss this idea at some length in
\S\ref{sec:rsd}, as it is emerging as a powerful route to measuring
the expansion history and testing GR growth predictions.

\subsection{Parameter Dependences and CMB Constraints}
\label{sec:dependences}

\begin{figure}
\begin{centering}
{\includegraphics[width=3.2in]{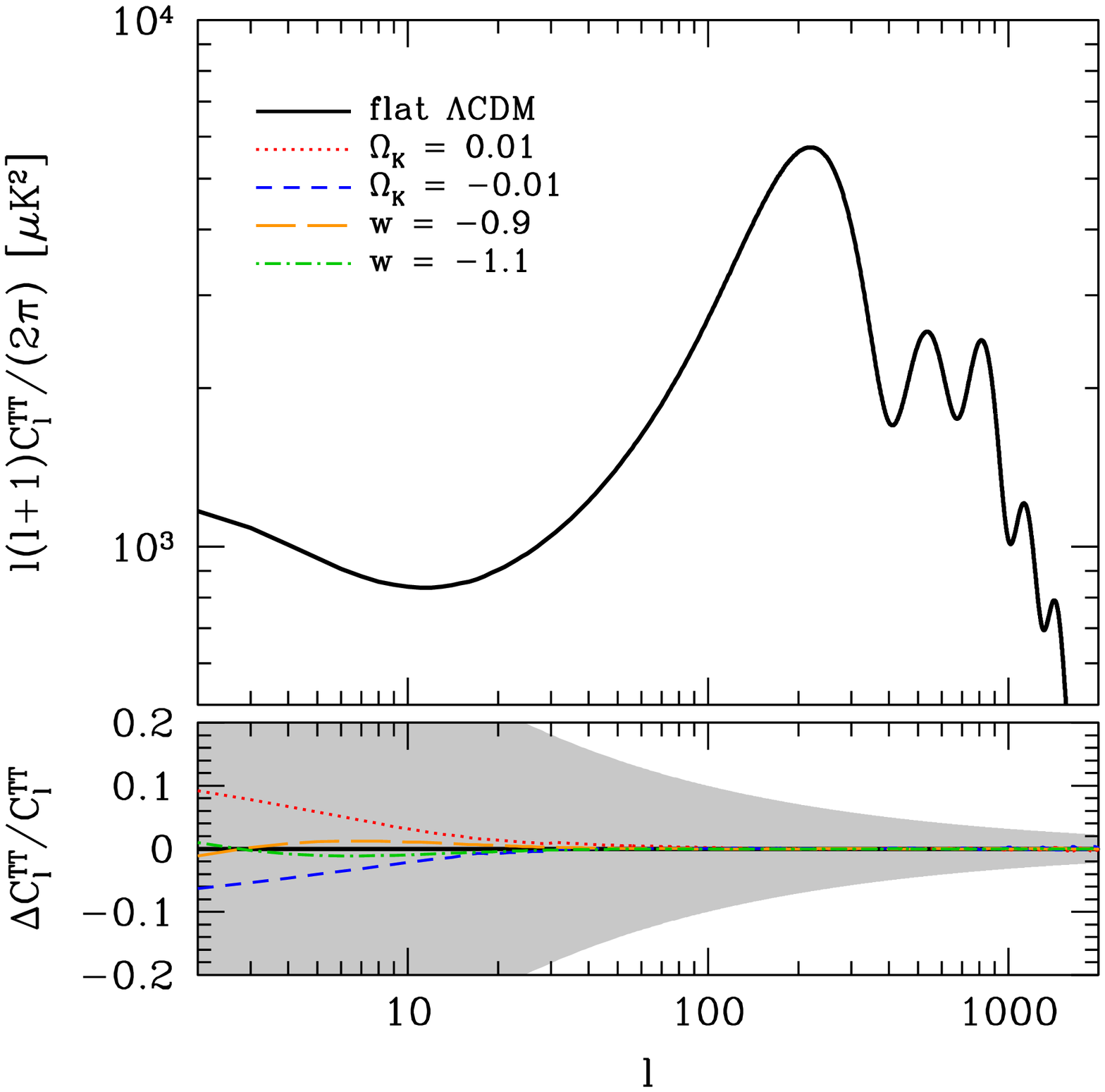}\hfill 
\includegraphics[width=3.2in]{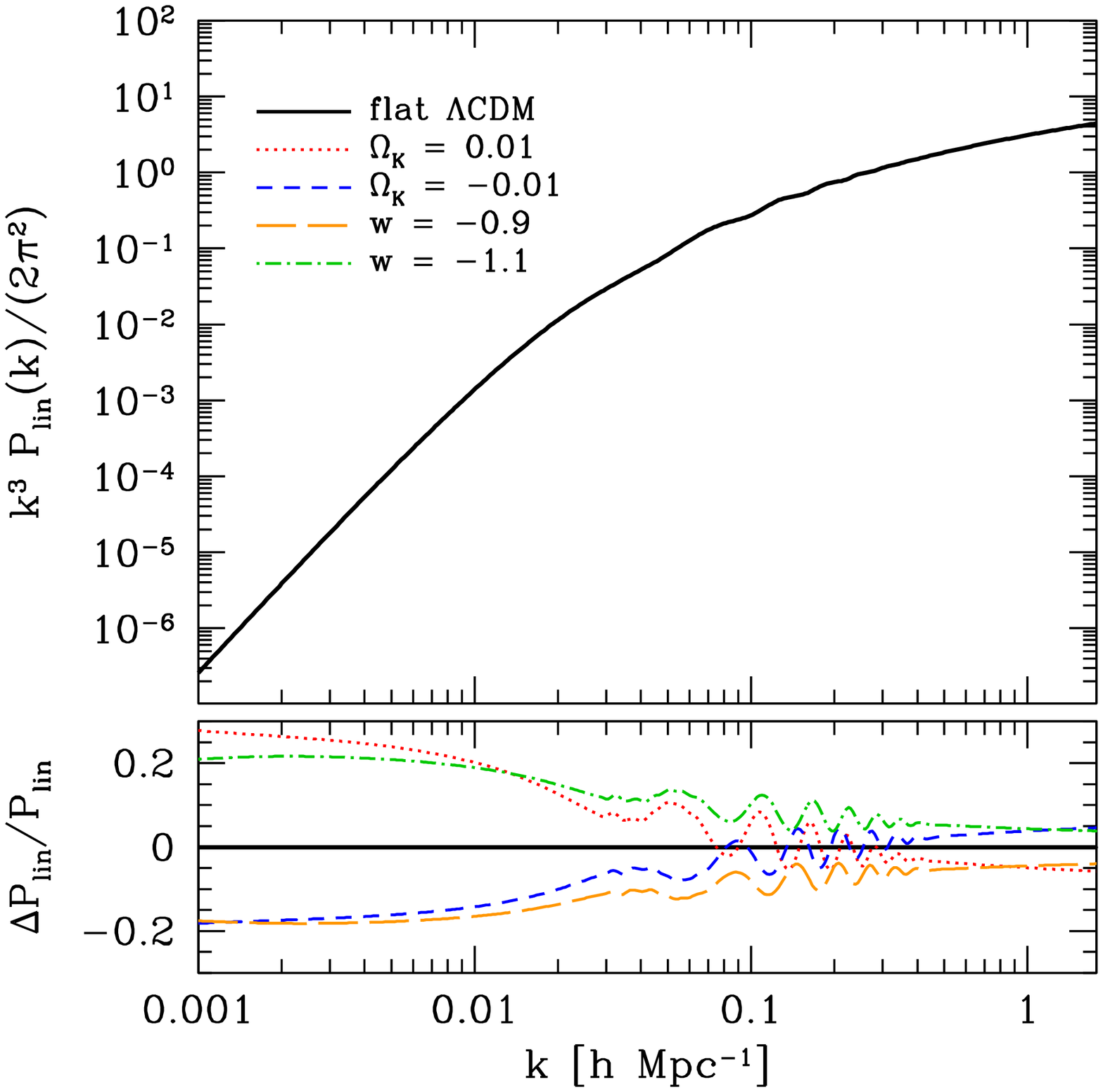}}
{\includegraphics[width=3.2in]{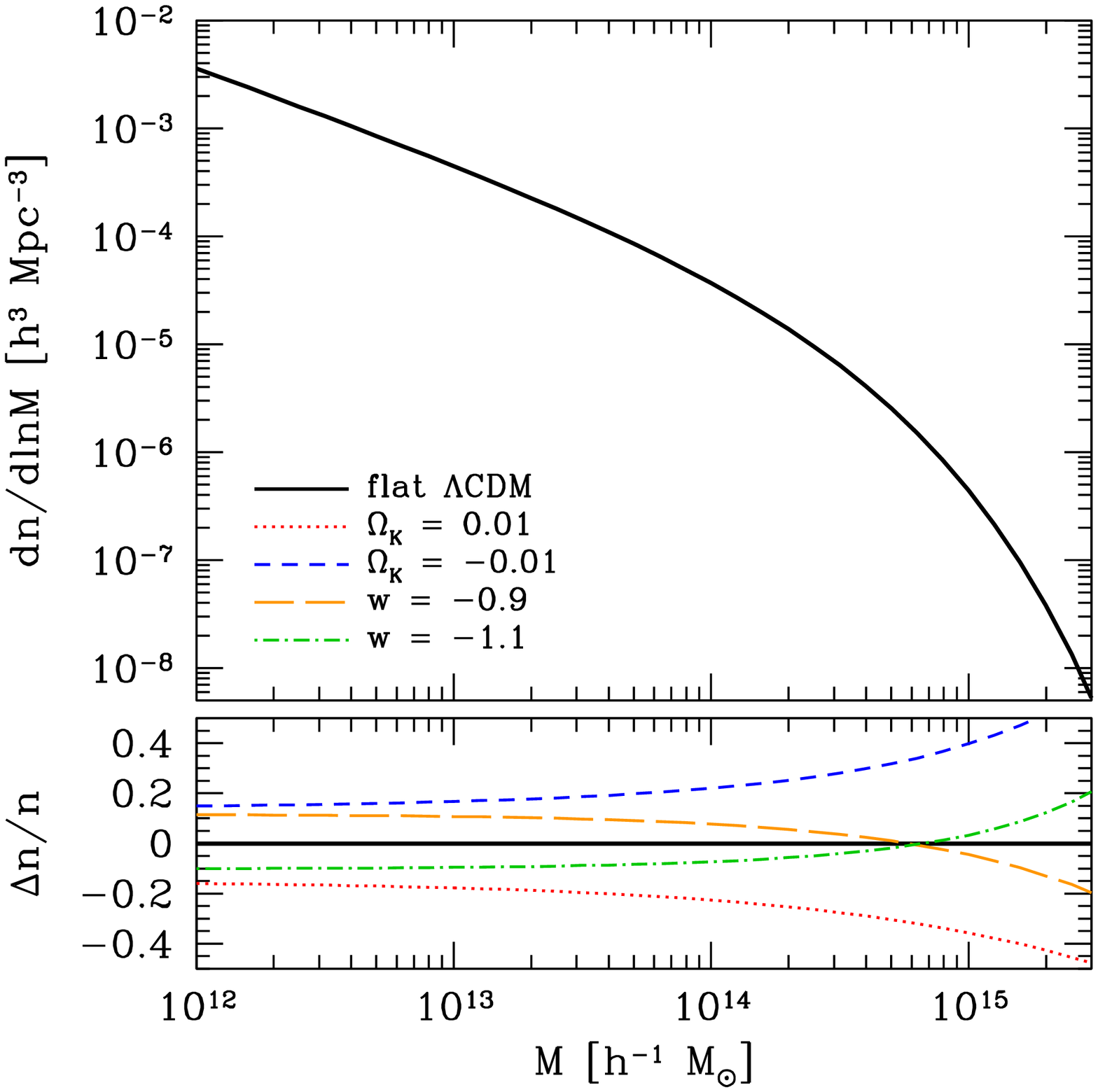}\hfill
\includegraphics[width=3.2in]{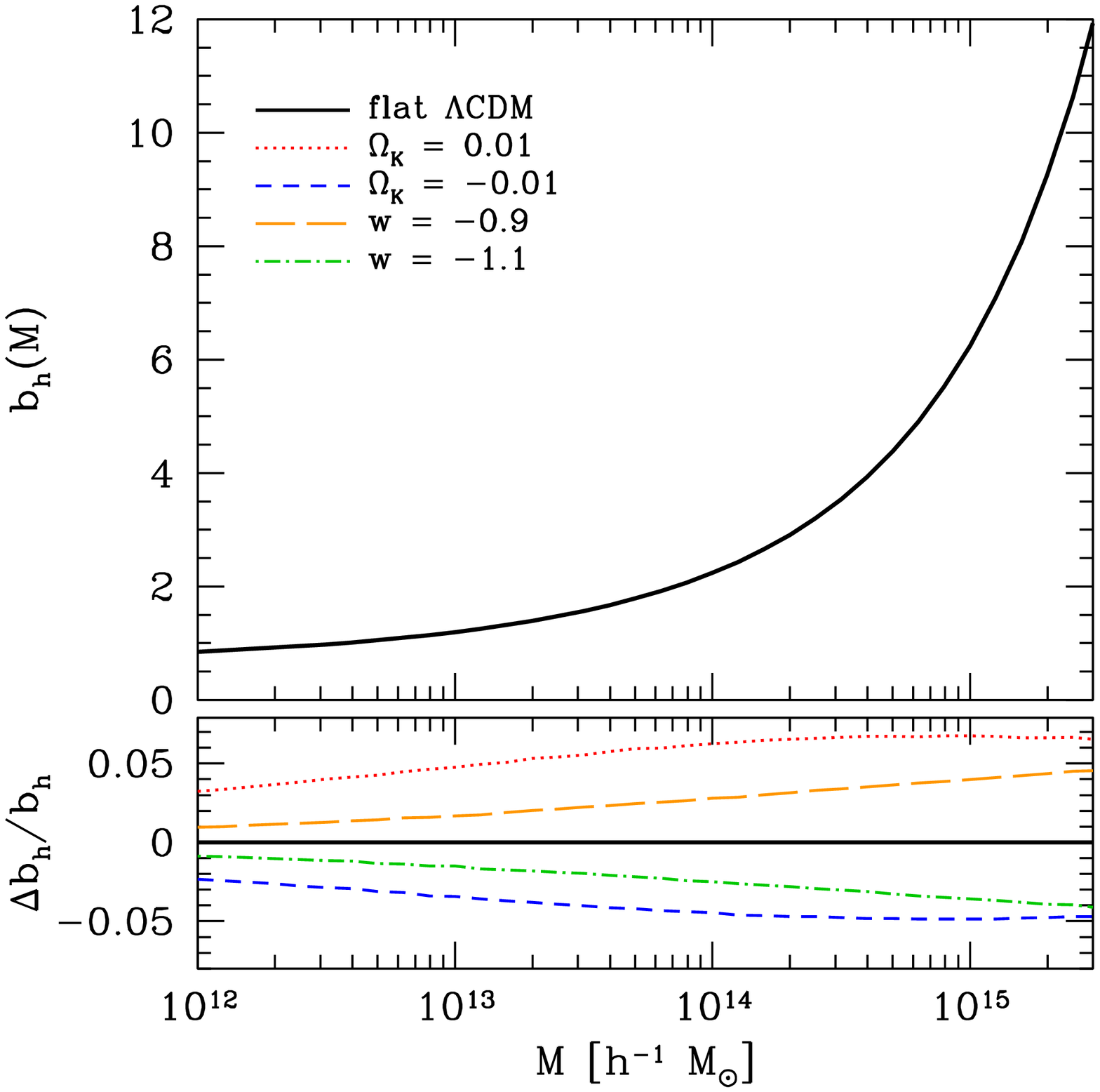}}
\end{centering}
\caption{\label{fig:structure} 
CMB angular power spectrum (upper left), variance of matter fluctuations
(upper right), halo mass function (lower left), and halo bias factor
(lower right).  Solid curves in the main panels show predictions of
the fiducial $\Lambda$CDM panel listed in Table~\ref{tbl:models}.
Curves in the lower panels show the fractional changes in these
statistics induced by changing $1+w$ to $\pm 0.1$ or $\ok$ to $\pm 0.01$
(see legend).  For each parameter change, we keep $\om h^2$, $\ob h^2$,
and $D_*$ fixed by adjusting $\om$, $\ob$, and $h$ 
(see Table~\ref{tbl:models}).  These compensating changes keep 
deviations in the CMB spectrum minimal, much smaller than the
cosmic variance errors indicated by the shaded region.
}
\end{figure}

Figure~\ref{fig:structure} illustrates the four statistics discussed above:
the CMB temperature angular power spectrum,
the matter variance $\Delta_{\rm lin}^2(k)$ computed from the linear theory
power spectrum at $z=0$, the $z=0$ halo mass function computed
from equations~(\ref{eqn:massfunction}) and~(\ref{eqn:fsigma}),
and the halo bias factor computed from equation~(6) of
\cite{tinker10} for overdensity 200 halos (relative to the mean
matter density).  Curves in the main panels
show a fiducial model with the likelihood-weighted mean parameters
for the seven-year \wmap\ CMB measurements 
(hereafter WMAP7; \citealt{larson11}) 
assuming a flat universe with a cosmological constant:
$\oco = 0.222$, $\obo=0.045$,
$\Omega_{\Lambda} = 0.733$, $h=0.71$, $n_s=0.963$, 
$\tau = 0.088$, and primordial power spectrum amplitude
$A_s(k=0.002\Mpc^{-1})=2.43\times 10^{-9}$.
(These parameters also assume no tensor fluctuations and
$dn_s/d\ln k = 0$.)
The CMB power spectrum shows the familiar pattern of acoustic
peaks, with the angular scale of the first peak corresponding approximately
to the sound horizon at recombination divided by the angular diameter
distance to the last scattering surface.  The matter variance
$\Delta_{\rm lin}^2(k)$ shows a slow change of slope starting at 
$k \approx 0.02 h\,{\rm Mpc}^{-1}$, corresponding to the horizon
scale at matter-radiation equality, and low amplitude wiggles at
smaller scales produced by BAO.  The halo mass function has an
approximate power-law form at low masses changing slowly to an 
exponential cutoff for $M \gg M^* = 3 \times 10^{12} \hmsun$.
The $b_h(M)$ relation is roughly flat for $M\la 5 M^*$ before rising
steeply at higher masses.  The $h$-dependences used for $k$, 
$dn/d\ln M$, and $M$ reflect the dependences that typically
arise when distances are estimated from redshifts and thus scale
as $h^{-1}$.

In the lower panels, we show the fractional change in these statistics
that arises when changing $1+w$ from 0 to $\pm 0.1$ and when changing
$\ok$ from 0 to $\pm 0.01$.  With any parameter variation, there is
the crucial question of what one holds fixed.  For this figure,
we have held fixed the parameter combinations that have the
strongest impact on the CMB power spectrum: $\Omega_m h^2$ and $\Omega_b h^2$,
which determine the heights of the acoustic peaks and the physical
scale of the sound horizon, and $D_* = D_A(z_*)$, 
which maps the physical scale
of the peaks into the angular scale.  We satisfy these constraints 
by allowing $h$ and $\Omega_m$ to
vary, maintaining $\ok=0$ for the 
$w$-variations and $w=-1$ for the $\ok$-variations, with $n_s$, $A_s$, and
$\tau$ fixed to the fiducial model values.  The parameter values
for these variant models appear in Table~\ref{tbl:models}.

%% CHANGED MICHAEL'S TABLE FORMAT TO DELUXETABLE
%\begin{table}[ht]
%\caption{Fiducial Model and Simple Variants}
%\centering
%\begin{tabular}{r r r r r r r}
%\hline\hline

\begin{deluxetable}{r r r r r r r}
\tablecolumns{7}
\tablecaption{Fiducial Model and Simple Variants\label{tbl:models}}
\tablehead{
$w$ & $\ok$ & $\Omega_c$ & $\Omega_b$ & $\Omega_\phi$ & $h$ & $\sigma_8$ 
%\\ 
}
\tablewidth{0pc}
%[0.5ex]
%\hline
\startdata
{\bf $-$1.0} & {\bf 0.00} & {\bf 0.222} & {\bf 0.045} & {\bf 0.733} & 
  {\bf 0.710} & {\bf 0.806} \\
$-0.9$ & 0.00 & 0.246 & 0.050 & 0.704 & 0.675 & 0.774 \\
$-1.1$ & 0.00 & 0.201 & 0.041 & 0.758 & 0.746 & 0.837 \\
$-1.0$ & 0.01 & 0.186 & 0.038 & 0.766 & 0.776 & 0.809 \\
$-1.0$ & $-0.01$ & 0.256 & 0.052 & 0.702 & 0.661 & 0.802 \\
%\hline
%\end{tabular}
\enddata
\tablecomments{
All models have $n_s=0.963$, $\tau=0.088$,
$A_s(k=0.002\,{\rm Mpc}^{-1})=2.43\times 10^{-9}$.
In addition, all models 
have the same values $\om h^2$, $\ob h^2$, and distance
to the last scattering surface $D_*$, so they produce nearly
indistinguishable CMB power spectra.
}
\end{deluxetable}

%\label{tbl:models}
%\end{table}

{}From the CMB panel, we can see that the changes in the angular power
spectrum induced by these parameter variations are small compared
to the cosmic variance error at every $l$, since we
have fixed the parameter combinations that mostly determine
the CMB spectrum.\footnote{The CMB cosmic variance error is
$\Delta\ln C^{TT}_l = [(2l+1)/2]^{-1/2},$ determined simply
by the number of modes on the sky at each angular scale $l$.}
The changes are coherent, of course, but
even considering model fits to the entire CMB spectrum
the $w$ changes would be undetectable at the level of errors forecast
for {\it Planck}, while the $\ok=\pm 0.01$ models would be distinguishable
from the fiducial model at about $1.5\sigma$.
The impact of these parameter changes must instead be sought
in other statistics at much lower redshifts.
Changes to the matter variance are $\sim 5\%$ at small scales,
growing to $\sim 20\%$ at large scales, with oscillations 
that reflect the shift in the BAO scale.  Fractional changes to the halo
space density at fixed mass can be much larger, especially at
high masses where the halo mass function is steep.  We caution, 
however, that the fractional change in mass at fixed abundance is 
significantly smaller, a point that we emphasize in \S\ref{sec:cl}.
The impact of a change in $w$ reverses sign at 
$M \approx 6\times 10^{14} \hmsun \approx 200 M^*$,
where the mass function begins to drop sharply.
Changes in bias factor at fixed mass are $\sim 5\%$
at high masses and smaller at low masses.

\begin{figure}
\begin{centering}
{\includegraphics[width=3.2in]{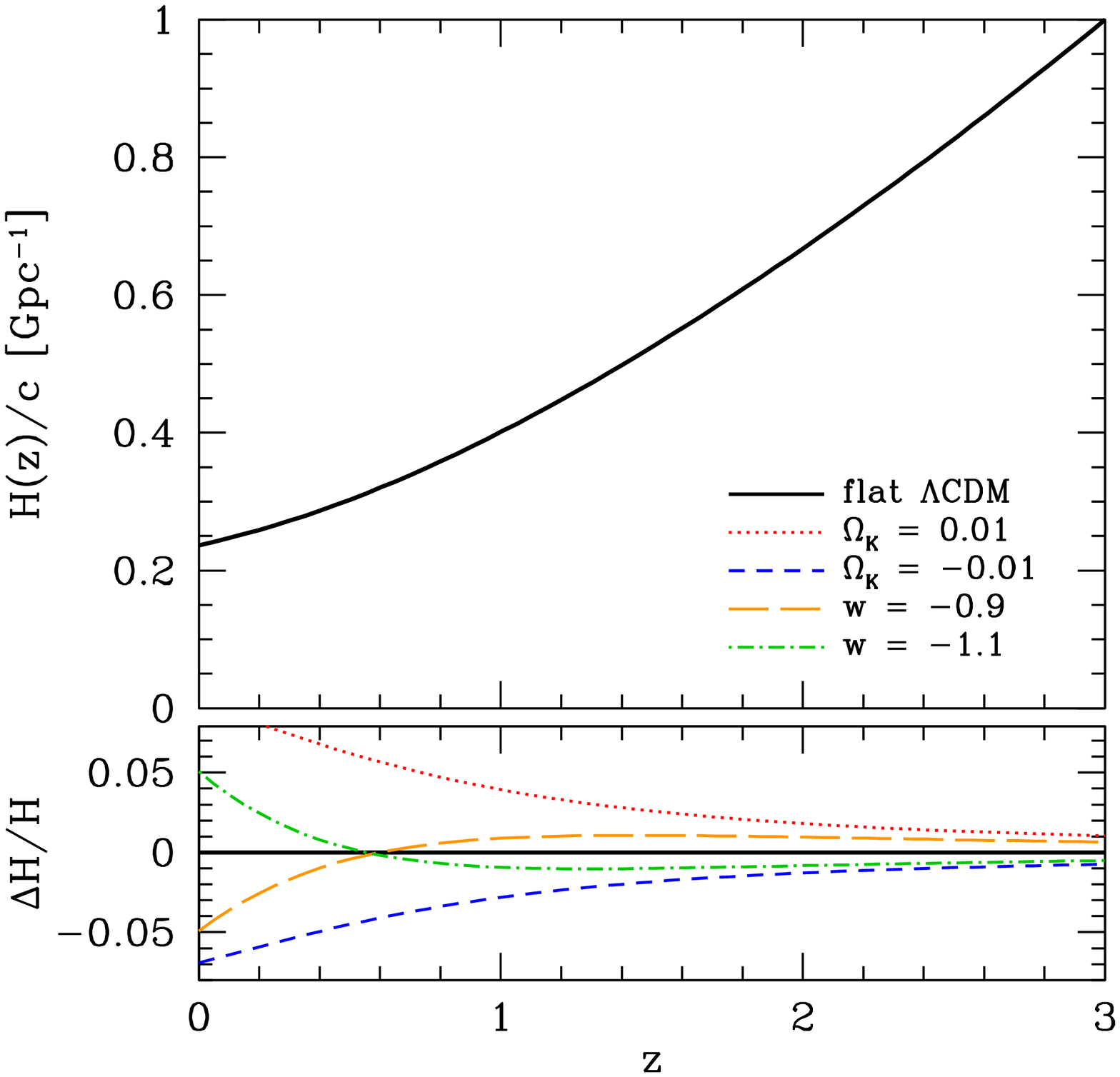}\hfill 
\includegraphics[width=3.2in]{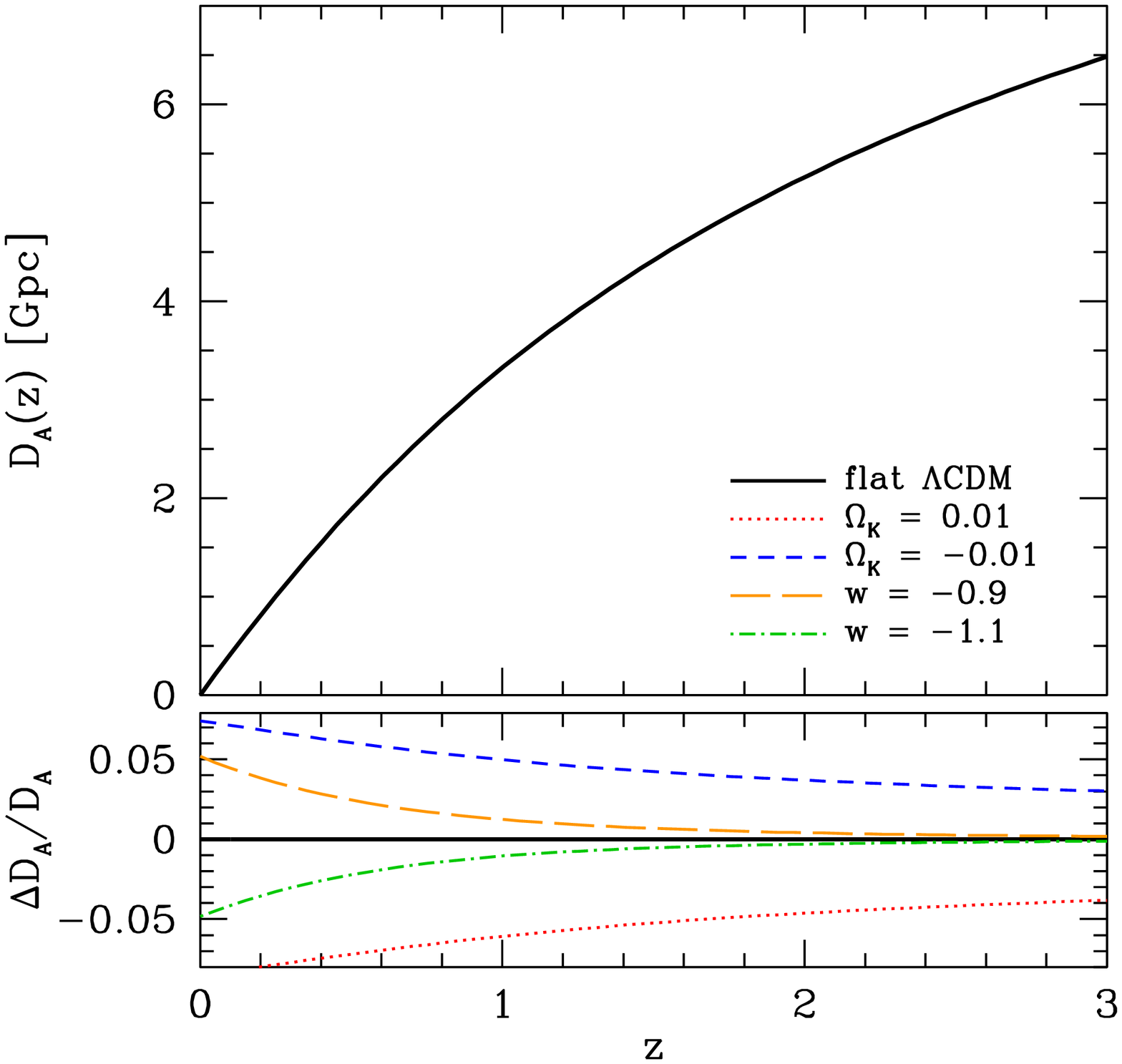}}
{\includegraphics[width=3.2in]{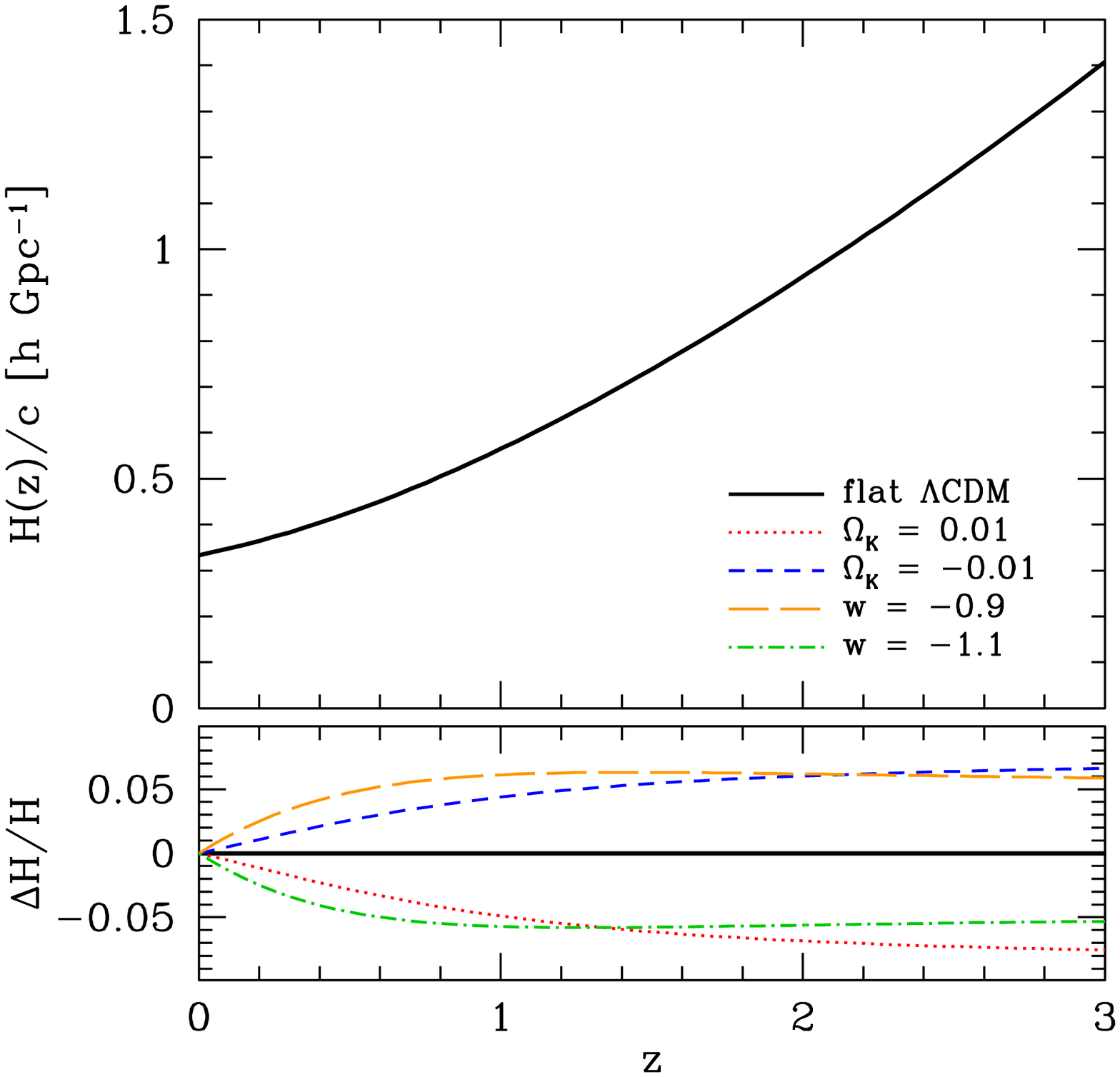}\hfill
\includegraphics[width=3.2in]{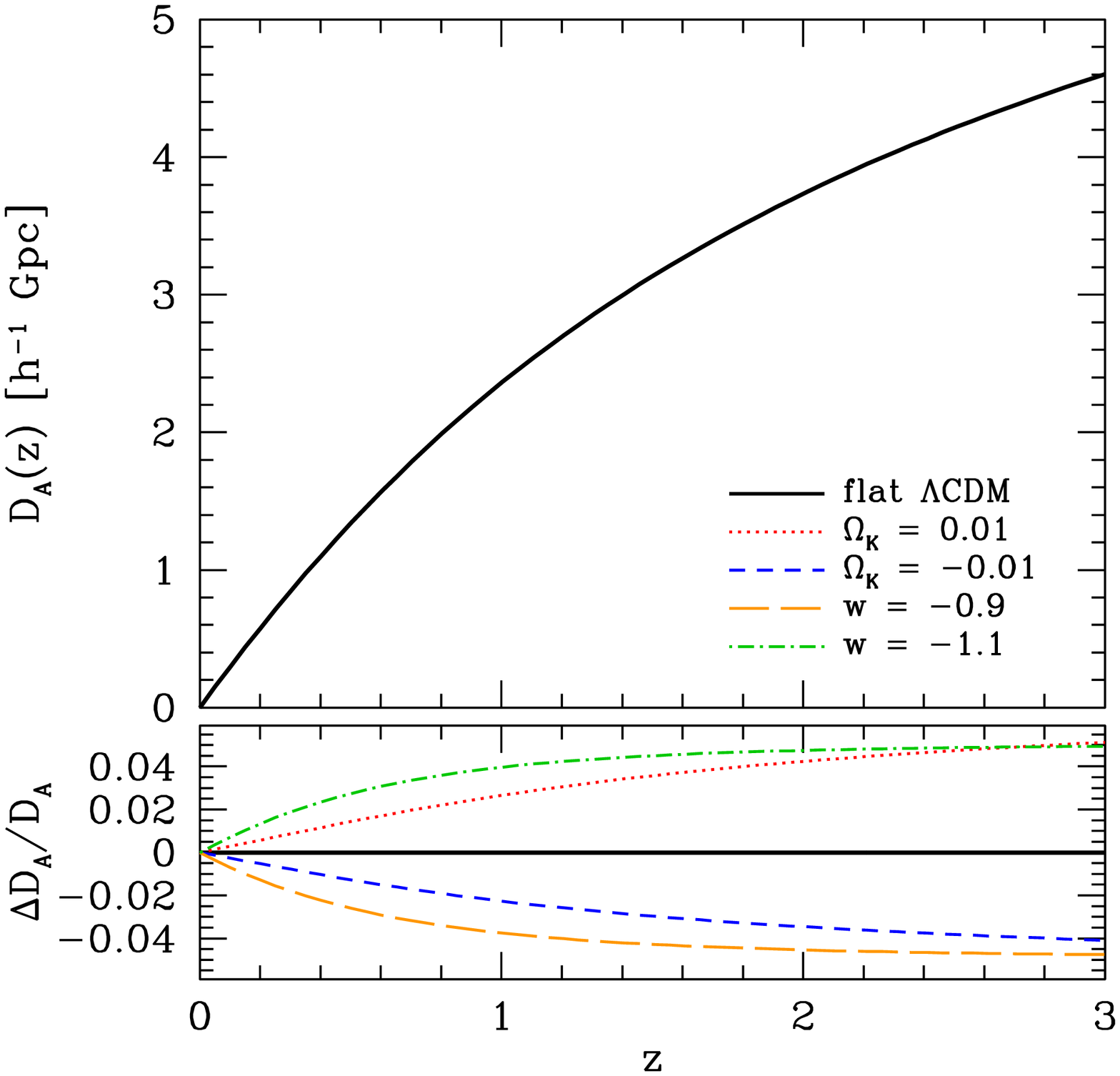}}
\end{centering}
\caption{\label{fig:hzdz} 
Evolution of the Hubble parameter (left) and the
comoving angular diameter distance
(right) for the fiducial $\Lambda$CDM model and for the variant models
shown in Figure~\ref{fig:structure}.  Upper panels are in absolute
units, relevant for BAO, while lower panels 
show distances in $h^{-1}$ Gpc,
relevant for supernovae or weak lensing.
}
\end{figure}

Figure~\ref{fig:hzdz} shows the redshift evolution and parameter
sensitivity of the Hubble parameter (eq.~\ref{eqn:friedmann}) and
the comoving angular diameter distance (eq.~\ref{eqn:adist}), for the
same fiducial model and parameter variations used in 
Figure~\ref{fig:structure}.  The upper panels show $H(z)$ and
$D_A(z)$ in absolute units, while the lower panels plot
them in $\hgpc$ units.
BAO studies measure in absolute units, but
supernova studies effectively measure $hD_A(z)$
because they are calibrated in the local Hubble flow.
Equivalently, supernova distances are determined in $\hmpc$ rather
than Mpc.\footnote{To be more precise, studies of supernovae
at redshifts $z_1$ and $z_2$ yield the distance ratio
$D(z_2)/D(z_1)$.  When the $z_1$ population is local, in the
sense that inferred distances have negligible cosmology
dependence except for the $H_0^{-1}$ scaling, then one
gets the distance $D(z_2)$ in $\hmpc$.}
Weak lensing predictions depend on distance ratios rather than
absolute distances, so in practice they also constrain $hD_A(z)$
rather than absolute $D_A(z)$.

In absolute units, model predictions diverge most strongly 
at $z=0$, and the impact of $\ok = \pm 0.01$ is larger than
the impact of $1+w = \pm 0.1$.  The impact of the $w$ change
on $H(z)$ reverses sign at $z\approx 0.6$, a consequence
of our CMB normalization.  Changing $w$ to $-0.9$
would on its own reduce the distance to $z_*$, and $H_0$ must 
therefore be lowered to keep $D_*$ fixed.  However, with 
$\om h^2$ fixed, lower $H_0$ implies a higher $\om$, which
raises the ratio $H(z)/H_0$, and at high redshift this effect
wins out over the lower $H_0$.
At $z>2$, $D_A(z)$ remains sensitive to $\ok$ but is
insensitive to $w$, while the sensitivity of $H(z)$ to $w$ is
roughly flat for $1 < z < 3$.
In $\hmpc$ units, models converge at $z=0$ by definition, and
the impact of $1+w = \pm 0.1$ is generally larger than the
impact of $\ok = \pm 0.01$.  The sensitivity of $hD_A(z)$ to
parameter changes increases monotonically with increasing redshift,
growing rapidly until $z=0.5$ and flattening beyond $z=1$.

\begin{figure}
\begin{centering}
{\includegraphics[width=3.2in]{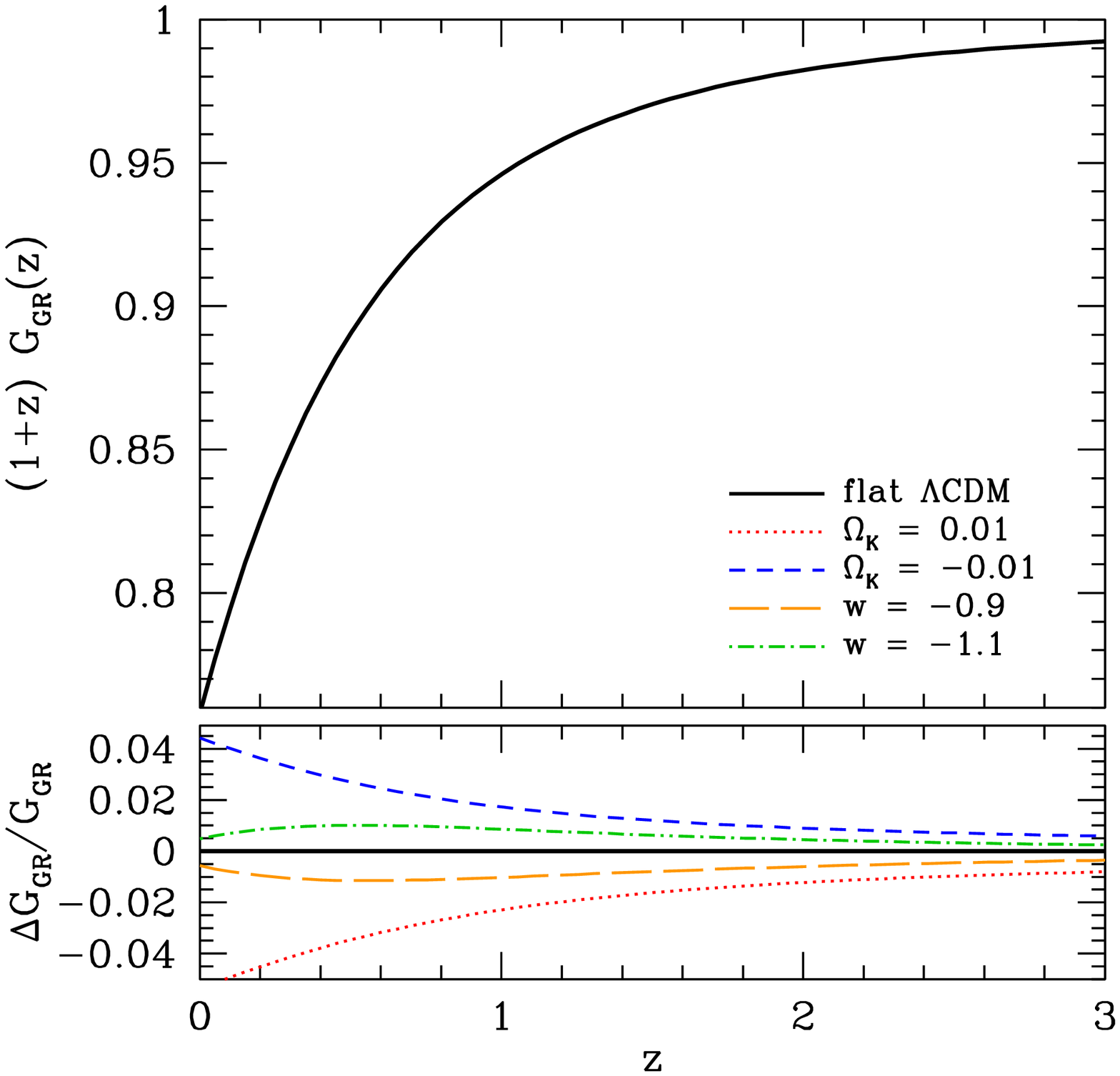}\hfill
\includegraphics[width=3.2in]{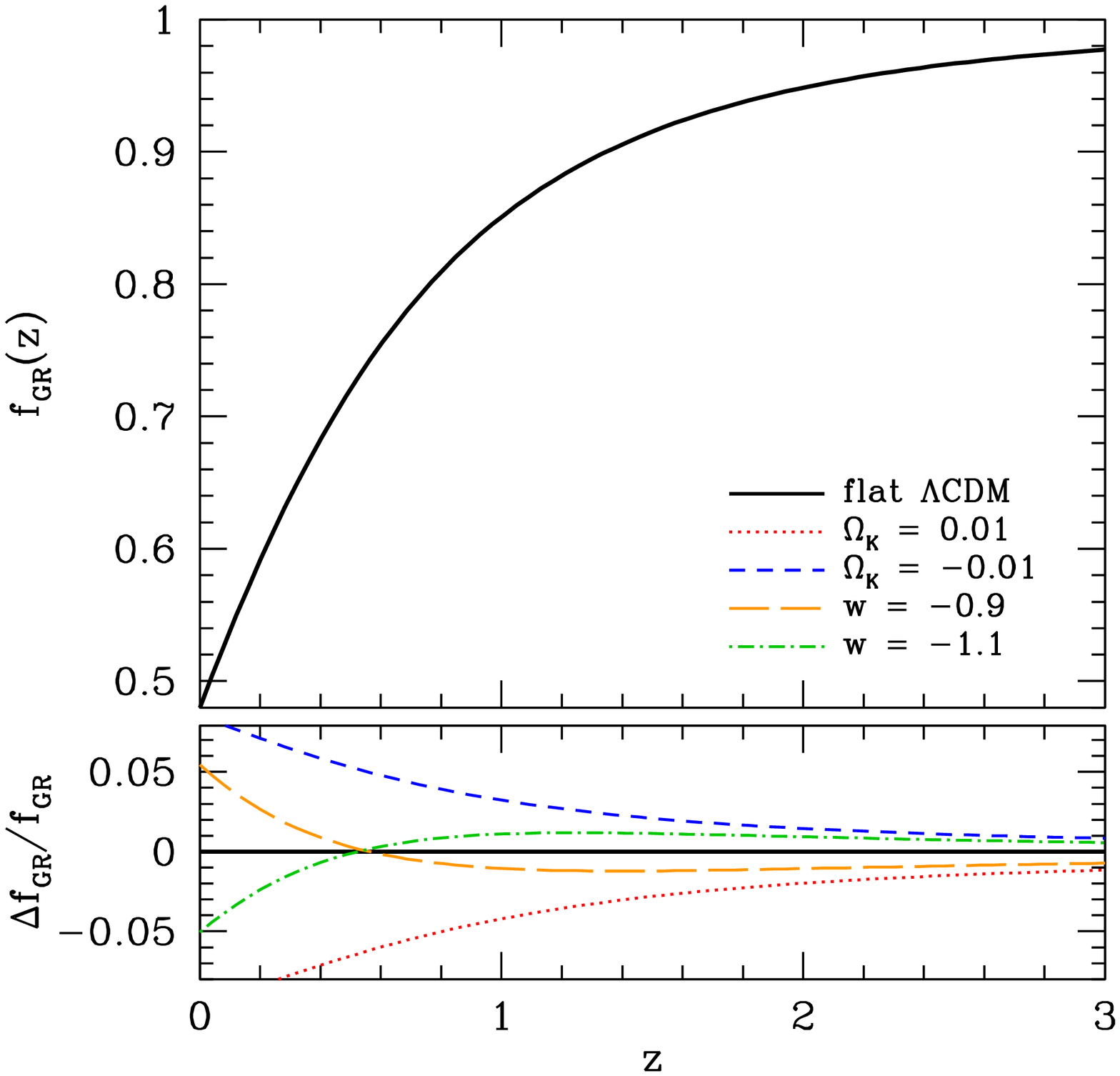}}
\caption{\label{fig:gz} 
Evolution of the linear growth factor $G(z)$ and
growth rate $f(z)$ for the 
models shown in Figure~\ref{fig:hzdz}, assuming GR.
The scaling in the left panel removes the $(1+z)$ evolution
that would arise in an $\om=1$ universe and normalizes $\Ggr(z)$
to one at $z=9$.
}
\end{centering}
\end{figure}

For structure growth, the issues of normalization are more subtle.
The normalization of the matter power spectrum is known better from
CMB anisotropy at $z_*$ than it is from local measurements at $z=0$,
and this will be still more true in the \planck\ era.
It therefore makes sense to anchor the normalization in the CMB,
even though the value at $z=0$ then depends on cosmological
parameters.  Figure~\ref{fig:gz} (left panel) plots 
$(1+z)\Ggr(z)$, where 
%$[\Ggr(z)/\Ggr(z=9)] \times (1+z)/10$,
$\Ggr(z)$ obeys equation~(\ref{eqn:lingrowth}) and is normalized
to unity at $z=9$.
In most models, dark energy is dynamically negligible at $z>9$,
making the growth from the CMB era up to that epoch independent of
dark energy.  In an $\om=1$ universe, $\Ggr(z) \propto (1+z)^{-1}$,
so the plotted ratio falls below unity when $\om(z)$ starts to fall
below one.  For $\ok=0.01$, $\om(z)$ is below that in our
fiducial model (see eqs.~\ref{eqn:friedmann} and~\ref{eqn:omegaz})
both because of the $\ok$ term in the Friedmann equation and because
we lower $\om(z=0)$ from 0.27 to 0.22 to keep $D_*$ fixed, thus depressing
$\Ggr(z)$ increasingly towards lower $z$.  For $w=-0.9$, however,
the depression of $\om(z)/\om(z=0)$ from the Friedmann equation
is countered by the higher value of $\om(z=0) = 0.30$ adopted to 
fix $D_*$, so the depression of $\Ggr(z)$ is smaller, and it actually
recovers towards the fiducial value as $z$ approaches zero.
The effects on the growth rate $f(z)$ (right panel) are similar
but stronger, with our adopted parameter changes producing larger
deviations from the fiducial model and the influence of $w$
actually reversing sign at $z < 0.5$.

\begin{figure}
\begin{centering}
{\includegraphics[width=3.2in]{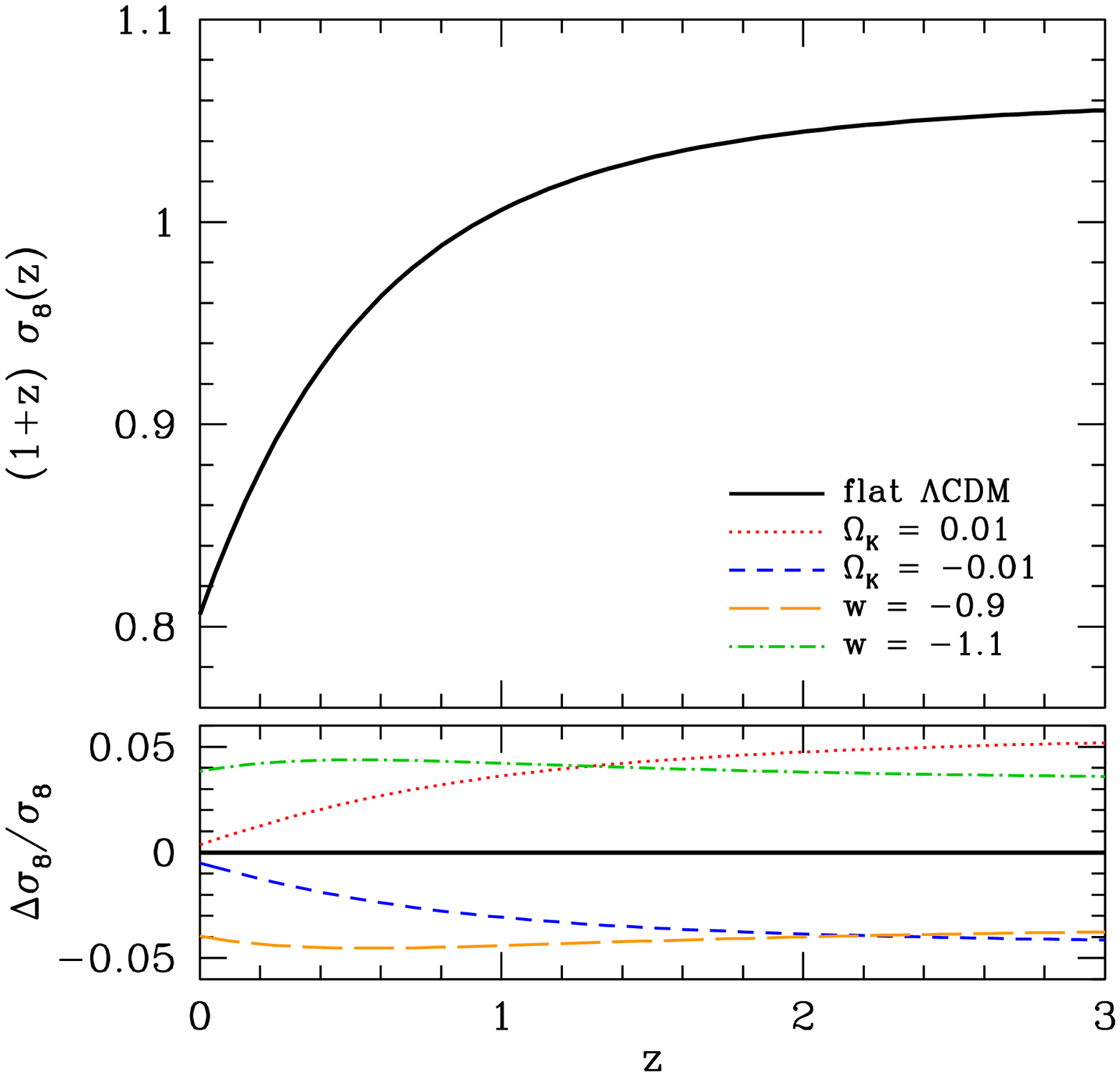}\hfill 
\includegraphics[width=3.2in]{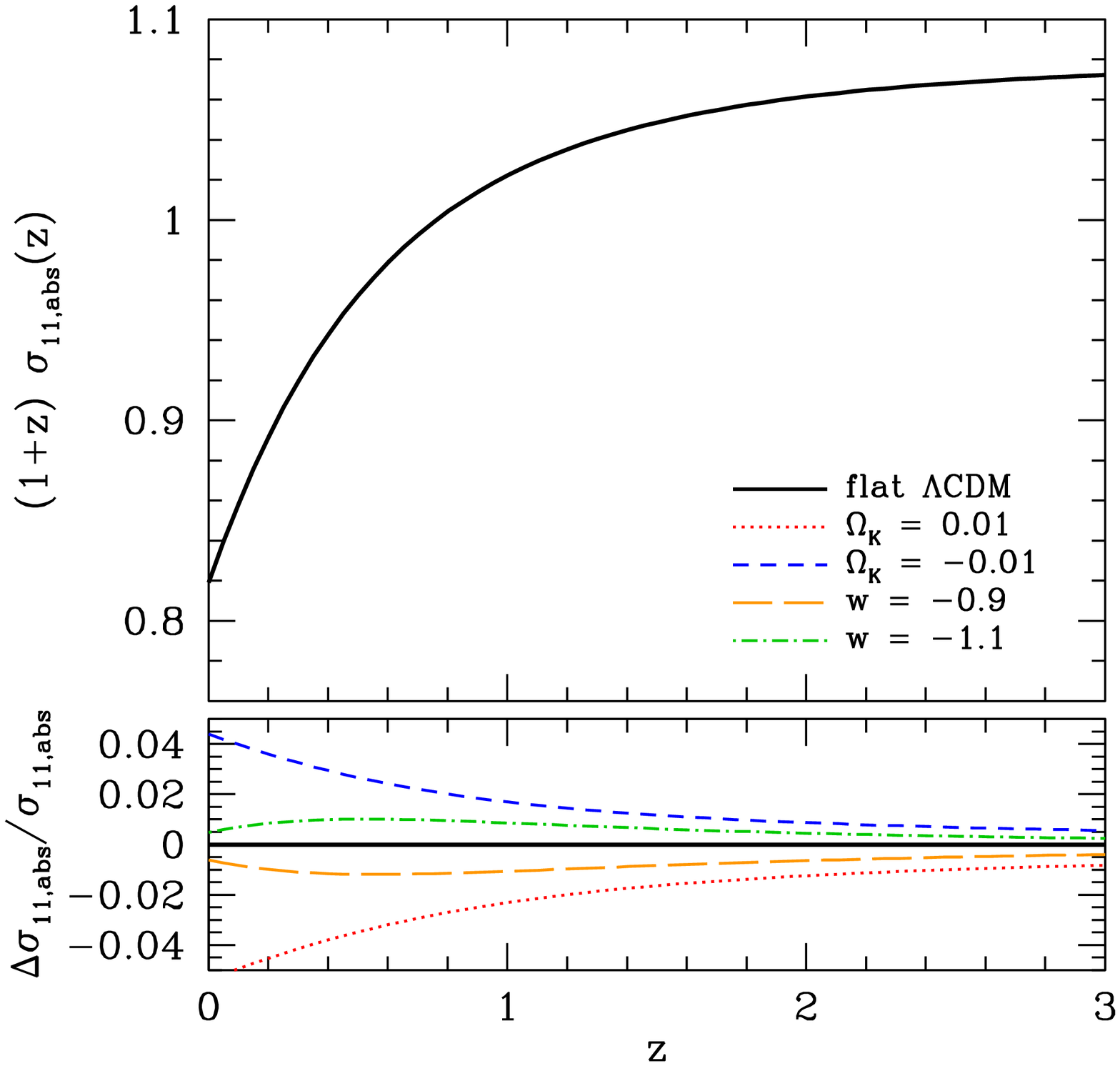}}
\end{centering}
\caption{\label{fig:sig8z} 
Evolution of the matter fluctuation amplitude for
the models shown in Figure~\ref{fig:gz}, characterized by
the rms linear fluctuation in comoving spheres of radius
$8\hmpc$ (left) or 11 Mpc (right).
All models are normalized to the WMAP7 CMB fluctuation amplitude.
}
\end{figure}

In practice, observations do not probe the growth factor
itself but the amplitude
of matter clustering, and in this case we must also account for the
changing relation between the CMB power spectrum and the matter
clustering normalization.  The left panel of Figure~\ref{fig:sig8z}
plots $\sigma_8(z) \times (1+z)$, where $\sigma_8(z)$ is the rms linear 
theory density contrast in a sphere of comoving radius $8\hmpc$
(eqs.~\ref{eqn:sigma2} and~\ref{eqn:tophattransform}).
The right panel instead plots $\sigma_{11,{\rm abs}}(z) \times (1+z)$,
where $\sigma_{11,{\rm abs}}$ refers to a sphere of radius 11 Mpc
(equivalent to $\sigma_8$ for $h=0.727$).
At high redshift these curves go flat as $\om(z)$ approaches one and
the growth rate approaches $\Ggr(z) \propto (1+z)^{-1}$.
In the CMB-matched models considered here, the impact of $w$ or
$\ok$ changes is complex, since changing these parameters alters
the best-fit values of $\om$ and $h$ as well as changing the
growth factor directly through equation~(\ref{eqn:lin_approx}).
The values of $\sigma_8(z)$ change by 4-5\% at all $z$
for $1+w = \pm 0.1$, but these changes mostly track the changes in $h$.
In absolute units, the changes to $\sigma_{11,{\rm abs}}(z)$ are $\la 1\%$,
tracking (by definition) the changes in $\Ggr(z)$ shown
in Figure~\ref{fig:gz}.
For $\ok = \pm 0.01$, $\sigma_8(z)$ changes by 4-5\% at high $z$
but converges nearly to the fiducial value at $z=0$,
while $\sigma_{11,{\rm abs}}(z)$ shows only 1\% differences
at high $z$ but diverges at low $z$.

All of these models have the WMAP7 \citep{larson11}
normalization of the power spectrum of inflationary fluctuations,
$A_s = 2.43 \times 10^{-9}$ at comoving scale
$k = 0.002\Mpc^{-1}$ at $z=z_*=1091$.
The primary uncertainty in this normalization
is the degeneracy with the electron optical depth $\tau$,
since late-time scattering suppresses the amplitude of the primary
CMB anisotropies by a factor $e^{-\tau}$ on the scales that determine
the normalization.  The WMAP7 constraints are
$\tau = 0.088 \pm 0.015$ ($1\sigma$), so the associated uncertainty 
in the matter fluctuation amplitude is 1.5\%.  
(Recall that the power spectrum amplitude is $\propto \sigma_8^2$,
so its fractional error is a factor of two larger.)
For \planck, \citet{holder03} estimate uncertainty 
$\sigma_\tau=0.01$ allowing for
complex reionization history, and we use this value in our own forecasts. While
there have been some changes in the situation since then (the polarized
foregrounds at large scales are worse than anticipated, and $\tau$ is
lower than the central value from the first-year WMAP results), this
expectation seems broadly consistent with more recent studies
\citep[e.g.,][]{mortonson08,colombo09}.\footnote{For
example, \citet{colombo09} find $\sigma_\tau\sim 0.006$, albeit
under somewhat optimistic assumptions regarding foregrounds and sky cuts.}
This will likely be the limiting factor for comparison of high-redshift
(CMB) measurements with low-redshift (e.g., WL) measurements of the growth
of structure (as opposed to measurements of evolution within the observed
low-$z$ range), {\em unless} other probes of reionization such as 21 cm
provide constraints on the reionization history
(see \S\ref{sec:wl_prospects} for further discussion).

Following \cite{albrecht09}, we parameterize departures from the GR growth
rate by a change $\Delta\gamma$ of the growth index (eq.~\ref{eqn:dlng.dlna})
and by an overall amplitude shift $G_9$ that is the ratio of the matter 
fluctuation amplitude at $z=9$ to the value that would be predicted
by GR given the same cosmological parameters and $w(z)$ 
history.\footnote{\cite{albrecht09} denote this quantity $G_0$ instead of
$G_9$, but we have reserved subscript-0 to refer to $z=0$ quantities.}
Some caution is required in defining $\Delta\gamma$, since 
equations~(\ref{eqn:dlng.dlna})-(\ref{eqn:gamma}) are not exact,
and their inaccuracies should not be defined as failures of GR!
For precise calculations, therefore, we adopt the \cite{albrecht09} 
expressions for growth factor evolution:
\begin{eqnarray}
f(z) &=& f_{\rm GR}(z)\left(1+\Delta\gamma\ln\om(z)\right) \label{eqn:fullfz}\\
G(z) &=& G_9 \times \Ggr(z) \times 
       \exp \left[\Delta\gamma \int_z^9 {dz' \over 1+z'} f_{\rm GR}(z')
                  \ln\om(z')\right], \label{eqn:fullgz}
\end{eqnarray}
where $\Ggr(z)$ and $f_{\rm GR}(z)$ follow
the (exact) solution to equation~(\ref{eqn:lingrowth}).

For practical purposes, one can use our definition of growth parameters
to calculate the normalized linear theory matter power spectrum at
redshift $z$, given an assumed set of cosmological parameter values
and a $w(z)$ history, as follows.  First, use CAMB \citep{lewis00} or
some similar program to compute the normalized linear matter power
spectrum at $z=9$.  Then multiply the power spectrum
by $G^2(z)/\Ggr^2(z=9)$, with $G(z)$ given by equation~(\ref{eqn:fullgz})
and $\Ggr(z)/\Ggr(z=9)$ given by the exact solution to 
equation~(\ref{eqn:lingrowth}), or by the approximate integral
solution~(\ref{eqn:lin_approx9}), computing $H(z)$ and $\om(z)$
from equations~(\ref{eqn:friedmann}) and~(\ref{eqn:omegaz}) given the
cosmological parameters and $w(z)$.  For reference, we note that CAMB
normalization with WMAP7 data yields, for a flat $\Lambda$CDM model,
\begin{eqnarray}
\label{eqn:cmbnorm}
%\sigma_8(z=9) \times (1+9) &=& 1.118 \left[{A_s
%   \over 2.43 \times 10^{-9}}\right]^{1/2} e^{2(n_s-1)} 
%   \left({\Omega_b h^2 \over 0.023}\right)^{-0.34} %\nonumber\\
%   \left({\Omega_m h^2 \over 0.13}\right)^{0.57} 
%   \left({h \over 0.71}\right)^{0.67+(n_s-1)/2}~, \\
\sigma_{11,{\rm abs}}(z=9) \times (1+9) &=& 1.134 
        \left[{A_s
        \over 2.43 \times 10^{-9}}\right]^{1/2} e^{2(n_s-1)} 
	\left({\Omega_b h^2 \over 0.023}\right)^{-0.34}
        \left({\Omega_m h^2 \over 0.13}\right)^{0.57}~, \\
\sigma_8(z=9) &=& \sigma_{11,{\rm abs}}(z=9) \times 0.9859
   \left({h \over 0.71}\right)^{0.67+(n_s-1)/2}~,
\end{eqnarray} 
where the primordial amplitude $A_s$ is defined at comoving
wavenumber $k = 0.002\Mpc^{-1}$.
%\begin{eqnarray}
%\label{eqn:cmbnorm}
%\sigma_8(z=9) \times (1+9) &=&
%    1.118 \left[{A_s(k=0.002\,{\rm Mpc}^{-1}) 
%    \over 2.43 \times 10^{-9}}\right]^{1/2} (75.85 h)^{(n_s-1)/2} \nonumber \\
%&&\times \left({\Omega_b h^2 \over 0.023}\right)^{-0.340}
%\left({\Omega_m h^2 \over 0.13}\right)^{0.574}
%\left({h \over 0.71}\right)^{0.674} .
%\end{eqnarray}
This formula, similar to that in \cite{hu04}, is found by varying
the parameters in CAMB calculations 
around the WMAP7 mean values
one at a time to evaluate logarithmic
derivatives; spot checks indicate that 
it is accurate to 0.2\% over the $2\sigma$ range of the WMAP7 errors,
and for the range of $w$ and $\ok$ variations in Table~\ref{tbl:models}.
For models other than flat \lcdm, 
one can use this formula to get $\sigma_8(z=9)$ in GR,
assuming that the effect of dark energy at $z>9$ is negligible,
then multiply by $G(z)/\Ggr(z=9)$ to get $\sigma_8(z)$.

For an analytic power spectrum, one can use the approximate
formula in equation~(25) of \cite{eisenstein99},
which includes suppression of small scale power by baryonic effects but does
not incorporate BAO.
This paper defines the power spectrum normalization in terms
of a parameter $\delta_H$, related to our growth factor
and normalization $A_s$ by
\begin{equation}
\label{eqn:deltaH}
\delta_H = {2 \over 5} A_s^{1/2} \left({G(z=0)\over\Omega_m}\right)
   \left({H_0 \over k_{\rm norm}}\right)^{(n_s-1)/2}~.
\end{equation}
Using the appropriate values for our fiducial WMAP7 flat $\Lambda$CDM
model [$G(z=0)=0.76$, $\om=0.267$, $A_s = 2.43\times 10^{-9}$ at
$k_{\rm norm} = 0.002\Mpc^{-1}$, $h=0.71$, $n_s=0.963$] with
this definition of $\delta_H$, the \cite{eisenstein99} formula
agrees with the result from CAMB to 2\% or better except
at the BAO scales, where deviations are up to 10\%.
One can also use this normalization for 
the more complex (but still analytic) formulae of
\cite{eisenstein98}, which do include BAO.
We caution that other papers and books (e.g., \citealt{dodelson03})
have different definitions of $\delta_H$.

There are, of course, degeneracies between the modified gravity parameters
$G_9$ and $\Delta\gamma$ and the $w(z)$ history, since both affect 
structure growth.  However, if $w(z)$ is pinned down well by $D(z)$ and $H(z)$
measurements, then measurements of matter clustering can be used to 
constrain $G_9$ and $\Delta\gamma$.
The clustering amplitude at a single redshift yields a degenerate
combination of these two parameters, but measurements at multiple redshifts
or direct measurements of the growth rate via redshift-space distortions 
can separate them in principle.
Of course, there is no guarantee that a modified
gravity prediction can be adequately described by $G_9$ and a constant
$\Delta\gamma$, and one might more generally consider 
(in eq.~\ref{eqn:fullgz}), for example, a functional
history $\gamma(z)$ analogous to $w(z)$, 
or a direct multiplicative change to the growth rate
$d\ln G/d\ln a$ rather than a change of the growth index $\gamma$.
However, any constraints
inconsistent with $G_9 = 1$, $\Delta\gamma=0$ after marginalizing
over $w(z)$ and cosmological parameters would be suggestive evidence
for a breakdown of GR.  Even if the measurements themselves are convincing,
one must be cautious in the interpretation, since apparent discrepancies
could arise from $w(z)$ histories outside the families considered in
marginalization or from other violations of the underlying assumptions.
To give two examples, ``early dark energy'' that is dynamically
significant at high redshift could cause an apparent $G_9 < 1$,
and decay of dark matter into dark energy could
cause an apparent $\Delta\gamma < 0$ because the value of
$\om(z)/\om(z=0)$ would be higher than in the standard picture.
In \S\ref{sec:gravity} we discuss other potential signatures of
modified gravity, such as scale-dependent growth, discrepancy
between masses inferred from lensing and from non-relativistic
tracers, and different accelerations in low and high density
environments, and we mention other parameterizations that
have been used to describe modified gravity models.

%Here $\oco=\omo-\obo$ is the density parameter of cold dark matter.
%Other curves in the figures show the effects of increasing $h$ by 10\%,
%increasing $\oco$ by 10\%, or decreasing $\Omega_{\Lambda,0}$ to 0.683
%(and thus changing $\oko$ to 0.05), or changing $w$ from $-1$ to $-0.9$,
%always keeping other parameters fixed at their fiducial values.

While they are not a substitute for full calculations, we find
the use of CMB-normalized models like those in this section to
be a valuable source of intuition for understanding the impact of distance
or structure growth measurements in a (realistic) situation where
CMB anisotropies impose tight parameter constraints.  To make
construction of such model sets easy, we note that for
small changes $|1+w| \la 0.1$ and $|\ok| \la 0.01$, the changes
to $h$ required to keep $D_*$ fixed are well approximated by
\begin{equation}
\label{eqn:cmb_dlnh}
\Delta\ln h \approx -0.5(1+w) + 8\ok + 0.9|\ok|.
\end{equation}
The changes to $\om$ and $\ob$ are then trivially found
by fixing $\om h^2$ and $\ob h^2$ to their fiducial model
values, and the dark energy density follows from
$\Omega_\phi = 1-\om-\ok-\Omega_r$.  The value of 
$\sigElev(z=9)$ is unchanged because $\om h^2$ and $\ob h^2$
are fixed, while the value of $\sigma_8(z=9)$ follows from
equation~(\ref{eqn:cmbnorm}) with the revised Hubble parameter.
To compute power spectrum normalizations at other redshifts
one uses equation~(\ref{eqn:lin_approx9}) with the new
$\om(z)$ implied by equations~(\ref{eqn:friedmann}) 
and~(\ref{eqn:omegaz}).  The changes to the normalization
at $z=0$ are approximately
\begin{equation}
\label{eqn:cmb_dlnsig8}
\Delta\ln \sigma_8(z=0) \approx -0.4(1+w) + 0.4\ok - 0.06|\ok|.
\end{equation}
The coefficients in equations~(\ref{eqn:cmb_dlnh}) and~(\ref{eqn:cmb_dlnsig8})
are chosen to reproduce the values in Table~\ref{tbl:models}; at
smaller $|1+w|$ or $|\ok|$ the best coefficients might be 
slightly different, but the changes themselves would be smaller.

\subsection{Overview of Methods}
\label{sec:overview}

We conclude our ``background'' material with a short overview of the
methods we will describe in detail over the next four sections.

Observations show that Type Ia supernovae have a peak luminosity
that is tightly correlated with the shape of their light curves ---
supernovae that rise and fall more slowly have higher peak luminosity.
The intrinsic dispersion around this relation is only about 0.12 mag,
allowing each well observed supernova to provide an estimated distance
with a $1\sigma$ uncertainty of about 6\%.  Surveys that detect
tens or hundreds of Type Ia supernovae and measure their light curves
and redshifts can therefore measure the distance-redshift relation
$D(z)$ with high precision.  Because the supernova luminosity is
calibrated mainly by local observations of systems whose distances are
inferred from their redshifts, supernova surveys effectively measure
$D(z)$ in units of $\hmpc$, not in absolute units independent
of $H_0$.

Baryon acoustic oscillations provide an entirely independent way
of measuring cosmic distance.  Sound waves propagating before
recombination imprint a characteristic scale on matter clustering,
which appears as a local enhancement in the correlation funtion
at $r \approx 150$ Mpc.  Imaging surveys can detect this feature
in the angular clustering of galaxies in bins of photometric
redshift, yielding the angular diameter distance $D(z_{\rm phot})$.
A spectroscopic survey over the same volume resolves the BAO
feature in the line-of-sight direction and thereby yields a more
precise $D_A(z)$ measurement.  Furthermore, measuring the
BAO scale in velocity separation allows a direct
determination of $H(z)$.  Other tracers of the matter distribution
can also be used to measure BAO.  Because the BAO scale is known
in absolute units (based on straightforward physical calculation
and parameter values well measured from the CMB), the BAO method
measures $D(z)$ in absolute units --- Mpc not $\hmpc$ --- so
BAO and supernova measurements to the same redshift carry
different information.

The shapes of distant galaxies are distorted by the weak gravitational
lensing from matter fluctuations along the line of sight.
The typical distortion is only $\sim 0.5\%$, much smaller than the
$\sim 30\%$ dispersion of intrinsic galaxy ellipticities,
but by measuring the correlation of ellipticities as a function
of angular separation, averaged over many galaxy pairs, one can
infer the power spectrum of the matter fluctuations producing
the lensing.  Alternatively, one can measure the average elongation
of background, lensed galaxies as a function of projected separation
from foreground lensing galaxies to infer the galaxy-mass
correlation function of the foreground sample, which can be
combined with measurements of galaxy clustering to infer the
matter clustering.  By measuring the projected matter power
spectrum for background galaxy samples at different $z$, weak
lensing can constrain the growth function $G(z)$.  However, the
strength of lensing also depends on distances to the sources
and lenses, so in practice the weak lensing method constrains
combinations of $G(z)$ and $D(z)$.

Clusters of galaxies trace the high end of the halo mass function,
typically $M \geq 10^{14} M_\odot$.  
Observationally, one measures the number of clusters as a function
of a mass proxy, which directly constrains
$dn/(d\ln M\,dV_c)$, where $dn/d\ln M$ is the halo mass function
(eq.~\ref{eqn:massfunction}) and $dV_c$ is the comoving
volume element at the redshift of interest (eq.~\ref{eqn:dvc}).
The mass function at high $M$ is sensitive to the amplitude
of matter fluctuations, and therefore to $G(z)$, though this
information is mixed with that in the cosmology dependence of
the volume element $dV_c \propto D_A^2 H^{-1}$.
Clusters can be identified
in optical/near-IR surveys that find peaks in the galaxy distribution
and measure their richness, in wide-area X-ray surveys that find
extended sources and measure their X-ray luminosity and temperature,
or in Sunyaev-Zel'dovich (SZ) surveys that find localized CMB
decrements and measure their depth.  The critical step in any
cluster cosmology investigation is calibrating the relation
between halo mass and the survey's cluster observable --- richness,
luminosity, temperature, SZ decrement --- so that the mass
function can be inferred from (or constrained by) the distribution
of observables.  We will argue in \S\ref{sec:cl} that the most
reliable route to such calibration is via weak lensing, making
wide-area optical or near-IR imaging a necessary component of
any high-precision cosmic acceleration studies with clusters.

Several of the ``alternative'' methods described in
\S\ref{sec:alternatives} may ultimately play an important role
in pinning down the origin of cosmic acceleration, even given
the high precision expected from Stage IV supernova, BAO, weak
lensing, and cluster surveys.  In some cases, such as 
redshift-space distortions, these alternatives are automatically
enabled by the same surveys conducted for BAO or weak lensing.
In other cases, such as direct measurement of $H_0$, the
required observational programs are different in character.

\vfill\eject

\section{Type Ia Supernovae}
\label{sec:sn}

\subsection{General Principles}
\label{sec:sn_method}

Supernovae (which we will often abbreviate to SN or SNe) are the most
straightforward tool for studying cosmic acceleration, and they are
the tool that directly discovered acceleration in the first place
(\citealt{riess98,perlmutter99}; both using local calibration samples
from the Cal\`an/Tololo survey, \citealt{hamuy96}).
Type Ia supernovae, defined observationally by the absence of
hydrogen and presence of SiII in their early-time spectra
\citep{filippenko97},
are thought to arise from thermonuclear
explosions of white dwarfs, though the evolutionary sequence or
sequences that lead to these explosions remains poorly understood.
The two broad classes of progenitor models are ``single degenerate,''
in which a white dwarf accreting from a binary companion is pushed
over the Chandrasekhar mass limit, and ``double degenerate,''
in which gravitational radiation causes an orbiting pair of
white dwarfs to merge and exceed the Chandrasekhar mass.
The observed supernova population could have contributions
from both channels
(see \citealt{livio00} for a review of Type Ia SN mechanisms).

To a rough approximation, Type Ia SNe are standard candles, with
rms dispersion of approximately 0.4 magnitudes in V-band at peak luminosity
(\citealt{hamuy96,riess96}).
This 0.4-mag scatter can be sharply reduced using an empirical correlation
between peak luminosity and light curve shape (LCS) --- supernovae with
higher peak luminosities decline more slowly after the peak.
This correlation, which we will refer to generically as the
luminosity-LCS relation,
was first quantified
by \cite{phillips93} based on a handful of objects including
the archetypes of low and high
luminosity Ia supernovae, 
SN 1991bg and SN 1991T, respectively.  Also important to the
refinement of distance determinations was the development of corrections for
the correlation between SN color and 
extinction \citep{riess96,tripp98,phillips99}
and $K$-corrections for redshifting effects \citep{kim96,nugent02}.
These were all quickly incorporated into analysis methods such as the
Multicolor Light Curve Shape (MLCS; \citealt{riess96}) technique
used by the High-$z$ Supernova Search \citep{schmidt98} and the
stretch-factor formalism used by the Supernova Cosmology Project
\citep{perlmutter97}.

With these corrections, the dispersion in well measured optical
band peak magnitudes is only $\sim 0.12$ magnitudes
(\citealt{Hicken09, Folatelli11}),
allowing each well measured supernova to provide a luminosity-distance
estimate with $\sim 6\%$ uncertainty.  The diversity of SN Ia light
curves is not fully understood, and peculiar SNe Ia appear to produce
$\sim$ 5\% non-Gaussian tails in the SN Ia distribution (\citealt{li11}).
For the bulk of the population,
the prevailing picture is that
the progenitor explosions produce varying amounts of Ni$^{56}$, whose
radioactivity powers the optical luminosity, and that the
correlation of peak luminosity with light curve shape arises from
radiative transfer effects
(\citealt{hoflich96,kasen07}).
Recent studies suggest that SN Ia are truly standard candles in the
near-IR, with peak luminosities at rest-frame $H$-band ($1.6\micron$)
that have only $\sim 0.1$ magnitude rms dispersion {\it independent}
of light curve shape, and with little sensitivity to uncertain reddening laws
(\citealt{mandel09,mandel11,barone-nugent12}).  
This small dispersion in near-IR peak luminosities
relative to optical is consistent with theoretical expectations
from radiative transfer models \citep{kasen06}.

To measure cosmic expansion with Type Ia SNe, one compares the corrected
peak apparent
magnitudes of distant supernovae to those of local calibrators at 
$0.03 < z < 0.1$, a ``sweet spot'' 
in which distances inferred from redshifts are insensitive to 
peculiar velocities and to the assumed densities of dark matter 
and dark energy.
Since the distances to the local calibrators are usually determined
from Hubble expansion, this method gives the luminosity distance $D_L$
in units of $h^{-1}$ Mpc.
More generally, the SN method yields relative distances in 
different redshift bins, even if one of those bins is not strictly local.
The $D_L(z)$ relation is sensitive to
dark energy through equations~(\ref{eqn:dcomove})
and~(\ref{eqn:friedmann}), and to space curvature through
equations~(\ref{eqn:adistapprox}) and~(\ref{eqn:ldist}).
A measurement of $N$ supernovae in a redshift bin with rms
observational errors $\sigma_{\rm obs}$ in peak magnitudes
yields an estimate of $D_L(z)$ with fractional statistical error
\begin{equation}
\label{eqn:sn_error}
\sigma_{\ln D} = {\left(\sigma_{\rm int}^2 + \sigma_{\rm obs}^2\right)^{1/2}
  \over \left(2\times 1.086 \times \sqrt{N}\right)},
\end{equation}
where $\sigma_{\rm int}$ is the rms intrinsic scatter, the factor 1.086
converts from magnitudes to natural logarithms, and the factor of two
converts from flux uncertainty to distance uncertainty.
As discussed in \S\ref{sec:sn_systematics} below, there are many
possible sources of systematic uncertainty, including flux calibration,
corrections for dust extinction, and possible redshift evolution of
the supernova population.  Of these, dust extinction looks like it may
ultimately be the
most difficult to control at the sub-percent level,
since even a 0.01-mag $E(B-V)$ color excess corresponds
to a 3\% suppression of $V$-band flux.
This consideration provides strong motivation for focusing Stage IV
supernova surveys on rest-frame near-IR photometry, where dust extinction
is a factor of 3 to 8 times smaller compared to the optical and where
the small scatter in peak luminosities may help minimize any evolutionary
effects.

\subsection{The Current State of Play}
\label{sec:sn_current}

Building on the initial discovery of cosmic acceleration, supernova
surveys have been a major area of activity in observational cosmology over the
last decade.  The largest high-redshift ($z \approx 0.4-1.0$)
data sets are those from the ESSENCE survey
(\citealt{WoodVasey07}; Narayan et al., in prep.;
$\sim$200 spectroscopically confirmed Type Ia SNe)
and the CFHT Supernova Legacy Survey
(SNLS; \citealt{Astier06,Conley11,Sullivan11}; 
$\sim$500 spectroscopically confirmed Type Ia SNe in the three-year data 
set SNLS3).  At very high redshifts, \hst\ surveys
\citep{Riess04,riess07,suzuki12}
have yielded $\sim$ 25 Type Ia SNe at $z>1.0$, 
which confirm the
expectation that the universe was decelerating at high redshift
and limit possible systematic effects from evolution of the
supernova population or intergalactic dust extinction.
At intermediate redshifts ($0.1 < z < 0.4$), the SDSS-II supernova survey
\citep{frieman08a,sako08} has discovered and monitored 500 spectroscopically
confirmed Type Ia SNe; only the first-year data set (103 SNe)
has so far been subjected to a full cosmological analysis \citep{kessler09},
but \cite{campbell12} present cosmological results from a sample
of 752 photometrically classified SDSS-II SNe with spectroscopic
host galaxy redshifts, and a joint analysis of the SNLS and SDSS-II
samples is in process (J.\ Frieman, private communication).
Finally, the last five years have also seen major efforts to
expand the sample of local calibrators and improve their measurements,
including rest-frame IR and rest-frame UV photometry
\citep{WoodVasey08,Stritzinger11,Contreras10,Hicken09b}.

The greatest cosmological utility from SNe Ia generally comes from the joint
use of
numerous samples that span a wide range in redshift.  To limit systematic
errors
introduced by combining disparate SN surveys, it is often valuable to
recompile a sample
from these surveys as homogeneously as possible.  This involves applying
consistent criteria for inclusion in the sample,
light curve fitting with a single algorithm, propagation of errors via
covariance matrices, consistent
use of $K$-corrections, and so forth.  While any such  ``survey of surveys'' is
not unique and may not be optimal for a specific application, these
compilations are popular because of 
their ease of use.  Recent examples include the
``Gold'' sample \citep{Riess04,riess07}, the ``Union'' and ``Union2'' samples
\citep{Kowalski08,Amanullah10}, the ``Constitution'' sample \citep{Hicken09b},
and the compilation of local, SDSS-II, SNLS3, and \hst\ supernovae
analyzed by \cite{Conley11}.

\begin{figure}
\begin{centering}
{\includegraphics[width=3.2in]{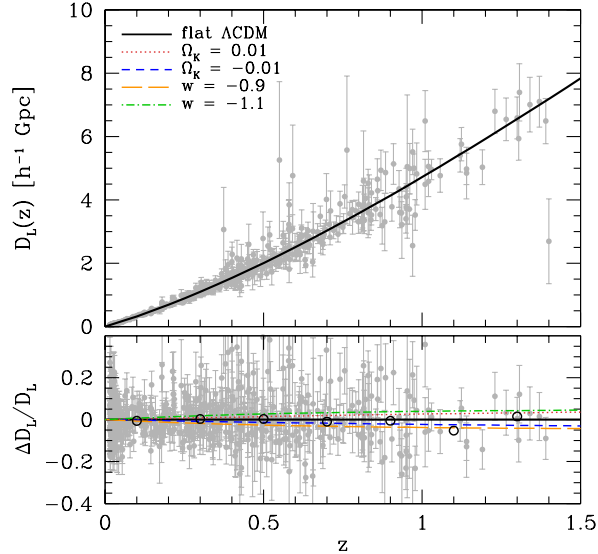}}
\caption{\label{fig:sn_hubble} 
Luminosity distance vs.\ redshift for our fiducial cosmological
model (solid curves), 
superposed on supernova measurements from the Union2
compilation \citep{Amanullah10}.  The lower panel shows
residuals from the fiducial model prediction for the SN data,
with open circles marking
medians of the data in $\Delta z = 0.2$ bins and broken curves showing 
the CMB-normalized variant models described in Table~\ref{tbl:models}.
Note that these distances are in $h^{-1}\,{\rm Gpc}$ units.
}
\end{centering}
\end{figure}

\begin{figure}
\begin{centering}
{\includegraphics[width=2.0in]{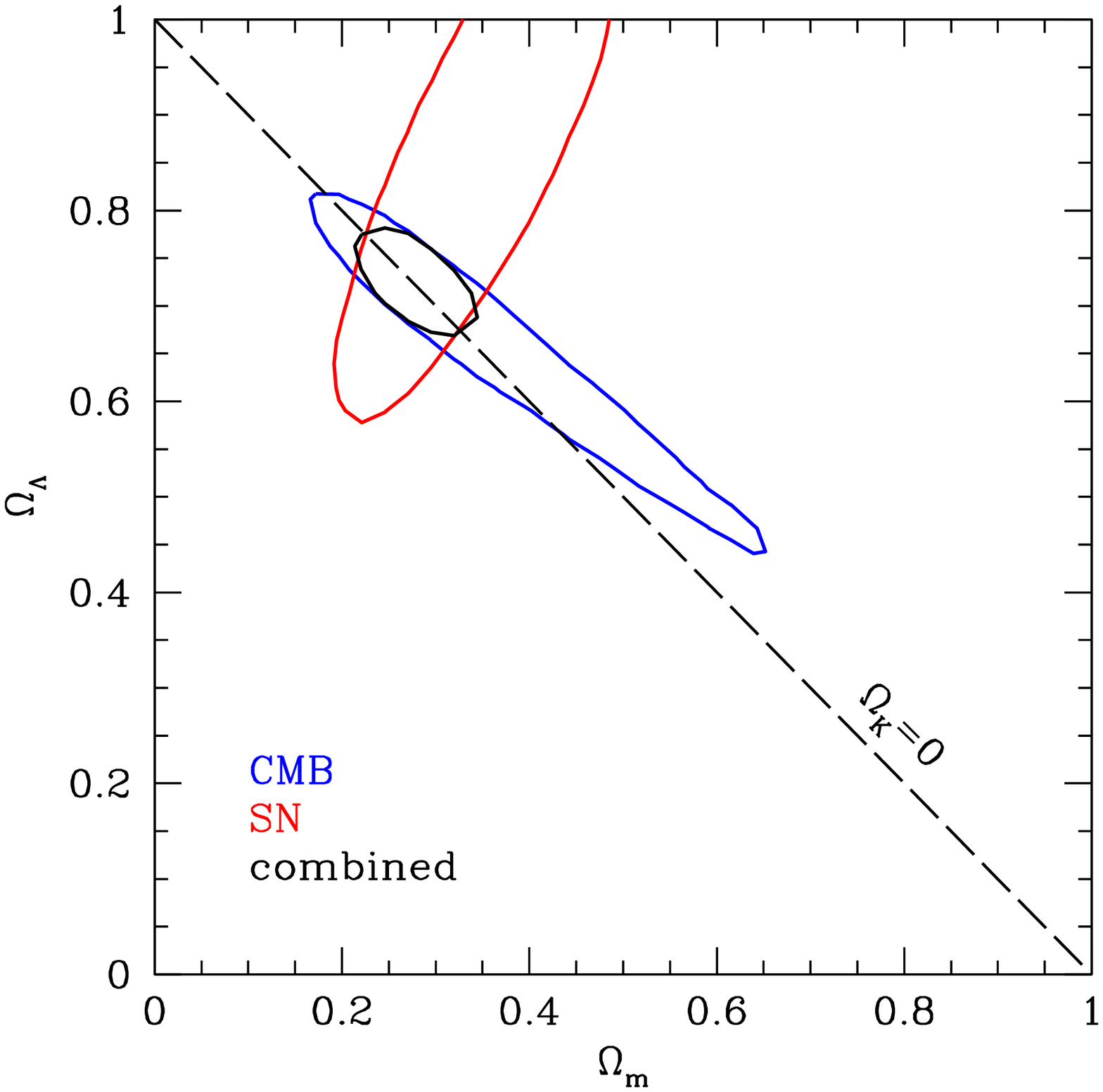}\hfill 
 \includegraphics[width=2.0in]{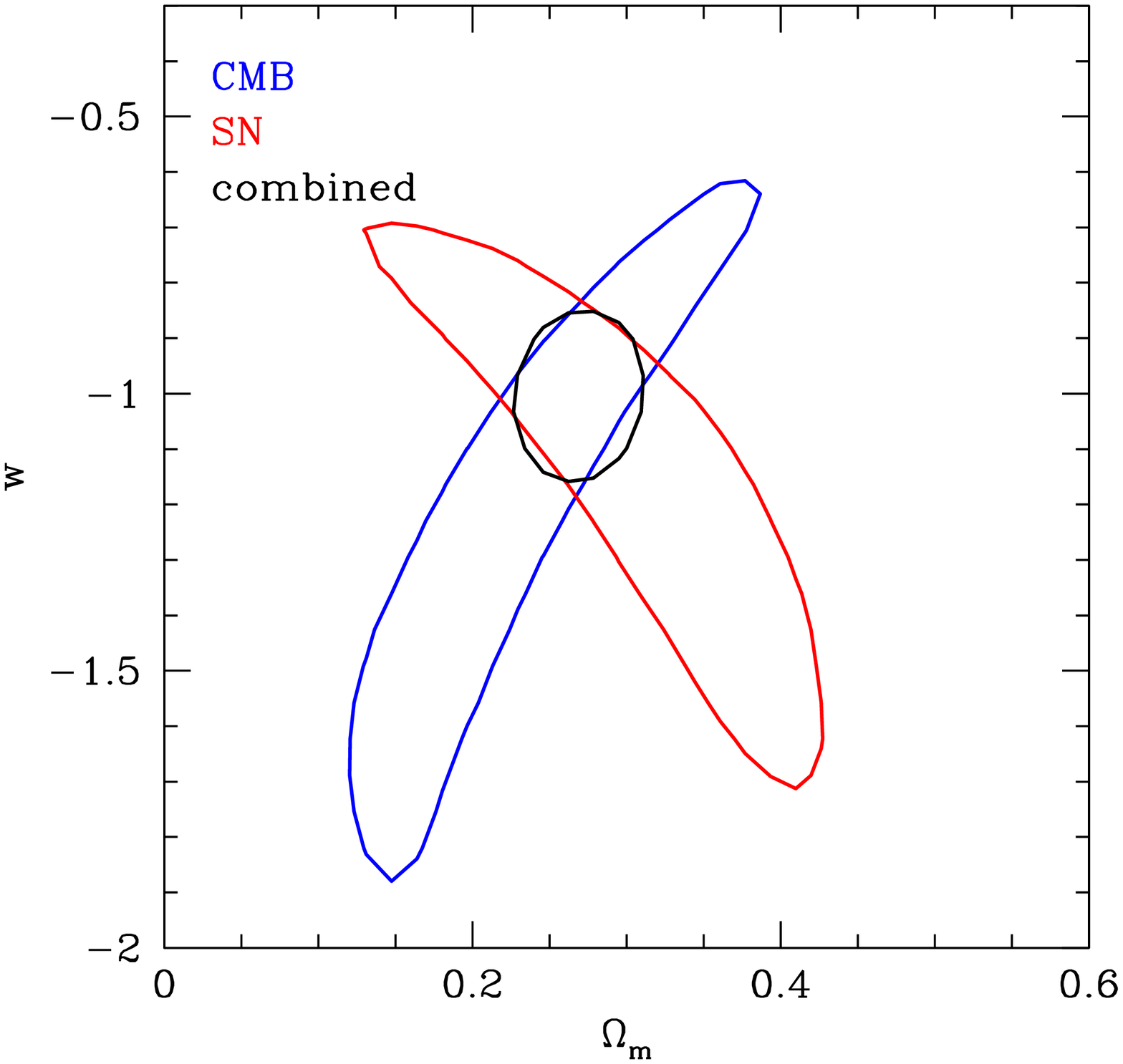}\hfill 
 \includegraphics[width=2.0in]{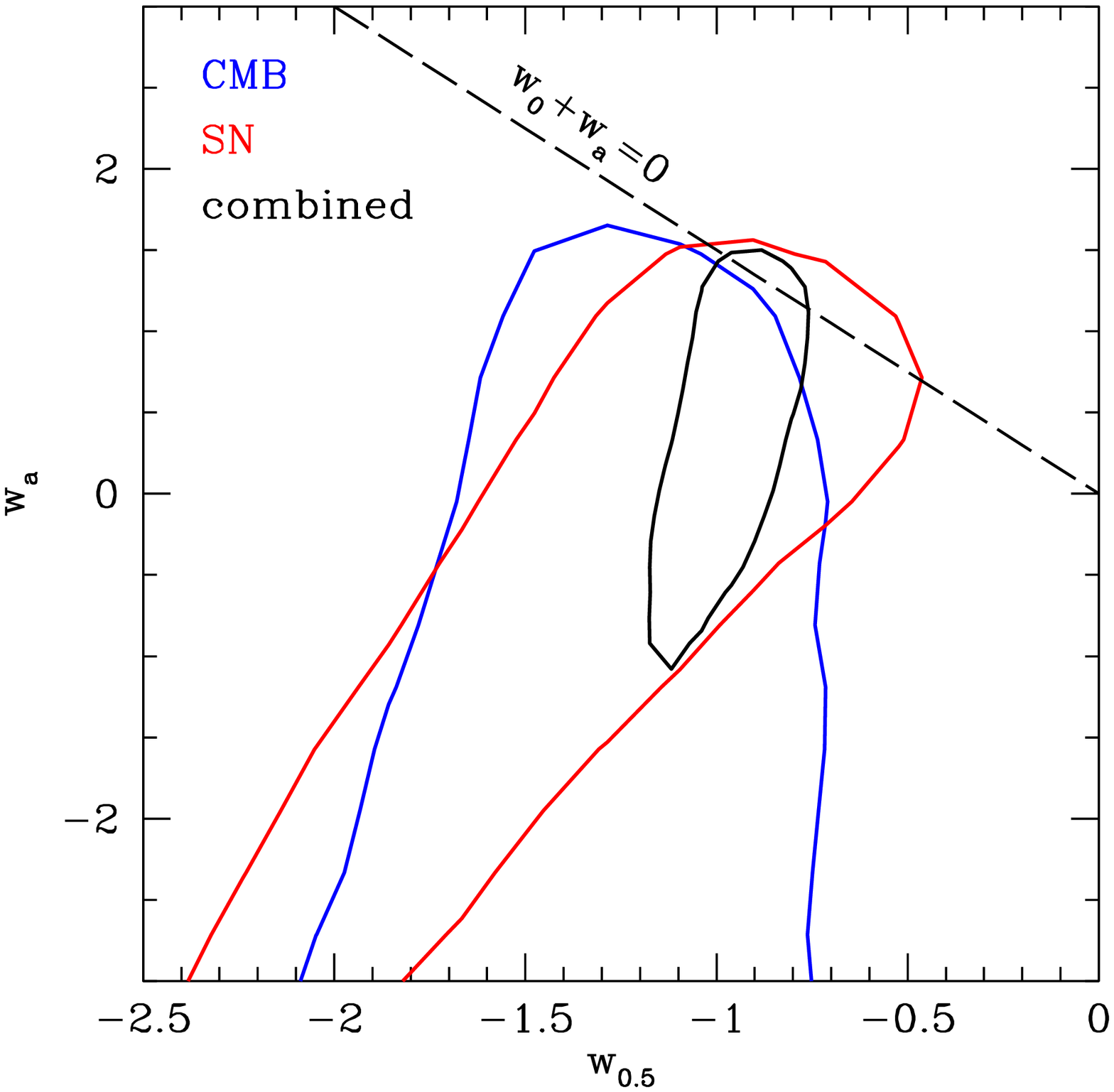}\hfill 
}
\end{centering}
\caption{\label{fig:sn_constraints} 
Constraints from WMAP7 CMB data, Union2 SN data, and the combination
of the two, in (a) the $(\om,\ol)$ plane assuming $w=-1$,
(b) the $(\om,w)$ plane assuming $\ok=0$, and
(c) the $(w_{0.5},w_a)$ plane assuming $\ok=0$, where 
$w_{0.5}$ is the value of $w$ at $z=0.5$.
Contours show 68\% confidence intervals.
In contrast to panels (a) and (b), the combined contour in
(c) is tighter than one would guess from the overlap of the
individual contours because the combined data set breaks
degeneracies among other parameters that are marginalized
over when inferring $w_{0.5}$ and $w_a$.
}
\end{figure}

Figure~\ref{fig:sn_hubble} plots luminosity distance measurements
from the Union2 compilation over the model predictions shown
previously in Figure~\ref{fig:hzdz} (multiplied by $1+z$ to convert
comoving angular diameter distance to luminosity distance).
The data are in good agreement with the fiducial cosmological
model, and the parameter changes in the bottom panel
($\ok=\pm 0.01$, $1+w=\pm 0.1$) are at the border of
detectability.  (Recall that other parameters are adjusted to
reproduce the CMB anisotropy of the fiducial model; see
Table~\ref{tbl:models}.)

Figure~\ref{fig:sn_constraints} illustrates model constraints
from the Union2 supernova data and WMAP7 CMB data, which we
have computed using CosmoMC \citep{lewis02}.  
We use the Union2 covariance matrix that includes
correlated systematic error contributions.
Panel (a) shows the ($\om,\ol$) plane assuming $w=-1$.
CMB and SN constraints are highly complementary in this plane because the 
former are most sensitive to the total energy density 
($\om+\ol$) and the latter to the difference between the densities
of ``attractive'' matter and ``repulsive'' dark energy.  Together the two
data sets yield tight constraints in this space,
$\om = 0.282 \pm 0.037$, $\ol = 0.723 \pm 0.030$, 
consistent with a flat universe.
Panel (b) shows the ($\om,w$) plane, where we now {\it assume}
spatial flatness and a constant value of $w$.  
Here again the SN and CMB data are highly complementary, yielding
a tight combined constraint
$\om = 0.270 \pm 0.023$, $w = - 1.007 \pm 0.081$, 
consistent with a cosmological constant.
Panel (c) shows the ($w_{0.5},w_a$) plane, where we have adopted
the 2-parameter dark energy model of equation~(\ref{eqn:w0wa});
$w_{0.5}$ is the value of $w$ at $z=0.5$, which is much better
determined than the value of $w_0$ and only weakly correlated with $w_a$.
Here we have assumed spatial flatness and marginalized over uncertainty
in $\om$.  CMB and SN data provide only weak constraints individually
in this model space, but the combination still provides a good 
constraint on $w_{0.5}$, with the error on $w_{0.5} = -1.008 \pm 0.132$ 
only degraded by $\sim 50\%$ compared to panel (b).  
Constraints on $w_a$, on the other hand, are very weak.
The $w$ and $w_{0.5}$ constraints in panels (b) and (c) would
degrade substantially if we allowed non-zero $\ok$; with this
level of flexibility, one must bring in additional data to
get useful constraints.  However, an $H_0$ or BAO constraint at the level
of current measurements is sufficient to remove most of the sensitivity
to $\ok$ \citep{mortonson10}.

Plots and constraints similar to 
Figures~\ref{fig:sn_hubble} and~\ref{fig:sn_constraints} appear
in many of the papers cited above.  The most up-to-date analysis
is that of \cite{Conley11}, who find 
$w=-0.91^{+0.16}_{-0.20}{\rm (stat)}^{+0.07}_{-0.14}{\rm (sys)}$
for SNe alone, assuming a flat universe with constant $w$ and
marginalizing over $\om$.
Combining this measurement with other data sets,
\cite{Sullivan11} find $w = -1.016^{+0.077}_{-0.079}$ in
combination with 7-year \wmap\ CMB constraints (similar to the
value and error bar quoted above), and
$w= -1.061^{+0.069}_{-0.068}$ after adding BAO and $H_0$ measurements.

There are several indications that current SN cosmology studies
are limited by systematic uncertainties associated with the linked
issues of dust extinction, SN colors, and photometric calibration.
In any cosmological analysis, one uses the color of a supernova
relative to a template expectation (derived from a training set)
to infer, and correct for, a correlation
between color and apparent magnitude arising from dust and/or 
intrinsic color variations.  
In the analysis of \cite{WoodVasey07}, different priors 
about host galaxy extinction change the inferred value of $w$
by amounts comparable to the statistical error.
When the ratio of extinction to reddening
is treated as a free parameter in the cosmological fits, the derived
values are typically quite far from those measured for Galactic
interstellar dust, e.g., $R_V \equiv A_V/E(B-V) = 1.5-2.5$
(\citealt{Hicken09,kessler09,Sullivan11})
instead of the mean $R_V=3.1$ found in the diffuse interstellar
medium of the Milky Way \citep{cardelli89}.
This difference could be a reflection of different kinds of dust along the
line of sight to the supernova (e.g., circumstellar dust), but it 
could also arise from intrinsic color differences among SNe Ia 
with similar light curve shapes, which would reduce the inferred $R_V$
if they are assumed to arise from reddening.
Supporting the latter idea, the distribution of SN colors shows
little dependence on host galaxy properties \citep{kessler09,Sullivan10},
while such dependence might be expected if the color distribution
is strongly affected by dust.  \cite{chotard11}, using spectroscopic
indicators of luminosity in nearby SNe, infer an extinction law
with $R_V = 2.8 \pm 0.3$, consistent with the Galactic value.
%The observed relation is likely to result from a mix of factors, and the
%demographics of this mix may shift with redshift and sample selection.
%These issues remain to be sorted out.

One of the main surprises in the first-year analysis of the SDSS-II
Supernova Survey \citep{kessler09} was the realization that the
two main algorithms developed by other groups
for global fitting of SN light curves and cosmological parameters ---
MLCS2k2 (\citealt{Jha07}) and SALT2 (\citealt{guy06}) --- initially gave
statistically
inconsistent cosmological results ($w=-0.76 \pm 0.07$ vs.\ $w=-0.96
\pm 0.06$,
quoting only the statistical errors) when applied to the same data
sets, a discrepancy that persisted even if the SDSS-II data themselves
were omitted from the fits.
\cite{kessler09} traced this discrepancy to two factors, one related to
calibration data
and the other to the treatment of SN colors.  For the calibration data,
ultraviolet flux measurements
in the local sample from the $U$-band
appear inconsistent with those from the $g$-band at only moderate redshift
and suggest a problem with the (observed frame) $U$-band 
calibration.\footnote{\cite{Conley11} provide further evidence for
an error in the local $U$-band calibration, and they omit these
data from their cosmological analysis.}
This problem translates into a difference between fitters because one
is trained with $U$-band
data and the other is not.
A more subtle difference arises from the determination of the correction to SN
brightness  from color measurements, specifically whether the correlation can
be assumed to be independent of redshift and survey and whether 
changes in color
are due solely to extinction.
While these systematic uncertainties will certainly be reduced by
larger multi-wavelength data sets and improved analysis methods,
the experience from these recent studies argues strongly for using
rest-frame IR photometry in precision cosmological studies to circumvent
uncertainties related to extinction.

\subsection{Observational Considerations}
\label{sec:sn_obs}

There are several steps to a supernova cosmology campaign:
discovery, monitoring, spectroscopic confirmation, and
calibration against low redshift samples.  In large area surveys, discovery and
monitoring are usually done together, through repeated imaging
of a large field of view in multiple bands.
A variety of image-differencing techniques can be used to identify
SNe (distinguished from other variable objects by their light curves)
and measure their magnitudes vs.\ time.
As a rule of thumb, a minimum rest-frame cadence of one observation per
$\sim$ 5 days\footnote{Observed-frame time intervals are larger by $1+z$.}
is needed
to get adequate measurements of light curve shapes and normalizations,
such that statistical errors are dominated by the intrinsic
dispersion of SN luminosities and not by observational errors.
The required cadence may be somewhat lower in the rest-frame IR,
where the dependence on light curve shape is weaker, but one must still
have enough data points to determine peak luminosity accurately.
At least two bands are needed to measure SN colors and thereby
infer dust extinction, though more are better,
and multiple colors may prove critical
to distinguishing different forms of extinction 
(interstellar, circumstellar, and intergalactic) 
from each other and from intrinsic color differences.

\begin{figure}
\begin{centering}
{\includegraphics[width=3.2in]{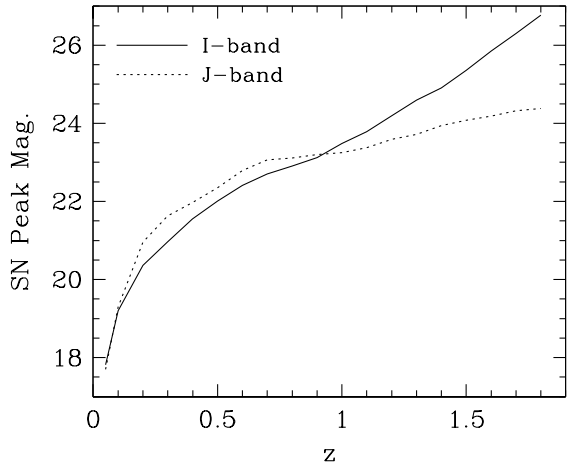}}
\caption{\label{fig:sn_magnitude} 
Peak apparent magnitude of a typical Type Ia supernova
as a function of redshift in observed-frame $I$-band
(solid) or $J$-band (dotted), from Table~7 of
\cite{tonry03}.  The $z>1.1$ portion of the $I$-band curve and
$z<0.4$ portion of the $J$-band curve rely on extrapolation of the 
template systems' spectral energy distributions beyond the observed range.
Magnitudes are on the Vega system.
}
\end{centering}
\end{figure}

Figure~\ref{fig:sn_magnitude}, based on Table 7 of \cite{tonry03},
plots the peak apparent magnitude of a typical Type Ia supernova
vs.\ redshift in observed frame $I$ and $J$ band.
As a rough rule of thumb, a survey with periodic and uniform exposures
targeting supernovae at
a given redshift should measure to a signal-to-noise ratio of  $\sim$ 15
at peak, so that it still
usefully measures the SN before or after peak when it is 1.5 magnitudes
fainter.  This depth ensures that
incompleteness for supernovae below the median luminosity does
not bias the results and that photometric errors do not
dominate over intrinsic scatter in cosmological analysis.
Ground-based surveys designed to observe SNe Ia to $z<0.8$ will typically
find $\sim$ 10 SNe Ia per square degree per month.

After discovering SNe, one must determine their type and redshift.
The most reliable approach is to
obtain their spectra to cross-correlate their spectral features with known
templates.
Spectral resolution $R\sim 250$ and S/N$\sim 5$ per resolution element
are adequate for these purposes,
but even at this level spectroscopic follow-up
is typically the most resource intensive step of a supernova campaign.
For the same telescope aperture, an epoch of spectroscopy requires
an order of magnitude more time than an epoch of photometry, and one
generally loses the parallelism afforded by photometric monitoring
with a large camera (which has several SNe per field of view at
a given time).
Spectroscopic follow-up of the SNLS3 sample, for example, used
more than 1600 hours of $8-10$m telescope time
(M.\ Sullivan, private communication).

In principle, photometric redshifts
can be used in place of spectroscopic redshifts, and if they
are accurate to a fractional distance error $\Delta D/D < 10\%$
they lead to only moderate degradation in statistical accuracy.
However, given the degeneracies among redshift, SN color, and dust
extinction, and the increased ambition of SN surveys to control systematics,
we are skeptical that cosmological SN surveys can
achieve the desired accuracy using only broad-band photometric
monitoring and spectroscopic follow-up of a small fraction of
the sample.  An intermediate approach that may work would be to 
%identify the significant 
measure the
cross-correlation of a supernova
SED with the SN Ia spectral features
using custom-designed optical filters that are matched to SN
spectroscopic features at different redshifts
(\citealt{scolnic10}).  It also may be possible to make use of subsamples of
SNe found in passive (non-star-forming) galaxies, which should host
only Type Ia SNe and which allow more accurate photometric redshifts
from host galaxies.  For type identification, one can also check for 
a second peak in the rest frame infrared light curve,
a morphological feature that is unique to SNe Ia.

Another intermediate approach is to obtain eventual spectroscopic
observations of all host galaxies in the cosmological analysis
sample but not attempt real-time spectroscopy of all candidate Type Ia
supernovae.  This scheme still yields precise redshifts, and it provides
host galaxy data that can be used to measure and remove
correlations between supernova and host galaxy 
properties (see \S\ref{sec:sn_systematics}).
While it still requires one faint-object spectrum per supernova,
the scheduling demands are much more flexible.
One can also apply data quality and other selection cuts before the 
spectroscopic observations to reduce the total number of spectra required,
though one must be careful not to let biases creep in at
this stage.
With good photometric monitoring and with subsequent spectroscopic redshifts
of apparent hosts, \cite{Kessler10} find that they can identify Type Ia SNe
with 70\% to 90\% confidence from the LCS and color alone,
and \cite{bernstein11des} forecast Type Ia purity as high as 98\%
for DES-like photometric observations.
A moderate amount of real-time supernova spectroscopy may then
suffice to assess efficiency and biases.  
The recent SDSS-II analysis by \cite{campbell12} puts this
approach into practice, illustrating its promise and its challenges.

Given the photometric and spectroscopic measurements for a selected
set of supernovae, one must fit the data set to infer cosmological
parameters.  Many of the algorithms in current use are descendants of the
Multicolor Light Curve Shape (MLCS; \citealt{riess96}) or
Spectral Adaptive Light Curve Template (SALT; \citealt{guy05})
methods.  In current implementations, MLCS fitters are ``trained''
on local supernovae to determine the relationships between
multi-band light curve shapes and peak absolute magnitudes,
and these relationships are applied to distant supernovae to
measure $D_L(z)$.  SALT-style fitters (which include the 
SiFTO [\citealt{conley08}] algorithm applied to SNLS3)
instead apply a global, simultaneous fit of parameters describing
cosmology and the relationship between supernova light curves
and absolute magnitude.  Of greater practical import, however,
is the different treatment of supernova colors in the two methods.
MLCS fitters attribute color differences at fixed peak luminosity
to dust reddening, and they adopt an explicit prior for the
distribution of reddening values.  SALT fitters allow scatter in
intrinsic colors at fixed peak luminosity and do not attempt
to separate intrinsic variations from dust reddening.
In reality, there certainly are intrinsic color variations
at some level, but there is also useful information in the fact
that dust reddening exists and has specific properties, in particular
that it cannot be negative.  An optimal approach should therefore
allow for both effects.  Bayesian fitting methods
(e.g., \citealt{mandel09,mandel11,march11a}) can in principle
incorporate a wide variety of parameterized relationships with
explicit priors, including dependences on redshift or host 
galaxy parameters, which are then marginalized over in cosmological
fits.  At the level of precision of current SN samples, 
the differences in fitting methods do matter
(e.g., \citealt{kessler09}), so this remains an area of active research.
Fortunately, the growing samples of well observed local and distant
SNe provide increasingly powerful data to guide this development.

The detailed spectra of SNe could potentially improve their luminosity
and/or color calibration relative to photometric light curves alone.
For example \cite{foley11} find a correlation between intrinsic
color and the ejecta velocity inferred from the line width
(see also \citealt{blondin12,foley12}). However, \cite{silverman12},
considering a variety of spectral indicators, find only marginal
evidence for a diagnostic that improves Hubble residuals, and
\cite{walker11} find similarly ambiguous results.
Given the substantial observing time required to measure good
spectroscopic diagnostics for high redshift SNe,
modest reductions in scatter are unlikely to win over simply
observing more supernovae.  However, spectral diagnostics merit
continued investigation to see whether matching spectral properties
between high and low redshift SNe can reduce susceptibility to
evolutionary systematics.

\subsection{Systematic Uncertainties and Strategies for Amelioration}
\label{sec:sn_systematics}

The largest current supernova surveys have $\sim 500$ Type Ia supernovae.
Future surveys hope to discover and monitor thousands of supernovae,
sufficient to yield statistical errors of 0.01 mag or smaller in narrow
redshift bins with $\Delta z \sim 0.1-0.2$.  Realizing the statistical
power of such surveys will require eliminating or limiting
several distinct sources of systematic error.  These include flux
calibration errors across a wide range of flux and redshift,
the systematics associated with SN colors and dust extinction,
the possible evolution of the supernova population with redshift,
and gravitational lensing.  
We discuss each of these issues in turn.\footnote{For detailed discussions
of systematics in the context of specific contemporary data sets, see, e.g.,
\cite{WoodVasey07}, \cite{Kessler10} and \cite{Conley11}.}

The leverage of SN studies comes from comparing SNe over a wide span
of redshift and thus an enormous range of flux; for example, the typical
peak $I$-band magnitude at $z=0.8$ is 23 mag
while the median peak $B$-band magnitude of the local calibrator sample
used in many analyses is 17 mag, implying a ratio of 250 in flux.
Maintaining sub-percent accuracy in relative flux calibration over such
a range would be challenging under any circumstances, and for SN
surveys it is complicated by the fact that
(a) local and distant SNe are usually observed with different telescopes
equipped with different filters,
(b) a given observed-frame filter intercepts a different portion of
the SN rest-frame spectral energy distribution (SED) at each redshift, and
(c) supernova SEDs are very different from those of the standard stars used
for flux calibration in most of astronomy.  
\cite{Conley11} identify calibration as the dominant systematic
in SNLS3, the only systematic in their analysis that makes
a major contribution to their total error budget.
Flux calibration uncertainties
can be reduced
by carefully designing photometric SN surveys with specialized
hardware (e.g., tunable lasers, NIST photodiodes and calibration sources;
\citealt{stubbs06}) to measure the system throughput {\it in situ} and by
choosing filter systems that provide a good
match in rest-frame SED sampling between low- and high-redshift samples.
The ACCESS rocket program should improve flux calibration with
sub-orbital flights that compare NIST photodiodes to calibration stars
(\citealt{Kaiser10}).
``Self-calibration'' that marginalizes over flux-calibration uncertainty
can further reduce this systematic error \citep{kim06},
but at the price of increasing statistical error.

As already noted in \S\ref{sec:sn_current}, uncertainties in dust
extinction, linked to uncertainties in intrinsic SN colors and in
photometric calibration, are already important systematics in SN
studies of cosmic acceleration.  These uncertainties can likely be reduced
with detailed, well calibrated, multi-wavelength observations of large
numbers of low redshift SNe, which can characterize the
separate dependence of SN colors on luminosity, light curve shape,
and time since explosion, and provide constraints on dust extinction 
laws that are
isolated from cosmological inferences.  The final analyses of data
from the SDSS-II supernova survey \citep{frieman08a} and the low-redshift
portion of the Carnegie Supernova Project \citep{hamuy06} should
allow advances on this front.  Analysis techniques that
eliminate the most highly reddened SNe can also
reduce extinction systematics if they can be applied in a way that
does not introduce selection biases; as an extreme example, one can
employ only SNe in early-type galactic hosts, which have low
amounts of interstellar dust.  Perhaps the most important strategy
for reducing extinction systematics is to work as far as possible
to red/near-IR rest-frame wavelengths, where extinction is low
compared to blue/visual wavelengths.  Most ground-based SN cosmology studies
to date
work at rest-frame $B$ ($0.4-0.5\micron$) or $V$ ($0.5-0.6\micron$)
wavelengths, which transform to observed-frame $I$-band
($0.7-0.9\micron$) at $z \approx 0.5-0.8$.  The high-redshift
portion of the Carnegie Supernova Project \citep{freedman09}
produced a SN Hubble diagram to $z\approx 0.7$ in {\it rest-frame}
$I$-band, where systematic errors due to uncertainty in the reddening laws
are roughly half that at $V$-band.
\cite{mandel09} find that the intrinsic dispersion of peak luminosities
is only $\sim 0.11$ mag at rest-frame $H$-band ($1.5-1.7\micron$), where
systematics due to extinction are only $\sim 1/6$ that at $V$-band.  However,
obtaining
rest-frame near-IR photometry for high-redshift supernovae requires
space observations due to the high backgrounds seen from the ground
(\S\ref{sec:sn_space}).

Locally observed SNe span a wide range in the age, metallicity,
and current star formation rate (SFR) of their host stellar populations.
This breadth of host conditions provides a laboratory for the investigation
of the evolution of SNe Ia as distance indicators.  Recently such an effect
was found and calibrated in the form of a modest, 0.03 mag dex$^{-1}$
relationship between host galaxy stellar mass (a likely tracer of metallicity)
and calibrated SN Ia magnitude (\citealt{Kelly10,Lampeitl10,Sullivan10};
see \citeauthor{Hicken09} [\citeyear{Hicken09}]
for an analysis with host morphology and \citeauthor{hayden12} 
[\citeyear{hayden12}] for an analysis that incorporates star formation
rate in an attempt to isolate metallicity).
At the level of precision enabled by current surveys, it is necessary
to correct for this effect \citep{Conley11}, 
but the uncertainty in the correction is not a limiting systematic.

Constraining evolutionary effects to a tenth of
$\sigma_{\rm int}$ ($\sim 0.01$ mag) or better is a challenge.
For example,
if there are two populations of Type Ia progenitors (e.g., single
and double degenerates) that have slightly shifted luminosity-LCS
relations, then evolution in the population ratio could produce
evolution in the mean relation at a fraction of $\sigma_{\rm int}$
(see, e.g., \citealt{sarkar08a}).
A strategy for limiting evolution systematics is to break
the SN sample into subsets defined by spectral features, light curve
shapes, or host properties and check for consistency of cosmological
results, since evolution is unlikely to affect all populations in
the same way.  A complementary path \citep{riess06} is to observe
supernovae at $z>2$, where predicted fluxes relative to
low-redshift samples are generally insensitive to dark
energy parameters; discrepancies would be an indication of evolutionary
effects or of unconventional dark energy models that could be
tested by other probes.  Finally, we note that any evolutionary
corrections may be weaker in the near-IR, both because of the narrower
range of luminosities and because of the weaker sensitivity to
metal lines (which may itself contribute to the narrower luminosity
range) and reddening laws.

Gravitational lensing by intervening large scale structure introduces
scatter in observed SN fluxes, at a level of $\sim 0.05$ magnitudes
for sources at $z=1$
(e.g., \citealt{frieman96,wang99}).
Flux conservation guarantees that the {\it mean} flux of the SN
population does not change.  However, some care is required
to ensure that selection effects or weighting schemes do not bias
results at the 0.01-mag level, especially as the magnification
distribution is highly non-Gaussian
(see, e.g., \citealt{sarkar08b}).
Since lensing effects are small and calculable, they are unlikely
to become a limiting systematic even for the most ambitious future
surveys.
Analyses that average fluxes of SNe in redshift bins 
or model the full flux distribution can 
minimize lensing systematics and may reduce some other systematic
effects as well \citep{wang00,amendola10,wang12}.

If rest-frame near-IR photometry can be obtained for large supernova
samples, we anticipate that flux calibration uncertainties will ultimately
set the floor on systematics. 
A detailed recent investigation of the \hst\ WFC3-IR system 
implies a limiting calibration uncertainty of $\sim 0.02$ mag
\citep{Riess12}.  A future mission designed with IR supernova
photometry as a key goal could presumably do better, so
$0.005-0.02$ mag seems a plausible bracket for calibration-limited
systematics.

\subsection{Space vs. Ground}
\label{sec:sn_space}

Space observations offer several key advantages for precision supernova
cosmology, a point emphasized early on by the {\it SNAP}
(SuperNova Acceleration Probe) collaboration
(e.g., \citealt{Aldering03}).
The first is the sharp and stable point-spread function (PSF) achievable
from space, which greatly increases sensitivity to faint, variable
point sources and the precision and accuracy of point-source photometry,
especially in the presence of a host galaxy background.
Adaptive optics can produce a sharp PSF from the ground, but it is
not likely to deliver photometry with 1\% precision and an image stable
enough to allow host subtraction
at random positions on the sky away from bright guide stars.
The second advantage is the greater accuracy and precision of flux calibration
achievable from space, with no time-variable atmospheric conditions
and (for a well chosen orbit) minimal variations in the telescope
environment.  The third is the vastly lower sky background in the
near-IR.  Typical sky backgrounds for ground-based observations are
16, 14, and 13  mag arcsec$^{-2}$ at $J$, $H$, and $K$
(Vega), while in space they are 6 to 8 mags fainter, limited by the zodiacal
light.

It is the last of these advantages that we regard as critical ---
no improvements in ground-based technology or observing strategy
will ever remove the IR sky background.  
We have already emphasized the key role of rest-frame near-IR
photometry in reducing systematics associated with dust extinction,
and possibly with evolution.  Obtaining rest-frame $J$-band ($1.2\micron$)
photometry of SNe at $z=0.8$ requires imaging at $\lambda = 2\micron$.
A 1.3-m space telescope --- the (unobstructed) 
aperture proposed for \wfirst\ ---
can make a S/N=15 measurement at the peak magnitude of a median
$z=0.8$ supernova at this wavelength in about 20 minutes.
A ground-based 4-m telescope with $0.8"$ seeing and a typical
IR sky background would require multiple nights, and even then
the accuracy of photometry would be compromised by variable sky background.

A space-based near-IR telescope also offers the option of discovering
and monitoring SNe at substantially higher redshifts, while working
at shorter rest-frame wavelengths.  However, for the reasons
discussed quantitatively in \S\ref{sec:bao} and \S\ref{sec:forecast},
we think that the most important role for a mission like \wfirst\ 
in SN studies is to provide the highest achievable accuracy and
precision at $z \leq 0.8$, as part of a combined dark energy program
that also includes ambitious BAO and weak lensing surveys.
At low redshifts, SNe can achieve a measurement precision
unmatched by other methods, but at higher redshifts they cannot
match the dark energy sensitivity of large BAO surveys unless they
can push statistical and systematic errors well below 0.01 mag
(see Table~\ref{tbl:baoforecast} in \S\ref{sec:forecasting}).
%Campaigns to detect and monitor
%high-redshift ($z>0.8$) SNe would be justified if they yield more
%leverage on dark energy parameters than other applications of the
%same observing time.
%(i.e., to weak lensing or BAO surveys).
The value of a high-$z$ SN program depends critically on whether
the systematics at high-$z$ are uncorrelated with those at low-$z$, in which
case the distant SNe provide new information even after the low-$z$ 
program has saturated its systematics limit, or whether the limiting
systematics are correlated across the full redshift range.
We discuss this point more quantitatively in
\S\ref{sec:results_simplewz} below.
For a given observing allocation, the maximally 
efficient use of \wfirst\  SN time may be in a combined
ground-space program, with ground-based photometry (in rest-frame
optical) providing high-cadence light-curve sampling and color
measurements and lower cadence space observations providing the
critical, well calibrated, dust-insensitive photometry used for
the SN distance determinations.  

\subsection{Prospects}
\label{sec:sn_prospects}

The next year or two should see the publication of final results from
the SDSS-II supernova survey, the five-year SNLS sample, and ESSENCE.
The measurements from these large
surveys should substantially reduce the statistical errors in the
SN Hubble diagram.  Perhaps more importantly, they should yield
significant reductions of systematic errors because of their
high sampling cadence, wide wavelength range, and greater attention to
photometric calibration.
Large campaigns to discover and monitor local supernovae
(e.g., PTF, LOSS, CSP, SN Factory) should also yield better
understanding of
potential systematics, as well as better local calibration.
A new \hst\ survey by the Higher-$z$ Team using WFC3 will find more high
redshift ($z>1.5$) SNe,
which provide additional leverage on the Hubble diagram and
constraints on evolution.

The largest new projects on the near horizon are the SN surveys
of PS1 (now underway) and DES (beginning observations in late 2012).
\cite{bernstein11des} discuss the DES strategy in some detail
and forecast discovery of up to 4000 Type Ia SNe out to 
redshift $z=1.2$.  For spectroscopic follow-up, DES aims 
to observe $\sim 10-20\%$ of their high-$z$ supernovae
but obtain nearly complete spectroscopic host galaxy redshifts for 
their cosmological sample.
A similarly detailed description of the PS1 strategy
is not yet available, but in principle PS1 should also be able
to discover thousands of Type Ia SNe.
In purely statistical terms, a sample of 2000 SNe out to $z=0.8$ can achieve
errors of 0.007 mag in redshift bins of $\Delta z = 0.2$, so both
PS1 and DES will almost certainly be limited by systematic rather
than statistical errors.

Looking further ahead, LSST is expected to yield samples of tens
or even hundreds of thousands of SNe \citep{lsst09}.
These photometric samples will certainly swamp spectroscopic
follow-up capabilities, and the LSST surveys will again be
systematics limited, though the enormous sample size
(allowing cross-checks and focus on the most favorable
subsamples) and the high-cadence monitoring with 
high photometric precision across the optical
spectrum should reduce systematics below those of PS1 and DES.
Finally, if \wfirst\ is completed and launched as per the Astro2010
recommendations, the access to the rest-frame near-IR should yield an
unmatchable advantage for SN cosmology and
the best achievable results in SN dark energy studies.

\vfill\eject

\section{Baryon Acoustic Oscillations}
\label{sec:bao}

\subsection{General Principles}
\label{sec:bao_method}

The baryon acoustic oscillation method relies on the imprint left by
sound waves in the early universe to provide a feature of known size
in the late-time clustering of matter and galaxies.
By measuring this
acoustic scale at a variety of redshifts, one can infer $D_A(z)$ and
$H(z)$.
The acoustic length scale can be computed as
the comoving distance that the sound waves could travel
from the Big Bang until recombination at $z=z_*$ \citep[see descriptions
by][]{hu96,eisenstein98}.  This is a simple integral
\begin{equation} r_s = \int_0^{t_*} {c_s(t)\over a(t)}  dt =
\int_{z_*}^\infty {c_s(z)\over H(z)}
dz. \label{eq:acoustic_scale}\end{equation}
The behavior of $H(z)$ at $z>z_*$
depends on the ratio of the matter density to radiation
density; in simple cosmologies, the radiation sector (photons and neutrinos)
is fixed and the ratio
is proportional to $\Omega_m h^2$.  The sound speed depends on the ratio
of radiation pressure to the energy density of the baryon-photon fluid,
determined by the baryon-to-photon ratio, which is proportional to
$\Omega_b h^2$.  Both the matter-to-radiation ratio and the baryon-to-photon
ratio are well measured by the relative heights of the acoustic peaks
in the CMB anisotropy power spectrum.  Analyses of \wmap\ data in the
usual $\Lambda$CDM cosmological models gives a 1.1\% inference of the acoustic
scale $r_s$ \citep{jarosik11}; \planck\ is expected to shrink this 
error bar to 0.25\%.  Note that the acoustic scale is determined in
absolute units, Mpc not $\hmpc$.

The acoustic scale is large, about 150 Mpc comoving, because
primordial sound waves travel at relativistic
speed, maxing out at $c/\sqrt{3}$ at early times when the
baryon density is negligible compared to radiation density.
The large size of
the acoustic scale protects this clustering feature from non-linear
structure formation in the low-redshift universe.  
As discussed below, both cosmological
perturbation theory and numerical simulations 
argue that the scale of the acoustic feature is
stable to better than 1\% accuracy, making it an excellent standard ruler.
The BAO method measures the cosmic distance scale using this
ruler.  Separations along the line of sight
correspond to differences in redshift that depend on the Hubble parameter
$H(z)r_s$.  Separations transverse to the line of sight correspond to
differences in angle that depend on the angular diameter distance
$D_A(z)/r_s$.

The challenge of the BAO method is primarily statistical: because this
is a weak signal at a large scale, one needs to map enormous volumes
of the universe to detect the BAO and obtain a precise distance
measurement.  Galaxy redshift surveys allow us to make these large
three-dimensional maps of the universe, although we will discuss other
methods as well.

At low redshift ($z\lesssim0.5$), the BAO method strongly complements
SN measurements because BAO provides an absolute distance scale
and a strong connection to the CMB acoustic peaks from $z=1000$,
while SN allow more precise measurements of relative distances and thus
offer a more fine-grained view of the distance-redshift
relation.  At higher redshift ($z\gtrsim0.5$), the large cosmic
volume and the direct access to $H(z)$ make the BAO method
an exceptionally powerful probe of dark energy and cosmic geometry.

\subsection{The Current State of Play}
\label{sec:bao_current}

\begin{figure}[t]
\begin{centering}
{
\includegraphics[width=3.2in]{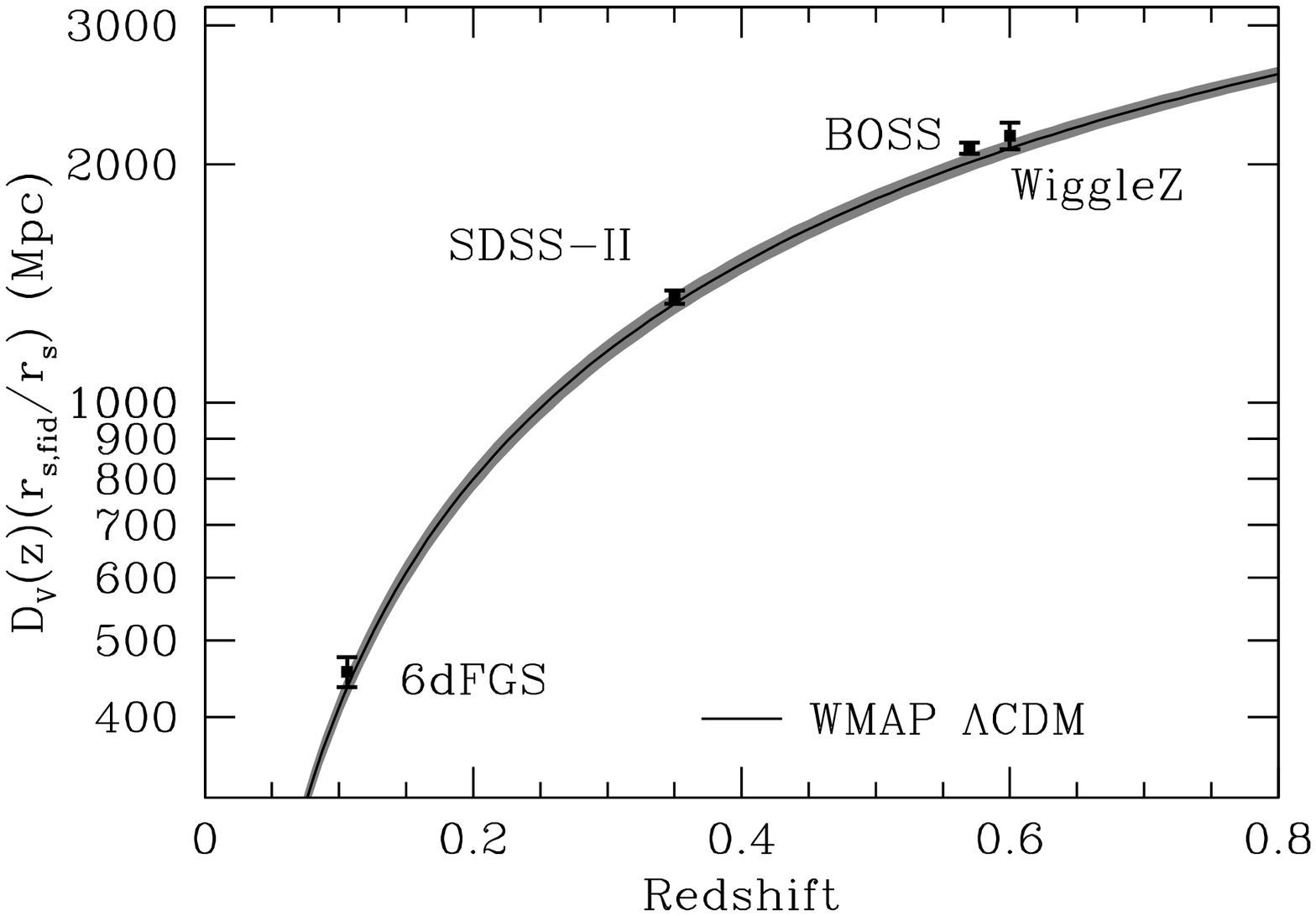}\hfill
\includegraphics[width=3.2in]{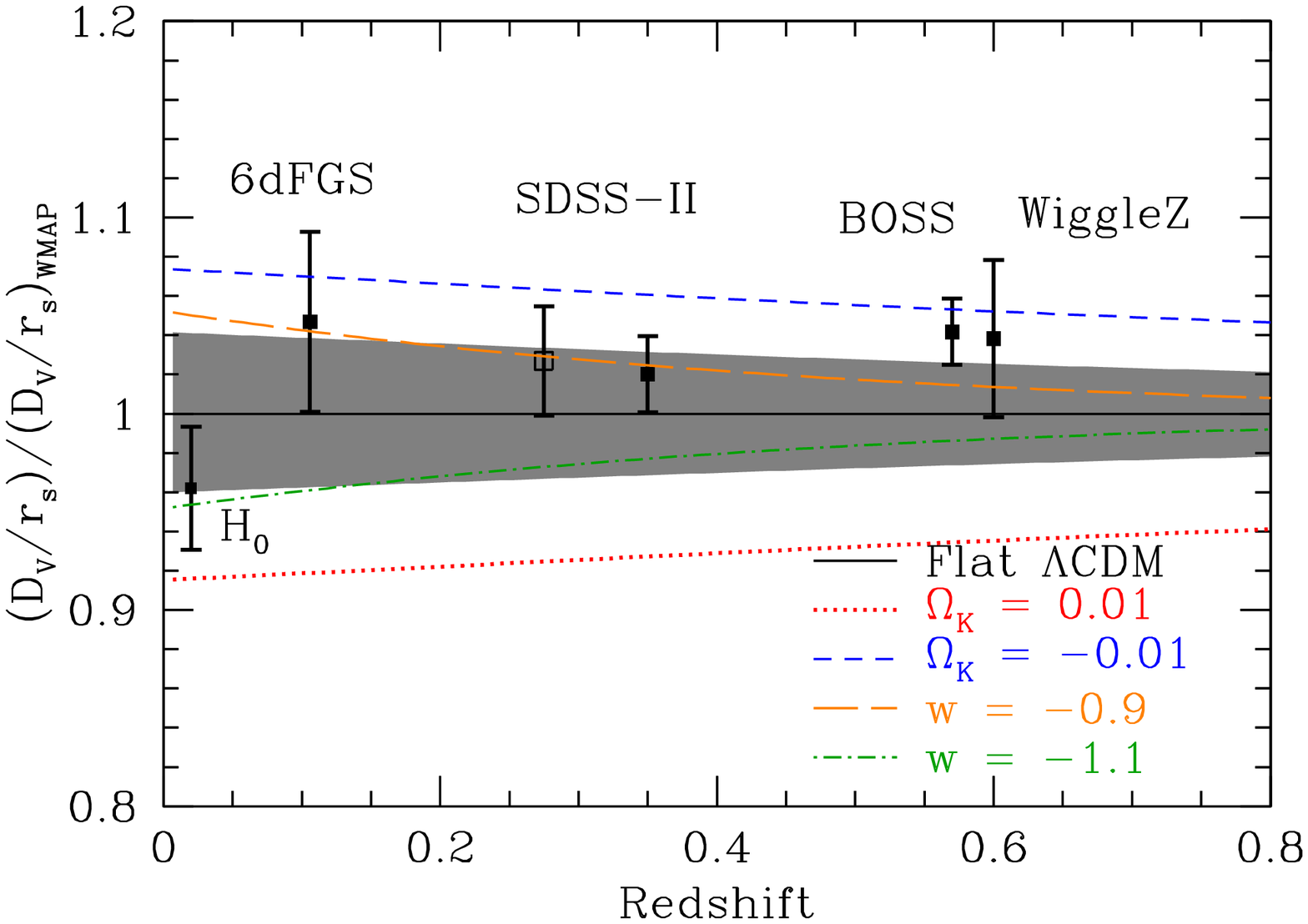}}
%\includegraphics[width=3.2in]{bao_logDV_current.eps}\hfill
%\includegraphics[width=3.2in]{bao_DVfid_review.eps}}
%%% lcdm_bao.eps
%%% w0wa_bao.eps
%%% w_bao.eps
\end{centering}
\caption{\label{fig:bao_DV}
(Left) The current BAO distance-redshift relation.  Individual
measurements are of the quantity $D_V(z)/r_s$.	We have multiplied
by the $r_s$ of the fiducial $\Lambda$CDM model to yield a distance;
the sound horizon is predicted to 1.1\% from WMAP7.  In increasing
redshift, data points are from the 6dFGS \protect\citep{beutler11},
SDSS-II \citep{padmanabhan12,xu12a},
BOSS \citep{anderson12}, 
and WiggleZ \protect\citep{blake11c}.
The WiggleZ paper 
also quotes correlated results from multiple redshift bins, but we have
chosen to plot only a single combined data point for each survey
so that the measurement errors are uncorrelated.
As described in the text, for a fixed choice of $w(z)$ and $\Omega_k$,
CMB data allows a prediction for $D_V(z)/r_s$.
The flat $\Lambda$CDM prediction from the best-fit WMAP7 model is the 
black line, and the grey region shows the
$1\sigma$ WMAP7 range.  This is not a fit to the data, but rather
the vanilla $\Lambda$CDM prediction from the CMB data.
(Right) The same plot after dividing by the $\Lambda$CDM prediction from WMAP7.
We have added an open point that shows the measurement from
\cite{percival10} using a combination of SDSS-II DR7 LRG and Main
sample galaxies and 2dFGRS galaxies; the \cite{padmanabhan12}
measurement from the DR7 LRG data alone has a smaller error
bar because of the increased precision afforded by reconstruction.
Also shown are the four alternative models from Table \protect\ref{tbl:models};
here we have suppressed the $1\sigma$ range that would surround each
line owing to uncertainties in the matter and baryon density.
Also shown is the direct $H_0$ value from \protect\citet{riess11}; here
we have assumed perfect knowledge of the sound horizon, which suppresses
a 1.1\% uncertainty term between this value and the BAO points.
These figures are adapted from the corresponding figures
in \cite{anderson12}.
We have omitted the very recent BAO detections from the BOSS \lya\ 
forest at $z\approx 2.3$ \citep{busca12,slosar13}, which are also 
consistent with $\Lambda$CDM predictions.
}
\end{figure}

\begin{figure}[t!]
\begin{centering}
{\includegraphics[width=2.1in]{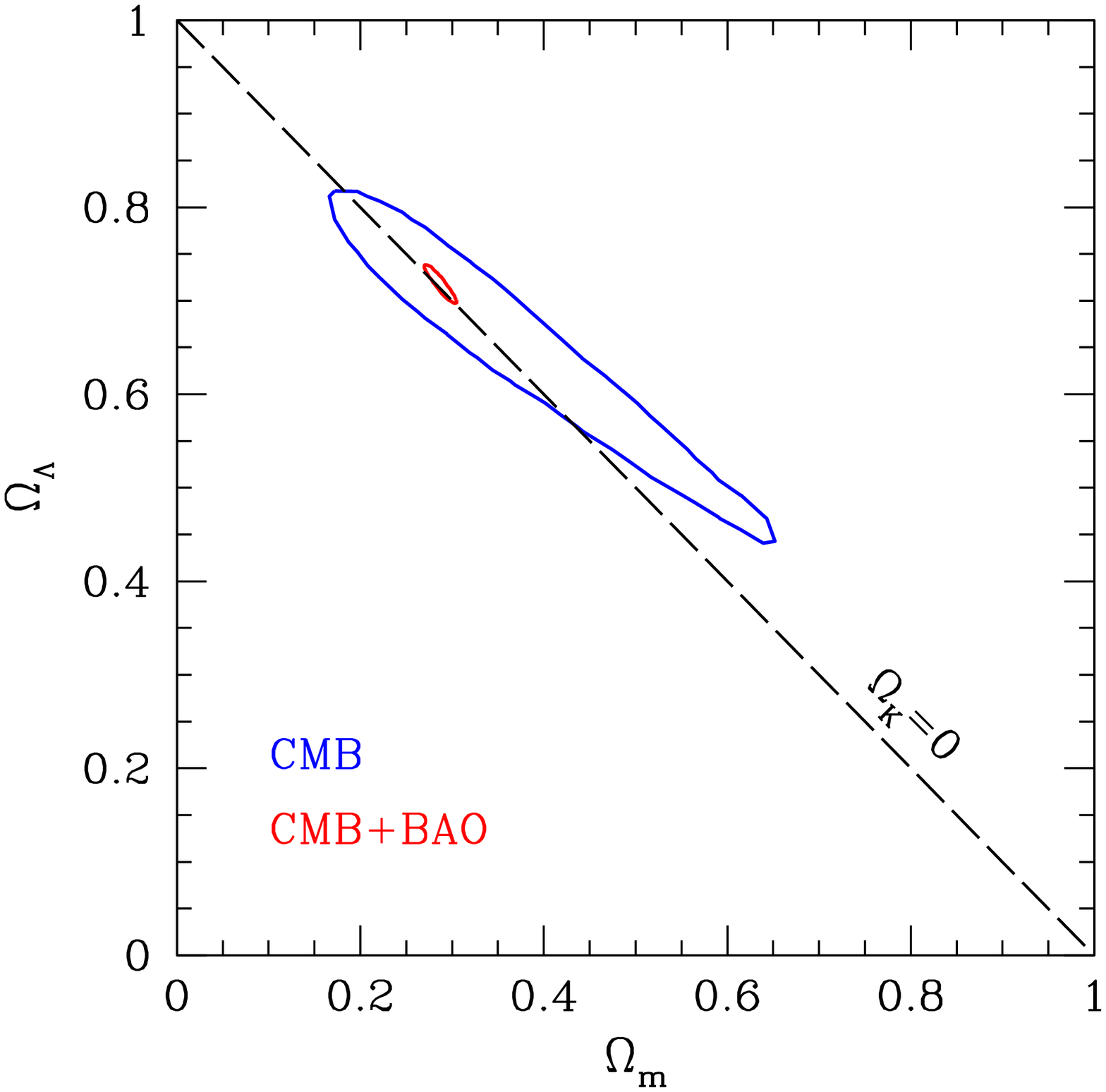}\hfill
\includegraphics[width=2.1in]{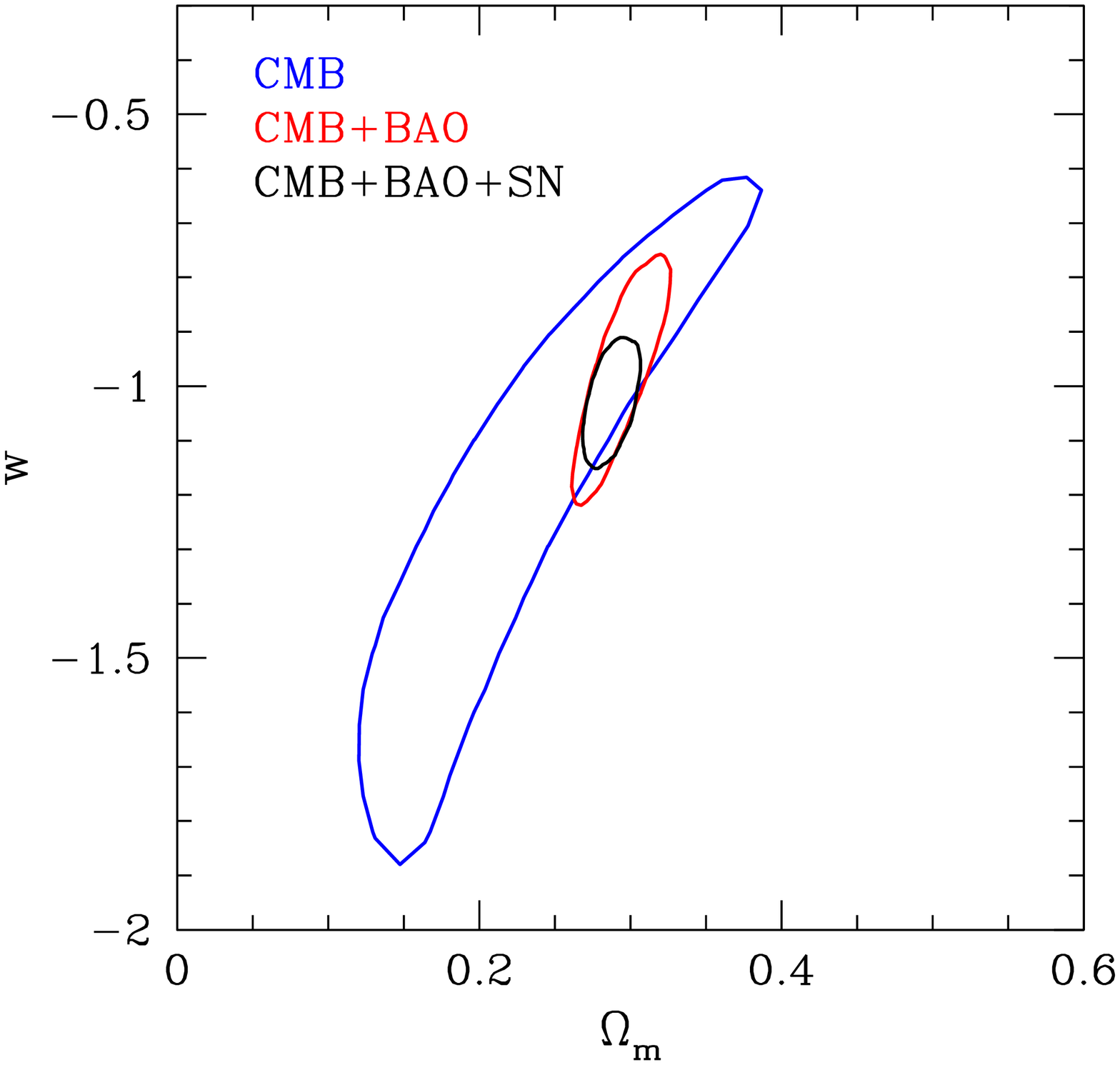}\hfill
\includegraphics[width=2.1in]{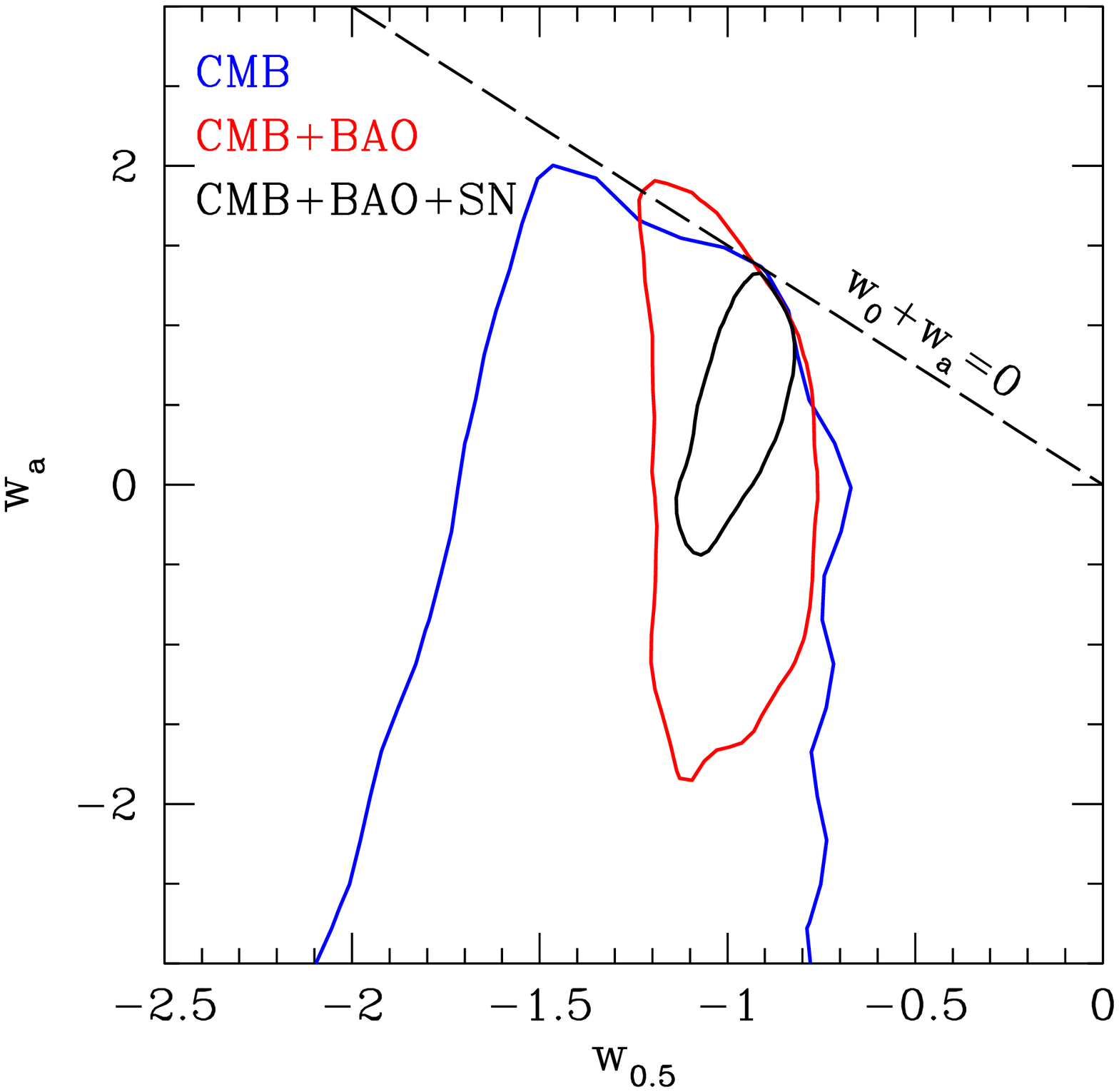}}
%% Note that "bao12" refers to the plots with the updated
%% BAO values from reconstructed SDSS-II and BOSS.
\end{centering}
\caption{\label{fig:bao_fisher}
Constraints from combinations of current BAO data (the four points
in the left hand panel of Fig.~\ref{fig:bao_DV}),
WMAP7 CMB data, and
Union2 SN data in (a) the $(\om,\ol)$ plane assuming $w=-1$,
(b) the $(\om,w)$ plane assuming $\ok=0$, and
(c) the $(w_{0.5},w_a)$ plane assuming $\ok=0$, where
$w_{0.5}$ is the value of $w$ at $z=0.5$.
Contours show 68\% confidence intervals.
We omit the CMB+BAO+SN combination from panel (a) because it is nearly
identical to the CMB+BAO combination.
}
\end{figure}

The acoustic oscillation phenomenon was identified as a potential effect
in the CMB sky in the late 1960s.  This was soon extended to the late-time
matter power spectrum by \citep{sakharov66,peebles70,sunyaev70};
of course, they were considering pure baryon cosmologies, where the effect is
very strong.  The introduction of adiabatic cold dark matter in the
mid-1980's made the predicted late-time acoustic peak very weak (particularly
in the $\Omega_m=1$ scenario), and the acoustic oscilllations were
primarily studied in the CMB context
\citep{bond84,bond87,jungman96,hu96,hu96a,hu97}.
A resurgence of interest in the
dynamics of the early universe post-{\it COBE} led to the identification
of the acoustic scale as a standard ruler, first in the CMB and then
in the matter power spectrum
\citep{kamionkowski94,jungman96,hu96,eisenstein98,meiksin99}.  Fisher matrix
forecasts for the combination of
CMB and large-scale structure identified the acoustic oscillations as
a critical feature in breaking the distance scale degeneracy between
$\Omega_m$ and $H_0$ in CMB model fits
\citep{tegmark97,goldberg98,efstathiou99,eisenstein98a}.  In particular,
the SDSS
Luminous Red Galaxy (LRG) sample \citep{eisenstein01}\footnote{
Alex Szalay and Jim Annis deserve particular credit for leading this
development in the early years.}
was proposed to maximize leverage on the
large-scale power spectrum, with BAO as one application.

After the discovery of cosmic acceleration
with Type Ia SNe, the focus on the
distance scale as a function of redshift became intense.
In 2003, several papers appeared discussing the acoustic scale as
a standard ruler for the measurement of dark energy in higher redshift
galaxy surveys \citep{eisenstein02,blake03,hu03b,linder03c,seo03}.
Compelling detections in 2005 intensified these
plans \citep{cole05,eisenstein05}, with several observational surveys
proposed and numerous
theoretical investigations.  The rapid development of the theory
led to the DETF featuring BAO as one of the
four leading methods for the study of dark energy \citep{albrecht06}.

Early results from the 2dF Galaxy Redshift Survey (2dFGRS)
\citep{percival01,efstathiou02,percival02} and the Abell/ACO cluster
sample \citep{miller99}
gave hints of the acoustic feature in the power spectrum.  However,
the first convincing detections of BAO came in 2005 from the
SDSS Data Release 3 (DR3) and final 2dFGRS
samples \citep{eisenstein05,cole05}.
\citet{eisenstein05} measured the large-scale correlation function
of SDSS LRGs in the
redshift range $0.16<z<0.47$ over 3,816 square degrees,
finding the acoustic peak with 3.4 $\sigma$ significance.
As they only measured the monopole of the correlation function,
\citet{eisenstein05} quoted the distance measurement as a blend of the
line-of-sight and transverse distance scale
\begin{equation} 
\label{eqn:dvdef}
D_V(z) = \left[D_A(z)\right]^{2/3}
\left[ cz \over H(z)\right]^{1/3}. 
\end{equation}
Comparing the size of the acoustic scale in SDSS to that in the CMB sky
from \wmap, they inferred the value of $D_V(0.35)$ divided by 
the distance to $z=1089$ with a $1\sigma$ uncertainty of 3.7\%.
(Recall that we use $D_A$ to denote the {\it comoving}
angular diameter distance.)
\citet{cole05} measured the power spectrum of 2dFGRS galaxies
in the redshift range $0<z<0.3$ over 1,800 square degrees.
The cosmological fitting analysis detected a baryon fraction
of $\Omega_b/\Omega_m = 0.185 \pm 0.046$, the non-zero result indicating
a detection of the BAO.  The distance precision of the
result was quoted as a 4.1\% measurement of $H_0$.

Since these first detections, the clustering of successively larger SDSS
spectroscopic samples has been analyzed by several groups using
different methods.
\citet{tegmark06} analyzed the DR4 LRG and main galaxy samples 
with a quadratic estimator for the power
spectrum and redshift-space distortion.
\citet{hutsi06} analyzed the monopole of the power spectrum of
the LRG data set with the \citet{feldman94} (FKP) method.
\citet{percival07} applied the FKP method to the combined DR5
LRG and main galaxy samples, along with the 2dFGRS sample, to measure the
acoustic scale at two different redshifts ($z=0.20$ and 0.35).
\citet{percival10} extended this analysis to the final SDSS-II sample (DR7),
obtaining
an aggregate distance precision of 2.7\% to $z=0.275$.
\citet{kazin10a} analyzed the DR7 LRG sample with the correlation
function (as did \citealt{martinez09}), achieving consistent results.
A recent reanalysis of the DR7 sample by 
\cite{padmanabhan12} and \cite{xu12a} improved the distance precision to 1.9\%
by applying the reconstruction technique described in 
\S\ref{sec:bao_theory_recon} below.

New BAO detections have recently been made in three other samples.
\citet{beutler11} report a $2.4\sigma$ detection from the
6-degree Field Galaxy Survey (6dFGS), which covered 17,000 deg$^2$
of sky, obtaining a 4.5\% distance measurement to $z=0.1$.
Stepping beyond $z=0.5$, the WiggleZ survey \citep{blake11b,blake11c} has
used the AAOmega instrument at the Anglo-Australian Telescope to
target emission-line galaxies at $0.4<z<1.0$.  The analysis of the
final data set of $\sim\!800$ deg$^2$ yields BAO detections in three
overlapping redshift slices centered on $z=0.44$, 0.60, and 0.73,
with an aggregate precision of 3.8\%.
\citet{anderson12} report the first BAO measurements from
the SDSS-III BOSS survey (discussed further below),
obtaining a 1.7\% distance precision to a sample with effective
redshift $z = 0.57$.

Combining SDSS-II, WiggleZ, and 6dFGS, \citet{blake11c} achieve a
$5\sigma$ detection of the acoustic peak; 
\citet{anderson12} combine the reconstructed SDSS-II measurement
and BOSS measurement to get a 6.7$\sigma$ detection.
Both studies find good agreement between the BAO and SN
distance-redshift relations.
These BAO measurements are displayed in Figure \ref{fig:bao_DV},
which shows $D_V$ as a function of redshift.
We can compare this $D_V(z)$ to the relation predicted
by WMAP7 under particular assumptions about dark energy and spatial
curvature.
A given value of $\Omega_mh^2$ and $\Omega_b h^2$ yields a sound horizon
$r_s$.	For any fixed choice of $\Omega_k$ and $w(z)$, the angular
acoustic scale in the CMB then breaks the $\Omega_m$--$H_0$ degeneracy,
which then specifies $D_V(z)$.
The left panel shows the WMAP7 prediction for flat $\Lambda$CDM, with the grey
band marginalizing over $1\sigma$ errors in $\Omega_m h^2$ and $\Omega_b h^2$,
while the right panel divides by this prediction.
The BAO measurements are all in good agreement with the
$+1\sigma$ edge of the WMAP7 band; in other words, they are
consistent with WMAP7 and $\Lambda$CDM but favor a value
of $\Omega_m h^2 \approx 0.139$ vs. $0.134$.
Intriguingly, the BAO data pull in the opposite direction
from the $H_0 = 73.8 \pm 2.4\hubunits$ measurement
of \cite{riess11}.  The discrepancy is only marginally significant
at present --- less than $2\sigma$ assuming $\Lambda$CDM 
--- but it illustrates how BAO and direct $H_0$ 
measurements can combine to reveal additional physics beyond
that in $\Lambda$CDM.

Curves in the right panel show how the
comparison to the data would vary with non-zero spatial curvature
or $w\ne-1$, using the CMB-normalized models introduced in
\S\ref{sec:dependences}.
One can see that small changes, particularly in spatial
curvature, make detectable differences in the prediction, so that 
comparison of the data to the prediction allows one to measure
$w(z)$ and $\Omega_k$.
Comparing variations in spatial curvature to variations of constant $w$,
one can see that variations in spatial curvature produce large offsets
but relatively small slopes.  SN determinations of relative distances
can only measure slopes on this graph, whereas absolute distance measurements
such as BAO can measure the offset.  This illustrates why, in fits to the
$w$--$\Omega_k$ model, the CMB+SN combination tends to measure $w$ better
while the CMB+BAO combination tends to measure $\Omega_k$ better.

Combining the WiggleZ, SDSS+2dFGRS \citep{percival10}, and 6dFGS
BAO measurements with WMAP7 and the Union-2 SN compilation,
\citet{blake11c} infer 
$\Omega_m=0.289\pm0.015$, 
$H_0= 68.7 \pm 1.9\hubunits$,
$\Omega_K = -0.004\pm0.006$, and 
$w = -1.03\pm0.08$ (where $w$ is assumed to be constant with redshift).
\citet{anderson12} obtain similar constraints when substituting
the BOSS and reconstructed SDSS-II BAO measurements and
the SNLS3 SN compilation:
$\Omega_m = 0.276 \pm 0.013$, 
$H_0 = 69.6 \pm 1.7\hubunits$,
$\Omega_K = -0.008 \pm 0.005$,
and $w = -1.09 \pm 0.08$.
One can think of these inferences approximately
as CMB acoustic peak heights measuring $\Omega_m h^2$, the BAO standard
ruler then
splitting $\Omega_m$ and $H_0$, the CMB angular acoustic scale
measuring $\Omega_K$, and the SNe measuring $w$.  
Figure \ref{fig:bao_fisher} displays our own constraints derived
from these data with CosmoMC \citep{lewis02}, with the same 
parameter space used for the SN and CMB constraints in
Figure~\ref{fig:sn_constraints}.  While Figure~\ref{fig:sn_constraints}
includes contours for SN alone, it makes little sense to consider 
BAO constraints independent of CMB data because the latter are
needed to calibrate the BAO ruler.  We therefore show contours
for CMB, CMB+BAO, and CMB+BAO+SN.  Consistent with our earlier
discussion, CMB+BAO provides much tighter constraints on $\ok$
and $\om$ in the $w=-1$ model than CMB+SN
(compare the left panels of Figs.~\ref{fig:sn_constraints}
and~\ref{fig:bao_fisher}), but CMB+SN provides better constraints
on $w$ (middle panels).  For a $w_0-w_a$ model with $\ok=0$
(right panel), the three data sets together yield a good
measurement of $w(z=0.5)$ but still only loose constraints on $w_a$.

%%% With WMAP consistently return $\Omega_m=0.29
%%% data sets indicate $\Omega_m=0.29 \pm 0.015$ and $H_0 = 69 \pm 1.7$
%%%
%%% \djecitet{Percival09} achieved an aggregate distance precision of 2.7\%
%%% to $z=0.275$
%%% and a value of $\Omega_m = 0.286\pm0.018$ and $H_0
%%% = 68.2 \pm 2.2$~km~s$^{-1}$~Mpc$^{-1}$.
%%% The relative distance between $z=0.20$ and $0.35$ was consistent
%%% with LCDM at 1.1~$\sigma$, albeit not highly constraining of $w$ due
%%% to the small path difference.  Combining with the WMAP 5-year
%%% measurements and the Union supernova sample \djecitep{cites} yielded
%%% a simultaneous measurement of $\Omega_K = -0.006\pm0.008$ and
%%% $w = -0.97\pm0.10$ for a constant equation of state.
%%%

Indications of the acoustic feature have also been found in a sample
of higher redshift LRG with photometric redshifts from the SDSS
\citep{padmanabhan07,blake07,crocce11,sawangwit11}.  These
analyses produced a 6.5\% measurement of the
angular diameter distance to $z=0.5$.  An analysis of the maxBCG cluster
catalog by \citet{hutsi10} also yields a 2-2.5 $\sigma$ detection.

Other analyses have focused on the anisotropic BAO signal, with the
intent of separating $D_A(z)$ and $H(z)$.	\citet{okumura08} performed
a correlation function analysis of the LRG sample from SDSS DR3,
achieving a weak indication of the radial BAO.	\citet{gaztanaga09a}
analyzed the correlations of the SDSS LRG sample, considering
only pairs very close to the line of sight.  They claimed a detection
of the BAO, thereby measuring $H(z)$; however, the proposed acoustic
peak is much higher amplitude than the predicted one, and  \cite{kazin10}
argue that it is likely to be noise (see \citealt{cabre11} for
a response to this criticism).
\citet{chuang11} analyzed the full SDSS DR7 LRG
sample with an anisotropic correlation function, finding separate
constraints on $D_A$ and $H$ at $z=0.35$.

%%% There are two spectroscopic surveys aimed at BAO currently operating.
%%% The WiggleZ survey \djecitet{Blake09} is using the AAOmega instrument
%%% at the Anglo-Australian Telescope to target emission-line galaxies
%%% at $0.5<z<1.0$.  Targets are selected from the combination of FUV and
%%% NUV imaging from the GALEX satellite with SDSS $r$ band imaging.
%%% The survey has completed its data taking, with a total of
%%% 400,??? galaxies
%%% over ??? square degrees.  The analysis of the final data set is
%%% expected soon and should achieve a BAO detection at $z\approx 0.7$
%%% with approximately 3\% distance precision.

The next generation of the SDSS large-scale structure survey is
BOSS, the Baryon Oscillation Spectroscopic Survey of SDSS-III.
BOSS is observing 1.5 million
luminous galaxies (mostly LRGs) out to $z=0.7$ over 10,000 square degrees
\citep{eisenstein11,dawson12}, with a selection that
triples the number density of LRGs at $z<0.4$ relative to SDSS-II
and extends to a new redshift range with a
dense sample at $0.5<z<0.7$.  The increased sampling should facilitate
accurate density-field reconstruction (\S\ref{sec:bao_theory_recon})
to boost the BAO performance.
BOSS is also surveying the $2<z<3$ universe using a grid of quasar
sightlines to provide a 3-dimensional view of the \lya\ forest,
with the goal of detecting BAO in the large-scale clustering
of neutral hydrogen at $z\sim2.5$.  This method was proposed by
\citet{white03} and \citet{mcdonald07}.  The clustering of \lya\
forest flux along
single lines of sight is well established as a probe of large-scale
structure (see \S\ref{sec:lyaf} for a discussion of the underlying theory),
and by using cross-correlations among multiple lines of
sight one can probe 3-dimensional structure.  Observationally
this method is attractive because quasars are very luminous and because
each quasar provides $\sim 50$ measurements of the large-scale density
field along its line of sight.	Since the BAO peak has an 
intrinsic rms width of $\sim 8\hmpc$, one need only
survey the \lya\ forest at modest resolution (a few hundred)
to retain full BAO information.
Furthermore, one does not need high signal-to-noise ratio spectra, as
one gets little gain from photon errors smaller than the intrinsic
variation in the small-scale forest.  
BOSS aims to measure 160,000 spectra of $z>2$ quasars over 10,000 deg$^2$.
Using the first third of this data set, \cite{busca12} and
\cite{slosar13} report the first BAO detections in the \lya\ forest,
and the first detections at any $z > 1$, with precision (2-3\% on an
isotropic dilation factor) that is 
roughly in line with theoretical expectations.
%BOSS has achieved the first detection
%of 3-d structure on $10-50\hmpc$ scales in the \lya\ forest
%\citep{slosar11}, and it aims to use 150,000 quasars
%over 10,000 square degrees to achieve the first
%measurements of BAO at $z>1$.

In summary, the BAO feature has been found in eight different samples ---
2dFGRS, SDSS LRG, SDSS Main, 6dFGS, WiggleZ, SDSS photometric, BOSS
galaxies, and the BOSS \lya\ forest
--- with analyses from several independent research teams and with
a variety of methods.  The best precision is now slightly below
2\%, with excellent agreement with the $\Lambda$CDM model.

\subsection{Theory of BAO}
\label{sec:bao_theory}

While the theory of supernova explosions is complicated, the use
of Type Ia SNe as distance indicators rests on empirically
determined correlations that are largely independent of that theory.
With BAO, on the other hand, we are using a standard ruler whose
length, imprint on the clustering of observable tracers, and
even very existence are derived from theory.
We therefore review both the long-established linear theory of
BAO and more recent work on non-linear evolution and galaxy bias,
and we discuss the implications of this work for analysis of BAO
data sets.

\subsubsection{Linear Theory}
\label{sec:bao_theory_linear}

\begin{figure}[p]
\centerline{\includegraphics[width=2.3in]{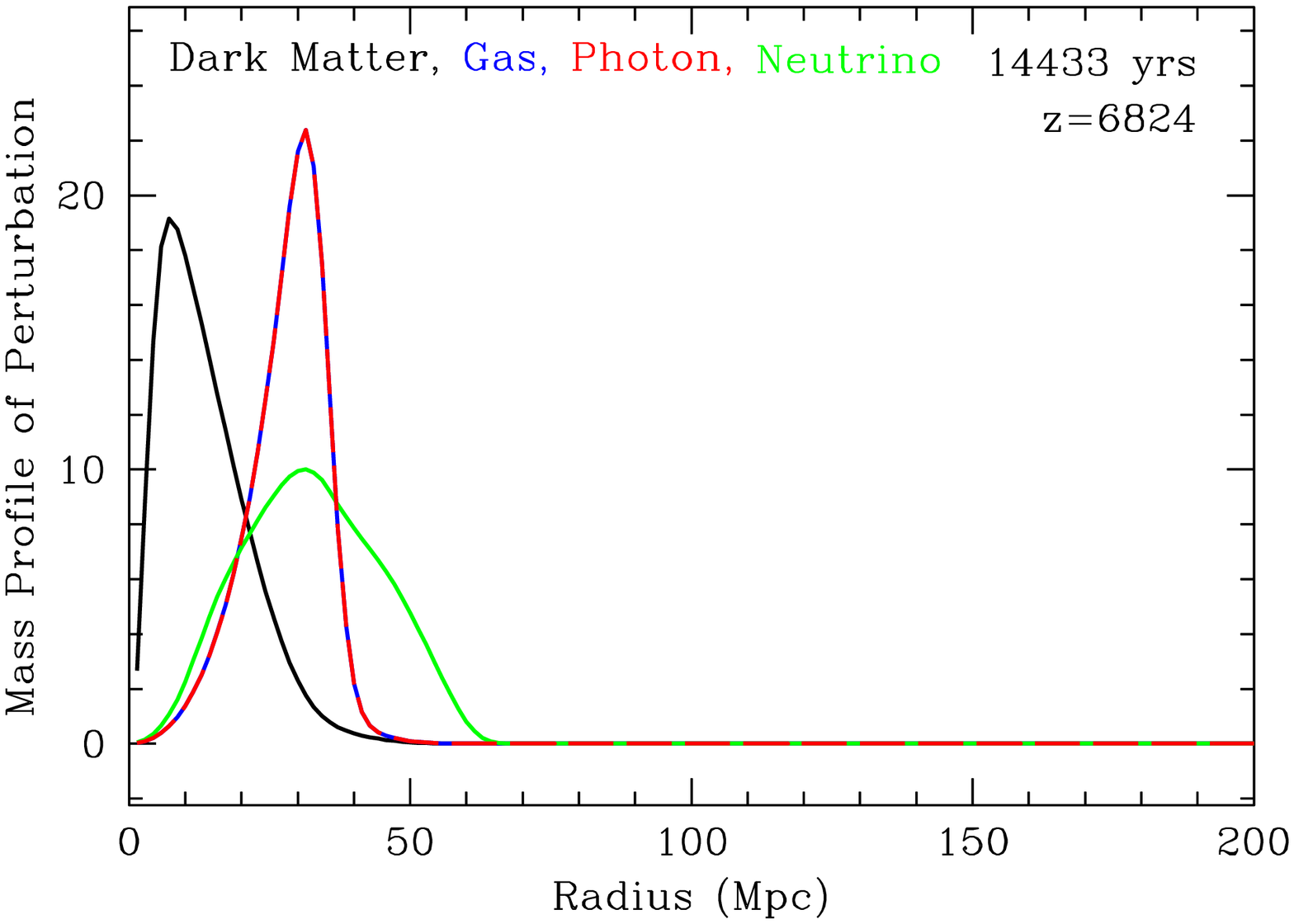}
    \quad  \includegraphics[width=2.3in]{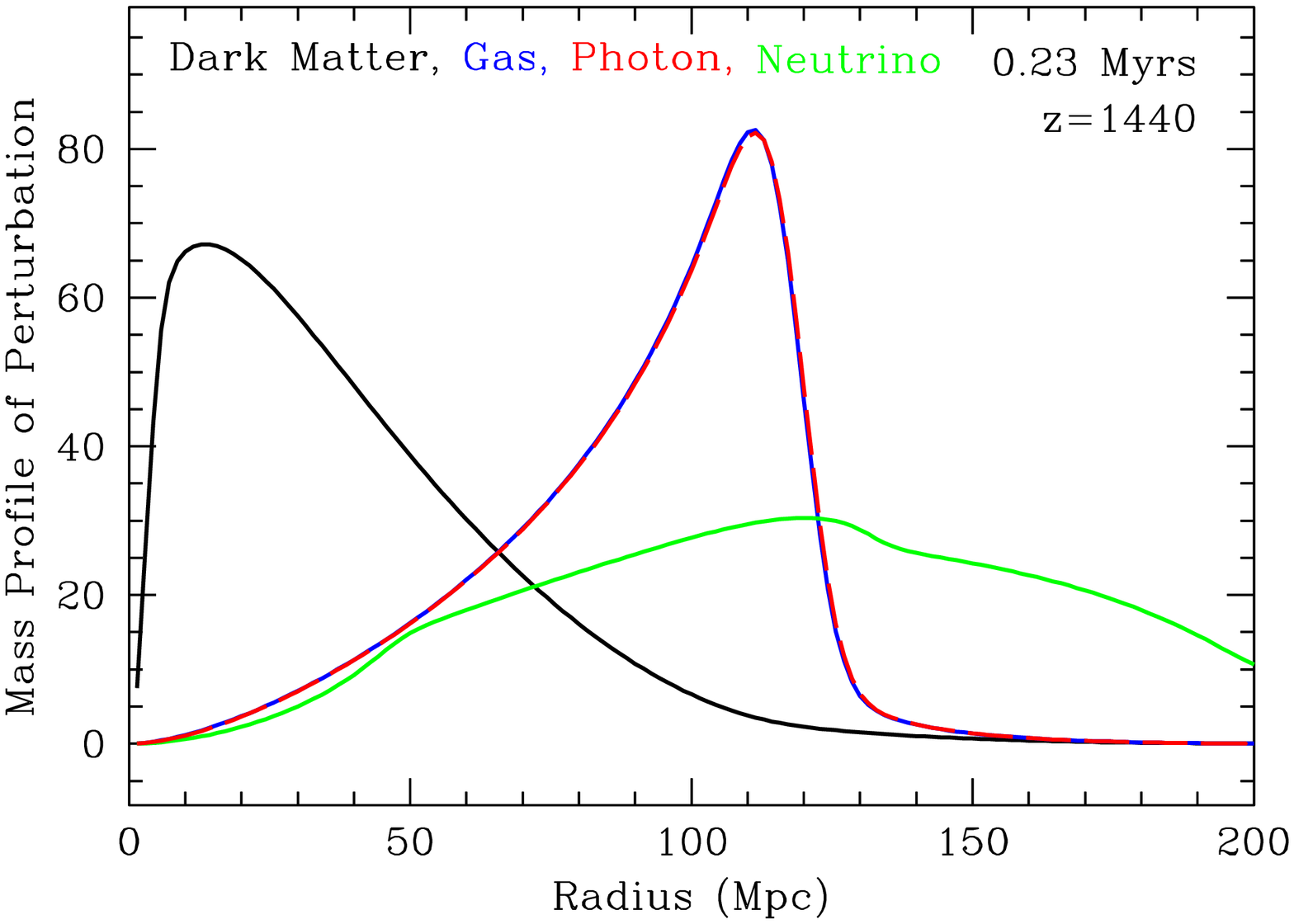}}
\centerline{\includegraphics[width=2.3in]{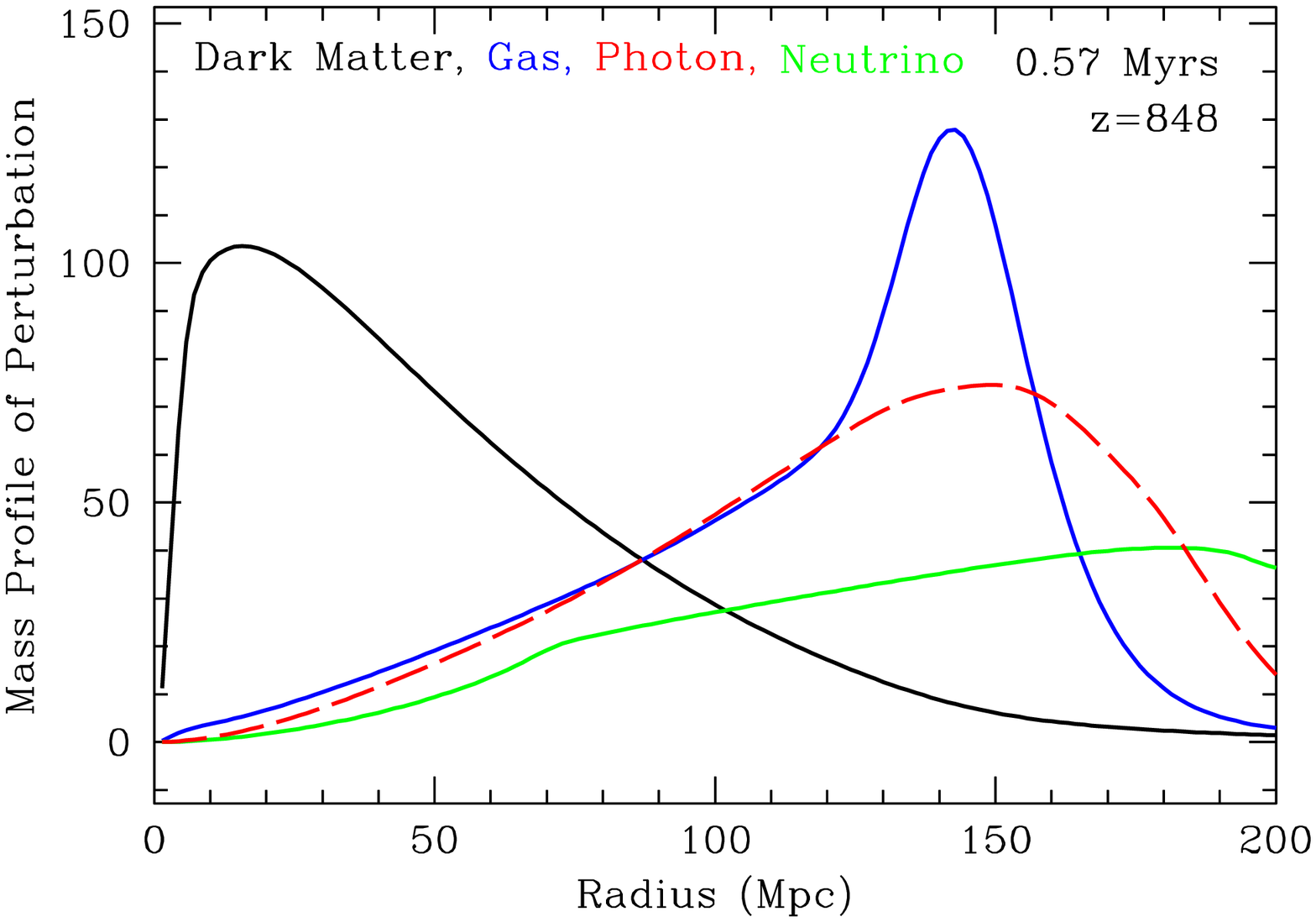}
    \quad  \includegraphics[width=2.3in]{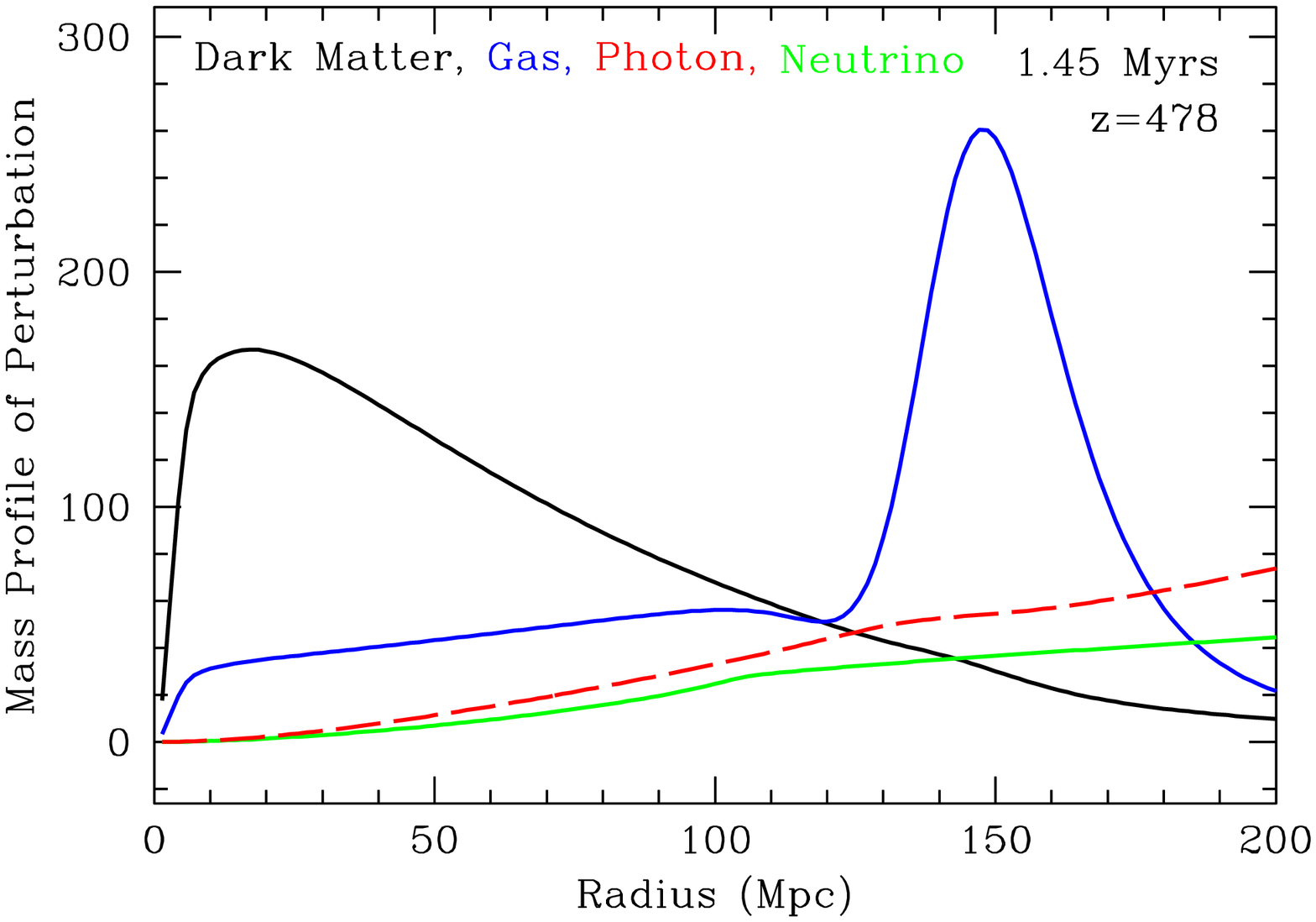}}
\centerline{\includegraphics[width=2.3in]{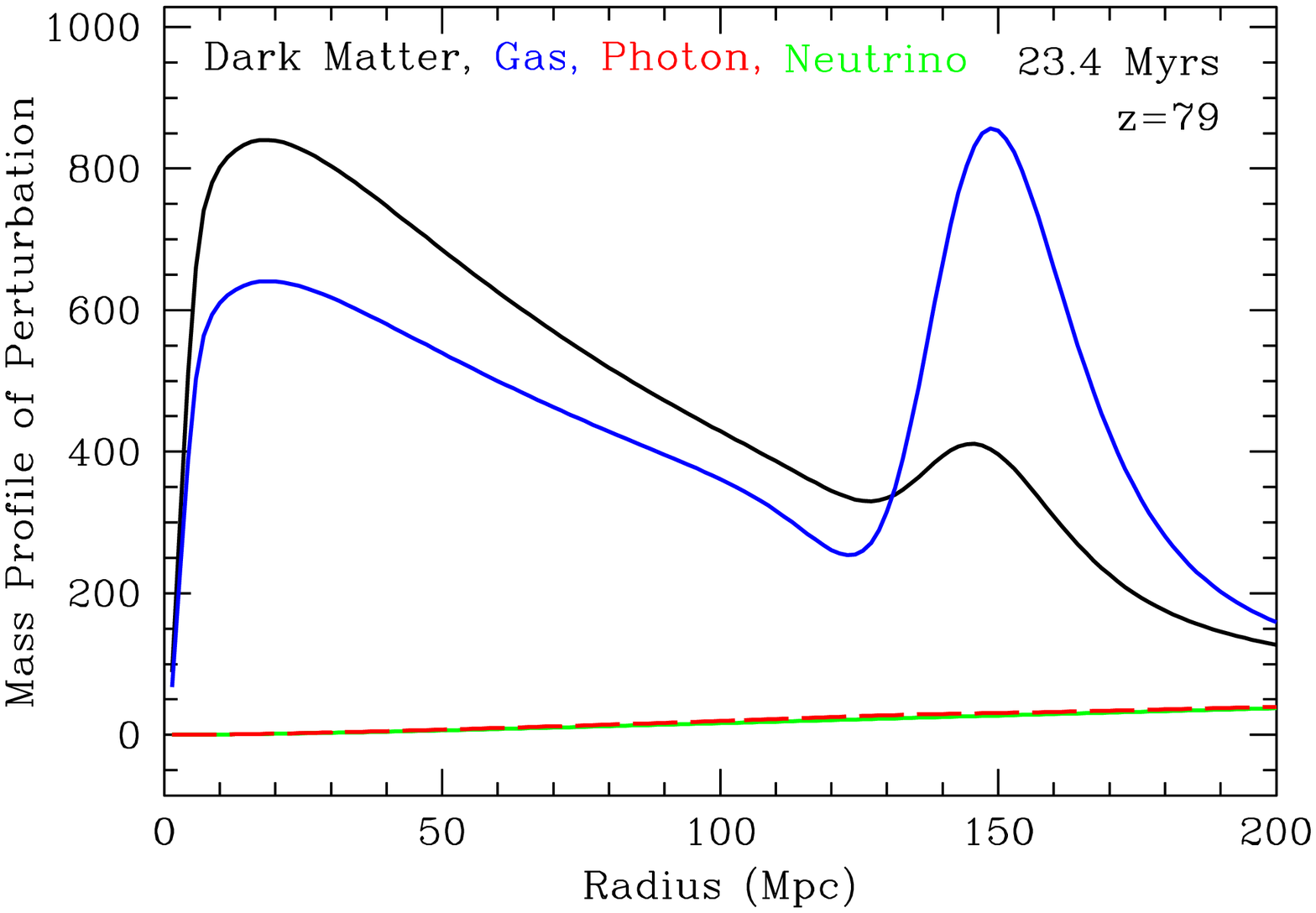}
    \quad  \includegraphics[width=2.3in]{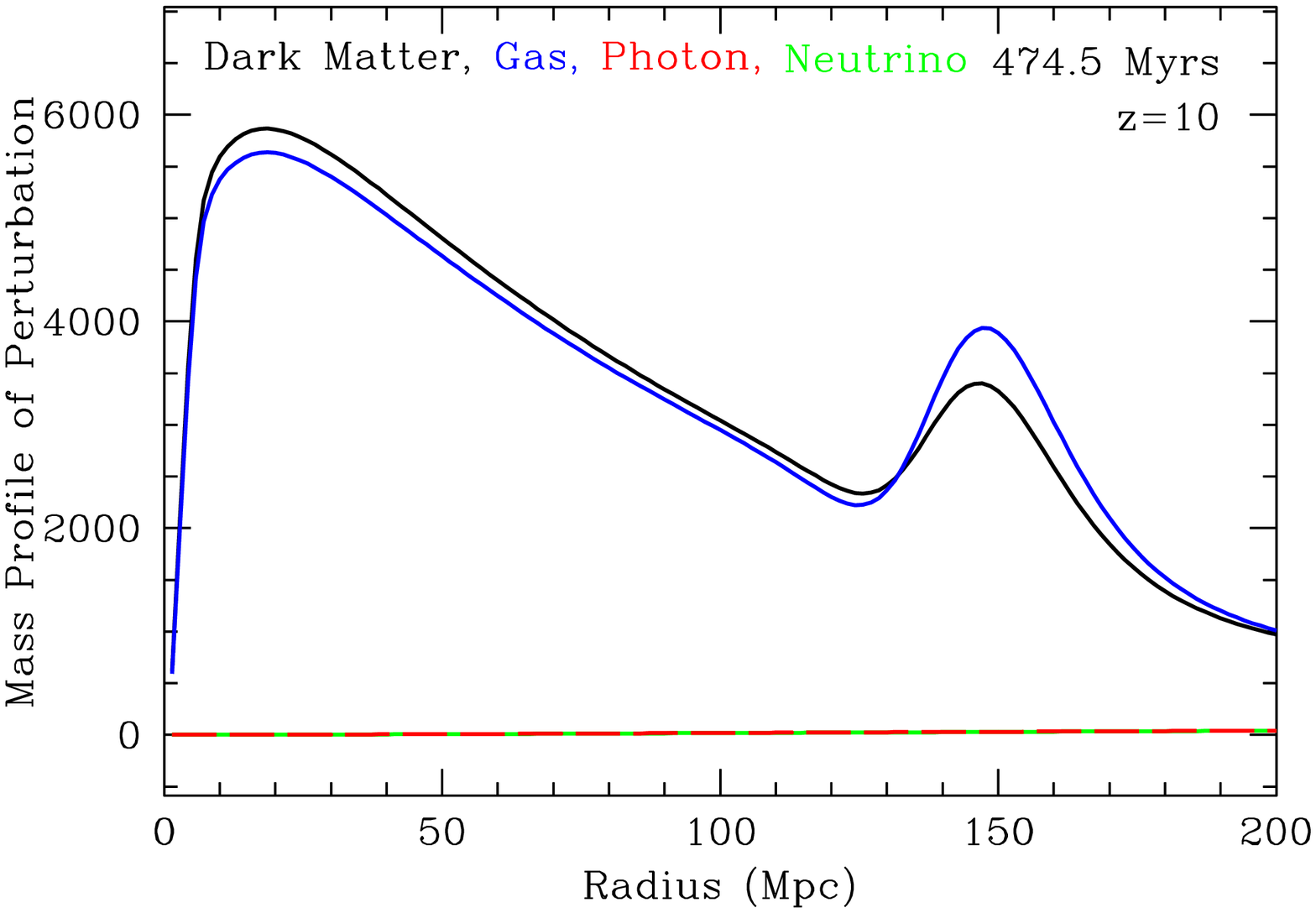}}
\caption{\label{fig:bao_linear}
The generation of the acoustic peak illustrated via the linear-theory
response to an initially point-like overdensity at the origin; this figure
is reproduced from \protect\citet{eisenstein07a}.
Each panel shows the radial perturbed mass profile 
in each of the four species: dark matter (black), baryons (blue),
photons (red), and neutrinos (green).  The redshift
and time after the Big Bang are given in each panel.
All perturbations are fractional for that species.
We have multiplied the radial density profile of the perturbation
by the square of the radius in order to yield the mass profile.
In detail, we begin with a compact but smooth profile at the origin, which
is why the mass profiles go to zero there.
As we are using linear theory, the normalization of the
amplitude of the perturbation (and thus the absolute scale of
the $y$-axis) is arbitrary.
a) Near the initial time, the photons and baryons are tightly coupled
in a spherical traveling wave.
b) The outward-going wave of baryons and relativistic species increases
the perturbation of the cold dark matter, similar to raising a wake.
c) At recombination, the photons decouple from the baryons.
d) With recombination complete, the CDM perturbation is near the origin,
while the baryonic perturbation is in a shell of 150 Mpc.
e) With pressure forces now small, baryons and dark matter are attracted
to these overdensities by gravitational instability.
f) Because most of the growth is drawn from the homogeneous bulk, the
baryon fraction converges toward the cosmic mean at late times.  Galaxy
formation is favored near the origin and at a radius of 150 Mpc.
These figures were made by suitable transforms of the transfer functions
created by CMBfast \protect\citep{seljak96,zaldarriaga00}.
}
\end{figure}

Prior to redshifts around $1000$, the universe is hot and dense enough
that the primordial gas is ionized.  The free electrons in this plasma
provide enough cross-section to the cosmic microwave background photons
via Thomson scattering to produce a mean free time well less than the
Hubble time.  The result is a close coupling between the electrons,
nuclei (baryons), and photons for sufficiently long wavelength
perturbations in the early universe.  The radiation pressure of the
photons is large compared to the gravitational forces in the
perturbations, with the result that perturbations in the baryon-photon
fluid oscillate as sound waves \citep{peebles70,sunyaev70}.
Diffusion of photons relative to baryons damps these oscillations
on comoving scales smaller than $\sim 8\hmpc$, the phenomenon 
known as Silk damping \citep{silk68}.

After recombination, the mean free time of the photons in the neutral
cosmic gas is long compared to the Hubble time.  The photons decouple
from the perturbations in the baryons and soon become smoothly
distributed.  The perturbations in the baryons are now subject to
gravitational instability, just like the dark matter perturbations.

As with normal sound waves, one can usefully view the BAO 
phenomenon from different linear basis sets.
We first consider the response to a density perturbation
at a particular initial location,
as illustrated in Figure \ref{fig:bao_linear};
see \citet{eisenstein07a} and \citet{eisenstein08} for further
description of this view.
Primordial perturbations of the adiabatic form 
predicted by standard inflation models consist of 
equal fractional density contrasts in all species.
The dark matter perturbation grows in place, slowly
at first in the radiation dominated epoch, then faster as the universe
becomes matter dominated.  The baryon-photon perturbation, on the other
hand, travels away from its origin as a sound wave.  At recombination,
the baryon part of the wave is left in a spherical shell centered on
the original perturbation.  Both the dark matter at the center and the
baryons on the shell seed gravitational instability, which grows to
form the halos in which galaxies form.	We therefore expect the
distribution of separations of pairs of galaxies (i.e., the two-point
correlation function generated by such perturbations) to 
show a small enhancement at the radius of
the shell, with galaxy concentrations in the central dark matter
clumps and in the shells induced by the baryons.

One can equally well view the BAO effect as a standing wave in Fourier space;
see \citet{hu96} and \citet{eisenstein98} for this explanation.
In Fourier space, the single acoustic scale gives rise to a harmonic
sequence of oscillations in the power spectrum.  This is easy to understand
physically.  The power spectrum encodes the response of the universe to
a plane wave perturbation.  Each crest in the initial wave produces a
planar sound wave that travels a distance equal to the acoustic scale.
If the wave deposits the baryon perturbation on another crest of the
dark matter perturbation, then one gets constructive interference;
if the sound wave ends in a dark matter trough, one gets destructive
interference.  The result is a harmonic relation between the
wavelength of the perturbation and the acoustic scale.

Mathematically, this correspondence can be seen by considering that
the correlation function and power spectrum are Fourier transform
pairs.  The Fourier transform of a delta function is a sinusoid,
and the smearing of a delta function simply provides a damping
envelope to that sinusoid.  In the case of the BAO, this smearing
is largely due to Silk damping
%%% {\bf explain (in a sentence) what this means}
in the early universe and to non-linear
structure formation at late times.  Both cause the higher harmonics
in the power spectrum to be reduced in amplitude or washed out.

While it is secondary to our pedagogical thread, we end
with some additional discussion of Figure \ref{fig:bao_linear} and
the evolution of the initial point-like density perturbation.  First,
because the perturbation is in the growing mode, only the density
perturbation is localized.  The velocity perturbation away from the
initial density perturbation has zero divergence but is non-zero;
hence it scales as $r^{-2}$ at large radius.  As the baryon-photon
and neutrino pulses expand, the gravity interior to the shell is
weaker than it would have been.  This causes the velocity perturbation
interior to grow less quickly, creating a non-zero divergence away
from the origin, which is why the CDM perturbation grows at non-zero
radius.  The size of this effect depends on the radiation to matter
density; this transformation of the CDM perturbation is the famous
$k^{-2}$ tail of the CDM transfer function \citep{peebles82a}.  The
non-zero velocity perturbation is also the reason why the neutrino
perturbation does not remain as a sharp peak.  Finally, we note
that this description of the behavior is the Green's function of
the system.  CMB Boltzmann codes typically compute the evolution
of individual Fourier standing waves; these are simply combined here to
generate the response to a point perturbation rather than a single
standing wave.

\subsubsection{Non-linear Evolution and Galaxy Clustering Bias}
\label{sec:bao_theory_nl}

%%% {\it
%%%	Broadening of the peak
%%%
%%%	Shift of the peak
%%%
%%%	Galaxy bias
%%% }
%%%
%%% {\bf Start by talking about broadening of peak and its
%%% effect on precision.  Then go to shift of peak.}

\begin{figure}[t]
\centerline{\includegraphics[width=3.2in]{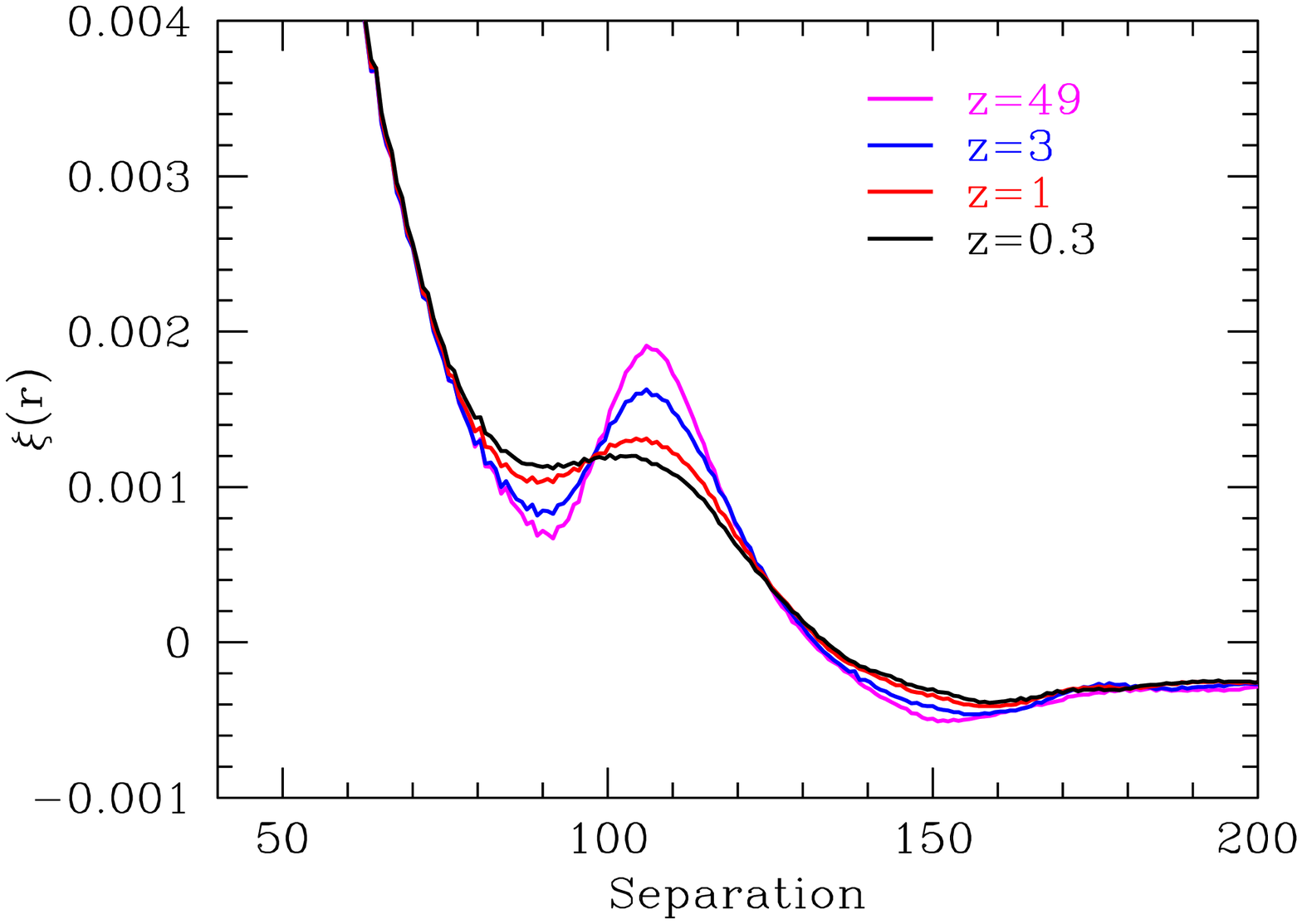}
    \quad  \includegraphics[width=3.2in]{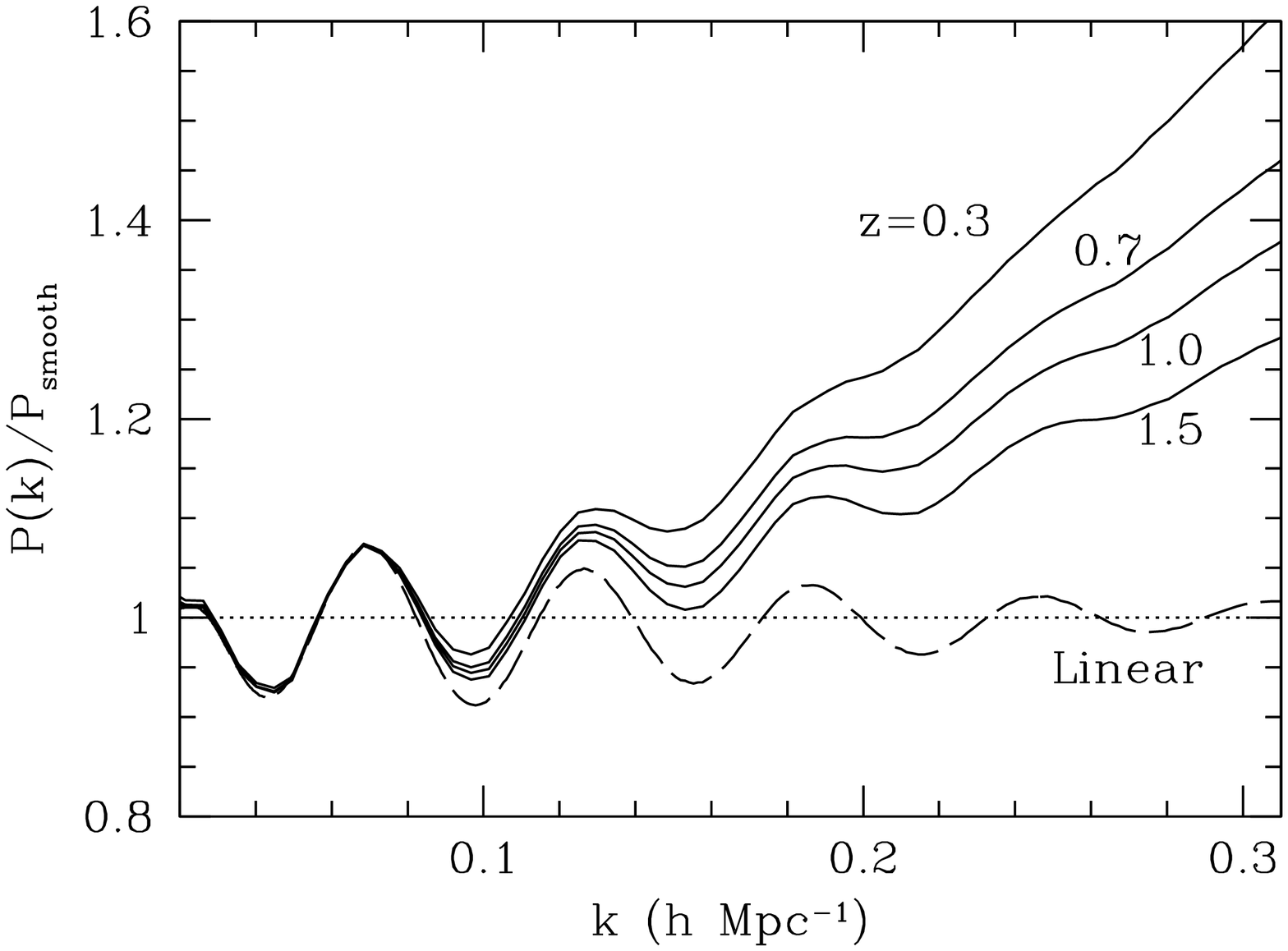}}
\caption{\label{fig:bao_sims}
The effects of non-linear clustering on the BAO.
(Left) Redshift-space matter correlation function at four different redshifts
from the simulations of \protect\citet{seo05}.
(Right) Real-space matter power spectra at four different redshifts from
the simulations of \protect\citet{seo08},
divided by a smooth power spectrum so as to reveal the acoustic oscillations.
The input linear theory is shown by the dashed line.
The effects of non-linear structure formation broaden the acoustic
peak in the correlation function.  In the power spectrum, this corresponds
to a damping of the higher harmonics.  Importantly, the boost of broad-band
power at late times visible in the power spectrum plot
corresponds largely to correlations at scales much smaller than the
acoustic peak.
}
\end{figure}

\begin{figure}[t]
\centerline{\includegraphics[width=3.2in]{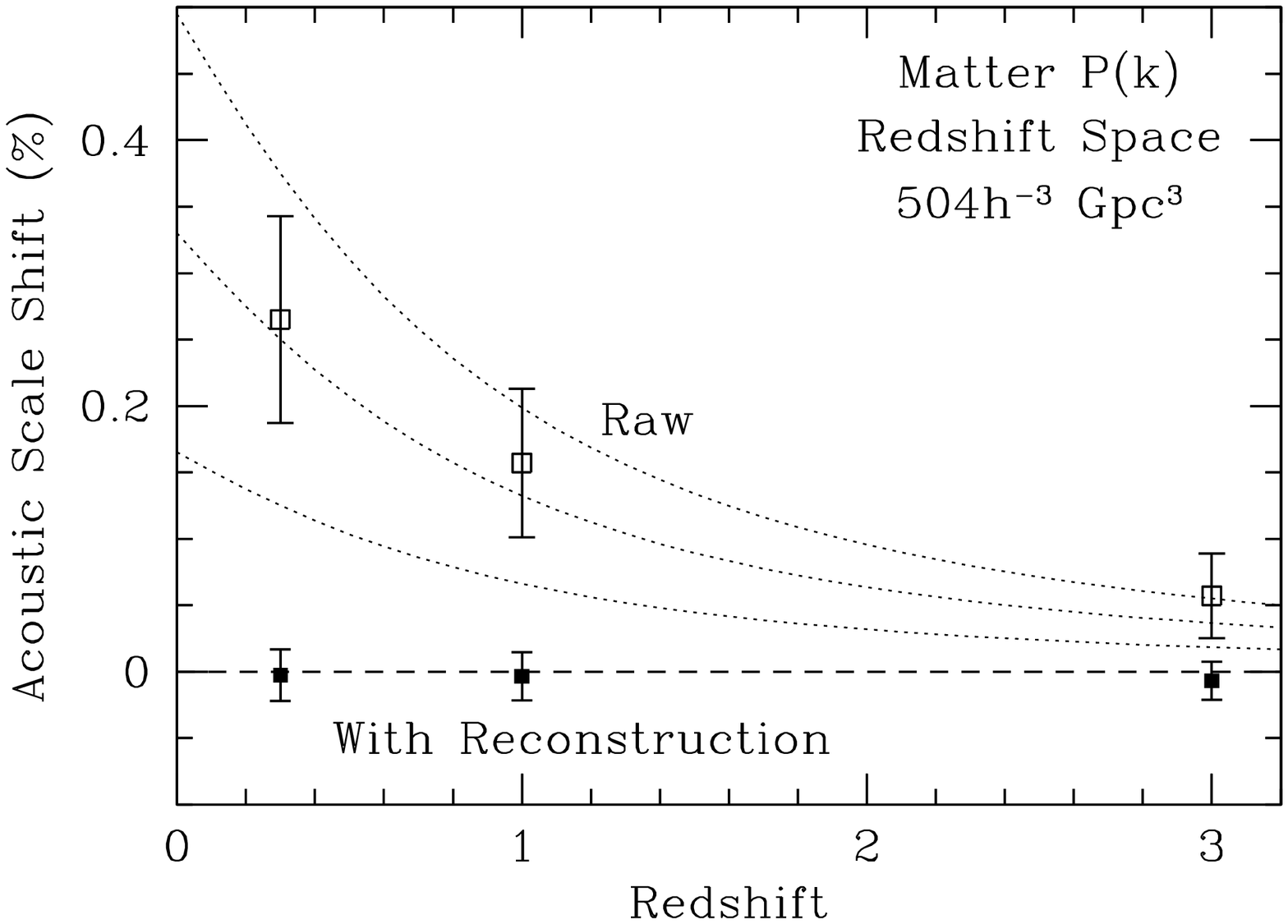}
    \quad  \includegraphics[width=3.2in]{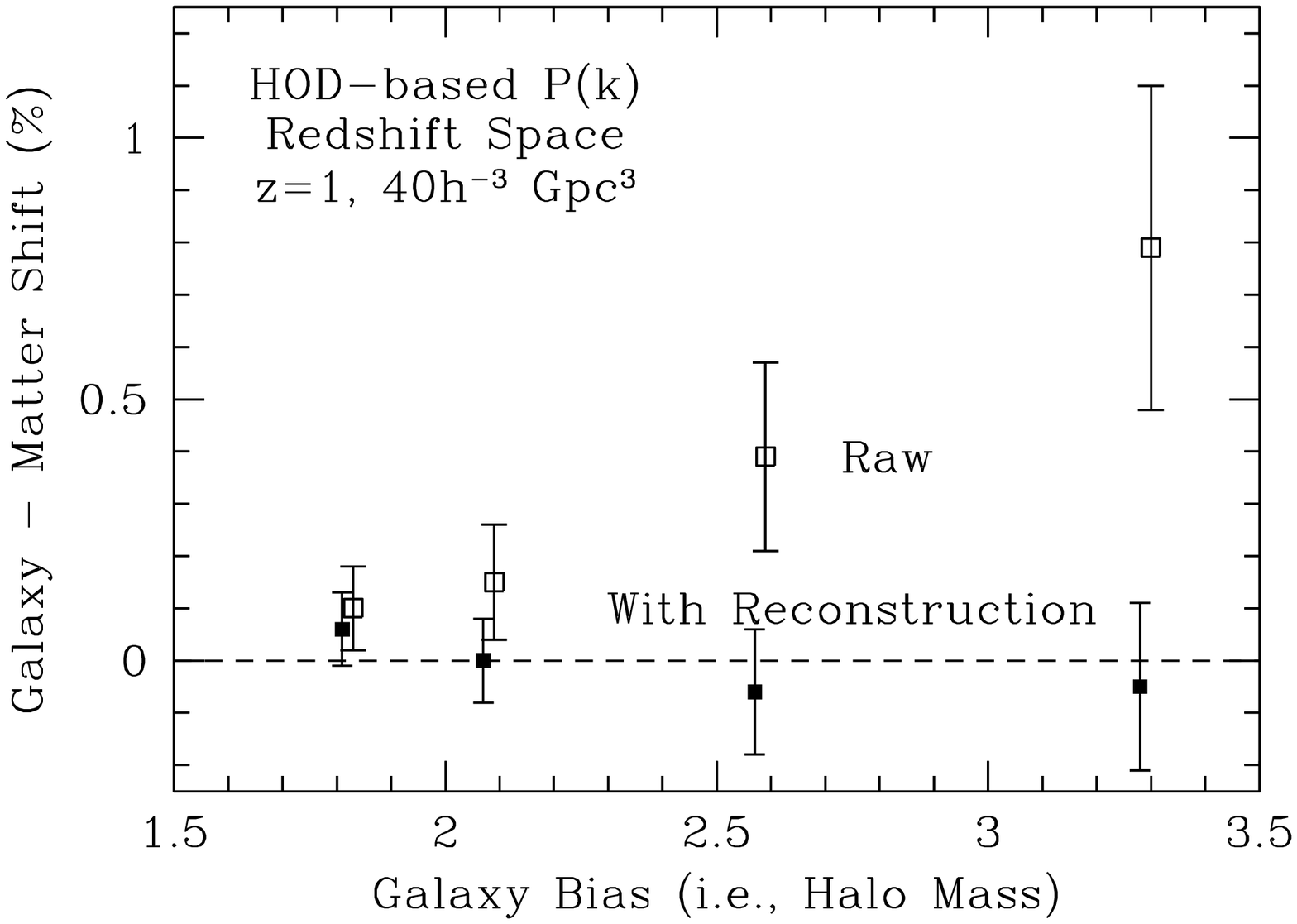}}
\caption{\label{fig:bao_alpha}
The shifts of the acoustic scale in cosmological N-body simulations.
(Left) Shifts of the acoustic scale in the redshift-space matter
power spectrum versus redshift from \protect\citet{seo10a}.  The
open symbols show the acoustic scale shifts prior to reconstruction;
the dashed lines show a scaling of the square of the linear growth function.
The solid symbols show the shifts after reconstruction is applied.
The error bars are derived from the variance among simulations.
(Right) Shifts of the acoustic scale in the redshift-space power spectrum
of mock galaxy distributions at $z=1$ from \protect\citet{mehta11}.  The
acoustic scale shift from the matter distribution in the same boxes has
been subtracted so as to decrease sample variance.  Galaxies
are placed via HOD prescriptions; increasing mass thresholds
leads to lower number densities and higher clustering bias.  The open
symbols show the shifts prior to reconstruction; the solid symbols,
after reconstruction.  The errors in the right panel 
are larger due to the smaller simulated
volume and the lower number density of tracers.
In all cases of both panels, reconstruction decreases the errors on the
acoustic scale and reduces the shift to be consistent with zero.
The left panel is based on 63 simulations, each using $576^3$ particles
in a $2\hgpc$ cube.  The right panel is based on 40
simulations, each with $1024^3$ particles in a $1\hgpc$ cube.
}
\end{figure}

The clustering of matter and galaxies undergoes substantial changes
at low redshift beyond the growth described by linear perturbation theory.
Small-scale structure grows non-linearly, peculiar velocities behave
differently from their linear prediction, and galaxies trace the
dark matter in a complicated manner.  We should worry that 
these effects might modify the location of the BAO feature relative
to the prediction of linear theory, thus distorting our standard ruler
\citep{meiksin99,seo05,angulo05,springel05,jeong06,huff07,angulo08,wagner08}.

Fortunately, the large scale of the acoustic peak insulates it from
most of non-linear structure formation \citep{eisenstein07a}.  A
typical pair of dark matter particles changes its comoving separation
by $10\hmpc$ (rms value) between high redshift and $z=0$.  These
motions broaden the acoustic peak, but the rms displacement is only
mildly larger than the $8\hmpc$ scale set by Silk damping.
The apparent displacement along the line of sight is larger in
redshift space, because the peculiar velocity is well correlated
with the displacement.
Figure \ref{fig:bao_sims} shows the correlation function and power
spectrum from N-body simulations; one can see that the acoustic
peak in the correlation function becomes broader at low redshift.
The corresponding effect in the power spectrum is the decreased
amplitude of the wiggles at higher wavenumber.  
Roughly speaking, one can think of the width of the evolved $\xi(r)$
peak as the quadrature sum of the initial width and the rms pairwise
displacement $\Sigma_{\rm NL}$ (see \citealt{orban11}, who examine
idealized BAO models numerically and analytically).  Equivalently,
the oscillations in $P(k)$ are damped by a factor 
$\exp(-k^2\Sigma_{\rm NL}^2)$.
As discussed in \S
\ref{sec:bao_theory_fitting}, the broadening of the BAO feature
does not significantly
bias the acoustic scale measurement provided one is using a suitable
template-fitting method.  However, it does degrade the precision
of the measurement for a given survey volume, as it is harder to 
centroid a broader feature.

To change the acoustic scale itself, one needs instead to move pairs
systematically closer or systematically further away.  This is a much
weaker effect than the rms motion of particles, as it depends on the
density variations in 150 Mpc spheres, which are percent level.  Moreover,
pairs of overdensities fall toward each other and pairs of underdensities
fall away from each other, and both situations count equally
toward a two-point statistic, causing a partial cancellation.

\citet{padmanabhan09} compute the change in the acoustic peak location
at second-order in gravitational perturbation theory.  \citet{crocce08}
have done similar calculations in renormalized perturbation
theory.
Both calculations reveal a second-order term
of the form $d\xi/dr$, which corresponds to moving the acoustic peak.
\citet{padmanabhan09} compute the size of this effect to be
around 0.25\% at $z=0$.

N-body simulations reveal a similar story.  \citet{seo10a} measure
the shift in the acoustic scale
in a large volume of simulations and detect a shift from $\alpha=1$
of $0.3\% \pm 0.015\%$ at $z=0.0$, with a scaling in redshift proportional
to the square of the linear growth function as expected for a second
order effect (left panel of figure \ref{fig:bao_alpha}).
\citet{padmanabhan09} validate their analytic
calculation with a similar set of simulations.

Redshift-space distortions have further effects on the BAO signal beyond
the extra broadening from the large-scale peculiar velocity.
Small-scale velocities, e.g., the Finger of God effect, blurs the
measurement of clustering along the line of sight, thereby broadening
the acoustic peak.  Moreover, the peculiar velocities create anisotropy in
the broadband clustering, which must be carefully accounted for when
extracting the acoustic scale (\S\ref{sec:bao_theory_fitting}).

Linear bias, with galaxy density contrast $\delta_g = b\delta_m$,
changes the amplitude of $\xi(r)$ or $P(k)$ but not the shape.  However,
any realistic bias relation must be at least somewhat non-linear,
which alters the relative weighting of overdense and underdense
regions and should shift the acoustic scale at second order.
Early work attempted
to measure this shift in simulations \citep{seo03,angulo08}, but
the volume of the simulations was insufficient to get a conclusive detection
of the effect.  More recently, \citet{padmanabhan09} explored
galaxy bias as the ratio of the second-order to first-order bias
term, finding shifts of a few tenths of a percent for reasonable bias cases.
\citet{mehta11} treated the problem numerically with halo occupation
distributions, finding shifts of 0.1\% to 0.8\% at $z=1$ depending
on the strength of the bias (right panel of figure \ref{fig:bao_alpha}).
For halo-based models or other prescriptions that tie galaxy bias
to the local density field, it therefore appears that bias-induced
shifts are small, and corrections of modest fractional accuracy
(e.g., to 20\% of the shift itself) will suffice to make them
negligible.  The relevant bias parameters should be tightly
constrained by smaller scale clustering measurements and higher order
statistics, enabling cross-checks of the model used for correction.

Non-local bias models that tie galaxy formation efficiency directly
to the environment on much larger scales 
(e.g., \citealt{babul91,bower93}) could perhaps induce larger
shifts of the acoustic scale.  However, such models require
fairly extreme physical effects, and they can be readily diagnosed
via their impact on clustering at scales below the BAO scale
\citep{narayanan00}.  A survey capable of measuring the acoustic
scale to the sub-percent statistical level will provide in its
millions of galaxies extensive opportunities to constrain even
very general bias models accurately enough to predict the
acoustic scale shift to within 10-20\% of its value, sufficient
to bring the systematic error below the statistical error.

\subsubsection{Reconstruction}
\label{sec:bao_theory_recon}

\begin{figure}[t]
\centerline{\includegraphics[width=2.3in]{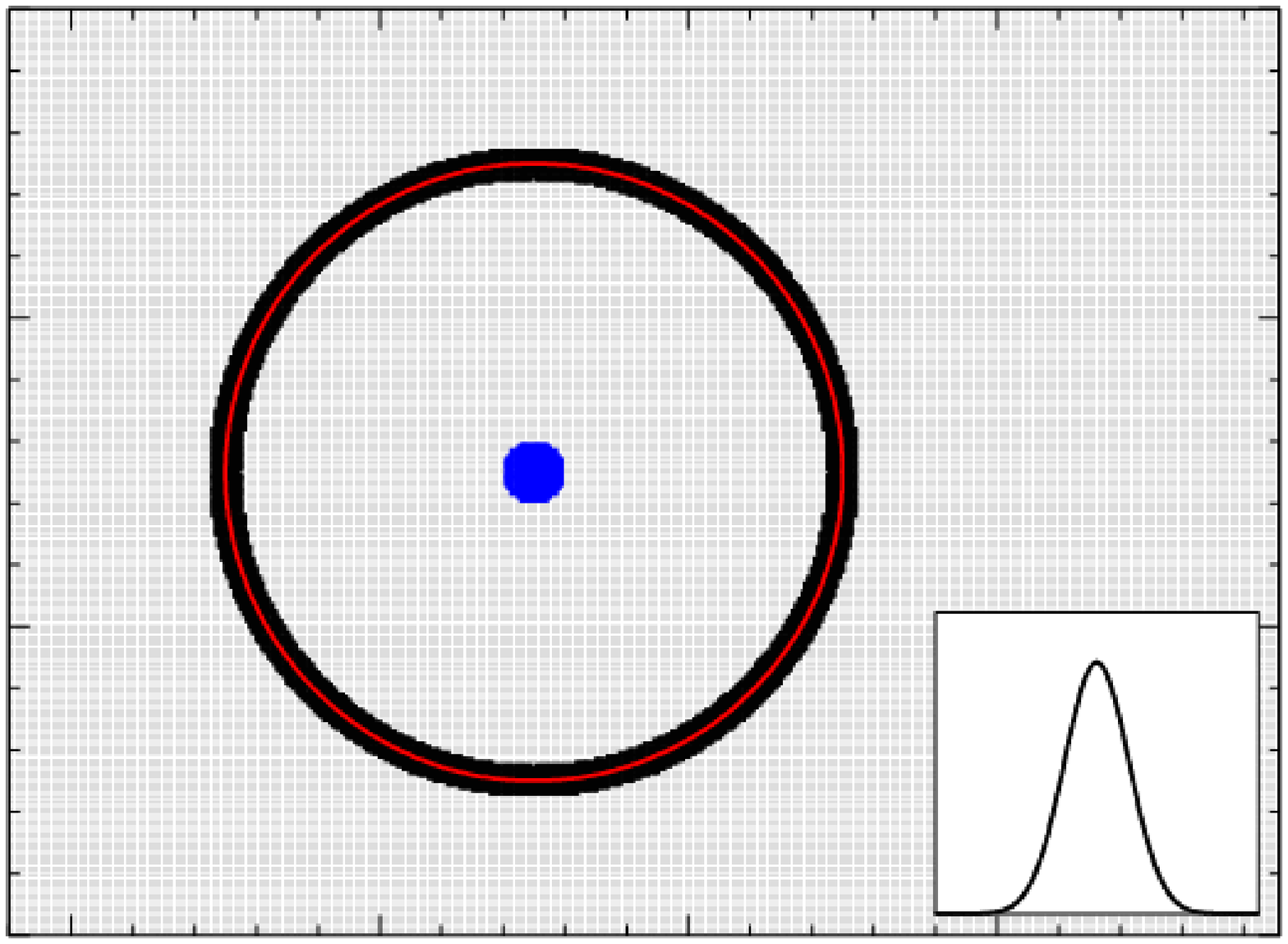}
    \quad  \includegraphics[width=2.3in]{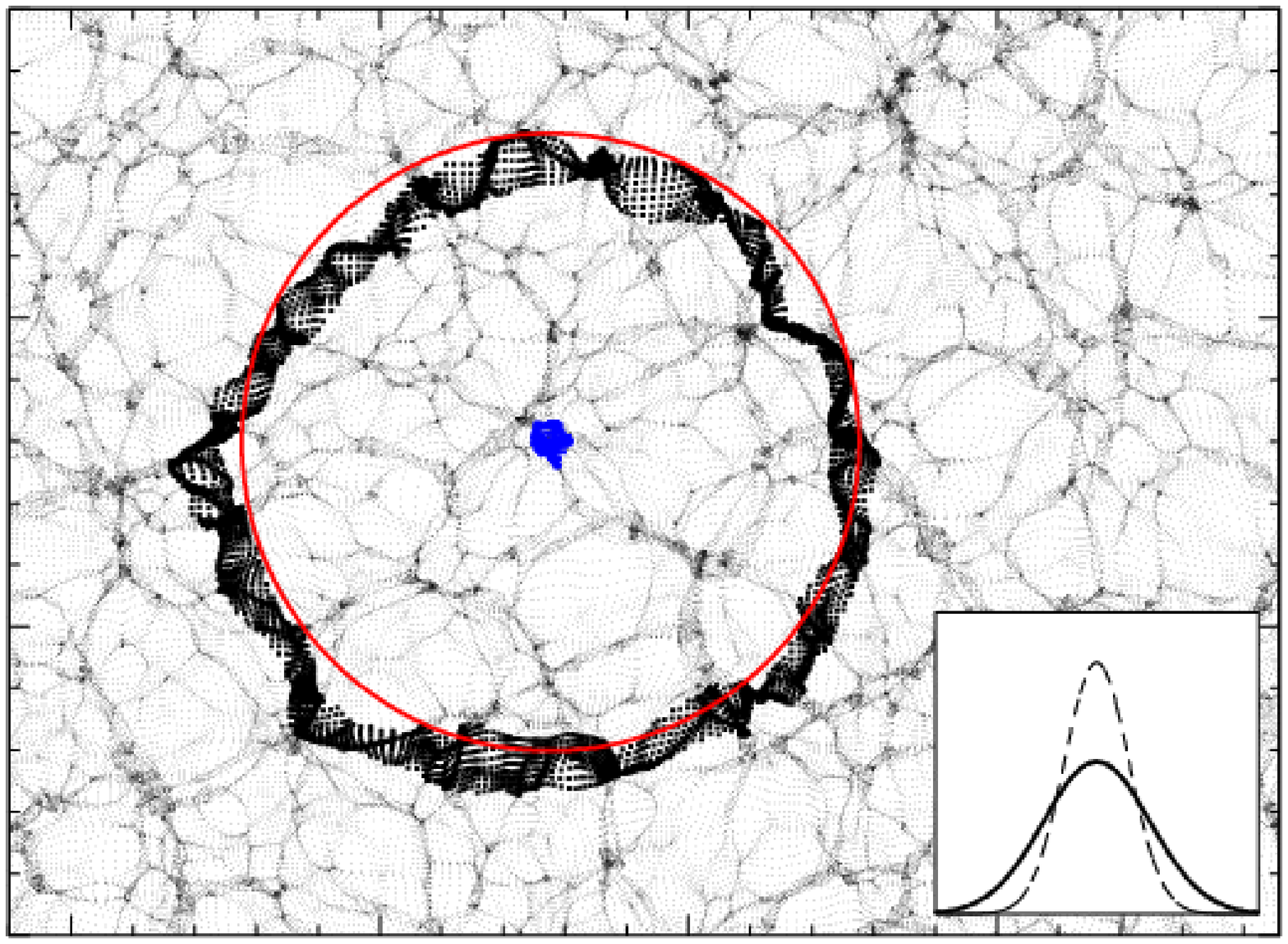}}
\centerline{\includegraphics[width=2.3in]{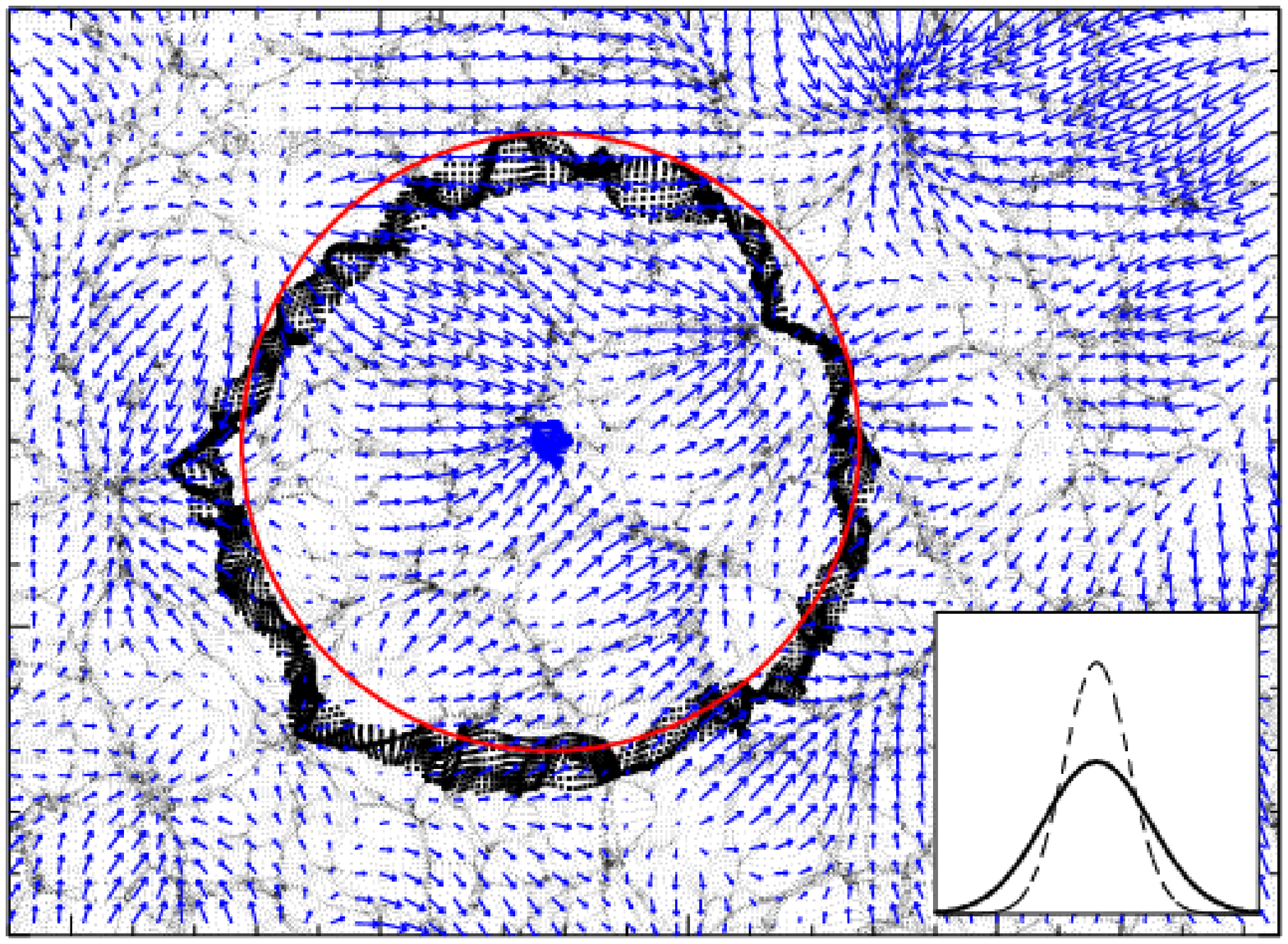}
    \quad  \includegraphics[width=2.3in]{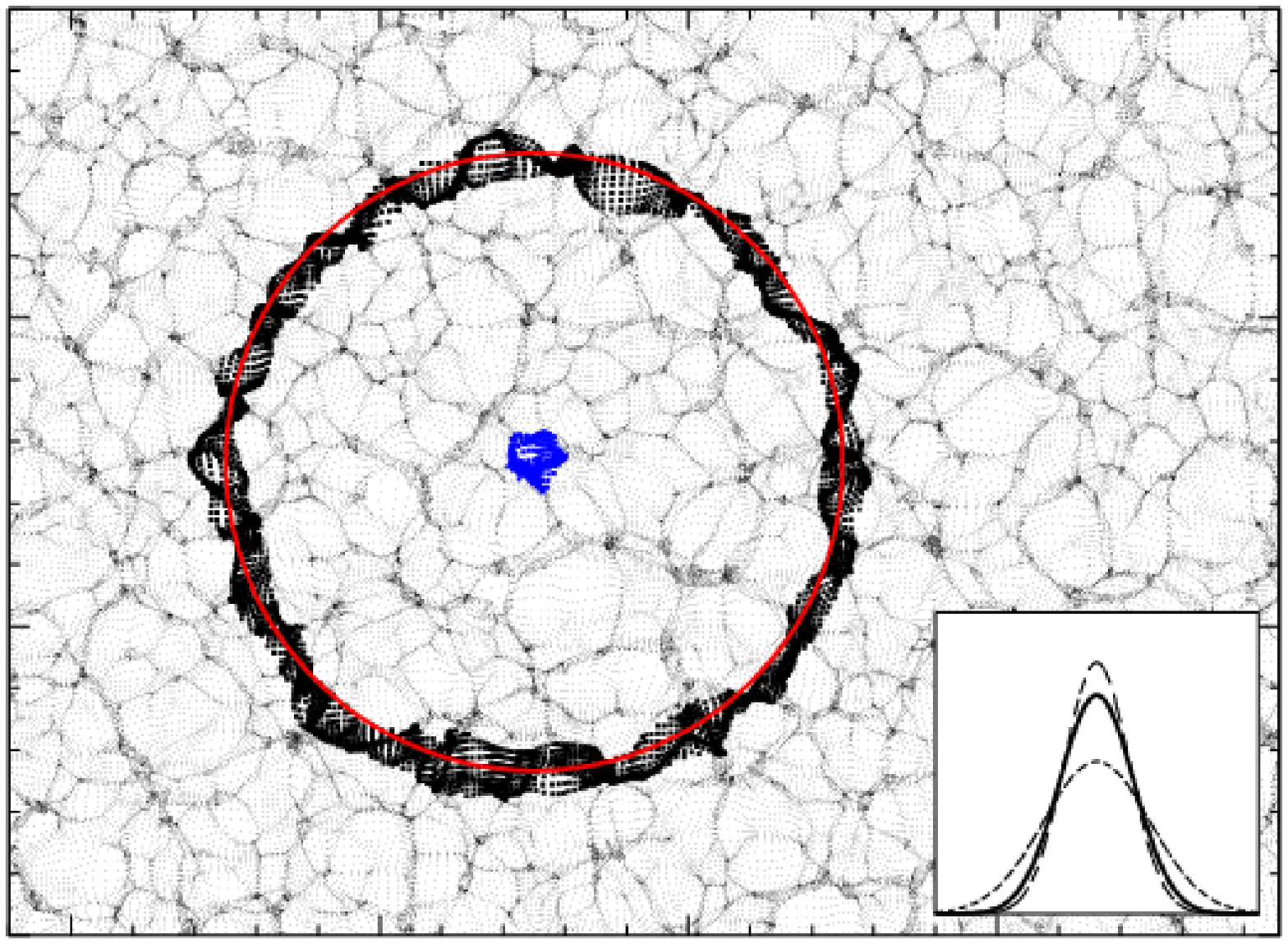}}
\caption{\label{fig:bao_recon}
A pedagogical illustration of how reconstruction can improve the
measurement of the acoustic scale; this figure is
from \cite{padmanabhan12}.  
Each panel shows a thin slice of a cosmological density field.
(Top Left) At early times, the density is nearly constant.  We mark
a set of points at the origin in blue and a ring of points at 150 Mpc
in heavy black.  We measure the distances between the black points and the
centroid of the blue point; the rms of these distances is represented
by the Gaussian in the inset.
(Top Right) At later times, structure has formed (in this calculation,
simply by the Zel'dovich approximation), and the points have moved.  
The red circle shows the initial radius of the ring, 
centered on the current centroid
of the blue points.  The fact that the black points no longer fall on
the red ring indicates that the acoustic peak has been broadened.  The
inset shows that the new rms of the radial distance (solid line)
is larger than the original (dashed line).
(Bottom Left) Arrows show the Zel'dovich displacements responsible for the
structure that has formed.  The idea of reconstruction is to estimate
these displacements and move the particles back.
(Bottom Right) We illustrate this by smoothing the density field by
a 10$h^{-1}$ Mpc filter and moving the particles back.	Because
the displacement field is imperfectly estimated, small-scale structure
remains.  But the black points now fall closer to the red ring, so
that the rms of the radial distance is close to the initial (inset).
The actual reconstruction algorithm of \cite{padmanabhan12}
is more complex, but this example shows the basic opportunity.
}
\end{figure}

By broadening and shifting the BAO feature in $\xi(r)$, non-linear
gravitational evolution degrades BAO precision and introduces a
possible systematic.  Is it possible to remove these effects by
``running gravity backwards'' to reconstruct the linear density field?
The \cite{zeldovich70} approximation --- in which particles follow
straight line trajectories in comoving coordinates at the rate
predicted by linear perturbation theory --- captures important 
aspects of non-linear evolution on large scales
(e.g., \citealt{weinberg90,melott94}).
\cite{eisenstein07} show that a simple reconstruction scheme
based on applying the (reversed) Zel'dovich approximation to the
smoothed non-linear density field is remarkably successful at
recovering BAO information, effectively shifting the low redshift
curves in Figure~\ref{fig:bao_sims} back towards the high redshift curves.
Figure~\ref{fig:bao_recon}, from \cite{padmanabhan12},
illustrates how reconstruction works in the idealized case of
an initial perturbation that exactly mimics the ``acoustic ring''
pattern.

\citet{seo07} and \citet{seo10a} investigate the effects of reconstruction
in more detail, showing that it noticeably
improves the scatter and decreases the shift of the recovered acoustic
scale from the matter density field of N-body simulations.  
The latest simulations demonstrate
that the non-linear shift of the scale has been removed to 0.02\%
or better (see Figure \ref{fig:bao_alpha}, left).
Moreover, comparing the initial conditions to final conditions on a
mode-by-mode basis shows that the linear density field has been
recovered to roughly double the pre-reconstruction wavenumber.  
\citet{padmanabhan09a}
analyze the method analytically, revealing the improvement while also
noting that the recovered density field is not exactly the linear one.

\citet{mehta11} extend this analysis to HOD-based mock galaxy catalogs
in simulations.  They consider a range of HOD prescriptions and find that the
reconstruction of the linear density field is not degraded by this form of
galaxy bias and that the shift of the acoustic scale after reconstruction
still vanishes, this time to 0.1\% precision
(Figure \ref{fig:bao_alpha}, right).  This success is not
surprising: the halo field traces the matter field fairly accurately
on the scales required for
reconstruction, so one is correctly estimating and removing the large
scale displacements.  Non-linear galaxy bias still alters
the weighting of convergent and divergent flows, but if the flows 
are being mostly removed, then it doesn't matter how they are weighted.

\begin{figure}[t]
\centerline{\hfil \includegraphics[width=2.3in]{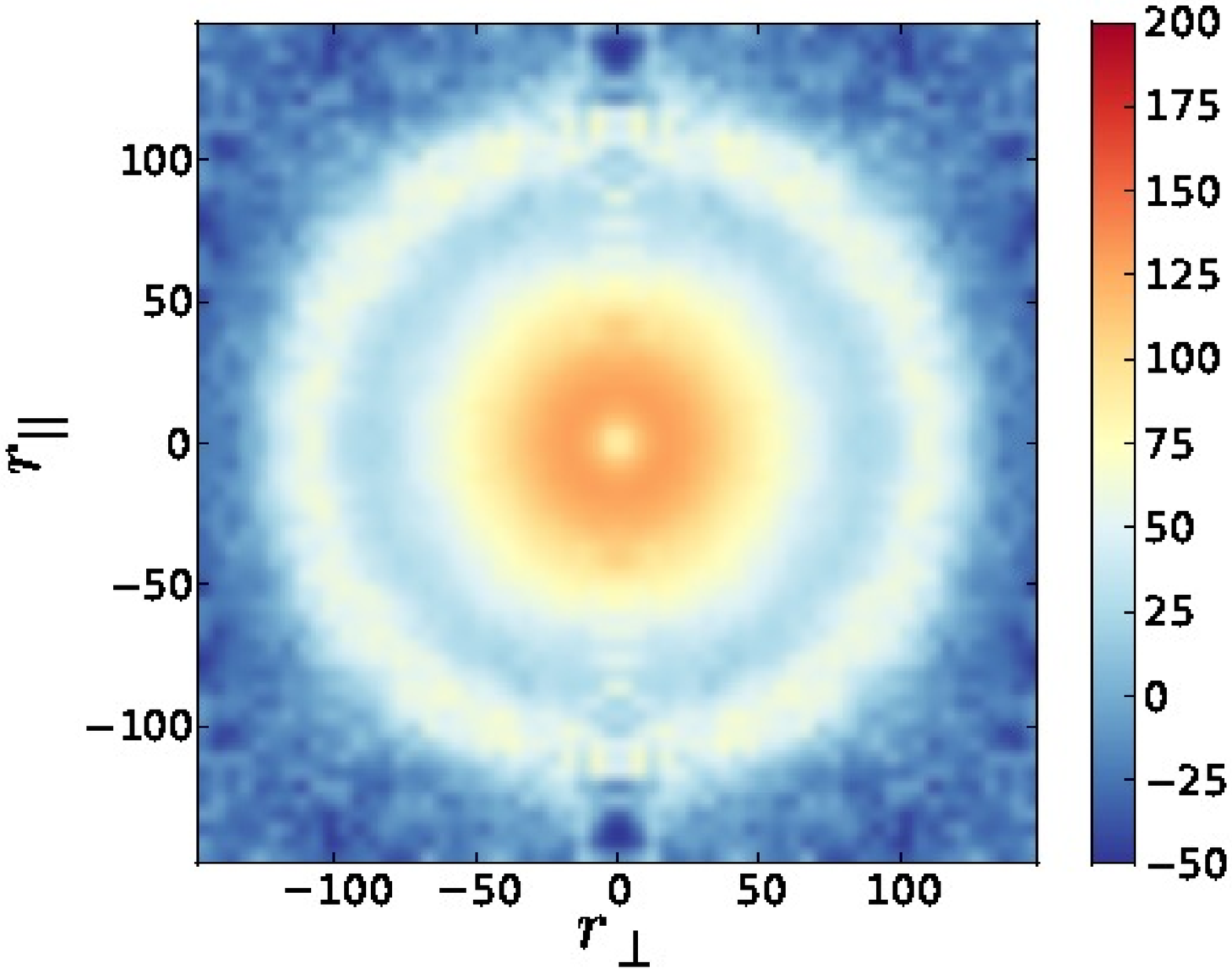}
    \hskip -0.20truein  \includegraphics[width=2.3in]{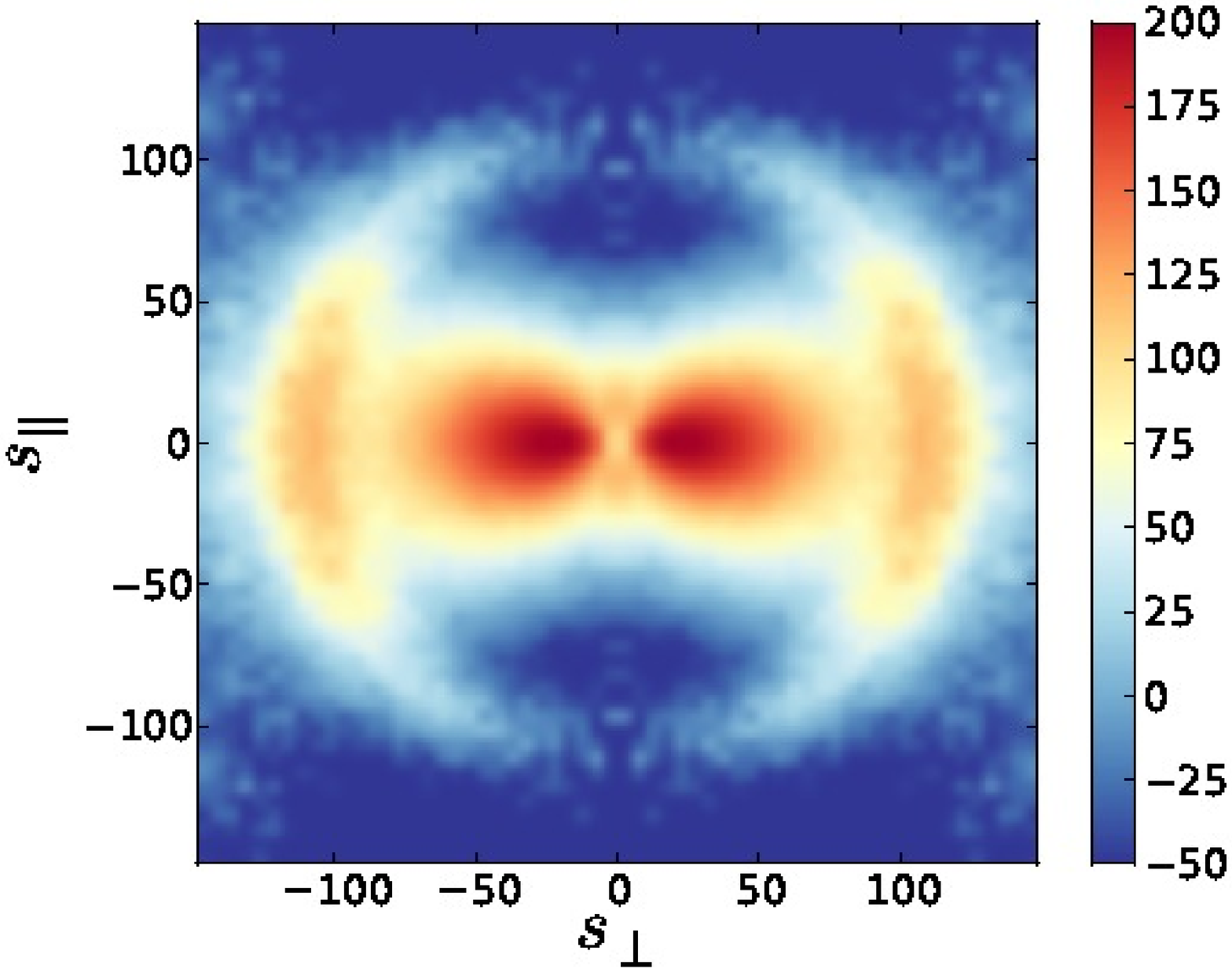}
    \hskip -0.20truein  \includegraphics[width=2.3in]{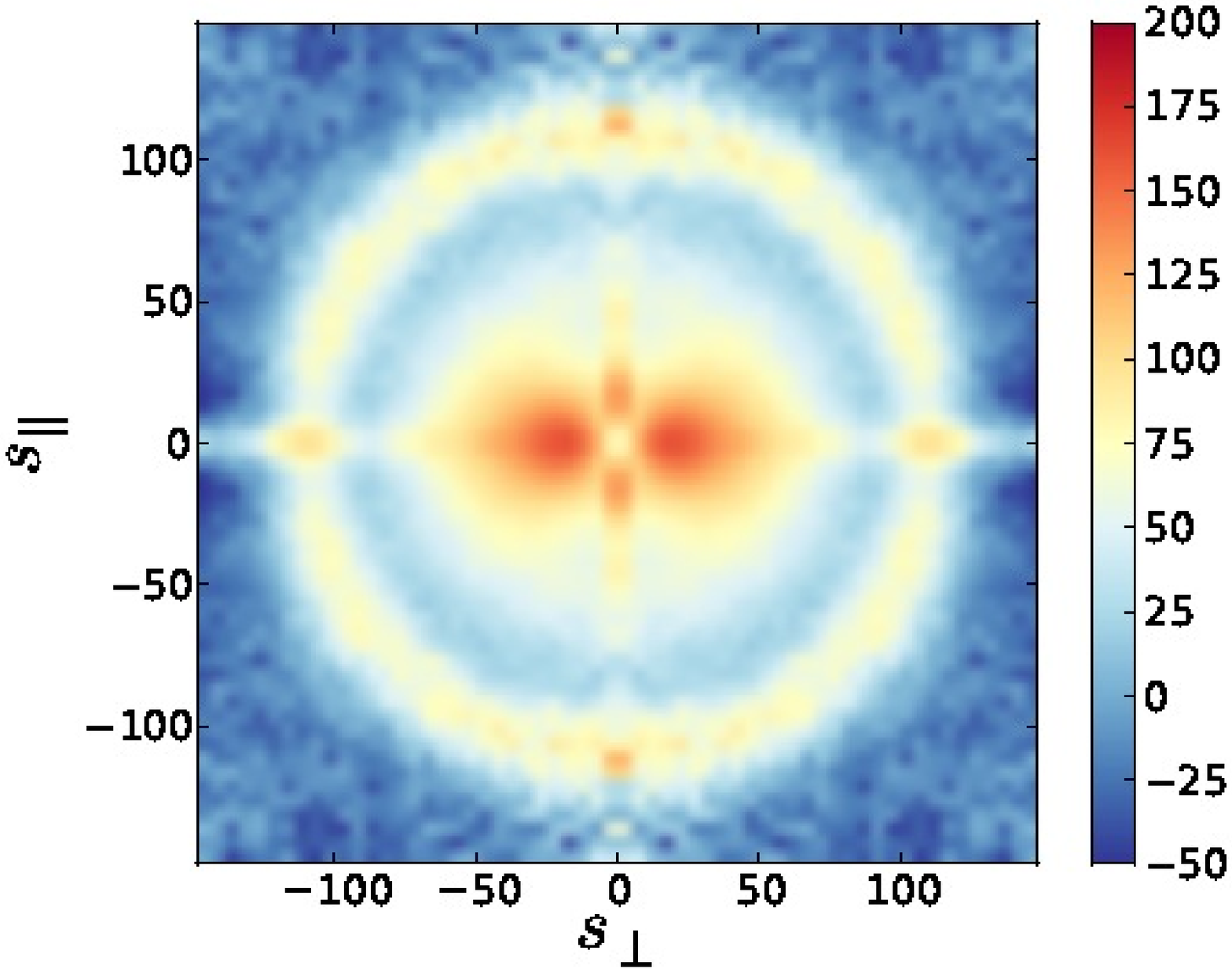}
	\hfil
}
\caption{\label{fig:bao_recon_zspace}
The ability of reconstruction to correct redshift-space
distortion of the BAO feature; figure panels taken from
\cite{padmanabhan12}.  
The left panel shows contours of the galaxy correlation function
from N-body mock catalogs, multiplied by $r^2$ to enhance the
large scale features, as a function of transverse ($r_\perp$)
and line-of-sight ($r_\parallel$) separation.
Apart from statistical fluctuations, this correlation function
is isotropic.  The middle panel shows contours of the galaxy
correlation function in redshift-space.  Peculiar 
velocities induce anisotropy, breaking the BAO ring into
two arcs.  The right panel shows the redshift-space correlation
function of the galaxy distribution after reconstruction,
including peculiar velocity correction at the level
of the Zel'dovich approximation,
which largely restores the isotropy of the BAO ring.
}
\end{figure}

Reconstruction is thus a powerful tool: one is achieving better
statistical precision for a given survey, typically by a factor of
1.5 to 2, equivalent to a factor of $2-4$ increase in survey size.
Meanwhile, one is mitigating the primary systematic error
from non-linear clustering and galaxy bias.  As an added benefit,
one can use the estimate of the large scale displacements
to remove large-scale redshift-space distortions,\footnote{The 
line-of-sight peculiar velocity is, in the Zel'dovich approximation,
equal to $f(z) H(z)$ times the line-of-sight component of the
3-d displacement.} 
as illustrated in Figure~\ref{fig:bao_recon_zspace}, 
which decreases
% (see fig.~4 of \citealt{padmanabhan12}), decreasing
that degradation of the BAO
accuracy and also pushes most of the BAO signal into the monopole and
quadrupole components of $\xi({\bf s})$ or $P({\bf k})$.
Without reconstruction, the redshift-space distortions contain significant
terms in the hexadecapole and beyond, and the quadrupolar squashing of
the Alcock-Paczynski effect couples to the quadrupole redshift-space distortion
to produce BAO signal in the hexadecapole.
To the extent that one is recovering the linear density field, one
can also hope that the large-scale density field is more Gaussian,
which is a major simplification for computing likelihood functions.
However, this last property has not been extensively tested.

The first applications of BAO reconstruction to observational
data appear in \cite{padmanabhan12} and \cite{anderson12}.
For the SDSS-II (DR7) LRG sample, reconstruction shrinks the
BAO distance error from 3.5\% to 1.9\%, equivalent to a factor
of three increase in survey volume.  It also improves the agreement
between the observed and predicted shapes of the correlation 
function in the BAO regime, thereby increasing the statistical significance
of the BAO detection from $3.3\sigma$ to $4.2\sigma$.
For the BOSS (DR9) sample, reconstruction produces 
little improvement in the BAO measurement, although this 
is consistent with the variation among DR9 mock catalogs
(see discussion by \citealt{anderson12}).

There is an extensive literature on reconstruction methods
for large-scale structure (e.g., 
\citealt{peebles89,weinberg92,nusser92,croft97,narayanan98,mohayaee06}).
Even simple methods appear adequate for BAO recovery, but better 
reconstruction is valuable for other applications of large-scale
structure \citep{reid10}.  Since BAO surveys are typically sparse,
an important area for continuing research is the performance of
methods in the presence of both galaxy bias and significant shot noise.
The effectiveness of reconstruction as a function of sampling
density might have important implications for survey design, favoring
different choices compared to the statistical considerations
discussed in \S\ref{sec:bao_obs} below.

\subsubsection{Fitting to Data}
\label{sec:bao_theory_fitting}

It is worth stressing that ``the acoustic scale'' is only an approximate
description of a more complicated physical situation.  For high
precision work, we cannot separate the concept of the acoustic scale
from the context of a Boltzmann code prediction for the
matter power spectrum and CMB anisotropy power spectrum.  The
sound horizon defined by equation~(\ref{eq:acoustic_scale}) does not 
correspond to the exact maximum of the
acoustic peak in the correlation function, nor do the harmonics
in the matter power spectrum have an identical scale to those in the
CMB anisotropy spectrum.  The differences arise from effects such as
the fact that photons decouple from the baryons earlier than the baryons
decouple from the photons, that the post-recombination matter growing mode
is largely set by the velocity perturbation at recombination rather
than the density perturbation, and that Silk damping alters the effective
redshift of recombination as a function of wavenumber.	These effects
are accurately calculated in the Boltzmann codes, resulting in precise
predictions for the matter and CMB power spectra.

When one wants to extract the acoustic scale (i.e., to measure distance
using the
BAO standard ruler) from a measurement of the two-point clustering, the
appropriate thing to do is to use the predicted clustering for the cosmology
one is testing as a template.  The optimal plan is then to fit that template
to the data over a range of scales using the correct covariance matrix or
likelihood.
Some early works instead used non-parametric models for the
acoustic peak, such as a Gaussian in configuration space or a damped
sinusoid in Fourier space \citep{blake03}, or simply identified the
maximum of the correlation function \citep{guzik07,smith08}.  We believe that,
because the acoustic scale is predicted only in the context of
an early universe model with parameters taken from fits to CMB data, there
is no extra value in avoiding the linear-theory model predictions.

Having said that, one does want to modify the template to allow for effects
of non-linear structure formation and perhaps to marginalize over broad-band
terms that might enter from scale-dependent clustering bias or velocity
bias or from errors in the calibration of one's survey.  This procedure
has been carried out by several different authors.  For example,
\citet{seo07} and and \citet{seo10a} fit the measured 
power spectrum to the form
\begin{equation}
\label{eqn:pmeasure}
P_{\rm measure}(k) = B(k) P_m(k/\alpha) + A(k)~,
\end{equation}
where $A(k)$ and $B(k)$ are smooth functions with parameters to be fit.
$P_m(k)$ is
the linear theory model with the acoustic oscillations additionally
damped by large-scale structure,
\begin{equation} 
\label{eqn:pkfit}
P_m(k) = \exp(-k^2 \Sigma_{\rm NL}) 
(P_{\rm lin}(k)-P_{\rm nw}(k)) + P_{\rm nw}(k), 
\end{equation}
where $\Sigma_{\rm NL}$ is a constant fit from simulations.
$P_{\rm nw}$ is the no-wiggle power spectrum from \citet{eisenstein98},
hand-crafted to edit out the acoustic oscillations.
$P_{\rm lin}$ is the exact linear theory power spectrum; 
note that $P_m(k)$ goes to this exact linear theory result
in the limit of negligible damping ($\Sigma_{\rm NL} \rightarrow 0$),
so the approximate form of $P_{\rm nw}(k)$ is acceptable.
The broadband terms
$A(k)$ and $B(k)$ will correct for non-linear power, 
the shot noise, scale-dependent bias, and any imperfections
in the survey.
The primary goal for the fit is to measure $\alpha$, 
a factor that dilates the scale of the
predicted clustering (and of the BAO feature in particular) 
relative to the observed clustering.
A value $\alpha=1$ indicates agreement with the
acoustic scale of the original model.  A value $\alpha\ne1$ indicates
that the acoustic scale of the linear-theory model is incorrect or that
the distance scale assumed in measuring the galaxy clustering was wrong.
Simple alterations of this prescription can be made
for fitting
the correlation function or mixed-space $\omega$ or wavelet statistics
\citep{xu10,arnaltemur11}.
This appproach thus allows one to fit for the scale of a standard
ruler without having to recompute a full predicted power spectrum
at every point in parameter space.

This fitting procedure is only compelling to the extent that 
the recovered value of $\alpha$
is stable (to within the statistical errors) as one varies the prescription
for the marginalization of parameters in $A(k)$ and $B(k)$.  Too little
freedom and one may be biased by broadband tilts and modulations that one
hasn't modelled properly; too much freedom and one will fit out the
acoustic signature and reduce the constraining power of the data.
Fortunately,
the separation between the acoustic scale and the typical non-linear scale
and Silk damping scale is large, i.e., the acoustic peak in the correlation
function is narrow.  This gives considerable freedom to fit away broadband
nuisance terms while not impacting the acoustic peak.  \citet{seo10a} show
stable results for $\alpha$ for various choices, e.g., polynomials of
different
order.	Similarly, $\alpha$ is robust to changes in the choice of
$\Sigma_{\rm NL}$,
so one is not sensitive to how one estimates that parameter in simulations
or mock catalogs.

An equivalent method has been used by \citet{percival07,percival10},
and in related works.
Here, one fits a spline to the measured power spectrum and divides
by that spline.  One does similarly for the template $P_m(k)$ and 
fits that to the residual spectrum of the data.  This is equivalent to
taking $B(k)$ to be a spline and setting $A(k)=0$.  Clearly the performance
depends on the number of spline points, but there is a broad stable region.

The definition of the acoustic scale as the distance a primordial
sound wave could travel before recombination
(eq.~\ref{eq:acoustic_scale}) 
is borne out in
such fits.  If one fits with the
power spectrum from a cosmology that is moderately wrong, then one infers
a different $\alpha$, but this change in $\alpha$ is proportional to the
ratio of the acoustic scales, as defined by the sound horizon integral
for each cosmology.  The stability of this scaling appears to be much
better than the statistical errors implied by the surveys that are defining
the range of interesting cosmological parameter space \citep{seo10a}.
In other words, one can use the acoustic scale integral to
adjust distance scale measurements of $D_A/r_s$ and $Hr_s$
between different cosmologies within the domain of interest.

Extending these approaches to the anisotropic case so as to extract $D_A$
and $H$ separately is more complicated and has not been fully developed.
The primary obstacle is to account for the anisotropic distortions
from peculiar velocities.
%\tbd{If we add the Padmanabhan ring figure, we should reference
%it here as showing the potential of reconstruction to improve
%angle-los separation.}
Examples of fit methodologies include \citet{okumura08},
\citet{padmanabhan09}, \citet{shoji09},
\citet{chuang11}, \citet{kazin11}, and 
\citet{xu12b}.
The ability of reconstruction to mitigate peculiar velocity
distortion (Fig.~\ref{fig:bao_recon_zspace}) may be a 
significant asset for disentangling $D_A$ and $H$.

With better modeling of non-linear structure and galaxy clustering bias,
one could of course extract additional cosmological information from the
two-point clustering of galaxies.  In particular, one can measure the
distance scale from the curvature (i.e., non-power-law form) of the
spherically-averaged power spectrum or correlation function.  This
physical scale arises from the size of the horizon at matter-radiation
equality, parameterized as $\Omega_m h^2$ in typical cosmologies.  However,
this curvature is a much broader feature and thus provides less leverage
on distance.  Most important, the width of the feature is comparable to
the scale itself, implying that one must control all extra broad-band sources
of power and scale-dependent galaxy biases in order to extract 
accurate distance information.  This is
much more challenging than the BAO application, but it is an important 
frontier
of the field of large-scale structure.	In particular, the application
of this approach to the quadrupole distortion known as the Alcock-Paczynski
effect will be discussed in \S \ref{sec:ap}.

\subsection{Observational Considerations}
\label{sec:bao_obs}

\subsubsection{Statistical Errors}
\label{sec:bao_obs_stats}

%%% {\it
%%%	Approximate formula for cosmic variance limit
%%%
%%%	Fisher matrix forecasts
%%%
%%%	Table of $H(z)$ and $D_A$ sensitivity, maybe a figure
%%% }
%%%
%%% {\bf My suggestion is to split the discussion so as to first
%%% say what the precision is for sample variance alone, then
%%% turn to shot noise in the next subsection.	We want to
%%% give some intuitive argument for the statistical error
%%% and its scaling in the sample variance limit and a formula
%%% for this scaling, before diving into quotation of Fisher
%%% matrix results.  You have good intuition about how errors
%%% depend on various things; we want to get that intuition
%%% down on paper.}

The primary challenge of the BAO method is that very large samples
of galaxies (or other tracers) are required to detect the acoustic
oscillations and hence measure a distance.  Like detecting
an emission line in a galaxy spectrum in order to measure a redshift,
one must have high enough signal-to-noise to detect the BAO peak
or one gets no useful distance information at all.
The minimum useful survey volumes are
of order $1h^{-3}$ Gpc$^3$, which yield a distance precision of about
5\%.

The two components of the statistical error are sample variance and
shot noise \citep{kaiser86}.  A given survey volume contains only a certain
number of Fourier modes; in the periodic box approximation
$dN_{\rm modes}/dk = 4\pi k^2 V/[(2\pi)^32]$, where the final
factor of 2 in the denominator 
accounts for the fact that the density field is real.
In a Gaussian random field, the real and imaginary parts of each mode
are independent with an intrinsic variance of $P(k)/2$.
In addition, each mode is imperfectly measured due to shot noise;
when treated in the Gaussian approximation (ignoring the 4-point
contributions from the Poisson distribution), this raises the variance
on the square of the complex norm to $[P(k)+1/n]^2$, where $n$ is the number
density of tracers.  The result is that the fractional
error bar on
the measurement of each mode is $\sigma_P/P = (nP+1)/nP$.  When combining
information from modes, we should sum the inverse variances, which are
\begin{equation} 
\label{eqn:pkerror}
{P^2\over\sigma_P^2} = \left(nP\over nP+1\right)^2 ~.
\end{equation}
We see that for $n\gg 1/P(k)$ we get unit information from each mode
but that the information drops rapidly for $n < 1/P(k)$.
We note that the relevant $P$ is
the redshift-space power spectrum; this can be substantially larger
than the real-space power spectrum for nearly radial large-scale
modes, thereby decreasing the shot noise impact on BAO estimation
of the Hubble parameter.

The mode-counting argument above neglects boundary effects,
effectively assuming that the survey volume is reasonably contiguous
with a high filling factor on scales of 150 Mpc.  In real
space, we can express this as the requirement that the number of pairs
of survey galaxies at 150 Mpc separation not be significantly diminished
compared to the case of a filled periodic box.  In Fourier space, we
must ensure that the survey window function not create aliasing between
modes in the crests and troughs of the acoustic oscillations.

Converting a power spectrum forecast into constraints on the distance
scale requires marginalizing over other cosmological parameters.  This
has been done with Fisher matrix analyses \citep{seo03,seo07}
or with Monte Carlo approaches \citep{blake03,glazebrook05,blake06}.
Several analyses have focused on the anisotropy of
the power spectrum in order to measure $H(z)$ and $D_A(z)$ separately.

\citet{seo07} constructed a fast approximation to the
full Fisher matrix calculation using an idealized treatment of the
acoustic oscillation, including non-linear structure formation and
redshift-space distortions.  This method allows forecasts for $H$ and
$D_A$ precision as a function of survey redshift, number density,
and volume (see their eq.~26). 
Tests with simulations \citep{seo10a,mehta11}
have shown this forecast to be
accurate to within 10-20\%, with a small trend toward over-optimism at
$nP<1$.  Whether this trend is intrinsic to shot noise or to the fact
that the low number density models used more massive mass thresholds
for halo bias is not clear at present.

\begin{deluxetable}{cccccccc}
\tablecolumns{8}
\tablecaption{BAO Forecasts for a Full-Sky BAO Survey\label{tbl:bao_cvl}}
\tablehead{
$z_{\rm min}$ &
$z_{\rm max}$ &
Volume &
\% Err $D_A(z)$ &
\% Err $H(z)$ &
$\Omega_\Lambda(z)$ &
S/N &
$\sigma_w$
}
\tablewidth{0pc}
\startdata
0.00  & 0.15  &   0.33 &   2.8 &  4.9  &    0.708 &   7.3 & 0.64  \\
0.15  & 0.32  &   2.62 &   0.95 & 1.7  &    0.616 &  18.2 & 0.088 \\
0.32  & 0.51  &   7.89 &   0.53 & 0.96 &    0.515 &  27.0 & 0.036 \\
0.51  & 0.73  &   16.5 &   0.35 & 0.63 &    0.413 &  32.9 & 0.021 \\
0.73  & 0.99  &   28.4 &   0.26 & 0.46 &    0.318 &  34.9 & 0.015 \\
0.99  & 1.28  &   42.9 &   0.21 & 0.36 &    0.236 &  33.3 & 0.013 \\
1.28  & 1.62  &   59.0 &   0.17 & 0.28 &    0.170 &  30.2 & 0.012 \\
1.62  & 2.00  &   75.8 &   0.14 & 0.24 &    0.119 &  25.2 & 0.013 \\
2.00  & 2.44  &   92.3 &   0.13 & 0.21 &    0.082 &  20.0 & 0.014 \\
2.44  & 2.95  &    108 &   0.12 & 0.18 &    0.056 &  15.5 & 0.016 \\
2.95  & 3.53  &    121 &   0.11 & 0.17 &    0.038 &  11.4 & 0.020 \\
3.53  & 4.20  &    133 &   0.10 & 0.15 &    0.025 &   8.3 & 0.025 \\
4.20  & 4.96  &    142 &   0.10 & 0.15 &    0.017 &   5.8 & 0.033 \\
\enddata
\tablecomments{
These forecasts assume a full-sky survey, use $nP=2$ at $k=0.2\invhmpc$, and
assume reconstruction improvements in the non-linear damping by a factor
of 2.  Statistical errors scale as $f_{\rm sky}^{-1/2}$.
The first and second columns
give the inner and outer edges of redshift bins;
the bins have equal width in $\ln(1+z)$.
The third column gives the comoving volume of the bin
in $h^{-3}$ Gpc$^3$, assuming $\Omega_m=0.25$.	The fourth and fifth
columns give
$1\sigma$ fractional errors (in percent) in $D_A(z)$ and $H(z)$, the
angular diameter
distance to and Hubble parameter at the bin center; note that
the errors on these two quantities are 40\% correlated.
We assume the sound horizon is known.
The sixth column gives $\Omega_\Lambda(z)$, i.e., the ratio of the
dark energy density to the critical density at that redshift
in a $\Lambda$-model.
Column 7 gives $\Omega_\Lambda(z) H/ 2\sigma_H$, which is the
significance at which one would detect the cosmological constant
at redshift $z$ using only the $H(z)$ BAO constraint and assuming
perfect knowledge of the matter density and curvature
(a good approximation, but not exact).
Column 8 shows the error on a constant value of $w$ that would
be obtained by comparing the BAO $H(z)$ measurement for this one
redshift to the value of $\rho_{\rm DE}$ at $z=0$, assuming
the latter is known perfectly.
}
\end{deluxetable}

Table \ref{tbl:bao_cvl} presents a summary of cosmic variance limited
BAO performance.  This is a tabulation of the \citet{seo07}
forecasts for a full-sky survey, using even binning in $\ln(1+z)$.  We
assume a shot noise level of $nP=2$ at $k=0.2\invhmpc$ (see
\S~\ref{sec:bao_obs_sampling}), and that reconstruction has decreased
the non-linear displacements by a factor of two in length scale,
i.e., reducing the quantity $\Sigma_{\rm NL}$ in 
equation~(\ref{eqn:pkfit}) by a factor of two below its
full non-linear value at each redshift.
Figure \ref{fig:bao_perform}, discussed further in the next section,
presents graphical summaries of the
main features of Table \ref{tbl:bao_cvl}.
One can see that the precision available in $D_A(z)$ and $H(z)$ is
excellent: of order 0.2\% per redshift bin at high redshift.
At low redshift, the precision is worse because there is far less
cosmic volume.	Of course, these statistical errors scale as 
$\fsky^{-1/2}$, where $\fsky$ is the fraction of sky surveyed.

\subsubsection{From BAO to Dark Energy}

\begin{figure}[t]
\centerline{\includegraphics[width=3.2in]{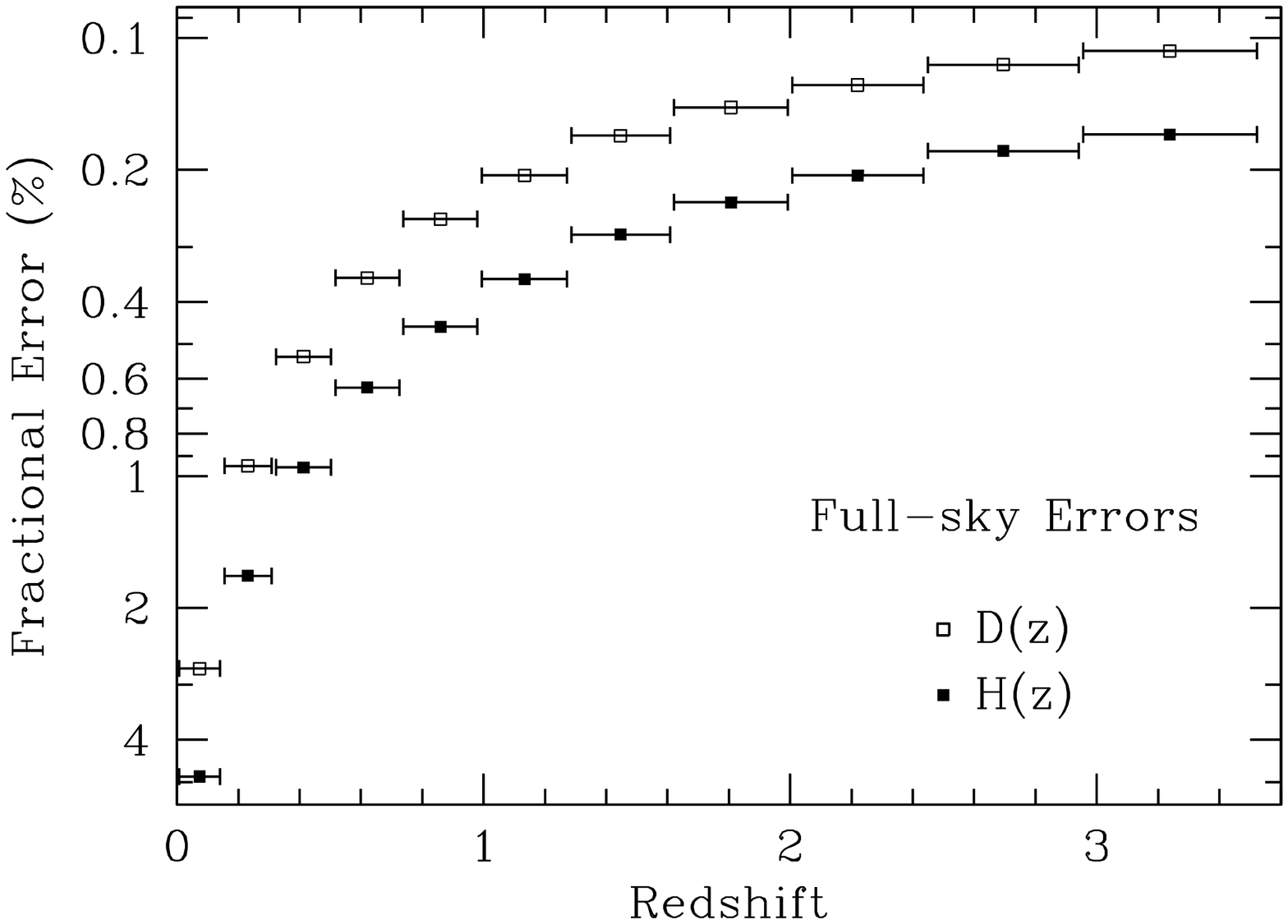}
    \quad  \includegraphics[width=3.2in]{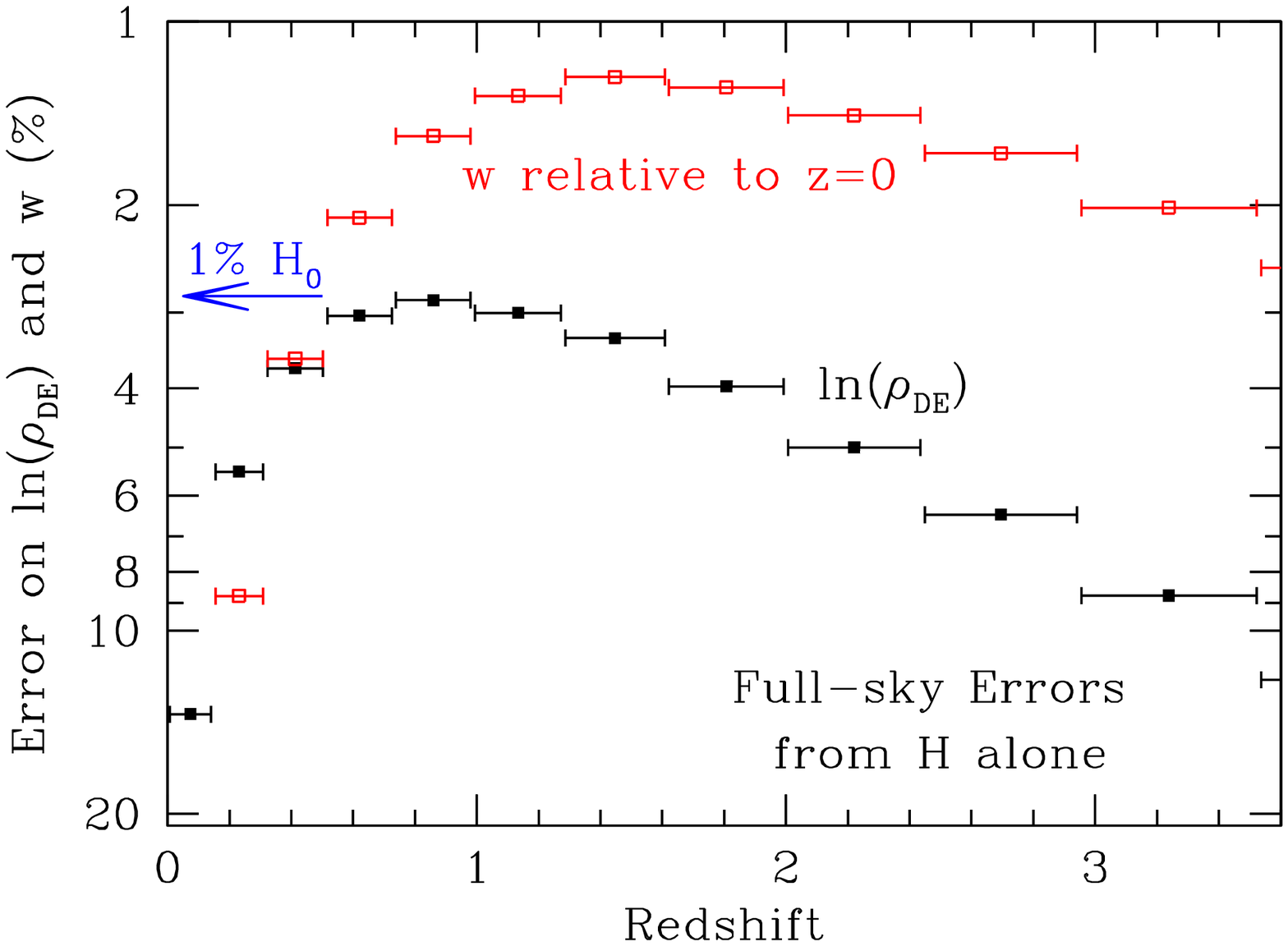}}
\caption{\label{fig:bao_perform}
Illustrative
BAO forecasts for a full-sky survey, from Table \protect\ref{tbl:bao_cvl}.
All errors scale as $f_{\rm sky}^{-1/2}$;
note that $y$-axes are inverted so that smaller errors appear higher
on the plot.
(Left) The fractional error on $D_A(z)$ and $H(z)$ in logarithmic
redshift bins, as open and solid points, respectively.
Note that performance of order 0.2\% per bin is
possible at high redshift.  Here we assume the sound horizon is known.
(Right) Illustration of the dark energy leverage available simply from
the $H(z)$ information in the previous panel.  Assuming perfect knowledge
of the matter density (i.e., $\Omega_m H_0^2$) and curvature, measurement
of $H(z)$ determines
the dark energy density.  The solid black points show the resulting fractional
errors on the dark energy density as a function of redshift, assuming
that the value is close to the cosmological constant.  Errors
of order 5\%, i.e., a 20-$\sigma$ detection of dark energy, are possible
at $0.5<z<2$, even if the dark energy density is simply constant.  
The evolution of dark energy can be expressed by comparing
the density at high redshift to that at $z=0$, assuming the latter is
known.	The open red points give the error on a power-law evolution in $1+z$,
expressed as the error on a constant $w$.  One sees that there is a
broad maximum in performance extending out to $z\approx 3$.
Of course, we must measure the dark energy density at $z=0$; the blue
arrow shows the fractional error on that density that would result from
a 1\% measurement of $H_0$ (which one might get from direct measures or
from a combination of BAO and supernovae), assuming perfect knowledge of
the matter density.  That the blue arrow is comparable
to or above the solid points indicates that we can reasonably expect
to be limited by our higher redshift data.  The open points are
optimistic in that we have assumed perfect knowledge of various inputs;
the intended lesson is that the large volume and larger redshift lever arm
at higher redshift can offset the fact that the dark energy makes
up a smaller fraction of the cosmic total.
} \end{figure}

We will explore how these performance estimates map to dark energy
parameter forecasts in \S\ref{sec:forecast}, but here we describe some
simplified treatments in order to build intuition.
Beginning at low redshift,
if we consider that CMB anisotropies give precise values for $\Omega_m
h^2$ and the acoustic scale $r_s$, then a BAO detection near $z=0$
is measuring a standard ruler and hence $H_0$ \citep{eisenstein98}.
Combining that with
$\Omega_m h^2$ yields $\Omega_m$.  No BAO measurement can be strictly
at $z=0$, but the inference of $\Omega_m$ and $H_0$ depends only on the
distance scale between $z=0$ and the survey redshift.  Even at $z=0.35$,
this brings in only a mild dependence on $w$ and $\Omega_k$
\citep{eisenstein05}.  Hence,
low-redshift BAO measurements offer a strong measurement of $\Omega_m$.
%%% For example, the SDSS DR3 result was
%%% $\Omega_m = 0.273 + 0.123(1+w_0)+0.137\Omega_K \pm 0.025$ (9\%),
%%% where $w_0$ is the value of the equation of state at $0<z<0.35$.
Determining $\Omega_m$ breaks a key degeneracy for the SN measurements
between $\Omega_m$ and $w$.

Moving to higher redshift ($z\gtrsim1$), we next consider the evolution
of the
density of dark
energy using only the $H(z)$ information from BAO
(right panel of Figure \ref{fig:bao_perform}).
If we know the matter density and spatial curvature perfectly,
then the Friedmann equation directly relates the measurement of $H(z)$
to the density of dark energy at that redshift.  Considering the null
hypothesis of the cosmological constant, we would achieve a detection
of the dark energy density with a significance of $\Omega_\Lambda
H/2\sigma(H)$,
where $\sigma(H)$ is the error on $H(z)$.
We next want to
consider the variation in the dark energy density.  Taking an example
in which one assumes the $z=0$ value is known perfectly, we can translate
the error at a given redshift to the error on the exponent of a power-law
variation,  which can in turn be rewritten as an error on a constant $w$
(eq.~\ref{eqn:uphi_vs_w}).
Of course, a full analysis must include the uncertainties on the
matter density, spatial curvature, and $z=0$ value of $\Omega_\Lambda$.

Despite the simplifications, Table \ref{tbl:bao_cvl} and Figure
\ref{fig:bao_perform}
offer some important
results for building intuition.  We find
that the sensitivity of BAO $H(z)$
to dark energy has a broad maximum over the range $0.6 < z < 3.5$.
This plateau arises
because the declining dynamical importance of dark energy is
compensated by the increasing statistical precision afforded by
larger comoving volume.  For $w=-1$, dark energy is only 10\%
of the total
density at $z=2$, but a cosmic-variance-limited BAO measurement can detect
that density at 20-$\sigma$ significance.  The large lever arm to $z=0$
translates this into a 1.3\% constraint on a constant $w$ model.
Of course, if $w>-1$, then $\rho_{\rm DE}$ is
higher at high redshift than it is for a cosmological constant,
increasing the statistical significance with which BAO can
detect it.

Meanwhile,
the transverse acoustic scale at $z\sim2$ and above can be compared
to the angular acoustic scale in the CMB to give a combination
constraint on early dark energy and the curvature of the universe
\citep{mcdonald07}.
This has considerable value in breaking degeneracies between
curvature and dark energy parameters at lower
redshift, and it should be considered an important consistency check
for the $\Lambda$CDM interpretation of the CMB.
A clear detection of non-zero curvature would have major implications
for inflation, and perhaps for quantum cosmology theories
\citep{gott82,guth12,kleban12}.
If the Alcock-Paczynski method can be applied at smaller
scales to obtain a precise determination of $H(z)D_A(z)$,
then the BAO values of $D_A(z)$ can also be used to improve
$H(z)$ determinations and thus the dark energy density
constraints (see \S\ref{sec:ap}).

\subsubsection{Sampling Density}
\label{sec:bao_obs_sampling}

%%% {\it
%%%	The $nP=1$ criterion
%%%
%%%	Can we say something about what the relevant criteria
%%%	  are in the case of Lyman-alpha forest and radio intensity
%%%	  mapping?
%%% }

The acoustic oscillations in the power spectrum are primarily at
wavenumbers $0.1-0.2\invhmpc$, so we want to design surveys with
$nP(k=0.2\invhmpc)\gtrsim 1$.  
Furthermore, if one wins sample size proportionally
to survey time, then $nP(k)=1$ is the optimal balance of survey depth
to sample volume at a given wavenumber \citep{kaiser86}.  One should beware
that this assumption rarely holds in surveys with multi-object
instruments:
the exposure time is driven by the faintest objects in the survey,
so that brighter galaxies are being overexposed in the chosen observation
time.
Also, the number density is often a function of
redshift, so one cannot hit the optimal density everywhere in the
survey.  Finally, one might care about distance precision differently
at different redshifts because of one's specific goals for testing
dark energy models.
See \citet{parkinson07} for a worked example of survey optimization.

For the concordance cosmology, the amplitude of the power spectrum
at $k=0.2\invhmpc$
is about $2700\sigma_{8,g}^2h^{-3}$~Mpc$^3$, where $\sigma_{8,g}^2$
is the variance of the fractional overdensity of the chosen tracer
at the survey redshift in spheres of $8h^{-1}$~Mpc radius.  This
implies that we seek number densities around
$n=(4\times10^{-4}h^3{\ \rm Mpc}^{-3})/\sigma_{8,g}^2$.  Fortunately,
this is well below the density of $L^*$ galaxies.

Higher galaxy bias is a good thing for the statistical
errors of a BAO survey.  The power spectrum amplitude scales as the
square of the bias, so an early-type galaxy is $3-4$ times more
valuable (in the sense of boosting $nP$) than a late-type galaxy.  Given
that there is no identified risk --- higher bias galaxies have
larger acoustic scale shifts, but this is correctable
(Figure~\ref{fig:bao_alpha}, right) --- it makes sense to use higher
bias tracers when possible.  
However, lower bias tracers can be more
effective if one can acquire their redshifts sufficiently quickly!

The balance of shot noise to sample variance is more complex in the
case of surveys with the \lya\ forest or HI intensity mapping.
However, the idea is the same: one wants to make a map in which the
pixel noise is dominated by sample variance, but not by much.
The power at $k=0.2\invhmpc$ corresponds roughly to density variance
of $8\hmpc$ spheres. Hence, we seek to measure the density of
individual regions of this size to a precision slightly better than
the intrinsic rms for such volumes for the chosen tracer (i.e.,
$b\sigma_8$).  If one is measuring too well, one would prefer to do
shallower measurements over a wider region.  In the case of the
\lya\ forest, this criterion concerns both the areal density
of the quasar sightlines and the signal-to-noise ratio of the
spectra \citep{mcdonald07,mcquinn11}.

\subsubsection{Spectroscopic vs. Photometric Redshifts}
\label{sec:bao_obs_photoz}

Photometric redshifts offer a cheap way to measure many galaxy redshifts and
hence to measure the BAO \citep{seo03,glazebrook05,dolney06,seo07}.
However, the larger errors are a challenge.  For velocity errors larger than
$1000\kms$ one is smearing out the acoustic scale along the line of sight
and failing to measure $H(z)$.	Note that this scale is set by the width of
the acoustic peak, not by the acoustic scale.
One only retains full information with rms precision below $300\kms$.

To measure $D_A(z)$, in principle precisions of $\sigma_z/(1+z)$ of
4\% are enough.  Worse precision causes catastrophic degradation
because the oscillations in angular power at the front and back of the
photometric redshift slab fall out of phase.  Redshift precision of
3-4\% yields poor constraints on the BAO per unit volume, with a rule of
thumb that one needs ten times more volume for a photometric redshift
survey than a spectroscopic survey \citep{blake05,seo07}.
Better redshift precision reduces this gap.
At $z<0.7$, current and ongoing spectroscopic surveys are already
covering 1/4 of the sky, so photometric redshift surveys are only competitive
at higher redshifts.  Extracting large-scale structure and BAO from
photometric redshift surveys requires very stringent calibration and
more extensive modeling than for spectroscopic surveys.
Photometric surveys with many narrow bands offer an
intermediate approach between imaging and spectroscopy, which
may be advantageous in some regimes \citep{benitez09}.

%%% {\bf Can we give some justification for this rule of thumb?
%%% What's the relevant scale for redshift precision?  Better
%%% than the BAO scale, or better than the width of the BAO peak?}

\subsubsection{Tracers of Structure}
\label{sec:bao_obs_tracers}

As we have seen, BAO surveys require surveys of very large volumes with
modest sampling density.  One wants to map a wide range of redshift so as
to measure the history of expansion.  The current generation of surveys
are mapping of order 1 million galaxies, and approaching the cosmic variance
limit at $z>1$ requires of order $10^8$ galaxies.

We have a lot of freedom in selecting the objects to trace the density
field.	Usually, we require isotropy of the selection but do not require
that the selection be unchanging as a function of redshift.  One is seeking
to minimize the observational cost for a given well-sampled survey volume.
There are many competing considerations \citep{glazebrook05}.
One desires a tracer with a strong spectroscopic signature to allow a
redshift determination to about $300\kms$ rms as fast as possible, with
few catastrophic redshift errors.
One desires a combination of density and clustering bias so that
$nP(k=0.2h{\rm\ Mpc}^{-1})\gtrsim1$.
One desires a higher clustering bias, so that the required number density
is lower; this allows one to use brighter objects and reduce exposure time.
For targeted surveys, one desires that the tracer can be readily selected,
so that one doesn't waste resources on undesired objects.
In more detail, one desires photometric redshifts good enough
that one can shape the $n(z)$ profile in a way that keeps $nP$ close
to unity at high redshift without being swamped by low luminosity
objects at low redshift.
And, of course, the observed wavelength of the spectroscopic feature 
determines a great deal about the instrumentation.

Luminous red galaxies are an effective choice at lower redshifts.
They have strong absorption features, notably the 4000\AA\ break,
and high surface brightnesses to allow rapid spectroscopy.  They
have a high bias ($b\sim2$) to reduce the required number density
and hence the number of spectroscopic fibers.   They are also easy
to select with photometric redshifts: essentially they are the
reddest galaxies at a given observed flux \citep{eisenstein01}.

As we work to $z=1$ and beyond, the advantages of using emission-line
galaxies increase.  Red galaxies are very faint in the optical at
$z\gtrsim1$ because of $K$ corrections, and the 4000\AA\ break moves
into the infrared where the forest of OH sky glow lines makes
spectroscopy more difficult.  But the star formation rates of normal
galaxies at $z>1$ are about ten times higher than today, and this
high star formation 
produces strong emission lines.  These emission lines can be detected
even when the stellar continuum cannot, and the galaxies with the
strongest lines can be measured in remarkably little time.  
Spectral resolution of a few thousand is desirable to work between
the OH sky glow lines and to resolve the [OII] doublet.
The challenge is primarily one of selection, how to use photometric data
to pick out the star forming
galaxies with the strongest lines.  Between the lower clustering
bias and the failure rate on weaker-lined systems, one needs to
survey many more blue galaxies than red.  Current expectations are
that the transition point from preferring red galaxies versus blue
is $z\approx 1$ \citep{glazebrook05}.
Slitless spectroscopy offers an alternative route to surveying
emission line galaxies, without prior target selection
(see \S\ref{sec:bao_space}).

Clusters of galaxies have been proposed as tracers \citep{angulo05,hutsi06}.
These can be readily
found from imaging data sets but have the disadvantage that their
$nP$ does not reach unity.  Also, acquiring spectroscopic redshifts
for the clusters imposes requirements similar in area and depth to
a red galaxy survey, so the gain in a highly multiplexed fiber
survey is much smaller than one would expect based on numbers alone.
Quasars have a similar problem of having $nP<1$, but they are extremely
luminous and easy to select.  This makes them a possible target for
a sparse, wide-field survey at $z>1$ \citep{sawangwit11a},
and they are readily added to a multi-fiber survey targeting
emission line galaxies or LRGs at other redshifts.

Using the \lya\ forest as a tracer is attractive because each
spectrum yields many density measurements (effectively about 50)
rather than just a single point in a map
\citep{white03,mcdonald07,norman09,mcquinn11}.
One wants to sample the
width of the acoustic peak, which is about $20h^{-1}$~Mpc FWHM.
This implies that one needs spectral resolution of only a few hundred
and moderate angular density of the lines of sight, preferably about
100 per square degree.	Quasars of this surface density are much 
brighter than the
Lyman-break galaxies that would be required to match the effective
sampling density.  As one has little gain from reducing the photon
noise errors to below the intrinsic variation of the forest on
$10\hmpc$ scales, one can afford to use low signal-to-noise ratio spectra.
The challenge here is systematics, as one must control the continuum
of the quasar and the spectrophotometry of the measurements to utilize
the spectral information.
It is also possible that theory systematics associated with the 
state of the IGM enter; so far, IGM uncertainties have not been shown
to affect BAO measurements from the forest, but the case has
not been investigated as thoroughly as it has for galaxies.
The recent detection of BAO in the \lya\ forest at a scale and
amplitude compatible with theoretical expectations
\citep{busca12,slosar13} is already encouraging, but much work
remains to demonstrate that systematic effects are below the
achievable level of statistical precision.

Star-forming galaxies also can be observed in the radio using the
21 cm line of neutral hydrogen.  This is a much weaker line, but
future generations of radio interferometers such as the Square
Kilometer Array offer phenomenal survey speed because one can
synthesize millions of simultaneous beams computationally.  Such
instruments could in principle achieve spectroscopic samples of
$10^9$ galaxies out to $z=2-3$ \citep{abdalla05}.

A different concept is that of 21 cm intensity mapping
\citep{peterson06,ansari08,chang08,loeb08,wyithe08,seo10}.  Here one does
not identify individual galaxies but instead measures the combined
emission of the 21 cm line from all galaxies in a volume of order
$10$ Mpc on a side.  The fluctuations in the map encode the large-scale
density field and hence the BAO.  Relative to an interferometer like
the SKA, one uses shorter baselines (around 300 meters) and a nearly
filled aperture to maintain surface brightness 
sensitivity.\footnote{An interferometer directly measures the 
Fourier transform (in the 
transverse direction) of the emission field; antennas separated by a 
distance $L$ measure Fourier modes with $k_\perp = 2\pi L/(\lambda D_A)$.}
Because one is not resolving individual galaxies above the instrumental noise,
one is using all of the neutral hydrogen even from low-mass galaxies.
In principle, one can map the BAO to the cosmic variance limit
out to $z\approx3$ with new interferometric arrays.  The challenge
is foreground subtraction, as the cosmic signal is several orders of
magnitude below the Galactic and extragalactic emission levels.
A first detection of large-scale structure in redshifted 21 cm has been
reported by \citet{chang10} by cross-correlating with an optical
galaxy redshift survey at $z=0.8$; cross-correlation removes
foregrounds that are not themselves correlated with the optical galaxies.
For intensity mapping to work on its own, one of course needs
to measure the auto-correlation signal.

Unlike for the case of galaxies, diffuse HI mapping does not
provide the {\em mean} level of emission (interferometers are not 
sensitive to this, and even if they were the Galactic emission would swamp
extragalactic HI); therefore $\delta_{\rm HI}$ is measured only up to a 
multiplicative constant. This does {\em not} present a problem for the
BAO technique because one is using the shape and not the amplitude of the
power spectrum. It does have an impact on redshift-space distortions
(\S\ref{sec:rsd}), as
without the mean level one cannot turn the observable $\beta$ into an 
estimate of the rate of growth of structure $f\sigma_8(z)$. This drawback 
is, however, also an opportunity to learn about astrophysics: measurement of
$\beta_{\rm HI}$ combined with independent knowledge of $f\sigma_8(z)$
would allow us to infer the mean HI signal and thus obtain the cosmic 
abundance of neutral gas as a function of time \citep{wyithe10}.

\subsection{Systematic Uncertainties and Strategies for Amelioration}
\label{sec:bao_systematics}

Given that we seek to measure the acoustic scale and hence cosmic distance
scale to a high level of precision,
it is important to consider the systematic errors that could cause
the inferred $D_A(z)/r_s$ and $H(z)r_s$ to be incorrect.
We consider three classes of systematic errors:
(1) observational errors, in which one mis-measures the large-scale structure
of the universe;
(2) astrophysical errors, in which our model of large-scale structure for
a given cosmology is incorrect; and
(3) cosmological errors, in which we mispredict the sound horizon given
our measurements because of new cosmological physics, either in the
early universe or at late times.

\subsubsection{Measurement Systematics}
\label{sec:bao_systematics_obs}

The measurement of large-scale structure requires the ability to
produce a well-calibrated density map of the universe.	The data need
not be homogeneous in quality so long as the inhomogeneities are
known well enough that one can correct for them statistically.

Observational errors involve imperfections in one's map of the
density field.	Examples of sources can be photometric miscalibrations
of the input catalog, mis-assessments of the incompleteness in the
input catalog, redshift failures or errors, incorrect tracking of
the target selection, failure to correct for deleterious interactions
between targets (e.g., fiber collisions), or imperfect assessment
of the redshift distribution of the map.  Another class of problems
involves understanding of the errors of the map, as one must assess
both the statistical properties of the density field and the point
sampling of it by galaxies.

Fortunately, these issues have been extensively studied in the general
context of the measurement of large-scale structure
\citep[e.g.,][]{tegmark98b}.  BAO measurement itself is
only a particular application of large-scale structure data, and it
turns out to be a relatively easy one because the acoustic peak
is narrow in scale and hence one has another differential opportunity
in the experimental design.  That is, one can compare the behavior
at 150 Mpc separation to the average of that at 120 and 180 Mpc, so
as to remove smooth errors.  The only way to produce a non-smooth
error is to have a sharply preferred scale in the systematic error.

For galaxy redshift surveys,
there is wide expertise in how to calibrate surveys and track their
selection functions, and there are many tests that can be employed
to look for specific problems.	Failing that, residual errors are
often intrinsically radial or angular in their nature, so one can
reject the purely radial and purely angular modes from a survey
\citep{vogeley96,tegmark98b}.  This is a small cost in information
content for an intrinsically 3-d field.  
A more targeted version of this 
idea is to use angular templates to remove systematic errors with
particular angular dependence, e.g., survey depth variations due to sky
brightness, seeing, or stellar density \citep{ho08,ross11,ross12}.
A further related idea is that for
a sharp scale in a systematic error to be a real threat, it must
be sharp for three dimensional spheres of separation.  For example,
even if a survey has an error that is modulated on a circular field
of view, the diameter of the field affects a range of 3-d separations
at a given redshift
simply because of the random orientations of pairs to the line-of-sight.

The BAO method is ultimately tied to the
separation of galaxies, which depend on astrometric positions and
redshifts.  These quantities can be exquisitely well measured, and
achieving 0.1\% precision on one's astrometric and wavelength scale
is easy.   The concern about systematic errors in the map is that
an erroneous tilt in the correlation function would cause one to
mismeasure the centroid of the acoustic peak.  This is a weaker
effect, and one can marginalize against such tilts if one wants,
using the techniques in \S \ref{sec:bao_theory_fitting}.

In short, it is very likely that a reasonable design for a galaxy
redshift survey will lead to sufficient accuracy for the BAO method.
The greater challenge for such surveys is to control the clustering analysis
for the broadband cosmological signals, which require a factor of
more than ten better accuracy.

On the other hand, the observational systematics for the \lya\
forest and 21 cm intensity mapping techniques are a serious concern.
Here we are trying to use every spectral pixel for our mapping data,
rather than differencing spectral pixels to measure a single redshift
per object.  Imperfections in the calibration of the spectra or the
subtraction of sky emission or Galactic foregrounds will appear as
cosmic structure.

For the \lya\ forest, we measure the absorption by assuming
that the quasar continuum is intrinsically smooth.  However, even
an unabsorbed spectrum would have variations due to the intrinsic
spectrum of the quasar and any errors in the removal of the sky
emissions or flat-fielding of the detector.  We do not know the
detailed unabsorbed quasar spectrum but instead need to estimate it from the
ensemble properties of quasars or from fitting to less absorbed
pixels.  The BAO signal is a very weak modulation on large scales.
Modeling errors far too weak to show up in any one spectrum could
inject correlations that bias the BAO scale or simply increase the
noise far above the expected sample variance.
The most detailed discussions of systematic errors in
large-scale \lya\ forest measurements are those of \cite{mcdonald06}
and \cite{slosar11}.
Measuring the scale of the BAO feature again appears much
easier than determining the broad-band shape or absolute amplitude of
the power spectrum.  Studies to date have not identified observational
problems that would prevent high precision BAO measurements, but the 
field is in its early days.

For 21 cm intensity mapping, we are looking for the correlations of the
extragalactic line emission as a function of wavelength (redshift) and sky
position. However, the Milky Way is producing synchrotron and free-free
emission that is three orders of magnitude higher than the extragalactic signal
and highly variable on the sky \citep{chang08,chang10}.
Fortunately
these emission mechanisms are smooth as a function of frequency, unlike
the cosmological signal where frequency maps to redshift, which should
enable foreground removal (e.g., \citealt{liu11}).  The challenge
here is primarily instrumental: undesired features in radio
interferometers such as far sidelobes or standing waves in the antenna are
strongly frequency dependent and can mimic a cosmological signal if not
suppressed. Moreover, the Galactic synchrotron radiation is polarized, and
Faraday rotation within the galaxy can lead to strongly
wavelength-dependent polarization amplitude 
(e.g., \citealt{haverkorn03}),
so the instrument and software must measure the
total intensity and remove Stokes $Q$ and $U$ from their maps --- a major
challenge given that radio antennas are inherently polarized. The problems
are similar to those of the 21 cm mapping of the epoch of reionization,
where several experiments are trying to achieve first detections. Projects
aimed at $z = 1$ are being started in order to investigate and hopefully
control the observational systematics.

\subsubsection{Astrophysical Systematics}
\label{sec:bao_systematics_astro}

Astrophysical systematic errors are principally due to non-linear
structure formation, redshift-space distortions, and galaxy clustering
bias.  These were discussed in \S \ref{sec:bao_theory_nl}.  Our understanding
of these effects in cold dark matter models has been greatly advanced
by numerical simulations and analytic theory over the past decades.
Fortunately, the acoustic scale is much larger than the scales of
non-linear structure formation and the hydrodynamic effects on galaxy
formation.  Gravitational forces are by far the dominant effect on
150 Mpc scales, and we can compute these at high accuracy.

Galaxy clustering bias based on halo occupations has been shown to
be manageable for the BAO method.  The raw shifts of the acoustic scale
are below 1\%, and they can be reduced below any reasonable detection limit
with reconstruction \citep{mehta11}, as shown in Figure \ref{fig:bao_alpha}.
Hence, the concern is now for a more complicated
clustering bias, e.g., one that couples more directly 
to the large-scale density
field or that features large-scale cooperative effects between galaxies
\citep{bower93}.
But clustering bias is no longer an arbitrary bogeyman.  We have many
observational probes that should test a bias model: two-point and
higher-point clustering over all scales, redshift-space distortion patterns,
cross-correlations between types of galaxies, galaxy-galaxy weak
lensing maps, and various measures of halo masses.  While the simplest
formulation of HOD is surely not the whole story of
clustering bias \citep[e.g.,][]{gao05,harker06,gao07}, the model has passed
significant tests.  An alternative mechanism
that couples to large-scale densities in a very different manner
so as to alter the BAO scale will almost certainly produce far more
detectable effects on smaller scales.

A possible complication to galaxy biasing at the BAO scale was
pointed out by \cite{tseliakhovich10} and \cite{yoo11}.
At the time of recombination, the pressure
of the photons causes the baryonic matter to have a relative velocity
compared to the dark matter, with a typical value $v_{\rm bc} \sim 30\kms$.  
This relative velocity is largely due to the same standing acoustic waves
that produce the BAO feature; it is coherent on scales of a few Mpc and
has a feature in its correlation function at 150 Mpc.  After
recombination, the sound speed in the baryons drops to $6\kms$, so the
relative velocity is supersonic. 
\cite{tseliakhovich10} 
argue that this boosts the effective Jeans mass
as small dark matter structures fail to retain baryons, thereby
suppressing the formation of the earliest galaxies ($M_{\rm halo}\sim
10^6 M_\odot$ at $z>10$).  This level of suppression depends on the
local $v_{\rm bc}$, which varies on large scales.  It is unclear whether
this varying suppression causes a detectable imprint on the properties
of much more massive galaxies at low redshift; it may be that it is
completely erased as galaxy-mass ($>10^{11} M_\odot$) halos form and
wipe out the small-scale initial conditions, or it may be that feedback
mechanisms such as early metal pollution allow some trace of $
v_{\rm bc}$ modulation to survive in structures at $z\ll 10$.  In the
latter case, it represents a potentially serious concern because (unlike
other systematic errors) the modulation contains the BAO scale 
\citep{yoo11}.  However, the form of the modulation is
predictable, and \cite{yoo11} find that the
measurements of the galaxy bispectrum would enable the detection and
removal of this effect.

Besides gravity, the only physical effect that we reasonably suspect
can modulate galaxy properties on large scales is radiation transport.
For example, it is predicted that reionization proceeds with bubbles
of scales of 10 Mpc for hydrogen and 100 Mpc for He II.  This may
affect the late-time galaxy density field in non-gravitational ways
\citep{mcquinn07,iliev08,mesinger08,zheng11,wyithe11}.	However, the scale
of the reionization bubbles
is not sharp enough to mimic the acoustic peak, e.g., any reasonable
variation in the luminosity of the ionizing sources will produce a
wide spread of bubble sizes.
Reionization effects could be a larger issue for the \lya\ forest
than for galaxy surveys because one is mapping the IGM directly.
Simulations of Gpc$^3$ volumes that incorporate models of these
reionization effects will be needed to see whether they can
detectably influence BAO measurements,
but the absence of a sharply preferred scale in reionization
should again provide protection if one marginalizes over broad band tilts.

\subsubsection{Cosmological Systematics}
\label{sec:bao_systematics_cosmo}

Cosmological effects that alter the sound horizon or the detailed
prediction of the linear power spectrum must contend with the fact
that the $z=1000$ universe and the acoustic oscillations themselves
are exquisitely well observed in the CMB anisotropies.	For example,
a change in the recombination history could alter the sound horizon,
but this produces correspondingly larger changes in the damping tail
of the primary anisotropies \citep{eisenstein04,debernardis09}.  Effects such
as particle decays that change the expansion history so as to alter
the sound horizon affect the gravitational potential of the
fluctuations and have large impact on the CMB anisotropies.

Models that combine adiabatic perturbations with smaller isocurvature
ones offer additional degrees of freedom to constrain in the CMB.
Most such combinations yield differences that can be detected in the
acoustic peak structure of the CMB before they affect late-time BAO
inferences.  However, \citet{mangilli10} show that a particular
combination of isocurvature modes may exist that can change the
sound horizon by a moderate amount before the CMB anisotropies are
observably altered.  This possibility 
bears more investigation, e.g., of other
late-time observable consequences of such a model
\citep{mangilli10,carbone11,zunckel11}.

Finally, we note that if the sound horizon or power spectrum template
predicted from $z=1000$ is wrong, then the effect on the BAO distance scale
will typically be a multiplicative error across all redshifts.	This
would alter the inference of $w(z)$, but with a particular redshift
dependence that one might choose to be suspicious of if one found it.

In short, while not all cosmological possibilities have been cataloged
for their effect on the BAO method, one should always judge such
possibilities in light of the CMB as well.  The combination of CMB
and BAO is likely to be self-diagnosing of new cosmological physics
at high redshift.  There may be exotica that can slip through
this net, but we don't view this potential confusion with dark
energy dynamics as a demerit of the method.  Large cosmological
surveys offer a rich spectrum of possible analyses with which to
corroborate our model of structure formation, and the discovery of
any discrepancy from vanilla $\Lambda$CDM will surely inspire a
vigorous search for alternative explanations.

\subsection{Space vs. Ground}
\label{sec:bao_space}

The principal challenge of the BAO method is obtaining the redshifts
of millions of faint galaxies.	Certainly we can obtain redshifts from
the ground for tracers at any redshift; the difficulty is in doing
this quickly and cheaply enough.

Most BAO work to date has used multi-fiber spectroscopic surveys at
optical wavelengths.  This is practical for surveys of order $10^7$
galaxies.  At $z>1$, one relies on finding very luminous line emitters,
and the desired number of galaxies to reach the cosmic variance limit
is of order $10^8$.  Routing optical fiber to $10^8$ objects is technically
very demanding.  We expect that fiber-fed optical galaxy redshift surveys
will do an excellent job out to $z=1$ and will make a start at
$1<z<1.5$, but will not approach the cosmic variance limit at $z>1$.

Photometric redshifts of either galaxies or clusters are an option
to sample a large volume at $z\approx 1$ with upcoming surveys and
probably at higher redshifts with deeper surveys like LSST.  Redshifts
$z<0.7$ will be done better with funded spectroscopic programs.
One is free to pick a subset of galaxies on which one has better
photometric redshift performance.  However, photometric redshifts
are best when relying on strong breaks, notably the 4000\AA\ and
Lyman break.  The former requires near-IR data at $z\gtrsim1$; the
latter requires space UV data at $z\lesssim3$.	As mentioned above,
photometric redshifts are not precise enough to capture the BAO $H(z)$
information, which is a large loss at higher redshifts.
We expect that the upcoming generation of imaging surveys will be the
first to map the BAO at $z\approx1$ over large areas of the sky.  This will
achieve an important constraint on $D_A(z)$.  Later spectroscopic surveys
will improve the $D_A(z)$ measurement and add $H(z)$.

A space mission offers the opportunity for slitless spectroscopy
\citep{glazebrook05a}.
This efficiently finds the strongest line emitters over a wide
instantaneous field.  Slitless spectroscopy of faint objects is
only practical in space, where the foreground (or ``sky'') emission
is low.  This is particularly attractive in the near-IR, where the
zodiacal background light is low and the H$\alpha$ line from $z>1$ galaxies
is very bright.  The UV with Ly$\alpha$ is another opportunity.

At $z>2$, the Ly$\alpha$ line (whether in emission or in the forest)
can penetrate the atmosphere.  This offers a renewed opportunity for
ground-based work, but only the Ly$\alpha$ forest is likely to be able
to approach the cosmic variance limit in the foreseeable future.
As described above, this method still has significant uncertainties about its
observational and thoeretical systematics.  Galaxy samples would again require
$>10^8$ objects to reach the cosmic variance limit, a factor of 100
more than planned surveys.
The Ly$\alpha$ forest gets undesirably
thick at $z>3.5$, and BAO surveys above this redshift might require
a space mission, such as the {\it Cosmic Inflation Probe} \citep{melnick09}.

A 21 cm facility such as the SKA capable of detecting individual
high-redshift galaxies is a multi-billion dollar project and hence
well in the future, albeit with a large cosmological payoff.  We
note that not all technical implementations of the SKA permit
full-sky mapping, and keeping this option does increase the cost of the
correlator.  The 21 cm intensity mapping technique is considerably
cheaper, but we do not know whether it can achieve the required
control of observational systematics.  Applying intensity mapping
to the reionization epoch could eventually measure the distance scale at $z>6$
\citep{mao08,rhook09}.

A space mission that would
target of order $10^8$ $1<z<2$ galaxies is the only robust
near-term path to approaching the cosmic 
cosmic variance limit for BAO over the enormous comoving
volume available in this redshift range.
Intensity mapping is an attractive opportunity,
but it needs substantially more development before it can be
realistically assessed.
Ground-based galaxy redshift surveys
and \lya\ forest surveys will explore $z>2$, though 
in the near-term approaching the cosmic variance limit depends on controlling
systematic errors in the \lya\ forest method, which are not yet
understood at the percent or sub-percent level.

\subsection{Prospects}
\label{sec:bao_prospects}

In contrast to essentially all of the other observational probes
that we consider in this review, we anticipate that even the most
ambitious BAO studies will remain limited by statistical errors
rather than systematic errors.  This assumption could prove
incorrect, either because we are overoptimistic about BAO systematics
or because we are too pessimistic about other methods.  
But it does imply a natural
long-term target for BAO investigations of cosmic acceleration:
survey a large fraction of the entire comoving volume out to
$z \approx 3.5$, beyond which the sensitivity to dark energy 
begins to decline (Table~\ref{tbl:bao_cvl}), with high enough
sampling density that the BAO measurements are limited by 
sample variance rather than shot noise.
No one survey will reach this goal on its own; rather, a variety
of projects can gradually map out the available volume by using
different facilities and techniques to target different redshift
ranges and areas of sky.
Surveys that cover the same redshift range with the same
technique are {\it not} redundant unless they cover the same 
region of the sky.
To zeroth order, the primary metric for a BAO survey is the 
comoving volume that is covered at adequate sampling density,
and it makes sense to choose redshift ranges according
to observational convenience (though of course one can further
optimize both survey strategy and instrument design).
Relative to the current state of the art described in
\S~\ref{sec:bao_current} --- roughly speaking,
analyses that have probed $f_{\rm sky}=0.4$ to $z=0.15$,
$f_{\rm sky}=0.25$ to $z=0.45$, and $f_{\rm sky}=0.08$
to $z=0.8$ --- BAO surveys have tremendous possibility for
growth, with correspondingly great opportunities for 
improved precision and redshift leverage on $D_A(z)$ and
$H(z)$ (Fig.~\ref{fig:bao_perform}).

With the completion of WiggleZ \citep{parkinson12}, the only
spectroscopic BAO survey currently operating is 
SDSS-III BOSS \citep{dawson12}.
BOSS is approximately midway through its five years of
spectroscopic observing and will conclude in mid-2014.  Forecasts for
the galaxy survey predict 1.0\% precision on $D_A$
at $z=0.35$ and $z=0.6$ and 1.8\% precision on $H(z)$ at these
redshifts.  The \lya\ forest survey is expected to yield 4.5\% precision
on $D_A$ and 2.6\% on $H(z)$ at $z=2.5$.
BOSS will provide a solid BAO anchor at low redshift, 
the first BAO measurements $z>2$,
and the first practical test of the \lya\ forest technique.

The Hobby-Eberly Telescope Dark Energy Experiment (HETDEX) is largely
funded and currently under construction.
HETDEX plans a survey of 800,000 \lya\ emission-line galaxies 
over 420 square degrees at redshifts $1.8<z<3.7$,
using a blind-pointing strategy with a large set of integral-field
spectrographs \citep{hill06}.  
The forecast precision on $D_A$ and $H$ is of order 2\% from BAO alone,
with additional gain possible if one can take advantage of the increased
linearity of the large-scale density field at high redshift to model the
full anisotropic clustering signal of the galaxies.

PanSTARRS and DES are two near-term imaging surveys with the depth
and area needed to probe BAO at $z\approx1$.  BAO analyses will 
likely focus on red galaxies as they afford more robust photometric 
redshifts and the two cameras employ red-sensitive detectors 
that achieve good depth in the $z$ and $y$ bands.
These projects will likely yield the first strong BAO constraints
at $z=1$.

A multitude of more ambitious projects are being planned.
On the imaging front, LSST should eventually yield an enormous sample of 
galaxies with good photometric redshifts, enabling photo-$z$
BAO studies to reach to $z=2$ and beyond.
Two near-term Spanish projects, PAU and JPAS,
aim to do shallower imaging with many
medium-band filters, designed to achieve high enough redshift precision
to recover $H(z)$ information out to $z\approx1$ 
\citep{benitez09,gaztanaga11}.
(PAU would use a new large-format camera built for the 
William Herschel Telescope while JPAS would use a new
telescope dedicated to the project.)
This medium-band strategy is intermediate between photometric
and spectroscopic approaches.

Returning to spectroscopy, eBOSS, part of a proposed (but not
yet fully funded) program of post-2014 surveys on the 
Sloan 2.5-meter telescope, would extend the BOSS survey
in several directions, using higher redshift LRGs 
(to $z=0.8$), emission line galaxies, and quasars, including
a denser set of $z>2$ quasars to improve measurements from
the \lya\ forest.  eBOSS would cover $1500-3000\mdeg^2$
depending on strategy details that are still to be decided.
The BigBOSS experiment \citep{schlegel11}
would use spectrographs fed by 5000 optical fibers over a 
3-degree field on the Mayall 4-meter telescope at Kitt Peak
to survey 14,000 deg$^2$.  For its five-year primary survey,
BigBOSS would target luminous red
galaxies to $z=1$ and emission line galaxies to $z=1.7$,
more than 10 million galaxies in total, with sampling density
$nP > 1$ out to $z\approx 1 - 1.2$.
BigBOSS would target high redshift quasars with a high enough
density to approach the sample variance limit for the \lya\ forest 
method at $2<z<3$.  The BigBOSS instrument could in principle
be moved to the Blanco telescope at CTIO to conduct a similar
survey of the southern hemisphere.  Alternatively, the DES
collaboration has considered a 4000-fiber instrument
(DESpec) that would use the DECam optical corrector on the
Blanco \citep{abdalla12};
this instrument could pursue a similar galaxy
redshift survey but would not (in its current design)
have the blue wavelength coverage needed to map the \lya\ forest.
The SuMIRe project proposed for
the Subaru 8-meter telescope would use optical/IR
prime focus spectrographs fed by 2400 fibers 
to carry out a large galaxy redshift
survey, mapping BAO in the redshift range $0.7 < z < 2.4$.
The current baseline program would survey 4 million [OII]-emitting
galaxies over 1420 deg$^2$.
Collectively, these ground-based optical/IR projects could 
cover a substantial fraction of the sky with fully sampled
galaxy surveys to $z \approx 1.2$, provide interesting BAO
measurements with lower sampling densities to $z \approx 1.7$,
and possibly measure BAO to something approaching the 
cosmic variance limit at $z=2-3$ using the \lya\ forest.

Both \euclid\ and \wfirst\ plan large BAO surveys as major
components of their dark energy science programs, using
slitless near-IR spectroscopy to measure redshifts of
strong H$\alpha$ emitters.
Current incarnations of these plans are described in
the \euclid\ Red Book \citep{laureijs11} and the \wfirst\
Science Definition Team's final report \citep{green12},
though technical specifications and survey strategies may
evolve to some degree prior to launch.
The present baseline strategy for \euclid\ has a
a survey area of approximately 14,000 deg$^2$ and
redshift range $0.7 < z < 2.0$, while the baseline strategy
for \wfirst\ adopts a smaller area (3,400 deg$^2$), fainter
flux limit, and higher redshift range ($1.3 < z < 2.7$).
Green et al.\ (\citeyear{green12}; see their Figs. 24 and 25)
attempt a side-by-side comparison of \euclid\ and \wfirst\
BAO performance, using common modeling assumptions that
include recent estimates of the luminosity function
\citep{sobral12} and clustering bias \citep{geach12}
of high-redshift H$\alpha$ emission line galaxies.
In their calculations, \euclid\ achieves a sampling
density $nP > 1$ out to $z \approx 0.9$, falling to
$nP \approx 0.15-0.3$ at $z = 1.3 - 2.0$.
\wfirst\ maintains $nP > 1$ out to $z \approx 2.4$,
falling to $nP \approx 0.5$ at $z=2.7$.
For \euclid\ they forecast fractional errors 
$\sigma_H/H$ of $1.3-1.8\%$ in bins of $\Delta z = 0.1$ 
out to $z=1.5$, rising to 2.5\% at $z=2.0$, while for \wfirst\
they forecast $\sigma_H/H = 1.7-1.8\%$ (again in $\Delta z = 0.1$ bins)
out to $z=2.4$, rising to 2.4\% at $z=2.7$.
These numbers should be
taken with a grain of salt, as they depend on uncertain hardware
and software performance and on details of survey strategy.
For example, a 2.4-meter implementation of WFIRST could potentially
survey 3-4 times larger area at similar depth \citep{dressler12}.
By the time these missions are launched, results from earlier
dark energy experiments or developments in modeling techniques
could well favor alternative strategies, e.g., with deeper
sampling but smaller sky area.  Furthermore, the \euclid\ 
and \wfirst\ dark energy programs are both limited by observing
time, and either could be more powerful with a longer mission.
It is clear, however, that these missions can dramatically
improve our knowledge of dark energy evolution at $z=1-3$.

Shifting wavelengths,
several 21 cm intensity mapping experiments for the range $0.8<z<3$ are
being planned, using different techniques. 
The CHIME project aims to build a 100 meter square filled interferometer 
using a cylindrical telescope array \citep{peterson06} 
and conduct a lengthy survey at $0.8<z<2.5$.  
If the foregrounds can be adequately controlled, CHIME
would be a powerful demonstrator of the 21 cm method and would yield
excellent cosmological information.  
Other projects include the FFT-based Omniscope \citep{tegmark10}
and the Baryon Acoustic Oscillation Broadband and Broad-beam (BAOBAB) 
interferometer array \citep{pober12}.  
Moving beyond intensity mapping, the
SKA could enable an HI-redshift survey of a billion galaxies, reaching
the sample variance limit over half the sky out to $z=3$ \citep{abdalla05},
which would be a good approximation to the ultimate BAO experiment.

\vfill\eject
\section{Weak Lensing}
\label{sec:wl}

%%%% BEGIN WL SECTION

The subtle distortion of shapes of distant galaxies by gravitational lensing
is a powerful probe of both the mass distribution and the global geometry of
the universe.  It has, however, turned out to be one of the most technically
difficult of the cosmological probes.  This section will cover the range of
applications of weak lensing (which we will sometimes abbreviate to WL), 
the recent and planned weak lensing surveys,
and the technical aspects of weak lensing image processing and control of
systematics.  By covering the latter subjects in some
detail (including some methods that we think have been under-appreciated or
under-utilized), we hope to stimulate further progress and be helpful
to readers
who are already experts in weak lensing.

This section is organized as follows: we begin with a qualitative
overview of weak lensing and its uses (\S\ref{ss:overview}).  We then
go into a mathematical treatment of the various statistics that can
be used and their dependences on the background cosmology and matter
power spectrum (\S\ref{sec:wl_gp}).  We then review the observational
results from recent weak lensing surveys (\S\ref{sec:wl_current}).
Section \ref{sec:wl_stat_errors} discusses the statistical errors
and cosmological sensitivity of cosmic shear surveys at a 
rule-of-thumb level; we expect this to be a useful entry point
for readers interested in understanding survey design.
We then turn to more technical aspects of survey design and analysis,
including source redshift estimation and the galaxy populations of 
optical/near-IR and radio surveys 
(\S\S\ref{sec:wl_galpop}-\ref{sec:wl_radio}), 
CMB lensing (\S\ref{sec:wl_cmblensing}),
the measurement of galaxy shapes (\S\ref{sec:wl_shapes}), 
and astrophysical uncertainties (\S\ref{sec:wl_astro}).
We summarize the major systematic errors and mitigation strategies
(\S\ref{sec:wl_systematics}).  We finally consider the advantages of a space
mission for weak lensing (\S\ref{sec:wl_space}) and prospects for the future
(\S\ref{sec:wl_prospects}).

Some of the material in this section is technical and in a first reading
may be either skipped or skimmed; but given that so much of the promise of
weak lensing depends on these issues, we felt compelled to include them.
The more technical
sections have been denoted with an asterisk (*).  They may be thought
of as analogous to, e.g., the ``Track 2'' material in \cite{misner73}.

% finish intro

\subsection{General principles: Overview}
\label{ss:overview}

The images of distant galaxies that we see are distorted by gravitational
lensing by foreground structures.  In rare cases, such as behind clusters, one
observes {\em strong lensing}: the deflection of light by massive structures
can result in multiple images of the same background galaxy.  More often,
however, images of galaxies are subjected only to {\em weak lensing}: a
small distortion of their size and shape, typically of the order of 1\%.
Since one does not know the intrinsic size or shape of a given galaxy, weak
lensing can only be measured statistically by examining the correlations of
shapes in deep and wide sky surveys. However, the payoff if these statistical
correlations can be measured is enormous: weak lensing provides a direct
measure of the distribution of matter, independent of any assumptions about
galaxy biasing.  Since this distribution can be predicted theoretically,
even in the quasilinear regime, and since its amplitude can be directly used
to constrain cosmology (unlike for galaxy surveys where one must marginalize
over the bias), weak lensing has great potential as a cosmological probe.

In principle, one may attempt to observe either the shearing of galaxies
(shape distortion) or their magnification (size distortion).  In practice,
the shape distortions have been used much more widely, since the mean shape
of galaxies is known (they are statistically round: as many galaxies are
elongated on the $x$-axis as on the $y$-axis) and the scatter in their
shapes is less than the scatter in their sizes.

A variety of statistical approaches have been used to extract information
from weak lensing shear (see later subsections for references).  
The simplest is the angular shear correlation
function, or its Fourier transform, the shear power spectrum.  These are
related to integrals over the matter power spectrum along the line of
sight, and as such in the linear regime at low redshift they scale as
$\propto\Omega_m^2\sigma_8^2$.\footnote{Warning: these scalings are altered
even at modest redshift, or in the nonlinear regime where the exponent
of $\sigma_8$ becomes closer to 3.}  Since the angular power spectrum is
rather featureless, more information can be extracted via {\em tomography}
--- the measurement of the shear correlation function as a function of the
redshifts of the galaxies observed, including the use of cross-correlations
between redshift slices.  Information on the relation between galaxies and
matter can be obtained via {\em galaxy-galaxy lensing}, i.e., the correlation
of the density field of nearby galaxies with the lensing shear measured on
more distant galaxies.  In the linear regime, the galaxy-galaxy lensing signal
scales as $\propto b\Omega_m\sigma_8^2$ and thus provides information on the
bias of the lensing galaxies, 
while in the nonlinear regime it probes individual galaxy halos and
hence places constraints on the halo occupation distribution
(\S\ref{sec:cmb_lss}).  Combination of this with the galaxy
clustering signal (which scales as $\propto b^2\sigma_8^2$) enables one
to eliminate the bias and measure $\Omega_m\sigma_8$.  The scaling of the
galaxy-galaxy lensing signal as a function of the source redshift, known
as {\em cosmography}, depends purely on geometric factors and hence can
be used to partially\footnote{The cosmography distance scale suffers from
three degeneracies, including the absolute-scale degeneracy that affects
supernova measurements; see \S\ref{sss:cosmography}.}
construct a distance-redshift relation.  Finally, the low-redshift matter
distribution is non-Gaussian, so higher-order statistics such as the
bispectrum or 3-point shear correlation function carry additional information.

For all of the applications of weak lensing to cosmology, deep wide-field
imaging is essential.  One can see this from a simple order-of-magnitude
estimate.  For a scatter in galaxy shapes of $\sigma_\gamma \sim 0.2$,
measuring a 1\%\ shear with unit signal-to-noise ratio requires $\sim 400$
galaxies ($0.2/\sqrt{400}\approx 0.01$).
Measuring the amplitude of density perturbations to 1\%\
accuracy requires that this be done over $\sim 10^4$ patches of sky, giving
a requirement of order $10^7$ galaxies, which for a density of 15 resolved
galaxies per arcmin$^2$ amounts to surveying 200 deg$^2$ of sky.  This is
the scale of the largest current surveys such as CFHTLS; in practice the
errors from these surveys are likely to be closer to several percent due to
``factors of a few'' that we have dropped here, and due to the inclusion
of systematic errors.  The eventual goal of the weak lensing community is
one or more ``Stage IV'' surveys (such as LSST on the ground and 
\euclid\ and \wfirst\ in space)
that would measure shapes of $\sim 10^9$ galaxies and achieve an additional
order of magnitude in precision.  Such surveys will have to face the daunting
task of reducing systematic errors by another order of magnitude.

There are unfortunately many sources of these systematic errors, and most of
the effort of the weak lensing community has been devoted to defeating them.
One is the measurement of galaxy shapes: while gravitational lensing by a
large-scale density perturbation can coherently align the images of many
galaxies, this can also arise from shaking of the telescope or optical
aberrations.  The accurate determination of the point-spread function (PSF)
of the telescope (usually based on observations of stars) and removal of its
effects is thus critical.  This problem gets much worse if one tries to model
galaxies with sizes similar to or smaller than the PSF.  High-resolution,
stable
imaging can help with this problem, motivating placement of future instruments
at the best ground-based sites or in space.  The determination
of redshifts for the large number of source galaxies is also a concern.  It
is not practical to obtain a robust spectroscopic redshift of every galaxy,
and hence ``photometric redshifts'' --- estimates of galaxies' redshifts
based on their broadband colors --- are used.  These must be calibrated with
well-known biases, scatters, and outlier distributions.  Finally, there are
astrophysical uncertainties: galaxies can suffer ``intrinsic alignments''
(non-random orientations), and the matter power spectrum may deviate from
pure CDM simulations at small scales.  Much of our discussion here
will be focused on the methodologies that have been developed to suppress
systematics at each stage of the observations and analysis.

\subsection{Weak lensing principles: Mathematical discussion}
\label{sec:wl_gp}

We will now go into greater detail on the mathematical aspects of weak
lensing, both the construction of the weak lensing field
and the various statistics that one can extract from it.
The modern theoretical formalism of weak lensing traces back
largely to the papers of \cite{blandford91}, \cite{miralda91},
and \cite{kaiser92}, though one can find roots in the much earlier 
papers of \cite{kristian66} and \cite{gunn67}.

\subsubsection{Deflection of light in cosmology}

Gravitational lensing gives a mapping from the intrinsic, unlensed image
of the sources of light on the sky --- the {\em source plane} --- to the
actual observable sky --- the {\em image plane}.  Our ultimate goal is
to extract information about the statistics and redshift dependence of this
mapping and use it to constrain cosmological parameters.  Our task here is
thus two-fold.  First, we must derive the mapping function that relates
the source to the image plane.  However,
since we do not know the intrinsic appearance of the sources, we cannot
directly infer the lens mapping from observations.  Therefore, our second
task will be to determine what properties of the lens map can be measured,
and with what accuracy.

In a fully general context, the lens mapping can be obtained by taking
an observer and following the geodesics corresponding to that observer's
past light cone.  We will make some simplifying approximations
here, namely that: (i) the spacetime is described by a
Friedmann-Robertson-Walker metric with
scalar perturbations and negligible anisotropic stresses (appropriate for
nonrelativistic matter, scalar fields, and $\Lambda$); (ii) deflection
angles are sufficiently small that we may use the flat-sky approximation;
(iii) the evolution of perturbations is slow enough that we may neglect
time derivatives of the gravitational potential $\Phi$ in comparison to
spatial derivatives (i.e., nonrelativistic motion); and (iv) such perturbations
are small enough that we may compute the lens mapping only to first order in
perturbation theory.\footnote{These approximations are sufficient to
analyze present power spectrum data, but corrections to
(iv) will become necessary in the future.}
Within these approximations, we may write the
angular coordinates $(\theta_1,\theta_2)$ of a light ray projected back to
comoving distance $D_C$ (see eq.~\ref{eqn:dcomove})
in terms of the position $(\theta_1^I,\theta_2^I)$
in the image plane as\footnote{The derivation of equation~(\ref{eq:wl:soln})
can be found in many works, though not always in the same notation.  See,
e.g., eq. (6.9) in the classic review by \cite{Bartelmann2001}.  The appendix
of \cite{Hirata2003} gives a shorter derivation in more similar notation.}
\begin{equation}
\theta_i(D_C) = \theta_i^I - 2 \int_0^{D_C} {\cal G}(D_{C1},D_C)
\frac{\partial \Phi}{\partial\theta_i}[D_{C1},\theta_i(D_{C1})]\, dD_{C1},
\label{eq:wl:soln}
\end{equation}
where ${\cal G}$ is the Green's function,
\begin{equation}
{\cal G}(D_{C1},D_C) = \int_{D_{C1}}^{D_C} [D_A(D_{C2})]^{-2} dD_{C2}
= \cot_K(D_{C1}) - \cot_K(D_C).
\label{eq:wl:green}
\end{equation}
Here $\cot_K(D_C)$ is the cotangentlike function,
\begin{equation}
\cot_K(D_C) = \left\{\begin{array}{ll} D_C^{-1} & {\rm ~flat} \\
K^{-1/2}\cot(K^{1/2}D_C) & {\rm ~closed} \\
|K|^{-1/2}\coth(|K|^{1/2}D_C) & {\rm ~open}, \end{array}\right.
\end{equation}
with the dimensional curvature $K$ defined in equation~(\ref{eqn:Kdef}),
and $\Phi$ is the Newtonian gravitational potential.

The potential derivative in equation~(\ref{eq:wl:soln}) is evaluated at the
position of the deflected ray $\theta_I(D_{C1})$, so it represents an implicit
solution to the light deflection problem.  However, in linear perturbation
theory (see our assumption iv above), we may evaluate it at the position
of the undeflected ray.  This is known as the {\em Born approximation}.
When we do this, it is permissible to pull the angular derivative out of
the integral and write
\begin{equation}
\theta^{\rm S}_i = \theta^{\rm I}_i + \frac{\partial \psi(D_C,\theta_i^{\rm
I})}{\partial\theta^{\rm I}_i},
\label{eq:wl:mapping}
\end{equation}
where $\psi$ is the {\em lensing potential}:
\begin{equation}
\psi(D_C,\theta_i) = - 2 \int_0^{D_C} [\cot_K(D_{C1}) - \cot_K(D_C)]
\Phi(D_{C1},\theta_i)\,dD_{C1}.
\label{eq:wl:soln2}
\end{equation}
Here it is important to remember that $D_C$ represents the distance to the
sources; one integrates over lens distances $D_{C1}$.

Equation~(\ref{eq:wl:mapping}) provides the mapping from the observed image
plane to the source plane, $\theta_i^{\rm S}(\theta_i^{\rm I})$.  In what
follows, we will assume that this mapping is one-to-one: this is known as
the regime of {\em weak lensing}.  In the small portion of sky covered
by very massive objects, the alternate regime of {\em strong lensing}
occurs, in which several points in the image plane map to the same point
in the source plane.  Strong lensing is an important probe of the matter
distribution in clusters, but we will not pursue it in this article;
we briefly discuss some applications of strong lensing
to cosmic acceleration in \S\ref{sec:strong_lensing}.

\subsubsection{Cosmic shear, magnification, and flexion}

We have now accomplished our first task: deriving the lens mapping from
the matter distribution.  However, we now need a way to classify the
observables in the lens mapping.  The potential $\psi$ is of course not
observable itself: like the Newtonian gravitational potential, its zero-level
is arbitrary.  Its angular derivative $\partial\psi/\partial\theta_i$ is
the {\em deflection angle}: the difference between the true position of a
source $\theta^{\rm S}_i$ and its apparent position $\theta^{\rm I}_i$.
However, since sources (in practice, galaxies) can be at any position,
we cannot measure the deflection angle either.

Let us now consider the second derivative of the lensing potential.  It is
simply the Jacobian of the mapping from image to source plane:
\begin{equation}
\frac{\partial \theta_i^{\rm S}}{\partial\theta_j^{\rm I}} = \delta_{ij}
+ \frac{\partial^2\psi}{\partial\theta_i\theta_j}
= \left( \begin{array}{cc} 1-\kappa-\gamma_+ & -\gamma_\times \\
-\gamma_\times & 1-\kappa+\gamma_+ \end{array}\right).
\label{eq:wl:jacobian}
\end{equation}
We have separated the 3 independent entries in the symmetric $2\times 2$
matrix of partial derivatives into 3 components: the {\em magnification}
(or {\em convergence})
$\kappa$ and the 2 components of {\em shear}, $\gamma_+$ and $\gamma_\times$.
The magnification has three effects:
\begin{itemize}
\item It makes the angular size of a galaxy look larger by a factor of
$1+\kappa$.
\item It makes the galaxy appear brighter by a factor of the
inverse-determinant of the Jacobian, $1+2\kappa$, since lensing conserves
surface brightness as dictated by Liouville's theorem.
\item It dilutes the number density of galaxies by a factor of $1-2\kappa$,
since the angular spacing between neighboring galaxies is increased by a
factor $1+\kappa$.
\end{itemize}
Magnification is a ``scalar'' in the sense that it is invariant under
rotations of the $(\theta_1,\theta_2)$ coordinate axes.

The shear stretches the galaxy along one axis and squeezes on the other:
the image of an intrinsically round galaxy appears elongated along the
$\theta_1$ axis if $\gamma_+>0$ and along the $\theta_2$ axis if $\gamma_+<0$.
The $\gamma_\times$ component stretches and squeezes along the diagonal
(45$^\circ$) axes.  The shear is a ``spin-2 tensor'' in the sense that
under a counterclockwise rotation of the coordinate axes by angle $\delta$,
it transforms as
\begin{equation}
\left( \begin{array}{c} \gamma_+ \\ \gamma_\times \end{array}\right)_{\rm new}
=
\left( \begin{array}{cc} \cos 2\delta & \sin 2\delta \\ -\sin 2\delta &
\cos 2\delta \end{array}\right)
\left( \begin{array}{c} \gamma_+ \\ \gamma_\times \end{array}\right)_{\rm
old}.
\end{equation}
If all galaxies were round, then each galaxy would provide a direct estimate
of the shear, since we could find the values of $(\gamma_+,\gamma_\times)$
that transformed an initially circular galaxy into the observed image.
In reality, galaxies come in many shapes, and any such estimate of the shear
components will have some standard deviation $\sigma_\gamma$ known as the
{\em shape noise}.  But in an ensemble average sense galaxies {\it are}
round --- there are as many galaxies in the universe elongated along
the $\theta_1$ axis as the $\theta_2$ axis. Thus,
if we take $N$ galaxies in the same region of sky,
we may expect that the shear components in that region can be measured with
a standard deviation 
of $\sim\sigma_\gamma/\sqrt N$.\footnote{As discussed 
further in \S\ref{sec:wl_sma} below,
a typical population of optically imaged galaxies (bulges and
randomly oriented disks) has an rms ellipticity $e_{\rm rms}\sim 0.4$
per component, which translates into an rms shear error
$\sigma_\gamma \approx 0.2$ via the shear response factor
(eq.~\ref{eq:wl:ea}).  Because there are two components
to shear one might expect to do a factor of $\sqrt{2}$ better in
statistical measurements,
but in the shear correlation function or power spectrum only one of the two
measurable components (the ``E-mode'' discussed in the next section)
contains a cosmological signal at leading order, so the relevant number for 
order-of-magnitude sensitivity estimates is generally 
$\sigma_\gamma \approx 0.2$.  Similarly, for galaxy-galaxy or
cluster-galaxy weak lensing, only the tangential shear contains cosmological information.}

Several caveats are in order at this point, and they form the basis for most
of the technical problems in weak lensing.  One is that a circular galaxy
re-mapped by the Jacobian (eq.~\ref{eq:wl:jacobian}) becomes an ellipse,
but since in the real sky one does not observe a population of galaxies with
homologous elliptical isophotes, there is no unique procedure to estimate
the shear.  Moreover, real telescopes, even in space, have finite resolution,
and the observed image is convolved with a PSF that smears the galaxy and
may introduce spurious elongation on some axis.  These two problems together
are referred to as the {\em shape measurement} problem.  A more fundamental
issue is that real galaxies are not randomly oriented: they have preferred
directions of orientation that are correlated with each other and with
large-scale structure, and thus contaminate statistical measures of the
cosmic shear field.  This is known as the {\em intrinsic alignment} problem.
Finally, as already mentioned above, relating the lensing potential
$\psi$ to the gravitational potential $\Phi(z)$, and hence to cosmological
parameters, requires accurate knowledge of the source galaxy
redshift distribution, presenting the {\em photometric redshift
calibration} problem.
We will discuss all of these problems in 
\S\S\ref{sec:wl_obs}--\ref{sec:wl_systematics}.

Measuring magnification $\kappa$ has proven more difficult than
measuring shear.  One might
imagine comparing the size, magnitude, or abundance of galaxies in some
region of sky to a typical or ``reference'' value, but there is a very wide
dispersion in galaxy sizes and magnitudes, and since some galaxies are too
faint to observe even in deep surveys one cannot measure such a thing as the
total number of galaxies.  Rather, one can measure the cumulative number
of galaxies brighter than some flux threshold, $N(>F)$.  If the number
counts have a power-law slope $\alpha$, i.e. $N(>F)\propto F^{-\alpha}$,
then magnification will perturb this distribution by a factor
\begin{equation}
N(>F,{\rm observed}) \propto [ 1 + 2(\alpha-1)\kappa ]F^{-\alpha}.
\label{eq:wl:mag}
\end{equation}
There are two competing effects here: in regions of higher
magnification the galaxies appear brighter, which gives the $2\alpha\kappa$
factor in equation~(\ref{eq:wl:mag}), but there is also the dilution of galaxy
number, which is responsible for the ``$-1$'' term.  Unfortunately,
for optical galaxies the observed number count slope is close to the critical
value $\alpha\approx 1$ for which magnification is not measurable.  Moreover,
the intrinsic clustering of galaxies gives large fluctuations in the number
density that greatly exceed those due to lensing effects.  For these reasons,
magnification has lagged behind shear as a cosmological probe, and the cosmic
magnification signal was not seen until \citet{scranton05} measured it using
cross-correlation of foreground galaxies and
background quasars. \citet{menard10} provide a more detailed analysis, using color
information to simultaneously detect lensing magnification and reddening of quasars
by dust correlated with intervening galaxies.

The most promising route to utilizing the cosmic magnification signal is to use
scaling relations that relate the size of a galaxy (as quantified by,
e.g., the half-light radius) to parameters that are magnification-independent
and can be measured in photometric surveys \citep{bertin06}, such as the
surface brightness, the Sersi\'c index, or (for AGN)
variability amplitude.
\citet{huff11mag} present a first application of this ``photometric
magnification'' method to galaxies, and
\citet{bauer11} an application to quasars.

After shear and magnification comes
the third derivative of the potential, i.e. the variation of shear
and convergence across a galaxy.  This effect
is called the {\em flexion}, and it
manifests itself via asymmetric banana and triangle-like distortions of an
initially circular galaxy \citep{goldbergbacon05}.  Flexion has been measured
by several groups (e.g. \citealt{leonard07,velander10,leonard11}), and there is
a growing literature on the theory of flexion measurement that parallels the
formalism required for shear measurement (e.g. \citealt{masseyfl07,schneider08,rowe12}).
However, because of the extra derivative it is sensitive mainly
to structure at the very
smallest scales, so it is primarily a tool for cluster lensing rather than
cosmological applications on larger scales.

\subsubsection{Power spectra and correlation functions*}

Just as for any other random field in cosmology, one may construct statistics
for the cosmic shear field.  The most popular are the power spectrum and
its real-space equivalent, the correlation function.

To construct the power spectrum, we take the Fourier transform of the
shear field,
\begin{equation}
\tilde\gamma_{+,\times}({\bf l}) = \int \gamma_{+,\times}({\theta}) e^{-i{\bf
l}\cdot\theta} d^2\theta
\;\;\;\;\leftrightarrow\;\;\;\;
\gamma_{+,\times}({\theta}) = \int \tilde\gamma_{+,\times}({\bf l}) e^{i{\bf
l}\cdot\theta} \frac{d^2{\bf l}}{(2\pi)^2}~.
\end{equation}
When considering the shear produced by a plane wave perturbation
of the lensing potential $\psi(\theta)$,
it is convenient to rotate the Fourier-space components from the coordinate
axis basis to a basis aligned with the direction of the wavevector, which
is a preferred direction in the problem.  The rotated components
are called the $E$-mode and $B$-mode:
\begin{equation}
\tilde\gamma_E({\bf l}) = \cos(2\phi_{\bf l}) \tilde\gamma_+({\bf l}) +
\sin(2\phi_{\bf l})\tilde\gamma_\times({\bf l})
{\rm ~~~and~~~}
\tilde\gamma_B({\bf l}) = \cos(2\phi_{\bf l}) \tilde\gamma_\times({\bf l})
- \sin(2\phi_{\bf l})\tilde\gamma_+({\bf l}),
\label{eq:wl_e_and_b}
\end{equation}
where $\tan\phi_{\bf l} = l_2/l_1$, with $l_1$ and $l_2$ being
the components of ${\bf l}$ in the pre-rotated coordinate system.
Thus the $E$-mode of the shear field corresponds to galaxies that are
stretched in the direction of the wave vector and squashed perpendicular
to it, whereas the $B$-mode corresponds to stretching and squashing at 
45$^\circ$ angles.  One may then define the power spectra:
\begin{equation}
\langle \tilde\gamma_E^\ast({\bf l}) \tilde\gamma_E({\bf l}')\rangle =
(2\pi)^2 C_{EE}(l) \delta^{(2)}({\bf l}-{\bf l}'),
\label{eq:wl:pedef}
\end{equation}
and similarly for $C_{EB}(l)$ and $C_{BB}(l)$.  Rotational symmetry 
of structure in the universe guarantees
that these depend only on the magnitude of ${\bf l}$ and not its direction,
and reflection symmetry guarantees that $C_{EB}(l)=0$.

In order to compute these power spectra, we need to express the Fourier
modes in terms of those of the lensing potential.  From the definition,
equation~(\ref{eq:wl:jacobian}),
the shear is seen to be the derivative of the deflection angle and hence
the second derivative
of the lensing potential,
\begin{equation}
\gamma_+(\theta) =- \frac12\left( \frac{\partial^2\psi}{\partial\theta_1^2}
- \frac{\partial^2\psi}{\partial\theta_2^2}\right)
{\rm~~and~~}
\gamma_\times(\theta) =
-\frac{\partial^2\psi}{\partial\theta_1\,\partial\theta_2}.
\end{equation}
Using the replacement $\partial/\partial\theta_i\rightarrow il_i$, we find
in Fourier space
\begin{equation}
\tilde\gamma_+({\bf l}) = \frac12(l_1^2-l_2^2)\tilde\psi({\bf l}) =
\frac12l^2\cos( 2\phi_{\bf l})\tilde\psi({\bf l})
 {\rm ~~and~~}
 \tilde\gamma_\times({\bf l}) = l_1l_2\tilde\psi({\bf l}) =
 \frac12l^2\sin(2\phi_{\bf l})\tilde\psi({\bf l}).
\end{equation}
Substitution into equation~(\ref{eq:wl_e_and_b}) implies that
\begin{equation}
\tilde\gamma_E({\bf l}) = \frac12l^2\tilde\psi({\bf l}) {\rm~~and~~}
\tilde\gamma_B({\bf l}) = 0.
\end{equation}
We thus arrive at the remarkable conclusion that cosmic shear possesses
only an $E$-mode; the $B$-mode shear must vanish, and we have $C_{BB}(l)=0$.
Confirming this prediction of vanishing $B$-mode provides 
a valuable, though not foolproof, test for systematics in WL surveys.

The $E$-mode shear power spectrum is simply $(l^2/2)^2$ times the lensing
potential power spectrum.  The latter may be found from the Limber
(small-angle)
approximation\footnote{See \citet{limber53} and \cite{limber54} for an
introduction to the theory. An exposition in terms of the power spectrum
is given by \citet{peebles73}.} in terms of the Newtonian potential power
spectrum, yielding
\begin{equation}
C_{EE}(l) = l^4 \int_0^{D_C} [\cot_K(D_{C1}) - \cot_K(D_C)]^2 \frac{
P_\Phi(k=l/D_{A1}) }{D_{A1}^2} dD_{C1}.
\end{equation}
(Here the power spectrum is evaluated at the redshift corresponding to
$D_{C1}$.)
We may put this in a more familiar form by recalling Poisson's equation, which
tells us that the potential and matter density perturbations are related by
\begin{equation}
P_\Phi(k) = \left[\frac32\Omega_mH_0^2 (1+z)\right]^2 k^{-4} P_\delta(k)~,
\end{equation}
yielding
\begin{equation}
C_{EE}(l) = \int_0^{D_C} [W(D_{C1},D_C)]^2
\frac{P_\delta(k=l/D_{A1})}{D_{A1}^2} dD_{C1}~,
\label{eq:wl:pee}
\end{equation}
where the {\em lensing window function}\footnote{Warning: many conventions
in use!} is\footnote{The Heaviside step function $\Theta$ is technically
unnecessary in equation~(\ref{eq:wl:window}), but it is convenient when considering
multiple populations of sources.}
\begin{equation}
W(D_{C1},D_C) = \frac32\Omega_mH_0^2 (1+z_1) D^2_{A1} [\cot_K(D_{C1}) -
\cot_K(D_C)] \Theta(D_C-D_{C1}).
\label{eq:wl:window}
\end{equation}
The window function
describes the contributions to lensing of sources at $D_C$
from lens structures at
distance $D_{C1}$.  Note that it vanishes as the lens approaches the source
($D_{C1}\rightarrow D_C$). In this equation, $D_{A1}$ is the comoving
angular diameter distance (eq.~\ref{eqn:adist}) to $D_{C1}$: in a curved
universe $D_{A1}\neq D_{C1}$. Note that in a flat universe, the window
function reduces to
\begin{equation}
W_{\rm flat}(D_{C1},D_C) = \frac32\Omega_mH_0^2 (1+z_1)
\frac{D_{C1}(D_C-D_{C1})}{D_C} \Theta(D_C-D_{C1}).
\end{equation}

One may also define the angular correlation function of the shear for two
galaxies separated by angle $\vartheta$.  Since the shear is a tensor, this
is more complicated than the correlation function for scalars.  Without loss
of generality, we may rotate the coordinate system so that the galaxies
are separated along the $\theta_1$-axis, and then take the $+$ and $\times$
components of the shear.  We then define the shear correlation functions,
\begin{equation}
C_{++}(\vartheta) = \langle \gamma_+(0) \gamma_+(\vartheta)\rangle
{\rm ~~and~~}
C_{\times\times}(\vartheta) = \langle \gamma_\times(0)
\gamma_\times(\vartheta)\rangle.
\end{equation}
As in the scalar case, these are related to the power spectra:
\begin{eqnarray}
C_{++}(\vartheta)\!\!\!  &=& \!\!\! \langle \gamma_+(0)
\gamma_+(\vartheta)\rangle
\nonumber \\
&=& \!\!\!
\int \frac{d^2{\bf l}}{(2\pi)^2} \int \frac{d^2{\bf l}'}{(2\pi)^2}
\langle \tilde\gamma_+({\bf l}) \tilde\gamma_+({\bf l}') \rangle
\exp(il'\vartheta\cos\phi_{{\bf l}'})
\nonumber \\
&=& \!\!\!
\int \frac{d^2{\bf l}}{(2\pi)^2}
\left[ \cos^2 (2\phi_{\bf l}) C_{EE}(l)
+ \sin^2 (2\phi_{\bf l}) C_{BB}(l)
\right]
\exp(il\vartheta\cos\phi_{{\bf l}})
\nonumber \\
&=& \!\!\!
\int_0^\infty \left\{ \frac{J_0(l\vartheta) + J_4(l\vartheta)}2C_{EE}(l)
+ \frac{J_0(l\vartheta)-J_4(l\vartheta)}2C_{BB}(l) \right\}\frac{l dl}{2\pi}~,
\label{eq:wl:cpp}
\end{eqnarray}
where $J_0$ and $J_4$ are spherical Bessel functions.
The expression for $C_{\times\times}$ is similar, but with $P_{EE}$ and
$P_{BB}$ switched.  The correlation
function $\{C_{++}(\vartheta),C_{\times\times}(\vartheta)\}$, if
measured over all scales, contains exactly the same information as the power
spectrum $\{C_{EE}(l),C_{BB}(l)\}$, as one can be derived from the other.
Therefore, the choice of which to measure is usually a technical one
based on the ease of data processing and handling of covariance matrices.
The condition for no $B$-modes, $C_{BB}(l)=0\,\forall l$,
is more complicated in correlation-function space.

An infinite number of other second-order statistics (i.e., expectation
values containing two powers of shear) can be constructed,
such as the aperture-mass variance \citep{schneideretal98}, ring statistics
\citep{schneiderkilbinger07}, and finite-interval orthogonal basis
decompositions (a.k.a.\ COSEBIs, \citealt{schneider10}).
These alternative statistics were introduced because they have useful
properties from the point
of view of data processing or systematics control --- e.g., for separation of
$E$ and $B$ modes,
or restriction to a particular range of scales --
but all of them are expressible as integrals over the power
spectrum or correlation function.

Formulae such as~(\ref{eq:wl:pee}) and~(\ref{eq:wl:cpp})
may be generalized to the full sky, as was first done for
CMB polarization \citep{kamionkowski97,zaldarriaga97}, but for cosmic shear most applications involve small
angular scales where the flat-sky approximation suffices.\footnote{The leading order
curved sky correction is the replacement of the scalar wavenumber $|{\bf l}|$ with
$\sqrt{l(l+1)}$, where here ``$l$'' is the spherical multipole number. Further corrections
are of the order of $1/l^2$ and are most important for the lowest multipoles.}

Having built the formalism to describe the statistics of weak lensing,
we can now consider the proposed ways of using it to measure cosmology.
Some methods will depend only on the expansion history of the universe,
while others are sensitive to the growth of perturbations.

\subsubsection{Method I: Cosmic Shear Power Spectrum*}

The conceptually simplest approach to using WL is to collect a sample of
source galaxies, obtain an estimator for the shear at each galaxy, measure the
correlation function 
or power spectrum, and do a comparison to equation~(\ref{eq:wl:pee}).
Of course not all galaxies are at the same redshift, but there is a
probability distribution of distances $p(D_C)$, and the observed mean shear
in a particular region of sky is then
\begin{equation}
\gamma_{+,\times}(\theta) = \int_0^{D_{C,\rm max}} p(D_C)
\gamma_{+,\times}(D_C,\theta)\, dD_C,
\end{equation}
where $D_{C,\rm max}$ is the comoving distance to the farthest galaxy in
the slice.
The power spectrum of this field can then be written as
\begin{equation}
C_{EE}(l) = \int_0^{D_{C,\rm max}} [W_{\rm eff}(D_{C1})]^2
\frac{P_\delta(k=l/D_{A1})}{D_{A1}^2} dD_{C1}.
\label{eq:wl:pee-cs}
\end{equation}
This is similar to equation~(\ref{eq:wl:pee}) with $W$ replaced by an effective
window function,
\begin{equation}
W_{\rm eff}(D_{C1}) = \int_0^{D_{C,\rm max}} p(D_C) W(D_{C1},D_C)\,dD_C,
\end{equation}
which is simply the usual window function appropriately weighted over the
source galaxies.

The cosmic shear power spectrum $C_{EE}(l)$ is sensitive to many
cosmological parameters.  Being an integral over the matter power spectrum,
it is $\propto\sigma_8^2$ in the linear regime, although its behavior in
the nonlinear regime is closer to $\propto\sigma_8^3$.  It also contains two
powers of $\Omega_m$, so we expect that the most important dependences in the
problem are that the WL power spectrum scales as $\sim\Omega_m^2\sigma_8^3$.
This is qualitatively correct, but the matter power spectrum and the mapping
between $D_A$ and $D_C$ at finite redshift contain sensitivities to all of
the cosmological parameters, and so a full answer to the question ``what
does the shear power spectrum constrain?'' requires us to actually do the
integral to obtain $C_{EE}(l)$.

The sensitivity to every parameter is both a virtue of the WL power spectrum
and its greatest fault: the featureless WL power spectrum contains too
many parameter degeneracies.
One way to break these degeneracies is to combine WL with other probes,
as discussed in \S\ref{sec:forecast}.  However, there are also
ways of using
WL that provide additional information and break these degeneracies
internally, as we now discuss.

\subsubsection{Method II: Power Spectrum Tomography*}

We can improve on the WL power spectrum constraints if we can split the source
galaxies into redshift slices.  In most practical cases, this would be done
with photometric redshifts.  In this case, instead of having a single power
spectrum, we have $N(N+1)/2$ power spectra and cross-spectra; if we denote
the slices by $\alpha,\beta\in\{1,2,...N\}$, then these spectra are
\begin{equation}
C_{EE}^{\alpha\beta}(l) = \int W_{{\rm eff},\alpha}(D_{C1}) W_{{\rm
eff},\beta}(D_{C1}) \frac{P_\delta(k=l/D_{A1})}{D_{A1}^2} dD_{C1},
\label{eq:wl:peetomo}
\end{equation}
where $W_{{\rm eff},\alpha}$ is the effective window function for the
$\alpha$ slice.  Note that because the window functions are multiplied,
this power spectrum depends only on the matter power spectrum at redshifts
closer than that of the nearby slice, i.e.\ at $z<\min\{z_\alpha,z_\beta\}$.
This makes sense because a given lens structure must be in front of both
sources to contribute to the shear cross-correlation.  
Lensing analysis that splits
samples by redshift and uses the redshift scalings to constrain
cosmology is known as {\em tomography}.

Like the shear power spectrum, the tomographic spectra are sensitive to
both the background geometry and the growth of structure: the shear power
spectrum at $l$ depends on the $D_C(z)$ relation, on $P_\delta(k=l/D_A;z)$
as a function of redshift, and on the curvature $K$.\footnote{There is also a
factor of $\Omega_m H_0^2$ in the window functions, but for now 
we will assume this combination has been measured accurately from the CMB.
Our forecasts in \S\ref{sec:forecast} marginalize over the uncertainty
in this combination, which matters at the precision of Stage IV experiments.}
With a single power spectrum
$C_{EE}(l)$ there is no hope of disentangling these functions with WL alone.
One might hope that having the tomographic cross-spectra as a function of
$z_\alpha$ and $z_\beta$ would allow the relevant degeneracies to be broken.
Unfortunately, such a program runs into three problems:
\begin{itemize}
\item A real WL survey has a maximum source redshift, and there is obviously
no sensitivity to structures farther than this.
\item There exist exact degeneracies among $\{ D_C(z), P_\delta(k=l/D_A;z),
K\}$ that lead to exactly the same lensing power spectra for all
$(l,z_\alpha,z_\beta)$.  The most obvious of these is the re-scaling
degeneracy: since lensing measures only dimensionless shears, it cannot
measure the absolute distance scale, only distance ratios.
Two other degeneracies are discussed
by \cite{bernstein06}; see also \S\ref{sss:cosmography}.
\item The broad, smooth nature of the lensing window functions $W_{{\rm
eff},\alpha}(D_C)$ implies near-degeneracies between power spectra
at adjacent redshifts.  For example, if one were to test a nonstandard
cosmology in which $P_\delta(k,z)$ had a rapid oscillation in $z$ superposed
on the expected evolution, the rapid oscillation would contribute little
to equation~(\ref{eq:wl:peetomo}) and would be easily buried by statistical or
systematic errors.
\end{itemize}
Despite these drawbacks, tomographic power spectra have far fewer parameter
degeneracies than the shear power spectrum alone.  More importantly, having
$N(N+1)/2$ power spectra provides many additional opportunities for internal
consistency tests and rejection of systematic errors.

\begin{figure}
\centerline{
\includegraphics[angle=-90,width=4in]{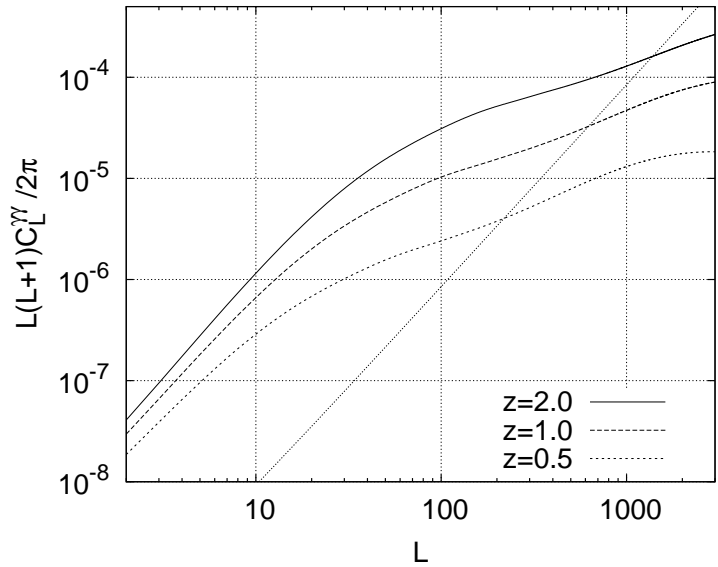}
}
\caption{\label{fig:wlps}The $E$-mode shear power spectra predicted for the
WMAP 7-year best fit
cosmology ($\Omega_m=0.265$, $\sigma_8=0.8$, $H_0=71.9\hubunits$). The curves
show power spectra for sources at $z=0.5$ (bottom), 1.0, and 2.0 (top). The
diagonal line
shows the shot noise contribution at a source density of $n_{\rm eff} = 10$
galaxies per arcmin$^2$; for this power spectrum measurement the
shot noise scales as $n_{\rm eff}^{-1}$. At small scales, where the noise power spectrum
exceeds the signal, it is not possible to measure individual structures in the weak lensing map.
However, with sufficient sky coverage, high-S/N measurement of the power spectrum or
correlation function is still possible (see \S\ref{sec:wl_stat_errors},
particularly eq.~\ref{eq:wl:staterr}).}
\end{figure}

Some examples of theoretical tomographic power spectra are shown in
Fig.~\ref{fig:wlps}.

\subsubsection{Method III: Galaxy-galaxy Lensing*}
\label{sec:wl_ggl}

A third way to use weak lensing is to look not just at the shear power
spectrum but at its correlation with the distribution of foreground galaxies.
This subject is known as {\em galaxy-galaxy lensing} (GGL), and it is a
powerful probe of the relation between dark matter and galaxies.  The angular
cross-power spectrum between the galaxies in one redshift slice $\alpha$
(the ``foreground'' or ``lens'' slice) and the $E$-mode shear in a more
distant slice $\beta$ (the ``background'' or ``source'' slice) is defined by
\begin{equation}
\langle \tilde \delta_g^{\alpha\ast}({\bf l}) \tilde\gamma^\beta_E({\bf l}')
\rangle =
(2\pi)^2 C_{gE}^{\alpha\beta}(l) \delta^{(2)}({\bf l}-{\bf l}'),
\end{equation}
where $\delta_g^\alpha$ is the 2-dimensional projected galaxy overdensity and
$\tilde \delta_g^\alpha$ is its Fourier transform, and $\alpha$ and $\beta$
represent redshift slices.  It can be computed via Limber's equation as
\begin{equation}
C^{\alpha\beta}_{gE}(l) = \int p_\alpha(D_{C1}) W_{{\rm eff},\beta}(D_{C1})
\frac{P_{g\delta}(k=l/D_{A1})}{D_{A1}^2}\,dD_{C1},
\label{eq:wl:cge}
\end{equation}
where $P_{g\delta}(k)$ is the 3-dimensional galaxy-matter cross-spectrum.
The real-space correlation function of galaxy density and shear is
\begin{equation}
C_{g+}^{\alpha\beta}(\vartheta) = -\int_0^\infty C^{\alpha\beta}_{gE}(l)
J_2(l\vartheta)\,\frac{l\;dl}{2\pi}.
\end{equation}
In the case where the foreground galaxy slice ($\alpha$) is narrow --
either due to use of spectroscopic foregrounds or high-quality photo-$z$s
-- the probability distribution 
in Limber's equation (eq.~\ref{eq:wl:cge}) becomes a
$\delta$-function, and the galaxy-matter cross-spectrum can be obtained.

One can also measure GGL by computing the mean tangential shear
(i.e., shear in the direction orthogonal to the lens-source vector)
of background galaxies around foreground galaxies as a function of
radius.  This view of the measurement is taken in many papers,
but it is (almost) mathematically equivalent to correlating the
shear field of the background galaxies with the density field
of the foreground galaxies.

From the perspective of dark energy studies, the principal advantage of
GGL over the shear power spectrum is observational: the shear is being
correlated with galaxies rather than itself.  
A spurious source of shear, e.g.\ from imperfections in the PSF model, 
is a source of systematic error in the shear power spectrum, but
in GGL it is only a source of noise because it is equally likely
to arise in regions of high and low foreground galaxy density.
The principal disadvantage
of GGL is that its interpretation requires assumptions about the galaxies,
which must ultimately be justified empirically.

Galaxy-galaxy lensing can be used in the linear, the weakly nonlinear,
and the fully nonlinear regimes:
\begin{itemize}
\item
{\em Linear regime}: In the linear regime, the galaxy-matter cross
spectrum is $P_{g\delta}(k) = bP_\delta(k)$, where $b$ is the galaxy bias
factor.
Thus $C^{\alpha\beta}_{gE}(l)$ is proportional to $b\sigma_8^2$, whereas the
galaxy power spectrum is proportional to $(b\sigma_8)^2$.  This provides a
way to measure the linear bias of the galaxies and hence obtain $\sigma_8$.
Unfortunately, one must reach very large scales (tens of Mpc) for linear
perturbation theory to be valid at the few percent level of accuracy, and
at these scales the signal-to-noise ratio of current GGL results is very low.
\item
{\em Weakly nonlinear regime}: At scales of order $\sim 10h^{-1}\,$Mpc,
nonlinear effects simply represent a correction to the linear theory, and one
might hope that a judicious combination of observables can remove them.
The key is to note that when stochasticity between galaxy and
matter densities is included, the GGL
signal is proportional to $br\sigma_8^2$, where $r$ is the galaxy-matter
cross-correlation coefficient,
so all we require to extract $b$
and $\sigma_8$ individually from GGL and galaxy clustering observables
is a theoretical prediction for the stochasticity.  This is a convenient
result because simulations show that $r=1$ is a much better
approximation in the weakly nonlinear regime than $b=\,$constant.  This type of
analysis is also best done in real space rather than Fourier space so that
the 1-halo contributions (see \S\ref{sec:cmb_lss})
to both clustering and lensing can be eliminated.
A specific outline for how to do this, including the next-order perturbation
theory corrections to $r=1$ and comparison to simulations, is presented
by \cite{baldauf10}.
\item
{\em Fully nonlinear regime}: GGL can be used on the scale of individual
halos ($k\sim 1$--10 Mpc$^{-1}$) to relate galaxy properties such as
luminosity, color, and stellar mass to the properties of the host dark
matter halo.  Such relations cannot be predicted {\it ab initio} because
of the complicated astrophysics involved.
Empirical constraints on these relations
are useful
for dark energy studies mostly because they enable us to test some of the
underlying assumptions of galaxy clustering models.
To gain some cosmological power beyond the weakly non-linear regime,
one can construct full galaxy HOD models and marginalize over
their parameters, using both GGL and galaxy clustering as
constraints \citep{yoo06,leauthaud11,leauthaud12}.
Within the weakly non-linear regime, this approach effectively
uses the HOD fits to compute the scale-dependence of $br$,
drawing on the information in the small-scale galaxy clustering
to improve the constraints.
\end{itemize}
\cite{yoo12} provide an extensive discussion of the cosmological
constraints that can be derived from the combination of GGL
and galaxy clustering, on small and large scales, 
in the simplified case where one isolates the population
of central galaxies, so that there is one galaxy per dark 
matter halo.

Cluster-galaxy lensing is similar to GGL, but one takes clusters of
galaxies rather than individual galaxies as the reference points
\citep{mandelbaum06,sheldon09}.
We will discuss this idea further in \S\ref{sec:cl}, arguing
that it offers the most reliable route to calibrating cluster
mass-observable relations and has the potential to sharpen
cosmological parameter constraints significantly.
Cluster-galaxy lensing may also be a useful tool for calibrating
uncertainties in shear calibration and photometric redshifts, since
the shear signal in the cluster regime is stronger and the
cluster photometric redshifts themselves are usually well determined.

\subsubsection{Method IV: Cosmography*}
\label{sss:cosmography}

The previous sections motivate us to ask whether there is a way to combine
the observational advantages of GGL with the model independence of the
shear power spectrum.  There is, although there is a large price to pay:
one can only obtain geometrical information.

The idea is to consider narrow slices of galaxies centered at redshifts
$z_\alpha<z_\beta<z_\gamma$ and measure the lensing of galaxies in
slices $z_\beta$
and $z_\gamma$ by galaxies in the foreground slice $z_\alpha$.  The ratio
of the galaxy-shear cross-spectra is, using equation~(\ref{eq:wl:cge}),
\begin{equation}
\frac{C^{\alpha\beta}_{gE}(l)}{C^{\alpha\gamma}_{gE}(l)} = \frac{\cot_K
D_C(z_\alpha) - \cot_K D_C(z_\beta)}
{\cot_K D_C(z_\alpha) - \cot_K D_C(z_\gamma)}~.
\label{eq:wl:cosmography}
\end{equation}
One can see that all dependence on the power spectra and the distribution
of galaxies has been cancelled, allowing a purely geometric test of
cosmology.  This is called the {\em cosmography} or {\em shear-ratio}
test \citep{jain03,bernstein04}.

One can see from equation~(\ref{eq:wl:cosmography}) that cosmography can
determine the $\cot_K D_C(z)$ relation up to any affine transformation,
i.e. transformations of the form
\begin{equation}
\cot_K D_C(z) \rightarrow a_0 + a_1 \cot_K D_C(z),
\label{eq:wl:affine}
\end{equation}
which leave the ratios of differences of $\cot_K D_C(z)$s unaffected.
(Recall that $\cot_KD_C = 1/D_C$ in a flat universe.)  It is clear that $a_1$
is the familiar overall rescaling degeneracy: cosmography measures only
dimensionless ratios and cannot distinguish two models with 
different $H_0$ but
the same values of $\Omega_m$, $w$, etc. Precisely the same degeneracy
afflicts the supernova $D_L(z)$ relation because the absolute magnitude of a
Type Ia supernova is not known {\it a priori}.
The $a_0$ degeneracy is trickier, arising from
the fact that $\infty$ is not a special distance in lensing
problems.\footnote{This is the same reason that the ``$\infty$''
setting on the focus knob for a camera is not special.}  Finally, since
only $\cot_K D_C(z)$ is measured, cosmography cannot by itself provide a
model-independent measurement of the curvature of the universe.  But aside
from these three degeneracies --- $a_1$, $a_0$, and $K$ --- the entire geometry
of the universe over the range of redshifts observed is measurable.

Unfortunately, the aforementioned degeneracies are similar in functional
form to the effects of $\Omega_m$ and $w$, and they have severely
limited the application of cosmography thus far.  This is particularly
true for observations restricted to low redshift: if one Taylor expands
the distance as $\tan_K D_C(z) = c_1z + c_2z^2 + c_3z^3 + ...$ then any
cosmological model is degenerate with one that has $(c_1,c_2)=(1,0)$,
and hence one must go through at least the $z^3$ term before cosmography
provides any useful information.  For example, at $(z_\alpha, z_\beta,
z_\gamma) = (0.25, 0.35, 0.70)$, the difference in the shear ratio
(eq.~\ref{eq:wl:cosmography}) between an $\Omega_m=0.3$ flat $\Lambda$CDM
cosmology and a pure CDM $\Omega_m=1$ cosmology is only 1\%!  In early
work \citep{mandelbaum05} cosmography was therefore used as a test for shear
systematics rather than a cosmological probe.

The outlook for cosmography is much brighter as we probe to larger redshifts,
or if we consider dark energy models with complicated redshift dependences
that cannot be mimicked by the degeneracy of equation~(\ref{eq:wl:affine}).
A particularly promising possibility is to use cosmography with lensing
of the anisotropies in the CMB ($z=1100$) to obtain a much longer lever
arm \citep{acquaviva08}.
In principle one can also apply the cosmography method
to strong gravitational lenses (see \S\ref{sec:strong_lensing} below).  
Here the challenge is that different sources
probe different locations in the lens, so one must be able to constrain
the lens potential extremely well to extract useful cosmographic
constraints.

\subsubsection{Method V: Non-Gaussian Statistics*}

The primordial density fluctuations in the universe were very nearly
Gaussian, as evidenced by the CMB.  In this case, the fluctuations are fully
described by the power spectrum, and this has become the common language
of CMB observations.  However, nonlinear evolution makes the matter
fluctuations and hence the lensing shear in the low-redshift universe
highly non-Gaussian on small and intermediate scales.
Therefore, many other statistical measures of the
shear field have been proposed, the most popular of which is the bispectrum.

The bispectrum is obtained by taking the product of three Fourier modes:
\begin{equation}
\langle \tilde\gamma^\alpha_E({\bf l}_1) \tilde\gamma^\beta_E({\bf l}_2)
\tilde\gamma^\gamma_E({\bf l}_3) \rangle
=(2\pi)^2 B^{\alpha\beta\gamma}_{EEE}(l_1,l_2,l_3) \delta^{(2)}({\bf
l}_1,{\bf l}_2,{\bf l}_3).
\end{equation}
Statistical homogeneity forces the three wave vectors involved to sum to
zero so the bispectrum is actually a function of the triangle configuration;
rotational and reflection symmetry then tell us that it depends only on the
side lengths $(l_1,l_2,l_3)$\footnote{The $EEB$ and $BBB$ bispectra flip sign
under reflections of the triangle, and some convention, e.g. that the sides
are given in counterclockwise order, must be imposed to avoid ambiguity.},
which must satisfy the triangle inequality.  Because there are 2 shear modes
($E$ and $B$), there are actually 4 types of bispectrum: $EEE$, $EEB$, $EBB$,
and $BBB$, but only $EEE$ can be produced cosmologically.  Limber's equation
expresses it in terms of the 3-dimensional matter bispectrum,
\begin{equation}
B^{\alpha\beta\gamma}_{EEE}(l_1,l_2,l_3) = \int W_{{\rm eff},\alpha}
W_{{\rm eff},\beta} W_{{\rm eff},\gamma}
\frac{B_\delta(l_1/\chi,l_2/\chi,l_3/\chi)}{D_{A1}^4} \,dD_{C1}.
\end{equation}
The bispectrum contains information equivalent to the shear 3-point
correlation function. The theory of transformations between the two and the implied
symmetry properties have been extensively studied \citep{zaldarriaga03I, schneider03I, takadajain03, schneider05I}.
Halo model based descriptions of the 3-point function are also available \citep[e.g.][]{cooray01}.

The original motivation to study the WL shear bispectrum was to break the
degeneracy between $\Omega_m$ and $\sigma_8$ \citep[e.g.][]{bernardeau97,hui99L,takadajain04}.  At low redshift, and on
large scales where perturbation theory applies, the WL power spectrum is
proportional to $\Omega_m^2\sigma_8^2$, whereas the bispectrum is proportional
to $\Omega_m^3\sigma_8^4$; it contains three powers of the
shear and hence three powers of $\Omega_m$, but the matter bispectrum is
generated by nonlinear interactions and is proportional to the square of
the matter power spectrum, i.e., to $\sigma_8^4$ rather than $\sigma_8^3$.
Unfortunately, this route to degeneracy breaking has proven difficult
because of the low
signal-to-noise ratio and high sampling variance of the bispectrum and
because the degeneracy directions of the power spectrum and
bispectrum are almost parallel in the $(\Omega_m,\sigma_8)$ plane.  A more
interesting application of the WL bispectrum in future surveys may be
as a constraint on modified gravity theories, though this has
not yet been well studied.

\subsection{The Current State of Play}
\label{sec:wl_current}

Weak lensing as a cosmological probe is only a decade old, although the
ideas go back much further. \citet{zwicky37} famously suggested gravitational
lensing as a tool to determine cluster masses (although the discussion focused
on strong lensing). We separately consider here the more recent history
of cosmic shear studies, and of galaxy-galaxy lensing as a cosmological
probe. Also the techniques and applications associated with lensing outside
the optical bandpasses are sufficiently different that we place them in a
separate section. Lensing by clusters is considered in the cluster section
(\S\ref{sec:cl}).

\subsubsection{Cosmic shear}

\citet{kristian67} described an initial attempt to measure statistical cosmic
shear using photographic plates taken on the Palomar 5 m telescope. He
even correctly identified intrinsic alignments as a systematic error,
and noted that the distance dependence could be used to separate them
from true cosmic shear. Interestingly, the objective of this analysis was
to search for cosmological-scale gravitational waves or other large-scale
anisotropies \citep{kristian66}. The author set a limit on the magnetic part
of the Weyl tensor\footnote{Equivalent to $\sim\omega^2h$, where $\omega$
is the gravitational wave frequency and $h$ is the strain.} of $\lesssim
200 H_0^{-2}$, which he describes as ``about the best that can be done with
this kind of measurement.'' Fortunately this has not remained the case --
indeed it was improved upon by two orders of magnitude by \citet{valdes83}.

The modern era of lensing studies was introduced by the availability of
arrays of large-format CCDs.  \citet{mould94} searched for cosmic shear and
reached percent-level sensitivity, but did not detect a signal. Cosmic shear
was finally detected in 2000 by several groups \citep{wittman00, bacon00,
vanwaerbeke00}, and in deeper but narrower data from \hst\ 
\citep{rhodes01,RRG02}. Over the same period, several additional square degrees
were observed with long exposure times in excellent seeing using ground-based
telescopes \citep{vanwaerbeke01,vanwaerbeke02,bacon03,hamana03}. The first
wide-shallow surveys were also carried out from the ground: the 53 deg$^2$
Red-Sequence Cluster Survey \citep{HYG02} and the 75 deg$^2$ CTIO survey
\citep{jarvis03}. These studies established the existence of cosmic shear,
but at a level far below that which would be expected in $\Omega_m\sim 1$
models normalized to the CMB.  The large error bars 
in early studies meant that 
only a single amplitude could be measured, yielding a constraint on the
combination $\sigma_8(\Omega_m/0.3)^\nu$, where the exponent $\nu$ varied
between 0.3 and 0.7 depending on the scale and depth. 
In the first detection of the cosmic shear bispectrum,
achieved with the VIRMOS-DESCART survey, \cite{penetal03} measured  
the skewness of the filtered shear signal and used it
in combination with the power spectrum to rule out 
large-$\Omega_m$, low-$\sigma_8$ solutions, finding 
$\Omega_m<0.5$ at 90\% confidence. The deep
COMBO-17 survey first detected the evolution of $\sigma_8$ as a function of
cosmic time \citep{bacon05}.

However, the early studies of cosmic shear were not free of trouble. As
one can see from Table~\ref{tbl:cosmicshear}, while most were broadly in
agreement with $\sigma_8$ in the 0.7--0.9 range, a detailed comparison shows
that the measurements were not all consistent. This discrepancy
stimulated discussions
about a number of possible ancillary issues with the data, such as the role
of intrinsic alignments, whether the source redshift distribution $N(z)$
was properly calibrated, and whether the models for the nonlinear power
spectrum and assumptions about the $P(k)$ shape parameter $\Gamma$ could be
leading to discrepancies. More seriously, most of the early measurements
contained $B$-mode signals at levels not far below the $E$-mode. This
was a clear signal of contamination of non-cosmological origin, probably
PSF correction residuals. Also, intrinsic alignments of galaxies were
detected at high significance even in the linear regime, at a level that
represented a potentially serious systematic error even for then-ongoing
surveys \citep{mandelbaum06}.

It was clear by 2006 that weak lensing was a {\em very} hard observational
problem and that a great deal of work lay ahead to turn it into a precision
cosmological probe. This resulted in a reduction in the rate of new cosmic
shear results, the reorganization of the field into larger teams, and detailed
looks at systematic errors ranging from optical distortions in telescopes
to intrinsic galaxy alignments.  Several
wide-field optical surveys were ongoing at the time, including the deep 170
deg$^2$ CFHT Legacy Survey (for which cosmic shear was a key science driver)
and the very deep multiwavelength COSMOS survey with high-resolution
optical imaging from \hst/ACS \citep{masseycosmos07,schrabback10}. 
The CFHTLS presented
some early results \citep{hoekstra06,semboloni06,fu08}, but following this
there was a rather bleak period of time. No new ground-based wide-field cosmic
shear results were published, and no new large surveys were undertaken with
\hst, nor do future large \hst\ weak lensing surveys
seem likely.\footnote{The premier lensing instrument on
the \hst\ (the Advanced Camera for Surveys) failed in January 2007. While
its wide-field channel was restored during the 2009 servicing mission,
the sky coverage possible with ACS is not competitive with next-generation
ground-based surveys, and it seems unlikely a major cosmic shear program will
be undertaken with \hst. Rather, the next major steps in space-based cosmic
shear will likely be the \euclid\ mission planned for 2020 and the
\wfirst\ mission planned for the early 2020s.}

In the past five years, however, great progress has been made in overcoming the
difficulties that at first appeared so daunting. The community made a massive
investment in algorithms to determine and correct for PSF ellipticities (we
will review some of these in \S\ref{sec:wl_shapes}), and in investigating
the physics that determines the PSF, including such complications as
atmospheric turbulence \citep{heymans11}. Equally important, these methods were
tested in public challenges on simulated data 
(STEP1, \citealt{heymans06}; STEP2, \citealt{massey07};
GREAT08, \citealt{GREAT08}; GREAT10, \citealt{GREAT10}; see
further dicussion in \S\ref{sec:wl_shapes}).
Progress was also made on astrophysical systematic errors. We learned
that large-scale intrinsic galaxy alignments are strongest for 
luminous red galaxies 
\citep{hirata07,mandelbaum11}, 
and that the linear alignment model,
once considered a crude analytical tool \citep{catelan01}, is in fact an
excellent description of the observations of early-type galaxies at $\ge
10h^{-1}\,$Mpc scales \citep{blazek11}.

As a result of this great effort by the community, the Stage II weak lensing
results are finally coming to fruition and yielding large data sets that pass
the standard systematics tests (e.g., $B$-modes consistent with zero). Two
groups \citep{lin11,huff11b} have performed a cosmic shear measurement
using the Sloan Digital Sky Survey deep co-added region --- a 120-degree long
stripe observed many times over the course of three years as part of the SDSS-II
supernova survey. These analyses used different methods to co-add their
data and correct for the PSF ellipticity, and they imposed different selection
cuts and hence had different redshift distributions, yet the results were
in agreement (and slightly more than 1$\sigma$ below the WMAP prediction
for $\sigma_8$). The largest of the Stage II weak lensing programs was the CFHT
Legacy Survey. After a thorough analysis, the lensing results and
cosmological implications were recently published 
\citep{heymans_cfht, benjamin_cfht, erben_cfht, kilbinger12, miller_cfht}.
They appear consistent with the standard $\Lambda$CDM cosmology with
WMAP-derived initial conditions, with the amplitude $\sigma_8$ measured to
$\pm 0.03$.

A summary of the current status of optical cosmic shear results is shown
in Table~\ref{tbl:cosmicshear}.

\begin{table*}
\caption{\label{tbl:cosmicshear}A summary of cosmic shear results from
the literature obtained in the optical. Note that some of these results
are independent analyses or extensions of previous data sets and hence are
not independent.}
{\tiny
\begin{tabular}{lcccl}
\hline\hline
Reference & Telescope/instrument & Area & Number of & Result \\
 & & (deg$^2$) & galaxies & \\
\hline
\citet{bacon00} & WHT/EEV-CCD & 0.5 & 27k & $\sigma_8=1.5\pm0.5$ (@
$\Omega_m=0.3$) \\
\citet{vanwaerbeke00} & CFHT/UH8K+CFH12K & 1.75 & 150k & Detection$^{\rm a}$ \\
\citet{wittman00} & Blanco/BTC & 1.5 & 145k & Detection$^{\rm b}$ \\
\citet{rhodes01} & HST/WFPC2 & 0.05 & 4k &
$\sigma_8(\Omega_m/0.3)^{0.48}=0.91^{+0.25}_{-0.30}$ \\
\citet{vanwaerbeke01} & CFHT/CFH12K & 6.5 & 400k &
$\sigma_8(\Omega_m/0.3)^{0.6}=0.99^{+0.08}_{-0.10}$ (95\%CL)$^{\rm c}$ \\
\citet{HYG02} & CFHT/CFH12K + Blanco/Mosaic II & 53 & 1.78M &
$\sigma_8(\Omega_m/0.3)^{0.55} = 0.87^{+0.17}_{-0.23}$ (95\%CL) \\
\citet{RRG02} & HST/WFPC2 & 0.36 & 31k & $\sigma_8=0.94\pm0.14$ (@
$\Omega_m=0.3$, $\Gamma=0.21$) \\
\citet{bacon03} & Keck II/ESI + WHT & 1.6 &  &
$\sigma_8(\Omega_m/0.3)^{0.68}=0.97\pm0.13$ \\
\citet{brown03} & MPG ESO 2.2m/WFI & 1.25 & & $\sigma_8(\Omega_m/0.3)^{0.49}
= 0.72\pm0.09^{\rm d,e}$ \\
\citet{jarvis03} & Blanco/BTC+Mosaic II & 75 & 2M  &
$\sigma_8(\Omega_m/0.3)^{0.57} = 0.71^{+0.12}_{-0.16}$ (2$\sigma$) \\
\citet{hamana03} & Subaru/SuprimeCam & 2.1 & 250k &
$\sigma_8(\Omega_m/0.3)^{0.37} = 0.78^{+0.55}_{-0.25}$ (95\%CL) \\
\citet{rhodesstis04} & HST/STIS & 0.25 & 26k &
$\sigma_8(\Omega_m/0.3)^{0.46}(\Gamma/0.21)^{0.18} = 1.02\pm0.16$ \\
\citet{heymans05} & HST/ACS & 0.22 & 50k & $\sigma_8(\Omega_m/0.3)^{0.65}
= 0.68\pm0.13$ \\
\citet{massey05} & WHT/PFIC & 4 & 200k & $\sigma_8(\Omega_m/0.3)^{0.5} =
1.02\pm 0.15$ \\
\citet{hoekstra06} & CFHT/MegaCam & 22 & 1.6M & $\sigma_8 = 0.85\pm0.06$ @
$\Omega_m=0.3$ \\
\citet{semboloni06} & CFHT/MegaCam & 3 & 150k & $\sigma_8=0.89\pm0.06$ @
$\Omega_m=0.3$ \\
\citet{benjamin07} & Various$^{\rm g}$ & 100 & 4.5M &
$\sigma_8(\Omega_m/0.3)^{0.59} = 0.74\pm0.04$ \\
\citet{hetterscheidt07} & MPG ESO 2.2m/WFI & 15 & 700k & $\sigma_8=0.80\pm0.10$
@ $\Omega_m=0.3$ \\
\citet{masseycosmos07} & HST/ACS & 1.64 & 200k & $\sigma_8(\Omega_m/0.3)^{0.44}
= 0.866^{+0.085}_{-0.068}$ \\
\citet{schrabback07} & HST/ACS & 0.4 & 100k &
$\sigma_8=0.52^{+0.11}_{-0.15}$(stat)$\pm0.07$(sys) @ $\Omega_m=0.3^{\rm f}$ \\
\citet{fu08} & CFHT/MegaCam & 57 & 1.7M & $\sigma_8(\Omega_m/0.3)^{0.64} =
0.70\pm0.04$ \\
\citet{schrabback10} & HST/ACS & 1.64 & 195k &
$\sigma_8(\Omega_m/0.3)^{0.51}=0.75\pm0.08$ \\
\citet{huff11b} & SDSS & 168 & 1.3M & $\sigma_8=0.636^{+0.109}_{-0.154}$ @
$\Omega_m=0.265^{\rm h}$ \\
\citet{lin11} & SDSS & 275 & 4.5M & $\sigma_8(\Omega_m/0.3)^{0.7} =
0.64^{+0.08\rm h}_{-0.12}$ \\
\citet{jee12} & Mayall+CTIO/Mosaic & 20 & 1M & $\sigma_8=0.833\pm0.034^{\rm i}$ \\
\citet{kilbinger12} & CFHT/MegaCam & 154 & 4.2M & $\sigma_8(\Omega_m/0.27)^{0.6} = 0.79\pm0.03$ \\
\hline\hline
\multicolumn5l{$^{\rm a}$Consistent with $\Omega_m=0.3$ ($\Lambda$ or open),
cluster normalized; $\Omega_m=1,\sigma_8=1$ excluded.}\\
\multicolumn5l{$^{\rm b}$Consistent with $\Lambda$CDM or OCDM, but not 
{\it COBE} normalized $\Omega_m=1$.}\\
\multicolumn5l{$^{\rm c}$Reanalysis by \citet{vanwaerbeke02} gives
$\sigma_8=0.98\pm0.06$ ($\Omega_m=0.3$, $\Gamma=0.2$, 68\%CL).}\\
\multicolumn5l{$^{\rm d}$Reanalysis by \citet{heymans04} to correct for
intrinsic alignments gives $\sigma_8(\Omega_m/0.3)^{0.6} = 0.67\pm 0.10$.}\\
\multicolumn5l{$^{\rm e}$
%\citet{brown05} 
Brown et al.\ (2005)
used a subset of this data to show
that the matter power spectrum increased with time.}\\
\multicolumn5l{$^{\rm f}$In the Chandra Deep-Field South; the authors warn that
this field was selected to be empty, hence $\sigma_8$ may be biased low.}\\
\multicolumn5l{$^{\rm g}$A combination of 4 previously published surveys.}\\
\multicolumn5l{$^{\rm h}$Both based on the same raw SDSS data, but with
analyses and reduction pipelines by 2 different groups.}\\
\multicolumn5l{$^{\rm i}$Other parameters fixed to WMAP 7-year values.}
\end{tabular}
}
\end{table*}

\subsubsection{Galaxy-galaxy lensing as a cosmological probe}

Like cosmic shear, galaxy-galaxy lensing is an old idea. The earliest
astrophysically interesting upper limit was that of \citet{tyson84},
who used the images of 200,000 galaxies measured by the now-obsolete
method of digitizing photographic plates to exclude extended isothermal
halos with $v_c>200\kms$ around an apparent magnitude-limited sample
of galaxies. Galaxy-galaxy lensing was observed at $\sim 4\sigma$ by
\citet{brainerd96}, 
the first clear detection of cosmological weak lensing.
Their analysis used a total of 3202 lens-source
pairs in a field of area 0.025 deg$^2$. Several other detections
followed this in deep surveys with limited sky coverage \citep{hudson98,
smith00w, hoekstra03cnoc}. However the full scientific exploitation of
the galaxy-galaxy lensing signal --- in contrast to cosmic shear --- favors 
wide-shallow surveys over deep-narrow surveys, since the S/N in the 
shape-noise limited regime scales as only $\bar n_{\rm source}^{1/2}$ 
rather than
$\bar n_{\rm source}$. Therefore, in the decade of the 2000s the leading
galaxy-galaxy lensing surveys became the 92 deg$^2$ Red-Sequence Cluster
Survey (RCS; \citealt{hoekstra04rcs, hoekstra05rcs,kleinheinrich06})
and eventually
the $10^4$ deg$^2$ SDSS (references below).  The availability of
spectroscopic redshifts in the latter allowed the signal from low-redshift
galaxies to be stacked in physical rather than angular coordinates, enabling
the detection of features as a function of transverse separation. 
The spectroscopic survey also provided detailed environmental
information, measures of star-formation history, and full 3-dimensional
clustering data (e.g., correlation lengths and redshift-space distortions)
for the lens galaxies.

The SDSS remains the premier galaxy-galaxy lensing survey today, 
for both galaxy evolution and cosmology applications, and it likely 
will remain so until DES and HSC results become available. The
SDSS Early Data Release, comprising only a few percent of the overall
survey, already detected the galaxy-galaxy lensing signal with high significance
\citep{fischer00,mckay01}. Some of the major results of cosmological importance
from the SDSS galaxy-galaxy lensing program have been:
\begin{itemize}
\item
The galaxy bias can be constrained directly by going to the very largest
scales and measuring galaxy-galaxy lensing in the 2-halo regime. By
dividing the galaxy clustering signal by the galaxy-mass correlation
function, \citet{sheldon04} found for $L_*$ galaxies a bias of $b =
(1.3\pm0.2)(\Omega_m/0.27)r$, where $r$ is the stochasticity (presumably $\sim
1$ at the largest scales), with no evidence of scale dependence.\footnote{Since
lensing measures $\delta\rho$ rather than $\delta\rho/\rho$, there is a factor
of $\Omega_m$ in this measurement.}$^,$\footnote{A re-analysis with the final
SDSS imaging data set and improved treatment of the stochasticity is underway.}
\item
The measurement of halo masses --- or more accurately, HOD parameters 
(see \S\ref{sec:cmb_lss}) ---
with galaxy-galaxy lensing also enables one to predict the galaxy bias,
by using the bias-mass relation $b(M)$. One can in principle use this
to constrain cosmological parameters, since the clustering of the lens
galaxies can be measured and hence one can obtain $\sigma_{8,\rm gal}\equiv
b\sigma_8$. The results of this analysis on 300,000 lens galaxies at $z\sim
0.1$ were presented by \citet{seljak05}. The direct constraints on $\sigma_8$
itself (with other parameters fixed) were uninteresting because 
the inferred bias at fixed halo mass is a decreasing
function of $\sigma_8$, and the observable product $\sigma_8b$ is almost
independent of $\sigma_8$. However, cosmological parameters that change
the shape of the power spectrum can be constrained quite well --- e.g., a
decrease in small-scale power can make halos rarer and hence
decrease $b$ without a compensating change in $\sigma_8$. This breaks
degeneracies internal to the CMB alone.
Combining with first-year \wmap\ data, \citet{seljak05} found that for the
case of three degenerate 
neutrinos one must have $\sum m_\nu < 0.54$ eV (95\%CL).
\item
The halo mass-concentration relation $c(M)$ 
(e.g., \citealt{bullock01}) is not in and of itself especially
useful as a dark energy probe; it depends somewhat on $\Omega_m$, but
also on baryonic physics. Nevertheless, testing it is important for
any cosmological application of the 1-halo regime, including cosmic shear
\citep{king11}, and galaxy-galaxy lensing is well suited to measuring it at
a range of halo masses.  (For clusters other techniques are available, 
such as strong lensing or X-ray measurements.) 
\citet{mandelbaum08} measured this
relation across the $10^{12}-10^{15}M_\odot$ range, finding $c_{200b}(M)
= (4.6\pm0.7)M_{14}^{-0.13\pm0.07}$, where $M$ is the halo mass in units
of $10^{14}h^{-1}M_\odot$. The normalization is 
about 2$\sigma$ below the theoretical predictions, 
but the discrepancy may well be a statistical accident, particularly given
that other methods have led to larger concentrations.
\item
\citet{reyes10} tested GR by comparing the galaxy-mass
correlation function, measured via weak lensing, to the galaxy-velocity
correlation function, measured via redshift-space distortions. The SDSS
luminous red galaxy sample was chosen due to its large volume. This measurement
requires an overlapping spectroscopic and WL survey. They find that
\begin{equation}
E_G = \frac{\Upsilon_{gm}(R)}{\beta\Upsilon_{gg}(R)}=0.39\pm 0.06,
\end{equation}
where $\Upsilon$ is a filtered correlation function 
(averaged over scales $R=10-50\hmpc$) and $\beta$ is the
redshift-space distortion parameter of equation~(\ref{eqn:pkmu}). 
The combination $E_G$
is equal to $\Omega_m/f(z)$ at the
redshift of the lenses, for which GR predicts $\Omega_m^{0.45}(z=0.32) =
0.408\pm 0.029$. This measurement establishes that the peculiar velocities
of galaxies are, to $\sim 15$\%\ precision, in agreement with expectations
based on the potential structure traced by lensing.
\end{itemize}
All of these measurements will become possible with much smaller error
bars once the Stage III WL experiments are operational. We look forward in
particular to much smaller error bars on $b/r$ and $E_G$ derived from the
largest scales, as well as improvements on $c(M)$.

\subsubsection{Lensing outside the optical bands}

All wavelengths of light are gravitationally lensed. The optical\footnote{By
``optical,'' we mean to include near-infrared wavelengths $\lambda>0.7$
$\mu$m at which stars are still the dominant source of luminosity, and
which are observed through traditional optical telescopes and with detector
technology based on the creation of electron-hole pairs in semiconductors.}
is not special in this regard --- rather, the emphasis on optical wavelengths
has been technological, as this is the cheapest band in which to observe and
resolve large numbers of galaxies at cosmological distances and obtain some
redshift information. However, advances in technology in other wavebands
have resulted in weak lensing being detected at several other wavelengths:
\begin{itemize}
\item In the radio, kilometer-scale interferometers are required to
resolve extragalactic sources, and at the present time one cannot obtain
a radio photo-$z$ because of the featureless synchrotron spectra. However,
\citet{chang04} detected cosmic shear of extended radio sources using the
Very Large Array FIRST survey.
\item Lensing of the CMB has been of interest for some time as it provides
the most distant possible source screen. The first search was carried out in
cross-correlation by \citet{hirata04a} using luminous red galaxies in SDSS as
the lenses and WMAP temperature anisotropies as the sources. The signal was
detected three years later with combinations of SDSS and NVSS data, and two
additional years of WMAP data \citep{smith07,hirata08}. Recently, the Atacama
Cosmology Telescope (ACT) 
carried out a cosmic shear autocorrelation analysis using
the CMB as a source and detected the signal at $4\sigma$ \citep{Das11}. While
apparently weak, this measurement shows that $\Omega_\Lambda>0$ using
CMB data alone, without assuming a flat universe \citep{blakes11}. The ACT
and South Pole Telescope (SPT) 
collaborations are next planning polarization surveys, which should
yield much higher $S/N$ detections of lensing and provide constraints on
the neutrino mass.
\end{itemize}

\subsection{Observational Considerations and Survey Design}
\label{sec:wl_obs}

\subsubsection{Statistical Errors}
\label{sec:wl_stat_errors}

The forecasting of statistical errors on the cosmological parameters is
much more involved for WL than for supernovae or BAO because of the complex
dependence of the observables on the underlying model. Nevertheless, some
intuition can be gained by making approximations to enable exact evaluation
of the integrals. Specifically, we assume (i) a single source redshift
$z_{\rm s}$; (ii) a power-law matter power spectrum,
\begin{equation}
P_\delta(k,z) = 4.2\times 10^{-4} \sigma_8^2 H_0^{-3} G^2(z) (k/H_0)^{-1.3}~,
\end{equation}
where the slope $k^{-1.3}$ is chosen to match that of the 
$\Lambda$CDM power spectrum at a scale of $\sim 10$ Mpc
and the normalization is chosen to give the correct
$\sigma_8$; (iii) evaluation of the normalization $(1+z)G(z)$ not at the
true lens redshift $z_{\rm l}$ (over which we integrate from 0 to 
$z_{\rm s}$) 
but at a ``typical'' lens redshift $z_{\rm s}/2$; and (iv) a flat
universe. Then equation~(\ref{eq:wl:pee}) gives
\begin{equation}
C_{EE}(l) = 1.1\times 10^{-3} \sigma_8^2 \left[ \left(1+\frac {z_{\rm
s}}2\right) G\left(\frac {z_{\rm s}}2\right) \right]^2 \Omega_m^2
[H_0D_C(z_{\rm s})]^{2.3} l^{-1.3}.
\label{eq:wl:ce1}
\end{equation}
The variance per logarithmic range in $l$ is
\begin{equation}
\Delta^2(l) \equiv \frac{l^2}{2\pi} C_{EE}(l) = 1.8\times 10^{-4} \sigma_8^2
\left[ \left(1+\frac {z_{\rm s}}2\right) G\left(\frac {z_{\rm s}}2\right)
\right]^2 \Omega_m^2 [H_0D_C(z_{\rm s})]^{2.3} l^{0.7};
\label{eq:wl:ce2}
\end{equation}
this is a measure of the shear variance at a particular angular scale
$\theta\sim l^{-1}$.
Recall that $(1+z)G(z)$ varies from $\approx 0.75$ at $z=0$ to one
at high redshift (see Fig.~\ref{fig:gz}).
Since $H_0$ enters only in the combination $H_0 D_C(z)$, and
$D_C(z) \propto H_0^{-1}$, we see again that the WL signal
depend on relative rather than absolute distances.

In practice, equation~(\ref{eq:wl:ce2}) 
is only a rough guide because of deviations
of $P_\delta(k)$ from a power law and the nonlinear enhancement of the
matter power spectrum on small scales. Nevertheless, we can see several
important features:
\newcounter{features}
\begin{list}{\arabic{features}. }{\usecounter{features}}
\item The typical shear, given by $\sqrt{\Delta^2(l)}$, 
is of order 1\% at cosmological
distances ($z_{\rm s}\sim 1$) and degree scales ($l\sim 100$).
The shear fluctuations are larger at smaller scales.
\item The shear power spectrum scales as $\propto\sigma_8^2$. Assuming
a known background cosmology and source redshift, a measurement of the
power spectrum to $X$\%\ determines $\sigma_8$ to an uncertainty of
$\frac12X$\%.  In the nonlinear regime the dependence of
the shear power spectrum is closer to $\sigma_8^3$, so in practice
the constraint on
$\sigma_8$ is better than equation~(\ref{eq:wl:ce2}) would suggest.
\item Alternatively, if one assumes perfect knowledge of the growth
of structure (hence $\sigma_8$, $\Omega_m$, and $G$), 
then the distance $D_C(z_{\rm s})$ to the sources can be
determined to an uncertainty of $\frac1{2.3}X$\%. Lensing thus acts as a
standard ``ruler.''
\item Measuring the shear power spectrum as a function of source redshift
$z_{\rm s}$ allows one to measure some combination of the growth function and
the distance as functions of redshift. However, one does not measure both
separately. In order to simultaneously constrain the functional forms
$G(z)$ and $D_C(z)$, lensing must be combined with another cosmological probe.
\item Systematic errors in any of the terms in equation~(\ref{eq:wl:ce2}) will
bias the cosmology results. In particular, a 1\% change in $z_{\rm s}$,
e.g. 1.00$\,\rightarrow 1.01$, changes the power spectrum by 2\%. 
(This is the result of a full calculation, not evident by simple
inspection of the equation.) Therefore,
careful estimation of the source redshift distribution is required for a
WL survey --- a challenge when relying on photometric redshifts 
for the vast majority of sources.
\end{list}

The statistical uncertainty on the shear power spectrum is determined by two
factors: sampling variance at low $l$ and shape noise at high $l$. Sampling
variance uncertainty is associated with the fact that there are only a
finite number $N$ of Fourier modes in the survey area, and consequently the
fractional uncertainty in the power can be no smaller than $\sqrt{2/N}$
(where the 2 arises because power is the variance of $\gamma_{\bf l}$,
not the rms amplitude).
If we measure the power spectrum in a bin of
width $\Delta l$, then the number of modes is $N = 2l\Delta l f_{\rm sky}$,
where $f_{\rm sky}$ is the fraction of the sky observed. This corresponds
to a sampling variance uncertainty
\begin{equation}
\frac{\sigma[C_{EE}(l)]}{C_{EE}(l)} = \frac1{\sqrt{l\Delta l \,f_{\rm sky}}}~,
\quad \hbox{sampling variance only}.
\end{equation}
If we measure modes up to some $l_{\rm max}$, there are $l_{\rm max}^2f_{\rm
sky}$ modes, and the sampling variance uncertainty in the normalization of
the power spectrum is 
$\sqrt{2}f^{-1/2}_{\rm sky} l^{-1}_{\rm max}$.
%$\sqrt{2/f_{\rm sky}}\,/l_{\rm max}$.

At high $l$, the errors on the WL power spectrum become dominated not by the
number of modes available but by how well each mode can be measured with a
finite number of galaxies. Individual galaxies are not round, and so a shear
estimator applied to a galaxy has an intrinsic scatter $\sigma_\gamma\sim
0.2$ rms in each component of shear ($\gamma_+$ or $\gamma_\times$), 
for typical galaxy populations with rms ellipticity $e_{\rm rms}\sim 0.4$
per component.  This
phenomenon is known as {\em shape noise}. Since it is uncorrelated between
distinct galaxies (at least as a first approximation), shape noise produces
a white noise ($l$-independent) power spectrum\footnote{We give the $E$-mode
noise here. There is an equal amount of shape noise power in the $B$-mode,
but the lensing $B$-mode is used only as a systematics test because it contains
no cosmological signal to first order.},
\begin{equation}
C_{EE}^{\rm shape}(l) = \frac{\sigma_\gamma^2}{\bar n_{\rm eff}},
\label{eq:wl-neff1}
\end{equation}
where $\bar n_{\rm eff}$ is the effective number of galaxies per steradian
(this is the true number of galaxies with a penalty applied for objects
where the observational measurement error on the shear becomes comparable
to $\sigma_\gamma$; see below). Since the cosmic shear $C_{EE}(l)$ is
decreasing with $l$, there is a transition scale $l_{\rm tr}$ where the shape
noise becomes comparable to the lensing signal. 
Using equation~(\ref{eq:wl:ce2}),
we estimate
\begin{equation}
l_{\rm tr} = 1300 \left( \frac{\sigma_8}{0.8} \right)^{1.54} \left[
\left(1+\frac {z_{\rm s}}2\right) G\left(\frac {z_{\rm s}}2\right)
\right]^{1.54} \left( \frac{\Omega_m}{0.3}\right)^{1.54} [H_0D_C(z_{\rm
s})]^{1.77} \left( \frac{\bar n_{\rm eff}}{20\,{\rm arcmin}^{-2}}
\right)^{0.77}.
\end{equation}
At angular scales smaller than $\theta\sim l_{\rm tr}^{-1}$, lensing cannot
detect (at S/N$\,>1$) a
typical fluctuation in the density field.\footnote{High-amplitude
features such as clusters may still be visible.} Statistical measurements are
still possible, however, and the power spectrum can be measured to an accuracy
of $\sqrt{2/N}\,C_{EE}^{\rm shape}(l)$ where $N$ is the number of modes. Thus,
in the shape-noise limited regime,
\begin{equation}
\label{eq:wl:staterr}
\frac{\sigma[C_{EE}(l)]}{C_{EE}(l)} = \frac1{\sqrt{l\Delta l \,f_{\rm
sky}}}\,\frac{C_{EE}^{\rm shape}(l)}{C_{EE}(l)}
= \frac1{\sqrt{l\Delta l \,f_{\rm sky}}}\,\left( \frac l{l_{\rm tr}}
\right)^{1.3}, \qquad l > l_{\rm tr}.
\end{equation}
One can see from this equation that the fractional uncertainty on
$C_{EE}(l)$ in bins of width $\Delta l/l\sim 1$ increases with $l$ for
$l>l_{\rm tr}$. Therefore we arrive at the important conclusion that the
power spectrum is best measured at the transition scale $l_{\rm tr}$:
on larger scales sampling variance degrades the measurement even though
individual structures are seen at high signal-to-noise ratio (SNR),
and on smaller scales shape
noise dominates. The aggregate uncertainty in the normalization of the
power spectrum is thus of order
\begin{equation}
\frac{\sigma({\rm normalization})}{\rm normalization} \sim \frac1{l_{\rm
tr}\sqrt{f_{\rm sky}}}.
\end{equation}
A full-sky experiment\footnote{In practice the Galactic Plane must be
avoided, so it is unlikely that optical astronomy would push beyond $f_{\rm
sky}\sim 0.7$ for any cosmological application.} reaching tens of galaxies
per arcmin$^2$ at redshifts of order unity would have $l_{\rm tr}\sim 1000$
and so could measure the normalization of the power spectrum to a statistical
precision of order 0.1\%. This would be an unprecedented measurement of
the strength of matter clustering.
However, as we will see below, there are substantial statistical
and systematic hurdles to such an experiment.

Finally, we consider galaxies measured at finite SNR. In the above
analysis, we assumed that each galaxy provided an estimate of the shear
with uncertainty $\sigma_\gamma$. At finite SNR there is also measurement
noise $\sigma_{\rm obs}$, so that each galaxy provides an estimate with error
$\sqrt{\sigma_\gamma^2+\sigma_{\rm obs}^2}$. Using inverse-variance weighting,
in the finite-SNR case the shape noise becomes equation~(\ref{eq:wl-neff1}),
with the effective source density 
\begin{equation}
\bar n_{\rm eff} = \frac1A \sum_{i=1}^{N_{\rm gal}}
\frac{\sigma_\gamma^2}{\sigma_\gamma^2+\sigma_{{\rm obs},i}^2} ~,
\label{eq:wl-neff2}
\end{equation}
where $A$ is the survey area and the sum is over the galaxies. This is always
less than $\bar n = N_{\rm gal}/A$. The effective source density $\bar
n_{\rm eff}$ is limited in part by the depth of the survey: $\sigma_{{\rm
obs},i}$ typically scales with integration time as $\propto t^{-1/2}$,
but once $\sigma_{{\rm obs},i}\ll \sigma_\gamma$ one no longer continues
to gain. How long does this take? In \S\ref{sss:sme}, we will show that
for nearly circular, Gaussian galaxies\footnote{For realistic non-Gaussian
profiles, the shape measurement error is usually worse by of order 20\%.}
\begin{equation}
\sigma_{\rm obs} = \frac1\nu \left( 1 + \frac{r_{\rm psf}^2}{r_{\rm gal}^2}
\right),
\end{equation}
where $r_{\rm psf}$ and $r_{\rm gal}$ are the half-light radii of the PSF
and the galaxy, respectively, and $\nu$ is the detection significance (in
$\sigma$s). Thus for galaxies with a similar size as the PSF, we expect
to reach $\sigma_{\rm obs}=0.1$ (measurement noise half of shape noise)
after integrating long enough to see the galaxy at 20$\sigma$.

In principle, the summation in equation~(\ref{eq:wl-neff2}) is over all objects
detected as extended sources, and any galaxy could be used if its detection
significance is high enough. In practice, this is dangerous: while one might
hope to obtain $\sigma_{\rm obs}=0.1$ on a galaxy with $r_{\rm gal}=0.5r_{\rm
psf}$ and a 50$\sigma$ detection, the ``ellipticity measurement'' on this
galaxy consists of measuring the small deviation of the image from the
PSF. Such a procedure tends to magnify systematic errors in the PSF model
and is usually unadvisable. Therefore, most WL surveys impose a cutoff on
$r_{\rm gal}/r_{\rm psf}$ or some similar property.

\subsubsection{The Galaxy Population for Optical Surveys}
\label{sec:wl_galpop}

The design of a WL survey must begin by considering the population
of galaxies.  We will focus here on the population in the 3-dimensional
space of redshift $z$, effective radius $r_{\rm eff}$, and apparent AB
magnitude in the $I$-band (a convenient choice for shape measurement with
red-sensitive CCDs from the ground).  The plots shown here are based on
the mock catalog of \cite{jouvel09}, which uses real galaxies from the
COSMOS survey but fills in missing information for individual galaxies
(e.g. redshifts or line fluxes) with photo-$z$s and models.

\begin{figure}
\centerline{
\includegraphics[angle=-90,width=4.5in]{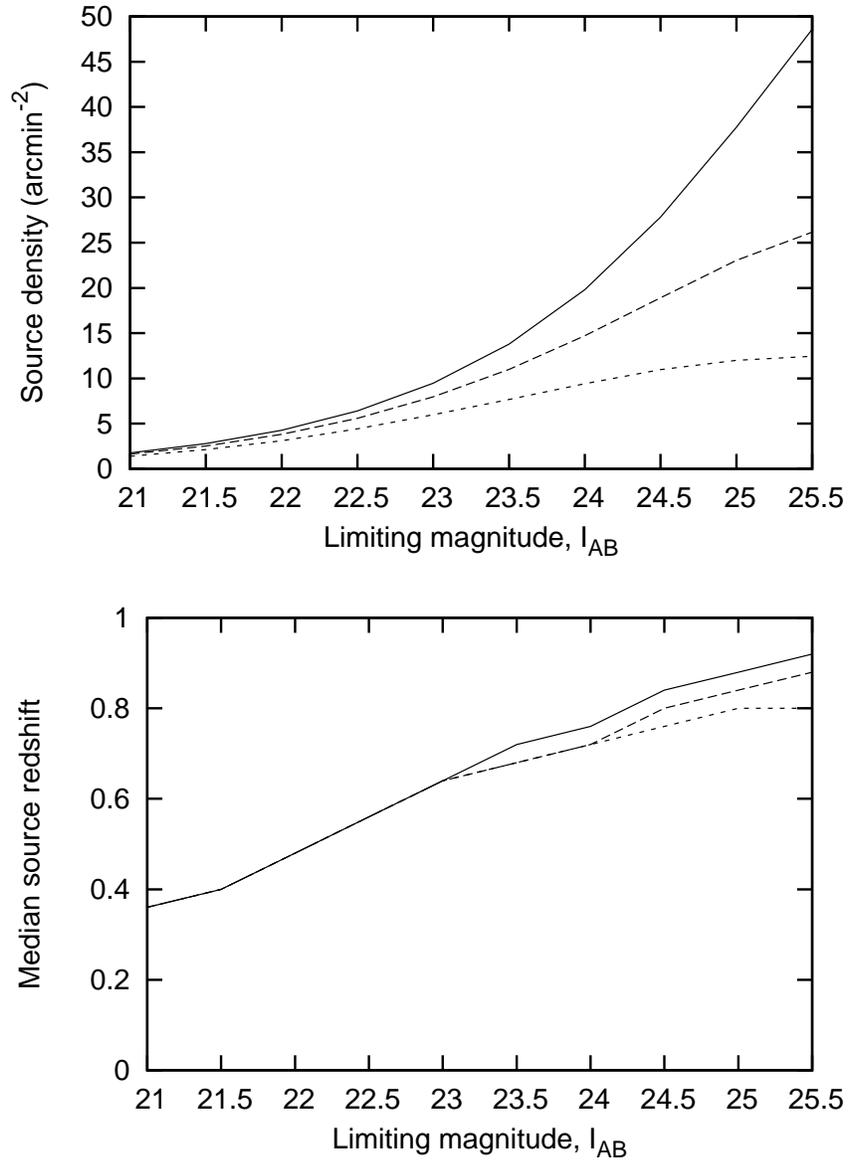}
}
\caption{\label{fig:wl_galdens}The mean surface density of galaxies (top panel)
and median redshift (bottom panel) as a function of limiting magnitude.
The three curves show different $r_{\rm eff}$ cuts: the top curve is a cut at
$0.15''$, which might be applied to a space-based survey; the middle curve
is a cut at $0.248''$, which would be an optimistic choice from the ground;
and the bottom curve is a cut at $0.35''$, a more conservative choice for
a ground-based survey with $\sim 0.7''$ seeing (FWHM). For galaxy-galaxy
lensing, one could make more aggressive cuts.}
\end{figure}

Figure~\ref{fig:wl_galdens} shows the mean surface density of galaxies and
the median source redshift as a function of limiting magnitude $I_{\rm
AB}$ for effective radius cuts of 0.15$''$, 0.248$''$, and
0.35$''$.  In general, one would like to use galaxies larger than the PSF
to avoid amplification of systematics when applying a PSF correction to
the shapes.  The ``effective radius'' (EE50, for 50\% encircled energy)
of a typical ground-based PSF
is $\sim 0.35''$ under good conditions, corresponding to a FWHM
of $\sim 0.7''$.  The 0.248$''$ cut is a factor of $\sqrt2$ smaller,
appropriate if one can make use of galaxies smaller than the PSF
%($\sigma_f^2=\frac12\sigma_G^2$)
or has sufficient \'etendue to do the
entire survey under the very best seeing conditions.  Measuring galaxies at
$r_{\rm eff}=0.15''$ is well beyond present ground-based cosmic shear survey
capabilities,
for both algorithmic and PSF-determination reasons, and will likely require
a space (or balloon) based platform.

\subsubsection{Photometric Redshifts and their Calibration}
\label{sec:wl_photoz}

Modern WL analyses all use photometric redshifts in some way.  They are
central to tomography and cosmography measurements, and they are also needed
in most schemes to remove the intrinsic alignment contamination.
In the case of GGL, photo-$z$s are used to select sources that are actually
behind the lens plane (sources in front of the lens are unlensed and dilute
the signal, whereas sources at the same redshift as the lens can contribute
intrinsic alignments).

One can characterize the photo-$z$ distribution using the joint probability
distribution for the photo-$z$ $z_p$ and the true redshift $z$ for some
sample of galaxies, $P(z_p,z)$.  In the case of lensing, we care about
the conditional probability distribution, $P(z|z_p)$.  This distribution
is sometimes characterized by its conditional bias and scatter,
\begin{equation}
\delta z(z_p) = z_p - \langle z\rangle|_{z_p}, {\rm ~~~} \sigma_z(z_p) =
\sqrt{\langle z^2\rangle|_{z_p} - \langle z \rangle|_{z_p}^2 } ~,
\end{equation}
but it is always non-Gaussian and in practice there are ``outliers'' or
``catastrophic failures'' with $|z-z_p|\sim{\cal O}(1)$.  The
conditional probability distribution is not symmetric: Bayes's theorem
tells us that
\begin{equation}
P(z|z_p) = \frac{P(z)}{P(z_p)} P(z_p|z),
\end{equation}
so a photo-$z$ that is is ``unbiased'' in the conventional sense of $\langle
z_p\rangle|_z = z$ may still have $\delta z(z_p)\neq 0$.
It is not required that photometric redshifts have $\delta z(z_p)=0$,
but one does need to know the value of $\delta z(z_p)$ to relate
observations to model parameters.  From the simplified example
discussed in \S\ref{sec:wl_stat_errors}, we can see that a systematic
error of $\sim 0.01$ in $\delta z(z_p)/z_p$ will lead to a
normalization error in the matter power spectrum of the order of 2\%.  Similarly, if 1\%
of galaxies in a source redshift bin $z_{\rm s}$ are actually outliers
with redshift $z \ll z_{\rm s}$, they will dilute the expected
lensing signal by 1\%, and the power spectrum by 2\%.

If the full distribution $P(z|z_p)$ is known, then the shear cross-power
spectra for any pair of redshift slices can be determined for a given
cosmological model.  However, the use of photo-$z$s to suppress intrinsic
alignments (\S\ref{sec:wl_ia}) does not work if the intrinsic alignments
of the outliers are significant, or even if the scatter is large enough
that galaxies can evolve significantly within a redshift bin, so there
is a strong motivation to reduce them to the minimum level possible.
Thus lensing programs must face two challenging problems: (i) obtaining a
low outlier rate, and (ii) determining $P(z|z_p)$ to sub-percent precision.

\begin{figure}[t]
\centerline{
\includegraphics[angle=-90,width=5in]{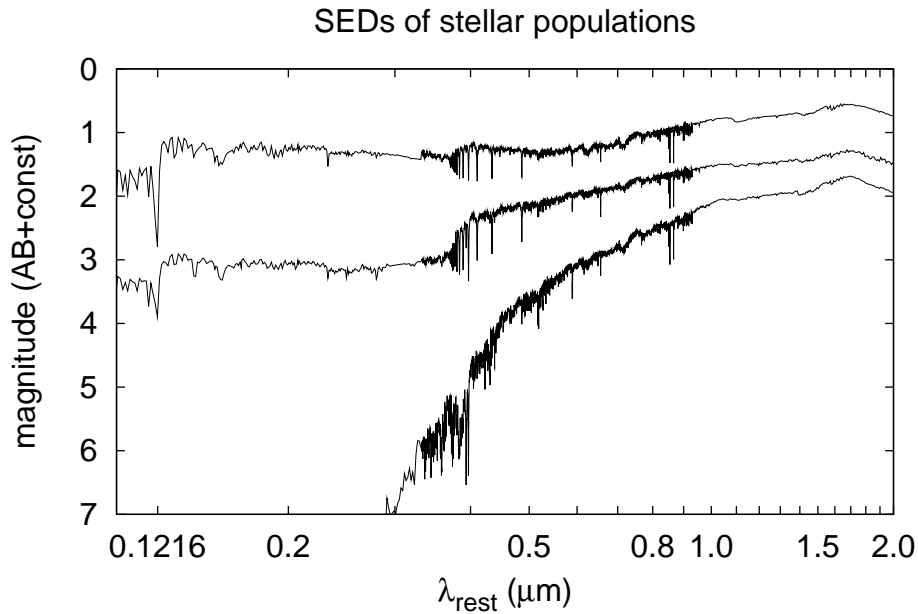}
}
\caption{\label{fig:photo-z}The SEDs of three stellar populations are shown:
a single burst at age 25 Myr (top); a continuous star-forming population of
6 Gyr age (middle);
and a single burst at 11 Gyr (bottom). All have solar metallicity. Blueward
of \lya\ 
they have been adjusted for an IGM transmission factor of 0.8 (appropriate
for $z=2.25$; see \citeauthor{mcdonald06} [\citeyear{mcdonald06}]),
but other corrections (dust, nebular emission) are not included. The models
are obtained from
\cite{bruzual03}. Note the break at $\sim 0.37-0.40\,\mu$m present in all
models, albeit with varying
shape, strength, and precise location.
}
\end{figure}

To understand how to reduce the outlier rate, we must investigate how
photo-$z$s work: they take several broad-band fluxes from a galaxy and try
to identify spectral features (see Fig.~\ref{fig:photo-z}).  
At low redshifts, the strongest feature in
the optical part of a galaxy spectrum is the break around 3800--4000\AA,
arising from metal line absorption in early-type galaxies and the Balmer
continuum (plus high-order lines) in late-type galaxies.
As the redshift of the galaxy increases, this feature moves to the red,
and above redshifts of $z\sim 1.3$ it is no longer useful for optical
photo-$z$s (depending on the SNR in $z$ and $y$ bands).  At $z\ge 2$, the
\lya\ break redshifts into the optical bands and can be used --
but it is possible to confuse it with the Balmer/4000\AA\ break.  This is
the principal example of a {\em photo-z degeneracy}.

The above discussion suggests that to reduce outliers across the whole range
of redshifts used for WL surveys ($z=0$ to $\sim 3$) one desires coverage
from blueward of the Balmer/4000\AA\ feature (i.e. a $u$-band) through the near-IR
($J+H$ bands), so that either the Balmer/4000\AA\ feature or \lya\ 
is robustly identifiable.  The optical bands can be easily observed from
the ground.  As one moves redward, however, the sky brightness as observed
from the ground increases rapidly, and obtaining the $J+H$ band photometry
matched to the depth of future surveys is only practical from space.

One is then left with the problem of measuring the photo-$z$ error
distribution.  The most direct and conceptually simplest 
way to do this is to collect
spectroscopic redshifts of a representative subsample of the sources used
for WL.  This is, however, very expensive in terms of telescope time: many
galaxies
have weak or absent emission lines (particularly if one restricts to the
optical range),
and so one searches for absorption features of faint ($i\sim 22-25$) galaxies.
Stage III/IV experiments may require ${\cal O}(10^5)$ redshifts to
calibrate photo-$z$s at the level of their statistical errors,
and we desire sub-percent failure
rates because the failures are likely concentrated at specific redshifts.
These failure rates are far below those that have actually been achieved by
spectroscopic surveys at the desired magnitudes.

An alternative idea \citep{newman08} is to use the 2-D angular 
cross-correlation of the photo-$z$ galaxies with a large area
spectroscopic redshift survey, which can target brighter galaxies and/or
the subset of faint galaxies that have strong emission lines.
For a bin of galaxies with photo-$z$ centered on $z_p$, the
amplitude of cross-correlation is proportional to
$b_{\rm spec}(z)b_{\rm phot}(z,z_p)P(z|z_p)$, where $b_{\rm spec}$
and $b_{\rm phot}$ are the clustering bias factors of the 
spectroscopic and photo-$z$ galaxies, respectively, at redshift $z$.
The auto-correlations of the spectroscopic and photo-$z$ samples
provide additional constraints on the bias factors, and one also
has the normalization condition $\int P(z|z_p)\,dz=1$ for 
each $z_p$ bin.  
The key uncertainty in this approach is constraining the full
redshift-dependent $b_{\rm phot}(z,z_p)$; if the bias varies with photo-$z$
error (e.g., because high-bias red galaxies and low-bias blue galaxies have
different photo-$z$ error distributions) then this dependence
must be modeled to extract $P(z|z_p)$ \citep{matthews10}.  
If one is using intermediate or small scale clustering, 
then one must also allow for scale-dependent bias
and for cross-correlation coefficients lower than unity between
different galaxy populations, which would lower the amplitude of
cross-correlations relative to auto-correlations.  
Finally, the approach requires a
spectroscopic sample that spans the full redshift range of the
photometric sample; a quasar redshift survey may provide sufficient
sampling density for probing the high-redshift tail of $P(z|z_p)$.
The cross-correlation technique has to date
not been used for WL surveys, but it has been used to measure other redshift
distributions --- see, e.g., the application to radio galaxies by \cite{ho08}.
Adding galaxy-galaxy lensing measurements to the galaxy clustering 
measurements may improve the robustness and accuracy of the
cross-correlation approach and allow some degree of ``self-calibration''
without relying on an external spectroscopic data set
\citep{zhang10}.

Overall, the problem of measuring $P(z|z_p)$ to the required accuracy
remains one of the greatest challenges for future WL projects.
Given the difficulty of assembling an ideal spectroscopic 
calibration sample, the treatment of photo-$z$ distributions
in Stage III and Stage IV WL analyses is likely to involve
some combination of direct calibration, cross-correlation
calibration, empirically motivated models of galaxy SEDs,
and marginalization over remaining uncertainties in parameterized
forms of $P(z|z_p)$.  Tomographic WL measurements themselves
have some power to constrain these distributions (at the cost
of some leverage on cosmological parameters), and weak lensing
by clusters that have well determined individual redshifts may
also be a valuable tool.

\subsubsection{Lensing in the Radio}
\label{sec:wl_radio}

An interesting alternative to shape measurement in the optical is to work in
the radio part of the spectrum, where late-type galaxies are observable via
their synchrotron emission.  In order to achieve the required resolution, one
needs to use a large interferometer: a fringe spacing of $1''$ is achievable
at 1 GHz with a baseline of 60 km.  One also needs a large collecting area
to obtain high-SNR images on a competitive number of galaxies; the SKA
could in principle measure billions of galaxies \citep{blake04}.  But let
us suppose such an interferometer were built.  What would it do for WL?
In principle, it could solve many problems at once:
\begin{itemize}
\item {\em Shape measurement}: An interferometer directly measures the
Fourier transform of the surface brightness of a galaxy, $\tilde I({\bf u})$,
thereby avoiding the difficulty of interpolating PSF properties from stars.
On long baselines, the ${\bf u}$-plane is usually sparsely sampled,
i.e., not all values of ${\bf u}$ are observed; the Fourier mode ${\bf
u}={\bf L}_\perp/\lambda$, where ${\bf L}_\perp$ is the interferometer
baseline projected into the plane of the sky.  At a given wavelength,
as the Earth rotates, each baseline thus traces out an ellipse in the
${\bf u}$-plane.  However, if one combines a finite range of $\lambda$
and many baselines, one could fill in the ${\bf u}$-plane, and
model-fitting shape measurement techniques can work even with significant
coverage gaps. Model-fitting methods were used in the analysis of the FIRST
survey at $\lambda =20$ cm \citep{chang04}, which resulted in a $3\sigma$
detection of cosmic shear.
\item {\em Redshifts}: Late-type galaxies contain atomic gas, and thus
radiate in the H{\sc~i} 21 cm line.  For nearby galaxies ($z<0.1$) this line
has a long history of being used as a redshift indicator.  A Stage IV radio
interferometer survey could collect hundreds of millions of spectroscopic
redshifts in this line out to $z\sim 1-2$, thereby obviating the need to
rely on photo-$z$s and calibrate photo-$z$ error distributions.
Conversely, it is not clear that one could make use of the many radio
galaxies {\it not} detected in H{\sc~i}, as even photometric redshifts
for large radio galaxy samples are difficult to obtain at high completeness.
\item {\em Shape noise reduction}: The radio part of the spectrum offers
interesting opportunities to reduce shape noise.  For example, if one
spatially resolved the H{\sc~i} disk of a galaxy, one could produce a
velocity map.  A perfect inclined disk has the long axis aligned with
the velocity gradient, and if sheared this alignment is destroyed, so a
measurement of the velocity gradient provides independent information on the
intrinsic shape of the galaxy \citep{morales06}.  Another idea is to use the
polarization of the synchrotron emission, which tends to be perpendicular
to the galactic disk and hence is an indicator of the position angle of the
intrinsic minor axis \citep{brown09}.  While promising, these ideas are new,
and their practical application to a WL survey may have to await at least
a partial SKA.
\end{itemize}

\subsubsection{Lensing of the CMB}
\label{sec:wl_cmblensing}

It is also possible to do lensing analyses on the CMB.  Here there are several
advantages: the source redshift is known exactly from cosmological parameters,
$z_{\rm src}=1100$; theory predicts exactly the statistical distribution of
hot and cold spots on the CMB, so there is no intrinsic alignment effect;
and the PSF (or ``beam'' shape) of microwave experiments tends to be far
more stable than in the optical.  The CMB is a diffuse field rather than
a collection of objects (galaxies), so reconstructing the shear requires
a different mathematical formalism than for galaxy lensing.  The basis for
this formalism is two-fold:
\begin{itemize}
\item
In the presence of lensing by a Fourier mode of the potential $\tilde\psi({\bf
l})$, the CMB anisotropy field is no longer statistically isotropic: different
temperature Fourier modes become correlated, $\langle\tilde T^\ast({\bf
l}_1)\tilde T({\bf l}_2)\rangle\propto\tilde\psi({\bf l}_2-{\bf l}_1)$.
These products of temperature modes can be used as estimators of the lensing
potential \citep{hu01}.
\item
A more dramatic effect occurs for CMB polarization.  The unlensed CMB
polarization is pure $E$-mode,\footnote{Primordial gravitational waves
can generate a $B$-mode on large scales, but such gravitational waves
are adiabatically damped on angular scales below a degree.  Thus the $\sim
10'$-scale $B$-mode should be dominated by lensing.} i.e., the polarization in
each Fourier mode is parallel to the wavevector rather than at a $45^\circ$
angle.  Lensing shear changes the direction of the wavevector ${\bf l}$
but not the polarization, so it can generate $B$-mode shear.
\end{itemize}
%The effective noise in the reconstructed shear field, $(\sigma_\gamma^2/\bar
%n)_{\rm eff}$, is shown in Fig.~??? for different model experiments.
%This noise is scale-dependent because
%it is much harder to reconstruct features in the lensing
%field on scales smaller than the CMB acoustic scale.

Until recently, because of SNR issues, lensing of the CMB had
been detected only in cross-correlation with foreground galaxies
\citep{smith07,hirata08,bleem12}. The advent of the arcminute-scale CMB experiments
ACT and SPT (primarily motivated by cluster cosmology using the SZ effect) has
enabled robust detections of the power spectrum of the CMB lensing field
\citep{Das11,vanengelen12}.

Because CMB lensing only provides a single source slice, it is unlikely to
ever replace galaxy lensing.  However, in combination with galaxy lensing,
it can provide the most distant source slice for tomography \citep{hu02b}
and cosmography \citep{acquaviva08}.

\subsection{Measuring Shears}
\label{sec:wl_shapes}

So far we have treated shear measurement as a black box: it takes in an
image of the galaxy and some knowledge of the instrument, and it returns 
$\hat\gamma_{+,\times}$, an unbiased estimator for the
true shear $\gamma$ with some uncertainty per component $\sigma_\gamma$.
This black box is very complicated on the inside, as one needs an accurate and
robust {\em shape measurement algorithm}, and even providing the necessary
inputs to such an algorithm, particularly an accurate determination 
of the PSF, has proven to be difficult.  
After a brief overview of these algorithms, we describe 
the idealized problem of measuring shear from an ensemble of galaxy images,
then turn to a more detailed discussion of the challenges that arise in 
practice.

There are two general strategies for shape measurement methods in common use
today. One class of methods is to measure moments of galaxies (in real or
Fourier space), and relate, e.g., the mean quadrupole moment of galaxies to the
shear.  These methods
started with {\em ad hoc} ``PSF correction'' prescriptions, but they
have recently
evolved toward methods that attempt to statistically close the hierarchy of
moments of galaxies and PSFs in a model-independent way. The other class
of methods is based on forward modeling: one adopts a model for a galaxy
(e.g., an elliptical Sersi\'c profile, or a linear combination of basis
images), simulates the observational procedure, and minimizes $\chi^2$. Both
approaches have their advantages and disadvantages. 
Much of the early WL work used
moments-based methods, but for years a generally applicable PSF correction
scheme seemed out of reach. Some of the more recent incarnations of the
Fourier domain moments-based methods work for arbitrary distributions of
galaxy and PSF profiles; however these are less mature in their practical
implementation, and they impose stringent requirements on input data quality
(e.g., sampling). The forward modeling methods can handle a much wider range
of observational defects (e.g., under some circumstances one may even be
able to measure a galaxy containing missing pixels), but they depend on a model
for the galaxy being observed; one must carefully assess the impact of an
insufficiently general model. Both strategies require exquisite knowledge
of the PSF.

Currently there are many algorithms in use in each category. The
prototype moments-based method was that of Kaiser, Squires, and Broadhurst
(KSB; \citealt{KSB95}; improved by \citealt{Luppino97, Hoekstra98}). Many
improvements of these methods have been made --- e.g., in computing better
conversion factors from shear to quadrupole moments\footnote{\cite{massey07}
\S3.1.1 give an excellent technical review of the methods derived
from KSB.}
\citep{Semboloni2006}.  Elliptical-weighted moments and the concept of
shear-covariance were introduced by \cite{bernstein02} and have been used
extensively in SDSS \citep{hirata03}. Further progress was made by moving to
moments in Fourier space, where the PSF ``correction'' becomes trivial (one
divides by the Fourier transform of the PSF, at least in the regions where it
is nonzero). This has culminated in the development of a shape measurement
method that is exact in the high-SNR limit \citep{bernstein10}. We discuss
this method and its development in \S\ref{sec:wl_sma}. An early example
of the model-fitting approach was {\sc im2shape} \citep{bridle02}. More
recently, Bayesian model fits have been introduced that are stable at lower
SNR \citep{miller07,kitching08}; these are currently being applied to the
CFHTLS. The ``shapelet'' basis \citep{refregier03I,refregier03II},
derived from
energy eigenstates of a 2D quantum harmonic oscillator,
is useful in both types of
methods.  The coefficients in a shapelet decomposition {\em are} moments, but
one may also fit a model galaxy parameterized by its shapelet coefficients.

The various shape measurement algorithms have been tested and compared
in blind simulations, such as the Shear Testing Program (STEP1/STEP2;
\citealt{heymans06, massey07}), GREAT08 \citep{GREAT08}, 
and GREAT10 \citep{GREAT10}.  In most of these
cases, the objective is to minimize both the shear calibration error $m$
(i.e. the error in the response to a given input shear) and the spurious
shear $c$ (i.e., the shear measured by the algorithm on an unlensed sample of
galaxies). The STEP2 simulations used typical ground-based PSFs and complex
galaxy morphologies and found that many of the measurement methods had shear
calibration errors $|m|$ of one-to-several percent, and spurious shear $|c|$
ranging from several $\times 10^{-4}$ to several $\times 10^{-3}$. This
level of performance should thus be considered typical of the more mature,
heavily used shear measurement algorithms, although recent methods have
done better. On the other hand, the algorithmic errors are only a portion
of the error budget in a WL experiment --- most importantly, the early
simulation tests did {\em not} require participants to recover the spatial
variability of the PSF. Such a test is currently ongoing as part of the
GREAT10 challenge \citep{GREAT10}.
Early results from GREAT10 are now available, but their
significance is still being digested.

In the remaining portions of this section we will discuss the mathematical
problem of shape measurement (\S\ref{sss:wl-sma-id}) and the basis for
{\em some} of the commonly used methods (\S\ref{sec:wl_sma}) and their
statistical errors (\S\ref{sss:sme}). We cannot of course do justice to
every method that has been suggested or used. We have chosen to highlight
the recent progress in Fourier-space methods, since in principle they provide
an exact solution in the limit of high SNR and are thus ripe for further
development and utilization \citep{bernstein10}.  There are some biases that
can result even for perfect shape measurement (or galaxies measured with
a $\delta$-function PSF), including the noise-related biases and selection
biases, which are probably present at some level for all known algorithms;
these are discussed in \S\ref{sss:nrsb}. Finally \S\ref{sec:wl_psf} describes
the determination of the PSF, which is taken as an input for any shape
measurement algorithm.

\subsubsection{The Idealized Problem}
\label{sss:wl-sma-id}

The idealized shape measurement problem is as follows: we have a galaxy in
the source plane whose surface brightness is $f_0({\bf x})$, where ${\bf x}$
is a 2-dimensional vector in the plane of the sky.  It is first sheared,
i.e., the galaxy in the image plane is $f({\bf x}) = f_0({\bf Sx})$, where
${\bf S}$ is the shearing matrix,
\begin{equation}
{\bf S} = \left( \begin{array}{cc} 1 - \gamma_+ & -\gamma_\times \\
-\gamma_\times & 1+\gamma_+ \end{array} \right).
\end{equation}
(We assume $|\gamma|\ll 1$ here and work to linear order in $\gamma$ for
simplicity, although higher-order corrections will be important for Stage
IV surveys.)
We do not observe the actual image on the sky, however --- we observe
it through an instrument with PSF\footnote{Here we use the term ``PSF''
to include not just the image of a point source produced by the telescope
optics but also pointing jitter and detector effects.  For example,
if the detector
has square pixels, the PSF is that delivered by the telescope convolved
with a square top-hat function.} $G({\bf x})$.  The resulting image is
\begin{equation}
I({\bf x}) = [f\star G]({\bf x}) = \int_{{\mathbb R}^2} f({\bf x}')
G({\bf x}-{\bf x}') d^2{\bf x}' = \int_{{\mathbb R}^2} f_0({\bf Sx}')
G({\bf x}-{\bf x}') d^2{\bf x}'.
\label{eq:wl:ireal}
\end{equation}
This equation may also be written in Fourier space: if we define
\begin{equation}
\tilde I({\bf u}) = \int_{{\mathbb R}^2} I({\bf x}) e^{-2\pi i{\bf u}\cdot{\bf
x}} d^2{\bf x}
\;\;\;\;\leftrightarrow\;\;\;\;
I({\bf x}) = \int_{{\mathbb R}^2} \tilde I({\bf x}) e^{2\pi i{\bf u}\cdot{\bf
x}} d^2{\bf u},
\end{equation}
then equation~(\ref{eq:wl:ireal}) simplifies to
\begin{equation}
\tilde I({\bf u}) = \tilde G({\bf u}) \tilde f_0({\bf S}^{-1}{\bf u}).
\label{eq:wl:ifourier}
\end{equation}
In practice, the image $I$ is only obtained at discrete values of ${\bf x}$,
i.e., at the pixel centers spaced by separation $\Delta$.  If the image is
oversampled, i.e., if the Fourier transform\footnote{It is important to recall
that the definition of oversampling required for equation~(\ref{eq:wl:sinc})
operates in Fourier space.  
The commonly used condition for oversampling that the FWHM should exceed
2 pixels is a good rule of thumb for smooth profiles such a Gaussian, but
it is not appropriate for general PSFs.}
of the
PSF is zero (or negligible) at wavenumbers above some $|{\bf u}|_{\rm max}$
with $|{\bf u}|_{\rm max}<1/(2\Delta)$, then it can be sinc-interpolated
to recover the full continuous function,
\begin{equation}
I({\bf x}) = \sum_{n_1n_2} I(n_1\Delta,n_2\Delta) \,{\rm
sinc} \,\frac{\pi(x_1-n_1\Delta)}{\Delta}\, {\rm sinc}\,
\frac{\pi(x_2-n_2\Delta)}{\Delta}.
\label{eq:wl:sinc}
\end{equation}
The pixelization thus represents no special difficulty, except that the sinc
function has noncompact support and must be smoothly truncated.  A second
implication of oversampling is that integrals of the form $\int P({\bf x})
I_1({\bf x}) I_2({\bf x})\,d^2{\bf x}$, where $P$ is a polynomial in the
coordinates and $I_1$ and $I_2$ are oversampled functions, can be replaced
without error by (infinite) sums over pixels: $\int \rightarrow \Delta^2\sum$.
Again, in practice such sums must be truncated.

We will also define a critical wavenumber $u_{\rm crit}$, which is the
smallest wave number for which there is a Fourier mode with $G({\bf u})=0$
with $|{\bf u}|=u_{\rm crit}$.  Then we have $G({\bf u})\neq 0$ for any
$|{\bf u}|<u_{\rm crit}$.  This critical wavenumber determines the region
within the Fourier plane over which deconvolution is possible, and over
which measurement of $\tilde f({\bf u})$ is possible.

A shape measurement algorithm is a functional $\hat\gamma_i[I;G]$,
$i\in\{+,\times\}$, that returns a shear estimate.  When averaged over a
population of galaxies with the same shear, such an algorithm will yield
an expectation value
\begin{equation}
\langle \hat\gamma_a \rangle = c_a + (\delta_{ab} + m_{ab}) \gamma_b +
{\cal O}(\gamma^2).
\label{eq:wl:bias}
\end{equation}
Here $c_a$ is called the {\em additive shear error} and $m_{ab}$ is the
{\em multiplicative shear error} or shear calibration error.  An ideal
algorithm will have $c_a=m_{ab}=0$.

Many WL surveys take multiple exposures of each field; if they are
oversampled, one may use equation~(\ref{eq:wl:sinc}) to reconstruct a continuous
function $I({\bf x})$ for each exposure.  If the PSFs in each exposure
differ (which they usually do), then to construct a stacked image, one
can either apply a convolution kernel to each input image to make the
PSFs the same or do a noise-weighted least squares fit to each Fourier
mode $\tilde f({\bf u})$.  If the individual exposures are undersampled
(as is likely for space-based data) and appropriately dithered, methods
are available in both Fourier space 
\citep{lauer99} and real space \citep{Fruchter2011,
Rowe2011} to reconstruct a fully-sampled and hence continuous image
$I({\bf x})$.\footnote{Much of the {\slshape HST}/COSMOS weak lensing work used
the ``Drizzle'' algorithm \citep{Fruchter2002}, which in general leads to
a slightly different PSF in each pixel. However, this did not represent a
limiting systematic for the $\sim 2$ deg$^2$ observed in COSMOS.}  In either
case, the problem is still one of measuring the shear from an ensemble of
images of different galaxies.  The one exception is that model-fitting
shape measurement techniques can operate either on the combined images
or via a direct fit to the raw input images. Even in this case, however,
with many exposures (as planned for LSST) object detection will have to be
carried out on the combined image in order to reach the full survey depth.

One would intuitively expect that shape measurement becomes more difficult
when the PSF is larger than the intrinsic size of the galaxy being measured.
This is indeed the case.  While the idealized problem of measuring shapes
in the presence of a PSF is well-defined for any nonzero galaxy size,
in practice both statistical and systematic errors blow up when the PSF
becomes significantly larger than the galaxy. The extent to which the
systematic errors in the high-SNR, $r_{\rm gal} < r_{\rm psf}$ regime can
be addressed will likely determine the constraining power of large-\'etendue
ground-based WL programs such as that planned for LSST.

\subsubsection{Shape Measurement Algorithms*}
\label{sec:wl_sma}

The most obvious --- but flawed --- way to construct a shape
measurement algorithm is to
simply use the quadrupole moment tensor of a galaxy: one could compute
\begin{equation}
Q_{ij}[I] = \int_{{\mathbb R}^2} I({\bf x}) (x_i-\bar x_i)(x_j-\bar x_j)
d^2{\bf x}, {\rm ~~~~~(incorrect)}
\label{eq:wl:q.wrong}
\end{equation}
where $\bar{\bf x}$ is the centroid and the $[I]$ implies that we compute
the quadrupole moment on an observed image.  It is easily seen from the
properties of convolutions that
$Q_{ij}[f] = Q_{ij}[I] - Q_{ij}[G]$, i.e., one may obtain the pre-PSF
quadrupole moment of a galaxy by subtracting the observed quadrupole moment
from that of a PSF.
Then one could construct
the ellipticities of the galaxy, which are simply the trace-free components
of the quadrupole moment normalized by the trace:
\begin{equation}
e_+[f] = \frac{Q_{11}[f]-Q_{22}[f]}{Q_{11}[f]+Q_{22}[f]}
{\rm ~~~and~~~}
e_\times[f] = \frac{2Q_{12}[f]}{Q_{11}[f]+Q_{22}[f]}.
\end{equation}
Since the quadrupole moment of $f$ is simply related to that of $f_0$ via
\begin{equation}
Q_{ij}[f] = (S^{-1})_{ik}(S^{-1})_{jl}Q_{kl}[f_0],
\label{eq:wl:qtrans}
\end{equation}
we may derive the transformation law for ellipticities under infinitesimal
shear:
\begin{eqnarray}
e_+[f] &=& e_+[f_0] + 2\gamma_+ - e_+[f_0] (\gamma_+e_+[f_0] + \gamma_\times
e_\times[f_0])
{\rm ~~and}\nonumber \\
e_\times[f] &=& e_\times[f_0] + 2\gamma_\times - e_\times[f_0]
(\gamma_+e_+[f_0] + \gamma_\times e_\times[f_0]).
\label{eq:wl:etrans}
\end{eqnarray}
It is then easily seen that the mean ellipticity of a population of
galaxies that has an initially isotropic distribution of ellipticities --
i.e., $P(e_+,e_\times)$ depends only on the magnitude $\sqrt{e_+^2+e_\times^2}$
and not on the direction $\arctan(e_\times/e_+)$ --- is
\begin{equation}
\langle e_a\rangle = \left( 2 - e^2_{\rm rms}\right) \gamma_a,
\label{eq:wl:ea}
\end{equation}
where $e_{\rm rms}^2$ is the mean square ellipticity per component ($+$
or $\times$).  Since we work to first order in $\gamma$, we may use the
mean square ellipticity of the observed sources in equation~(\ref{eq:wl:ea}).
So the galaxy ellipticity divided by $2-e_{\rm rms}^2$ is a shear estimator
satisfying our desired conditions: by comparison to equation~(\ref{eq:wl:bias})
there is no additive or multiplicative bias.

The problem with this procedure is that the unweighted quadrupole moment,
equation~(\ref{eq:wl:q.wrong}), involves an integral over the entire sky, with a
weight that increases $\propto x^2$ as one moves away from the centroid of
the galaxy.  Therefore its measurement noise is infinite.  It also fails
to converge if the wings of the PSF decline as $G({\bf x})\propto |{\bf
x}|^{-\alpha}$ for $\alpha\le 4$, i.e., it fails to converge for all PSFs
realized in modern optical telescopes.  Therefore equation~(\ref{eq:wl:q.wrong})
needs modification.

A conceptually simple approach is to do a model fit to each galaxy.
If one fits a model of an exponential or de Vaucouleurs profile galaxy
with homologous elliptical isophotes, then one can obtain the quadrupole
moment $Q_{ij}[f]$ analytically from the model and hence the ellipticity
of the galaxy.  Modern model-fitting techniques can even fit more general
radial profiles, or simultaneously fit bulge + disk models.  Model fitting
is also robust against many types of nastiness that occur in real data,
such as dead pixels, cosmic rays, or nonlinear detector effects.  However,
model fitting assumes that the galaxy actually obeys the model --- and
especially at $z>1$, the appearance of galaxies is not simple and they
are not describable by simple analytical functions.  At present, our best
approach to understand what happens when simple model fits are confronted with
complex galaxies is with simulations.  One can even imagine ``re-calibrating''
these methods using the simulations, e.g. by subtracting the simulated $c_i$
from each shear and multiplying by the matrix inverse of $\delta_{ij}+m_{ij}$
(see eq.~\ref{eq:wl:bias}); but of course one is then relying on the galaxy
population in the simulation to closely trace reality.

One could also attempt to do a regularized deconvolution of the galaxy.
The most popular such technique is a basis function technique: one writes the
galaxy image as $f({\bf x}) = \sum_n b_n \psi_n({\bf x})$, where $\{\psi_n\}$
are a finite basis set and $b_n$ are the fit coefficients; this then becomes
a model-fitting problem.  A common choice is the ``shapelet'' basis, where
the $\{\psi_n\}$ are the energy eigenmodes of the 2-D quantum harmonic
oscillator (polynomials times Gaussians); this requires $(N+1)(N+2)/2$
eigenfunctions to represent the $0...N$ energy levels \citep{refregier03I,
refregier03II}.  This basis is complete in the limit of large $N$, and the
Gaussian endows the basis coefficients with simple transformation properties
under translation and shear.  Real galaxies often require very large $N$
to be well-represented, however, especially for cuspy profiles.

A final class of ideas has been to note that any ellipticity
formula that is {\em shear-covariant} in the sense of transforming via
equation~(\ref{eq:wl:etrans}) enables us to use equation~(\ref{eq:wl:ea}).  For example,
suppose that we had the galaxy image $f$ before PSF convolution, and did
an unweighted least-squares fit, in the sense of minimizing
\begin{equation}
c = \int_{{\mathbb R}^2} [f({\bf x}) - f_{\rm model}({\bf x}|{\bf
p})]^2\,d^2{\bf x}.
\label{eq:wl:c}
\end{equation}
Here $f_{\rm model}$ is an elliptical Gaussian fit to the image with free
amplitude $A$, centroid $\bar x_i$, and second moment matrix $Q_{ij}^{\rm
elfit}$ (6 parameters).  Then $Q_{ij}^{\rm elfit}$ and the ellipticities
constructed from it would be shear-covariant --- even if the galaxy's true
radial profile does not resemble a Gaussian!\footnote{This is easily seen
because the measure $d^2{\bf x}$ in equation~(\ref{eq:wl:c}) is shear-invariant.}
Early work on implementing this idea in the presence of a PSF attempted to
determine the second moment matrix of the image on the sky $Q_{ij}^{\rm
elfit}[f]$ from the observed image and the PSF.  For example, Gaussian
galaxies and PSFs satisfy $Q_{ij}^{\rm elfit}[f] = Q_{ij}^{\rm elfit}[I]
- Q_{ij}^{\rm elfit}[G]$, and so ``non-Gaussianity corrections''  were
introduced \citep{bernstein02, hirata03} that yielded shear calibration
errors of a few percent.  But these methods were heuristic, and moreover
they suffer from a fundamental limitation: $Q_{ij}^{\rm elfit}[f]$ depends
on very high-wavenumber Fourier modes ${\bf u}$ of the image, which are
not preserved by the PSF, i.e. $\tilde G({\bf u})=0$.  It is therefore
mathematically impossible to determine $Q_{ij}[f]$ from the data in a
model-independent manner.

To understand this point more fully, and illustrate a solution, let us
imagine that we are doing an unweighted least-squares fit of a parameterized
image $f_{\rm model}({\bf p})$, using equation~(\ref{eq:wl:c}).  For convenience,
we will write the parameters as ${\bf p} = \{ A, \sigma_{\rm gal}, \bar x_1,
\bar x_2, e_+, e_\times \}$, where $\sigma_{\rm gal} = (\det {\bf Q})^{1/4}$
is a characteristic scale length of the galaxy, so that they have simple
transformation properties under rotations.  Written in Fourier space,
it becomes
\begin{equation}
c = \int_{{\mathbb R}^2} [\tilde f({\bf u}) - \tilde f_{\rm model}({\bf
u}|{\bf p})]^2\,d^2{\bf u},
\end{equation}
and its minimum is given by the simultaneous solution of the 6 equations
\begin{equation}
0 = \int_{{\mathbb R}^2} \tilde f({\bf u}) \frac{\partial \tilde f_{\rm
model}({\bf u}|{\bf p})}{\partial p_\alpha}\,d^2{\bf u},
\label{eq:wl:min}
\end{equation}
where $p_\alpha$ is any of the 6 parameters.  The problem occurs because
$\partial \tilde f_{\rm model}({\bf u}|{\bf p})/\partial p_\alpha$ has support
at $|{\bf u}|>u_{\rm crit}$, where we cannot determine $\tilde f({\bf u})$.

A solution to this problem has been proposed by
\cite{bernstein10}\footnote{See also \cite{kaiser00}, which contains many
of these ideas but seems to have been promptly forgotten by most of the WL
comminity!}, which is in principle exact in the low-noise limit and has been
applied to simulations (but not yet to actual data).  The key is to work in
the Fourier domain, where the effect of the PSF is simple and the effect
of the shear is as simple as in real space.  We present the solution here
in its most general form, and refer the reader to \citet{bernstein10} for
implementation details.  The solution is to replace equation~(\ref{eq:wl:min}) with
\begin{equation}
0 = t_\alpha = \int_{{\mathbb R}^2} \tilde f({\bf u}) \tilde W_\alpha({\bf
u}|{\bf p}) \,d^2{\bf u},
\label{eq:wl:nl}
\end{equation}
where $W_1...W_6$ are weight functions.  These should be envisioned to be
qualitatively similar to the derivatives in equation~(\ref{eq:wl:min}); but
the only rules that we will impose are that: (i) the Fourier transforms
$\tilde W_\alpha({\bf u}|{\bf p})$ have compact support, confined to $|{\bf
u}|<u_{\rm crit}$; and (ii) they are rotation and translation-covariant,
e.g., changing the centroid parameter by $\delta\bar{\bf x}$ simply translates
the function $W_\alpha({\bf x})\rightarrow W_\alpha({\bf x}-\delta\bar{\bf
x})$, and there is a similar transformation when rotating the ellipticity
components.\footnote{Note that $W_{\bar{\bf x}}$ must transform as a vector
under rotations, and $W_{\bf e}$ as a spin-2 tensor.}  We do {\em not} require
the $W_\alpha$ to be shear-covariant: indeed, since a large shear can map
any mode to another mode with $|{\bf u}|>u_{\rm crit}$, such a requirement
would be inconsistent with rule (i).  Now we may write equation~(\ref{eq:wl:nl}) as
\begin{equation}
0 = \int_{|{\bf u}|<u_{\rm crit}} \tilde I({\bf u}) \frac{\tilde W_\alpha({\bf
u}|{\bf p})}{\tilde G({\bf u})} \,d^2{\bf u}.
\end{equation}
The combination $\tilde W_\alpha/\tilde G$ is well-defined, and $\tilde
I({\bf u})$ is the Fourier transform of the {\em observed} image, so the
parameters ${\bf p}$ can be measured from the data.

By rule (ii), we have rotation covariance, so the mean of the ellipticities
$\langle{\bf e}\rangle$ over an isotropic population of galaxies is
zero --- even if the PSF is anisotropic.  Thus there is no additive bias
(except for selection and noise effects --- see warnings below).  However,
dropping shear covariance has come at a price: the ellipticities
$(e_+,e_\times)$ no longer transform according to equation~(\ref{eq:wl:etrans}),
and the responsivity coefficient $\langle{\bf e}\rangle={\cal R}\gamma$ must
be determined.  Fortunately, we can evaluate the effect of an infinitesimal
shear on equation~(\ref{eq:wl:nl}): if ${\bf S} = {\bf 1} + \delta{\bf S}$,
then to first order in $\delta{\bf S}$,
\begin{equation}
\delta \tilde f({\bf u}) = -u_i [\delta{\bf S}]_{ij}\frac{\partial \tilde
f({\bf u})}{\partial u_j},
\end{equation}
and so
\begin{equation}
0 = \delta t_\alpha =
- \int_{{\mathbb R}^2} u_i [\delta {\bf S}]_{ij}\frac{\partial \tilde f({\bf
u})}{\partial u_j} \tilde W_\alpha({\bf u}|{\bf p}) \, d^2{\bf u}
+ \left\{ \int_{{\mathbb R}^2} \tilde f_0({\bf u}) \frac{ \partial \tilde
W_\alpha({\bf u}|{\bf p}) }{\partial p_\beta} \,d^2{\bf u} \right\}
\delta p_\beta.
\end{equation}
The integral in braces $\{\}$ is simply a $6\times 6$ matrix, which we
denote $E_{\alpha\beta}$.  Using integration by parts, the tracelessness
of $\delta{\bf S}$ in the first integral, and the substitution $\tilde
f\rightarrow \tilde I/\tilde G$, we then find
\begin{equation}
\delta p_\beta = -[{\bf E}^{-1}]_{\beta\alpha} [\delta{\bf S}]_{ij}
 \int_{|{\bf u}|<u_{\rm crit}} \frac{\tilde I({\bf u})}{\tilde G({\bf u})} u_i
 \frac{\partial \tilde W_\alpha({\bf u}|{\bf p})}{\partial u_j}\, d^2{\bf u},
\end{equation}
which is well-defined.  This equation tells us how the parameters for each
galaxy vary under an infinitesimal shear; their ensemble average gives
${\cal R}$.  Note that once shear covariance has been dropped, it is only
possible to know the responsivity factor ${\cal R}$ if one has a sample of
real galaxies to observe, since one needs the sample of real galaxies to
compute the matrix $E_{\alpha\beta}$.

A related approach to solving the shear calibration problem was suggested by
\cite{Mandelbaum11shera}. They noted that given a high-resolution image of
a galaxy (e.g., a space-based image) with PSF $G_1$, it is often possible to
construct a lower resolution but sheared image of the same galaxy 
with PSF $G_2$ 
in a model-independent way. One can thus directly test {\em any}
shear estimator on the sheared images, and extract the shear calibration
factor. Conceptually, the criterion for this to work is that all of the
Fourier modes of the image observable using PSF $G_2$ must be within the band
limit of $G_1$ with enough ``padding'' to make sure that the shear (which
also shears the Fourier plane!) does not bring unobserved high-wavenumber
modes not seen with $G_1$ into the region seen by $G_2$. Mathematically,
the criterion for this to be possible are that there exist two critical
wavenumbers $u_{\rm c}$ and $u_{\rm d}$ such that (i) all the power in the
low-resolution PSF is below $u_{\rm c}$, i.e. $\tilde G_2({\bf u}) = 0$ for
$|{\bf u}|\ge u_{\rm c}$; (ii) the high-resolution transfer function $\tilde
G_1({\bf u})$ is far from zero, i.e. $1/\tilde G_1({\bf u})$ is well-behaved,
at all $|{\bf u}|<u_{\rm d}$; and (iii) $u_{\rm c}>(1-\gamma)u_{\rm d}$. Then
one can use the Fourier-domain multiplication:
\begin{equation}
\tilde I_2^{(\gamma)}({\bf u}) = \tilde G_2({\bf u}) \tilde T({\bf S}^{-1}{\bf
u}) \tilde I_1({\bf S}^{-1}{\bf u}),
\end{equation}
where $\tilde T({\bf u})=1/\tilde G_1({\bf u})$ for wave vectors $|{\bf
u}|<u_{\rm d}$.
As implemented, this method requires a higher-resolution image of a fair
subsample of galaxies, which is not always available. It may however be
quite useful in the Stage III ground-based programs, where one might use
HST data for the ``high resolution'' image; see \cite{Mandelbaum11shera}
for a preliminary application of HST data to shear calibration in SDSS.

\subsubsection{Shape Measurement Errors*}
\label{sss:sme}

The statistical uncertainty in ellipticity estimation depends on the method
used and the radial profile of the galaxy, as well as the sizes of the
galaxy and PSF and the SNR.  Rules of thumb can be obtained
by considering nearly circular Gaussians.  
Propagating instrument noise through the
elliptical Gaussian fitting method, \cite{bernstein02} find, in the absence
of a PSF,
\begin{equation}
\sigma(e_+[f]) = \sigma(e_\times[f]) = \frac{\sqrt{16\pi n}\,\sigma_f}{F}
= \frac2\nu,
\label{eq:wl:noise-nopsf}
\end{equation}
where $n$ is the flux noise variance per unit area, $F$ is the galaxy flux,
$\sigma_f$ is the $1\sigma$ width of the galaxy (note: the effective radius
of a Gaussian is $1.177\sigma_f$), and $\nu$ is the detection SNR in an
optimal filter.  In the presence of a circular Gaussian PSF, the ellipticity
is diluted by
\begin{equation}
{\bf e}[I] = \frac{\sigma_f^2}{\sigma_I^2}{\bf e}[f] =
\frac{\sigma_f^2}{\sigma_f^2 + \sigma_G^2}{\bf e}[f],
\label{eq:wl:dil}
\end{equation}
where $\sigma_G$ is the PSF width and $\sigma_I$ is the width of the
PSF-convolved galaxy image.  Furthermore, the detection SNR is reduced
because the galaxy is smeared out into an aperture with more noise, so it
follows that equation~(\ref{eq:wl:noise-nopsf}) should be modified by replacing
$\sigma_f\rightarrow\sigma_I$; and, if we want the uncertainty in the
pre-PSF galaxy ellipticity, we must divide out the $\sigma_f^2/\sigma_I^2$
factor from equation~(\ref{eq:wl:dil}).  This gives
\begin{equation}
\sigma(e_+[f]) = \sigma(e_\times[f]) = \frac{\sqrt{16\pi
n}\,\sigma_I^3}{F\sigma_f^2}  = \frac2\nu \frac{\sigma_I^2}{\sigma_f^2}.
\end{equation}
This provides a large advantage for making the PSF smaller than the galaxy:
since the noise variance $n$ scales with observing time as $t^{-1}$, the
time required to measure the shape of a galaxy scales as
\begin{equation}
t \propto \sigma_I^6 \propto \left( 1 + \frac{\sigma_G^2}{\sigma_f^2}
\right)^3;
\label{eq:wl:scale}
\end{equation}
in the limit of a poorly resolved galaxy ($\sigma_f\ll \sigma_G$) a factor
of 2 improvement in the PSF provides a factor of 64 gain in speed.  However,
as the PSF becomes smaller than the galaxy this advantage saturates.

Equation~(\ref{eq:wl:dil}) also illustrates another property of shape
measurement: {\em systematic} errors as well as statistical errors are
inflated by having large PSFs.  For example, if there is a systematic
error in the ellipticity of the observed image $I$, it propagates to the
estimated pre-PSF ellipticity ${\bf e}[f]$ with a multiplying factor of
$(\sigma_f^2+\sigma_G^2)/\sigma_f^2$.  Therefore there is a systematics
advantage to having $\sigma_G\ll\sigma_f$.

The shear uncertainty is a factor of $\sim2$ smaller than the
ellipticity uncertainty owing to the responsivity factor $2-e_{\rm rms}^2$
(eq.~\ref{eq:wl:ea}).  It does however have a minimum value: the ellipticity
of an individual galaxy has an RMS variation of $e_{\rm rms}\sim0.4$ per
component, so there is a limiting ``shape noise'' contribution to the shear
measurement uncertainty of $\sigma_\gamma \approx 0.2$.
There are some ideas for how to circumvent this limit using
the color- or scale-dependence of ellipticity \citep{lombardi98,jarvis08} or taking
advantage of the non-Gaussianity of the ellipticity distribution \citep{kaiser00, bernstein02},
but there are no clear routes to large improvement for
optical galaxies.  For galaxies imaged in the HI 21cm line,
one might be able to use kinematic signatures to distinguish
random orientation from lensing shear \citep[e.g.][]{morales06}.

\subsubsection{Noise Rectification and Selection Biases*}
\label{sss:nrsb}

Two pernicious biases can arise even for the ``exact'' shape measurement
algorithms described above: the noise rectification and selection biases.

{\em Noise rectification bias} arises whenever a nonlinear transformation,
such as ellipticity measurement, is applied to noisy data.  If we
Taylor-expand the mean of the ellipticity measured on the true image $I_{\rm
obs}$ around the noiseless image $I$, we find
\begin{equation}
\langle {\bf e}[I_{\rm obs}] \rangle = {\bf e}[I] + \frac12 \sum_{ab}
\frac{\partial^2 {\bf e}}{\partial I({\bf x}_a) \partial I({\bf x}_b)}
{\rm Cov}[I({\bf x}_a),
I({\bf x}_b)] + ...,
\label{eq:wl:noiserect}
\end{equation}
where the sum is over pairs of pixels in the image, and ${\bf x}_a$ and
${\bf x}_b$ are positions of those pixels.  The bias is proportional to
the noise variance, i.e., to $(S/N)^{-2}$ at leading order.

One might at first think that the pixel covariance is described by
uncorrelated white noise, which is statistically shear-invariant and thus
leads to no bias, but in the presence of a PSF correction [i.e., dividing by
$\tilde G({\bf u})$] this is no longer the case.  The noise rectification
bias was first recognized in the context of WL by \cite{kaiser00}, who
showed that because the centroiding of a galaxy is more accurate on the
``short'' than the ``long'' axis of the PSF there is a preference for the
measured second moment of the galaxy to be elongated along the PSF, even if
the PSF correction method is perfect in the deterministic case.  This was
generalized by \cite{bernstein02} to incorporate other noise-related biases
and by \cite{hirata04} to include the effect on shear calibration errors.
Equation~(\ref{eq:wl:noiserect}) provides a unified framework for computing
all of these biases to order $(S/N)^{-2}$.  At low $S/N$ higher-order
terms in the expansion may become important, and the expansion itself may
break down, e.g., as fitting algorithms jump to alternate $\chi^2$ minima.
It is our judgment that it is best to stay away from this ``nonperturbative
noise'' regime.
For a recent investigation of noise rectification bias in the 
context of current shape-measurement algorithms, see
\cite{melchior12}.

{\em Selection biases} are well-known in astronomy.  In our case, they
will affect the shear if there is a bias in favor of detecting
galaxies in some orientations rather than others, producing
an additive shear error,
or if selection depends on the magnitude of the ellipticity, which leads to 
a multiplicative shear error because galaxies are preferentially
selected when their intrinsic ellipticity is aligned with the shear
\citep{hirata03}.  A similar bias results if galaxies are weighted by various
properties (e.g., ellipticity uncertainty) that are not shear-invariant.
The formalism of \S\ref{sec:wl_sma} can in principle handle this problem
if instead of computing $\langle{\bf e}\rangle$ we compute $\langle w{\bf
e}\rangle$ where the weight $w=0$ for galaxies that are rejected.  However,
the assessment of selection biases in practice has been addressed through
simulations such as the STEP program.

A problem related to selection biases is {\em blending}: the superposition
of images of two galaxies.  If the galaxies are at the same redshift, they
are affected by the same shear, and an ideal shape measurement algorithm
that measures the blend should recover the ``correct'' answer --- indeed,
existing WL surveys must contain many sources that are actually blended
with their own satellite galaxies.  But if the deblending algorithm is
not shear-invariant there can be a bias in the shear.  Another issue,
particularly for ground-based Stage IV experiments that will aim for high
source densities at modest resolution and very small statistical errors,
is accidental blending of galaxies at different redshifts (and hence
different shears).

The general strategy for dealing with these categories of biases is:
(1) make choices (e.g., S/N cuts) that keep them small to the extent
possible;
(2) compute corrections using simulations and/or analytic estimates,
and apply them to the measurements;
(3) test the accuracy of these corrections in the data
by looking for the expected scalings with S/N, source size,
and so forth; and
(4) marginalize over remaining uncertainties in the corrections.

\subsubsection{Determining the PSF and Instrument Properties}
\label{sec:wl_psf}

Shape measurement algorithms are only as useful as their inputs: in this case
a map of the PSF $G({\bf x})$ at each point in the field.  Determining the
PSF to sub-percent accuracy is one of the major challenges in WL.  Errors in
the PSF model introduce correlated structure into the ellipticity field
of the galaxies, since residual anisotropy in the PSF determination is
interpreted as shear by a shape measurement algorithm.

Fortunately, Nature has provided us with stars, which under typical observing
conditions can be treated as point sources.  Unfortunately, there is only a
finite density of stars in high Galactic latitude fields, typically of order 1
per arcmin$^2$, so one must interpolate the PSF to the position of a galaxy.
This is a demanding challenge; any error in the interpolated PSF is likely to
have spatial structure.  It is also not an easy problem, as the PSF
is an entire function $G({\bf x};\theta)$ at every 2-d position $\theta$
on the sky, and in contrast to shape measurement, interpolation from stars is
underconstrained.\footnote{As a reminder, here ${\bf x}$ is used to refer
to the location within the image of each star, i.e., of order $\sim 1''$,
whereas the independent variable $\theta$ accounts for variation across the
entire field, of order $\sim 1^\circ$.}  To date, most of the methods applied
to real data are heuristic.  For example, the SDSS analyses
fit a low-order polynomial,
\begin{equation}
G({\bf x};\theta) = \sum_{i=0}^N \sum_{j=0}^{N-i} \sum_{k=1}^M a_{ijk}
\theta_1^i \theta_2^j G^{(k)}({\bf x}),
\end{equation}
where the $\{G^{(k)}\}$ are the top $M=3$ principal components of the stellar
images, $N=2$ is the interpolation order, and $a_{ijk}$ are coefficients.
Small scale structure in the PSF variation
may not be well represented by this approach
unless $N$ is large, but if the required number of polynomial
coefficients $(N+1)(N+2)/2$ exceeds the number of stars in each frame
then the method falls apart.  If the small-scale structure is repeatable,
for example if 
it is associated with low-order aberrations in the telescope or the
topography of the focal plane, then one may make progress by applying PCA
to the angular dependence in instrument-fixed coordinates \citep{jarvis04},
choosing the top $K$ modes out of the space of $(N+1)(N+2)/2$ polynomials.
Recent work has focused on improved interpolation schemes that outperform
polynomials (e.g., \citealt{Berge11}).

For space-based data, one can either build a physical model of the PSF
\citep{rhodes06} or use PCA \citep{jee07}. However, for ground-based data
where the PSF has a large
contribution from atmospheric turbulence, the more empirical interpolation
schemes have been the methods of choice.

Once one has the PSF, one needs a method of quality assessment.  We need
to be able to determine, or at least bound, the power spectrum of the
residual PSF systematics that leak into cosmic shear results.  (For GGL,
this job is easier because residual PSF anisotropy adds noise but does not
correlate with the positions of the galaxies.)  One way is to do null tests:
one can compute the correlation function of ellipticities of the stars and
(supposedly) PSF-corrected galaxies, or search for $B$-mode shear. 
The latter is not foolproof, as a PSF systematic of $E$-mode type
can arise from some aberrations.  A very attractive (but underutilized)
test is to mask some of the stars in the PSF fitting and compare the
interpolated PSFs at their locations to the observed stellar images. There
are also
methods for using combinations of these correlation functions to test for
``overfitting''
-- the phenomenon in which a too-general PSF model begins to fit noise or
small-scale structure in the stellar images, with the effect that the
interpolated PSF
is actually worse \citep{rowe10}.

Even when this is done, there remain two other errors that have received
increasing attention recently, which may cause the PSF of a galaxy to differ
from that of a star:
\begin{itemize}
\item {\em Color dependence}:
Real PSFs depend on the wavelength of light: a diffraction-limited telescope
has a PSF size that scales $\propto\lambda$, seeing through a Kolmogorov
atmosphere gives a size $\propto\lambda^{-1/5}$, aberrations introduce
$\lambda$-dependence into not just the size of the PSF but its morphology
and radial profile, real detectors have response functions that depend on
wavelength, and in ground-based data atmospheric dispersion acts like a prism
and causes a centroid shift with wavelength.  Since galaxies have different
SEDs than the stars used to fit the PSF (the galaxies are usually redder),
the PSF measured from the stars is not always appropriate to the galaxies
\citep{cypriano09}.  Moreover, each galaxy contains a range of SEDs due to
differing stellar ages and metallicities and dust columns; and for each of
these SEDs there is a different PSF, with the resulting images superposed
on the focal plane \citep{Voigt11}.  This represents a major challenge:
while the centroid
wavelength of galaxies in a typical filter ($\lambda/\Delta\lambda\sim
5$) varies by several percent, Stage IV surveys will require sub-percent
shear calibration accuracy, and as yet no WL survey has published a
color correction to its PSF at all.  While this area requires much more work,
in a problem of this complexity prevention is the first step to a cure.
For example, one can employ an atmospheric dispersion compensator on the
ground, and one can use narrower filters.  For smooth spectra or 
spectra averaged
over moderate redshift ranges, the variation of wavelength centroid scales
as $\propto\Delta\lambda^2$.  The advantage for narrower filters
must of course be weighed against the
slower survey speed with smaller $\Delta\lambda$. If many observations of
a given field are taken under varying conditions, as planned for LSST,
one has the intriguing possibility of using the different seeing, focus,
or hour angle dependences of these errors to solve them out and distinguish
them from shear. A final strategy is to use ``calibration samples'' of galaxies
observed in multiple filters to correct the effect for a lensing survey with a
single, wide band \citep{semboloni12}.
\item {\em Detector effects}:
A CCD image can be altered during readout due to finite {\em charge-transfer
inefficiency}.  This results from photoelectrons that temporarily bind
to defects (``traps'') in the material, causing each sky object to leave a
trail along the readout direction.  If not corrected, this can appear as a
PSF anisotropy, and it is greater for faint objects than for bright
objects because some of the traps saturate, so stellar images tend to
underestimate the effect.  Charge transfer inefficiency is primarily a concern 
for space-based data, since the principal cause of traps is radiation damage.
Remedies applied to HST data have included empirical corrections to the
galaxy ellipticities as a function of row and magnitude, or pixel-level
corrections that begin by ``undoing'' the charge transfer inefficiency
before subsequent processing \citep{massey10, Rhodes10cti}.
\item
Near-infrared detectors (as planned for \wfirst) suffer from different potential
detector-induced systematic errors than CCDs. In particular, since the image
is read ``in place'' on the detector rather than transferred out as on a
CCD, there is no charge transfer inefficiency. Instead the major concerns
have been {\em interpixel capacitance} (the sensitivity of the voltage on
each pixel to the charge collected in neighboring pixels); {\em persistence}
(a charge trapping and release phenomenon in which each exposure contains a
small amount of residual signal from previous exposures); and {\em reciprocity
failure} (a detector exposed to twice the signal for half the time does not
produce the same response, which becomes an issue for comparing the images
of stars and galaxies of very different magnitudes).
\end{itemize}

\subsection{Astrophysical systematics}
\label{sec:wl_astro}

The principal advantage of weak lensing is that --- despite its technical
difficulty --- it is directly sensitive to mass. It is thus less affected
by astrophysical uncertainties than other probes of cosmic structure such
as the galaxy power spectrum or X-ray cluster counts. 
However, it is not entirely
free of astrophysical contamination. The two major sources of uncertainties
in this case are intrinsic galaxy alignments, which can mimic the coherent
distortion of galaxies by gravitational lensing, and the prediction of the
matter power spectrum.

\subsubsection{Intrinsic Alignments*}
\label{sec:wl_ia}

We have thus far
assumed that the intrinsic ellipticities of galaxies
are independent, adding noise but not spurious signal to
cosmic shear measurements.
However, the orientations
of galaxies are determined by physical processes --- mergers, torquing
by tidal fields from the host halo and large scale structure, etc. ---
that could produce correlated {\em intrinsic
alignments}.  We first describe here the general formalism for
the impact of intrinsic alignments, then consider
what observations and theory have taught us about them.
We conclude by discussing prospects for intrinsic alignment removal.

The field of intrinsic galaxy ellipticities
is a tensor function ${\bf e}({\bf r},\hat{\bf n})$
of position ${\bf r}$ and viewing direction $\hat{\bf n}$.
In this sense it is very similar to CMB polarization.  In principle it also
depends on the type of galaxy under consideration and on the observational
details --- for example, the $B$ and $I$-band
images of a galaxy could have different ellipticities.  We may also discuss
either the unweighted intrinsic ellipticity field ${\bf e}_{\rm unwt}$
or the field weighted by the galaxies,
\begin{equation}
{\bf e}_{\rm wt}({\bf r},\hat{\bf n}) \equiv [1+g({\bf r})] {\bf e}_{\rm
unwt}({\bf r},\hat{\bf n}),
\end{equation}
where $g = (n_{\rm gal}-\bar n)/\bar n$ is the galaxy overdensity.  In what
follows, we use ${\bf e}$ to denote the galaxy-weighted field ${\bf e}_{\rm
wt}$, since this is most closely related to what one observes in a survey.

Like any other field, ${\bf e}$ can be Fourier transformed to give $\tilde{\bf
e}({\bf k},\hat{\bf n})$, with a power spectrum
\begin{equation}
\langle \tilde e_a^\ast({\bf k},\hat{\bf n}) \tilde e_b({\bf k}',\hat{\bf
n}')\rangle = (2\pi)^3 \delta^{(3)}({\bf k}-{\bf k}')
  P_{e;ab}({\bf k},\hat{\bf n},\hat{\bf n}')~,
\end{equation}
where $a,b$ are spin-2 tensor indices.  Here we break from the train
of reasoning in CMB polarization studies: instead of doing a multipole
decomposition of ${\bf e}$, we note that in the Limber approximation (which
we use exclusively here) there is only one relevant viewing direction --
the direction to the observer --- so $\hat{\bf n}=\hat{\bf n}'$.
Moreover,
the Fourier wave vectors that we care about are perpendicular to the line of
sight, so ${\bf k}\cdot\hat{\bf n}=0$.  We will thus write this particular
configuration as simply $P_{e;ab}(k)$.  An $E$/$B$ mode decomposition is
also possible if we rotate the coordinate basis so that the $E$-component of
ellipticity is aligned along the direction of ${\bf k}$ and the $B$-component
is at a 45$^\circ$ angle; we then have two ellipticity power spectra,
$P_{e}^{EE}(k)$ and $P_{e}^{BB}(k)$ (the $EB$-term vanishes by parity).
One can also write correlations of the ellipticity with scalar fields such
as the galaxy or matter density.  In this case, only the $E$-mode can be
correlated, and we write $P_{e\delta}$, $P_{eg}$, etc.

The measured shear on the sky is a superposition of the WL shear and the
intrinsic ellipticity (converted to shear using the algorithm-specific
responsivity factor ${\cal R}$):
\begin{equation}
\gamma_{\rm obs}(\theta) = \gamma(\theta) + \frac1{\cal R} \int p(D_C){\bf
e}(\theta,D_C) \,dD_C.
\end{equation}
Limber's equation can then be used to obtain the observed shear power
spectrum between the $\alpha$ and $\beta$ redshift slices.  The $E$-mode
contains three terms:
\begin{equation}
C_{EE}^{\alpha\beta}(l;{\rm obs}) = C_{EE}^{\alpha\beta}(l;GG) +
C_{EE}^{\alpha\beta}(l;GI) + C_{EE}^{\alpha\beta}(l;II),
\end{equation}
where the $GG$ term is the gravitational lensing shear contribution, $II$
is the intrinsic ellipticity contribution, and $GI$ is the cross-correlation.
The $GG$ term is the desired signal and is given by equation~(\ref{eq:wl:peetomo}).
The other terms are
\begin{equation}
C_{EE}^{\alpha\beta}(l;II) = \frac1{{\cal R}^2} \int p_{\alpha}(D_{C1})
p_{\beta}(D_{C1}) \frac{P_e^{EE}(k=l/D_{A1})}{D_{A1}^2} dD_{C1}
\label{eq:wl:ii}
\end{equation}
and
\begin{equation}
C_{EE}^{\alpha\beta}(l;GI) = \frac1{\cal R}\int \left[
W_{{\rm eff},\alpha}(D_{C1}) p_{\beta}(D_{C1}) +
W_{{\rm eff},\beta}(D_{C1}) p_{\alpha}(D_{C1}) \right]
\frac{P_{e\delta}(k=l/D_{A1})}{D_{A1}^2} dD_{C1}.
\label{eq:wl:gi}
\end{equation}
There is also an $II$ contribution to the $B$-mode power spectrum similar
to equation~(\ref{eq:wl:ii}).  Since there is no $B$-mode gravitational shear,
there is no $GG$ or $GI$ contribution to the $B$-mode power spectrum.

Several generic features can be noted from these equations:
\begin{itemize}
\item
The $II$ contribution to the cross-spectrum is only nonzero if the two
redshift distributions overlap, since it arises from intrinsically
aligned galaxies at the same redshift.  Therefore, if
low-scatter photo-$z$s are
available, it can be eliminated.  This is one motivation for doing tomography
instead of simply measuring the shear power spectrum on a magnitude-limited
sample of galaxies.
\item
The $GI$ contribution is more troublesome.  It arises from the lensing of
the more distant slice by the same matter field that controls the intrinsic
ellipticity of the nearby slice.  Inspection of the properties of the
window function $W_{\rm eff}$ shows that equation~(\ref{eq:wl:gi}) is nonzero
for {\em all} redshift distributions, unless either $P_{e\delta}(k)=0$
or the redshift distributions are $\delta$-functions at the same redshift,
which would dramatically enhance $II$.  Thus we are led to conclude that
every tomographic power spectrum can suffer intrinsic alignment contributions.
\item
$B$-mode ``shear'' can be generated by intrinsic alignments, and if nonzero
$C_{BB}^{\alpha\beta}(l)$ is observed this is one possible explanation
(PSF model errors are another).  However, it is possible for the intrinsic
alignment contribution to the $E$-mode to be much larger because (i) there
is no theoretical reason from galaxy formation to expect $P_e^{EE}(k)\sim
P_e^{BB}(k)$ --- indeed, we will see below that $P_e^{EE}(k)\gg P_e^{BB}(k)$
may be natural --- and (ii) $C_{EE}^{\alpha\beta}(l)$ can also contain a $GI$
term, which for broad redshift distributions usually dominates over $II$.
Therefore a nondetection of $B$-mode shear does not rule out significant
intrinsic alignment contamination.
\end{itemize}

Before we discuss removal of intrinsic alignments, it is helpful to consider
the physics underlying their power spectra.  One can distinguish two cases:
early-type galaxies, which are triaxial and whose intrinsic ellipticity is
presumably related to the direction of the most recent merger or the direction
of anisotropic collapse (depending on one's idea of how these galaxies are
formed), and late-type galaxies, whose ellipticity is determined by the
disk angular momentum (perhaps acquired via tidal torquing during collapse,
reshuffled by disk-halo interactions, and perturbed by minor mergers).
The detailed physics of these processes remains elusive, but some predictions
can still be made by traditional galaxy biasing arguments.  For example, if
one considers the formation of early-type galaxies in a particular region
of the universe, one could argue that at linear order in the large-scale
density field a galaxy's formation sequence can be sensitive only to the
density and tidal field coming from the linear modes, and to small-scale
structure.  Since only the tidal field has the correct symmetry properties
to be related to an ellipticity, it follows that the ellipticity should be
proportional to the tidal field,
\begin{equation}
e_+ = \frac{C_1}{4\pi G\bar\rho a^2} (\partial_1^2-\partial_2^2)\Phi,
{\rm~~and~~} e_\times = \frac{C_1}{4\pi G\bar\rho a^2}
(2\partial_1\partial_2)\Phi ~,
\label{eq:wl:ia-lrg1}
\end{equation}
where $C_1$ controls the strength of alignment and
$\partial_1$ and $\partial_2$ denote derivatives along
two orthogonal axes on the sky.
This implies that the ellipticity traces the density field, and in particular
\begin{equation}
P^{EE}_e(k) = C_1^2P_\delta(k), {\rm~~} P^{BB}_e(k)=0,  {\rm~~and~~}
P_{e\delta} = C_1P_\delta(k).
\label{eq:wl:ia-lrg}
\end{equation}
Equations~(\ref{eq:wl:ia-lrg1}, \ref{eq:wl:ia-lrg}) are known as the {\em
linear alignment model} \citep{catelan01}.  
Note that they predict only $E$-mode intrinsic
alignments, because the alignments are linearly sourced by a scalar
field.\footnote{If one interprets the model of equation~(\ref{eq:wl:ia-lrg1})
as applying to the unweighted ellipticity field, then converting to a
galaxy-weighted field introduces a $B$-mode.  However it is much smaller
than the $E$-mode signal and vanishes in the linear regime.}

Observations of LRGs in the SDSS have shown that the galaxy-ellipticity
correlation\footnote{This has been measured as the line-of-sight integral of
the correlation function, $w(r_p)$ where $r_p$ is the transverse separation,
which contains the same information as the power spectrum.} $w_{ge}(r_p)$
has the same power-law slope as the galaxy correlation function $w_g(r_p)
\propto r^{-0.7}$ \citep{mandelbaum06,hirata07}, with an amplitude that
increases rapidly with LRG luminosity.  This is a quantitative success of
the linear model.  However, on small scales it is not clear how accurate
equation~(\ref{eq:wl:ia-lrg}) should be.

For late-type galaxies, it is less clear what to expect.  The oldest and
most widely discussed model is that disk galaxies acquired their angular
momentum from tidal fields acting on nonspherical protogalaxies, an effect
that would make the resulting intrinsic ellipticity quadratic in the tidal
field: this is known as the {\em quadratic alignment model} \citep{pen00}.
This model produces both $E$ and $B$-mode $II$ signals, but to leading order
it predicts $P_{e\delta}(k)=0$ and hence gives no $GI$ signal \citep{hirata04}.
However, one should be cautious about this argument for several reasons, most
importantly because there has not yet been any quantitative observational
confirmation of the scale and configuration dependence predicted by the
quadratic model, and additionally because perturbation theory arguments show
that the nonlinear evolution of the tidal field can generate a linear
type alignment \citep{hui08}.  What is clear from observations is that the
alignments of late-type galaxies on large scales, at least as measured by
$w_{ge}(r_p)$, are consistent with zero and are certainly much less than
for LRGs \citep{hirata07, mandelbaum11}.

Detailed assessments of the intrinsic alignment contamination have been
made on the basis of SDSS, 2SLAQ, and WiggleZ observations of $w_{ge}(r_p)$
\citep{hirata07,mandelbaum11,joachimi11}.  These studies show
that for surveys of modest depth
($z_{\rm med}\sim 0.7$) the $GI$ contamination may be up to several percent
of the expected cosmic shear signal
for late-type galaxies if it is near current upper limits, and it could be
tens of percent for LRGs.  As one probes to higher source redshifts the
level of contamination becomes increasingly uncertain, because there are
not yet galaxy surveys at $z\ge 1$ that are capable of probing intrinsic
alignments at interesting levels.  The $II$ contamination for broad redshift
distributions is found to be much less than $GI$ for linear alignment models.

Finally, we consider the methods used to remove intrinsic alignments.
One starts with prevention: in the recent COSMOS analysis, \cite{schrabback10}
suppressed $II$ by throwing out the auto-power spectra of each of their
redshift slices with itself, keeping only the cross-spectra.  They also
suppressed $GI$ by not including LRGs in the foreground redshift slice,
since LRGs contribute the most to the contamination.  However, it is not
clear that sample selection alone
will provide sufficient $GI$ rejection for Stage III surveys
and beyond.  Two general approaches to $GI$ rejection have been proposed,
model-independent and model-dependent.

The model-independent $GI$ rejection method is to note that if we have narrow
redshift bins, and denote the foreground and background slices by $z_\alpha$
and $z_\beta$ respectively, then the $GI$ signal depends only on intrinsic
alignments at $z_\alpha$ (i.e., in the nearer bin).  Then at fixed $z_\alpha$
the $GI$ signal is proportional to
\begin{equation}
C_{EE}^{\alpha\beta}(l;GI) \propto \cot_KD_C(z_\alpha) - \cot_KD_C(z_\beta),
\end{equation}
which becomes small if $z_\beta-z_\alpha$ is small.  This is a different
redshift dependence than the $GG$ signal, which is a linear function
of $\cot_KD_C(z_\beta)$ but remains finite as $z_\beta\rightarrow
z_\alpha$. Hence it could be projected out \citep{hirata04} --- e.g., one
could take the $\alpha\beta$ shear cross-spectrum at several background
bins and extrapolate to $z_\alpha$.  An alternative implementation of this
idea is nulling \citep{hirata04,joachimi08,joachimi09}, 
constructing a synthetic redshift slice by weighting of
the different $z_\beta$ whose window function
\begin{equation}
W_{\rm eff,syn}[D_C(z_\alpha)] = \sum_\beta w_\beta
W[D_C(z_\alpha),D_C(z_\beta)] = 0.
\end{equation}
Clearly some of the weights $w_\beta$ must be negative.  This class of
techniques assumes nothing about the physics of intrinsic alignments,
but because of the extrapolations or negative weights it can amplify
observational systematics, and to date it has not been successfully
implemented on real data.

A model-dependent alternative, less demanding in terms of observational
systematics, is to construct the $3\times 3$ symmetric matrix of power
spectra of the matter, galaxies, and intrinsic ellipticity,
\begin{equation}
{\bf P}(k) = \left(\begin{array}{ccc} P_\delta & P_{g\delta} & P_{e\delta} \\
P_{g\delta} & P_g & P_{eg} \\
P_{e\delta} & P_{eg} & P_e^{EE} \end{array} \right)(k).
\end{equation}
This has six free functions of wavenumber, of which one ($P_\delta$) can be
predicted from cosmological parameters.  However, since the tidal field
is determined by the matter distribution, if galaxy alignments are really
determined by the tidal field then they should not additionally care
where the other galaxies are: the conditional probability distribution
Prob$(e|\delta,g)=$Prob$(e|\delta)$.  In this case, and in the limit
of a Gaussian field, one should have the restriction\footnote{This is
a slightly more general relation than assuming that the galaxies are
linearly biased with no stochasticity, in which case one could replace
$P_\delta(k)/P_{g\delta}(k)\rightarrow 1/b$.}
\begin{equation}
%Pem/Peg = be*re/(be*bg*re*rg) = 1/(bg*rg).
P_{e\delta}(k) = \frac{P_{\delta}(k)}{P_{g\delta}(k)} P_{eg}(k).
\label{eq:wl:ed}
\end{equation}
This relation was assumed by the DETF in their WL parameter forecasts
\citep{albrecht06}, and if valid it is very useful because it relates the $GI$
contamination ($P_{e\delta}$) to theory ($P_\delta$), GGL ($P_{g\delta}$),
and galaxy-ellipticity correlations at the same redshift ($P_{eg}$).
Unfortunately, its accuracy is unclear in the nonlinear regime, since for
non-Gaussian density fields, Prob$(e|\delta,g)=$Prob$(e|\delta)$ no longer
implies equation~(\ref{eq:wl:ed}); an investigation of this in simulations should
be a high priority.  Nevertheless, equation~(\ref{eq:wl:ed}) may be usable
if the $GI$ correlation for late-type galaxies turns out to be far below
current upper bounds, in which case even a crude correction could reduce
it to below statistical error bars.
Further discussions of this approach and an application to observational
data may be found in \cite{bernstein09}, \cite{joachimi10}, 
and \cite{kirk10}.

Intrinsic alignments also represent a contaminant to GGL if the
``lens'' and ``source'' redshift distributions overlap; some of the
``sources'' may then be physically associated with the lens and show an
alignment that is a result of galaxy formation physics rather than lensing
\citep{BernsteinNorberg02,hirata04X}.  However, in this case the availability
of good photo-$z$s solves the problem, since for GGL there are only $II$
alignments, which can be eliminated by restricting cross-correlations
to non-overlapping redshift slices.
Consistency checks between GGL and cosmic shear may provide
a useful route to diagnosing the impact of intrinsic alignments
on the latter.

\subsubsection{Theoretical uncertainties in the matter power spectrum*}
\label{sec:wl_matterpk}

An important systematic error in weak lensing is the prediction of the
cosmic shear power spectrum, which --- although far more theoretically
tractable than galaxy clustering --- is not free of uncertainty. WL gets
most of its information from the nonlinear regime, where the only way to
accurately predict the power spectrum is using large $N$-body simulations.
At the present time, most WL constraints have used physically
motivated fitting formulae calibrated to
$N$-body simulations (e.g., \citealt{peacock96,smith02}),
but these have limited accuracy because of the limited
resolution and box size of the simulations and
the limited ranges of cosmological parameters that have been explored.
The situation has improved dramatically in recent years thanks to Moore's law
and the fact that the ``interesting'' region of parameter space
has shrunk considerably.  Much improved nonlinear matter power spectrum
calculations have been obtained from the ``Coyote Universe'' simulations
\citep{heitmann09, heitmann10, lawrence10}.  Given this progress, and
given that the $N$-body problem is perfectly well-defined mathematically,
we expect that the theoretical uncertainty in the power spectrum for pure
dark matter models will not be a limiting systematic for WL.

The situation is more complicated when one goes beyond pure dark matter.
Baryons make up $\sim17$\%\ of the matter in the universe, and on small
scales they do not trace the dark matter.  Hydrodynamic simulations can follow
them, but one cannot hope to model the processes of cooling, star formation,
metal enrichment and feedback from first principles.  On quasilinear scales,
$k\sim$few$\times 0.1h\,$Mpc$^{-1}$, the largest uncertainty appears to come
from clusters, where redistribution of the radial distribution of baryons
affects the 1-halo contribution to the power spectrum. Observations of
clusters --- in particular measurement of cluster concentrations --- may help
to constrain this effect \citep{rudd08}. It has been proposed to either
``self-calibrate''
the cluster profile effect \citep{zentner08} or incorporate information
from cluster-galaxy lensing \citep{mandelbaum08}, although this has not yet
been necessary
for present cosmic shear experiments.

A second uncertainty is associated
with the {\em missing baryon problem} --- the fact that most of the baryons
that should be in galaxy-sized halos (assuming a cosmic baryon:CDM ratio)
are not observed in the stellar, H{\sc~i}, and molecular gas components.
If these baryons have been ejected from the host halo, e.g. via galactic
winds or AGN feedback, then they could reduce the matter power spectrum.
These effects were discussed in an idealized ``nightmare scenario'' by
\cite{levine06};
more recently, the detailed hydrodynamic simulations of \cite{vandaalen11} 
have shown a suppression of the matter power spectrum by 1\%
at $k=0.3 h/$Mpc and 10\% at $k =1h/$Mpc. 
These effects are large compared to the statistical errors
of Stage IV WL experiments.
Worrisomely, \cite{vandaalen11} find that the predictions for
the matter power spectrum depend significantly on the treatment
of star formation and AGN feedback in the simulations. In their simulations,
AGN feedback has the effect of reducing the baryon content of the haloes,
consistent with X-ray observations of intrahalo gas: the matter power suppression quoted above
is thus within the range of ``reasonable'' rather than ``extreme/unrealistic'' models.

\cite{semboloni11} show that the results of the simulations can be
captured by a parameterized halo model for the baryons, so one
may be able to use this approach to marginalize over uncertainties,
but at the price of reducing the cosmological information derived
from WL measurements on these scales. Moreover, their mitigation procedure
involves tuning the halo model to the \cite{vandaalen11} simulations. Therefore
one should worry that the removal of baryonic physics-induced bias seen by \cite{semboloni11}
might not be realized in practice, if the simulation captures the qualitative features of AGN feedback
but does not quantitatively reproduce the correct functional form. \cite{zentner12} avoid this issue
by fitting cosmic shear power spectra based on the \cite{vandaalen11} simulations using
a mitigation procedure tuned to the \citet{rudd08} simulations. However they find that this
procedure, while successful on simulated Stage III (DES) survey data, is not adequate for the
more ambitious task of Stage IV data analysis.

In summary, it is clear that better predictions for baryonic effects in the matter power spectrum,
ancillary observations of baryonic gas to constrain the range of outcomes realized in the real universe,
and optimal methods for incorporating these effects 
with minimal damage to cosmological constraints are critical
areas for further investigation.

A final issue is the accuracy of the leading-order mapping from
$P_\delta(k,z)$ to the shear power spectrum, equation~(\ref{eq:wl:pee}).
Next-order perturbation theory arguments \citep{krause10} suggest that
the correction is small, only a few $\sigma$ for Stage IV experiments.
Ultimately, this correction should be computed with ray-tracing simulations
that solve the full deflection equation.

\subsection{Systematic Errors and their Amelioration: Summary}
\label{sec:wl_systematics}

Summarizing results from our earlier discussion,
the principal systematic errors in weak lensing measurements are:
\begin{itemize}
\item {\slshape PSF correction and shape measurement biases 
(\S\S\ref{sec:wl_sma}--\ref{sec:wl_psf})}: 
For the
typical case of a PSF of a similar size to the galaxy, the correction of
the galaxy ellipticity for PSF effects is of order unity, and the desired
accuracy is $<10^{-3}$. This requires both very accurate knowledge of the
PSF and appropriate algorithms to correct for it.
\item {\slshape Redshift distribution uncertainties (\S\ref{sec:wl_photoz})}: 
Using source galaxies
at a higher redshift increases the WL power spectrum, and hence there
is a degeneracy between $z_{\rm source}$ and the inferred cosmological
parameters. If the source redshift distribution (or distributions, in
the case of a tomographic analysis using photometric redshifts) are not
well-calibrated, there is a resulting error in the inferred cosmology.
\item {\slshape Intrinsic alignments (\S\ref{sec:wl_ia})}: 
The ellipticities of nearby
galaxies may be correlated with each other due to formation in a common
environment. This ``$II$ effect'' adds to the observed shear power spectrum
and represents a systematic error. There can also be cross-correlation of the
intrinsic ellipticity of a nearby galaxy at $z=z_1$ with the lensing signal
on a more distant galaxy at $z=z_2>z_1$, since both depend on the tidal
field at $z=z_1$. This latter ``$GI$ effect'' contaminates all tomographic
cross-power spectra and hence is more difficult to remove than $II$.
\item {\slshape Matter power spectrum uncertainties (\S\ref{sec:wl_matterpk})}: 
Predicting $P_m(k,z)$ 
from a set of cosmological parameters is a nontrivial task.  This can
now be done accurately for dark matter only cosmologies (assuming that the
dark matter behaves as simple CDM), but on small scales the influence of
baryonic physics (cooling, feedback, etc.) is difficult to model.
\end{itemize}
These errors, and the steps to remove them, are not independent --- for
example, marginalizing out the intrinsic alignment effects can amplify
systematic errors in photometric redshifts \citep{bridle07}. The development
of systematic error budgets and requirements for future surveys thus requires
a global analysis of all of the statistical and systematic uncertainties
and their possible degeneracies \citep{bernstein09}.

We have described numerous strategies for suppressing most of these effects, 
but a few features stand out. {\em First}, exquisite knowledge of the PSF
must be achieved through some
combination of good engineering (designing a stable telescope and instrument
and putting it in the best possible environment), good choice of observing
strategy (more dithers and repeat visits), and good algorithms (one needs
to generate a homogeneous catalog with well-understood ellipticity errors
and selection effects). {\em Second}, precise photo-$z$s over
the entire range covered by the survey are desirable for characterizing the
redshift distribution, and they are {\it required} if one is to even attempt
a model-independent removal of intrinsic alignments --- something that has
not yet successfully been done.
To achieve these photo-$z$ requirements, one wants optical and near-IR
photometry to distinguish Balmer/4000\AA\ breaks from \lya\ breaks
and spectroscopic samples that span the full range of the WL samples.
Cross-correlation against large redshift surveys can be an important
tool in photo-$z$ calibration \citep{newman08}.

\subsection{Advantages of a Space Mission}
\label{sec:wl_space}

A space platform offers two critical advantages for weak lensing: (i)
the availability of a small and stable PSF, and (ii) the low sky brightness
in the near-IR, which allows deeper observations. For this reason, weak lensing
has been highlighted as an important science objective for the 
\euclid\ and \wfirst\ space missions.

The small PSF enables the telescope to resolve many more galaxies (see
Fig.~\ref{fig:wl_galdens}). 
The space-based PSF size is normally determined by the diffraction
limit: for an ideal Airy disk with an unobstructed circular aperture (off-axis
telescope), the 50\% encircled energy radius {\tt EE50} 
is $0.535 \lambda/D$. This
worsens for obstructed telescopes, reaching $1.25 \lambda/D$ in the extreme
case of blocking 25\%\ of the area of the telescope entrance. Nevertheless,
for typical $\lambda$ (of order 0.8 $\mu$m for a visible mission and 1.5
$\mu$m for a near-IR mission) and reasonable telescope size ($D\ge1.1$ m)
the {\tt EE50} radius is several times smaller than the typical $\sim 0.3-0.4$
arcsec from a good ground-based site. 
There are additional contributions to the PSF size
-- charge diffusion, the pixel tophat, aberrations, and pointing jitter --
but on a space weak lensing mission these would be designed to be subdominant
to diffraction.

A perhaps more important advantage is the {\em stability} of the PSF on a space
mission, which allows for better characterization. The dominant contribution
to a ground-based PSF is from atmospheric turbulence, which varies rapidly as
a function of time and field position. This is eliminated in space. Moreover,
contributions to the optical distortions from temperature variations and
gravity loading can be reduced or (in the latter case) eliminated, particularly at the L2 Lagrange
point, in a high Earth orbit during periods where shadow is avoided, and/or
by using temperature-controlled optics. The three dominant
contributions to PSF ellipticity on a space mission are (i) astigmatism,
which causes the ellipticity of the PSF to vary with focus position; (ii)
coma from misaligned optics, which at second order leads to ellipticity; and
(iii) anisotropic pointing jitter. Of these, (i) and (ii) are functions of
mirror positions, whose time and field position dependence are controlled
by a small number of parameters. The pointing jitter is the least stable --
it may be different in every exposure --- but it has a controlled position
dependence, no color dependence (at least with all-reflective optics),
and can be monitored with the same fine guidance sensors used to point
the telescope. Therefore a space mission offers the possibility of a PSF
whose entire structure is determined by a small number of parameters that
can be tracked as a function of time \citep{ma08b}. This means it provides
the best possibility of providing accurate PSF knowledge at every point in
every exposure. The diffraction PSF has the unfortunate feature of having
a size that is highly color-dependent ($\propto \lambda/D$), and in the
presence of aberrations the ellipticity is color-dependent as well.  However,
in contrast to ground-based observations, the color dependence is controlled by
the same wavefront error that determines the PSF morphology.

As already noted, optimal photo-$z$ performance across the entire relevant
range of redshifts can be obtained only with continuous coverage from
blueward of the 4000 \AA\ break (at $z=0$) through the near-IR. In particular
the Balmer/4000 \AA\ feature is always detected except at very high redshift
($z>3$), which reduces the number of objects with no breaks identified and
provides cleaner separation of the \lya\ versus Balmer/4000 \AA\
breaks. Collecting photometric data points in the bluer bands (starting at the
$\sim 3200$ \AA\ atmospheric cutoff) is quite reasonable from the ground, and in
this area there is no major advantage to a space mission. However, as we move
to the red the space mission begins to look much more attractive. 
From the ground, the near-IR sky brightness (relevant for broadband imaging) 
is dominated by
the decay of OH radicals, which are produced in vibrationally excited states at
$\sim 90$ km altitude in the Earth's upper atmosphere \citep{leinert98}. The
typical sky brightness rises from 18.5 mag AB arcsec$^{-2}$ in the $Z$
band through 15.4 mag AB arcsec$^{-2}$ in the $H$ band.\footnote{See the
WFCAM website, {\tt http://casu.ast.cam.ac.uk/surveys-projects/wfcam}, and
beware of Vega to AB conversions, which are significant in the near-IR.} 
In space
in the 1--2 $\mu$m region the dominant background is instead scattering of
sunlight off of interplanetary dust particles (the ``zodiacal light''). The
typical brightness is $\sim 23$ mag AB arcsec$^{-2}$ near the ecliptic poles
and 21.5 mag AB arcsec$^{-2}$ in the ecliptic plane \citep{leinert98}. Thus
in the $H$ band the sky brightness is a factor of 300--1000 lower in space,
which means that a space telescope with even $\sim 1$ m$^2$ collecting area
would outperform the best ground-based telescopes in terms of near-IR imaging
survey speed. Note also that because of the altitude of the OH emitting layer,
airplane or balloon based platforms cannot access the low background available
in space.

\subsection{Prospects}
\label{sec:wl_prospects}

The next several years promise to be very exciting for weak lensing as
we enter the Stage III era. Two major wide-field ground-based imagers are
coming online in the 2012/13 timeframe: the Dark Energy Camera\footnote{\tt
http://www.darkenergysurvey.org/} (DECam) at CTIO in the Southern Hemisphere,
and the Hyper Suprime Cam (HSC) on Subaru in the Northern Hemisphere
\citep{miyazaki06}. These will provide great leaps in \'etendue, roughly
35 m$^2$ deg$^2$ for DECam and 70 m$^2$ deg$^2$ for HSC (versus 8 m$^2$
deg$^2$ for CFHT/MegaCam). The Dark Energy Survey (DES; using DECam) plans
to observe 5000 deg$^2$ in the $grizy$ bands over five years to $\sim 24$th
magnitude (10$\sigma$ $r$ band AB, shallower in $y$). The HSC plans a somewhat
deeper and narrower survey, also in $grizy$ (2000 deg$^2$, 25th magnitude
$10\sigma$). These projects together will measure the shapes of roughly 300
million galaxies and provide accurate photometric redshifts out to $z\sim
1.3$; this represents a $1\frac12$ order of magnitude increase relative
to current data sets. We expect that the use of several revisits and shape
measurements in multiple bands, as well as incorporating the lessons from
Stage II WL projects such as the CFHTLS and SDSS, will provide additional
control over systematic errors in shape measurement. With careful attention
to the source redshift distribution as well, and the photo-$z$ capability
provided by $y$-band imaging, the Stage III cosmic shear projects (DES and
HSC) should reach the 1\%\ level of precision on the amplitude $\sigma_8$,
as well as providing high-$S/N$ measurements of its increase as a function
of cosmic time. If the stochasticity issue turns out to be tractable,
a similar level of precision will be reached by using galaxy-galaxy lensing
to constrain the bias of galaxies and infer $\sigma_8$ indirectly
from galaxy clustering.

The Stage III projects will also mark the completion of the research program of
extrapolating the amplitude measured from the CMB forward in time and comparing
it to the value of $\sigma_8$ measured via WL, and using the agreement of the
amplitudes to measure $w(z)$ or test GR. There is a fundamental limitation to
this type of comparison coming from reionization: while Planck will measure the
CMB power spectrum to very high accuracy, one needs the optical depth $\tau$
to convert this into a normalization of the initial perturbations. This seems
unlikely to be measured from the CMB $E$-mode to significantly better than 0.01
due to cosmic variance, foregrounds, and modeling uncertainties
\citep{holder03,mortonson08,colombo09}.\footnote{In
the more distant future, 21 cm measurements may improve our understanding
of reionization to the point where this limitation is removed; however such
an advanced understanding is not anticipated in the immediate future.}

The completion of the DES and HSC will not, however, mark the end of the road
for cosmic shear. Because of the reionization degeneracy, the next step will
be to make highly accurate measurements of the {\em shape} of the signal
(dependence on scale and redshift) rather than its amplitude. This is a
scientific matter of critical importance: if DES/HSC find a convincing
deviation from the expected amplitude of low-redshift structure, one does not
know whether this reflects a breakdown of GR at late times (a phenomenon that
might be linked to cosmic acceleration) or something that happened to alter the
growth of structure between $z\sim 10^3$ and $z\sim$a few 
(such as massive neutrinos, though early dark energy would also be
a possible explanation).  What is needed next is the survey that
measures the rate at which the growth of structure is suppressed internally
to the low-redshift data.  In our \S\ref{sec:forecast} forecasts we
describe deviations from the GR-predicted growth rate using the
parameter $\Delta\gamma$ (see eqs.~\ref{eqn:dlng.dlna} and~\ref{eqn:fullfz}),
though other choices are possible.  Even the Stage III surveys may
make only preliminary measurements in this direction. 
\cite{albrecht09} estimated that DES could measure $\Delta\gamma$ to
a $1\sigma$ accuracy of only 0.2 using the evolution of the WL signal,
and our fiducial Stage III forecast in \S\ref{sec:results} yields
$\sigma_{\Delta\gamma} = 0.148$ (see Table~\ref{tbl:forecasts2}).  
Clusters calibrated
by stacked weak lensing might enable a significantly tighter constraint
(\S\ref{sec:forecast_cl}, Fig.~\ref{fig:contours_cl}),
and redshift-space distortions could also enable good
measurements of $\Delta\gamma$ (\S\ref{sec:rsd}).
It is not clear, however, that any method
will achieve percent-level measurements of the rate of low redshift
structure growth in the 2010 decade.  Reaching this goal is one
of the major drivers for Stage IV projects using WL and other probes
of structure growth.  It requires
highly accurate, low-systematics shape measurements, of galaxies across a
wide range of redshifts, including $z>1$ where the angular radii of galaxies
are small and the shape measurement challenges are immense.

Fortunately, the Stage IV WL experiments are already being planned, although
their first light is not expected until $\ge2020$. There are several
approaches. One is the Large Synoptic Survey Telescope (LSST), which would
feature a giant-\'etendue (290 m$^2$ deg$^2$) telescope dedicated to optical
surveys of the Southern Hemisphere. Over a ten-year operating period, LSST
would acquire hundreds of images of every point on the sky, which should
go a long way toward identifying and removing any residual sources of
PSF systematics. The incorporation of 6 bands ($ugrizy$) will likely lead
to the best photometric redshifts practical from the ground over such a
wide area. LSST will survey the entire extragalactic sky available from the
south, perhaps 12,000--15,000 deg$^2$. The usable density of source galaxies,
particularly at high redshift, is not certain as it depends on both advances
in measuring galaxies small compared to the PSF and the quality of photometric
redshifts in the notorious $1.3<z<2$ range. However, by achieving high S/N
on almost every resolved galaxy, LSST is likely to represent the ``ultimate
experiment'' for ground-based optical weak lensing.

An alternative approach is to exploit the small and stable PSF and availability
of the near-IR bands from space, as planned for ESA's \euclid\ mission 
(scheduled 
launch in 2020) and the NASA \wfirst\ mission (launch date to be determined;
see below).
\euclid\ will be a 1.2 m telescope with a 0.5 deg$^2$ focal plane
that will survey 15,000 deg$^2$ in a parallel WL+BAO mode, with shape
measurements performed in a broad red band (0.55--0.92 $\mu$m).
\euclid\ will have only 3--4 observations of
each galaxy, but this is predicted to be acceptable given the much greater
stability of {\it Euclid}'s PSF relative to anything possible on the ground. At
$\sim 30$ galaxies per arcmin$^2$, \euclid\ would measure shapes for $\sim 1.6$
billion galaxies.
\euclid\ will also obtain near-IR photometry in three bands,
which will be combined with ground-based optical photometry
(from LSST where available) for photometric redshifts;
the IR imaging is underresolved and will not be used for shape measurements.
\wfirst, in the ``DRM1'' configuration described by \citet{green12},
would be a $1.3\,$m, unobstructed (i.e., off-axis secondary) 
infrared space telescope capable of surveying 1400 deg$^2$/yr in 
a combined WL+BAO mode (4-band imaging and slitless spectroscopy
with resolution $\lambda/\Delta\lambda \approx 600$).
The baseline WL program has 5--8 exposures in each of three
shape measurement filters (J, H, and K), with an effective
source density $n_{\rm eff} = 40\,$arcmin$^{-2}$,
and in a shorter wavelength
filter (Y) that provides additional information for photometric redshifts.
(LSST or other ground-based data are again required to provide
optical photometry.)
Multiple bands provide control of color-dependence of the PSF,
and the degree of data redundancy is much higher than in \euclid\
because of the larger number of exposures and the ability to 
correlate shape measurements in different bands --- the WL signal
should be achromatic, but many systematics would not be.
However, this greater redundancy, and the fact that the telescope
is shared with other science programs, comes at the expense of
what will likely be a smaller survey.  The \cite{green12} design
reference mission calls for 2.4 years of high-latitude imaging
and spectroscopy (out of a 5-year mission lifetime), which is
sufficient to cover 3400 deg$^2$.\footnote{An additional 0.45
years would be devoted to an imaging and spectroscopic survey
for supernovae.} 
As mentioned in \S\ref{sec:forward}, the transfer of two 2.4-m
on-axis space telescopes from the U.S. National Reconnaissance Office (NRO)
to NASA opens an alternative route to \wfirst, with initial
ideas described by \cite{dressler12}.  While this implementation
may not increase the survey area\footnote{In wide-field 
ground-based surveys, an increase in
telescope aperture, e.g. 2 m $\rightarrow$ 4 m, increases the \'etendue,
resulting in a faster survey at the same seeing-limited resolution. For
space-based surveys, the natural choice when receiving a larger telescope
is to maintain the same sampling of the PSF (and hence the same $f$-ratio
if the detector properties remain fixed), which results in 
each pixel subtending a smaller number of arcseconds. The \'etendue
and hence survey speed to reach the same extended-source sensitivity are
unchanged if the pixel count is held fixed, but the 
angular resolution is improved as $\sim\lambda/D$. This is of course an
enormous advantage for weak lensing.}, the superior angular resolution and 
light-gathering power of this hardware
make it the only plausible option (at least in the optical-NIR bands) to reach
source galaxy densities of $\sim 70$ galaxies/arcmin$^2$ over thousands
of $\mdeg^2$.  A detailed study of a 2.4-m implementation of \wfirst\ is
ongoing, with a report planned for April 2013.

By the end of the 2020s, we should have a rich data set from all three of these
projects (LSST, \euclid, and \wfirst) --- 
and perhaps also from a large-scale radio
interferometer such as the SKA. 
These surveys represent very different approaches to
the Stage IV WL problem and will provide for multiple cross-checks of final
results and internal cross-correlations of different data sets. The total
number of galaxies with accurately measured shapes will probably reach $\sim
4$ billion, with most observed by at least two instruments and some with all
three. Robust measurements of the suppression of the growth of structure to
$\sigma_{\Delta\gamma} \approx 0.03$ --- 
a factor of several better
than Stage III --- should then be possible (see Table~\ref{tbl:forecasts1}), 
as well as tests of other possible
deviations from $\Lambda$CDM that we have not yet imagined. But a great deal
of work will be necessary before then to ensure that the systematic errors
are controlled at this level.

\begin{figure} [t]
\begin{center}
{\includegraphics[width=3.2in]{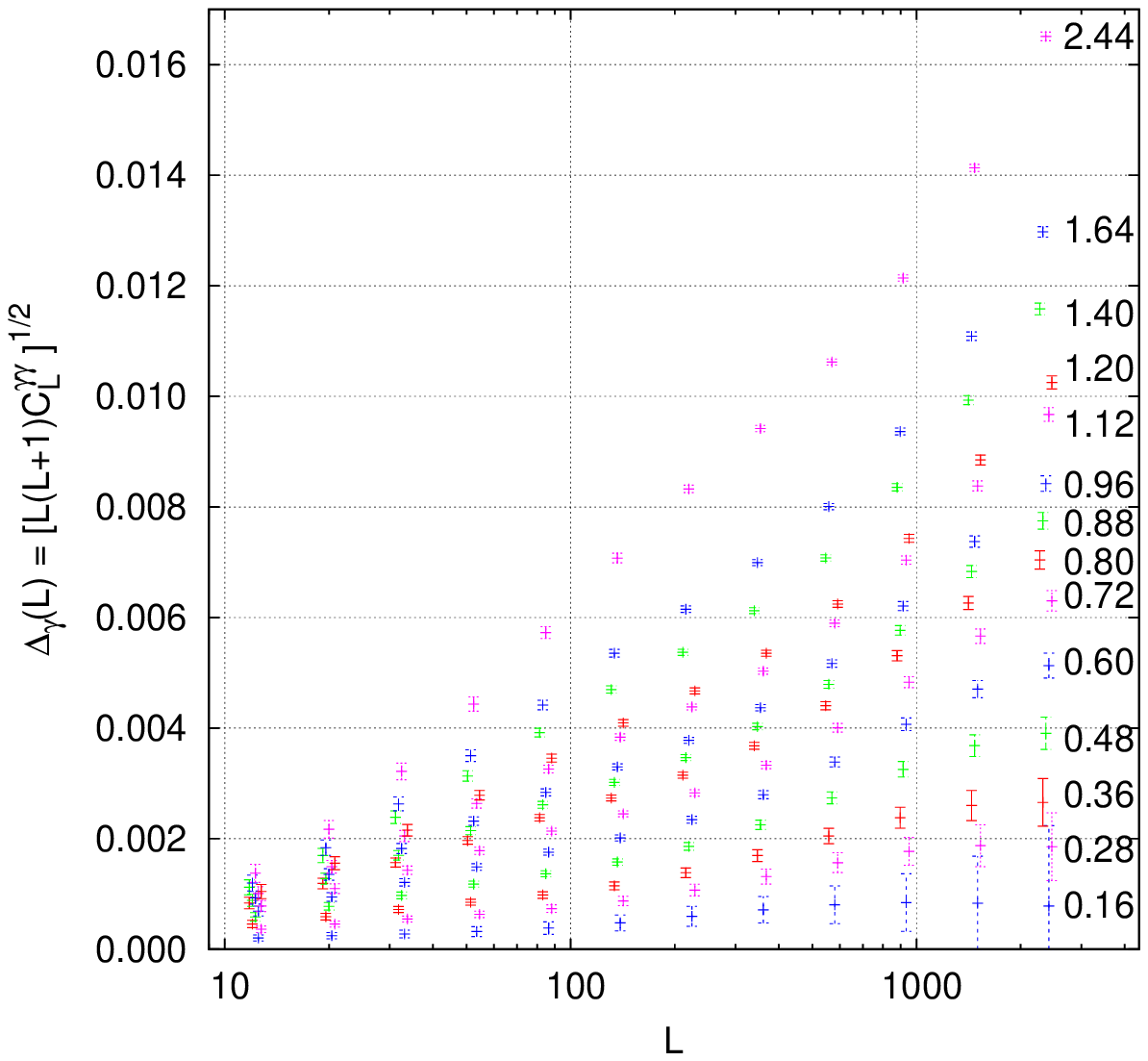}
\includegraphics[width=3.2in]{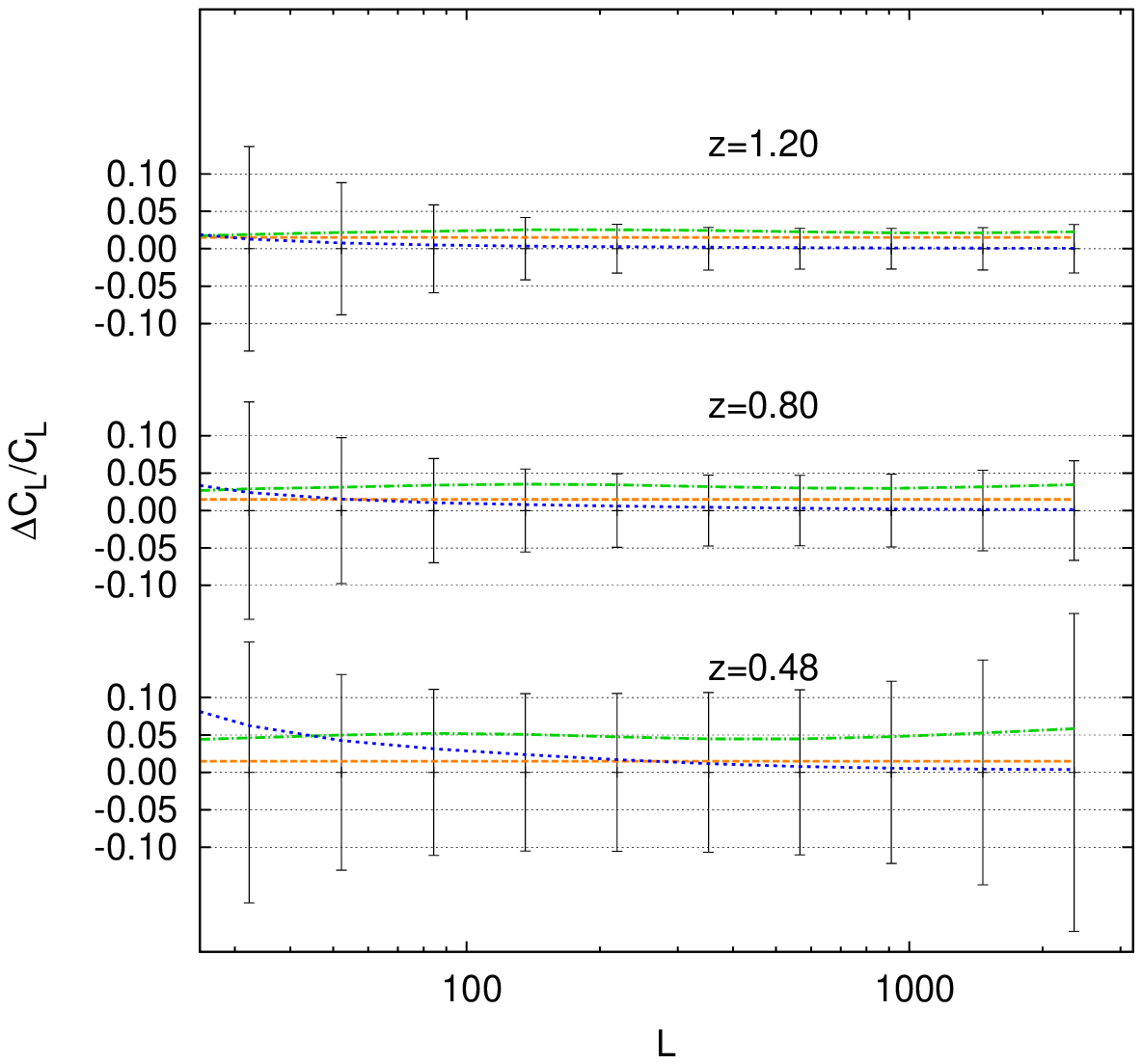}
}
\caption{(Left) The predicted cosmic shear power spectrum and statistical 
errors in each of 14 photo-$z$ bins assuming a \lcdm\ cosmological
model with the parameters of \cite{albrecht09} and the
survey parameters of our fiducial Stage IV WL program.
(Right) Impact of systematic errors relative to statistical errors.
For three of the photo-$z$ bins from the left panel, error bars show 
the $\pm 1\sigma$ statistical errors 
(in bins of width $\Delta\log l = 0.2\,$dex),
with vertical offsets between bins for clarity.
Solid, dashed, and dotted curves show, respectively, the effect
of a multiplicative 
shear calibration bias of $2\times 10^{-3}\times\sqrt{14}$ (orange),
a mean photo-$z$ offset of $2\times 10^{-3}\times\sqrt{14}$ (green), 
or an additive
shear bias of $3\times10^{-4}\times\sqrt{14}$ (blue) {\em per $z$-bin}.
(The $\sqrt{14}$ is inserted here since an actual survey would combine all 14 bins.)
The power in the additive shear bias was equally distributed in $\ln l$
for the purposes of this plot.
}
\label{fig:wlpkerrors}
\end{center} \end{figure}

For our forecasts in \S\ref{sec:forecast} we adopt a fiducial
Stage IV WL program that assumes an effective source density 
$n_{\rm eff} = 23$ arcmin$^{-2}$ over $10^4\,$deg$^2$, for
a total of $8.3\times 10^8$ shape measurements in 14 bins
of photometric redshift (see \S\ref{sec:fiducial} for details).
We incorporate (and marginalize over) a multiplicative shear
calibration uncertainty of $2\times 10^{-3}$ and a mean photo-$z$
uncertainty of $2\times 10^{-3}$; these are aggregate values,
and are larger by $\sqrt{14}$ in each photo-$z$ bin.
LSST and \euclid\ both anticipate a larger number of shape
measurements and thus smaller statistical errors than our
fiducial program.  The baseline \wfirst\ DRM1 survey has a factor
of two fewer shape measurements, but the mission's technical
requirement for shape systematics is a factor of two better.
Thus, our fiducial program is conservative relative to the
stated goals of all three experiments, though highly ambitious
relative to the current state of the field.

For this fiducial Stage IV program,
Figure~\ref{fig:wlpkerrors} shows the predicted shear power spectrum
and $1\sigma$ statistical errors, in each of the 14 photo-$z$ bins.
In addition to these auto-spectra, the data allow measurements
of $N_{\rm bin}(N_{\rm bin}-1)$ cross-spectra among the bins,
providing additional statistical power and tests for intrinsic
alignment and other systematics.
In a given photo-$z$ bin, the errors in different $l$ bins are
independent.  However, the errors from one photo-$z$ bin to another
are correlated because the same foreground structure can lens galaxies in
multiple background redshift shells.
We compare the statistical errors to the impact of cosmological
parameter changes in \S\ref{sec:forecast_multiprobe}.
The aggregate statistical precision on the overall amplitude of the WL
power spectrum, i.e., on a constant multiplicative factor applied to 
the auto- and cross-correlations in all photo-$z$ bins, is $\approx 0.21\%$.
The corresponding error on $\sigma_8$, treated as a single parameter
change, is about three times smaller because the power spectrum
scales as $\sigma_8^3$ in the regime where it is best measured.
The right panel compares the statistical errors in four representative
photo-$z$ bins to the effects of a multiplicative shear calibration
bias of $2\times 10^{-3}$, a mean photo-$z$ bias of $2\times 10^{-3}$,
or an additive shear bias of $3\times 10^{-4}$.
We see that systematic errors of this magnitude would be
smaller than the statistical errors in an individual photo-$z$
bin, but the overall impact would be larger than the aggregate
statistical errors. Thus, for our fiducial assumptions the Stage IV program is
systematics limited rather than statistics limited, but not by an enormous 
factor.  Even though our assumptions for this fiducial program are
arguably conservative, it would achieve powerful constraints
on the cosmic expansion history and the history of structure growth,
as discussed in \S\ref{sec:forecast}.

Is there a future for WL beyond Stage IV, both in terms of science
motivation and technical capability? It seems unlikely that there would be
a follow-on experiment that consists of simply a super-size LSST, \euclid,
or \wfirst, particularly given that these experiments will come within a
factor of a few of the cosmic variance limit at several tens of galaxies
per arcmin$^2$. Rather the more distant future would have to involve new
technology and a new science case not subject to the usual limitations. An
example might be to look for lensing by primordial gravitational waves, which
is not practical using galaxies as sources \citep{dodelsongw03} but is at least
in principle possible using highly-redshifted 21 cm radiation as the source,
even for tensor-to-scalar ratios as low as $10^{-9}$ \citep{book11}. But we
have now entered the speculative realm of post-2030 science and technology,
where our ability to forecast the future is of limited reliability. We
thus conclude our discussion of weak lensing here.

\vfill\eject

\section{Clusters of Galaxies} \label{sec:cl}

\subsection{General Principles} \label{sec:cl_method}

Galaxy clusters have a long and storied history as cosmological probes.  They
provided the first line of evidence for the existence of dark matter
\citep{zwicky33,smith36}, and cluster mass-to-light ratio measurements 
suggested that the matter density in the universe
was sub-critical ($\Omega_m < 1$) as far back as the early 1970's
\citep[see][and references therein]{gott74}.
The evidence for low $\Omega_m$ was substantially strengthened
by baryon fraction measurements \citep{white93b,evrard97}, and by the
discovery of massive clusters at high $(z\approx 0.8)$ redshift
\citep[e.g.,][]{henry97,eke98,donahue98}.  Today, clusters remain an important
cosmological tool, capable of testing cosmology in a variety of ways.  Here
we focus on cluster abundances as a tool for 
constraining the growth of structure in the matter distribution.
Tight geometrical constraints from BAO and supernovae
in turn yield tight predictions for structure growth assuming
GR to be correct.  Deviations from these predictions, revealed
by weak lensing or by clusters, would constitute direct evidence
for modified gravity as the driver of accelerated expansion.
The excellent review by \cite{allenetal11} discusses other 
cosmological applications of clusters and examines recent cluster
abundance results in detail (see also the earlier review by \citealt{voit05});
we summarize recent work in 
\S\ref{sec:cl_current} but devote most of our attention to methods
for Stage III and Stage IV cluster surveys.
Other recent reviews in the field include
\cite{kravtsov12}, who review the physics of cluster formation with
emphasis on the insights gained from hydrodynamic cosmological
simulations, and \cite{kneib12}, who review strong and weak 
lensing by clusters.

The basic idea of cluster abundance studies is to compare the
predicted space density of massive halos (Figure~\ref{fig:structure})
to the observed space density of clusters, which can be
identified via optical, X-ray, or CMB observables that should
correlate with halo mass.  In optical searches, the basic observable
is the richness, the number of galaxies in a specified luminosity
and color range within a fiducial radius (typically taken to be
the estimated virial radius of the halo).  In X-ray searches, 
the luminosity $L_X$, temperature $T_X$, and inferred gas mass $M_{\rm gas}$ 
all provide observable indicators of halo mass.  In CMB searches, clusters
can be characterized by the central or integrated value of the
flux decrement $Y_{\rm SZ}$ produced by the 
Sunyaev-Zel'dovich (1970; hereafter SZ)
effect: Compton up-scattering of CMB photons by hot electrons
in the intracluster medium.  The product $Y_X = T_X M_{\rm gas}$
defines an X-ray observable that should scale with $Y_{\rm SZ}$,
and numerical simulations predict that $Y_X$ tracks
halo mass more closely than temperature or gas mass alone
\citep{kravtsov06}.

The first applications of this approach were made by
\citet{peeblesetal89} and \citet{evrard89}, who 
used observed cluster abundances to argue against 
an $\Omega_m=1$ CDM cosmological model 
(see also \citealt{kaiser86b,kaiser91}, who compared the
observed evolution of X-ray clusters to predictions of a
self-similar model with $\om=1$).
Halo abundance is sensitive to the amplitude of the matter
power-spectrum $\sigma_8$ and the matter density $\Omega_m$.  
The mean matter content in a sphere of
comoving radius $8\hmpc$ is $\approx 2\times 10^{14}\ \msun$.  Thus, 
cluster-mass halos form from
the gravitational collapse of fluctuations on 
about this scale, and their abundance 
naturally tracks $\sigma_8$.  Moreover,
because the total mass of each collapsed volume scales linearly with
$\Omega_m$, the number of halos at a given mass
can be raised either by raising $\sigma_8$, so that fluctuations are larger, 
or by raising $\Omega_m$, so that the mass associated with each perturbation is larger.
The quantity most tightly constrained by cluster abundances is 
a combination of the form $\sigma_8\Omega_m^q$, 
with $q \approx 0.4$ \citep{white93}.  
The degeneracy between $\sigma_8$ and $\Omega_m$ can be broken
by measuring abundances at a variety of masses.  This argument also holds at
higher redshift, so one can think of cluster abundances as 
primarily constraining
$\sigma_8(z)\Omega_m^q$, modulated by the additional cosmological
dependence of the volume element $dV_c(z) \propto D_A^2H^{-1} d\Omega dz$,
and by any intrinsic dependence of cluster observables on the 
distance--redshift relations.
Note that, as elsewhere in this article, $\Omega_m$ always refers to the $z=0$
value unless $\Omega_m(z)$ is written explicitly.

\begin{figure} [t]
\begin{center}
\includegraphics[width=2.8in]{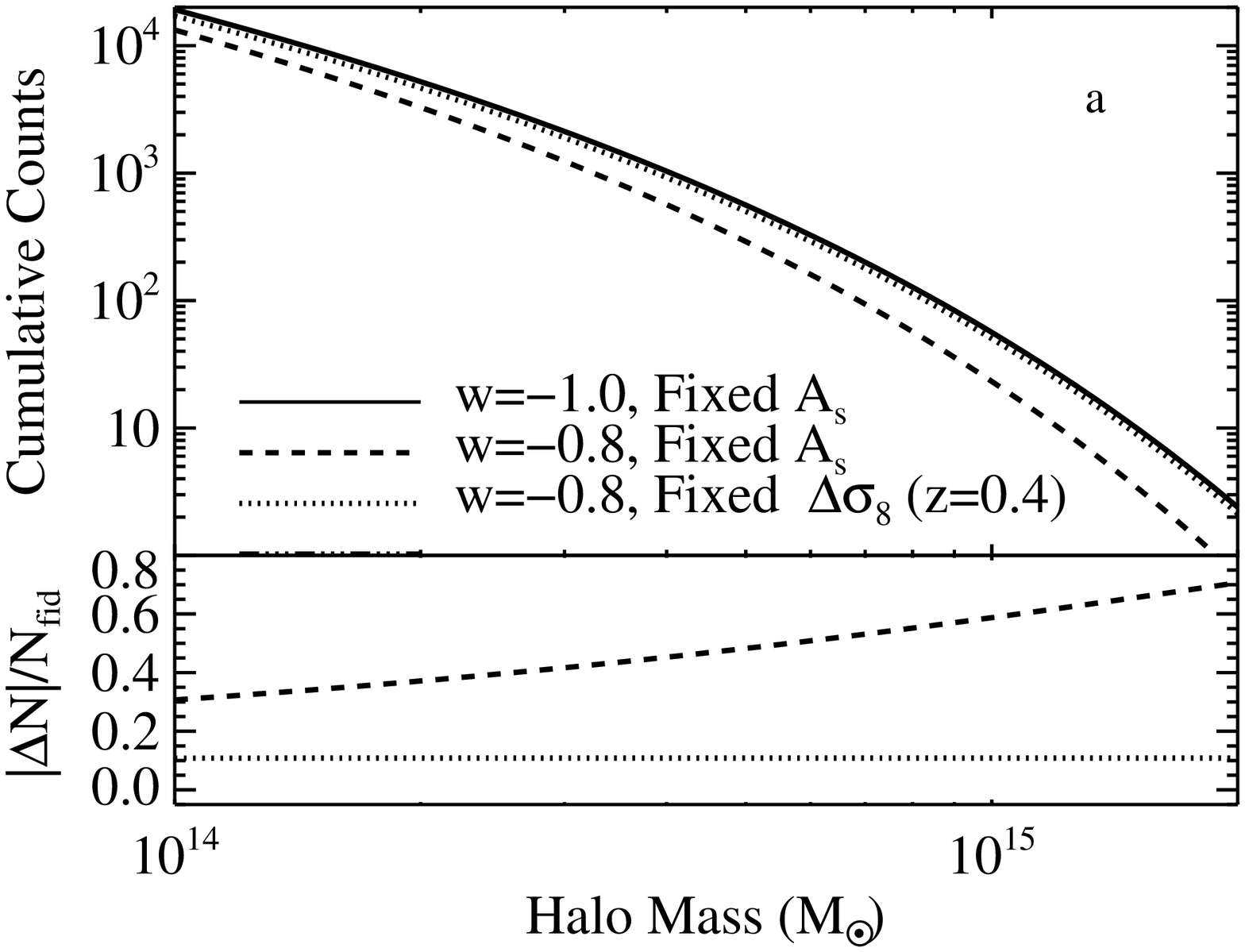}
\includegraphics[width=2.8in]{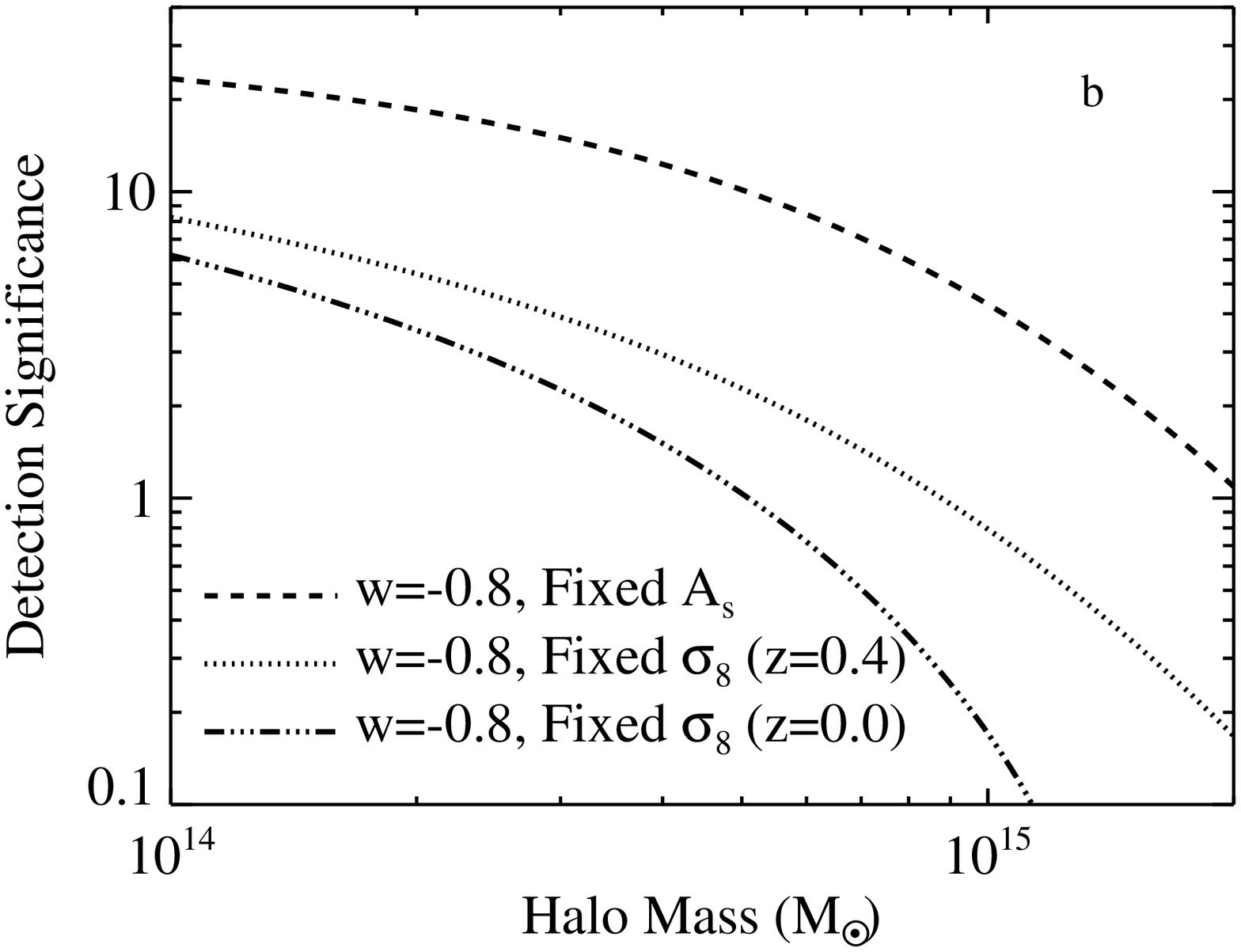} 
\caption{(a) Cumulative halo counts as a function of limiting mass for a 
$10^4\mdeg^2$
survey in a redshift slice $z=0.4\pm 0.05$.  The solid line shows the
fiducial model from Table~\ref{tbl:models}.  The dashed line corresponds to
$w=-0.8$ with the amplitude of the primordial matter power spectrum held fixed.
The dotted line has $w=-0.8$, but holds $\sigma_8(z=0.4)$ fixed.  Residuals
relative to the fiducial model are shown in the bottom panel.  The small,
nearly constant offset of the dotted line is sourced by the dark energy
dependence of the comoving volume element $dV_C$.  (b)
The significance with which this hypothetical halo sample could
distinguish the fiducial model from the alternatives in panel (a)
as a function of mass threshold, using the statistical error
of equation~(\ref{eq:nerr}).  
The dot-dashed line shows an additional model in which
$\sigma_8(z=0)$ is held fixed.
Even though the high mass end of the halo mass
function depends most strongly on cosmology, the statistical power of the
cluster abundances is dominated by the low mass end because of the
much lower measurement errors.
}
\label{fig:halomf}
\end{center} \end{figure}

We illustrate these ideas in Figure~\ref{fig:halomf}.  Panel (a) shows the
expected halo abundance as a function of the limiting mass in a redshift 
slice $z=0.35-0.45$ subtending $10^4\ \deg^2$.  
Plots for other redshift slices are qualitatively similar.  
For this plot, and throughout the
rest of this section unless otherwise noted, halo mass refers to the
mass enclosed within a sphere whose mean interior overdensity is
$\Delta=200$ relative to the mean matter density of the universe.  
The solid line is the abundance in our fiducial model (see
Table~\ref{tbl:models}), while the dashed line shows the corresponding halo
abundance when setting $w=-0.8$ and holding $\Omega_m$ and 
the primordial power spectrum amplitude $A_s(k=0.002\,{\rm Mpc}^{-1}$) fixed.
Unlike in Figure~\ref{fig:structure}, this choice does not leave the
CMB observables fixed, but it better illustrates the intrinsic sensitivity of 
cluster
abundances.  
For $w=-0.8$, dark energy becomes dynamically important earlier
than for $w=-1$, suppressing growth and lowering $\sigma_8(z=0.4)$
from 0.66 to 0.62.  This sharply reduces the halo
abundance, by $\approx 30\%$ at a threshold of $10^{14} M_\odot$ and by $\approx60\%$
at $10^{15} M_\odot$.  If we raise $A_s$ 
so as to hold $\sigma_8(z=0.4)$ fixed, then 
the $w=-1$ and $w=-0.8$ models differ by a nearly constant
factor of 1.1, which is the ratio of the comoving volumes of the redshift
slices in the two cases.  This volume effect is clearly weaker than 
the overall scaling of halo abundances with $\sigma_8$.

%%%%%%%%%%%%%%%%%%%%%%%%%%%%%%%%%%%%%%%%%%%%%%%%
%%%%%%%%%%%%%%%%%%%%%%%%%%%%%%%%%%%%%%%%%%%%%%%%

%%%%%%%%%%%%%%%%%%%%%%%%%%%%%%%%%%%%%%%%%%%%%%%%
%%%%%%%%%%%%%%%%%%%%%%%%%%%%%%%%%%%%%%%%%%%%%%%%

While the mean halo abundance becomes more sensitive to $\sigEightz$
at higher masses, 
the statistical precision with which one can measure $\sigEightz$
{\it decreases} with increasing mass because of the larger Poisson
fluctuations for rarer clusters.  
This point is illustrated in Figure~\ref{fig:halomf}b,
which shows the statistical significance at which a $10^4\mdeg^2$,
$z=0.35-0.45$ cluster survey would distinguish the models
shown in panel (a).
For reference, we also show the case in which $\sigEight$ is held fixed
at $z=0$, which reduces model differences because the growth
and volume element effects act in opposite directions.
We discuss statistical errors in cluster abundances, including the role of sample variance,
in \S\ref{sec:cl_numbers}.  The key conclusion from
Figure~\ref{fig:halomf}b is that lower mass clusters
allow stronger model discrimination.

Cluster cosmology requires making 
an explicit
link between the theoretically predicted
population of halos as a function of mass and an observed
population of clusters.  This problem
is complicated by the fact that the halo population is usually  
characterized
using dark matter simulations, whereas clusters are identified using
baryonically-sourced signatures such as the presence of galaxy overdensities,
extended X-ray emission, or SZ decrements (see \S\ref{sec:cl_finding}).
The lower mass limit probed by cluster abundance experiments is partly set
by the detection thresholds intrinsic to each method, but also by the
difficulty of characterizing the relation between low mass halos and poor
clusters.  Different researchers adopt varying definitions of halos and of
clusters.  Within a reasonable range, such variation is acceptable, provided
each study is self-consistent and the halo--cluster relation is accurately
characterized.  In recent years, numerical studies have mostly
shifted from the friends-of-friends algorithm used in earlier
work (e.g., \citealt{efstathiou88}) to spherical overdensity
definitions (e.g., \citealt{tinker08}), thus avoiding the tendency
of the friends-of-friends method to occasionally link distinct
mass concentrations via narrow bridges \citep[see][and references therein
for a more detailed discussion]{more11}.
Halo boundaries are typically drawn at overdensities
$\Delta\approx100-500$, where clusters are in approximate dynamical 
equilibrium and where mass predictions are fairly robust to baryonic 
physics.  
The overdensity $\Delta$ can be quoted relative to the mean matter density of
the universe at the cluster redshift or relative to the critical density at
that redshift.
In this section, we will adopt $\Delta = 200$ with respect
to the mean density as our definition unless otherwise specified.

\begin{figure} [t]
\begin{center}
\includegraphics[width=3.2in]{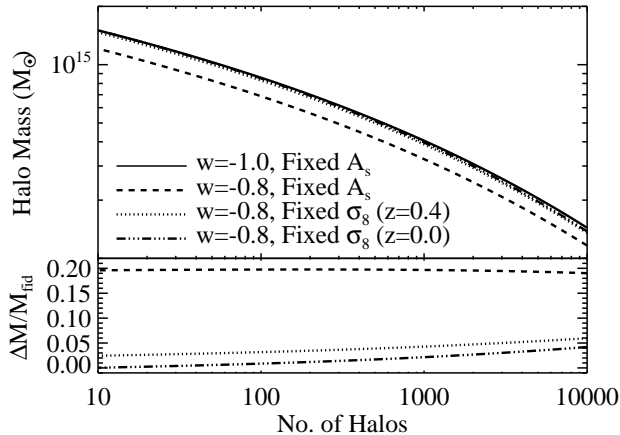}
\caption{Halo mass thresholds as a function of cumulative number counts, i.e.
flipping the $x$ and $y$ axes of Figure \ref{fig:halomf}a.  The
$x$-axis shows the number of halos predicted in a $10^4\mdeg^2$ survey in a
redshift slice $z=0.4\pm 0.05$.  The lower panel shows the fractional change
in mass threshold relative to the fiducial cosmological model.  
}
\label{fig:flip_halomf} \end{center} \end{figure}

The principal challenge to precision cosmology with clusters is not
cluster identification {\it per se},
but the accurate calibration of the relation
between cluster observables (e.g., richness, X-ray luminosity, SZ decrement)
and halo masses.
Figure~\ref{fig:flip_halomf} illustrates this point by
flipping the $x$ and $y$ axes of panel (a) in Figure~\ref{fig:halomf}, 
thus plotting 
the mass threshold at fixed cluster abundance for the different cosmological
models. Changing from $w=-1$ to
$w=-0.8$ while holding $A_s$ fixed
changed the predicted abundances by $30-60\%$, but the corresponding change in
mass threshold is only about 20\%.  For fixed $\sigma_8(z=0.4)$, the 15\%
change in abundance corresponds to a $2.5\%-6\%$ change in mass threshold.
These, then, are the levels of accuracy in mass calibration that must be
attained to distinguish between the two $w=-0.8$ models and our fiducial
$w=-1$ model.  The issue of mass calibration will arise repeatedly
in this section, especially in 
\S\ref{sec:cl_mass_calibration} and~\S\ref{sec:cl_core_calibration}.

%%%%%%%%%%%%%%%%%%%%%%%%%%%%%%%%%%%%%%%%%%%%%%%%
%%%%%%%%%%%%%%%%%%%%%%%%%%%%%%%%%%%%%%%%%%%%%%%%

%%%%%%%%%%%%%%%%%%%%%%%%%%%%%%%%%%%%%%%%%%%%%%%%
%%%%%%%%%%%%%%%%%%%%%%%%%%%%%%%%%%%%%%%%%%%%%%%%

In principle, cluster abundances are sensitive to $\sigEightz$, $\Omega_m$,
and the comoving volume element $dV_C$, as well as any inherent sensitivity
of the relation between cluster mass and cluster observables on the 
distance--redshift relations.
To simplify our discussion,
we will usually assume that a combination of other data
sets (CMB, SN, BAO, WL, etc.)
will determine both $\Omega_m$ and $dV_C(z)$ at higher
precision than that achievable from cluster abundances.  
Consequently, we will focus on
the sensitivity of cluster abundances to $\sigEightz$
while holding $\Omega_m$, $dV_C(z)$, and the angular and luminosity
distances fixed.
In practice, we expect our assumption should be a good one as far as
the comoving volume element and the distances are concerned.  
However, the sensitivity attainable with clusters is high
enough that holding $\Omega_m$ fixed may be incorrect in detail.  
We will discuss this point in \S\ref{sec:cl_prospects} and 
again in \S\ref{sec:forecast_cl}.

Many cluster cosmology papers quote masses in $h^{-1}\msun$ because
observational mass estimates (and, to some extent, theoretical 
predictions) scale inversely with $h$.
However, at non-zero redshift many other parameters also come into
play, and $h$ is itself one of the parameters constrained by
dark energy experiments.
Thus, we have elected to quote masses in $\msun$
rather than $h^{-1}\msun$.  In a similar vein, we will switch
most of our subsequent
discussion from $\sigEight$ to $\sigElev$, the rms fluctuation on a scale
of $R=11\,\Mpc$ (equal to $\sigEight$ for $h=0.727$).
For some observables (e.g., the X-ray estimated gas mass $M_{\rm gas}$,
the inferred cluster mass is sensitive to the angular diameter
distance $D_A(z)$ and this dependence itself provides useful
cosmological constraints; this point is discussed by
\cite{allenetal11} but we will not address it further here.
Our primary focus is
the statistical precision with which cluster abundances constrain
$\sigElevz$, and the level at which systematic uncertainties must be
controlled to achieve these statistical limits.  
In \S\ref{sec:cl_prospects} we compare the precision potentially
attainable with clusters to forecasts (described in \S\ref{sec:forecast})
from fiducial Stage III and Stage IV CMB+SN+BAO+WL programs.

%%%%%%%%%%%%%%%%%%%%%%%%%%%%%%%%%%%%%%%%%%%%%%%%
%%%%%%%%%%%%%%%%%%%%%%%%%%%%%%%%%%%%%%%%%%%%%%%%
%%%%%%%%%%%%%%%%%%%%%%%%%%%%%%%%%%%%%%%%%%%%%%%%
%%%%%%%%%%%%%%%%%%%%%%%%%%%%%%%%%%%%%%%%%%%%%%%%
%%%%%%%%%%%%%%%%%%%%%%%%%%%%%%%%%%%%%%%%%%%%%%%%
%%%%%%%%%%%%%%%%%%%%%%%%%%%%%%%%%%%%%%%%%%%%%%%%
%%%%%%%%%%%%%%%%%%%%%%%%%%%%%%%%%%%%%%%%%%%%%%%%
%%%%%%%%%%%%%%%%%%%%%%%%%%%%%%%%%%%%%%%%%%%%%%%%

\subsection{The Current State of Play} \label{sec:cl_current}

Most cluster cosmology studies of the past decade have been based on X-ray
catalogs, with typical cluster samples numbering in the several tens to few
hundreds of clusters.  The vast majority of these catalogs rely on \rosat\
data --- either from the \rosat\ All-Sky Survey \citep[RASS;][]{voges99} or
from serendipitous detections in pointed observations --- though there are
also samples selected based on \xmm\ and \chandra\ imaging.  
Table~\ref{tbl:Xray_catalogs} summarizes some of the main 
X-ray catalogs that have been employed in these studies.  The recently approved
XXL survey will add $\approx 50\ \deg^2$ of imaging, contributing $\approx 600$
clusters out to $z=1$ and above.
The next big step forward for X-ray samples is the \erosita\ mission, 
which should identify
$\approx\,$80,000 galaxy clusters at high confidence 
(see \S\ref{sec:cl_space}).

%%%%%%%%%%%%%%%%%%%%%%%%%%%%%%%%%%%%%%%%%%%%%%%%
%%%%%%%%%%%%%%%%%%%%%%%%%%%%%%%%%%%%%%%%%%%%%%%%

\begin{deluxetable}{c c c c} 
\tablecolumns{4} 
\tablecaption{X-ray Cluster Catalogs 
\label{tbl:Xray_catalogs}} 
\tablehead{ Catalog/Reference & Type of Survey & 
  No.  of Clusters & Redshift Limit\\
} \tablewidth{0pc} %[0.5ex] %\hline
\startdata 
BCS \citep{ebeling00} & Wide/Shallow & 107 & 0.3 \\ 
NORAS \citep{bohringer00} & Wide/Shallow & 378 & 0.3 \\ 
HIFLUGCS \citep{reiprich02} & Wide/Shallow & 63 & 0.2 \\ 
WARPS \citep{perlman02} & Narrow/Deep & 34 & 0.8 \\ 
SHARC \citep{burke03} & Narrow/Deep & 48 & 0.7 \\ 
$160\ \deg^2$ \citep{mullis03} & Narrow/Deep & 201 & 0.7 \\ 
REFLEX \citep{bohringer04} & Wide/Shallow & 447 & 0.3 \\ 
$400\ \deg^2$ \citep{burenin07} & Narrow/Deep & 287 & 0.8\\ 
MACS \citep{ebeling10} & Wide/Shallow & 34 & 0.6 \\ 
MCXC \citep{piffaretti10} & Compilation & 1783 & 0.8 \\ 
XCS \citep{lloyd-davies10} & Narrow/Deep & 1022/3669$^*$ & 0.8 \\
%\hline
%\end{tabular} 
\enddata 
\tablecomments{ All cluster catalogs included above
are drawn from \rosat\ data, except for XCS, which is a serendipitous cluster
search in \xmm\ archival data \citep[see][for the first data release]{mehrtens11}.  
Wide/shallow survey catalogs refer to cluster
searches in the \rosat\ All-Sky Survey (RASS), whereas narrow/deep catalogs
are drawn from pointed \rosat\ or \xmm\ observations.  MCXC is a compilation
of various X-ray cluster catalogs.  The characteristic high redshift limit shown is
not the redshift of the highest redshift cluster in the sample, but rather a redshift
that contains $\gtrsim 90\%$ of the galaxy clusters.  The highest cluster redshifts
can be significantly higher than the redshift quoted, as expected for flux
limited surveys.\\
$^*$1022 is the number of galaxy clusters with $\geq 300$ photons, 
allowing for $T_X$ estimates.  3669 is the number of $4\sigma$ cluster 
candidates.
}
\end{deluxetable}

%%%%%%%%%%%%%%%%%%%%%%%%%%%%%%%%%%%%%%%%%%%%%%%%
%%%%%%%%%%%%%%%%%%%%%%%%%%%%%%%%%%%%%%%%%%%%%%%%

The largest existing cluster samples are optically selected, using
either spectroscopic or photometric galaxy catalogs.  The former benefit
from much finer spatial resolution along the line of sight.  They tend to be
shallow, with typical $z\lesssim 0.2$
\citep{merchan02,kochanek03,miller05,merchan05,berlind06,yang07,li08,
blackburne11}, though high redshift spectroscopic catalogs do exist
\citep{gerke05,coil06}.  Photometric cluster catalogs hail back as far as the
original \citet{abell58} catalog, which contained upwards of 2500 systems and
served as the primary basis of cluster studies for decades.
Though many
recent photometric catalogs have focused on narrow but deep survey data
\citep[$z\lesssim 1$, e.g.,][]{gonzalez01,gladders05,milkeraitis10,adami10},
the SDSS has led to the publication of several moderately deep
($z\lesssim 0.5$) and wide catalogs, which can contain upwards of 50,000
clusters \citep[e.g.][]{koester07,wen09,hao10,szabo10}.  Extensions that
reach out to $z\approx 1$ over $1000\mdeg^2$ or more 
from current or near future
photometric surveys --- such as RCS-2, DES, Pan-STARRS, and HSC --- 
will expand samples to the hundreds of thousands.

One limiting factor that affects these optical cluster finding experiments is
that the $4000\,$\AA\ break in the spectrum of early-type galaxies shifts into
the near-IR at $z\approx 1$, making optical detection challenging above this
redshift.  This difficulty can be overcome with IR adaptations of optical
cluster finding techniques.  Today, there are two independent efforts aiming
to detect galaxy clusters using IR data: the IRAC Shallow Cluster Survey
\citep[ISCS;][]{eisenhardt08} and the \spitzer\ Adaptation of the Red-Sequence
Cluster Survey \citep[SpARCS;][]{wilson06}.  Both surveys have discovered and
spectroscopically confirmed candidate galaxy clusters out to redshift
$z\lesssim 1.5$
\citep[e.g.,][]{stanford05,brodwin06,eisenhardt08,muzzin09,wilson09,demarco10},
with some recent detections reaching $z\lesssim 2$ \citep{stanford12,zeimann12}.
Additionally, some of these systems have also been detected in X-rays and/or
SZ \citep{brodwin10,andreon11,brodwin12}.
These early results are encouraging and suggest that IR detection of high redshift
clusters can play an important role in the future of cluster cosmology.

While detections of the SZ effect in known galaxy clusters 
date back as early as 1976 \citep{gull76},
it is only recently that instrumentation advances have made large
scale SZ searches feasible.  The first three successful cluster SZ surveys ---
using
the South Pole Telescope (SPT), the Atacama Cosmology Telescope (ACT), and the
\planck\ satellite --- are all currently ongoing. 
All three projects have released SZ-selected cluster samples
\citep{vanderlinde10,marriage10,planck_esz,williamson11,reichardt12}.  
These samples tend to
be of very massive clusters (see Figure \ref{fig:selection})
and, in the case of ACT and SPT,  extend to $z\approx 2$, with the upper limit
set by the lack of massive galaxy clusters above this redshift.
For ACT and SPT, this redshift coverage is limited
only by the abundance of such massive objects at high redshift.  \planck\ is
limited in part by its relatively large beam, but it has the important benefit
of being an all sky survey, which results in a larger cluster yield overall.
Based on the sensitivity estimates shown in Figure~\ref{fig:selection} below,
we anticipate $\sim 700$ clusters in 2500 deg$^2$ for SPT 
and $\sim 11,000$ over the full sky for \planck.  
We emphasize, however, that these numbers
can easily shift by factors of $\sim 2-3$ depending on the 
signal-to-noise cut adopted for cluster identification.
In contrast to optical and X-ray techniques,
there is not likely to be a major leap forward
in SZ capabilities in the next few
years, so the SPT, ACT, and \planck\ 
samples will probably remain the largest SZ cluster
samples available for the next decade.   
That said, the limiting masses of SZ cluster samples will go down as these
and other facilities conduct deeper
surveys focused on CMB polarization (e.g.,  ACTPol and SPTPol).

Existing cluster cosmology constraints have come primarily from X-ray data
\citep[see, e.g.,][]{henry00,reiprich02,schuecker03,allen03, pierpaoli03},
reflecting the fact that X-ray observables can be related to mass via
simulations and/or analytic approximations and by 
hydrostatic modeling for well observed clusters.  
All three of the most recent X-ray analyses
yielded tight, consistent cosmological constraints, which can be summarized as
$\sigma_8(\Omega_m/0.25)^{0.45} = 0.80 \pm 0.03$
\citep{henry09,vikhlinin09,mantz10}.  Cosmological analyses from optical
samples have typically been less constraining because of uncertain mass
calibration \citep[see, e.g.,][]{bahcall03,gladders07,wen10}.  However, recent
work that uses stacked weak lensing analysis for mass calibration
\citep{johnston07,mandelbaum08,sheldon09} 
has allowed optical samples to
achieve the same level of precision as X-ray samples \citep{rozo10},
with comparable levels of systematic error.
Constraints from SZ selected samples are emerging
\citep{vanderlinde10,sehgal10,reichardt12}, and while they are currently weak
because of the relatively large uncertainty in the 
SZ-mass scaling relation, 
the extensive follow-up campaigns that are currently underway will reduce this
scaling uncertainty and bring these constraints
to a level comparable to those from optical and X-ray cluster 
catalogs \citep[e.g.][]{high12,hoekstra12,planck12,rozo12e}.

%%%%%%%%%%%%%%%%%%%%%%%%%%%%%%%%%%%%%%%%%%%%%%%%
%%%%%%%%%%%%%%%%%%%%%%%%%%%%%%%%%%%%%%%%%%%%%%%%

\begin{figure} 
\begin{center}
\includegraphics[width=4.2in]{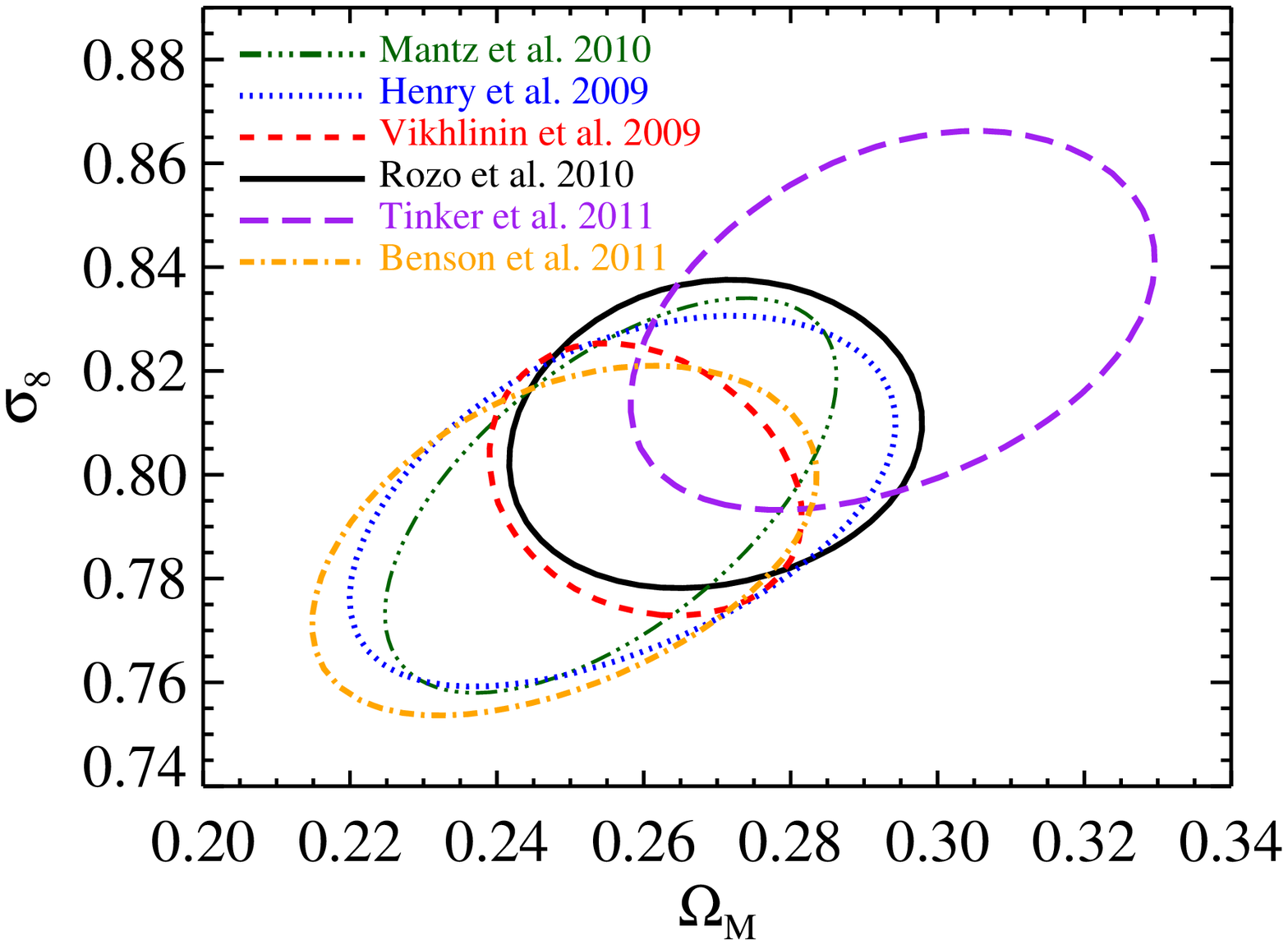}
\caption{Comparison of the $68\%$ confidence regions derived 
from galaxy cluster abundances and WMAP CMB data by various groups.  
The first three error ellipses --- using quoted uncertainties from 
\cite{mantz10}, \cite{henry09}, and \cite{vikhlinin09} ---
all come from X-ray selected cluster samples.  
The \cite{rozo10} ellipse comes from an optically selected cluster sample
with stacked weak lensing mass calibration. 
The \cite{tinker11} constraint uses the same optical clusters and
mass calibration, but relies on galaxy clustering and
mass-to-number ratios to derive cosmological constraints, making it essentially
an independent cross-check.
The \cite{benson12} ellipse comes from the SPT selected cluster sample.
}
\label{fig:comparison} 
\end{center} 
\end{figure}

%%%%%%%%%%%%%%%%%%%%%%%%%%%%%%%%%%%%%%%%%%%%%%%%
%%%%%%%%%%%%%%%%%%%%%%%%%%%%%%%%%%%%%%%%%%%%%%%%

Regardless of the wavelength of choice, current
cluster abundance constraints are limited not by the number of clusters but
by uncertainty in mass calibration. 
Figure~\ref{fig:comparison} shows the cluster 
abundance constraints from several recent analyses.
Because the current X-ray and optical mass
calibrations are fundamentally different (hydrostatic vs.\ weak lensing), the
excellent agreement illustrated in Figure \ref{fig:comparison} provides a strong test of
systematic uncertainties.  However, the results from the 
\citet{Planck11b} have sounded a cautionary
note, as the optical mass estimates used to derive cosmological 
parameters in \citet{rozo10}
appear to be inconsistent with SZ data \citep[see also][]{draper11}.
\citet{biesiadzinski12} have attributed this inconsistency to miscentering,
while \citet{angulo12} point out the importance of systematics covariance.
\citet{rozo12b,rozo12c,rozo12d} 
argue that the optical, X-ray, and SZ data can be reconciled by
considering, in addition to these effects, the systematics of 
X-ray temperature measurements indicated by the offsets among 
estimates from different groups, and departures from
hydrostatic equilibrium at the level predicted by hydrodynamic cosmological
simulations (e.g., \citealt{nagai07}).
Regardless of how this issue is ultimately resolved, it is clear that 
further tightening cosmological constraints will require a significant
improvement in our ability to estimate cluster masses.

On this last count, 
we note that Figure \ref{fig:comparison} also includes cosmological
constraints from an analysis by \citet{tinker11} that does not 
rely on cluster abundances.
\cite{tinker11} use a halo occupation model (see \S\ref{sec:cmb_lss})
fit to SDSS galaxy clustering, which yields a prediction for the
mass-to-number ratio of clusters\footnote{Analogous to mass-to-light ratio,
but with galaxy number instead of integrated luminosity.} as a function
of $\sigma_8$ and $\om$.
While this analysis uses the same weak lensing mass calibration as
\cite{rozo10}, the method is less sensitive to the mass scale and is
entirely independent of abundance uncertainties, making it
a largely independent measurement and a 
powerful systematics cross-check.
The same approach can be adapted to future, deeper photometric
surveys.
We also note that stacked weak lensing measurements for clusters
can be extended far beyond the virial radius \citep{sheldon09}, 
into the regime where they measure the large scale cluster-mass
cross-correlation function, and that these large scale measurements
can also be used to constrain cosmological parameters \citep{zu12a}.

%%%%%%%%%%%%%%%%%%%%%%%%%%%%%%%%%%%%%%%%%%%%%%%%
%%%%%%%%%%%%%%%%%%%%%%%%%%%%%%%%%%%%%%%%%%%%%%%%
%%%%%%%%%%%%%%%%%%%%%%%%%%%%%%%%%%%%%%%%%%%%%%%%
%%%%%%%%%%%%%%%%%%%%%%%%%%%%%%%%%%%%%%%%%%%%%%%%
%%%%%%%%%%%%%%%%%%%%%%%%%%%%%%%%%%%%%%%%%%%%%%%%
%%%%%%%%%%%%%%%%%%%%%%%%%%%%%%%%%%%%%%%%%%%%%%%%
%%%%%%%%%%%%%%%%%%%%%%%%%%%%%%%%%%%%%%%%%%%%%%%%
%%%%%%%%%%%%%%%%%%%%%%%%%%%%%%%%%%%%%%%%%%%%%%%%

\subsection{Observational Considerations} \label{sec:cl_obs}

\subsubsection{Expected Numbers and Cosmological Sensitivity}
\label{sec:cl_numbers}

\begin{figure} [t]
\begin{center}
\includegraphics[width=2.8in,
height=2.4in]{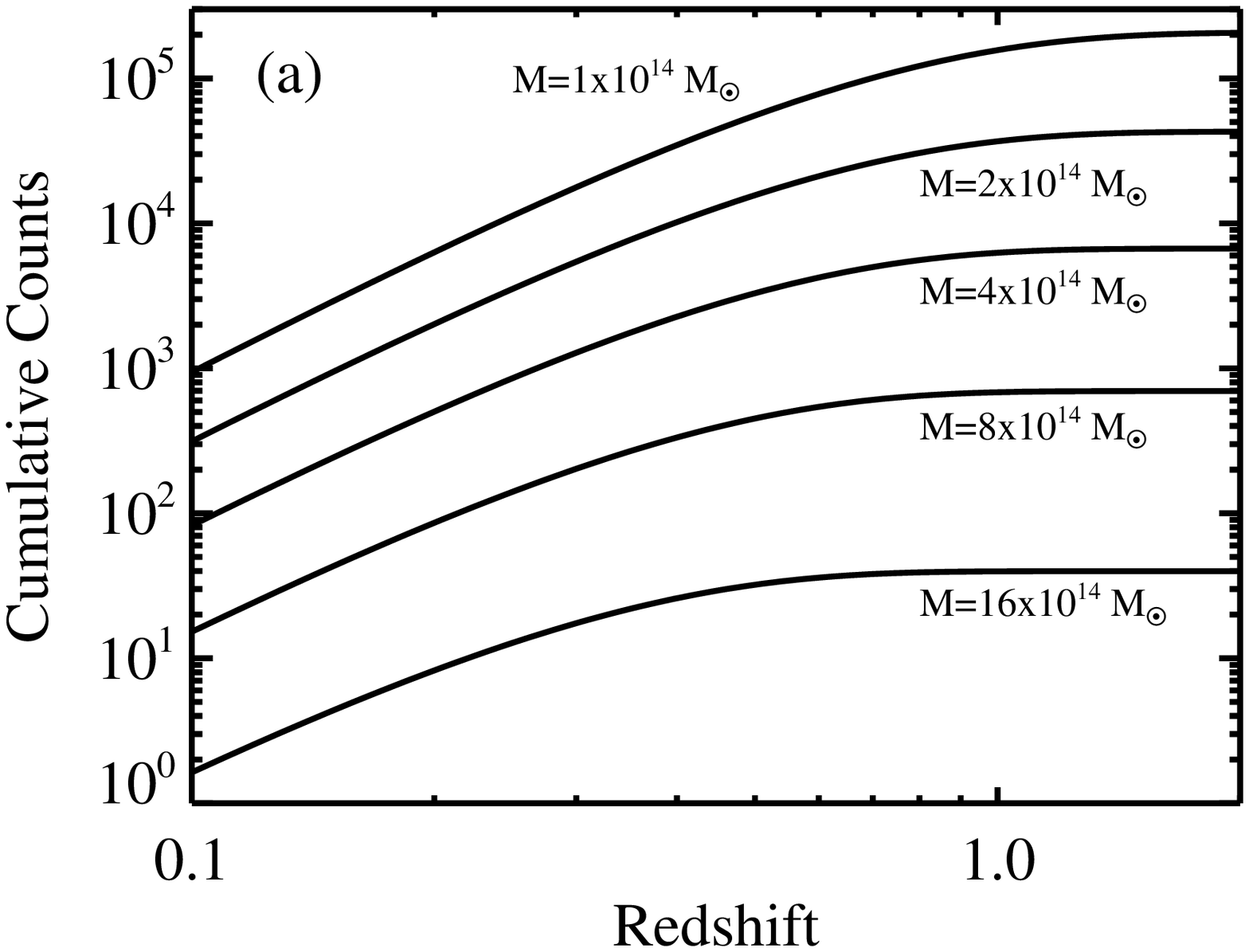} \includegraphics[width=2.8in,
height=2.4in]{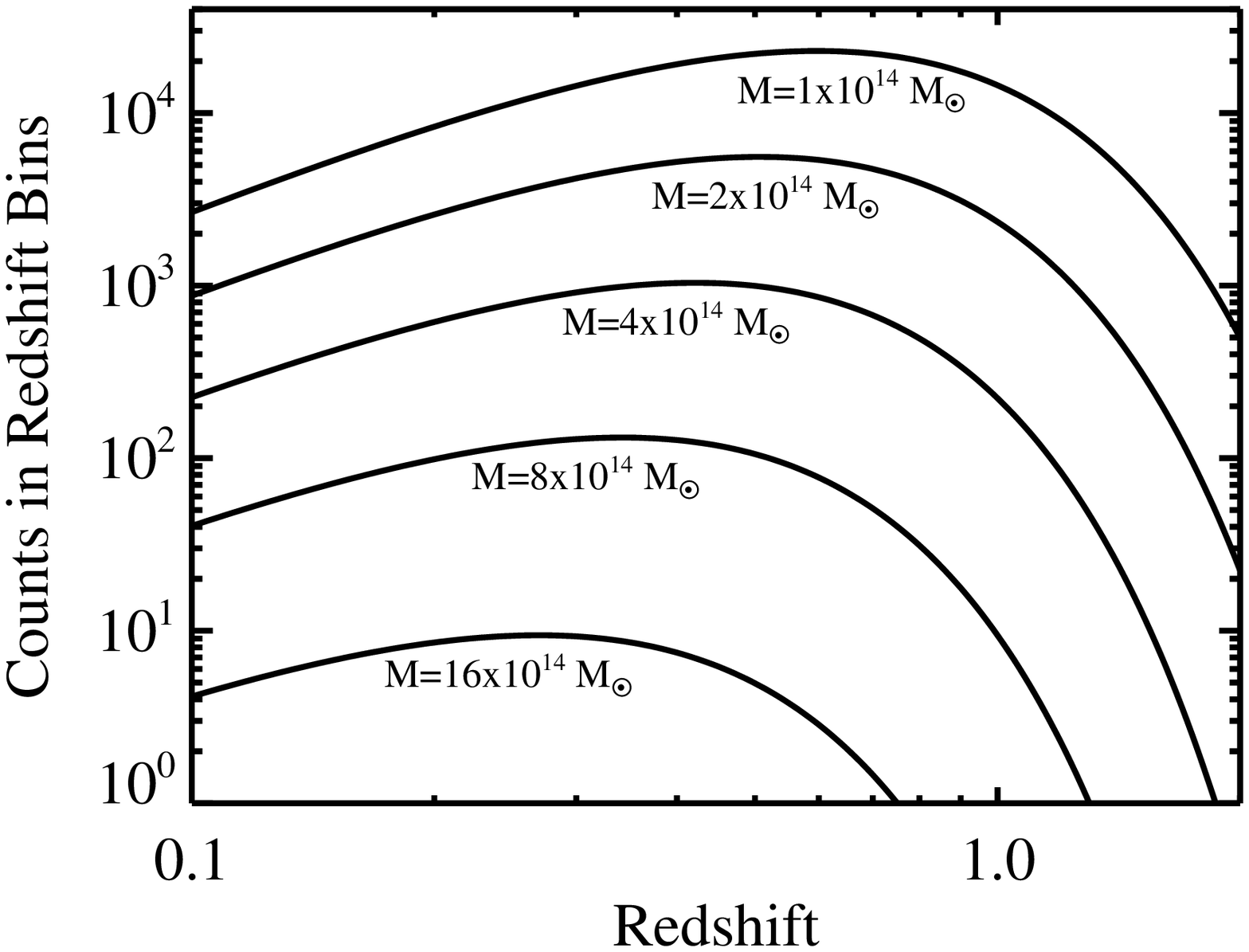} \includegraphics[width=2.8in,
height=2.4in]{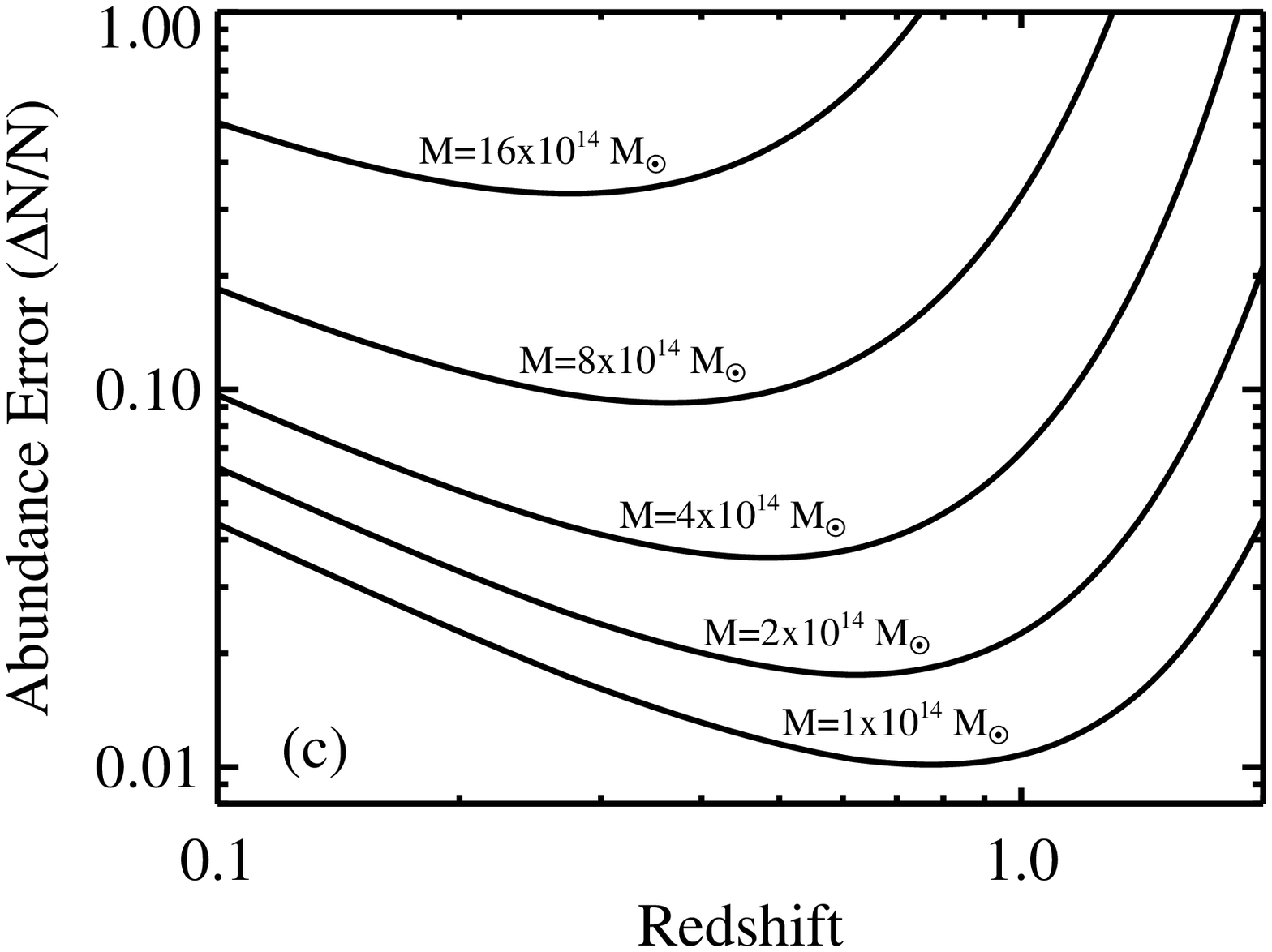} \includegraphics[width=2.8in,
height=2.4in]{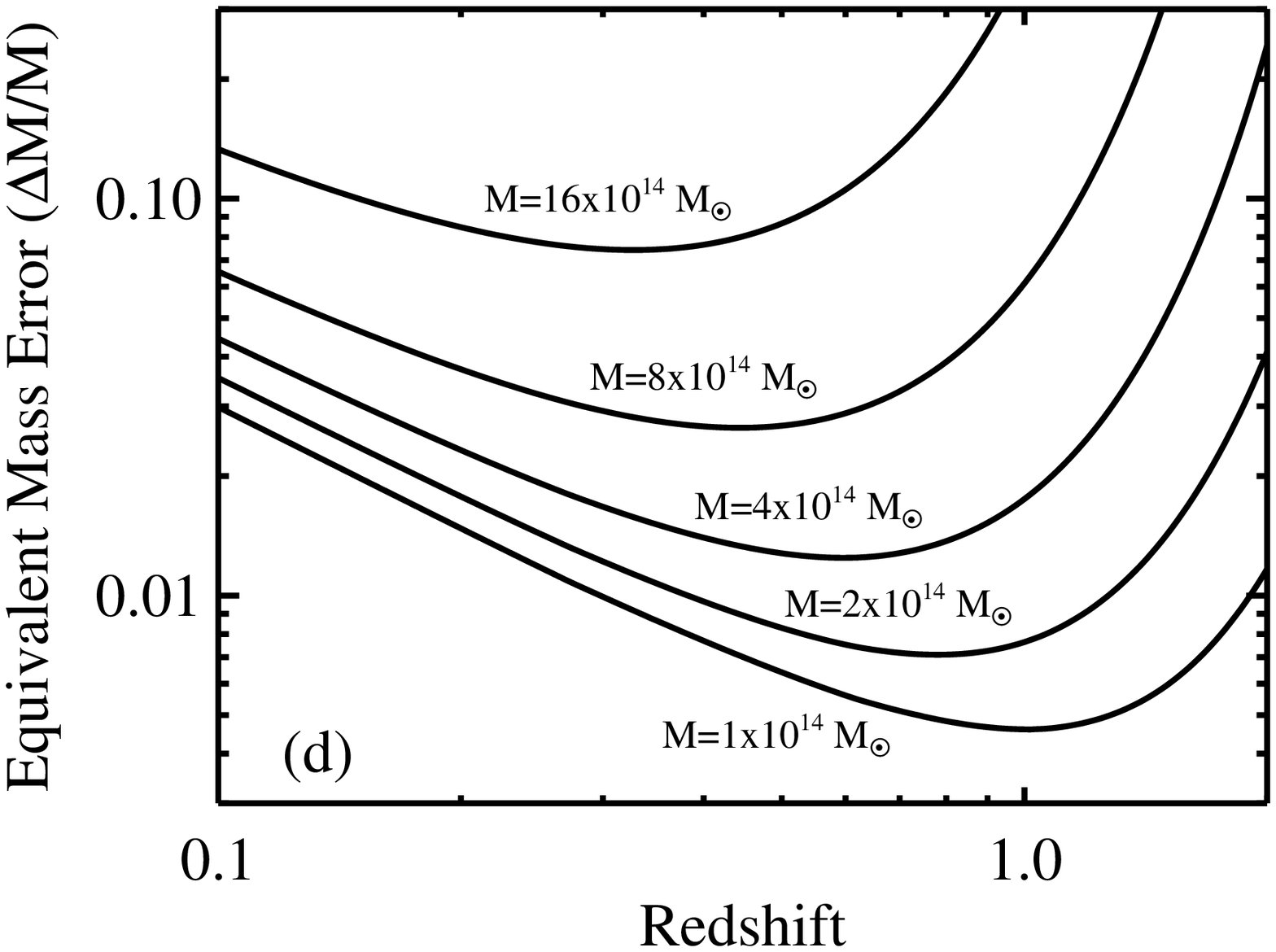} 
\end{center}
\caption{(a) Cumulative
halo number counts above the indicated mass thresholds $M$
as a function of the limiting survey redshift.
We assume the fiducial cosmological model from
Table~\ref{tbl:models}, and survey area of $10^4\ \deg^2$.  (b)
Counts above the mass threshold
in redshift bins $z=z_c\pm 0.05$.  (c) Statistical error
in the number of clusters above the mass threshold from 
equation~(\ref{eq:nerr}), 
again in redshift bins $z=z_c\pm 0.05$.  (d)
The mass accuracy required to ensure that cosmological constraints are limited
by the statistical precision in the number of galaxy clusters rather than by
uncertainties in mass estimation.  
} \label{fig:cumcounts} \end{figure}

Figure \ref{fig:cumcounts}a shows the expected cluster
counts in our fiducial cosmological model for a variety of 
limiting masses, as a function of the limiting redshift $z$ 
of a $10^4\mdeg^2$ survey.  (Note that these are
lower limits on mass but upper limits on redshift.)
Panel (b) shows number counts in redshift bins of width $\pm 0.05$;
e.g.,
at $z=0.15$, we show the halo counts in the redshift bin $[0.1,0.2]$.  We
maintain this redshift binning convention throughout.  Together, these two
figures give a broad-brush sense for the typical sample sizes and redshift
distribution of galaxy clusters as a function of limiting mass and redshift.

%%%%%%%%%%%%%%%%%%%%%%%%%%%%%%%%%%%%%%%%%%%%%%%%
%%%%%%%%%%%%%%%%%%%%%%%%%%%%%%%%%%%%%%%%%%%%%%%%

%%%%%%%%%%%%%%%%%%%%%%%%%%%%%%%%%%%%%%%%%%%%%%%%
%%%%%%%%%%%%%%%%%%%%%%%%%%%%%%%%%%%%%%%%%%%%%%%%

Assuming halo masses can be adequately measured, the statistical error in
cluster abundances is the sum in quadrature of Poisson noise and sample
variance \citep{hu03}, 
\begin{equation}
\label{eq:nerr} 
(\Delta N)^2 = N + \bar b^2 N^2 \sigma^2(V).
\end{equation}
Here, $N$ is the mean number of halos in the volume of
interest, $\bar b$ is the mean bias of the halos, and $\sigma^2(V)$ is the
variance of the matter density field over the survey 
volume.\footnote{For example, in our fiducial cosmology at $z=0.6$,
the matter variance in a volume of $\Delta z = 0.1$ and area
10,000 deg$^2$ is $\sigma(V) \approx 0.2\%$, and the mean halo bias is 
$\approx 3.0$ and $\approx 5.7$ for mass thresholds of $10^{14}M_\odot$ and
$4\times 10^{14}M_\odot$, respectively.}
Figure \ref{fig:cumcounts}c shows the fractional error $\Delta N/N$
for the fiducial model, again for redshift bins $z=z_c\pm 0.05$ where $z_c$ is
the central redshift of the bin.  
Sample variance becomes larger than Poisson variance below a transition mass
$\sim 4\times 10^{14}\ \msun$ at $z=0.1$ and $\sim  10^{14}\ \msun$
at $z=1$.
However, the statistical error is never more than a factor $\sim 2$
above the $N^{-1/2}$ Poisson expectation (see Figure~\ref{fig:errors} below),
and total statistical errors should scale with survey area roughly as 
$(A/10^4\mdeg^2)^{-1/2}$.
For any mass threshold the statistical error first
decreases with redshift, as the number of clusters grows with the increasing
comoving volume per $\Delta z$.  
This trend flattens when the clusters become exponentially
rare, at which point further increase in redshift leads to a precipitous drop
in the number of clusters and a corresponding rise in Poisson errors.
These competing effects lead to
the characteristic U-shape of the curves in Figure \ref{fig:cumcounts}c.

Figure~\ref{fig:cumcounts}d converts these statistical abundance errors to
equivalent errors in mass by dividing $\Delta N/N$ by the logarithmic slope of
the cumulative halo mass function, $\alpha= - d\ln N/d\ln M$,
which ranges between 2 and 5 depending
on redshift and mass.  While observational samples are not thresholded 
exactly in mass,
the sensitivity of 
cluster abundances to an overall shift in the mean mass at fixed
observable is well captured by this heuristic argument.
In order for clusters to saturate the statistical limit in the abundances,
the uncertainty in mass calibration must be
smaller than this $\Delta M/M$.
For a $10^4\mdeg^2$ survey and $M\geq 8\times 10^{14}\ \msun$,
a mass accuracy of $3\%-10\%$ (depending on $z$) suffices. By $M\approx
2\times 10^{14}\ \msun$, however, the accuracy requirement has 
sharpened to $\lesssim 1\%$.   
(This last number 
agrees well with the more detailed analysis of 
\citeauthor{cunha10} [\citeyear{cunha10}] for a mass threshold
of $10^{14.2}\,\msun$; see in particular the top panels in their Figure 2.)
Achieving such accuracy is a tall order,
and current studies are clearly limited by the systematic
uncertainty in cluster masses rather than abundance statistics.
Note that the required accuracy scales roughly as
$(A/10^4\mdeg^2)^{-1/2}$, and it applies to the overall mass scale (i.e.,  the
mean of the mass--observable relation) rather than the mass of any individual
system.

%%%%%%%%%%%%%%%%%%%%%%%%%%%%%%%%%%%%%%%%%%%%%%%%
%%%%%%%%%%%%%%%%%%%%%%%%%%%%%%%%%%%%%%%%%%%%%%%%

\begin{figure} [t]
\begin{center}
\includegraphics[width=2.8in,
height=2.4in]{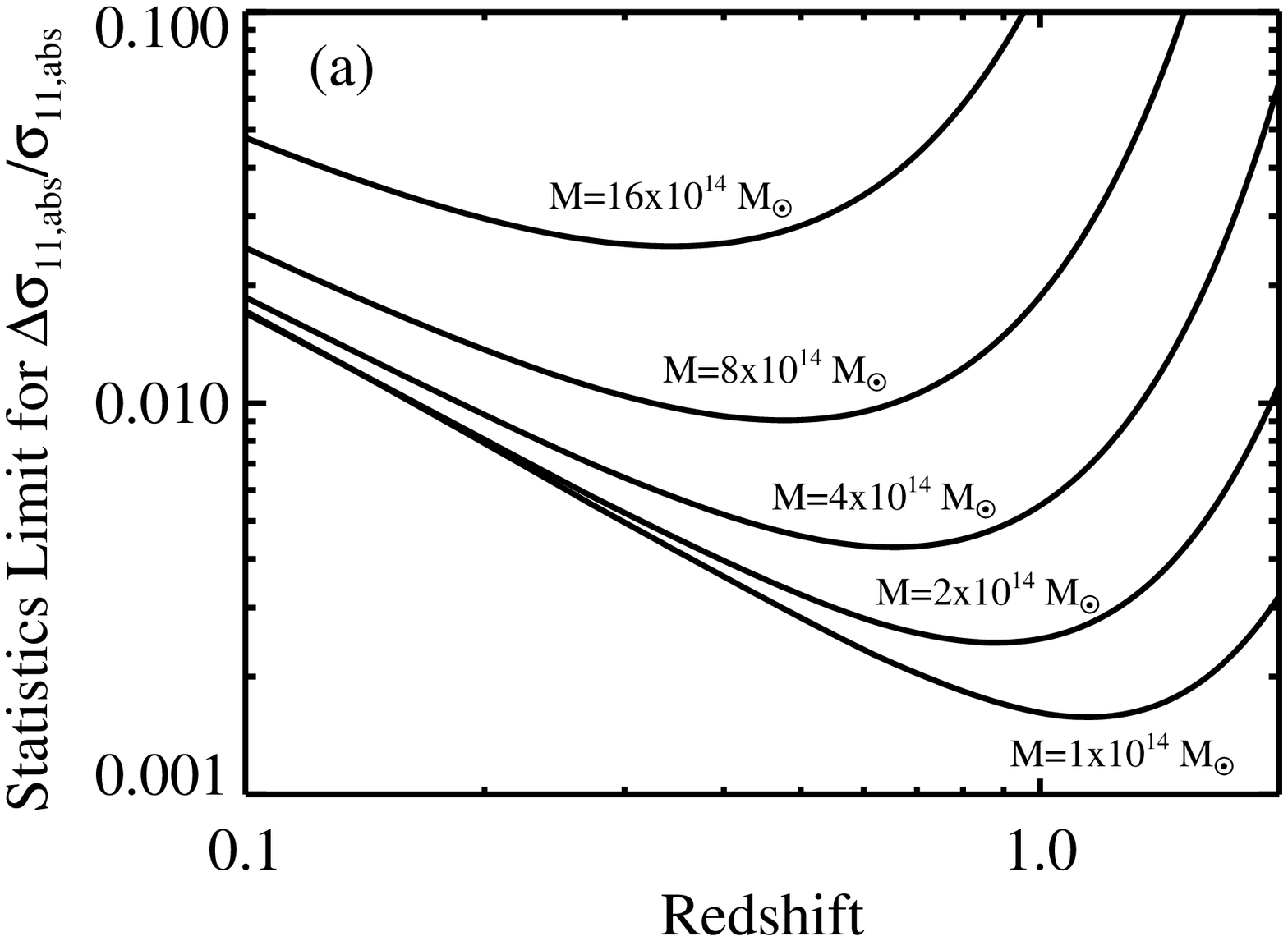} \includegraphics[width=2.8in,
height=2.4in]{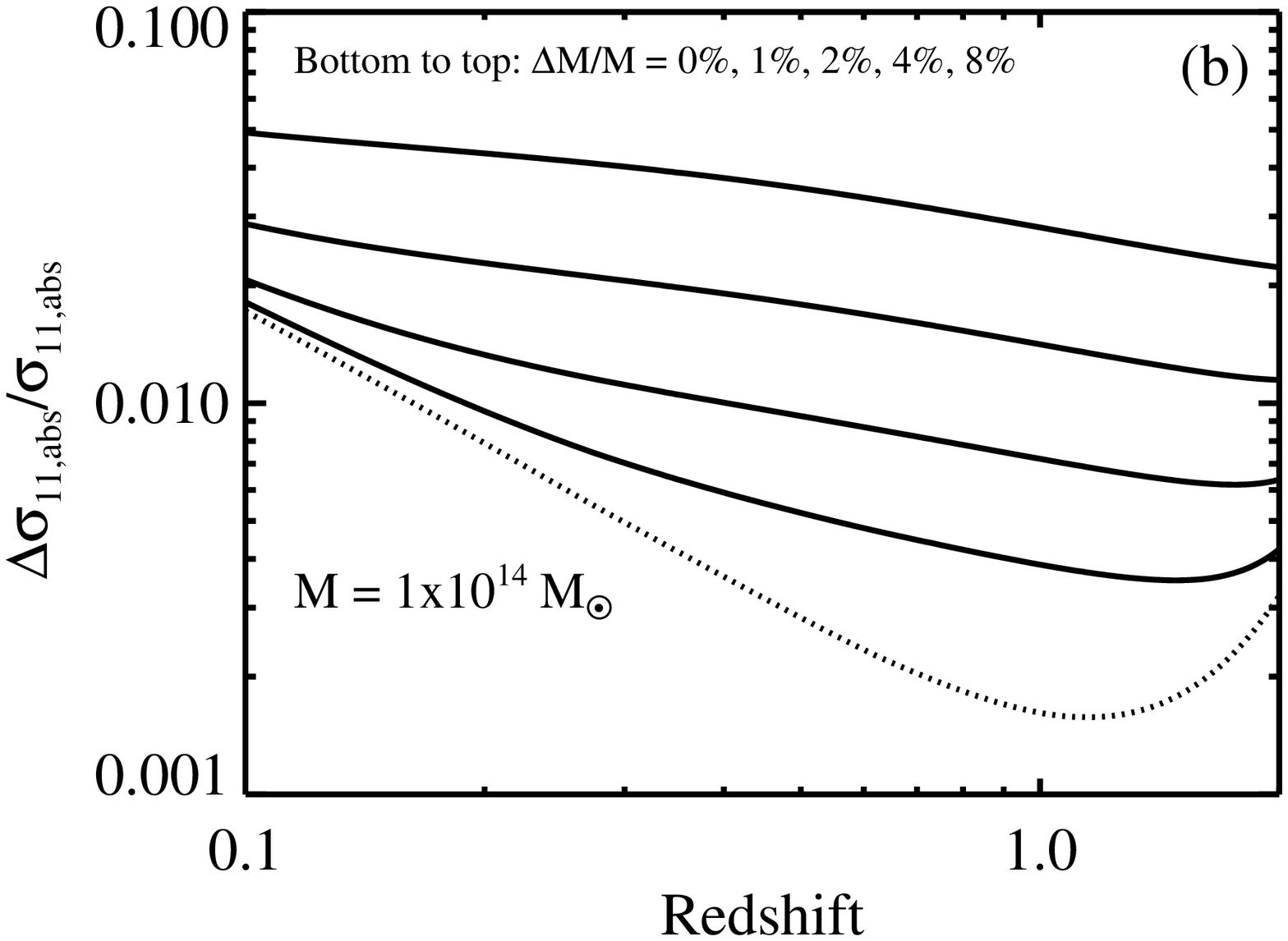}
\includegraphics[width=2.8in,
height=2.4in]{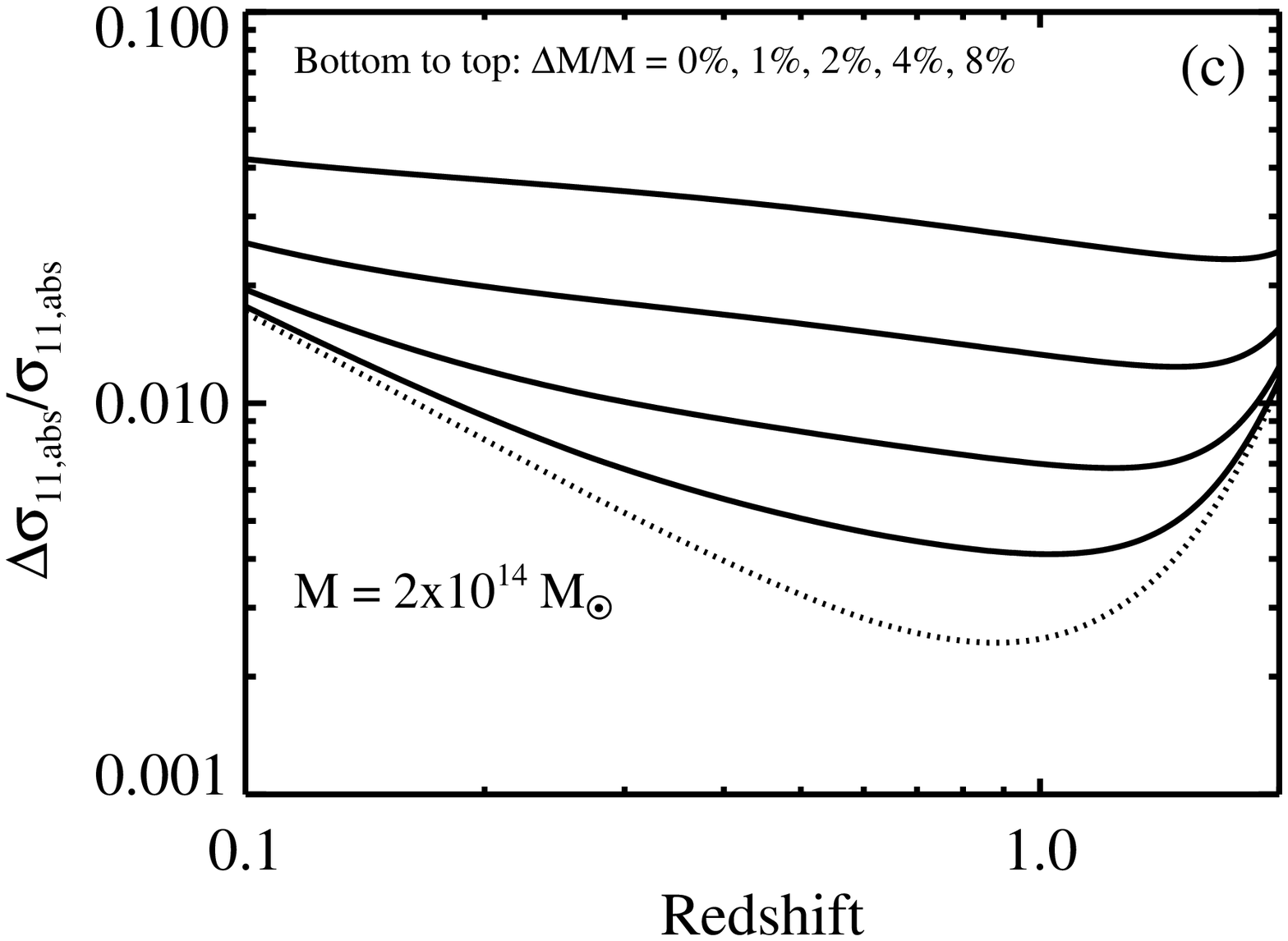}
\includegraphics[width=2.8in,
height=2.4in]{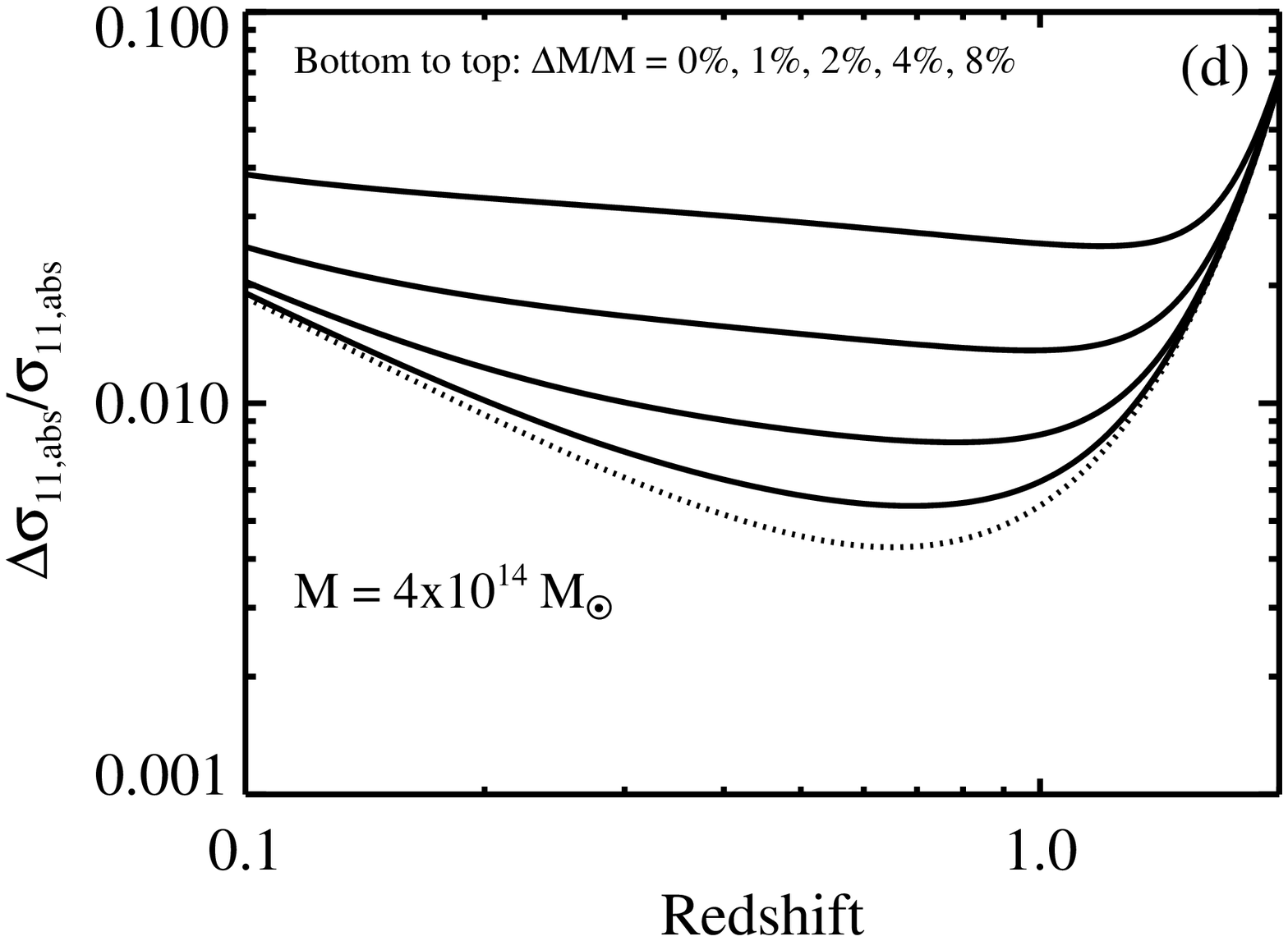} 
\end{center}
\caption{(a)
Statistical error on $\sigElevz$ as a function of redshift, in
redshift bins $z=z_c\pm 0.05$, for different mass thresholds as labeled.  We
assume a $10^4\ \deg^2$ survey area, and the fiducial cosmological model.  We
also assume that $\Omega_m$, the shape of the matter power spectrum, and the
comoving volume element $dV_C$ are perfectly known from independent data
(CMB+SN+BAO+WL).  
Panels (b)-(d) refer to specific mass thresholds as labeled.
In each panel the solid curves show the effect
of different mass calibration uncertainties as labeled while the dotted curve
assume the perfect mass calibration values (i.e., number statistics limited) 
from panel (a).
For reference, the uncertainty in $\sigElevz$ that we forecast for a 
fiducial CMB+SN+BAO+WL program is $\sim 1\%$ for Stage IV data sets
and $\sim 2-3\%$ for Stage III data sets
(see \S\ref{sec:cl_prospects} and~\S\ref{sec:forecast_cl}).
} \label{fig:serr} \end{figure}

%%%%%%%%%%%%%%%%%%%%%%%%%%%%%%%%%%%%%%%%%%%%%%%%
%%%%%%%%%%%%%%%%%%%%%%%%%%%%%%%%%%%%%%%%%%%%%%%%

Figure \ref{fig:serr} translates the errors on cluster abundance from Figure
\ref{fig:cumcounts} to errors on the matter power spectrum amplitude
$\sigElev(z)$, again for a $10^4\deg^2$ survey with $z=z_c\pm 0.05$ bins.
For simplicity, we assume
that $\Om$, the comoving volume element $dV_c(z)$, and the power
spectrum shape are perfectly known from independent data (CMB+SN+BAO+WL),
so that $\sigElev(z)$ is the single cosmological parameter controlling the
cluster abundance.  As discussed in \S\ref{sec:cl_method},
if the uncertainty in $\Omega_m$ is non-negligible, then it is the combination
$\sigma_8(z)\Omega_m^q$ that is constrained instead.
Panel (a) shows the case where mass calibration errors are
negligible.  The errors on $\sigElev(z)$ roughly track the abundance errors
$\dNN$ in Figure \ref{fig:cumcounts}, but because the sensitivity 
of the abundance to $\sigElev(z)$ at fixed mass increases with increasing
redshift, the best constraint on $\sigElev(z)$ comes at a higher redshift
than the one at which $\dNN$ is minimized.
The remaining panels show the impact of 1\%, 2\%, 4\%, and 8\% mass
calibration errors for three different threshold masses.  

The basic features
in Figure \ref{fig:serr} are simple to understand at a quantitative level,
starting from the knowledge that
cluster abundances constrain the combination
$\sigElevz\Omega_m^q$ with $q\approx 0.4$.  
Since the mass of a collapsed volume scales linearly with $\Omega_m$, 
a shift of the mass scale by a constant factor is 
nearly degenerate with a change of $\Omega_m$ by the same factor.
Together these scalings imply $\sigElevz \propto M^q$, where $M$ is
the mass scale at fixed abundance, making
$\Delta \ln \sigElevz \approx q \Delta \ln M$ for a survey
limited by mass calibration uncertainty $\Delta\ln M$.
For a survey limited by halo statistics,
the corresponding effective mass error is 
$(\Delta \ln M)_{\rm eff} = \alpha^{-1} \Delta \ln N$ where 
$\alpha= - d\ln N/d\ln M \approx 2-5$ 
is the slope of the cumulative halo mass function, so in this case
$\Delta \ln \sigElevz \approx q \alpha^{-1} \Delta \ln N$.  
Combining the two limits we arrive at
\begin{equation}
\Delta \ln \sigElevz \approx q\times 
        \max\left[ \Delta \ln M, \alpha^{-1} \Delta \ln N \right].
\label{eq:sigElapprox}
\end{equation}
The above expression fits the data in Figure \ref{fig:serr} with better than $30\%$
accuracy (typically $\lesssim 15\%$).

\begin{figure} 
\begin{centering}
\includegraphics[width=3.5in,height=2.9in]{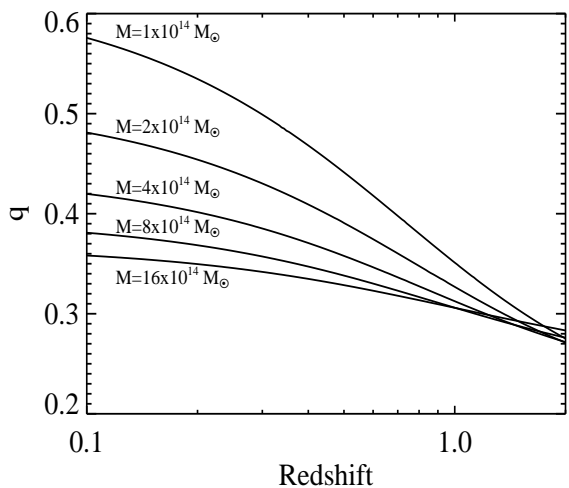}
\caption{The degeneracy exponent $q$ as a 
function of redshift for a series of threshold masses.
The parameter $q$ is the exponent in $\sigElevz\Omega_m^q$ that holds the 
abundance of galaxy clusters above the quoted threshold mass at the 
appropriate redshift bin fixed for small, oppositely
directed changes in $\sigElevz$ and $\om$.
} 
\label{fig:qval} 
\end{centering}
\end{figure}

Figure~\ref{fig:qval} plots the value of the degeneracy exponent $q$
as a function of limiting mass and redshift.
In the Press-Schechter (\citeyear{press74}) theory of the halo mass
function, the cumulative abundance is set by the probability that a
point in a Gaussian field of variance $\sigma^2(M)$ exceeds the
critical threshold $\delta_c\approx 1.69$ for spherical collapse
(see \S\ref{sec:cmb_lss}), so that 
$N\propto \left[1-\mbox{erf}(\delta_c/\sqrt{2}\sigma(M)) \right]$.
Putting in the $\sigma(M,z)$ relation for a $\Lambda$CDM power
spectrum yields a logarithmic derivative
$d\ln N/d\ln \sigma \equiv \alpha_\sigma \approx 5-9$ depending
on mass and redshift.  Because cluster abundances
are degenerate in $\Omega_m/M$, 
the logarithmic derivative of cluster abundances relative to 
$\Omega_m$ is the same 
as the slope $\alpha$ of the mass function (but with opposite sign), 
so locally the cumulative mass function scales as 
\begin{equation}
N(m) \propto \left[\sigElevz\right]^{\alpha_\sigma}\Omega_m^{-\alpha} 
   = \left[\sigElevz \Omega_m^{-\alpha/\alpha_\sigma} \right]^{\alpha_\sigma}.
\end{equation}
We see that halo abundances are degenerate in $\sigElevz \Omega_m^q$ 
with $q=-\alpha/\alpha_\sigma \approx 3/7 \approx 0.4$.
We plot the ratio $\alpha/\alpha_\sigma$ --- computed using the 
\citet{tinker08} mass function rather than the Press-Schechter
mass function --- in Figure~\ref{fig:qval}.

%%%%%%%%%%%%%%%%%%%%%%%%%%%%%%%%%%%%%%%%%%%%%%%%
%%%%%%%%%%%%%%%%%%%%%%%%%%%%%%%%%%%%%%%%%%%%%%%%

%%%%%%%%%%%%%%%%%%%%%%%%%%%%%%%%%%%%%%%%%%%%%%%%
%%%%%%%%%%%%%%%%%%%%%%%%%%%%%%%%%%%%%%%%%%%%%%%%

A cluster abundance analysis becomes limited by mass scale uncertainty
rather than halo abundance statistics when 
$\Delta\ln M > \alpha^{-1}\Delta \ln N$.
If we approximate the error as Poisson, $\Delta\ln N = N^{-1/2}$,
then an experiment is limited by mass uncertainty if 
the sample size is $N \geq (\alpha \Delta \ln M)^{-2}$.   
Current systematic uncertainties in mass calibration are $\approx 10\%$,
which for $\alpha \approx 3$ corresponds to $N\approx 10$.  
Thus, cluster abundance studies are limited by uncertainty in 
the overall mass scale even
for samples with as few as $\approx 10-20$ galaxy clusters.
For cluster samples with $N\approx 10^3\ (10^4)$,
the accuracy required in mass estimation for an experiment 
to be dominated by halo statistics is $\approx 1\%\ (0.3\%)$.
So that one may apply the rule-of-thumb estimates 
derived in this section, 
Figure \ref{fig:errors} plots the mass-function slope $\alpha$
and the ratio of the total error $\Delta \ln N$ 
to the Poisson uncertainty $N^{-1/2}$.  
Note that abundance errors including sample variance almost never 
exceed twice the Poisson error and are often much closer.
Using Figures~\ref{fig:qval} and \ref{fig:errors} 
along with equation (\ref{eq:sigElapprox}), one 
can quickly estimate how well 
an experiment with given number of galaxy clusters $N$ can
constrain $\sigElevz$.  

%%%%%%%%%%%%%%%%%%%%%%%%%%%%%%%%%%%%%%%%%%%%%%%%
%%%%%%%%%%%%%%%%%%%%%%%%%%%%%%%%%%%%%%%%%%%%%%%%

\begin{figure} 
\centerline{
\includegraphics[width=2.8in, height=2.4in]{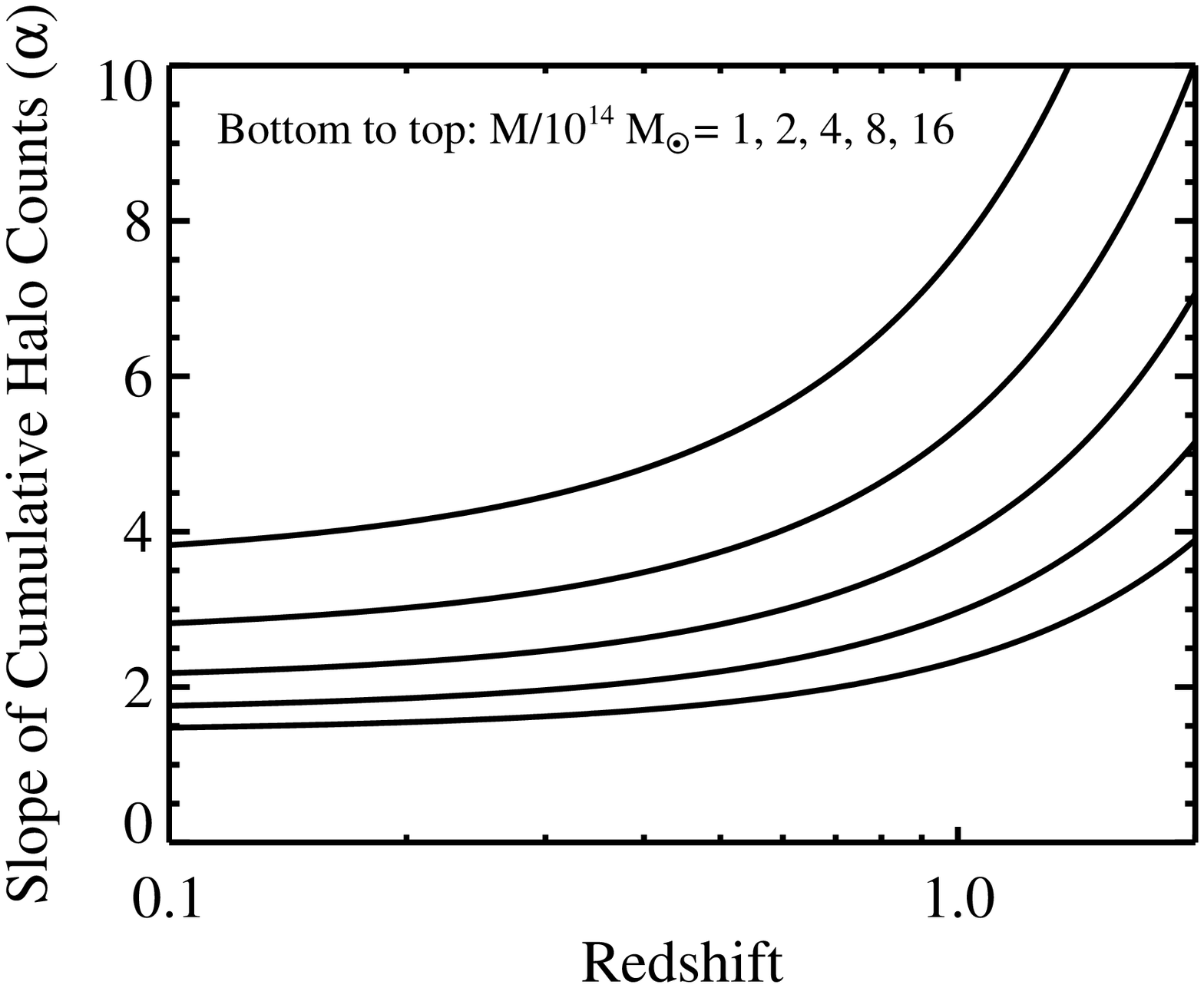}
\includegraphics[width=2.8in, height=2.4in]{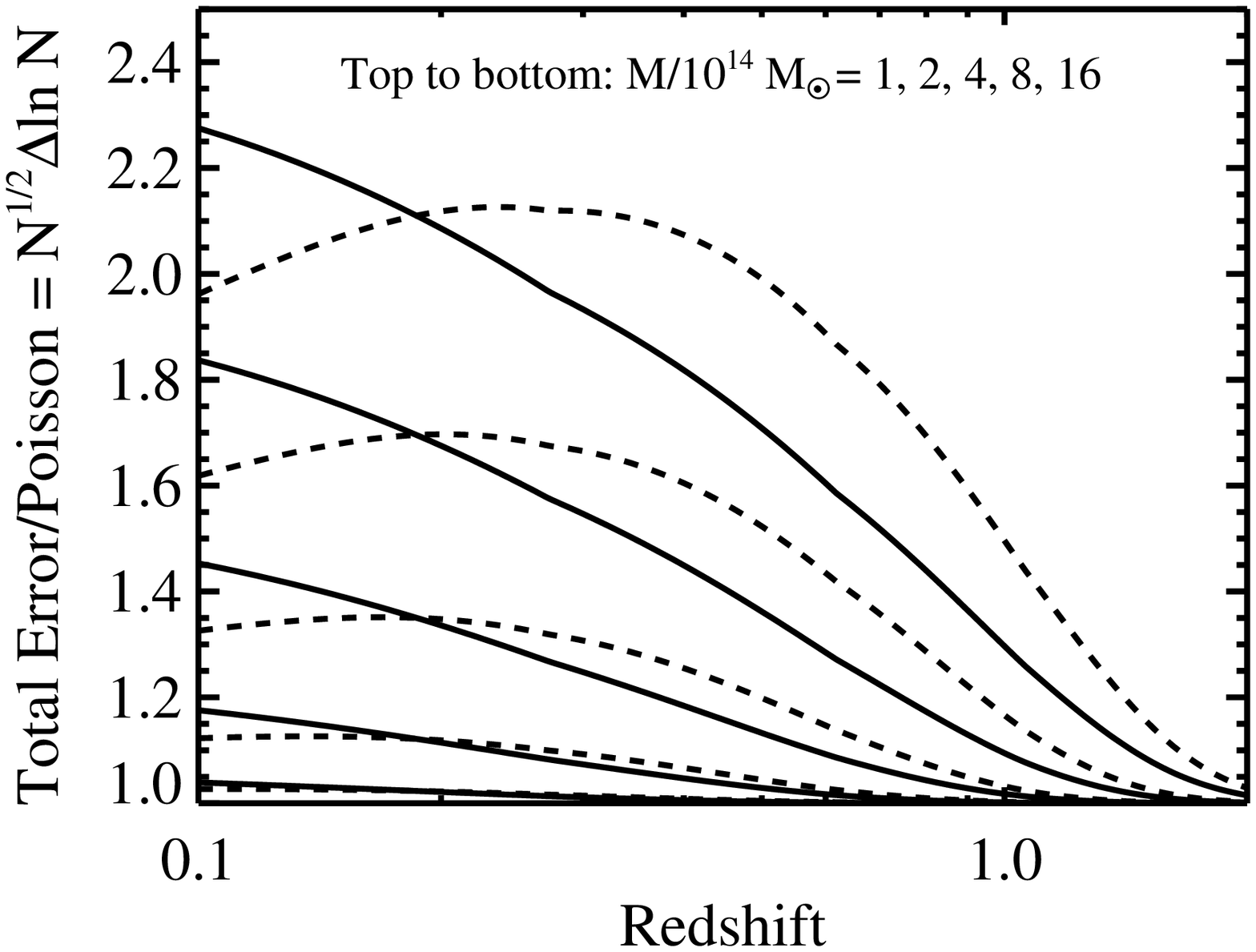}
}
\caption{{\it Left:} Logarithmic derivative
$\alpha=-d\ln N/d\ln M$ of the cumulative halo counts, 
as a function of redshift, for five mass thresholds as labeled.
{\it Right:} The ratio of the total (Poisson + sample variance) 
error in the halo counts
$\Delta \ln N$ to the Poisson error $N^{-1/2}$.   Solid
lines assume a survey area of 10,000$\mdeg^2$, while 
dashed lines correspond to $100\mdeg^2$.  In conjunction with
Fig.~\ref{fig:qval} and
equation~(\ref{eq:sigElapprox}), these figures allow one 
to quickly estimate how well 
$\sigElevz$ can be constrained at each redshift by a galaxy cluster sample
with $N$ clusters.
} 
\label{fig:errors} 
\end{figure}

%%%%%%%%%%%%%%%%%%%%%%%%%%%%%%%%%%%%%%%%%%%%%%%%
%%%%%%%%%%%%%%%%%%%%%%%%%%%%%%%%%%%%%%%%%%%%%%%%

If $\Om$ and $dV_c(z)$ are not perfectly known, then cluster abundances will
constrain a combination of cosmological parameters rather than the matter
fluctuation amplitude alone.  Predicted abundances are proportional to
$dV_c(z)$, so for an experiment dominated by uncertainty in the mass scale,
uncertainty in the volume element will affect the interpretation
if $\Delta\ln dV_c \ga \alpha\Delta\ln M$, the effective abundance uncertainty.
SN and BAO surveys should typically yield uncertainties 
below this limit, so we expect 
regarding $dV_c(z)$ as known to be an adequate approximation for our purposes,
though it may fail for sufficiently powerful cluster surveys.
Since a pure shift in $\Om$ is equivalent to a shift in mass scale,
uncertainties in $\Om$ are relevant if $\Delta \ln \Om \gtrsim \Delta \ln M$,
where we have again assumed the experiment in question is dominated by the
mass error $\Delta \ln M$.
If the uncertainty in $\Om$ is larger than this critical
scale, then clusters will effectively constrain $\sigElevz\Omega_m^{q}$
rather than $\sigElevz$ alone.  Equation (\ref{eq:sigElapprox}) will
still hold, but one must replace $\Delta \ln \sigElevz$ by 
$\Delta \ln \left[ \sigElevz\Omega_m^{q} \right]$.
Current fractional uncertainties in $\Om$ from CMB and other
observables are $\sim 10\%$, comparable
to mass calibration systematics.  Future studies will reduce $\Om$
uncertainties, but they may remain significant compared to improved
mass calibration errors in cluster surveys.

We have focused our discussion here on
cumulative cluster abundances --- i.e., space densities of clusters above a
mass threshold --- while observational analyses usually examine the
differential distribution as a function of observable mass-proxies.
Differential distributions are useful for breaking degeneracies (e.g., among
$\sigElev$, $\Om$, and $dV_c$), and for constraining ``nuisance parameters''
such as the scatter of the observable-mass relation.  However, for
single-parameter constraints on $\sigElev(z)$, we expect that our analysis
of the cumulative abundance uncertainties provides an accurate guide, as it
makes use of the single number best determined by the data for any given mass
threshold and redshift range.
We anticipate that observational analyses will continue to concentrate
mainly on differential distributions, but cumulative distributions
are more amenable to the kind of rule-of-thumb estimates that 
we try to develop throughout this section, so they
provide a more intuitive way of understanding the cosmological
information content of cluster surveys.

%%%%%%%%%%%%%%%%%%%%%%%%%%%%%%%%%%%%%%%%%%%%%%%%
%%%%%%%%%%%%%%%%%%%%%%%%%%%%%%%%%%%%%%%%%%%%%%%%
%%%%%%%%%%%%%%%%%%%%%%%%%%%%%%%%%%%%%%%%%%%%%%%%
%%%%%%%%%%%%%%%%%%%%%%%%%%%%%%%%%%%%%%%%%%%%%%%%
%%%%%%%%%%%%%%%%%%%%%%%%%%%%%%%%%%%%%%%%%%%%%%%%
%%%%%%%%%%%%%%%%%%%%%%%%%%%%%%%%%%%%%%%%%%%%%%%%
%%%%%%%%%%%%%%%%%%%%%%%%%%%%%%%%%%%%%%%%%%%%%%%%
%%%%%%%%%%%%%%%%%%%%%%%%%%%%%%%%%%%%%%%%%%%%%%%%

\subsubsection{Cluster Finding} \label{sec:cl_finding}

Each of the three main methods for finding galaxy clusters --- optical, X-ray,
and SZ --- has its own virtues and deficiencies.  The principal advantage of
optical surveys is sheer statistics, reflecting the low mass threshold for
optical detection; clusters with masses as low as $5\times 10^{13}\ \msun$ are
capable of hosting significant galaxy overdensities.   
Near-future surveys
(RCS-2, DES, HSC, Pan-STARRS) should find $\approx 10^5$ systems in areas of
$10^3-10^4\mdeg^2$ out to $z\approx 1$.
On a longer time scale ($\approx 10$ years),
surveys with LSST should increase the available cluster samples by another
factor of $5-10$, due both to larger area ($\approx $20,000 deg$^2$)
and to deeper imaging, which should allow cluster detection out to 
$z\approx 1.5$.  Finally, cluster searches in the IR are capable of finding
galaxy clusters out to $z\approx 2$, but large survey areas 
to this depth will
only be achievable with the advent of \euclid\ and/or \wfirst.
With the stacked weak lensing mass calibration that we advocate in
\S\ref{sec:cl_mass_calibration}, the calibration accuracy scales with
cluster number as $N^{-1/2}$, so enormous samples are statistically
advantageous even if mass uncertainties dominate the error budget.

The main drawback for optical cluster detection is projection effects, i.e.,
chance alignments of multiple low mass halos along the line of sight that are
misidentified as a single massive galaxy cluster.  While this systematic has
been drastically suppressed in modern surveys with multi-band photometry and
photometric redshift estimators, one still expects 
$5\%-20\%$ of photometrically selected clusters to suffer from serious 
projection
effects \citep{cohn07,rozo11}.  The importance of projection effects
increases with decreasing mass, so we expect it is projection effects
rather than survey depths that will ultimately set the detection mass
threshold for optical cluster finding in future surveys.

Unlike optical studies, X-ray cluster searches are nearly free from projection
effects.  This robustness to the presence of structures along the line of
sight reflects the fact that X-ray emission scales as density-squared, 
which 
enhances the relative contrast of a cluster in the sky, and it is the principal
reason that X-rays are considered the cleanest method for selecting
galaxy clusters.  The main difficulty for X-ray selection is a
technological one, specifically, the need for space-based observatories.  A
dramatic leap forward in capabilities will happen with the launch of \erosita,
which should detect $\approx 10^5$ galaxy clusters over the full sky out to
$z=1$ and beyond, ensuring that X-rays will continue to play a critical
role in the development of cosmologically relevant cluster samples over the
coming decade.  On a longer time scale, further improvements would require
X-ray observatories that
reach lower flux limits with higher angular resolution, both of which are
needed to detect large numbers of systems at $z\gtrsim 1$.

The primary advantage of SZ searches is that they do not suffer from
cosmological dimming. The SZ signature arises from
up-scattering of CMB photons by the hot intra-cluster plasma, and because
the number of up-scattered photons does not depend on the distance to the
cluster the signal is roughly redshift independent.  In practice, the SZ signal is not
exactly redshift independent because of residual sensitivity to the relative
size of the cluster and the beam of the telescope.  Unfortunately, achieving
sufficient sensitivity to detect low mass clusters in SZ is technologically
very challenging.  For instance, the current SPT, ACT, and \planck\ surveys
are expected to be complete at all redshifts
above mass thresholds of $7\times 10^{14}\,\msun$, $10^{15}\,\msun$,
and $2\times 10^{15}\,\msun$ respectively
\citep{vanderlinde10,marriage10,planck_esz};
while these limits will go down, they will not reach thresholds
comparable to those of X-ray or optical cluster selection.
Consequently,
while these experiments are currently the best avenue to probe the $z\approx
1$ massive cluster population, on a $3-5$ year time scale
the focus of cluster detection is likely to shift towards optical and
X-ray.  To our knowledge, there are no current plans to develop a new
generation of SZ survey instruments that would dramatically improve upon the
capabilities of current experiments for cluster detection, at least compared to
the differences in optical (e.g., DES vs. SDSS) and X-ray (\erosita\ vs \rosat).
However, both SPTpol and ACTpol 
should lead to significantly lower mass thresholds
for SZ cluster detection than the current SPT and ACT cluster samples.

\begin{figure} 
\begin{center}
\includegraphics[width=2.8in, height=2.4in]{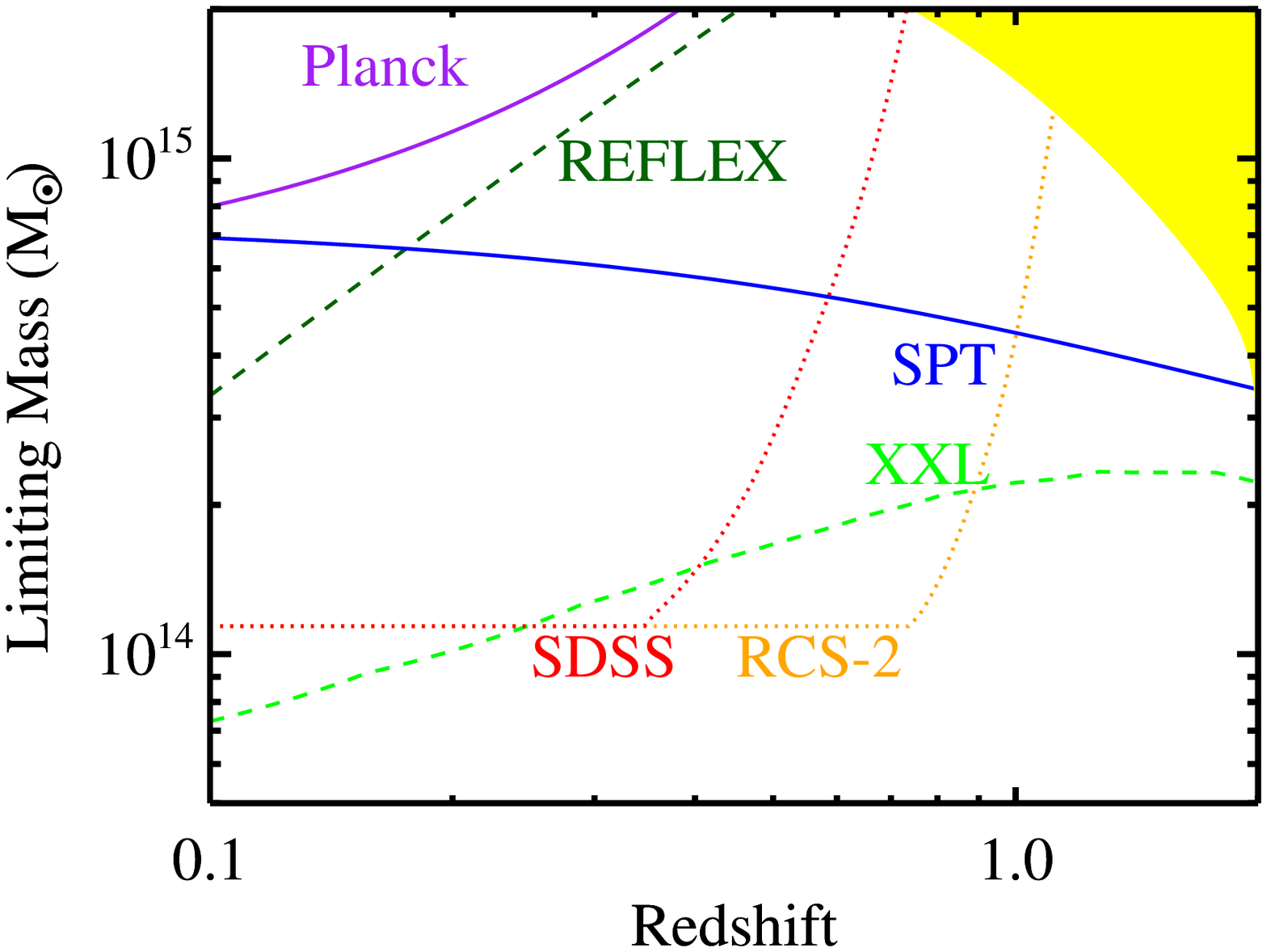} 
\includegraphics[width=2.8in,height=2.4in]{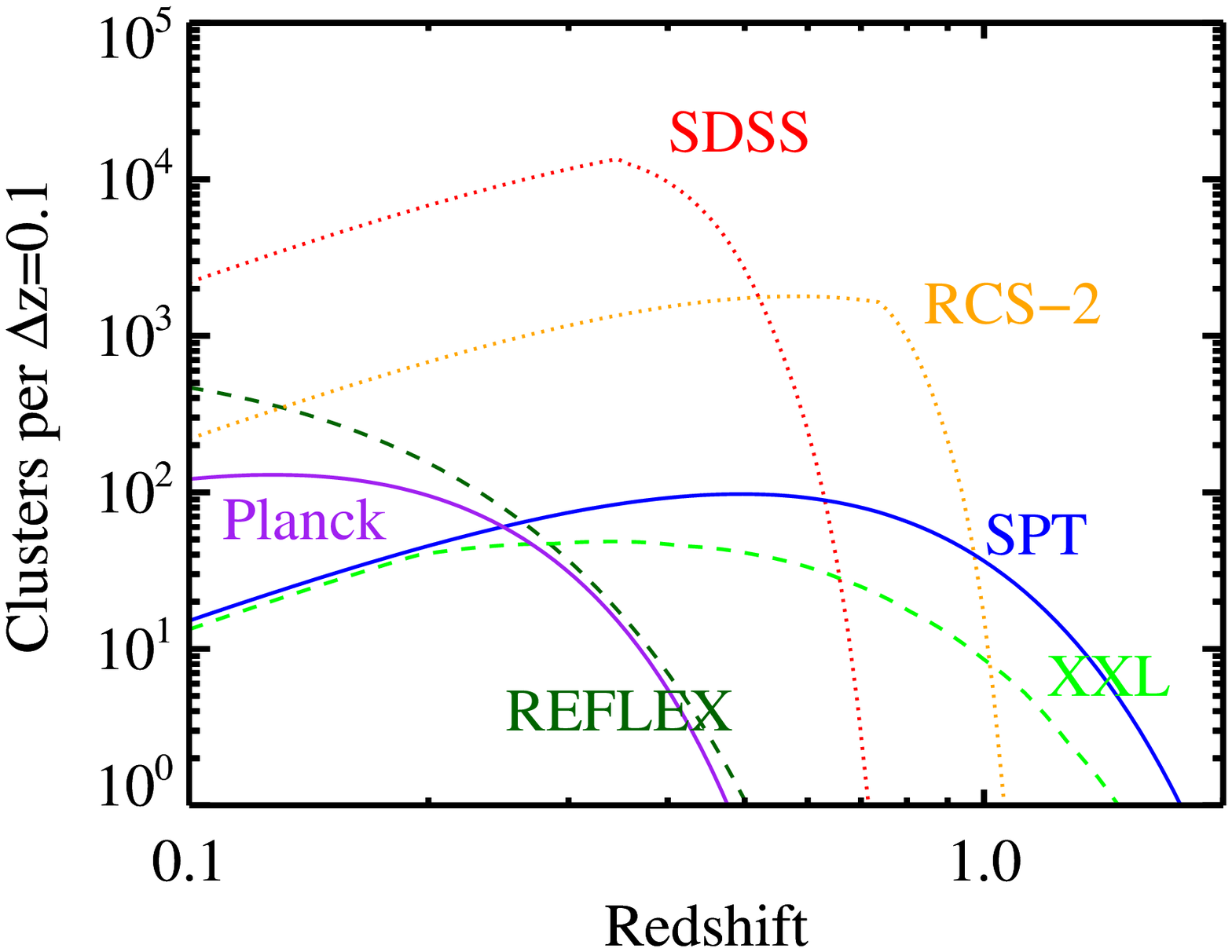}
\includegraphics[width=2.8in, height=2.4in]{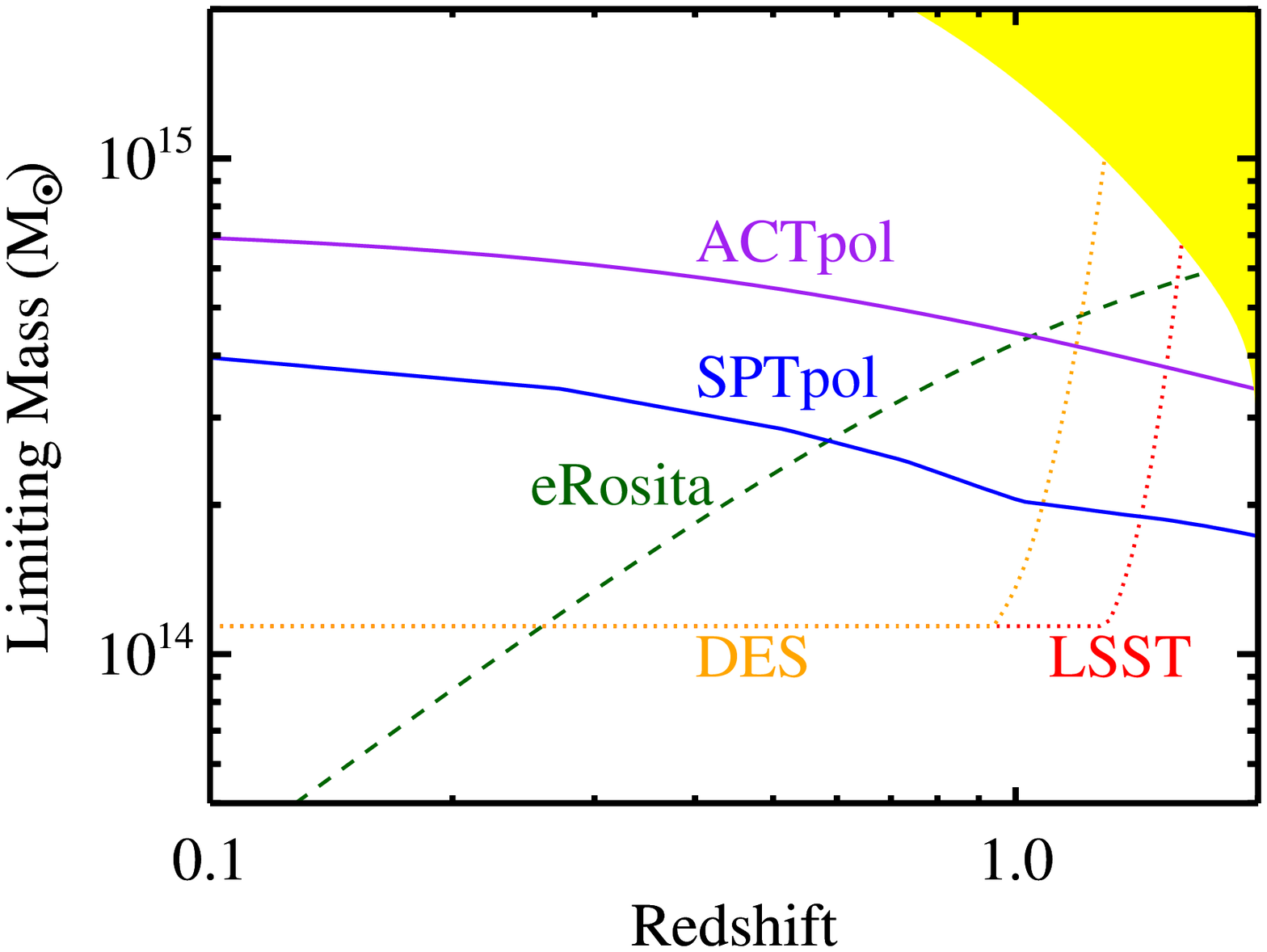} 
\includegraphics[width=2.8in,height=2.4in]{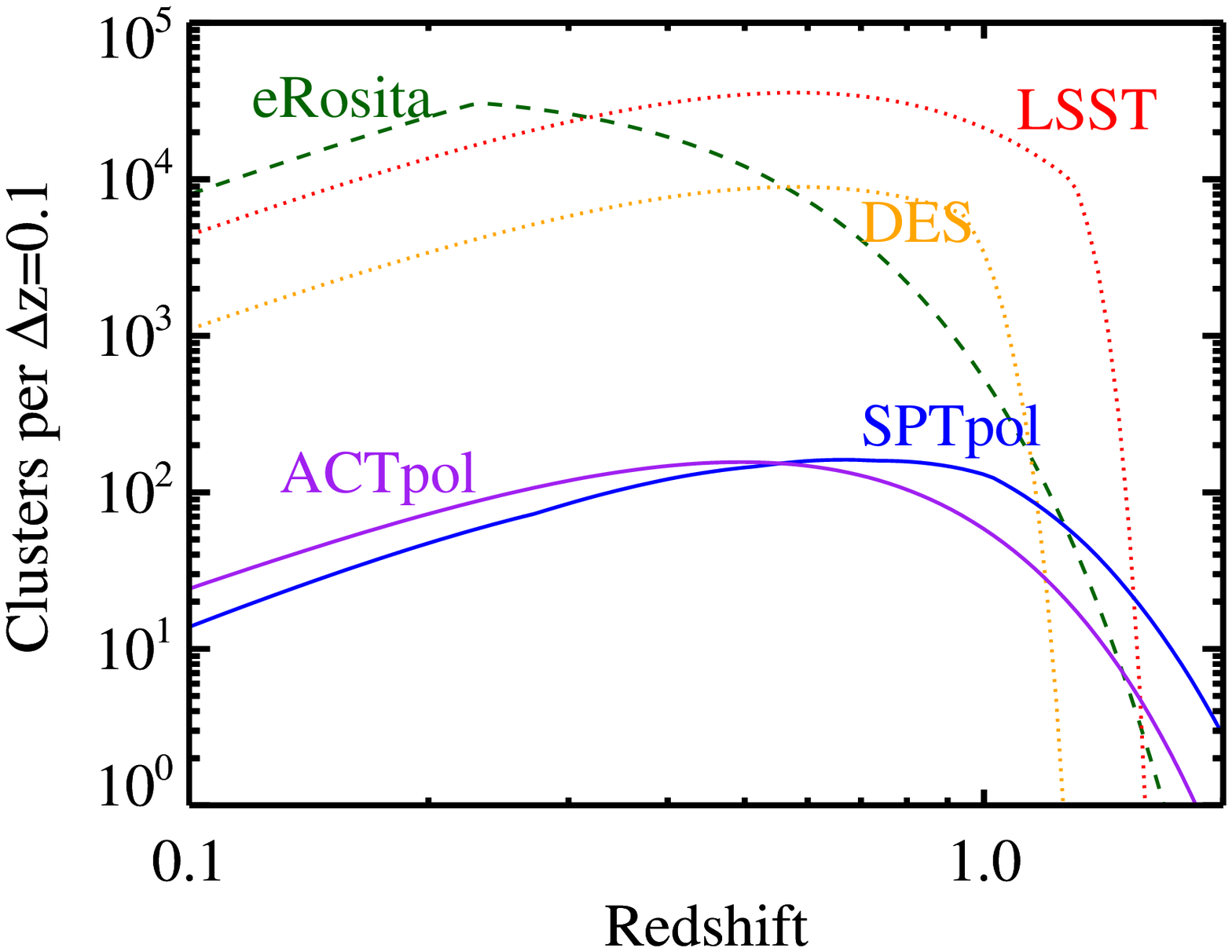}
\end{center}
\caption{Selection function for several representative 
cluster samples, as labelled.  
The top panels show surveys that are completed
or currently ongoing.  
The bottom panels show future surveys.
Left panels show the limiting mass as a function of redshift, 
while right panels show
the number of galaxy clusters above the limiting mass 
in redshift bins of width $\Delta z =0.1$.
The yellow region in the left panels corresponds 
to the area in parameter space where one
expects fewer than one galaxy cluster above the mass and redshift under consideration.
For the abundance plot, we consider the appropriate area for each of the surveys: 
30,000 deg$^2$ for the \erosita\ and \planck\ cluster samples, 
10,000 deg$^2$ for the REFLEX sample,
20,000 deg$^2$ for LSST, 
10,000 deg$^2$ for SDSS, 
5,000  deg$^2$ for DES, 
1,000  deg$^2$ for RCS-2,
2500  deg$^2$ for SPT, 
600   deg$^2$ for SPTpol, and 
4000 deg$^2$ for ACTPol.
The current ACT survey (not plotted)
is similar to SPT, with a somewhat higher mass threshold
and a $1000\mdeg^2$ survey area.
Different line types are used only to aid visual discrimination.
} 
\label{fig:selection} \end{figure}

Figure~\ref{fig:selection} showcases the difference of the cluster
populations from the various selection methods, where we have limited ourselves
to wide surveys ($1000\mdeg^2$ or higher) 
and have shown only a handful of representative
selection functions.   The top row shows the selection
functions for existing or ongoing surveys, while the
bottom-row shows the selection for future surveys.
The left panels shows the limiting
mass as a function of redshift for each of 
the surveys considered, while the right panels
shows the number above the limiting mass 
in a redshift bin of width $\Delta z=0.1$, accounting
for survey area.  We emphasize that in practice cluster samples never have 
a sharp mass threshold; the curves shown in Figure \ref{fig:selection}
are only roughly indicative of the mass and redshift ranges probed.
The number of clusters detected depends in detail on the selection cuts
applied, and small changes in threshold translate to larger changes
in abundance, so factor-of-two deviations from the projections
in Figure~\ref{fig:selection} would not be particularly surprising.

For the optical detection threshold we have 
assumed that projection effects limit 
useful cluster catalogs to a minimum richness $\lambda=15$
in the algorithm of \citet{rykoff11}, which counts galaxies
of luminosity $L\geq 0.2L_*$.  
To account for mass-richness scatter, we choose an effective
mass threshold that yields approximately the same space density
as this richness threshold.
The sharp upturn occurs when
$0.2L_*$ matches the magnitude limit of the survey.
In SZ, we 
see that the SPT mass threshold (kindly provided by the SPT collaboration,
and normalized to a total cluster yield of $\approx 700$ clusters at full depth)
is only mildly sensitive to redshift. 
The gentle decrease in limiting mass with increasing $z$ reflects
the fact more distant clusters subtend smaller angles that better
match the SPT beam size, and that clusters are hotter at fixed
mass with increasing redshift.
For \planck, conversely, the decreasing angular size of clusters
reduces sensitivity at higher redshift because the beam itself is large.
The curve shown is a rough estimate of the \planck\ Early SZ sample
\citep{planck_esz}, though the final selection will go considerably
lower in mass, because of both deeper data and lower S/N cuts.
The SPTpol curve is similar to SPT, but it reaches lower masses
over a smaller area, while the ACTpol curve
reaches similar noise levels to SPT 
\citep[$\approx 20\mu\mbox{K}$,][]{niemack10} over a larger area.
(ACTpol also plans a separate survey, deeper and narrower than SPTpol.)
Turning to X-rays, the 
REFLEX, XXL, and \erosita\ curves all show the 
increase of mass threshold with redshift characteristic of 
flux-limited surveys.  
The XXL selection is that of \citet{valageas11} scaled to match the 
observed density of C1 clusters in the XMM--LSS field \citep{pacaud07},
while the \erosita\ threshold represents a flux limit 
$\approx 4\times 10^{-14}\,{\rm erg}\,{\rm s}^{-1}$, corresponding
to $\approx 50$ photon counts \citep{pillepich11}.  The mass limit is higher by a factor of $\approx 3$
for clusters reaching $300$ photon counts.

%%%%%%%%%%%%%%%%%%%%%%%%%%%%%%%%%%%%%%%%%%%%%%%%
%%%%%%%%%%%%%%%%%%%%%%%%%%%%%%%%%%%%%%%%%%%%%%%%

%%%%%%%%%%%%%%%%%%%%%%%%%%%%%%%%%%%%%%%%%%%%%%%%
%%%%%%%%%%%%%%%%%%%%%%%%%%%%%%%%%%%%%%%%%%%%%%%%

Current wide X-ray samples are largely limited to massive systems at 
moderate redshifts, but narrow/deep samples reaching $z\approx 1$ and above
do exist. By comparison, the SDSS reaches lower mass over large areas
of the sky, but it only extends to $z\approx 0.5$.  RCS-2 reaches $z\approx 1$,
but over a smaller (though still quite significant) area.
The \planck\ SZ survey is largely limited to massive, moderate redshift
systems, while the SPT SZ survey 
has the best current sensitivity to high redshift clusters.  
In the near future, DES will extend the range of
optical identification to $z\approx 1$ over a large area, but 
\erosita\ should ultimately produce a larger sample.
While DES has a lower mass threshold over the range $0.3 < z < 1$,
the larger (all-sky) area of \erosita\ leads to a larger cluster total,
and \erosita\ should continue to detect clusters at $z>1$ where the
DES sensitivity declines rapidly.  On a longer
time scale, LSST will push the optical selection limit to $z\approx 1.5$, 
increasing the number of $z>1$ galaxy clusters by one to two orders 
of magnitude.

Another proposed method for detecting galaxy clusters is to search for
peaks in the weak lensing shear field.  
However, while massive halos produce local shear peaks,
shear peak statistics are known to suffer from severe projection effects:
many peaks arise from the superposition of multiple
halos along the line of sight.  Consequently, shear peak selection is not a
particularly effective method for selecting clusters of galaxies.  
That said, the shear peak abundance is an observable 
that can be predicted from numerical
simulations in much the same way as the halo mass function, and
this approach may well yield useful 
cosmological constraints
\citep[e.g.,][]{marian09,dietrich10}.  For the remainder of this review,
however, we focus on abundances of clusters identified by optical,
X-ray, or SZ methods.  We emphasize that stacked weak lensing
mass calibration of clusters
identified by other methods is {\it not} equivalent to shear peak
statistics, since cluster methods use the additional information 
afforded by 
baryonic density peaks to drastically reduce the impact of projection
effects on cluster selection.

%%%%%%%%%%%%%%%%%%%%%%%%%%%%%%%%%%%%%%%%%%%%%%%%
%%%%%%%%%%%%%%%%%%%%%%%%%%%%%%%%%%%%%%%%%%%%%%%%
%%%%%%%%%%%%%%%%%%%%%%%%%%%%%%%%%%%%%%%%%%%%%%%%
%%%%%%%%%%%%%%%%%%%%%%%%%%%%%%%%%%%%%%%%%%%%%%%%
%%%%%%%%%%%%%%%%%%%%%%%%%%%%%%%%%%%%%%%%%%%%%%%%
%%%%%%%%%%%%%%%%%%%%%%%%%%%%%%%%%%%%%%%%%%%%%%%%
%%%%%%%%%%%%%%%%%%%%%%%%%%%%%%%%%%%%%%%%%%%%%%%%
%%%%%%%%%%%%%%%%%%%%%%%%%%%%%%%%%%%%%%%%%%%%%%%%

\subsubsection{Calibrating the Observable--Mass Relation}
\label{sec:cl_mass_calibration}

The biggest challenge for cluster cosmology is characterizing the
observable--mass relation $P(X|M,z)$, where $X$ is a cluster observable that
is correlated with mass (e.g., richness, $Y_{SZ}$, $L_X$) and $P(X|M,z)$
is the probability that a halo of mass $M$ at redshift $z$ is detected as a
cluster with observable $X$.
This relation is usually described by parameters that specify
the mean relation, the rms scatter, and perhaps a measure
of skewness or kurtosis, all of which can evolve with redshift.
There are three general approaches to determining these parameters:
simulations, direct calibration, and statistical calibration.  

In the
simulation approach, one relies on numerical simulations to calibrate the
observable--mass relation \citep[e.g.][]{vanderlinde10,sehgal10}.  The main
difficulty that simulation methods face is our incomplete understanding of
baryonic physics, particularly galaxy formation feedback processes.  These
difficulties can be minimized by defining new X-ray observables that are
expected to be robust to these details, and through careful exploration of the
sensitivity of the observable--mass relation to the physics that goes into the
simulations \citep[e.g.,][]{nagai07,rudd09,stanek10,fabjan11,battaglia11}.  
The simulations themselves are steadily improving thanks to increased
computer power, more sophisticated algorithms, and the availability
of better data to test the input physics.  Despite these trends,
we think it unlikely that simulations
will achieve the $\sim 0.5-2\%$ 
level of accuracy required for cluster abundance experiments
to become statistics dominated in the next ten years.

The second approach to calibrating the observable--mass relation is the direct
method, in which a small subset of galaxy clusters have X-ray hydrostatic 
mass estimates and/or weak lensing mass estimates that are taken to represent
``true'' masses.
The observable--mass relation is directly calibrated on this
small subset of galaxy clusters, then applied to the general cluster
population \citep[][]{vikhlinin09,mantz10}.  
Unfortunately, hydrostatic mass estimates are 
themselves problematic 
because non-thermal pressure support (bulk motions, magnetic fields,
cosmic rays) is expected to bias them 
at the $\approx 10\%-20\%$ level \citep{lau09,meneghetti10},  
and it is not clear that
these biases can be predicted at the required level of accuracy.
We therefore suspect that hydrostatic estimates will play a steadily
decreasing role in future cluster abundance experiments.
Weak lensing mass estimates of individual clusters can in
principle be unbiased in the mean, but they are typically available
only for the most massive galaxy clusters in a given sample because
of limited signal-to-noise ratio.
In addition, even if the WL shape noise is small,
halo orientation and large scale structure
introduce irreducible noise in the mass estimates of individual 
clusters at the $20\%-30\%$ level \citep{becker11}. 
Nonetheless, ambitious efforts to achieve accurate weak lensing
masses for substantial samples ($\approx 50$) of X-ray or SZ-selected
clusters are likely to play a key role in improving cluster cosmological
constraints over the next few years
\citep{hoekstra12,vonderlinden12}.

The final approach to calibrating the observable--mass relation is statistical:
instead of relying on precise
mass estimates of a subsample of galaxy clusters, the 
relation is calibrated using additional observables for the
full sample that correlate
with mass.  One such statistical method uses the spatial clustering
of the clusters themselves, as characterized by the variance of
counts-in-cells \citep{lima04} or by the cluster correlation function
or power spectrum \citep{schuecker03,majumdar04,huetsi08}.  
Because the bias of halo clustering depends on mass
(Figure~\ref{fig:structure}), the amplitude and scale-dependence
of clustering provides information about the mass-observable relation.
Operationally, one parameterizes this relation, then uses standard
likelihood methods to
jointly fit for both cosmology and the $P(X|M,z)$ parameters
\citep{hu06,holder06}.  
These types of analyses are often referred to as ``self-calibration'' because
they do not require ``direct''
mass calibration data.  However, we think the descriptor ``statistical
mass calibration'' is more accurate.

The other statistical method we consider is stacked weak lensing, wherein one
measures the mean tangential shear of background galaxies around galaxy
clusters in a bin of fixed observable.  In other words, the stacked weak
lensing signal is the cluster--shear correlation function,  
which can be inverted to yield the mean 3-d mass profile of clusters
in the bin \citep{johnston07}.  Because this
measurement allows one to stack many clusters, one can easily obtain high
signal-to-noise measurements even for low mass clusters and large angular
distances \citep{mandelbaum08,sheldon09}.  
Since the underlying
halo population is randomly oriented relative to the line of sight,
stacked weak lensing mass calibration does not suffer from orientation biases
so long as the cluster identification itself does not preferentially
select halos oriented along a particular direction or aligned with
line-of-sight structure.
However, orientation biases in the cluster selection method will
probably exist to some degree, and they must
be calibrated carefully on simulations.  Finally, because this method relies on stacking
all galaxy clusters, it only provides information about the mean of the mass--richness relation,
so additional data are required to provide 
tight constraints on the scatter.\footnote{The distinction between
statistical calibration via stacked weak lensing and direct calibration
using weak lensing mass measurements is not a sharp one, and both methods
share the virtue that the relation between mass and weak lensing signal
is governed by well understood gravitational physics.  By
``stacked weak lensing'' we mean to emphasize the case where
(a) the WL measurements come from a large area imaging survey
that overlaps the cluster catalog (and may have been used to create it)
rather than from cluster-by-cluster follow-up observations, and
(b) the S/N of the mass measurement for any individual cluster
may be $\leq 1$, though the S/N for the ensemble is high.}

\begin{figure}  [t]
\begin{center}
\includegraphics[width=2.8in,height=2.4in]{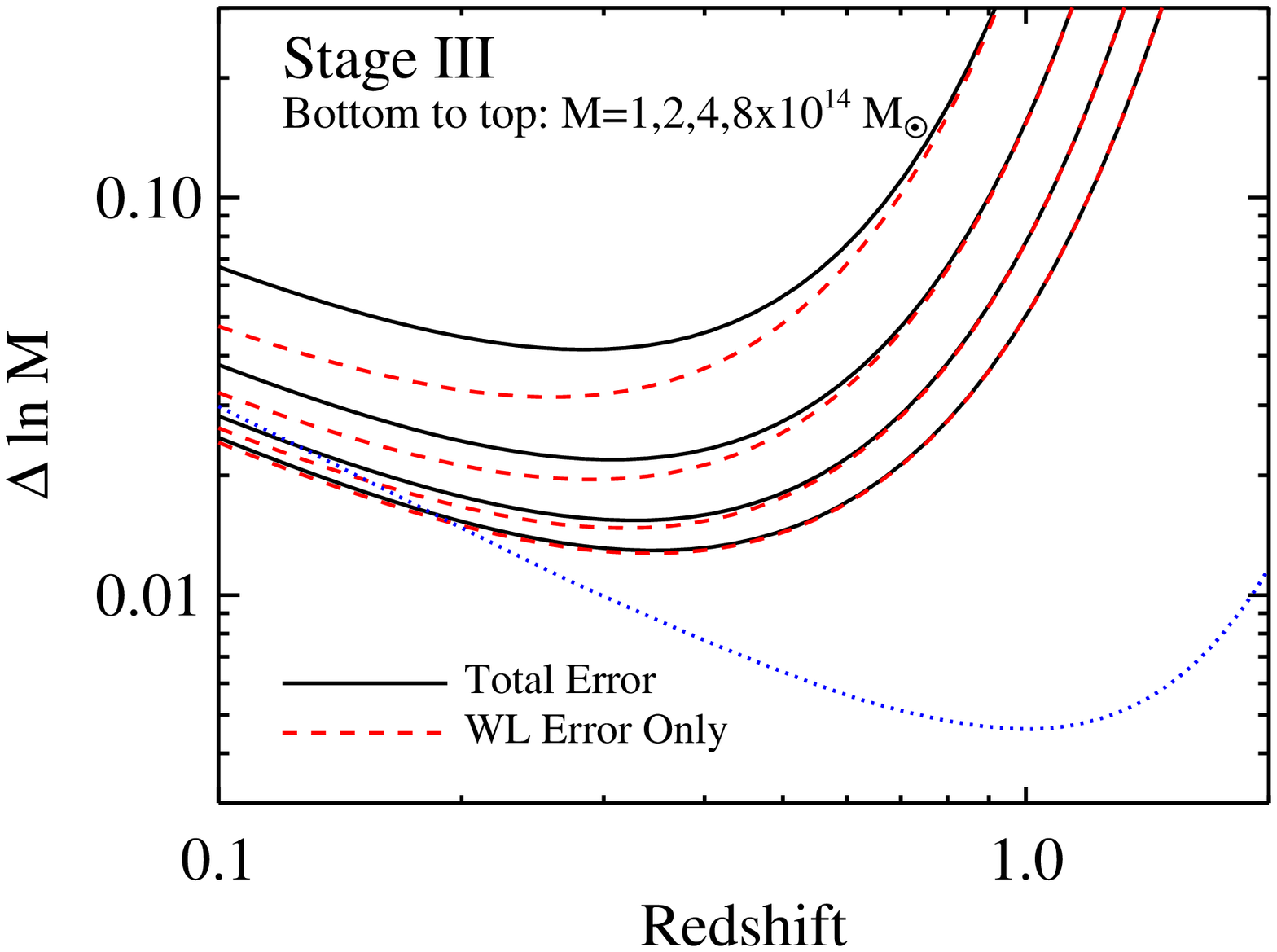} 
\includegraphics[width=2.8in,height=2.4in]{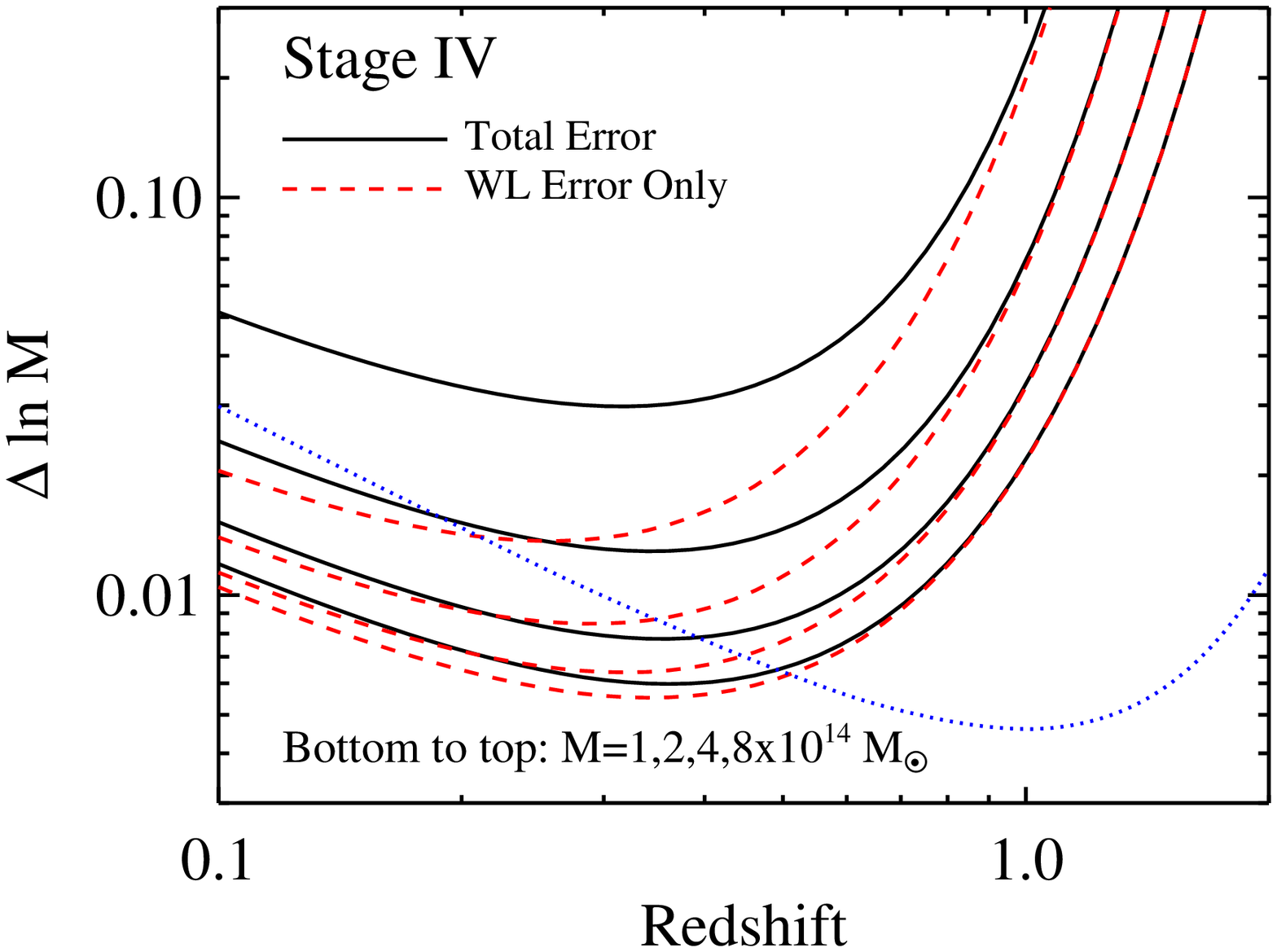} 
\end{center}
\caption{Mass uncertainty from
stacked weak lensing calibration as a function of redshift, assuming only WL 
shape noise (dashed red curves) 
and including sample variance due to intrinsic scatter between
WL mass and halo mass (solid curves).
For Stage III data (left) we assume
$\sigma_e=0.4$ and $\bar n_g =10$ galaxies/$\arcmin^2$, while for
Stage IV (right) we assume
$\sigma_e=0.3$ and $\bar n_g =30$ galaxies/$\arcmin^2$.  
For both cases we 
assume a $10^4\mdeg^2$ survey, and the redshift bin width is $z=z_c\pm 0.05$.
Each curve corresponds to a different mass threshold as labeled.  
The blue dotted line
shows the mass error corresponding to a statistics-limited cluster survey 
with a threshold
mass of $10^{14}\,\msun$, as per Figure \ref{fig:cumcounts}.  
The intersection between the blue dotted line and the lowest
solid black line marks the redshift at which a cluster 
abundance experiment with a threshold
mass of $10^{14}\ \msun$ transitions
from being dominated by the statistical error in cluster abundance
(at low redshift) to the 
error in the weak lensing mass calibration (at high redshift).
}
\label{fig:wl_merr} \end{figure}

Figure \ref{fig:wl_merr} shows the error in mass calibration that can be
achieved using stacked weak lensing for both ``Stage III'' (left panel) and
``Stage IV'' (right panel) observations, calculated via the methodology
described by \citet{rozo10b}.  Briefly, we assume a source redshift distribution
appropriate for DES-like survey depth, and we sum over all annuli within 
the radius $2R_{200}$, which is a rough approximation for the location 
of the one-to-two halo transition of the matter correlation function
using the \citet{hayashi08} model.  
(Other studies, e.g.\ \citeauthor{tavio08} [\citeyear{tavio08}],
also find that one-halo regime of the mass profile
extends well beyond $R_{200}$.) 
For our Stage III estimates we
assume an intrinsic shape noise $\sigma_e=0.4$ and 
source galaxy surface density $\bar n_g = 10\,\arcmin^{-2}$,
while for Stage IV we assume $\sigma_e=0.3$ and $\bar n_g = 30\,\arcmin^{-2}$.
Note that the corresponding tangential shear error is 
$\sigma_\gamma \approx \sigma_e/\sqrt{2}$.
These values correspond roughly to expectations for
DES data and \euclid/\wfirst\ data, respectively; 
the lower $\sigma_e$ for the latter reflects higher image
quality, though the partition of this improvement between $\sigma_e$
and $\bar{n}_g$ is somewhat arbitrary.
LSST falls between these two cases but closer to Stage IV.  
We assume that clusters have NFW mass profiles (\citealt{navarro96}),
and we include the decrease 
in background source density with increasing cluster redshift.
In all cases, the redshift distribution is set to 
\begin{equation}
F(z) \propto z^2 \exp \left[ - (z/z_*)^2 \right] 
\end{equation}
with $z_* = 0.5$.  This is appropriate for DES and 
underestimates the redshift depth for LSST, which will 
result in a slight overestimate of the statistical 
uncertainties for Stage IV experiments, particularly 
at the highest redshift bins.

In each panel of Figure~\ref{fig:wl_merr},
dashed red curves show the error from shape noise
alone, while solid curves include the intrinsic scatter between
noiseless WL mass estimates and true three-dimensional halo masses, 
a consequence of non-spherical mass distributions,
which we add in quadrature to the shape noise assuming
an intrinsic scatter per cluster of $\sigma_{\rm wl}=0.3$ \citep{becker11}.
The two curves separate when the number of sources is high enough
to measure individual clusters with S/N$\sim 3$.
We assume the stacked weak lensing signal uses all halos within a redshift bin
$z=z_c\pm 0.05$ and above a given mass threshold as labeled.  
The forecast mass errors are marginalized over concentration.
The improvement in precision with decreasing mass is driven by the rapid 
increase in the number of halos as the mass threshold decreases. 
For mass thresholds $1-2 \times 10^{14} \msun$, calibration at the 1-2\%
level is achievable in principle with Stage III data and at the sub-percent
level with Stage IV data.  These are errors per $\Delta z = \pm 0.05$
bin, so if one assumes a smooth, parameterized evolution of $P(X|M,z)$
it may be possible to constrain the overall normalization more
tightly.  Conversely, some forms of WL systematics (e.g., uncertainty
in the shear calibration or source redshift distribution) could introduce
mass calibration errors correlated across redshift bins.  
The results in Figure~\ref{fig:wl_merr}
are broadly consistent with those from the more detailed treatment by
\citet{oguri11b}.

%%%%%%%%%%%%%%%%%%%%%%%%%%%%%%%%%%%%%%%%%%%%%%%%
%%%%%%%%%%%%%%%%%%%%%%%%%%%%%%%%%%%%%%%%%%%%%%%%

%%%%%%%%%%%%%%%%%%%%%%%%%%%%%%%%%%%%%%%%%%%%%%%%
%%%%%%%%%%%%%%%%%%%%%%%%%%%%%%%%%%%%%%%%%%%%%%%%

Comparing Figures \ref{fig:cumcounts}d and \ref{fig:wl_merr}, we see that
Stage IV weak lensing data can in principle calibrate the mean
relation well enough that a $10^4\mdeg^2$ cluster survey would
be limited by the statistical uncertainty in abundance for
$z \la 0.5$, though mass calibration error would dominate
at higher redshift.   
(The abundance error and weak lensing calibration error both
scale with area as $A^{-1/2}$.)
The statistics limit for $M=10^{14}M_\odot$ 
from Figure \ref{fig:cumcounts}d is shown
in Figure \ref{fig:wl_merr} as the blue dotted line.
Stage III weak lensing data fall short of this
goal by a factor $\sim 3$, but they can still achieve
powerful constraints on $\sigElevz$ (see Figure~\ref{fig:wl_serr} below).

The general trends in Figure~\ref{fig:wl_merr} can be 
understood using simple arguments.  For a singular isothermal sphere 
(SIS) of velocity dispersion $\sigma_V\propto M^{1/3}$, the tangential
shear is $\gamma(\theta) = \theta_E/2\theta$, where $\theta$
is the angular distance to the cluster center, and $\theta_E$ is the 
Einstein radius.
The Einstein radius is related to the velocity dispersion via \citep{fort94}
\begin{equation}
\label{eqn:einsteinring}
\theta_E = 4\pi \left( \frac{\sigma_v}{c} \right)^2 \frac{D_{ls}}{D_s} 
    \approx 0.07\ \mbox{arcmin} \left( \frac{\sigma_V}{550\kms} \right)^2 
    \left({D_{ls}/D_s \over 0.5}\right)~,
\end{equation}
where $D_s$ is the distance to the source, 
$D_{ls}$ is the distance to the source as seen from the lens,
and we have scaled to a typical value of their ratio.
We have also scaled equation~(\ref{eqn:einsteinring}) to the (1-dimensional)
velocity dispersion of a $2\times 10^{14} M_\odot$ cluster at $z=0.5$.
Each source galaxy gives a low S/N estimate of 
$\gamma$ and hence of $\theta_E=2\theta\gamma$.
The variance of this estimate is
$\Var(\hat \theta_E)=2\theta^2 \sigma_e^2$, 
where $\sigma_e=\sqrt{2}\sigma_\gamma$ is the WL shape noise.
The number of source galaxies in a logarithmic angular interval
d$\ln\theta$ is $2\pi {\bar n}_g\theta^2 d\ln\theta$, so each 
such interval contributes equally to the S/N on $\theta_E$, from
$\theta_{\rm min}$ where the weak lensing approximation fails
to $\theta_{\rm max}$, the angular extent of the cluster.
The variance of the estimate for an individual cluster is thus
\begin{equation}
\Var(\hat \theta_E) = 
\frac{2 \theta^2 \sigma_e^2}{2\pi{\bar n}_g \theta^2
     \ln \left(\theta_{\rm max}/\theta_{\rm min}\right)}~,
\end{equation}
and the variance for $N$ clusters is smaller by $N$.
As representative values we take $\theta_E=0.07\,\arcmin$, 
$\theta_{\rm min} = 5\theta_E = 0.35\,\arcmin$ 
(so $\gamma_{\rm max} = 0.1$), and
$\theta_{\rm max} = 6.5\,\arcmin$, the angle subtended by a radius $R=2R_{200}$
at $z=0.5$ (for $M = 2\times 10^{14} M_\odot$), 
yielding $\ln(\theta_{\rm max}/\theta_{\rm min}) \approx 3$.
Since $\theta_E \propto \sigma_V^2 \propto M^{2/3}$,
$\Delta\ln M = 1.5\Delta\ln \theta_E$, with 
$\Delta\ln \theta_E = \theta_E^{-1}[\Var({\hat \theta_E})]^{1/2}$.
Putting these results together yields a total shape noise error at $z=0.5$ of
\begin{eqnarray}
\label{eqn:cl_wlmasserr}
\Delta\ln M & \approx & 1.5 N^{-1/2} 
    \left[{ \sigma_e^2 \over \pi \bar{n}_g \theta_E^2 
            \ln(\theta_{\rm max}/\theta_{\rm min}) }\right]^{1/2} \\
    & \approx &
    6\times 10^{-3} \left({N\over 4000}\right)^{-1/2} 
    \left({\sigma_e \over 0.3}\right)
    \left({M \over 2\times 10^{14}\ \msun}\right)^{-2/3}
    \left({\bar n_g \over 30\,\arcmin^{-2}}\right)^{-1/2}
    \left({D_{ls}/D_s \over 0.5}\right)^{-1}~.  \hspace{0.2in}
\end{eqnarray}
This error estimate is 25\% smaller than the value plotted in 
Figure~\ref{fig:wl_merr} (which shows $\Delta\ln M \approx 0.008$
at $z=0.5$ from shape noise alone),
in part because the
surface density of sources {\it behind} the clusters is lower than
$\bar n_g$,  and in part because marginalizing over the NFW
concentration parameter further increases the mass error. 
Including the dependence of $N$ on mass threshold,
equation~(\ref{eqn:cl_wlmasserr}) implies
\begin{equation}
\Delta \ln M_{\rm shape} \propto \theta_E^{-1}N^{-1/2} \propto M^{-2/3+\alpha/2}
~,
\end{equation}
where $\alpha$ is the mass function slope shown in Figure~\ref{fig:errors}.
For $\alpha \geq 4/3$, which is always satisfied for 
$M\geq 10^{14}\,\msun$,
the increase in abundance at lower
masses outweighs the lower S/N per cluster, yielding 
higher precision at lower mass threshold as
seen in Figure \ref{fig:wl_merr}.
To obtain the total noise, one simply adds the intrinsic weak-lensing 
noise $\sigma_{\rm wl}N^{-1/2}$ in quadrature to the shape noise.

Multi-wavelength
studies of galaxy clusters also allow for statistical mass calibration from
cross-correlation studies.  Just as the clustering of clusters
is a mass-dependent observable, so too are the abundance functions of
different observables.  Consequently, overlapping surveys allow for the
possibility of measuring the abundance of galaxy clusters as a function of
{\it two} observables $X_1$ and $X_2$.  While an overall shift in the normalization
of the multi-variate observable--mass relation $P(X_1,X_2|M)$ is still degenerate
with cosmology, the addition of the clustering signal --- which depends on cluster masses
directly --- allows one to jointly calibrate $P(X_1,X_2|M)$ while still improving the cosmological
constraints relative to those derived from a single observable \citep{cunha09}.
The improvement is driven by the fact that
using two cluster observables simultaneously allows one to 
better constrain the scatter of the observable--mass relation \citep[see also][]{stanek10}.
Given the large overlap
between many of the currently ongoing or near future cluster surveys 
(e.g., DES fully overlaps
with SPT), we
expect this type of analysis to become increasingly important in the coming
decade.

It remains to be seen whether statistical calibration of the mean
observable-mass relation via clustering can compete with stacked
weak lensing calibration, but we suspect that the answer is no based on
the following approximate argument.  If the cluster bias factor is
measured with uncertainty $\Delta\ln b$, then the corresponding mass
scale uncertainty is $\Delta\ln M \approx \eta^{-1} \Delta\ln b$,
where $\eta \equiv d\ln b/d\ln M \approx 0.4-0.5$ is the logarithmic 
slope of the bias-mass relation for cluster mass halos.  We have
computed $\Delta\ln b$ for an optimally weighted measurement of
cluster pairs in a wide radial bin, $20\Mpc < R < 100\Mpc$ (comoving),
considering {\it only} Poisson pair count errors, not sample variance
errors.  For our usual $\Delta z = 0.1$ redshift bin over $10^4\mdeg^2$,
centered at $z=0.5$, we find that the corresponding 
$\Delta\ln M$ rises from 6\% for
a $10^{14}M_\odot$ threshold to $\sim 50\%$ for a $4\times 10^{14}M_\odot$
threshold, much worse than our estimated errors for Stage III stacked
weak lensing calibration shown in Figure~\ref{fig:wl_merr}.
Cross-correlation with a much denser galaxy sample might evade
this argument by allowing higher precision bias measurements,
but sample variance will set a floor to these errors, 
and the bias of the cross-correlation sample must also be known.
Our expectation is that clustering may well help constrain the scatter
given mass constraints from weak lensing, but that it will
prove insufficiently powerful to pin down the mass scale of
clusters on its own.

In practice, the distinction between simulation, direct, and statistical mass
calibration is somewhat artificial. One can use simulation and direct mass
calibration to place priors on the observable--mass relations, then use
statistical methods to arrive at the final constraint.  
High quality observations
of individual clusters can provide important information about the
scatter of the observable--mass relation, a quantity that is only indirectly
constrained via statistical calibration methods.  Conversely, we expect
that only statistical methods, and particularly stacked weak lensing, are
likely to achieve the $\approx 1\%$ mass scale 
accuracy demanded by Stage IV experiments.
To the extent that this is true, optical imaging of galaxy clusters will be a
necessary component of all future cluster surveys, not just for redshifts, but
also for cluster mass calibration.
Conversely, imaging surveys conducted for WL studies of cosmic
acceleration will automatically enable cluster studies.

With spectroscopic follow-up data or an overlapping galaxy
redshift survey, one can also try to calibrate cluster observable-mass
relations using virial mass estimators \citep{heisler85},
``hydrostatic'' estimators for the galaxy population
\citep{carlberg97}, or ``velocity caustics'' that mark
the boundary between galaxies bound to the cluster potential and
galaxies above the escape velocity
\citep{regos89,diaferio99,rines03}.
The key systematic issue for this approach is the possible influence
of galaxy formation physics on the velocity field and velocity
dispersion profile, though \cite{diaferio99} argues that these effects
should be small for velocity caustics.  These approaches can 
again be applied in either a ``direct'' mode for individual clusters
or a ``statistical'' mode using velocity distributions
measured for large samples.  Studies to date have not established
the robustness of these methods at the few-percent level needed for
future progress, but with the large spectroscopic surveys underway
or planned for dark energy measurements the approach merits further
investigation \citep[e.g.,][]{white10,saro12}.
\cite{zu12} show that the mean radial infall profile for clusters
can be extracted from measurements of the redshift-space
cluster-galaxy cross-correlation function, which may provide
a practical route to implementation.  
Even if the calibration precision from redshift-space distortions
is lower than that from stacked weak lensing, comparison of the
two enables tests of modified gravity models that
predict differences between the potentials affecting lensing
and non-relativistic motions (see \S\ref{sec:gravity}).

%%%%%%%%%%%%%%%%%%%%%%%%%%%%%%%%%%%%%%%%%%%%%%%%
%%%%%%%%%%%%%%%%%%%%%%%%%%%%%%%%%%%%%%%%%%%%%%%
%%%%%%%%%%%%%%%%%%%%%%%%%%%%%%%%%%%%%%%%%%%%%%%%
%%%%%%%%%%%%%%%%%%%%%%%%%%%%%%%%%%%%%%%%%%%%%%%%
%%%%%%%%%%%%%%%%%%%%%%%%%%%%%%%%%%%%%%%%%%%%%%%%
%%%%%%%%%%%%%%%%%%%%%%%%%%%%%%%%%%%%%%%%%%%%%%%%

\subsection{Systematic Uncertainties and Strategies for Amelioration}
\label{sec:cl_systematics}

If $X$ is a cluster observable correlated with mass, and $P(X|M,z)$
the mass-observable relation discussed in \S\ref{sec:cl_mass_calibration},
then the expected number of clusters in a volume $V$ at redshift $z$
above a threshold $X_{\rm min}$ is
\begin{equation} 
\label{eq:abundance}
N(X_{\rm min},z) = \int_{X_{\rm min}}^\infty dX {dN \over dX} =  
                 \int_{X_{\rm min}}^\infty dX \int_0^\infty dM\,V(z)\,
		 {dn(z)\over dM}P(X|M,z),
\end{equation}
where $dn(z)/dM$ is the halo mass function at redshift $z$.
From equation~(\ref{eq:abundance}) we can identify several sources
of potential systematic uncertainties: errors in cluster redshifts,
incompleteness and contamination that produce extended non-Gaussian
tails of $P(X|M,z)$, the form and calibration of the ``core'' of
$P(X|M,z)$, and the theoretical prediction of $dn/dM$ itself.
We discuss each of these categories in turn.

%%%%%%%%%%%%%%%%%%%%%%%%%%%%%%%%%%%%%%%%%%%%%%%%
%%%%%%%%%%%%%%%%%%%%%%%%%%%%%%%%%%%%%%%%%%%%%%%%
%%%%%%%%%%%%%%%%%%%%%%%%%%%%%%%%%%%%%%%%%%%%%%%%
%%%%%%%%%%%%%%%%%%%%%%%%%%%%%%%%%%%%%%%%%%%%%%%%
%%%%%%%%%%%%%%%%%%%%%%%%%%%%%%%%%%%%%%%%%%%%%%%%
%%%%%%%%%%%%%%%%%%%%%%%%%%%%%%%%%%%%%%%%%%%%%%%%

\subsubsection{Redshift Uncertainties}
\label{sec:cl_photoz}

Equation~(\ref{eq:abundance}) implicitly assumes that all clusters
are assigned the correct redshifts.  As cluster samples grow to the
tens and even hundreds of thousands, obtaining spectroscopic redshifts 
for all systems becomes impractical, and photometric
redshifts are essential.  Fortunately, clusters contain many galaxies
with uniform (red-sequence) colors, allowing precise
and accurate photo-$z$'s.  
\citet{lima07} estimated the level at
which the bias and scatter of photometric redshift errors must be
controlled in a Stage III dark energy experiment so as to not degrade
cosmological information, finding that the rms scatter must be
held to $\sigma_z \leq 0.03$ and that any bias in the mean photo-$z$
must be held below $\Delta z = 0.003$.
Current cluster photometric redshift estimates have a
dispersion of $\approx 0.01$ \citep[e.g.][]{koester07}, so controlling the
scatter at the $0.03$ level is not particularly problematic.  The bias on the
mean is more challenging, but current catalogs
do achieve close to the necessary accuracy.  For instance, the bias
of the SDSS maxBCG catalog, measured by comparing cluster photo-$z$'s
to spectroscopic redshifts, is $\approx 0.004$ \citep{koester07}.
We expect these successes will still hold as we push to higher redshifts,
so cluster photometric redshift errors 
are unlikely to be a significant source of systematic
uncertainty in abundance studies, at least for samples below $z\approx 1$.
Above this redshift, the $4000$\AA\ break feature in the spectrum of early-type
galaxies red-shifts into the IR, and the photometric redshift accuracy 
will become
more difficult to control at the required level unless near IR data are
available.
X-ray and SZ cluster samples require deep multi-band optical imaging
and/or spectroscopic follow-up to achieve these errors.  In particular, while
the use of iron lines in X-ray spectroscopy has proven to a reliable indicator
of cluster redshift \citep[e.g.][]{yu11}, the accuracy achieved by these methods is only of order 
$\approx 0.03$, with a not-insignificant outlier fraction, and even then this requires
a significant number of photon counts.
Nevertheless, for high redshift systems without IR data 
this information is often
the only indicator of a cluster's redshift, and it
can therefore play a critical role.

\subsubsection{Contamination and Incompleteness: The Tails of $P(X|M,z)$}
\label{sec:cl_tails}

Equation~(\ref{eq:abundance}) assumes a one-to-one match between
halos and observable clusters.  In practice, any observed cluster
catalog suffers some degree of contamination, the presence 
of systems whose true halo mass is far below the value suggested
by the observable $X$.
Cluster catalogs are also affected by incompleteness, halos
whose corresponding observable $X$ is anomalously low so that
they are assigned masses far below their true masses, or perhaps
fail to make it into the catalog at all.
Thus, we can think of contamination and incompleteness as
characterizing the extended non-Gaussian tails of $P(X|M,z)$.   

Significant levels of contamination and incompleteness can be
tolerated provided that they are well calibrated.
A contamination fraction $C$ increases the estimated cluster abundance by a
factor $(1+C)$ relative to the true value, while an incompleteness
fraction $I$ reduces the estimated abundance by a factor $(1-I)$.
To prevent them becoming the limiting factor in cluster abundance
measurements, the product $(1+C)(1-I)$ must be determined to
a fractional accuracy that is smaller than the uncertainty in the 
cluster space density, roughly $N^{-1/2}$ if limited by cluster 
statistics or $\alpha\Delta \ln M$ if limited by mass calibration 
uncertainty.

Contamination can also impact mass calibration \citep{cohn07,erickson11}. 
In the simplest case, if $\bar M$ is the mean mass of a sample of 
clusters selected by some range of observable and
contaminating clusters have mass $M \ll \bar M$, they dilute
the sample and reduce the mean mass inferred from calibration
by a factor $(1+C)$.
Incompleteness, on the other hand, should not affect the estimated
mean mass of a galaxy cluster sample, provided that the reason 
a cluster of given $X$ 
fails to be detected is not correlated with its halo mass.
Keeping the impact of contamination uncertainty sub-dominant
requires that the contamination level be known to
$\Delta C \approx \Delta\ln(1+C) \leq \Delta\ln M$.
This is a stiffer requirement than that on the product
$(1-I)(1+C)$, by a factor of $\alpha \approx 3$, so it will be
more difficult to achieve in practice.

Different cluster finding techniques are sensitive to different sources
of contamination and incompleteness.  In X-rays, the principal contaminants
are X-ray
point sources (AGNs), which can be effectively removed from cluster
catalogs by demanding that galaxy clusters be detected as spatially extended
emission.  With this cut, the fraction of galaxy clusters where AGNs have a
significant impact on the cluster emission is 
$\lesssim 5\%$ \citep{burenin07,mantz10}.  
The few percent contamination level of today's X-ray cluster surveys
is not an important systematic relative to mass calibration uncertainty.
However, the demands will be stiffer for \erosita, so whether AGN
contamination will continue to be a negligible systematic in the
future remains to be seen.
Incompleteness (in the sense of clusters that reside in non-Gaussian tails) 
is a source of possible concern, since \erosita\ will
probe significantly lower cluster masses than current X-ray surveys,
and the regularity of the intracluster medium could break down at
lower halo masses because of greater importance of radiative cooling
or galaxy and AGN feedback.
However, \chandra\ studies of group-scale 
systems show that the scaling relations of galaxy clusters extend down to 
$M\approx 4\times 10^{13}\ \msun$ \citep{sun09}, 
so \erosita\ should be able to use the vast majority of
all X-ray selected groups and clusters for cosmological investigations.
As usual, the largest open question is accuracy of the mass calibration.

Because SZ clusters work in the low S/N limit, 
with typical detections being
$\approx 5\sigma$, SZ cluster samples typically can contain a
few false detections --- sources 
that do not correspond to massive galaxy clusters
but rather reflect
the stochastic nature of the CMB and/or instrumental noise.  
However both of these
sources of stochasticity can be very well characterized, so 
we do not expect them to be a limiting systematic: their 
impact on $P(X|M,z)$ is calculable.  Radio emission
by point sources and/or dusty star forming galaxies can 
systematically reduce the SZ signal of clusters, but these 
effects are expected to fall 
below the $10\%$ level \citep[e.g.,][]{vanderlinde10}.  
Further study of the ongoing SZ surveys 
will better illuminate the impact that such sources 
can have on cosmological constraints from SZ 
cluster samples.   
Contamination by intrinsic CMB fluctuations and point sources
are both mitigated by multi-frequency observations, 
since the SZ effect has a distinct spectral signature.
While contamination and incompleteness of SZ samples remains an area
of active research, we think these effects are unlikely to
compete with mass calibration as a limiting uncertainty.

For optical cluster searches the primary source of contamination 
is projection effects --- two or more small halos lining up to
produce the apparent galaxy overdensity of a larger halo.
These projections can arise from truly random superpositions
or from galaxies or groups that lie in the same filamentary
structure but not within the virial radius of a common halo.
Even with galaxy spectroscopic redshifts, projection effects in
the optical can produce contamination levels of 5\%-20\%
depending on the richness threshold \citep{cohn07,rozo11};
in a direct comparison of optical and X-ray catalogs, 
\cite{andreon11} conclude that the contamination of the
former is 10\% or less.
The principal reason that projection effects are more
important in optical catalogs than in X-ray or SZ catalogs is that optical
catalogs tend to reach significantly lower mass thresholds at 
high redshift, which results in higher surface densities of clusters
and therefore stronger projection effects.  
In fact, projection effects may well set the lower mass threshold
at which cosmological analyses with optical clusters are possible.
We anticipate that
incompleteness and contamination can be adequately modeled 
through the use of realistic mock catalogs constructed
using numerical simulations, provided they are constructed
to match the clustering data of the survey under consideration.
These mock catalogs can be
analyzed using the same algorithms applied to the observational data,
allowing one to quantitatively characterize the impact of projection
effects.  Many of the most recent optical analysis
draw on such detailed mock catalogs, but greater accuracy
will be needed for next generation surveys.

The impact of contamination on weak lensing mass calibration is
somewhat subtle, and probably weaker than the naive expectation
of depressing the estimated mass by $(1+C)$ through dilution.
When superposed galaxy groups masquerade as a single more massive cluster,
their projected mass distributions are also superposed,
and the lensing signal from this blend may be close to
the signal that would come from a cluster of the combined richness.
The net impact must again be evaluated with detailed mock catalogs.

%%%%%%%%%%%%%%%%%%%%%%%%%%%%%%%%%%%%%%%%%%%%%%%%
%%%%%%%%%%%%%%%%%%%%%%%%%%%%%%%%%%%%%%%%%%%%%%%%
%%%%%%%%%%%%%%%%%%%%%%%%%%%%%%%%%%%%%%%%%%%%%%%%

\subsubsection{Calibrating the Core of $P(X|M,z)$}
\label{sec:cl_core_calibration}

In addition to characterizing extended tails of the mass-observable
relation, one must calibrate the ``core'' of $P(X|M,z)$, where
scatter arises from physical variations in cluster properties
at fixed halo mass, from observational noise, and from low
level contamination that produces small random fluctuations in 
the observable.  These effects are typically assumed to produce
a log-normal form of $P(X|M,z)$, i.e., Gaussian scatter in $\ln X$
at fixed $M$.  The calibration task is then to determine the
mean relation 
$\langle \ln X | M,z\rangle$ 
and the variance
$\Var(\ln X|M,z)$,
and to characterize any deviations from
log-normal form that are large enough to affect the predicted
abundance.  As the notation indicates, the relation can evolve
with redshift, and the scatter and non-Gaussianity may depend
on halo mass at fixed redshift.  

We consider each of the relevant terms in turn, starting with the
mean observable--mass relation.
We have already expressed our view that statistical calibration
methods, and stacked weak lensing in particular, are the most
promising route to meeting the stringent demands of next-generation
cluster surveys.  \cite{cunha09b} and \cite{oguri11} show
that this approach allows the mass and redshift dependence of
$\langle \ln X | M,z\rangle$ and $\Var(\ln X|M,z)$ to
be parameterized in an extremely flexible way while 
retaining enough information to yield strong
cosmological constraints.

If the mean mass-observable relation is calibrated using stacked weak
lensing, then the systematic effects discussed for WL in 
\S\ref{sec:wl_systematics} are also sources of uncertainty
for cluster studies.  In particular, errors in the source galaxy
redshift distribution and/or shear calibration will shift
the inferred cluster mass scale.  
For these systematics to be insignificant, the rule of thumb is that the
uncertainty in the mean inverse critical surface 
density $\avg{\Sigma_{\rm crit}^{-1}}$ of the source galaxies and the 
error in the shear calibration
must be smaller than the mass errors plotted in 
Figure~\ref{fig:wl_merr}, divided by 1.5.  
The 1.5 factor comes in because an error in 
$\avg{\Sigma_{\rm crit}^{-1}}$ or shear calibration
uniformly biases the recovered cluster density profile
and therefore biases the estimate of $R_{200}$.
A bias $b$ in the mass at a fixed
aperture becomes roughly a bias $b^{1.5}$ in the estimated virial mass.
Typically, a systematic error $\Delta\bar{z}$ in the mean redshift of sources
produces a corresponding error $\sim \Delta\bar{z}/2$ in 
$\avg{\Sigma_{\rm crit}^{-1}}$.
Recent work suggests that controlling 
photometric redshifts at the level required for weak lensing
mass calibration of galaxy clusters is possible \citep{sheldon11}.
Importantly, because cluster weak lensing depends on the mean tangential shear
around cluster centers, some forms of cosmic shear systematics 
are automatically averaged away and therefore not relevant for weak lensing 
mass calibration of galaxy clusters.  
For instance, errors that 
are coherent on scales larger than
cluster diameters (typically a few arcmin) but incoherent
on still larger scales will be averaged out in a stacked lensing measurement.
Moreover, because the weak lensing signal about galaxy clusters 
is stronger than cosmic shear, uncertainties that appear for very
low shear values (e.g., additive biases) are less important.
All in all, the demands on
weak lensing systematics for stacked weak lensing 
calibration of galaxy clusters are likely to be 
lower than those for cosmic shear.

There are some systematics specific 
to stacked cluster lensing, the most significant of which is
cluster mis-centering.
If the observationally determined center of a cluster
does not match the location of the center of the dark matter halo
that one would select in simulations, then the observed mean tangential shear
about the assigned center will differ from the theoretical expectation.
Cluster mis-centering should not be problematic in X-ray
experiments with high angular resolution, as gas in hydrostatic equilibrium
traces the underlying gravitational potential.  While a few exceptions will 
arise, such as the famed Bullet Cluster \citep{clowe06}, 
the frequency of these systems is low.
For similar reasons, centroiding of SZ systems is expected to be fairly
robust.
The mis-centering problem is most difficult in the optical, 
where the center is typically chosen to be a specific galaxy but the
choice of galaxy is not necessarily obvious;
X-ray studies of SDSS maxBCG clusters suggest that the 
mis-centered fraction is about 30\% \citep{andreon11}.
Mis-centering
is currently one of the dominant systematics in stacked cluster lensing,
introducing uncertainties at the $\approx 5\%-10\%$ 
level \citep{johnston07}.  There are ongoing efforts aimed
at improving cluster centering \citep[][Rykoff et al. in preparation]{george12}.
\citet{oguri11b} find that marginalizing over parameters that
describe mis-centering
does not significantly dilute the cosmological power of cluster abundance
studies, so it may be that future analyses will simply treat mis-centering
via an additional set of nuisance parameters.  
Alternative weak lensing estimators can be constructed
to avoid mis-centering biases in the inner
regions of clusters \citep{mandelbaum10}.
Other potential biases that affect stacked cluster lensing are modulation 
of the source population by
lensing magnification, non-linear shear corrections, and 
source density modulation due to obscuration by cluster members 
(see \citealt{rozo10b,hartlap11}).
These effects can also have impact on cosmic shear experiments.

Turning to scatter, we can show that the magnitude of the variance 
$\Var(\ln X|M,z)$ is degenerate
with the mass scale through a simple argument.  
Suppose the observable of interest is a mass 
estimator $X=M_{\rm obs}$, where the
subscript indicates the observationally estimated cluster mass.  
The observed abundance is
\begin{equation}
\frac{dn}{d\ln M_{\rm obs}} = \int d\ln M\ \frac{dn}{d\ln M} P(M_{\rm obs}|M,z).
\end{equation}
For a power-law mass function $dn/d\ln M = A M^{-\alpha}=A\exp(-\alpha \ln M)$ 
and log-normal scatter of variance
$\sigma^2 = \langle (\ln M_{\rm obs} - \ln M)^2 \rangle$, one
can readily compute the observed abundance by completing the square, finding
\begin{equation}
\label{eqn:cl_sigshift}
\frac{dn}{d\ln M_{\rm obs}} = A\exp\left( -\alpha \ln M_{\rm obs} - \frac{1}{2}\alpha^2 \sigma^2 \right) ~.
\end{equation}
From equation~(\ref{eqn:cl_sigshift})
it is evident that a shift in mass $\Delta \ln M$ is 
degenerate with a shift in the variance 
$\Delta \sigma^2 = 2\alpha^{-1} \Delta \ln M$.
(For a more rigorous argument that arrives at the same conclusion, see 
\citealt{lima05}.)
Thus, if the mass scale is controlled with an accuracy $\Delta \ln M$, then the scatter must be controlled with an accuracy
$\Delta \sigma^2 = 2\alpha^{-1} \Delta \ln M$. 
If we further set $\Delta\sigma^2 = 2\sigma\Delta\sigma$, we arrive
at $\Delta\sigma = \alpha^{-1} \sigma^{-1} \Delta \ln M$.  
The {\it fractional} accuracy with which $\sigma$ must be known
to avoid competing with $\Delta\ln M$ scales as $\sigma^{-2}$, so
the requirement is much less demanding if the scatter is smaller
to begin with.
As an illustrative example, we set $\alpha=3$ and $\sigma=0.2$,
which is roughly appropriate for SZ and likely slightly optimistic
for optical.   We find that the uncertainty due to errors in the scatter becomes
comparable to that from errors in the mass when $\Delta \sigma \approx 1.7\Delta \ln M$.  
For Stage III experiments with weak lensing calibration,
yielding $\Delta \ln M \approx 2\%$, the scatter needs to be known 
at the $\Delta \sigma \approx 0.04$ level,
a value in agreement with the more rigorous estimate by 
\citet{rozo10b} and likely to be achievable in the near future
\citep[see, e.g.,][]{rykoff11}.
If Stage IV experiments reach $0.5\%$ precision, the corresponding 
uncertainty in the scatter must be below $0.01$ (absolute, not fractional),
which is difficult to achieve from an {\it ab initio} calculation
but may be possible with statistical calibration methods.

Finally, we must consider the possibility that, in addition to
extended tails reflecting contamination and incompleteness,
the core of $P(X|M,z)$ deviates from log-normal form.
This problem was considered by \cite{shaw08}, whose discussion we
paraphrase here.
An observable-mass relation can be approximated by
\begin{equation} 
P(\ln X|M) = G(x) -
\frac{\gamma}{6}\frac{d^3G}{dx^3} + \frac{\kappa}{24}\frac{d^4G}{dx^4} + ...
\end{equation} 
known as the Edgeworth expansion.  
Here $G$ is a Gaussian of
zero mean and unit standard deviation, $x=(\ln X - \avg{\ln X})/\slnx$,
$\gamma$ is the skewness of the distribution, and $\kappa$ is the
kurtosis.  
For a power-law mass function $dn/d\ln M \propto M^{-\alpha}$,
it is straightforward to check that the resulting cluster
abundance is
\begin{equation} 
{dn \over dX} = \int dM {dn \over dM} P(X|M)
     = \left({dn \over dX}\right)_0 
       \left[1+ \frac{\alpha^3\sigma^3}{6}\gamma +
       \frac{\alpha^4\sigma^4}{24}\kappa + ...  \right] ,
\end{equation} 
where
$(dn/dX)_0$ is the abundance for a purely log-normal distribution.  
(Note that this $\alpha$ is also the logarithmic slope of the 
cumulative halo mass function $d\ln N/d\ln M$ that appears in our
earlier discussion.)

Setting $\alpha=3$ and assuming $10\%$ scatter for X-ray masses, a $3\%$
correction to the abundance --- equivalent to a $1\%$ correction in the mass
--- requires extreme non-Gaussianity with $\gamma \approx 7$ or 
$\kappa \approx 90$.  
Numerical simulations, on the other hand,
predict distributions of X-ray observables
that are close to log-normal (see, e.g., \citealt{stanek10}, Fig. 8;
\citealt{fabjan11}, Fig. 3).
We therefore do not expect
X-ray studies to be sensitive to departures from a log-normal $P(X|M,z)$.
For $\slnx = 0.2$, typical for SZ and perhaps achievable for optical,
a 3\% abundance change arises from $\gamma \approx 0.8$ or 
$\kappa \approx 6$, still quite large deviations from Gaussianity.  
For $\slnx=0.4$ these numbers drop to 0.1 and 0.35, respectively,
so with this level of scatter a moderate degree of non-Gaussianity
can have noticeable impact on the predicted abundances.
For example, a Poisson distribution for a cluster
with $\avg{N}=10$ galaxies corresponds to a skewness $\gamma\approx 0.3$. 
This discussion demonstrates the value of finding improved optical richness
estimators that have lower scatter relative to mass
\citep{rozo09b,rykoff11}.

\begin{figure} 
\begin{centering}
\includegraphics[width=3.5in,height=2.9in]{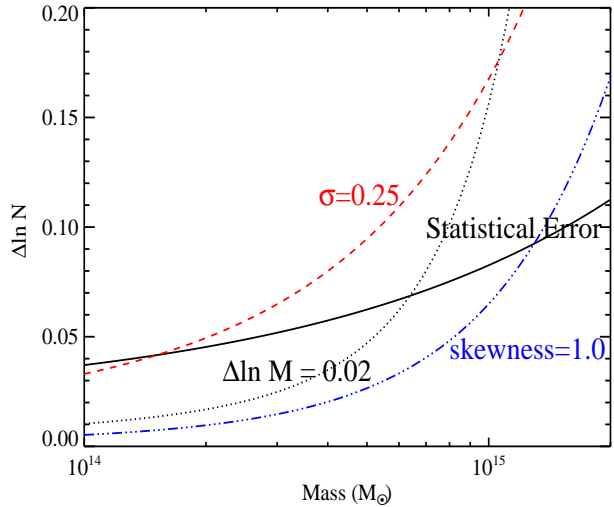} 
\caption{Relative change in cluster abundances at $z=0.6$ 
as a function of mass due
to a 2\% bias in the mass ($\Delta \ln M=0.02$), raising the log-normal scatter 
$\sigma$ from $0.2$ to $0.25$,
or introducing skewness $\gamma=1$ in $P(X|M,z)$
(solid, dashed, and dot-dashed curves, respectively).
The statistical error in number counts for $A=10^4\mdeg^2$ is
shown by the dotted line. The sensitivity of $\Delta \ln N$ to 
systematic errors in the mass, scatter, or skewness can be estimated using
the rule-of-thumb
approximations in equations~(\ref{eq:mean})-(\ref{eq:gamma}).
}
\label{fig:cl_scatter} 
\end{centering}
\end{figure}

Figure~\ref{fig:cl_scatter} shows the impact that various elements of $P(X|M,z)$
can have on the recovered cluster counts.  For illustrative purposes, 
we assume that $X$ is an observed mass and show the change in the observed 
mass function due to changes in $P(M_{\rm obs}|M,z)$.  
For our reference model, we assume $M_{\rm obs}$
is unbiased and has log-normal scatter $\sigma=0.2$, 
and we compute the cumulative
cluster counts above $M_{\rm obs}$ for our fiducial cosmology at $z=0.6$ 
in a redshift bin of width $\Delta z =0.1$.
Results at other redshifts are qualitatively similar.

%%%%%%%%%%%%%%%%%%%%%%%%%%%%%%%%%%%%%%%%%%%%%%%%
%%%%%%%%%%%%%%%%%%%%%%%%%%%%%%%%%%%%%%%%%%%%%%%%

%%%%%%%%%%%%%%%%%%%%%%%%%%%%%%%%%%%%%%%%%%%%%%%%
%%%%%%%%%%%%%%%%%%%%%%%%%%%%%%%%%%%%%%%%%%%%%%%%

Solid, dashed, and dot-dashed curves show
the change in the 
cumulative number counts $\Delta \ln N$
if $M_{\rm obs}$ is biased by $2\%$ 
($\Delta \ln M =0.02$), if the scatter is increased
from $\sigma = 0.2$ to $\sigma=0.25$, or
if the skewness is increased from $\gamma=0$
to $\gamma=1$ using the Edgeworth expansion.  For reference, we also show the
statistical error on the cluster counts for $A=10^4\mdeg^2$ as a dotted line.  
The details of $P(X|M,z)$ affect the recovered cluster counts, 
and the impact is larger at higher masses than at lower masses.  
Moreover, the relative impact of skewness
to scatter and of scatter to bias is mass dependent, 
with lower masses being more robust
to uncertainties in the scatter and skewness.  
This is as expected: the shallower the slope of the
mass function, the less important the details of $P(X|M,z)$.
The systematic offsets in Figure \ref{fig:cl_scatter} 
are well approximated (to
$\approx 10\%$ and $30\%$ for scatter and skewness respectively) by
the rule-of-thumb calculations we have described above,
specifically
\begin{eqnarray}
\Delta \ln N_{\rm predicted} & = & \alpha \Delta \ln M ~, \label{eq:mean} \\
\Delta \ln N_{\rm predicted} & = & \alpha^2 \sigma \Delta \sigma ~, \\
\Delta \ln N_{\rm predicted} & = & \frac{1}{6}\alpha^3\sigma^3 \Delta \gamma~. 
\label{eq:gamma}
\end{eqnarray}
Given the values of $\Delta\ln N$ and $\alpha$ expected for a survey
(Fig.~\ref{fig:errors}; typical values $\Delta\ln N \approx N^{-1/2}$
and $\alpha \approx 3$), one can
use equations~(\ref{eq:mean})-(\ref{eq:gamma})
to infer the uncertainties $\Delta\ln M$,
$\Delta\sigma$, and $\Delta\gamma$ required to keep
a cosmological analysis limited by abundance statistics.

%%%%%%%%%%%%%%%%%%%%%%%%%%%%%%%%%%%%%%%%%%%%%%%%
%%%%%%%%%%%%%%%%%%%%%%%%%%%%%%%%%%%%%%%%%%%%%%%%
%%%%%%%%%%%%%%%%%%%%%%%%%%%%%%%%%%%%%%%%%%%%%%%%
%%%%%%%%%%%%%%%%%%%%%%%%%%%%%%%%%%%%%%%%%%%%%%%%
%%%%%%%%%%%%%%%%%%%%%%%%%%%%%%%%%%%%%%%%%%%%%%%%
%%%%%%%%%%%%%%%%%%%%%%%%%%%%%%%%%%%%%%%%%%%%%%%%

\subsubsection{Theoretical Systematics}
\label{sec:cl_theoretical_systematics}

Predicting observed cluster counts via equation~(\ref{eq:abundance}) 
requires knowledge of the halo mass function $dn/dM$ for any
cosmological model under consideration.  If the fractional
uncertainty in $dn/dM$ exceeds the observational error in
cluster counts $\Delta \ln N$, or if the equivalent mass scale 
uncertainty exceeds the mass calibration error $\Delta\ln M$, then
cosmological constraints will be limited by theoretical uncertainty
rather than by observational errors.  
The study of \cite{tinker08} finds agreement in $dn/dM$ at the $\la 5\%$
level among multiple simulations by different groups for a
$\Lambda$CDM cosmological model with WMAP3 parameters.
This is roughly the level required for large area surveys
of $M > 4\times 10^{14} M_\odot$ clusters in $\Delta z = 0.1$ bins, 
though higher
accuracy is needed for lower mass thresholds 
(for detailed discussion see \citealt{cunha10,wu10b}).
The formula~(\ref{eqn:fsigma}) describes Tinker et al.'s $z=0$ 
results accurately, but at redshifts $z=0.5-2.5$ they find
deviations of $\sim 10-30\%$ from this ``universal'' 
prescription.  While these deviations are themselves
numerically calibrated, their existence suggests that the mass
function may depend on the dark energy model even when expressed
in terms of the $\sigma(M)$ relation as in equation~(\ref{eqn:fsigma}).
In addition, consistency in halo definitions is clearly critical.  For instance,
\cite{bhattacharya11} find that mass functions in
their suite of wCDM simulations --- which are calculated using 
friends-of-friends halo finders --- deviate by up 
to 10\% from a fitting formula calibrated on their $\Lambda$CDM
simulation suite.
It seems likely that Stage III and certainly Stage IV
experiments will need to move to emulator based methods with
comprehensive N-body libraries \citep[e.g.,][]{lawrence10}
rather than simple fitting formulae.

While further N-body work is needed to interpret future surveys,
dark matter evolution is straightforward in principle,
and the problem should yield to sufficient applications of
computational force.  Baryonic evolution is potentially
a thornier issue.
Some X-ray studies suggest a depletion of baryonic mass 
(stars + hot gas) relative to the universal $\Omega_b/\Omega_m$
ratio by $20-30\%$ within the $\Delta = 500\rho_c$ radius,
with systematically larger depletion in less massive clusters
\citep[e.g.,][]{giodini09}.  
For $\Omega_b/\Omega_m = 0.17$,
a 20\% deviation in baryonic mass is a 3.4\% deviation in total
mass, and thus comparable to or larger than the statistical mass
calibration errors achievable with stacked weak lensing
(Figure~\ref{fig:wl_merr}), as well as the precision required to 
achieve the statistical limits of large cluster surveys
(Figure~\ref{fig:cumcounts}d).  
Hydrodynamic simulations can explain baryon depletions 
comparable to those observed \citep{young11}, but the magnitude
and even the sign of the baryonic effects depend on the star formation
and feedback physics (e.g., \citealt{stanek09,cui11}). 
Furthermore, because the baryons influence the dark matter profile, they
can have substantial impact ($\sim 15\%$) on the total mass
within a high overdensity threshold (e.g., the $\Delta=500\rho_{\rm crit}$
threshold frequently adopted in X-ray analyses; see \citealt{stanek09}).
In all of these simulations the corrections are smaller at larger radii,
so defining halo boundaries at lower overdensity 
(such as the $\Delta=200\bar\rho$ convention used here) 
is beneficial in this respect.

It may be possible to calibrate baryonic effects well enough with simulations
and detailed observations of selected systems to remove them as a 
source of systematic uncertainty, but this problem will
require concerted effort, particularly when Stage IV experiments
get underway.  By the same token,
if stacked weak lensing is the primary mass calibration tool,
then one must also develop robust theoretical models for
predicting the weak gravitational lensing signal, which in turn
requires that the halo--mass correlation function be characterized
at the same level as $\Delta \ln M$.  Current analytical
models are accurate only at the $\approx 10\%-20\%$ 
level \citep{hayashi08}, so this is another area that requires 
further theoretical study.  

A final caveat related to the halo mass function is that primordial
non-Gaussianity could alter its form
\citep[e.g.,][]{weinberg92a,dalal08,grossi09,loverde11,damico11}
and thereby change the cluster abundances predicted for a given
dark energy model \citep[e.g.,][]{cunha10b,pillepich11}.
Of course, evidence for non-Gaussian initial conditions would
be exciting in its own right, with important implications for 
early-universe physics.  However, it appears that the levels of
non-Gaussianity that would have significant impact on cluster 
abundances are already ruled out by other constraints, unless one 
allows the magnitude of the non-Gaussianity to be scale-dependent
\citep[e.g.,][]{hoyle11,paranjape11}.
Given the strong theoretical prior for Gaussian initial conditions
and the multiple observational probes that could detect and
characterize primordial non-Gaussianity if it exists, we think
it unlikely that non-Gaussianity will limit the power of cluster
abundances as a probe of dark energy and modified gravity.

%%%%%%%%%%%%%%%%%%%%%%%%%%%%%%%%%%%%%%%%%%%%%%%%
%%%%%%%%%%%%%%%%%%%%%%%%%%%%%%%%%%%%%%%%%%%%%%%%
%%%%%%%%%%%%%%%%%%%%%%%%%%%%%%%%%%%%%%%%%%%%%%%%
%%%%%%%%%%%%%%%%%%%%%%%%%%%%%%%%%%%%%%%%%%%%%%%%
%%%%%%%%%%%%%%%%%%%%%%%%%%%%%%%%%%%%%%%%%%%%%%%%
%%%%%%%%%%%%%%%%%%%%%%%%%%%%%%%%%%%%%%%%%%%%%%%%

\subsection{Space vs.  Ground} \label{sec:cl_space}

As discussed in \S\ref{sec:cl_current}, X-ray observations, possible
only from space, have played a central role in nearly all cluster cosmological studies to date.  The \rosat\ All-Sky Survey has been the
basis for many of the cluster samples used in these studies
(Table~\ref{tbl:Xray_catalogs}).
Pointed observations with a variety of telescopes, especially \xmm\ and \chandra, have been the basis of
mass calibration for X-ray observables and the source of most empirical knowledge about the physics of the intracluster
gas.  Ongoing \xmm\ surveys will expand the dynamic range and
size of X-ray catalogs over the next few years.  The most important
advance will come with the \erosita\ mission \citep{merloni12}, which should
produce the definitive all-sky survey of massive
($M \ga 4\times 10^{14} M_\odot$) clusters out to $z\approx 1$,
with an extended tail of higher redshift clusters reaching $z\approx 2$.  
Follow-up X-ray studies at higher angular resolution will help better 
assess point-source contamination
and will improve the mass calibration of the \erosita\ catalog.
For comparable numbers of clusters, X-ray catalogs offer significant
advantages over SZ or optical catalogs because of the low scatter
expected between X-ray observables and halo mass, which
reduces sensitivity to uncertainties in the width and form
of the observable-mass relation (\S\ref{sec:cl_core_calibration}).

For SZ searches, ground-based telescopes have higher
sensitivity than space observatories because of their larger collecting
area and higher angular resolution.  The larger
beam size of the \planck\ observatory ($\approx 5\ \mbox{arcmin}$) relative to SPT and ACT 
($\approx 1\ \mbox{arcmin}$)
reduces its ability to detect high redshift
systems.
Nonetheless, the all-sky nature of \planck\ observations is an important
asset, and the \planck\ catalog of high mass clusters will be useful
both for direct cosmological constraints and for cross-correlation
studies with clusters identified at other wavelengths.  Thus, we view
the \planck, SPT, and ACT surveys
as highly complementary.
Any future CMB space mission designed to probe inflation physics
and primordial gravity waves would also 
produce a much more sensitive all-sky SZ cluster catalog, 
provided it achieved high angular resolution. 

Turning to optical searches, 
space observatories provide little advantage for cluster 
detection at $z\lesssim 1$, since
cluster detection does not gain much from the improved image resolution 
achievable from space.
However, as discussed in \S\ref{sec:cl_finding}, space-based near-IR 
imaging is highly desirable for extending (rest-frame) optical
cluster catalogs to $z\approx 2$. In the near future, such searches 
will rely on \spitzer\ data, as in the case of 
ISCS \citep{eisenhardt08}, SpARCS \citep{wilson06}, and the recently approved
$100\ \deg^2$ \spitzer--SPT Deep Field.  
Additional IR data is or will soon be available from surveys like 
VHS, UKIDSS, and \wise, which may 
allow for high redshift cluster finding \citep{gettings12}.
The VIKING survey, covering
$\approx 1500\ \deg^2$, should be sufficiently deep
to allow for robust cluster detection at $z>1$.
In the longer term, IR imaging from \euclid\ and/or \wfirst\ could make a key contribution to high redshift
cluster surveys.
High redshift cluster detection should also be feasible with extremely deep optical imaging from the ground,
like that planned for LSST, which should reach $z\approx 1.5$.

In the long run, however, the most important contribution of space observations
to cluster cosmology will come via weak lensing mass calibration
rather than cluster finding.  The statistical error of WL
mass calibration scales as ${\bar n_g}^{-1/2}$, where ${\bar n_g}$ is
the source surface density.  As can be seen from
Figure \ref{fig:wl_merr},
a surface density ${\bar n_g} \approx 30\ \arcmin^{-2}$ is 
required to reduce mass calibration error below the statistical 
abundance error, 
and even then only for $z \la 0.5$.
This source density is expected for an optical space mission
like \euclid, but it is probably higher than can be achieved
by ground-based observations, even with the depth and image
quality of LSST. 
The cluster counting error and
mass calibration error both scale with survey area as $A^{-1/2}$,
so the area effect cancels out if the cluster and WL surveys
overlap completely.  If the cluster survey covers a larger area (e.g., the all-sky \erosita\ catalog), then the 
WL source density required to saturate the halo statistics limit is even higher.
Reaching the calibration accuracy allowed by the source galaxy
statistics also requires excellent control of shape measurement
systematics, generally expected to be lower from a
space-based platform, and photo-$z$ systematics, which probably
require space-based IR imaging to achieve the stringent demands implied by Figure~\ref{fig:wl_merr}.
More generally, if the error in WL mass calibration sets the
ultimate limit of cluster measurements of fluctuation growth,
as we have speculated it will, then the achievable error 
on $\sigElevz$ scales as ${\bar n_g}^{-1/2}$, or as
$(\Delta\gamma)_{\rm sys}^{-1}$ if the WL measurements
are themselves limited by a shear measurement
systematic $(\Delta\gamma)_{\rm sys}$.

%%%%%%%%%%%%%%%%%%%%%%%%%%%%%%%%%%%%%%%%%%%%%%%%
%%%%%%%%%%%%%%%%%%%%%%%%%%%%%%%%%%%%%%%%%%%%%%%%
%%%%%%%%%%%%%%%%%%%%%%%%%%%%%%%%%%%%%%%%%%%%%%%%

\subsection{Prospects} \label{sec:cl_prospects}

We expect cluster abundance studies to undergo substantial and steady
improvements over the next decade and beyond.
In the near term ($\lesssim 3$ years), 
we anticipate advances in X-ray, SZ, and optical cluster studies.
The XMM Cluster Survey (XCS) and XMM XXL Survey
will yield much larger X-ray cluster samples at $z\gtrsim 0.3$. 
\planck\ will produce the definitive all-sky
SZ catalog of massive clusters out to $z\lesssim 0.7$, while
SPT and ACT will probe
$z\gtrsim 0.7$ cluster populations over thousands of square degrees for
the first time.  
In the optical, continuing studies with the SDSS will lead to
improved cluster finders and richness estimators, as well as
improved weak lensing calibration thanks to better centering and better
source photometric redshifts.
On a comparable time scale, the RCS-2 
survey will obtain $g$, $r$, and $z$ imaging to a nominal depth of
$r \approx 24.8$ (roughly 2 magnitudes deeper than SDSS) over
$1000\mdeg^2$, yielding the first large area optical cluster
catalog extending to $z\approx 1$.
Relative to the results shown in Figure~\ref{fig:comparison},
these X-ray, SZ, and optical studies will improve the low 
redshift $\sigma_8$--$\Omega_m$ constraint and
extend it, at somewhat lower precision, to $z\approx 0.5-1$.
At the same time, improved calibration and cross-checks among surveys
will test for and reduce remaining sources of systematic error.

In the medium term ($\approx 3-8$ years), several new optical surveys
will cover thousands of $\mdeg^2$ with greater depth than SDSS
and larger area and/or more photometric bands than RCS-2.
These include the Kilo-Degree Survey (KIDS, $1500\mdeg^2$ in 
$ugriz$),
DES ($5000\mdeg^2$ in $grizY$),
PS1 (15,000$\mdeg^2$ in $grizY$),
and the Hyper-Suprime Camera survey 
(HSC, $1500\mdeg^2$ in $grizY$).
These surveys should significantly improve the cosmological constraints
relative to RCS-2, thanks
to higher cluster numbers, lower statistical errors in weak lensing
mass calibration, and better control of photometric redshift
uncertainties.  The VIKING survey will cover
$1500\mdeg^2$ at near-IR wavelengths ($ZYJHK_s$)
at sufficient depth to allow cluster identification and accurate
photometric redshifts at $z=1-2$.  In addition, all of these surveys
will overlap with \planck, and often with either the ACT or SPT surveys,
which can further enhance the utility of both sets of catalogs.
DES in particular is designed to cover the entire footprint
of the SPT SZ survey.

With launch expected 2013-2014, \erosita\ will produce the ultimate
all-sky catalog of massive clusters
(see \S\ref{sec:cl_space}).
The optical imaging surveys will allow weak lensing calibration 
of the \erosita\ mass-observable relations, with multiple independent
surveys affording larger overlap area and thus more precise
calibration.  This combination of X-ray selection and optical WL
calibration offers bright prospects for the coming decade of
cluster cosmology.  Optical surveys will further extend
this leverage by probing cluster abundances to masses below those
probed by \erosita.

On a longer timescale, LSST plans to image 20,000$\mdeg^2$ of
high-latitude sky 
in six bands ($ugrizY$), with each single pass comparable in
depth to the medium-term surveys described above and co-added
data reaching $2.5-3$ magnitudes deeper. 
The increased depth of LSST should allow one to cleanly select galaxy clusters
out to $z\approx 1.5$.
While the greater dynamic range of the cluster
catalogs will be an asset in itself, LSST's most 
important contribution to cluster 
cosmology will be in the form of
improved WL mass calibration, both for \erosita\ and for LSST's own clusters.
\euclid\ could provide even better WL calibration
over a similar sky area, while \wfirst\ should achieve a 
high WL source density but over a smaller survey area.  The IR sensitivity
of \euclid\ and/or \wfirst\ should also enable
cluster searches at $z\approx 2$ and beyond.

\begin{figure} 
\begin{centering}
\includegraphics[width=2.8in, height=2.4in]{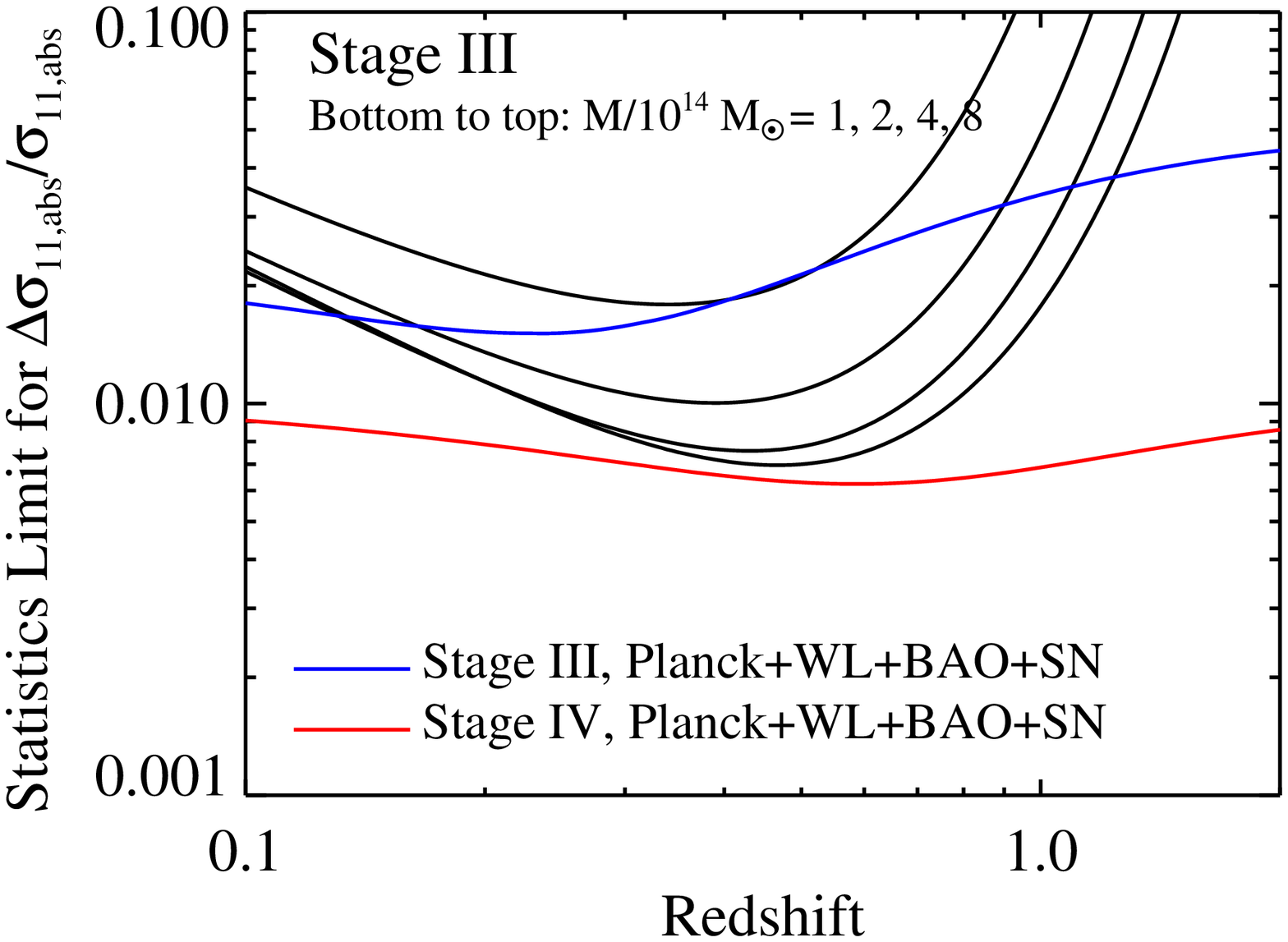} 
\includegraphics[width=2.8in, height=2.4in]{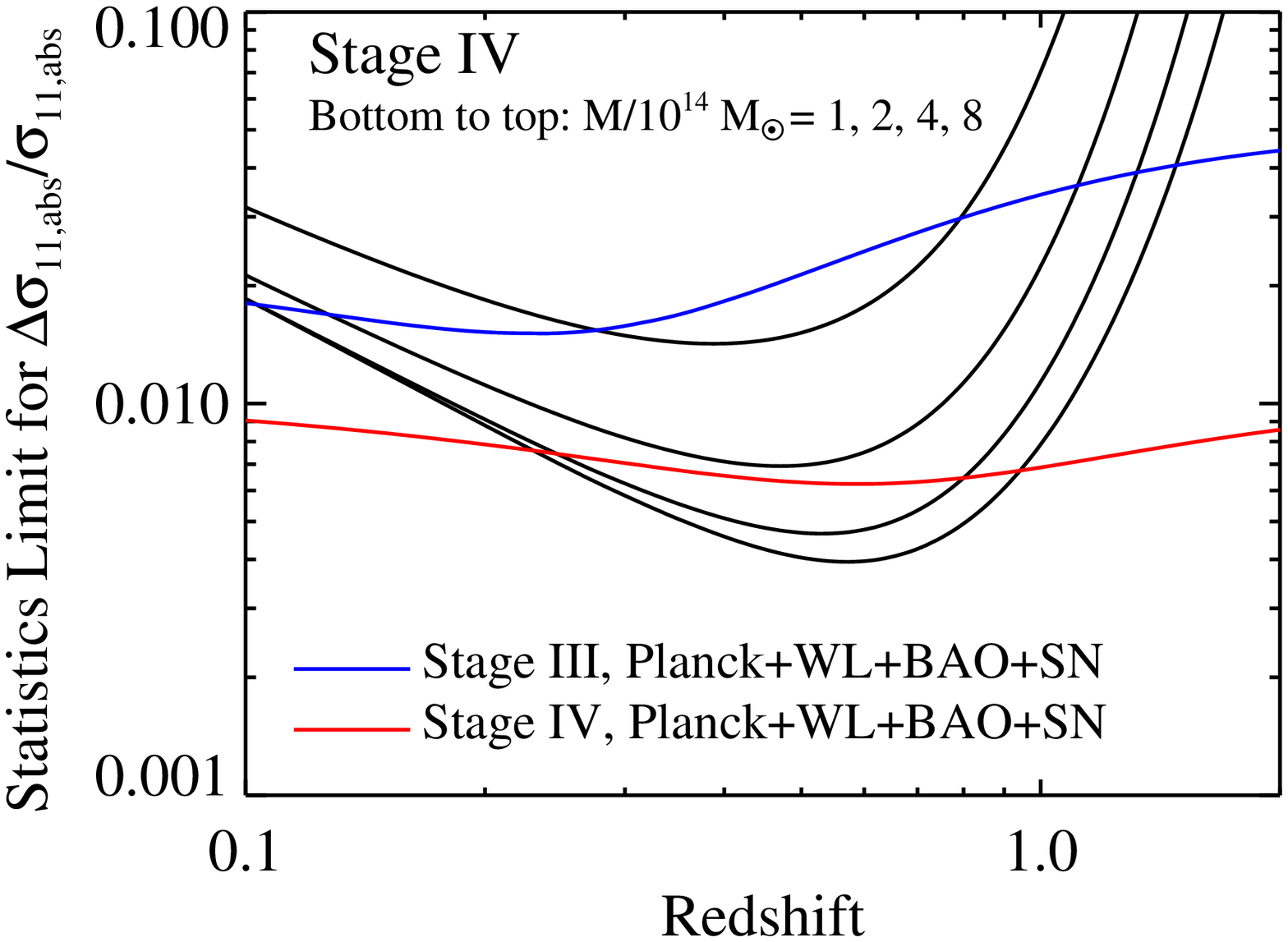} 
\caption{
Error on $\sigElev(z)$ achievable by 
measuring cluster abundances in a redshift bin $z=z_c\pm 0.05$ in a 
$10^4\mdeg^2$, assuming mass calibration via stacked weak lensing
with Stage III or Stage IV source densities.  We assume that all geometric
cosmological parameters --- most significantly the comoving volume element 
and the matter density parameter $\Omega_m$ --- are held fixed, being
effectively constrained by a joint CMB+SN+BAO+WL experiment.
Also shown for comparison are the forecast
constraints on $\sigElevz$ derived from such 
a joint analysis using our fiducial Stage III and 
Stage IV surveys, assuming a $w_0$--$w_a$ parameterization
of dark energy and allowing deviations from GR parameterized by
$G_9$ and $\Delta\gamma$ (see \S\ref{sec:forecast_cl} for details).  
} 
\label{fig:wl_serr} 
\end{centering}
\end{figure}

We have argued throughout this section that mass calibration
will be the likely limiting factor in cluster studies of 
cosmic acceleration, and that stacked weak lensing is the
most promising avenue to achieve accurate mass calibration.  
Figure~\ref{fig:wl_serr} combines information 
from Figures~\ref{fig:serr} and~\ref{fig:wl_merr}, showing
the fractional error on $\sigElevz$ in $\Delta z = 0.1$ bins
that can be achieved with a $10^4\mdeg^2$ cluster survey,
using the WL mass calibration errors we have forecast for
Stage III (left panel) or Stage IV (right panel) source densities.
With Stage III lensing calibration, errors on 
$\sigElevz$ are below 1\% at $z\approx 0.5$ for cluster
mass thresholds of $1-2\times 10^{14} M_\odot$, and
$\sim 1.5\%$ for a mass threshold of $4\times 10^{14} M_\odot$.
With Stage IV lensing calibration, the peak sensitivity is
better than 0.5\% for the lower mass thresholds and better
than 1\% for the $4\times 10^{14} M_\odot$ threshold.

%%%%%%%%%%%%%%%%%%%%%%%%%%%%%%%%%%%%%%%%%%%%%%%%
%%%%%%%%%%%%%%%%%%%%%%%%%%%%%%%%%%%%%%%%%%%%%%%%

%%%%%%%%%%%%%%%%%%%%%%%%%%%%%%%%%%%%%%%%%%%%%%%%
%%%%%%%%%%%%%%%%%%%%%%%%%%%%%%%%%%%%%%%%%%%%%%%%

The additional red and blue curves in 
Figure~\ref{fig:wl_serr} show the forecast constraints on
$\sigElevz$ for a fiducial Stage III (blue) or Stage IV (red)
program combining SN, BAO, WL, and CMB data as discussed in 
\S\ref{sec:forecast}.  These forecasts
assume a $w_0-w_a$ dark energy model and allow departures from 
GR-predicted growth described by an overall multiplicative
offset $G_9$ and a growth index deviation $\Delta\gamma$
(see \S\ref{sec:parameterizations}).  The fiducial
programs are defined in \S\ref{sec:fiducial}.
If WL systematics are controlled at the level assumed in these
fiducial programs then they
should be negligible for cluster mass calibration relative to statistical 
errors, so we have not included them in computing $\Delta\ln M$.

From Figure~\ref{fig:wl_serr} we see that
a $10^4\mdeg^2$ cluster survey with Stage III WL calibration
data can easily exceed the $\sigElevz$ precision expected from
the Stage III CMB+SN+BAO+WL program, by as much as a factor of $\approx 3$
for a threshold of $10^{14} M_\odot$.  Similarly, cluster
constraints with Stage IV WL calibration improve on the fiducial
Stage IV $\sigElevz$ precision without clusters by a factor of
$\approx 2$.  
The visual impression that clusters can outperform the fiducial
program only at $z \approx 0.4-0.8$ but perform worse at high and
low redshifts is artificial, since the CMB+SN+BAO+WL curves assume
a smooth growth model while the cluster constraints 
in Figure \ref{fig:wl_serr}
are those that can be achieved from galaxy clusters 
within each individual redshift bin.
For Figure~\ref{fig:wl_serr} we have assumed that errors on $\Om$
and $dV_c(z)$ are negligible.  While the assumption for $dV_c(z)$
should prove reasonably accurate, the forecast CMB+SN+BAO+WL errors on
$\Om$ (4\% and 1\% for Stage III and Stage IV, respectively)
are larger than our assumed
WL mass calibration errors for $M \la 2\times 10^{14} M_\odot$
(see Figure~\ref{fig:wl_merr}).  In practice, therefore,
the fractional errors in Figure~\ref{fig:wl_serr} would apply
not to $\sigElevz$ but to the parameter combination
$\sigElevz\Omega_m^q$, with $q \approx 0.4$. 
We return to these points in \S\ref{sec:forecast_cl} below,
where we discuss the improvements in constraints on 
the dark energy equation of state and on $G_9$ and
$\Delta\gamma$ achievable with clusters.

If some alternative mass calibration method proves better 
than stacked weak lensing, then the situation
could be even better than Figure~\ref{fig:wl_serr} suggests.
This would be especially true for Stage III, where the WL
source density is the clear limiting factor on the overall error.
For our assumed Stage IV source density, the uncertainty from WL
mass calibration is already close to the statistical uncertainty
in cluster counts at $z \la 0.6$.  Conversely, the situation would be worse
than Figure~\ref{fig:wl_serr} suggests if some other systematic
uncertainty --- e.g., contamination, miscentering, theory, or WL photo-$z$
calibration --- makes it impossible to achieve the statistical
limits of the WL mass calibration.  

In summary, our analysis indicates
that cluster abundances with masses calibrated by stacked weak lensing
could provide strong tests of cosmic acceleration models,
beyond those afforded by the 2-point WL statistics described
in \S\ref{sec:wl}.
However, achieving this potential requires that mass calibration
uncertainties 
be controlled at the $1-3\%$ level
for Stage III and at the $0.5-1.5\%$ level for Stage IV.
We see no obvious show stoppers, but the challenge is a
demanding one.

\vfill\eject

\section{Alternative Methods}
\label{sec:alternatives}

In \S\S\ref{sec:sn}-\ref{sec:cl}, we have reviewed in detail
the four observational methods that have been most widely 
discussed, and applied, as probes for the origin of cosmic
acceleration.  We now review more briefly some of the other
techniques for testing cosmic acceleration models.
In some cases, surveys conducted for SN, BAO, or WL studies
will automatically provide the data needed for these alternative
methods.  For example, redshift-space distortions (\S\ref{sec:rsd})
and the Alcock-Paczynksi effect (\S\ref{sec:ap}) can be measured
in galaxy redshift surveys designed for BAO measurements, and 
synoptic surveys designed for Type Ia supernovae will discover
other transients that might provide alternative distance indicators
(\S\ref{sec:other_di}).  Just as cluster investigations will
increase the cosmological return from WL surveys, these methods
will increase the return from BAO or SN surveys.
The potential gains are large, but they are uncertain
because the level of theoretical or observational systematics
for these methods has not yet been comprehensively explored.
In \S\ref{sec:forecast_alt} we will examine how precisely our
fiducial Stage III or Stage IV CMB+SN+BAO+WL programs predict
the observables of these methods, setting targets for the precision
and accuracy they should achieve to make major contributions
to cosmic acceleration studies.

Some of the other methods described below require completely
different types of observations or experiments, falling outside
of the ``survey mode'' that characterizes the methods we have
discussed so far.  
Compared to the combined SN+BAO+WL+CL approach, 
these methods may yield more limited
information or be sensitive only to certain classes of acceleration
models, but they can provide high-precision tests of the
standard $\Lambda$CDM model, and they could yield
surprising results that would give strong guidance to the
physical origin of acceleration.

\subsection{Measurement of the Hubble Constant at $z\approx 0$}
\label{sec:h0}

As emphasized by \cite{hu05}, a precise measurement of the
Hubble constant, when combined with CMB data, 
allows a powerful test of dark energy models
and tightened constraints on cosmological parameters.
In effect, the CMB and $H_0$
provide the longest achievable lever arm for measuring the
evolution of the cosmic energy density, from $z\approx 1100$
to $z=0$.  The sensitivity of $H_0$ to dark energy is illustrated in
Figure~\ref{fig:hzdz}, which shows that a change $\Delta w = \pm 0.1$
alters the predicted value of $H_0$ by 5\% in $\ok=0$ models
that are normalized to produce the same CMB anisotropies.
More generally, a low redshift determination of the Hubble constant
combined with \planck-level CMB data constrains $w$ with an
uncertainty that is twice the fractional uncertainty in $H_0$,
assuming constant $w$ and flatness.
The challenge for future $H_0$ studies is to achieve the
percent-level statistical and systematic uncertainties needed
to remain competitive with other cosmic acceleration methods.
\cite{freedman10} review recent progress in $H_0$ determinations
and prospects for future improvements; the past five years
have seen substantial advances in both statistical precision
and reduction of systematic uncertainties.

One of the defining goals of the {\it Hubble Space Telescope} was to
measure $H_0$ to an accuracy of 10\%.
The $H_0$ Key Project achieved this goal, with a final estimate
$H_0 = 72 \pm 8\hubunits$ \citep{freedman01}, where the error
bar was intended to encompass both statistical and systematic
contributions.  This estimate used Cepheid-based distances to
relatively nearby ($D < 25\Mpc$) galaxies observed with WFPC2 to
calibrate a variety
of secondary distance indicators --- Type Ia supernovae, Type II
supernovae,
the Tully-Fisher relation of disk galaxies, and the fundamental
plane and surface-brightness fluctuations of early-type galaxies.
These secondary indicators were
in turn applied to galaxies ``in the Hubble flow,''
meaning galaxies at large enough distance
$(D \approx 40-400\Mpc)$ that their peculiar
velocities $v_{\rm pec}$ did not contribute significant uncertainty
when computing $H_0 = v/d$.
The Cepheid period-luminosity ($P-L$) relation was calibrated
to an adopted distance modulus of $18.50\pm 0.10$ mag for
the Large Magellanic Cloud (LMC).  The 
%uncertainty of the LMC distance itself and the 
uncertainty in adjusting the LMC optical
$P-L$ relation to the higher characteristic metallicities of
calibrator galaxies was an important contributor to the final
error budget; \cite{freedman01} adopted a $\pm 0.2$ mag/dex
uncertainty in the metallicity dependence, implying a $\sim 0.07$ mag
systematic uncertainty in the correction (3.5\% in distance).
Another important systematic was the uncertainty
in differential measurements of Cepheid fluxes and colors
over a wide dynamic range along the distance ladder.
The uncertainty of the LMC distance itself was also a significant
fraction of the error budget.

A number of subsequent developments have allowed substantial
improvements in the measurement of $H_0$
(see \citealt{riess09,riess11,freedman10,freedman12}).
The recent determination of $H_0 = 73.8 \pm 2.4\hubunits$
by \cite{riess11} yields a $1\sigma$ uncertainty of only
3.3\%, including all identified sources of systematic uncertainty
and calibration error.
One important change in this analysis is a shift to Cepheid
calibration based on the maser distances to NGC 4258
\citep{herrnstein99,humphreys08,humphreys13}
%Greenhill et al. 2009
and  on parallaxes to Galactic Cepheids
measured with \hipparcos\ \citep{vanleeuwen07} and
with the \hst\ fine-guidance sensors \citep{benedict07}.
These calibrations circumvent the statistical and systematic
uncertainties in the LMC distance, and they directly calibrate
the $P-L$ relation in the metallicity range typical of
calibrator galaxies, albeit with a sample of only $\sim 10$ stars
reaching an error-on-the-mean of 2.8\% in the case of Milky Way parallaxes.
A second improvement is more than doubling the sample
of ``ideal'' Type Ia SNe --- with
modern photometry, low-reddening, typical properties,
and caught before maximum ---
from the two available to \cite{freedman01} to eight.
Of all secondary distance indicators,
Type Ia supernovae have the smallest statistical errors,
%constraining $H(z)$ at $z<0.1$ to 0.5\% precision,
and probably the smallest systematic errors, and they can be
tied to large samples of supernovae observed at distances
that are clearly in the Hubble flow. 
Riess et al.\ (\citeyear{riess09}, \citeyear{riess11})
use Type Ia supernovae exclusively in their $H_0$ estimates.
Third, Cepheid observations at near-IR wavelengths (1.6$\mu$m)
have reduced uncertainties associated with extinction and the
dependence of Cepheid luminosity on metallicity
\citep{riess12c}.
Finally, relative calibration uncertainties of Cepheid photometry
obtained with different
instruments and photometric systems along the distance ladder have been
mitigated by the use of a single instrument, \hst's WFC3,
for a large fraction of the data.  

Extending the trend towards longer wavelength calibration, 
\cite{freedman11} and \cite{scowcroft11} argue that 3.6$\mu$m
measurements --- possible with {\it Spitzer} and eventually
with \jwst\ --- minimize systematic uncertainties in the
Cepheid distance scale because of low reddening and weak
metallicity dependence.  \cite{monson12} calibrate the 
3.6$\mu$m $P-L$ zero-point against Galactic Cepheid samples,
including the \cite{benedict07} parallax sample,
and thereby infer the distance modulus to the LMC
as a test of the optical Cepheid $P-L$ relation and its metallicity 
correction.
\cite{freedman12} use these Milky Way parallaxes to calibrate the
optical Cepheid $P-L$ relation and then recalibrate the Key Project
data set to infer $H_0$; this determination 
still relies on optical, WFPC2 Cepheid data (and the associated
metallicity corrections and flux and color zero-point uncertainties),
and the Key Project SN Ia calibrator sample includes several SNe with 
photographic photometry or high extinction.
\cite{sorce12} use the Tully-Fisher (\citeyear{tully77}) relation
in the {\it Spitzer} 3.6$\mu$m band, normalized to the \cite{monson12}
LMC distance, to recalibrate the SN Ia absolute magnitude scale and 
thereby infer $H_0$.  \cite{suyu12} infer $H_0$ from gravitational lens
time delays for two well constrained systems, an approach that 
sidesteps the traditional distance ladder 
entirely (see \S\ref{sec:strong_lensing} for further discussion).
These four recent $H_0$ determinations 
(\citealt{riess11,freedman12,sorce12,suyu12}) agree with each
other to better than 2\%.  While the data used for the first three
are only partly independent, this level of consistency is
nonetheless an encouraging indicator of the maturity of the field.
With the precision of $H_0$ measurements already at a level that
allows critical tests of dark energy models in combination
with CMB, BAO, and SN data (see, e.g., \citealt{anderson12}),
a key challenge for the field is convergence on error budgets that
neither underrepresent the power of the data nor understate 
systematic uncertainties.

Over the next decade, it should be possible to reduce
the uncertainty in direct measurement of $H_0$ to approach the
one-percent level.  One crucial
step will be the 1\% to 5\% parallax calibration of hundreds of
long-period Galactic Cepheids within 5 kpc by the \gaia\ mission,
setting the fundamental calibration of the multi-wavelength $P-L$ relation 
and, to some degree, its metallicity dependence on a solid geometrical
base with distance precision easily better than 1\%.
New Milky Way parallax measurements using the spatial scanning
capability of \hst\ may achieve this precision even sooner.
Discovery of additional galaxies
with maser distances (like NGC 4258) may also improve the
Cepheid calibration or, if they are in the Hubble flow, may provide a
direct
determination of the Hubble constant
(see \citealt{reid12a} for a recent measurement of UGC 3789
and \citealt{greenhill09} for additional candidates).
The other key step will be the Cepheid calibration of
more Type Ia supernovae, which occur at a rate of one per $2-3$
years in the range $D<35\Mpc$ accessible to \hst\ with WFC3.
\jwst\ could increase this range to $D < 60\Mpc$, quadrupling
the rate of usable supernovae.  Ultimately a sample of 20 to 30
calibrations of the SN Ia luminosity is needed to reduce the sample size
contribution to uncertainty in $H_0$ below 1\%.
With firmer $P-L$ calibration
and a larger Type Ia sample, the remaining uncertainty in $H_0$
is likely to be dominated by systematic uncertainty
in the linearity of the photometric systems observing nearby and
distant Cepheids.
This may be minimized by the careful construction of ``flux ladders,''
analogous to distance ladders but
used to compare the measurements of disparate flux levels.  Additional
contributions to the
determination of $H_0$ with few percent precision could come from
``golden'' lensing systems, infrared Tully-Fisher distances,
surface brightness fluctuation measurements further
into the Hubble flow, Sunyaev-Zel'dovich effect measurements, and local
volume measurements of BAO.

We discuss the potential contribution of $H_0$ measurements to dark
energy constraints in \S\S\ref{sec:results} and~\ref{sec:forecast_alt}
below.  Already, the combination of
the 3\% measurement of \cite{riess11} with CMB data alone
yields $w = -1.08 \pm 0.10$, assuming a flat universe with
constant $w$.  The limitation of $H_0$ is, of course, that it is a
single number at a single redshift, so while it can test any well
specified dark energy model, it provides
little guidance on how to interpret deviations from
model predictions.  However, precision $H_0$ measurements
can significantly increase the constraining power of
other measurements: for our fiducial Stage IV program described in
\S\ref{sec:forecast}, assuming a $w_0-w_a$ model for dark energy, a 1\%
$H_0$ measurement would raise the DETF Figure of Merit by 40\%.
A direct measurement of $H_0$ also has the potential to reveal
departures
from the smooth evolution of dark energy enforced by the
$w_0-w_a$ parameterization.  In essence, the dark energy model
transfers the absolute distance calibration from moderate
redshift BAO measurements down to $z=0$, but unusual
low redshift evolution of dark energy can break this link,
shifting $H_0$ away from its expected value.
A precise determination of $H_0$, coupled to a $w(z)$ parameterization
that allows low-redshift variation, could reveal recent evolution
of dark energy and definitively answer the basic question,
``Is the universe {\it still} accelerating?''

\subsection{Redshift-Space Distortions}
\label{sec:rsd}

As discussed in \S\ref{sec:cmb_lss}, peculiar velocities make
large scale galaxy clustering anisotropic in redshift space
\citep{kaiser87}.
In linear theory, the relation between the real-space matter
power spectrum $P(k)$ and the redshift-space galaxy power
spectrum $P_g(k,\mu)$ at redshift $z$ follows equation~(\ref{eqn:pkmu}):
$P_g(k,\mu)=[b_g(z)+\mu^2 f(z)]^2 P(k),$
where $b_g(z)$ is the galaxy bias factor, $f(z)$ is the logarithmic
growth rate of fluctuations, and $\mu$ is the cosine of the
angle between the wavevector {\bf k} and the line of sight.
The strength of the anisotropy is governed by distortion parameter
$\beta = f(z)/b_g(z)$, which has been measured for a variety of
galaxy redshift samples (e.g., 
\citealt{cole95,peacock01,hawkins03,okumura08}).
By modeling the full redshift-space galaxy power spectrum one
can extract the parameter combination $f(z)\sigma_8(z)$, the
product of the matter clustering amplitude and the growth rate
(see \citealt{percival09}, who provide a
clear review of the physics of redshift-space distortions and recent
theoretical developments).  
Like any galaxy clustering measurement, statistical errors for
redshift-space distortion (RSD) come from the combination of 
sample variance --- determined by 
the finite number of structures present in the survey volume ---
and shot noise in the measurement of these structures
(see \S\ref{sec:bao_obs_stats}).
Optimal weighting of galaxies based on their host halo masses
can reduce the effects of shot noise
below the naive expectation from Poisson statistics
\citep{seljak09,cai11,hamaus12}.
However, sample variance has a large impact on RSD measurements because 
filaments and walls extend for many tens of Mpc with specific orientations,
so even in real space one would find isotropic clustering only
after averaging over many such structures.
\cite{mcdonald09} show that one can partly beat the limits
imposed by sample variance by analyzing multiple galaxy
populations with distinct bias factors in the same volume, which
allows one to extract information from the $b_g$-dependence of the
amplitude $\delta_g({\bf k},\mu)$ of each individual mode, rather than
just the variance of the modes. \cite{bernstein11} provide a
nicely pedagogical discussion of this idea.

Anisotropy of clustering in galaxy
redshift surveys thus offers an alternative to weak lensing and
cluster abundances as a tool for measuring the growth of structure.
While WL and clusters constrain the amplitude of matter
clustering and yield growth {\it rate} constraints from measurements
at multiple redshifts, redshift-space distortions directly measure
the rate at which structure is growing at the redshift of observation;
the coherent flows responsible for RSD are the same flows that are
driving the growth of fluctuations (eq.~\ref{eqn:divv}).
Recent observational analyses include the measurement of 
\cite{guzzo08} from the VIMOS-VLT Deep Survey (VVDS),
$f(z) = 0.91 \pm 0.36$ at $z\approx 0.8$,
the measurement of \cite{samushia11}
from SDSS DR7, obtaining $\sim 10\%$ constraints on $f(z)\sigma_8(z)$
at $z=0.25$ and $z=0.37$, the measurement of \cite{blake11a} from
the WiggleZ survey, obtaining $\sim 10\%$ constraints in each of four 
redshift bins from $z=0.1$ to $z=0.9$, 
the measurement of \cite{reid12} from BOSS, 
obtaining a $\sim 8\%$ constraint at $z=0.57$, 
and the local measurement of \cite{beutler12b} from the 6dFGS,
obtaining a $\sim 13\%$ constraint at $z = 0.067$.
All of these measurements assume \lcdm\ geometry when
inferring $f(z)\sigma_8(z)$, and their derived growth parameters
are consistent with \lcdm\ predictions.

Redshift-space distortions can be measured with much higher precision
from future redshift surveys designed for BAO studies.
These measurements can improve constraints on dark energy models
assuming GR to be correct, and they can be used to constrain 
(or reveal) departures from GR by testing consistency of the 
growth and expansion histories.  The key challenge in modeling 
RSD is accounting for non-linear effects,
including non-linear or scale-dependent bias between galaxies and
matter, at the level of accuracy demanded by the measurement
precision.  The linear theory formula (\ref{eqn:pkmu}) is an inadequate
approximation even on scales of $50\hmpc$ or more
\citep{cole94,hatton98,scoccimarro04} because of a variety of
non-linear effects, including the ``finger-of-God'' (FoG) distortions 
in collapsing and virialized regions, which are opposite in sign 
from the linear theory distortions.  
Their effects are commonly modeled by
adding an incoherent small scale velocity dispersion to the
linear theory distortions, but this model is physically incomplete,
and it typically leaves $5-10$\% systematic errors in $\beta$ 
estimates \citep{hatton98}.  Higher order perturbation theory can be
used to refine the large scale predictions \citep{scoccimarro04}, but
this does not capture the small scale dispersion effects, which 
are themselves significantly different for galaxies vs. dark matter.
\cite{tinker06} and \cite{tinker07} advocate an approach based
on halo occupation modeling, which has the virtue of adopting an
explicit, self-consistent physical description that can encompass
linear, quasi-linear, and fully non-linear scales.  However, the
model is complicated, and it is presently implemented 
using numerically calibrated fitting formulas that may not
generalize to all cosmologies.  Following similar lines,
\cite{reid11} present a simpler and more fully analytic scheme for
computing redshift-space clustering of halos, which may prove
sufficiently accurate for the large scales probed by future surveys.
\citet{hikage11} suggest using galaxy-galaxy lensing to estimate the
radial distribution of tracer galaxies in their dark matter halos and
combining with the virial theorem to predict the FoG profile.
Other recent discussions of analytic or numerically calibrated models
of non-linear RSD, from different perspectives,
include \cite{taruya10}, \cite{jennings11b}, \cite{seljak11},
\cite{jennings12a}, and \cite{okumura12}; this is a highly
active area of current research.

Linear theory RSD depends only on the growth parameters $f(z)\sigma_8(z)$,
but testing non-GR models such as $f(R)$ gravity with RSD may 
require full numerical simulations to capture non-linear effects
in these models (e.g., \citealt{jennings12b,li12}).
While analytic models are convenient when fitting data to extract
parameter values and errors, there is no problem of principle in
using brute-force numerics to compute RSD predictions, for either
GR or modified gravity models.
The fundamental
question is the limit on the accuracy of predictions that
will be imposed by uncertainties in galaxy formation physics,
such as the relative velocity dispersions of galaxies and dark
matter within halos.
These limits are poorly understood at present.

Since the number of Fourier modes in a 3-dimensional volume increases
as $k^3$, the precision of clustering measurement is generally higher
on smaller scales, at least until one hits the shot noise limits of
the tracer population.  Forecasts of cosmological constraints from
RSD remain uncertain because it is not clear 
how small a scale and how high a precision one can go to before
being limited by theoretical modeling systematics.
The impact of theoretical systematics is often characterized
implicitly in terms of a maximum wavenumber $\kmax$ used in
the modeling.  As one goes to larger $\kmax$ the non-linear effects
are larger, and the demands on modeling them accurately become
more stringent because the statistical precision is higher.
One can think of $\kmax$ as representing the crossover scale
where theoretical uncertainties become comparable to the statistical
uncertainty, a scale that depends on the survey volume as well
as the modeling accuracy itself.  Most forecasts (including
ours below) assume that modeling is perfect up to $\kmax$
but uses no information from higher $k$.  In practice, analyses
may continue to high $k$ but marginalize over systematic uncertainties,
leading to an ``effective'' value of $\kmax$ that determines
the strength of the RSD constraints.

Plausible assumptions suggest promising prospects for future
RSD experiments.
For example, assuming a maximum $k$ equal to $0.075 \invhmpc$
at $z=0$ and tracking the non-linear scale $k_{\rm nl}$ at
higher redshifts, \cite{white09} predict $1\sigma$ errors on
$f(z)\sigma_8(z)$ of a few percent per $\Delta z = 0.1$ redshift
bin out to $z=0.6$ from the SDSS-III BOSS survey, and for a 
space-based survey that achieves a high galaxy density\footnote{We caution
that the space densities assumed by \cite{white09} are much
higher than those that \euclid\ or \wfirst\ is likely to achieve,
so these RSD precision forecasts may prove overoptimistic.
Recent forecasts in the specific context of \euclid\ and \wfirst\
appear in \cite{majerotto12} and \cite{green12}, respectively.}
out to $z=2$ they predict errors per $\Delta z = 0.1$
that drop from $\sim 1\%$ at $z=0.8$ to $\sim 0.2\%$ at $z=1.9$.
These forecasts incorporate the \cite{mcdonald09} 
method for beating sample variance.
\cite{reid11} examine BOSS RSD forecasts in more detail, considering the
impact of modeling uncertainties.  They forecast a $1\sigma$
error on $f(z)\sigma_8(z)$ at $z=0.55$ of 1.5\% using correlation
function measurements down to a comoving scale $s_{\rm min}=10\hmpc$,
rising to 3\% if the minimum scale is $s_{\rm min}=30\hmpc$.
(The corresponding wavenumber scale is 
$k_{\rm max} \approx 1.15\pi/s_{\rm min}$.)
These forecasts assume marginalization over a nuisance parameter
$\sigma_v$ characterizing the small scale velocity dispersion.
They improve by a factor of $\sim 1.5$ if $\sigma_v$ is assumed to
be known perfectly, demonstrating the potential gains from a method
(like that of \citealt{tinker07}) that can use smaller scale
measurements to pin down the impact of velocity dispersions.
More generally, small scale clustering may be useful to pin down
the nuisance parameters of large scale RSD models and therefore
improve the precision of the cosmological parameter 
measurements \citep{tinker06}.

At the percent level there is another potential systematic error in RSD
if the selection function has an orientation dependence (e.g., due to fiber
aperture or self-extinction by dust in the target galaxy) and galaxies are
aligned by the large-scale tidal field. This exactly mimics RSD, even in
the linear regime \citep{hirata09}, but fortunately
the effect seems to be negligible for present surveys.
Orientation-dependent selection is predicted to be a larger effect for
high-$z$ \lya\ emitters \citep{zheng11},
since there the radiation can resonantly scatter in the IGM
and must make its way out through the large-scale velocity flows
surrounding the galaxy; at very high redshift ($z=5.7$) simulations
predict an order unity effect. The implications for \lya\ emitters
at more modest redshift will become clear with the HETDEX survey.

In \S\ref{sec:forecast_rsd} we show that our fiducial Stage IV
program (CMB+SN+BAO+WL)
constrains $\sigma_8(z)f(z)$ to a $1\sigma$ precision of
2\% at $z=0.5$ and 1\% at $z \ga 1$ if we assume a $w_0-w_a$
dark energy model with $G_9$ and $\Delta\gamma$ as parameters
to describe departures from GR.  Thus, RSD
measurements with this level of precision or better can significantly
improve the figure-of-merit for dark energy constraints and sharpen
tests of GR, even in a combined program that includes powerful
weak lensing constraints.  Much weaker RSD
measurements could still make a significant contribution to Stage III
constraints.  Forecasts of the contribution of redshift-space
distortions to constraints from specific Stage IV experiments
(a BigBOSS-like ground-based survey and a \euclid- or 
\wfirst-like space-based survey)
are presented by \cite{stril09}, \cite{wang10}, and \cite{wang12b}.

\begin{figure} 
\begin{centering}
\includegraphics[width=4.0in]{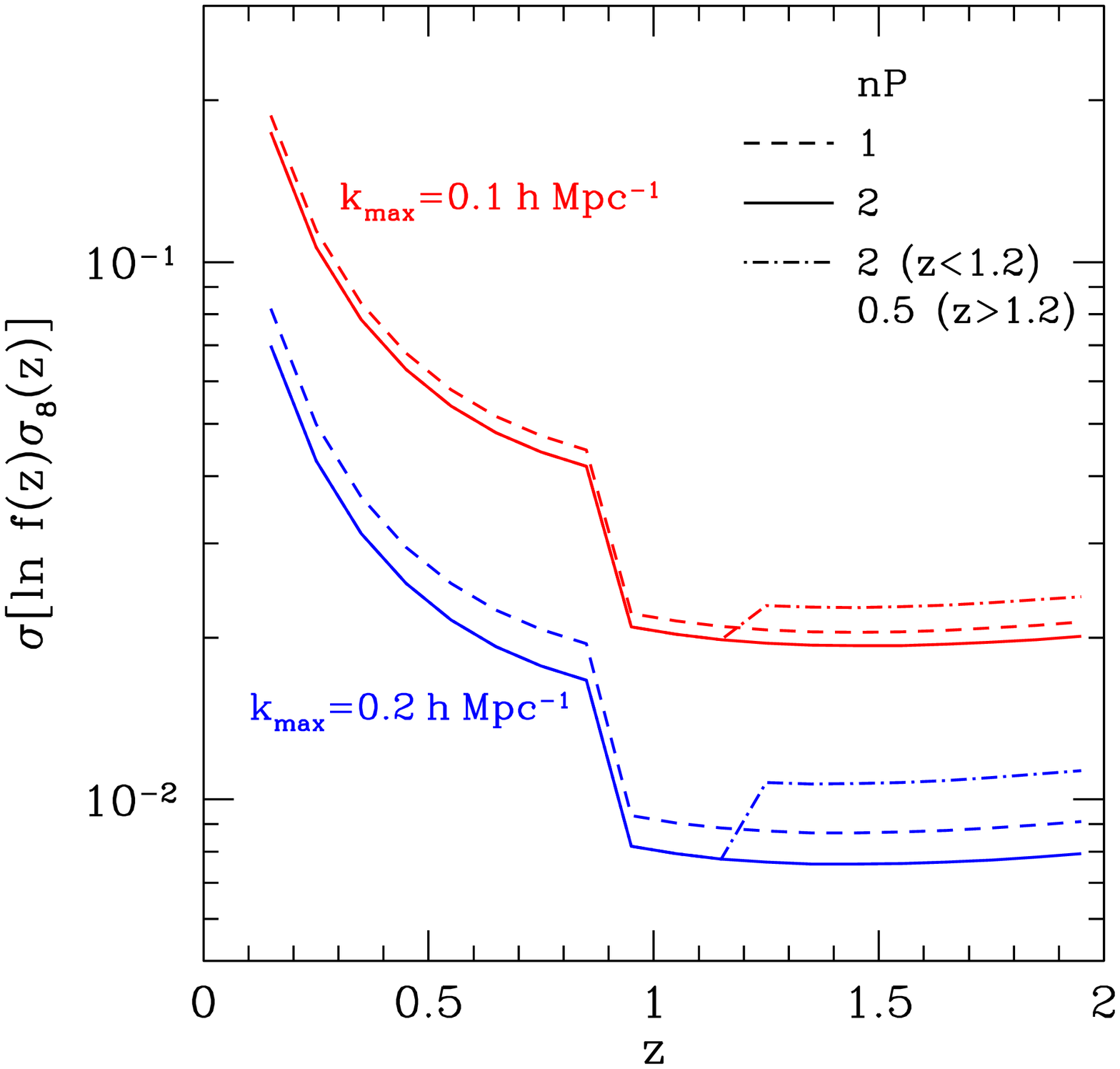}
\caption{
Forecast errors 
on $f(z)\sigma_8(z)$ 
per $\Delta z = 0.1$ redshift bin 
from an RSD survey with $\fsky=0.25$, computed with the
code of \cite{white09}.  For simplicity, we assume linear theory
up to $\kmax=0.1\invhmpc$ (upper, red curves) or
$\kmax=0.2\invhmpc$ (lower, blue curves) and no information
from smaller scales.  Solid and dashed curves show sampling
densities $nP=2$ and $nP=1$, respectively, and dot-dashed 
curves show a case where $nP$ drops from 2 to 0.5 at $z=1.2$.
The sharp drop at $z=0.9$ reflects an assumed change in bias
factors from a high value (appropriate to absorption-line galaxies)
at low $z$ to a lower value (appropriate to emission-line galaxies)
at high $z$.
} 
\label{fig:rsderr}
\end{centering}
\end{figure}

To provide guidance for our forecast discussions in \S\ref{sec:forecast},
we have used the publicly available code of \cite{white09}
to predict errors on $f(z)\sigma_8(z)$ for different assumptions
about survey parameters and modeling limitations.
The solid curves in Figure~\ref{fig:rsderr} show predicted fractional
errors per $\Delta z = 0.1$ redshift bin assuming a survey with
$\fsky = 0.25$ and a sampling density that yields $nP(k=0.2\invhmpc) = 2$,
the same assumptions that we make for our fiducial Stage IV BAO
program (see \S\ref{sec:fiducial}).
In contrast to BAO forecasts, which depend only on the combination
$nP$, RSD forecasts also depend separately on the bias evolution
$b_g(z)$ because more strongly biased tracers 
exhibit weaker anisotropy (lower $\beta$) and therefore provide less leverage
on $f(z)\sigma_8(z)$.  Here we have assumed strongly biased tracers
(such as luminous red galaxies) at $z < 0.9$ and weaker bias
(appropriate to emission line galaxies) at $z>0.9$; specifically,
we adopt $b_g(z)\sigma_8(z) = 1.3$ at $z<0.9$ (motivated by \citealt{reid12})
and $b_g(z)\sigma_8(z) = 0.6$ at $z>0.9$, corresponding to $b_g=1.5$ 
at $z=1.5$ (see \citealt{orsi10} and \citealt{geach12}).
This change in assumed bias factor produces the sharp drop in the
forecast error at $z=0.9$.  Maintaining $nP=2$ would, of course,
require a corresponding jump in galaxy density for $z>0.9$.
Upper and lower curves correspond to $\kmax=0.1\invhmpc$ and
$0.2\invhmpc$, respectively.  In either case,
the forecast error drops with increasing redshift out to $z=0.9$
because of the larger comoving volume per $\Delta z = 0.1$ bin, then
stays roughly constant from $z = 0.9 - 2$.

For $\kmax=0.2\invhmpc$ and $nP=2$, 
the error per bin is about 0.8\% from $z = 0.9 - 2$.
Lowering the sampling density from $nP=2$ to
$nP=1$ degrades the fractional error by about 12.5\%,
equivalent to a 25\% reduction in $\fsky$.
The dot-dashed curve shows the case in which we assume $nP=2$ for
$z<1.2$ and $nP=0.5$ for $z>1.2$, where emission line galaxy redshifts
become increasingly difficult to obtain from the ground and the
dominant samples may eventually come from slitless spectroscopic
surveys with \euclid\ and \wfirst.  In this case, the high
redshift error increases by $\sim 40\%$.  Reducing $nP$ has less 
impact for $\kmax=0.1\invhmpc$ because structure at this larger
scale is more fully sampled, leaving sample variance as the dominant
source of measurement uncertainty.

Figure~\ref{fig:rsderr} highlights the critical role of
modeling uncertainty in determining the ultimate cosmological
return from RSD measurements.  If we assume $\kmax=0.1\invhmpc$
instead of $\kmax=0.2\invhmpc$, then the errors for $nP=2$
are larger by a factor $\sim 2.5$, only slightly less than
the factor $(0.2/0.1)^{3/2} = 2.83$ suggested by a pure mode
counting argument.  In both cases we have assumed that $\kmax$
is constant with redshift in comoving coordinates, in contrast
to \cite{white09} who assume that it scales with $k_{\rm nl}$,
and this difference largely accounts for the substantially larger
errors that we forecast at high redshift.  It is not clear 
which assumption is more appropriate, since it is not clear 
whether the scale at which modeling uncertainties dominate
will be set by non-linearity in the matter clustering,
which tracks $k_{\rm nl}$, or by non-linearity in the biased
galaxy clustering, which stays roughly constant in comoving coordinates
because of compensation between $b_g(z)$ and $\sigma_8(z)$.
While our $\kmax=0.2\invhmpc$, $nP=2$ case yields 0.8\% errors
per bin at $z>0.9$, the actual demand on modeling accuracy is tighter
by $\sqrt{N_{\rm bin}}\sim 3.3$ because a systematic modeling error would
be likely to affect all bins coherently.

Our forecasts here include only $P(k)$ modeling, not the additional
gains that are potentially available by applying the \cite{mcdonald09}
method to tracer populations with different bias factors in the
same volume.  High redshift surveys may not yield galaxy samples
with a wide range of bias, but at $z<1$ this approach could reduce
errors significantly relative to those presented here.

For the calculations in Figure~\ref{fig:rsderr}, we have set the
small scale velocity dispersion $\sigma_v=0$, i.e., we have assumed
pure linear theory up to $k=\kmax$.  Setting $\sigma_v=300\kms$
produces only mild degradation of the errors, much smaller than the
difference between $nP=1$ and $nP=2$.  However, marginalizing over
$\sigma_v$, or more generally over parameters that describe non-linear
effects, could degrade precision significantly unless smaller scale
data can be used to constrain these parameters.  Conversely, modeling
to higher $\kmax$ can yield substantially tigher errors.  In experiments
with $\sigma_v$ fixed to $300\kms$, we find that the $\kmax^{3/2}$ scaling
holds at the factor-of-two level up to $\kmax \sim 1\invhmpc$, so the
potential gains are large. 
This analysis thus confirms the key point of this section:
RSD analyses of the same redshift surveys conducted for BAO
could provide powerful constraints on dark energy and stringent
tests of GR growth predictions, but exploiting this potential will require
development of theoretical modeling methods that are accurate
at the sub-percent level in the moderately non-linear regime.

%but as one goes to larger $\kmax$ the
%non-linear effects are larger and the demands on modeling them
%accurately become more stringent.  We conclude from this analysis
%that the data from BAO surveys will allow very tight constraints 
%on structure growth via RSD, with limitations that will likely
%come from theoretical modeling uncertainty.

\subsection{The Alcock-Paczynski Test}
\label{sec:ap}

The translation from angular and redshift separations to comoving
separations depends on $D_A(z)$ and $H(z)$, respectively.
Therefore, even if peculiar velocities are negligible, clustering in
redshift space will appear anisotropic if one adopts an incorrect
cosmological model --- specifically, one with an incorrect value of the
product
$H(z)D_A(z)$.  Alcock \& Paczynski (\citeyear{alcock79}; hereafter AP)
proposed an idealized cosmological test using this idea, based on a
hypothetical population of intrinsically spherical galaxy clusters.
The AP test can be implemented in practice by using the
amplitude of quasar or galaxy clustering to identify equivalent
scales in the angular and redshift dimensions
\citep{ballinger96,matsubara96,popowski98,matsubara01}
or by using anisotropy of clustering in the \lya\ forest
\citep{hui99,mcdonald99}.
AP measurements provide a cosmological test in their own right, and they
allow high-redshift distance measurements to be translated into
constraints on $H(z)$, which is a more direct measure of energy density.
Recently \cite{blake11d} have measured the AP parameter $H(z)D_A(z)$
from galaxy clustering in the WiggleZ survey, obtaining $10-15\%$
precision in each of four redshift bins out to $z=0.8$, and
\cite{reid12} have obtained $\sim 6\%$ precision at $z=0.55$
from BOSS galaxies, improving to $\sim 3.5\%$ if they assume
that the growth rate (and hence the peculiar velocity distortion)
has the value predicted by \lcdm.  Both results are consistent
with flat-$\Lambda$ geometry for WMAP7 values of $\om$.

\begin{figure}
\begin{centering}
\includegraphics[width=3.2in]{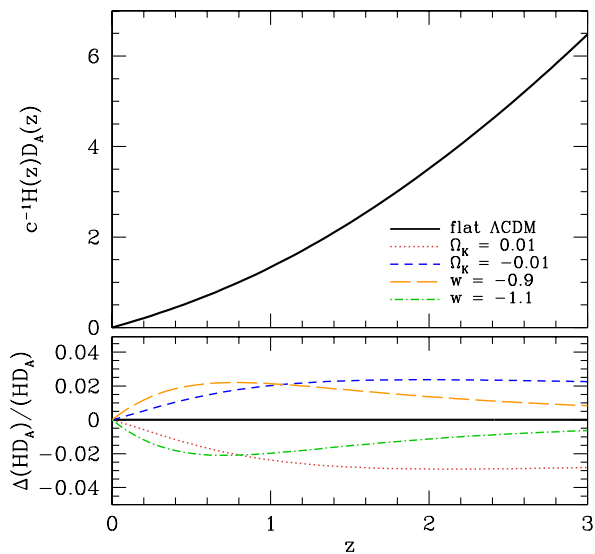}
\caption{\label{fig:models_ap}
Evolution of the parameter combination $H(z)D_A(z)$ 
constrained by the Alcock-Paczynski test, for the same
suite of CMB-normalized models shown in Figure~\ref{fig:hzdz}.
}
\end{centering}
\end{figure}

Figure~\ref{fig:models_ap} shows the evolution of $c^{-1} H(z)D_A(z)$
for the same set of CMB-normalized models shown earlier in 
Figures~\ref{fig:hzdz}-\ref{fig:sig8z}.
At low redshift, the model dependence resembles that of $H(z)/ch$
(lower left panel of Figure~\ref{fig:hzdz}), but 
deviations are reduced in amplitude
because of partial cancellation between $H(z)$ and 
$D_A(z) \propto \int_0^z dz' H^{-1}(z')$.  At high redshift
$\Omega_k = \pm 0.01$ has a larger impact than $1+w = \pm 0.1$.
Note that negative space curvature (positive $\Omega_k$)
tends to increase $D_A$, but because the CMB normalization lowers
$\Omega_m$ (see Table~\ref{tbl:models}) and thus $H(z)/H_0$, the
net effect is to decrease $H(z) D_A(z)$.
In \S\ref{sec:forecast_ap} we show that our fiducial Stage IV
program predicts $H(z) D_A(z)$ with an accuracy of $\sim 0.15-0.3\%$,
assuming a $w_0-w_a$ dark energy model.  AP measurements at this level
could significantly improve dark energy constraints.  
For Stage III, the predictions are considerably weaker, $\sim 0.5-0.9\%$.

Like redshift-space distortion measurements, the AP test is 
automatically enabled by redshift surveys conducted for BAO.  
In practice the two effects must be modeled together
(see, e.g., \citealt{matsubara04}), and the principal systematic
uncertainty for AP measurements is the uncertainty in modeling
non-linear redshift-space distortions.  
At present, it is difficult to forecast the likely precision of
future AP measurements because there have been no rigorous tests
of the accuracy of redshift-space distortion corrections at the
level of precision reachable by such surveys.  
If one assumes that redshift-space distortions 
can be modeled adequately up to $k \sim k_{\rm nl}$ then
the potential gain from AP measurements is impressively large.
For example, \cite{wang10} find that using the full galaxy
power spectrum in a space-based emission-line redshift survey
increases the forecast value of the DETF FoM by a factor of $\sim 3$
relative to the BAO measurement alone; this gain is not broken down
into separate contributions, but we suspect that a large portion
comes from AP.\footnote{Since RSD and AP both depend on modeling the 
broadband $P(k,\mu)$, it is difficult even in principle to separate 
the two types of constraints in an observational analysis.}
In the context of HETDEX, \cite{shoji09} show that
the AP test substantially improves the expected cosmological constraints
relative to BAO analysis alone, even after marginalizing over
linear RSD and a small-scale velocity dispersion parameter.
The appendix to that paper presents analytic formalism for
incorporating AP and RSD constraints into Fisher matrix analyses.

Halo occupation methods (see \S\ref{sec:cmb_lss}) provide a useful
way of approaching peculiar velocity uncertainties in AP measurements.
Observations and theory imply that galaxies reside in halos, and on
average the velocity of galaxies in a halo should equal the halo's
center-of-mass velocity because galaxies and dark matter feel the
same large scale acceleration.  However, the dispersion of satellite
galaxy velocities in a halo could differ from the dispersion of dark
matter particle velocities by a factor of order unity, and central 
galaxies could have a dispersion of velocities relative to the
halo center-of-mass \citep{vandenbosch05,tinker06}.
To be convincing, an AP measurement must show that it is robust
(relative to its statistical errors) to plausible variations in
the halo occupation distribution and to plausible variations in
the velocity dispersion of satellite and central galaxies.
Alternatively, the AP errors can be marginalized over uncertainties
in these multiple galaxy bias parameters, drawing on constraints 
from the observed redshift-space clustering.

BAO measurements from spectroscopic surveys in some sense already
encompass the AP effect, since they use the location of the BAO
scale as a function of angular and redshift separations to 
separately constrain $D_A(z)$ and $H(z)$.  
The presence of a feature at a particular scale makes the separation
of the AP effect from peculiar velocity RSD much more straightforward
(see fig.~6 of \citealt{reid12}).
However, the addition of a 
high-precision AP measurement from smaller scales could significantly
improve the BAO cosmology constraints.
BAO measurements typically constrain $D_A(z)$ better than $H(z)$
because there are two angular dimensions and only one line-of-sight
dimension.  However, at high redshift $H(z)$ is more sensitive
to dark energy than $D_A(z)$, since $H(z)$ responds directly
to $\uphi(z)$ through the Friedmann equation~(\ref{eqn:friedmann})
while $D_A(z)$ is an integral of $H_0/H(z')$ over all $z' < z$.
An AP measurement would allow the BAO measurement of $D_A(z)$
to be ``transferred'' to $H(z)$, thus yielding a better measure
of the dark energy contribution.
\cite{blake11d} have recently implemented a similar idea by using
their AP measurements in the WiggleZ survey 
to convert SN luminosity distances into $H(z)$ determinations.

The AP test can be implemented with measures other than the
power spectrum or correlation function.
One option is to use the angular distribution of small scale pairs
of quasars or galaxies \citep{phillipps94,marinoni10}, though peculiar
velocities still affect this measure, in a redshift- and cosmology-dependent
way \citep{jennings11}.  A promising recent suggestion is to use
the average shape of voids in the galaxy distribution;
individual voids are ellipsoidal, but in the absence of peculiar
velocities the {\it mean} shape should be spherical
\citep{ryden95,lavaux10}.
Typical voids are of moderate scale ($R\sim 10\hmpc$) and have 
a large filling factor $f$, so the achievable precision in a
large redshift survey is high if the sampling density is 
sufficient to allow accurate void definition.
A naive estimate for the error
on the mean ellipticity of voids with rms ellipticity $\epsilon_{\rm rms}$
in a survey volume $V$ is
$\sigma_{\bar\epsilon} \sim 
\epsilon_{\rm rms} (fV / \frac{4}{3}\pi R^3)^{-1/2} 
 \approx 6\times 10^{-4} (\epsilon_{\rm rms}/0.3)
   (fV/1\,h^{-3}\,{\rm Gpc}^3)^{-1/2} (R/10\hmpc)^{-3/2},$
if the galaxy density is high enough to make the shot noise 
contribution to the ellipticity scatter negligible on scale $R$.
Peculiar velocities have a small, though not negligible, impact on
void sizes and shapes \citep{little91,ryden96,lavaux11},
so one can hope that the uncertainty in this impact will be
small, but this hope has yet to be tested.  Assuming statistical errors
only, \cite{lavaux11} estimate that a void-based AP
constraint from a \euclid-like redshift survey would provide several
times better dark energy constraints than the BAO measurement from
the same data set, mainly because the scale of voids is so much
smaller than the BAO scale.
\cite{sutter12} have recently applied the AP test to a void catalog
constructed from the SDSS DR7 redshift surveys, though with this
sample the statistical errors are too large to yield a significant
detection of the predicted effect.

\subsection{Alternative Distance Indicators}
\label{sec:other_di}

In \S\S\ref{sec:sn} and~\ref{sec:bao} we have discussed the two most
well established methods for measuring the cosmological distance
scale beyond the local Hubble flow: Type Ia supernovae and BAO.
These two methods set a high bar for any alternative distance
indicators.  Type Ia supernovae are highly luminous, making them
relatively easy to discover and measure at large distances.
Once corrected for light curve duration, local Type Ia's have a
dispersion of $0.1-0.15$ mag in peak luminosity despite sampling
stellar populations with a wide range of age and metallicity,
and extreme outliers are apparently rare \citep{li11}.
Thus far, surveys are roughly succeeding in achieving the
$\sqrt{N}$ error reduction from large samples, though
progress on systematic uncertainties will be required to continue
these gains.
The BAO standard ruler is based on well understood physics,
and it yields distances in absolute units.
``Evolutionary'' corrections (from non-linear clustering
and galaxy bias) are small and calculable from theory.

Core collapse supernovae exhibit much greater diversity than
Type Ia supernovae, which is not surprising given the greater
diversity of their progenitors.  However, Type IIP supernovae,
characterized by a long ``plateau'' in the light curve after
peak, show a correlation between expansion velocity (measured
via spectral lines) and the bolometric luminosity of the 
plateau phase, making them potentially useful as standardized
candles with $\sim 0.2$ mag luminosity scatter
(\citealt{hamuy02}; see \citealt{maguire10} for a recent 
discussion).
%\footnote{Type IIP supernovae can also be used for
%distance measurements via the Expanding Photosphere Method
%\citep{schmidt94}, but the high data quality required for
%this method makes it impractical beyond the local Hubble flow.}
Unfortunately, as distance indicators Type IIP supernovae appear to be
at least slightly inferior to Type Ia supernovae on every score:
they are less luminous, the scatter is larger, the fraction of
outliers may be larger, and they arise in star-forming environments
that are prone to dust extinction.  With the existence of cosmic
acceleration now well established by multiple methods, we are
skeptical that Type IIP supernovae can make a significant contribution
to refinement of dark energy constraints.

The door for alternative distance indicators is more open
beyond $z=1$, the effective limit of most current SN and BAO surveys.
Gamma-ray bursts (GRBs) are highly luminous, so they can be 
detected to much higher redshifts than optical supernovae;
the current record holder is at $z\approx 8.2$
\citep{tanvir09,salvaterra09}.  GRBs are extremely diverse
and highly beamed, but they exhibit correlations
\citep{amati06,ghirlanda06} between equivalent isotropic
energy and spectral properties (such as the energy of peak 
intensity) or variability.  These correlations can be used
to construct distance-redshift diagrams for those
systems with redshift measured via spectroscopy of afterglow
emission or of host galaxies (e.g., \citealt{schaefer07};
see \citealt{demianski11} for a recent review and discussion).  While
GRBs reach to otherwise inaccessible redshifts, we are again
skeptical that they can contribute to our understanding
of dark energy because of statistical limitations and 
susceptibility to systematics.  It has taken detailed observations of 
many hundreds of Type Ia supernovae, local and distant, 
to understand their systematics and statistics.
The number of GRBs with spectroscopic redshifts is $\sim 100$,
and the spectroscopic sample may be a biased subset of the full 
GRB population because of the requirement of a bright optical
afterglow or identified host galaxy.

Quasars are another tool
for reaching high redshifts, drawing on empirical correlations
between line equivalent widths and luminosity
\citep{baldwin77} or between luminosity and the broad line region
radius $R_{\rm BLR}$ \citep{bentz09}. 
For example, \cite{watson11} have recently proposed 
reverberation mapping (which measures $R_{\rm BLR}$)
of large quasar samples to constrain dark energy models.
The high redshift quasar population is systematically 
different (in black hole mass and host galaxy environment)
from lower redshift calibrators.
Quasar spectral properties appear remarkably stable
over a wide span of redshift \citep{steffen06}, 
and the dependence of $R_{\rm BLR}$ on luminosity is 
driven primarily by photoionization physics.  However, in
the event of a ``surprising'' result from quasar distance
indicators, one would have 
to be prepared to argue that subtle (e.g., 10\% or smaller) changes
with redshift were a consequence of cosmology rather than evolution.
Of course, quasar distance indicators also face the same challenges
of photometric calibration, k-corrections, and dust
extinction that affect supernova studies.

Radio galaxies have been employed as a standard (or at least
standardizable) ruler for distance-redshift studies,
drawing on empirically tested theoretical models that
connect the source size to its radio properties
\citep{daly94,daly02}. 
Analysis of 30 radio galaxies out to $z=1.8$ gives
results consistent with those from Type Ia supernovae
\citep{daly09}.
The number of radio galaxies to which
this technique can be applied is limited, and the model assumptions
used to translate observables into distance estimates
are fairly complex (see \citealt{daly09}, \S 2.1).  
We therefore expect that both statistical and systematic
limitations will prevent this method from becoming competitive
with supernovae and BAO.

In \S\ref{sec:forecast_distance} we show forecast distance
errors for our fiducial Stage III and Stage IV experimental programs,
presenting a target for alternative methods.  If one assumes
a $w_0-w_a$ model then the constraints are very tight, with
errors below $\sim 0.25\%$ at Stage IV and $\sim 0.5\%$ at
Stage III.  However, with a general $w(z)$ model the constraints
become much weaker outside the redshift range directly
measured by Type Ia SNe or BAO.  In particular, our Stage IV
forecasts presume large BAO surveys at $z>1$, and if these do
not come to fruition there is much more room for alternative
indicators at high redshift.

\subsection{Standard Sirens}
\label{sec:sirens}

Gravitational wave astronomy opens an entirely different route
to distance measurement, with an indicator that is
grounded in fundamental physics \citep{schutz86}.  The basic concept is
illustrated by considering a nearly Newtonian binary system of two black
holes with total mass $M$ and reduced mass $\mu$, in a nearly circular
orbit at separation $a$. The gravitational wave luminosity of such a source is
\begin{equation}
L_{\rm GW} = \frac{32G^4\mu^2M^3}{5c^{5}a^5}.
\end{equation}
If one can measure the angular velocty of the orbit\footnote{The 
observed frequency
of the gravitational wave is $2\omega$ because the source is a
quadrupole, producing two crests and two troughs per orbit.} $\omega =
\sqrt{GM/a^3}$, its rate of change due to inspiral as the binary loses
energy
$\dot\omega/\omega = 96G^{3}\mu M^{2}/5c^{5}a^{4}$, and the orbital velocity
$v=\sqrt{GM/a}$ (using relativistic corrections to the emitted waveform),
one has enough information to solve for $a$, $M$, and $\mu$.
One can therefore calculate $L_{\rm GW}$ from the measured observables
and compare to the measured energy flux to infer distance.
In practice, one would
need to solve for other dimensionless parameters such as the eccentricity
and the orientation of the orbit, black hole spins, and source position
on the sky.  The solution is not trivial (!),
but gravitational waveforms from relativistic
binaries encode this information in higher harmonics and modulation of
the signal due to precession \citep{arun07}.
Because of the analogy between gravitational wave observations and
acoustic wave detection, this approach is often referred to
as the ``standard siren'' method.

There are several practical obstacles to gravitational wave cosmology.
First, of course, gravitational waves from an extragalactic source must
be detected.  The most promising near-term possibility is nearby ($z\ll 1$)
neutron star binaries, which should be detected by the ground-based Advanced
LIGO detector (to start observations in $\sim 2014$) and upgraded
VIRGO detector, and which could be used to measure $H_0$. 
The space-based
gravitational wave detector \lisa\ (possible launch in the 2020 decade)
is designed to allow high S/N measurements of 
the mergers of massive black holes at the centers of
galaxies at $z\sim{\cal O}(1)$, which 
would enable a full Hubble diagram $D_L(z)$ to be constructed.  
A second complication is that gravitational wave observations yield a
distance but do not give an independent
source redshift.  One thus needs an identification of the host galaxy,
and given the angular positioning accuracy of gravitational wave observations
this will generally require identification of an electromagnetic
transient that accompanies the gravitational wave burst.
Possibilities include GRBs
resulting from neutron star mergers \citep{dalal06} and the 
optical, X-ray, or radio signatures of the response of
an accretion disk to a massive black hole merger \citep{milosavljevic05,
lippai08}.  However, both the event rates and the characteristics of the
electromagnetic signatures are poorly understood at present.  
One can also make identifications statistically using large
scale structure \citep{macleod08}.  
A third complication, important for the high S/N observations expected
from \lisa, is that weak lensing magnification becomes a dominant
source of noise at $z \ga 1$, inducing a scatter in distance
of several percent per observed source 
\citep{markovic93, holz05, jonsson07}.  By taking advantage of the
non-Gaussian shape of the lensing scatter, one can reduce the error
on the mean by a factor $\sim 2-3$ below the naive $\sigma/\sqrt{N}$
expectation \citep{hirata10,shang11}, so samples of a few dozen
well observed sources could yield sub-percent distance scale errors.

\citet{nissanke10} forecast constraints on
$H_0$ from next-generation ground-based gravitational wave detectors,
including Monte Carlo simulations of parameter recovery from neutron
star-neutron star and neutron star-black hole mergers. They find that
$H_0$ can be constrained to 5\% for 15 NS-NS mergers with GRB
counterparts and a network of three gravitational wave 
detectors. While the event rate is
highly uncertain, tens of events per year are quite possible and could
lead to percent-level constraints on $H_0$ a decade or so from now.
\cite{taylor12} discuss the prospects for a network of
``third generation'' ground-based interferometers, which could
detect $\sim 10^5$ double neutron star binaries.

It remains to be seen whether standard sirens can compete
with other distance indicators in the LIGO/VIRGO and \lisa\ era.
Looking further ahead, \cite{cutler09} show that an interferometer
mission designed to search for gravitational waves from the
inflation epoch in the $0.03 - 3\,$Hz range, such as
the {\it Big Bang Observer} ({\it BBO}, \citealt{phinney04})
or the Japanese {\it Decigo} mission \citep{kawamura08}, 
would detect hundreds of thousands of compact star binaries
out to $z \sim 5$.  The arc-second level resolution would be
sufficient to identify the host galaxies for most binaries
even at high $z$, enabling redshift determinations through
follow-up observations.  With the S/N expected for {\it BBO},
the error in luminosity distance for most of these standard 
sirens would be dominated by weak lensing magnification, and
the $\sqrt{N}$ available for beating down this scatter
would be enormous.  Indeed, \cite{cutler09} argue that the
dispersion of distance estimates as a function of separation
could itself be used to probe structure growth via the strength of the
WL effect.  For the {\it BBO} case, \cite{cutler09} forecast 
errors of $\sim 0.1\%$, 0.01, and 0.1 on $H_0$, $w_0$, and $w_a$
when combining with a \planck\ CMB prior, assuming a flat universe.
Because the distance indicator is rooted in fundamental physics,
there are no obvious systematic limitations to this method
provided the calibration of the gravitational wave measurements
themselves is adequate.  

\subsection{The \lya\ Forest as a Probe of Structure Growth}
\label{sec:lyaf}

The \lya\ forest is an efficient tool for mapping structure at
$z\approx 2-4$ because each quasar spectrum provides many independent
samples of the density field along its line of sight.
(At lower redshifts \lya\ absorption moves to UV wavelengths
unobservable from the ground, and at higher redshifts the forest
becomes too opaque to trace structure effectively.)
The relation between \lya\ absorption and matter density is non-linear
and to some degree stochastic.  However, the physics of this relation
is straightforward and fairly well understood, in contrast to 
the more complicated processes that govern galaxy formation.
We have previously discussed the \lya\ forest as a method of measuring
BAO at $z>2$, which requires only that the forest provide a linearly
biased tracer of the matter distribution on $\sim 150\,$Mpc scales.
However, by drawing on a more detailed theoretical description 
of the forest, one can use \lya\ flux statistics to infer the 
amplitude of matter fluctuations and thus measure structure growth
at redshifts inaccessible to weak lensing or clusters.

The \lya\ forest is described to surprisingly good accuracy by
the Fluctuating Gunn-Peterson Approximation
(FGPA, \citealt{weinberg98}; see also \citealt{gunn65,rauch97,croft98}),
which relates the transmitted flux $F=\exp(-\tau_{{\rm Ly}\alpha})$
to the dark matter overdensity $\rho/\bar{\rho}$, with the latter
smoothed on approximately the Jeans scale of the diffuse intergalactic
medium (IGM) where gas pressure supports the gas against gravity
\citep{schaye01}.  Most gas in the low density IGM follows a power-law
relation between temperature and density, $T=T_0(\rho/\bar{\rho})^\alpha$
with $\alpha \la 0.6$, which arises from the competition between 
photo-ionization heating and adiabatic cooling
\citep{katz96,hui97}.  The \lya\ optical depth is proportional
to the hydrogen recombination rate, which scales as
$\rho^2 T^{-0.7}$ in the relevant temperature range near $10^4\K$.
This line of argument leads to the relation 
\begin{equation}
\label{eqn:fgpa}
F = \exp(-\tau_{{\rm Ly}\alpha}) \approx 
    \exp\left[-A(\rho/\bar{\rho})^{2-0.7\alpha}\right] ~.
\end{equation}
The constant $A$ depends on a combination of parameters that are 
individually uncertain (see \citealt{croft98,peeples10}), and the
value of $\alpha$ depends on the IGM reionization history,
so in practice these parameters must be inferred empirically from the
\lya\ forest observables.  However, even after marginalizing over
these parameters there is enough information in the clustering statistics
of the flux $F$ to constrain the shape and amplitude of the
matter power spectrum (e.g., \citealt{croft02,viel04,mcdonald06}).
\lya\ forest surveys conducted for BAO allow high precision measurements
of flux correlations on smaller scales, so they have the statistical
power to achieve tight constraints on matter clustering.

There are numerous physical complications not captured by 
equation~(\ref{eqn:fgpa}).  On small scales, absorption is smoothed
along the line of sight by thermal motions of atoms.  Peculiar
velocities add scatter to the relation between flux and density,
though this effect is mitigated if one uses the redshift-space 
$\rho/\bar{\rho}$ in equation~(\ref{eqn:fgpa}).
Gas does not perfectly trace dark matter, so
$(\rho/\bar{\rho})_{\rm gas}$ is not identical to $(\rho/\bar{\rho})_{\rm DM}$.
Shock heating and radiative cooling push 
some gas off of the temperature-density relation.
All of these effects can be calibrated using hydrodynamic cosmological
simulations, and since the physical conditions are not highly non-linear
and the effects are moderate to begin with, uncertainties in the
effects are not a major source of concern.  

A more serious obstacle
to accurate predictions is the possibility that inhomogeneous IGM
heating --- especially heating associated with helium reionization, which
is thought to occur at $z\approx 3$ --- produces spatially coherent
fluctuations in the temperature-density relation that appear as
extra power in \lya\ forest clustering, or makes the relation more
complicated than the power law that is usually assumed 
\citep{mcquinn11a,meiksin11}.
Fluctuations in the ionizing background radiation can also produce extra 
structure in the forest, though this effect should be small on 
comoving scales below $\sim 100\Mpc$ \citep{mcquinn11a}.  On the observational
side, the primary complication is the need to estimate the unabsorbed
continuum of the quasar, relative to which the absorption is
measured.  (In our notation, $F$ is the ratio of the observed flux
to that of the unabsorbed continuum.)
For statistical analysis of a large sample, the continuum does
not have to be accurate on a quasar-by-quasar basis, and there
are strategies (such as measuring fluctuations relative to a running
mean) for mitigating any bias caused by continuum errors
(see, e.g., \citealt{slosar11}).
Nonetheless, residual uncerainties from continuum determination can
be significant compared to the precision of measurements.

A discrepancy between clustering growth inferred from the \lya\ 
forest and cosmological models favored by other data would face 
a stiff burden of proof, to demonstrate that the \lya\ forest
results were not biased by the theoretical and observational
systematics discussed above.  However, complementary clustering statistics
and different physical scales have distinct responses to systematics and
to changes in the matter clustering amplitude, so it may be
possible to build a convincing case.  
For example, the bispectrum \citep{mandelbaum03,viel04b}
and flux probability distribution (e.g., \citealt{lidz06})
provide alternative ways to break the degeneracy between 
mean absorption and power spectrum amplitude and to
test whether a given model of IGM physics is really an adequate
description of the forest.
\lya\ forest tests will assume
special importance if other measures 
indicate discrepancies at lower
redshifts with the growth predicted by GR combined with simple dark energy
models.  Growth measurements at $z\sim 3$ from the \lya\ forest
could then play a critical role in distinguishing between modified
gravity explanations and models with unusual dark energy history.
Because it probes high redshifts and moderate overdensities,
the \lya\ forest can also constrain the primordial power spectrum
on small scales that are inaccessible to other methods.  The 
resulting lever arm may be powerful for detecting or constraining
the scale-dependent growth expected in some modified gravity
models, as discussed further in the next section.

\subsection{Other Tests of Modified Gravity}
\label{sec:gravity}

We have concentrated our discussion of modified gravity on
tests for consistency between measured matter fluctuation
amplitudes and growth rates --- from weak lensing, clusters, and
redshift-space distortions --- with the predictions of dark energy
models that assume GR.  However, ``not General Relativity'' is a 
broad category, and there are many other potentially observable
signatures of modified gravity models.  For an extensive
review of modified gravity theories and observational tests,
we refer the reader to \cite{jain10}.  
We follow their notation and discussion in our brief summary here.

In Newtonian gauge, the spacetime metric with scalar perturbations
can be written in the form
\begin{equation}
\label{eqn:metric}
ds^2 = - (1+2\Psi)dt^2 + (1-2\Phi) a^2(t) dl^2,
\end{equation}
which is general to any metric theory of gravity.
If the dominant components of the stress-energy tensor have
negligible anisotropic stress, then the Einstein equation 
of GR predicts that $\Psi=\Phi$, i.e., the same gravitational
potential governs the time-time and space-space components of
the metric.  We have made this assumption implicitly in the
WL discussion of \S\ref{sec:wl}.
Anisotropic stress should be negligible
in the matter-dominated era, and most proposed forms of dark
energy (e.g., scalar fields) also have negligible anisotropic
stress.  Therefore, one generic form of modified gravity test
is to check for the GR-predicted consistency between $\Psi$
and $\Phi$.  For example, if the Ricci curvature scalar $R$ in
the GR spacetime action $S \propto \int d^4x\sqrt{-g}R$
is replaced by a function $f(R)$, then $\Psi$ and $\Phi$ are
generically unequal.  In the forecasts of \S\ref{sec:forecast}
we focus on GR-deviations
described by the $G_9$ and $\Delta\gamma$
parameters that characterize structure growth 
(\S\ref{sec:parameterizations}), but an alternative approach
parametrizes the ratios of $\Psi$ and $\Phi$ to their 
GR-predicted values (see \citealt{koivisto06,bean10,daniel10}, and references
therein).  The $G_9$ and $\Delta\gamma$ formulation is well matched to 
observables that can be measured by large surveys, but the
potentials formulation is arguably closer to the physics
of modified gravity.

The main approach to testing the consistency of $\Psi$ and $\Phi$
exploits the fact that the gravitational accelerations of
non-relativistic particles are determined entirely by $\Psi$
but the paths of photons depend on $\Psi+\Phi$.
Thus, an inequality of $\Psi$ and $\Phi$ should show up 
observationally as a mismatch between mass distributions
estimated from stellar or gas dynamics and mass distributions estimated
from gravitational lensing.  
(In typical modified gravity scenarios, it is then the lensing measurement
that characterizes the true mass distribution.)
The approximate agreement between
X-ray and weak lensing cluster masses
already rules out large disagreements between
$\Psi$ and $\Phi$.  A systematic statistical approach to this
test, employing the techniques discussed in \S\ref{sec:cl}, could
probably sharpen it to the few percent level, limited by the
theoretical uncertainty in converting X-ray observations to absolute
masses.  To reach high precision on cosmological scales, the most promising
route is to test for consistency between growth measurements from
redshift-space distortions, which respond to the non-relativistic
potential $\Psi$, and growth measurements from weak lensing.
Implementing an approach suggested by \cite{zhang07},
\cite{reyes10} present a form of this test that draws on
redshift-space distortion measurements of SDSS luminous red galaxies
by \cite{tegmark06} and galaxy-galaxy lensing measurements
of the same population.  The precision of the test in \cite{reyes10}
is only $\sim 30\%$, limited mainly by the redshift-space distortion
measurement, but this is already enough to rule out some otherwise viable
models.  In the long term, this approach could well be pushed to the
sub-percent level, with the limiting factors being the modeling
uncertainty in redshift-space distortions and systematics in 
weak lensing calibration.  Similar tests on the $\sim$kpc scales
of elliptical galaxies have been carried out by
\cite{bolton06} and \cite{schwab10}.

Some modified gravity models allow $\Psi$ and $\Phi$ to depend
on scale and/or time, yielding an ``effective'' gravitational
constant 
$G_{\rm Newton} \longrightarrow G_{\rm Newton}(k,t)$
(where $k$ denotes Fourier wavenumber).
Scale-dependent gravitational growth will alter the shape of the matter
power spectrum relative to that predicted by GR for the same
matter and radiation content.  Precise measurements of the galaxy
power spectrum shape can constrain or detect such scale-dependent
growth.  Uncertainties in the scale-dependence of galaxy bias
may be the limiting factor in this test, though departures from
expectations could also arise from non-standard radiation or matter
content or an unusual inflationary
power spectrum, and these effects may be difficult to disentangle from
scale-dependent growth.  The lever arm for determining the
power spectrum shape can be extended by using the \lya\ forest
or, in the long term, redshifted 21cm maps to make small-scale
measurements.  Time-dependent $G_{\rm Newton}$ would alter
the history of structure growth, leading to non-GR values
of $G_9$ or $\Delta\gamma$, but it could also be revealed by
quite different classes of tests.  For example, the consistency
of big bang nucleosynthesis with the baryon density inferred from
the CMB requires $G_{\rm Newton}$ at $t\approx 1\,$sec to equal
the present day value to within $\sim 10\%$ \citep{yang79,steigman10}.
Variation of $G_{\rm Newton}$ over the last 12 Gyr would also
influence stellar evolution, and it is therefore constrained by
the Hertszprung-Russell diagram of star clusters \citep{deglinnocenti96}
and by helioseismology \citep{guenther98}.

Departures from GR are very tightly constrained by high-precision
tests in the solar system, and many modified gravity models require
a screening mechanism that forces them
towards GR in the solar system and Milky Way environment.\footnote{To
give one example, the Shapiro delay of radio waves passing near the Sun, 
measured to agree with GR to five decimal places \citep{bertotti03},
is reduced in theories of gravity that contain scalar fields,
but the effect could be suppressed by scalar self-interaction in
dense environments.}
Screening may be triggered by a deep gravitational potential,
in which case the strength of gravity could be significantly different in other
cosmological environments.  For a generic class of theories,
the value of $G_{\rm Newton}$ would be higher by 4/3 in 
unscreened environments, allowing order unity effects
(see, for example, the discussion of $f(R)$ gravity by
\citealt{chiba03} and DGP gravity by \citealt{lue06}).
\cite{chang11} suggest tests with evolved stars, which could
be screened in the dense core and unscreened in the diffuse
envelope; the stars should be located in isolated dwarf galaxies 
so that the gravitational potential of
the galaxy, group, or supercluster environment does not trigger screening
on its own.  \cite{hui09} and \cite{jain11} 
propose testing for differential acceleration
of screened and unscreened objects in low density environments
(e.g., stars vs. gas, or dwarf galaxies vs. giant galaxies),
in effect looking for macroscopic and order unity violations
of the equivalence principle.
\cite{jain12} provides a systematic, high-level review of these
ideas and their implications for survey experiments, emphasizing
the value of including dwarf galaxies at low redshifts
within large survey programs.

Evidence for modified gravity could emerge from some
very different direction, such as high precision laboratory or
solar system tests, tests in binary pulsar systems, or gravity
wave experiments.  In many of these areas, technological advances
allow potentially dramatic improvements of measurement
precision --- for example, the proposed {\it STEP} satellite
could sharpen the test of the equivalence principle by 
five orders-of-magnitude \citep{overduin09}.
Modified gravity or a dark energy
field that couples to non-gravitational forces could also lead to 
time-variation of fundamental ``constants'' such as the fine-structure
constant $\alpha$.  Unfortunately, there are no ``generic''
predictions for the level of deviations in these tests,
so searches of this sort necessarily remain fishing expeditions.
However, the existence of cosmic acceleration suggests that
there may be interesting fish to catch.

\subsection{The Integrated Sachs-Wolfe Effect}
\label{sec:isw}

On large angular scales, a major contribution to CMB anisotropies
comes from gravitational redshifts and blueshifts of photon
energies \citep{sachs67}.
In a universe with $\Omega_{\rm tot}=\om=1$, potential fluctuations
$\delta\Phi \sim G\,\delta M/R$ stay constant because (in linear
perturbation theory) $\delta M$ and $R$ both grow in proportion
to $a(t)$.  In this case, a photon's gravitational energy
shift depends only on the difference between the potential at
its location in the last scattering surface and its potential
at earth.  However, once curvature or dark energy becomes 
important, $\delta M$ grows slower than $a(t)$, potential
wells decay, and photon energies gain a contribution from
an integral of the potential time derivative
($\dot\Psi + \dot\Phi$ in the notation of \S\ref{sec:gravity})
known as the Integrated Sachs-Wolfe (ISW) effect.
In more detail, one should distinguish the early ISW effect,
associated with the transition from radiation to matter
domination, from the late ISW effect, associated with the
transition to dark energy domination.
The ISW effect depends on the history of dark energy, which
determines the rate at which potential wells decay.
It can also test whether anisotropy is consistent with 
the GR prediction --- in particular whether the $\Psi$
and $\Phi$ potentials are equal as expected.

As an observational probe, the ISW effect has two major shortcomings.
First, it is significant only on large angular scales, where
cosmic variance severely and unavoidably limits measurement
precision.  (On scales much smaller than the horizon,
potential wells do not decay significantly in the time it takes
a photon to cross them.)
Second, even on these large scales the ISW contribution is small
compared to primary CMB anisotropies.  The second shortcoming
can be partly addressed by measuring the cross-correlation
between the CMB and tracers of the foreground matter distribution,
which separates the ISW effect from anisotropies present at the
last scattering surface.  
The initial searches, yielding upper limits on $\Omega_\Lambda$,
were carried out by cross-correlating {\it COBE} CMB maps with the
X-ray background (mostly from AGN, which trace the distribution
of their host galaxies) as measured by {\it HEAO} \citep{boughn98,boughn04}.
The \wmap\ era, combined with the availability of large optical
galaxy samples with well-characterized redshift distributions, led to
renewed interest in ISW and to the first marginal-significance
detections \citep{fosalba03,scranton03,afshordi04,boughn04,fosalba04,nolta04}

Realizing the cosmological potential of the ISW effect requires
cross-correlating the CMB with large scale structure tracers over a range
of redshifts at the largest achievable scales, and properly treating
the covariance arising from the redshift range and sky coverage of each
data set. \citet{ho08} used 2MASS objects ($z<0.2$), photometrically
selected SDSS LRGs ($0.2<z<0.6$) and quasars ($0.6<z<2.0$), and NVSS radio
galaxies, finding an overall detection significance of 3.7$\sigma$.
\citet{giannantonio08} used a similar sample
(but with a different SDSS galaxy and quasar selection, and with the
inclusion of the {\it HEAO} X-ray background maps) and found a 4.5$\sigma$
detection of ISW. Both of these measurements are consistent
with the ``standard'' $\Lambda$CDM cosmology.\footnote{The \citet{ho08}
measurement is almost $2\sigma$ above $\Lambda$CDM, but
we attribute no special significance to this!}
\cite{zhao10} utilize the \cite{giannantonio08} measurement
in combination with other data to test for late-time transitions
in the potentials $\Phi$ and $\Psi$, finding consistency with GR.

\cite{giannantonio08} estimate that
a cosmic variance limited experiment could achieve a $7-10\sigma$ ISW
detection.  
Because of the low S/N ratio, the ISW effect does not add usefully
to the precision of parameter determinations within standard
dark energy models, but it could reveal signatures of 
non-standard models.
Early dark energy --- dynamically significant at the
redshift of matter-radiation equality --- can produce observable
CMB changes via the early ISW effect (\citealt{doran07};
\citealt{deputter09} discuss the related problem of constraining
early dark energy via CMB lensing).
Perhaps the most interesting application of ISW
measurements is to constrain, or possibly reveal, inhomogeneities
in the dark energy density (see \S\ref{sec:parameterizations}),
which produce CMB anisotropies via ISW and are confined to
large scales in any case (see \citealt{deputter10}).  
However, it is not clear whether even exotic models can produce
an ISW effect that is distinguishable from the $\Lambda$CDM
prediction at high significance.  Measuring the ISW cross-correlation
requires careful attention to angular selection effects in the 
foreground catalogs, but these effects should be controllable,
and independent tracers allow cross-checks of results.
Since the prediction of conventional dark energy models is
robust compared to expected statistical errors, a clear deviation
from that prediction would be a surprise with important implications.

\subsection{Cross-Correlation of Weak Lensing and Spectroscopic Surveys}

Our forecasts in \S\ref{sec:rsd} incorporate an ambitious Stage IV
weak lensing program, and in \S\ref{sec:forecast_rsd} we consider the
impact of adding an independent measurement of $f(z)\sigma_8(z)$
from redshift-space distortions in a spectroscopic galaxy survey,
finding that a $1-2\%$ measurement can significantly improve constraints
on the growth-rate parameter $\Delta\gamma$ relative to our fiducial
program.  However, some recent papers 
\citep{bernstein11,gaztanaga11,cai11}
suggest that a combined analysis of overlapping weak lensing and
galaxy redshift surveys can yield much stronger dark energy and
growth constraints than an after-the-fact combination of independent
WL and RSD measurements.

The analysis envisioned in these papers involves measurement of all
cross-correlations among the WL shear fields (and perhaps magnification
fields) in tomographic bins, the angular clustering of galaxies in photo-$z$
bins of the imaging survey, and the redshift-space clustering of
galaxies in redshift bins of the spectroscopic survey, as well as 
auto-correlations of these fields.  While the forecast gains emerge
from detailed Fisher-matrix calculations, the essential physics
\citep{bernstein11,gaztanaga11} appears to be absolute calibration
of the bias factor of the spectroscopic galaxies via their weak
lensing of the photometric galaxies (galaxy-galaxy lensing, 
\S\ref{sec:wl_ggl}).  This calibration breaks degeneracy in the
modeling of RSD, and it effectively translates the spectroscopic
measurement of the galaxy power spectrum into a normalized measurement
of the matter power spectrum.  While the second technique can also
be applied to galaxy clustering in the photometric survey,
using photo-$z$'s, the clustering measurement in a spectroscopic
survey is much more precise because there are more modes in 3-d
than in 2-d.  The cross-correlation approach is more powerful if
the spectroscopic survey includes galaxies with a wide range
of bias factors \citep{mcdonald09}, e.g., a mix of massive
absorption-line galaxies and lower mass emission-line galaxies.

These studies are still in an early phase, and it remains to be
seen what gains can be realized in practice.  
The largest synergistic impact arises when the WL and RSD surveys are 
comparably powerful in their individual measurements of growth
parameters \citep{cai11}.  Stochasticity in
the relation between the galaxy and mass density fields depresses
cross-correlations relative to auto-correlations, a potentially
important theoretical systematic, though stochasticity is expected
to be small at large scales, and corrections can be computed with
halo-based models that are constrained by small and intermediate-scale
clustering.  Conversely, it may be possible to realize these gains
even when the weak lensing and spectroscopic surveys do not overlap,
by calibrating the bias of a photometric sample that has the same target
selection criteria as the spectroscopic sample.
\cite{gaztanaga11} also investigate the possibility of implementing
these techniques in a narrow-band imaging survey with large numbers
of filters, which is effectively a low-resolution spectroscopic survey.
They find that most of the gains of a spectroscopic survey are
achieved if the rms photo-$z$ uncertainty is $\Delta z/(1+z) \la 0.0035$,
while larger uncertainties degrade the results.

\subsection{Strong Gravitational Lenses}
\label{sec:strong_lensing}

Strong gravitational lenses can be employed to measure expansion 
history in a variety of ways.  One is to simply count the number
of lenses as a function of angular separation and (when available)
redshift --- the predicted counts depend on dark energy because
the probability of lensing goes up when path lengths are larger
\citep{kochanek96}.  This approach requires a large sample of
lenses with well understood statistical characteristics and
selection biases.  Observationally it is a stiff challenge, but
the more fundamental difficulty is that the predicted counts
also depend on the evolution of the mass distribution, in particular
that of the early-type galaxies that dominate the lensing statistics.
For example, \cite{oguri12} derive constraints from a
sample of 19 strong lenses found from 50,000+ source quasars 
in the SDSS, and while the analysis yields a robust detection
of dark energy, the cosmological parameter values are
significantly degenerate with uncertainties in the 
evolution of the galaxy velocity function.
The latter can be addressed
to some degree with measurements of galaxy velocity dispersions, 
but even then one is left with uncertainty in the exact relation
between the observable stellar velocity dispersion and the
potential well depth that is important for lensing.
With a sufficiently large sample of lenses one can ``self-calibrate''
by using the full distribution as a function of angular
separation and redshift, though one still relies on 
direct galaxy counts for an overall normalization.
While future imaging surveys will yield much larger samples
of lens candidates (which can then be confirmed spectroscopically),
we suspect that the systematic uncertainties in interpreting
lens counts will be too large for this method to be 
competitive as a dark energy probe --- rather, it will probe
the dynamical history of galaxy assembly.
The number of giant cluster arcs as a function of curvature
radius and redshift offers an alternative form of the 
``lens counting'' test that depends on both geometry 
and structure growth (e.g., \citealt{horesh11}),
but one must again understand the detailed mass assembly
history of the central regions of massive halos
to derive dark energy constraints \citep{killedar12}.

A second approach is to use angular positions
of multiple sources with different (known) redshifts to obtain
distance ratios, a strong-lensing version of the cosmography test
discussed in \S\ref{sss:cosmography}.
Clusters of galaxies are the best targets for this approach,
as they sometimes have large numbers of lensed sources
(see \citealt{soucail04} and \citealt{jullo10} for
representative observational applications).
The difficulty is that the sources with different 
redshifts also probe different locations in the
lens potential, so to derive cosmological results 
one needs strong constraints on 
substructure in the lens cluster and in the foreground
and background mass distributions.
\cite{meneghetti05} and \cite{daloisio11} 
discuss some of the theoretical issues in this approach
and their implications for data requirements.
Obtaining strong cosmological constraints from
cosmographic analysis of cluster lenses will be
observationally demanding, and mass modeling uncertainties
may be a limiting systematic.

The most promising of the strong-lensing approaches uses
time delay measurements, which attach an absolute scale
to the angular separations and redshifts measured for 
a gravitational lens.
Predicted time delays scale as $H_0^{-1}$, and more generally
with the angular diameter distance relation $D_A(z)$.
The critical systematic is again the uncertainty in the total mass and mass
distribution of the lens, which enters predictions of the angular
positions and time delays.  
Given the observed positions, a time-delay measurement determines
a degenerate combination of the distance scale ($H_0^{-1}$) and
the surface mass density of the lens within the Einstein radius
\citep{kochanek03}, so a dynamical measurement, usually the 
stellar velocity dispersion, is required to break the degeneracy
and isolate the cosmological information.
When the sources are resolved
(i.e., galaxies rather than AGN in the optical, or jets mapped
with VLBI), or when the multiplicity of images is unusually high,
then reproducing the data places more stringent constraints on the mass
distributions (e.g., \citealt{suyu11}).
In the best cases, distance scale constraints at the $\sim 5\%$
level, after marginalizing over uncertainties in the mass distribution, are
achievable for an individual lens (e.g., \citealt{suyu10,suyu12}).
Careful measurements and modeling
of a number of well constrained lenses could lead to high aggregate
precision, adding significant power to tests of dark energy models
\citep{coe09,linder11}.
The critical question from the point of view of dark energy is
whether the mass modeling uncertainties are independent from
system to system, so that the net uncertainty drops as $1/\sqrt{N}$,
or whether global uncertainties in the approach to mass modeling
impose a systematic floor.
The answer to this question should become clearer as the
current generation of cosmological time-delay analyses proceeds.

\subsection{Galaxy Ages}
\label{sec:ages}

In principle the age of the universe is an observable that 
can probe cosmic expansion history.  The integral that determines age,
\begin{equation}
\label{eqn:age}
t(z) = \int_z^\infty {dz' \over 1+z'} H^{-1}(z') ~,
\end{equation}
is similar to the integral for comoving distance
(eq.~\ref{eqn:dcomove}), except that it extends from 
$z$ to $\infty$ instead of 0 to $z$.
The conflict between the ages of globular clusters and
the value of $t_0$ in a decelerating universe was one of
the significant early arguments for cosmic acceleration, and
a number of authors have employed ages of high-redshift galaxies or clusters
of galaxies as a constraint on dark energy models 
(e.g., \citealt{lima00,jimenez03,capozziello04}).

\cite{jimenez02} proposed using {\it differential} ages
of galaxies at different redshifts to measure, in effect, $H(z)$.
Observational studies thus far have concentrated on the massive
ellipticals, as these have the fewest complications (dust, ongoing star
formation). 
This differential approach removes some of the uncertainties
in the population synthesis models, but it relies on identifying
a population of galaxies at one redshift that is just an
aged version of a population at a higher redshift,
or accounting for the evolutionary corrections that arise
from mergers, low-level star formation, and movement of galaxies
into or out of the passive population.
The state-of-the art observational study is the analysis
of $\sim 11,000$ early-type galaxies from several large surveys
by \cite{moresco12}.
They report measurements of $H(z)$ in eight bins out to $z\approx 1$,
with uncertainties of $\sim 5-15\%$ per bin including estimated
systematic errors.
At low ($z<0.3$) redshift, \cite{moresco11} analyzed a sample of 
14,000 early type galaxies from the SDSS, obtaining 
$H_0 = 72.6 \pm 2.9\,{\rm (stat)} \pm 2.3\,{\rm (syst)}\,\hubunits$,
compatible with other estimates of $H_0$ and competitive in precision.
These papers provide extensive discussions of systematic
uncertainties and argue that they can be well controlled.
Nonetheless, the reliance on population synthesis 
and galaxy evolution models means that this method will
face a stiff burden of proof if it finds a discrepancy
with simple dark energy models or other observational analyses,
particularly if the differences are at the sub-percent
level that is the target of future dark energy experiments.

\subsection{Redshift Drift}
\label{sec:drift}

The redshift of a comoving source changes as the universe expands.
\cite{sandage62} was the first to propose this ``redshift drift'' as
a cosmological test, but he noted that it appeared far beyond
the capabilities of existing experimental techniques.
\cite{loeb98} repopularized the idea, noting that high-resolution
spectrographs on large telescopes could potentially measure the
effect in absorption-line spectra of high-redshift quasars.
\cite{quercellini10} provide an extensive review of the
redshift-drift method and other forms of ``real-time cosmology''
experiments.  The expected change in redshift over a time 
interval $\Delta t$, expressed as an apparent velocity shift, is
\begin{equation}
\label{eqn:zdrift}
\Delta v \equiv {c\Delta z\over 1+z} = 
c H_0\Delta t \left[1 - {H(z)/H_0\over 1+z}\right]~,
\end{equation}
which vanishes for a coasting universe with $H(z)/H_0 = (1+z)^{-1}$.
For our fiducial cosmological model, the predicted change over a
$\Delta t = 10$ year observational span is
$+1.32\cms$, $-1.21\cms$, and $-3.66\cms$ for sources at
$z=2$, 3, and 4, respectively.  \cite{corasaniti07} estimate
that observations of 240 quasars over a span of 30 years
using the CODEX spectrograph proposed for the European 
Extremely Large Telescope
could measure $H(z)$ over $z=2-5$ with an aggregate precision
of $\approx 2\%$ (see \citealt{liske06} and \citealt{balbi07}
for similar discussions and \citealt{liske08} for
detailed calculations of CODEX performance).
The BAO component of the fiducial Stage IV program that we
present in \S\ref{sec:fiducial} (which assumes 25\% sky coverage
and errors that are 1.8 times those of linear theory 
sample variance) yields errors of $0.6-0.7\%$ in $H(z)$
per bin of 0.07 in $\ln(1+z)$ at $z=2-3$ (see Table~\ref{tbl:baoforecast}).
The fiducial BAO program would thus
be much more powerful than the redshift-drift approach,
but it is not yet clear that high-redshift BAO surveys of
this volume will prove practical.

\subsection{Alternative Methods: Summary}
\label{sec:alt_summary}

Of the methods described in this section, redshift-space distortion
is the one that seems almost guaranteed to have a broad-ranging
impact in future studies of cosmic acceleration.  Because of its
direct sensitivity to $f(z)$ and its use of non-relativistic
tracers, RSD is a valuable complement to weak lensing and clusters
as a probe of structure growth.  The redshift surveys ongoing or
planned for BAO measurements will automatically enable RSD measurements
with high statistical precision over a significant range in redshift.
The primary systematics for RSD are associated with theoretical
modeling uncertainties, and these are likely to diminish with time
as more sophisticated techniques are developed, numerical simulations
become more powerful, and galaxy clustering measurements provide
more detailed tests of the models.
The AP method may also amplify the return from BAO redshift surveys,
perhaps by a large factor, depending on the success of modeling
non-linear velocity distortions.

The $z=0$ Hubble constant is only a single number, but it is an
important one, as it determines the current critical density and
the current expansion rate, with the maximal leverage against the
CMB at $z \approx 1100$.  Ongoing technical developments seem to
offer a realistic path to achieving $H_0$ measurements with 1-2\% uncertainty,
tight enough to make significant contributions to dark energy constraints
even in the era of Stage IV experiments.  \jwst\ and \gaia\ represent
dramatic improvements in technological capabilities for $H_0$ studies,
perhaps allowing progress to the sub-percent level.

The impact of the other methods discussed here is more difficult
to predict.  In some cases, interesting results at the few-percent
level are already in hand but it is still too early to say whether
the limiting systematics can be reduced to the percent
or sub-percent level required for long-term progress.
In other cases, such as precision gravity tests, there are clear
paths to radically improved measurements, but these will provide
useful clues or tests only for a limited class of cosmic acceleration
models.

The one method that looks like it could ultimately outperform
even Stage IV experiments is that of standard sirens applied to
compact star binaries (\S\ref{sec:sirens}), because of the
opportunity to measure hundreds of thousands of distances
by a technique that has no obvious limiting systematics.
This approach requires extraordinary technological advances,
which are unlikely to be achieved in the next decade,
and perhaps not in the next two decades.  However, the
most precise mapping of the expansion history of the universe
might ultimately come from spacetime ripples rather than
electromagnetic waves.

\vfill\eject

\section{A Balanced Program on Cosmic Acceleration}
\label{sec:forecast}

Having discussed many observational methods individually, we now
turn to what we might hope to learn from them in concert.
To the extent that this report has an underlying editorial theme,
it is the value of a balanced observational program that 
pursues multiple techniques at comparable levels of precision.
In our view, there is much more to be gained by doing a good job
on three or four methods than by doing a maximal job on one at
the expense of the others.  This is {\it not} a ``try everything''
philosophy --- moving forward from where we are today, an observational
method is interesting only if it has reasonable prospects of achieving
percent- or sub-percent-level errors, both statistical and systematic,
on observables such as $H(z)$, $D(z)$, and $G(z)$.  The successes
of cosmic acceleration studies to date have raised the field's
entry bar impressively high.

A balanced strategy is important both for cross-checking of systematics
and for taking advantage of complementary information.
Regarding systematics, the next generation of cosmic acceleration
experiments seek much higher precision than those carried out to date,
so the risk of being limited or biased by systematic errors is
much higher.  Most methods allow internal checks for systematics ---
e.g., comparing distinct populations of SNe, measuring angular 
dependence and tracer dependence of BAO signals, testing for $B$-modes and
redshift-scaling of WL --- but conclusions about cosmic acceleration
will be far more convincing if they are reached independently
by methods with different systematic uncertainties.
Two methods only provide a useful cross-check of systematics if they
have comparable statistical precision; otherwise a result found
only in the more sensitive method cannot be checked by the less
sensitive method.

Regarding information content, we have already emphasized the 
complementarity of SN and BAO as distance determination methods.
SN have unbeatable statistical power at $z \la 0.6$, while
BAO surveys that map a large fraction of the sky with adequate
sampling can achieve higher precision at $z \ga 0.8$.  Overlapping
SN and BAO measurements provide independent physical information
because the former measure relative distances and the latter
absolute distances ($\hmpc$ vs.\ Mpc), and the value of $h$ is
itself a powerful dark energy diagnostic in the context of CMB
constraints (see \S\ref{sec:h0} and \S\ref{sec:forecast_h0}).
WL, clusters, and redshift-space distortions provide independent
constraints on expansion history, at levels that can be competitive
with SN and BAO, and they provide sensitivity to structure growth.
Without structure probes, we would have little hope of clues that
might locate the origin of acceleration in the gravitational sector
rather than the stress-energy sector, and we would, more generally,
reduce the odds of ``surprises'' that might push us beyond
our current theories of cosmic acceleration.

The primary purpose of this section is to present quantitative
forecasts for a program of Stage IV dark energy experiments and
to investigate how the forecast constraints depend on the performance
of the individual components of such a program.  
Our forecasts are analogous to those of the DETF \citep{albrecht06},
updated with a more focused idea of what a Stage IV program might
look like, and updated in light of subsequent work on parameterized
models and figures of merit for dark energy experiments, most directly
that of the \jdem\ Figure-of-Merit Science Working Group 
(FoMSWG; \citealt{albrecht09}).
In \S\ref{sec:fiducial} we summarize our assumptions about the
fiducial program.  In \S\ref{sec:forecasting} we describe the
methodology of our forecasts, in particular the construction
of Fisher matrices for the fiducial program.
In \S\ref{sec:results} we present results for the fiducial program
and for variants in which one or more components of this program are
made significantly better or worse.  We also compare these results
to forecasts of a ``Stage III'' program represented by experiments
now underway or nearing their first observations.

We have elected to focus on SN, BAO, and WL as the components of 
these forecasts, for two reasons.  First, it is more straightforward
(though still not easy) to define the expected statistical and
systematic errors for these methods than for others.  Second,
the most promising alternative methods --- clusters, redshift-space
distortions, and the Alcock-Paczynksi effect --- will be enabled by
the {\it same} data sets obtained for WL and BAO studies.
It is therefore reasonable to view these as auxiliary methods
that may improve the return from these data sets (perhaps by substantial
factors) rather than as drivers for the observational programs
themselves.
In \S\S\ref{sec:forecast_cl} and~\ref{sec:forecast_alt}
we present forecasts for how well the fiducial CMB+SN+BAO+WL
programs predict the observables of these and other alternative
methods, providing a target for how well they must perform to 
add new information beyond that in our primary probes.  
In some cases we find that plausible levels of performance 
could substantially improve tests of cosmic acceleration models.
In \S\ref{sec:forecast_aggregate} we focus on the precision with which
our fiducial program measures fundamental observables, and
we discuss aggregate precision as a useful, nearly model-independent
way of characterizing the power of an experiment and 
the level of systematics control required to realize it.
Section \ref{sec:forecast_multiprobe} provides a high-level
summary, discussing the potential yield from programs
that combine CMB, SN, BAO, and WL measurements with
additional constraints from clusters, redshift-space distortions,
and direct $H_0$ determinations.

\subsection{A Fiducial Program}
\label{sec:fiducial}

As discussed in \S\ref{sec:forward}, Astro2010 and the European 
Astronet report have placed high priority on ground- and space-based
dark energy experiments.  
The Stage III experiments currently underway will already allow
much stronger tests of cosmic acceleration models, and Stage IV
facilities built over the next decade should advance the field much
further still.
Our Stage IV program corresponds roughly to the goals recommended
by the Cosmology and Fundamental Physics panel report of Astro2010.

For SN studies, we anticipate that Stage IV efforts will be limited
not by statistical errors but by systematics associated with 
photometric calibration, dust extinction, and evolution of the SN population.
For our fiducial program, we assume that SN surveys will achieve
net errors (statistical + systematic) of 0.01 mag for the mean
distance modulus in each of three redshift bins of width $\Delta z = 0.2$ 
extending from $z = 0.2$ to a maximum redshift $\zmax = 0.8$
(see discussion in \S\ref{sec:sn_systematics}).
We also assume the existence of a local SN sample at $z=0.05$
with the same 0.01 mag net error.
High quality observations could  yield a smaller systematic
error in the local sample, but we suspect that the most challenging
systematic for this local calibration will be transferring it
to the more distant bins.
%\tbd{Revised to reflect new SN forecasts.}
We treat the bin-to-bin errors
as uncorrelated, though this is clearly an approximation to systematic
errors that are correlated at nearby redshifts and gradually decorrelate
as one considers differing redshift ranges and observed-frame 
wavelengths.
%\footnote{We have considered a model with random errors
%of 0.007 mag with a $\Delta z=0.2$ correlation length, 
%\tbd{[state more precisely]} 
%and we obtain nearly identical results.}
Even with 0.15 mag errors per SN, achieving this level of statistical
error requires only 225 SNe per bin, and we expect that the error
per SN can be reduced by working at red/IR wavelengths and by selecting
sub-populations based on host galaxy type, spectral properties,
and light curve shape.  For purely ground-based efforts, we consider
our 0.01 mag floor for systematic errors to be somewhat optimistic, given
the challenges of dust extinction corrections and photometric calibration.
However, a space-based program at rest-frame near-IR wavelengths,
enabled by \wfirst, could plausibly achieve better than 0.01 mag systematics.
We suspect that it will be hard to push calibration and evolution
systematics below 0.005 mag even with \wfirst, and pushing statistical
errors below this level begins to place severe demands on spectroscopic
capabilities, unless purely photometric information can be used to
identify populations with scatter below 0.1 mag per SN.
We also consider the impact of increasing $\zmax$
beyond 0.8, though we argue that this is beneficial mainly
when one is hitting a systematics floor at lower $z$ and 
high-$z$ observations have uncorrelated systematics.

For BAO, the primary metric of statistical constraining power is
the total comoving volume mapped spectroscopically with a sampling
density high enough to keep shot-noise sub-dominant.
There are several projects in the planning stages that could map
significant fractions of the comoving volume available out to $z\approx 3$.
These include the near-IR spectroscopic components of \euclid\ and \wfirst,
ground-based optical facilities such as BigBOSS, DEspec, and SuMIRe PFS,
and radio intensity-mapping experiments (see \S\ref{sec:bao_prospects}).  
For our fiducial program,
we assume that these projects will collectively map 25\% of the
comoving volume out to $z=3$, with errors a factor of 1.8 larger
than the linear theory sample variance errors.\footnote{This is
equivalent to assuming linear theory sample variance over a fractional
volume $25\%/1.8^2 = 7.7\%$.}
We specifically assume full redshift coverage from $z=0-3$ with 
$\fsky=25\%$ sky fraction, but other combinations of redshift
coverage and $\fsky$ that have the same total comoving volume
yield similar results.  
The factor 1.8 accounts for imperfect sampling (hence non-negligible shot-noise)
and for non-linear degradation of the BAO signal.
It approximates the effects of sampling with $nP=2$ and
using reconstruction (\S\ref{sec:bao_theory_recon}) to remove
50\% but not 100\% of the non-linear Lagrangian displacement of tracers.
We implicitly assume that theoretical systematics associated with
location of the BAO peak will remain below this level, an assumption 
we consider reasonable but not incontrovertible based on the
discussion in \S\ref{sec:bao_systematics}.

For WL, the primary metric of statistical constraining power is the
total number of galaxies that have well measured shapes and good enough
photometric redshifts to allow accurate model predictions and removal
of intrinsic alignment systematics.  For our fiducial case, we assume
a survey of $10^4\,$deg$^2$ achieving an effective surface density of
23 galaxies per arcmin$^2$ with $z_{\rm med}=0.84$,
corresponding to $I_{\rm AB}<25$ and $r_{\rm eff}>0.25\arcsec$.
The effective galaxy number is $8.3 \times 10^8$.
\euclid\ plans a 14,000 deg$^2$ imaging survey and can likely
achieve this surface density or slightly higher.
LSST will survey a still larger area,
and it might or might not achieve this effective surface density, 
depending on how low a value of $r_{\rm eff}/r_{\rm PSF}$ 
it can work to before shape measurements are systematics dominated.
The \wfirst\ design reference mission (\citealt{green12}; DRM1)
would achieve $n_{\rm eff} \approx 40\,\arcmin^{-2}$ but would
only image 3400 deg$^2$ in its 2.4-year high-latitude survey,
thus measuring about $4.8\times 10^8$ galaxy shapes.  
An extended \wfirst\ mission, or an implementation of \wfirst\
using one of the NRO 2.4-m telescopes \citep{dressler12}, 
could potentially reach
$10^4$ deg$^2$.  Even individually, therefore, any one of these
projects may well exceed the number of shape measurements assumed
in our fiducial program, and collectively they will almost
certainly do so.
We compute constraints from cosmic shear in 14 bins of photometric
redshift and from the shear-ratio test described in 
\S\ref{sss:cosmography},
but we do not incorporate higher order lensing statistics or
galaxy-shear cross-correlations.
We include information up to multipole $l_{\rm max}=3000$, beyond which
statistical power becomes limited at this surface density and 
systematic uncertainties associated with non-linear evolution and
baryonic effects become significant.

Forecasting the systematic uncertainties in Stage IV WL experiments
is very much a shot in the dark.  Systematic errors are already 
comparable to statistical errors in surveys of $100\,$deg$^2$,
so lowering them to the level of statistical errors in a $10^4\,$deg$^2$
survey that has higher galaxy surface density requires more than
an order of magnitude improvement.
We therefore consider a ``fiducial'' and an ``optimistic'' case
for WL systematics.  For the fiducial case, we incorporate (and
marginalize over) aggregate uncertainties 
of $2\times 10^{-3}$ in shear calibration and
$2\times 10^{-3}$ in the mean photo-$z$, with
errors in each redshift bin larger by $\sqrt{14}$ but uncorrelated
across bins.  We also incorporate intrinsic alignment uncertainty
as described by Albrecht et al.\ (\citeyear{albrecht09}, \S 2h of
Appendix A), 
which includes marginalization over both GI and II components
(see \S\ref{sec:wl_ia}).
For our ``optimistic'' case we adopt no specific form of the
systematic errors but simply assume that they will double the
statistical errors throughout.  
At an order of magnitude level, we can see that the optimistic
case corresponds to a global fractional error
$\sigma \sim 2 N_{\rm mode}^{-1/2} \sim 
2 \fsky^{-1/2} \lmax^{-1} = 1.3 \times 10^{-3}$,
significantly lower than the fiducial case assumption of
$2\times 10^{-3}$ errors for shear and photo-$z$ calibration
(which, roughly speaking, combine in quadrature to make
a $2.8 \times 10^{-3}$ multiplicative uncertainty).
However, at scales and redshifts where the statistical errors
are large, multiplying them by two can be a larger change than
adding the shear-calibration and photo-$z$ systematics.  As a result,
there will be some measures (e.g., the error on $\ok$) for which our
``optimistic'' program performs slightly worse than our fiducial program.
Of course, WL experiments that achieved the statistical limits
of several $\times 10^9$ source galaxies --- possible in principle ---
would be several times more powerful than even our 
optimistic scenario.
% In addition, galaxy-galaxy lensing measurements combined with galaxy
% clustering could lead to significantly stronger cosmological constraints
% if the theoretical systematics associated with non-linear galaxy
% bias can be adequately controlled, perhaps using parameterized
% models (such as HOD prescriptions) constrained by small and
% intermediate scale clustering.

\subsection{Forecasting Constraints}
\label{sec:forecasting}

The fiducial program outlined above provides a baseline for evaluating
improvement in the determination of the cosmological parameters relative 
to current constraints. We use a Fisher matrix analysis to quantify this
improvement and to study the complementarity of the main probes of 
cosmic acceleration. Since our knowledge of the exact design of future 
surveys and the systematic errors they will face is inherently imperfect,
we also consider the effect of varying the precision of each technique
in our forecasts, including both pessimistic and optimistic cases for 
SN, BAO, and WL data.

Quantifying the impact of each probe on our understanding of cosmic 
acceleration requires metrics for evaluating progress. 
The precision with which the dark energy equation of state (and its
possible time dependence) can be measured is a common choice;
while not the only quantity of interest, it is clearly a 
central piece of the puzzle.
One of the main quantities we use below is the DETF figure of merit 
defined in equation~(\ref{eqn:fom}), 
${\rm FoM} = [\sigma(w_p)\sigma(w_a)]^{-1}$.
The FoM indicates how well an experiment determines the dark energy equation 
of state parameter and its derivative $dw/da$ at
the pivot redshift $z_p$, and it thereby indicates 
the ability to detect deviations from the standard $\Lambda$CDM model 
with $w_p=-1$ and $w_a=0$.
When one considers experiments of increasing power,
$\sigma(w_p)$ and $\sigma(w_a)$ tend to shrink in concert,
so the DETF FoM scales roughly as an inverse variance and
therefore increases linearly with data volume when statistical
errors dominate.  If the error of every individual measurement
(e.g., each $D_L$ or $H(z)$ measurement) goes down by $\sqrt{2}$,
then the FoM doubles.

While the DETF FoM is relatively simple to evaluate for a particular 
experiment, it omits much of the information that will be available 
from future experiments, including some potentially important clues 
to the nature of cosmic acceleration. For example, the true dark energy 
dynamics may be considerably more complicated than what the two-parameter 
linear model can accommodate, so that constraints on $w_0$ and $w_a$ 
may yield incomplete or misleading results. Additionally, the 
equation of state alone is insufficient to describe the full range of 
possible alternatives to the standard cosmological model. 
For example, modified gravity theories can mimic the effect of any 
particular equation of state evolution on the Hubble expansion rate and 
the distance-redshift relation while altering the rate of growth 
of large-scale structure 
(e.g., \citealt{lue04,song07}). 
Including such possibilities requires extra 
parameters that describe changes in the growth history that are 
independent of equation of state variations,
as discussed in \S\ref{sec:parameterizations}.
Other standard parameters of the cosmological model, such as the 
spatial curvature and the Hubble constant, are important due to 
degeneracies with the effects of cosmic acceleration that can 
limit the precision of constraints on the dark energy equation of state.

To include more general variations of the equation of state as well as 
altered growth of structure from modifications to GR on large scales, we adopt 
the \jdem\ FoMSWG parameterization \citep{albrecht09}.
The equation of state in this parameterization
is allowed to vary independently in each of 36 bins of width $\Delta a = 0.025$
extending from the present to $a=0.1$ ($z=9$). Specifically, the equation 
of state has a constant value of $w_i$ at $(1-0.025i) < a < [1-0.025(i-1)]$,
for $i=1,\ldots,36$.
At earlier times, the equation of state is assumed to be $w=-1$, although
the impact of this assumption is typically quite small since dark 
energy accounts for a negligible fraction of the total density at $z>9$
in most models.
Modifications to the linear growth function of GR $\Ggr(z)$ are included 
through the parameters $G_9$ and $\Delta \gamma$ as 
defined in equations~(\ref{eqn:fullfz}) and~(\ref{eqn:fullgz}).
These parameters describe the change relative to GR in the normalization 
of the growth of structure at $z=9$ and in the growth rate at $z<9$, 
respectively. Adding these to the binned $w_i$ values and the standard 
$\Lambda$CDM parameters, the full set is
\begin{equation}
\label{eqn:forecastparams}
{\bf p} = ( w_1, \ldots, w_{36}, \ln G_9, \Delta \gamma, \om h^2, 
\ob h^2, \ok h^2, \op h^2, 
\ln A_s, n_s, \Delta \mathcal{M} )~,
\end{equation}
where the primordial amplitude $A_s$ is defined at $k=0.05 \Mpc^{-1}$.
$\Delta \mathcal{M}$ is an overall offset in the absolute magnitude
scale of Type Ia supernovae.
The Hubble constant is determined by these parameters through
$h^2 = \om h^2+\ok h^2+\op h^2$.
We compute our forecasts at the fiducial parameter values chosen by 
the FoMSWG to match CMB constraints from the 5-year release of 
WMAP data \citep{komatsu09}; 
these are listed in Table~\ref{tbl:fiducial}. 
These parameters are similar but not identical to those of 
the model used in \S\ref{sec:observables} (Table~\ref{tbl:models}),
which is based on WMAP7.
Note that spatially 
flat $\Lambda$CDM and GR are assumed for the fiducial model.

%\begin{table}[t]
%\caption{Fiducial Model for Forecasts}
%\centering
%\begin{tabular}{r r r r r r r r r r r r}
%\hline\hline
%$w_1$ & \ldots & $w_{36}$ & $\ln G_9$ & $\Delta \gamma$ & $\om h^2$ & 
%$\ob h^2$ & $\ok h^2$ & $\op h^2$ & $\ln A_s$ & $n_s$ & $\Delta \mathcal{M}$ 
%\\ [0.5ex]
%\hline
%$-1$ & \ldots & $-1$ & 0 & 0 & 0.1326 & 0.0227 & 0 & 0.3844 & $-19.9628$ &
%0.963 & 0 \\
%\hline
%\end{tabular}
%\label{tbl:fiducial}
%\end{table}

\begin{deluxetable}{rrrrrrrrrrrr}
\tablecolumns{12}
\tablecaption{Fiducial Model for Forecasts\label{tbl:fiducial}}
\tablehead{
$w_1$ & \ldots & $w_{36}$ & $\ln G_9$ & $\Delta \gamma$ & $\om h^2$ &
$\ob h^2$ & $\ok h^2$ & $\op h^2$ & $\ln A_s$ & $n_s$ & $\Delta \mathcal{M}$
}
\tablewidth{0pc}
\startdata
$-1$ & \ldots & $-1$ & 0 & 0 & 0.1326 & 0.0227 & 0 & 0.3844 & $-19.9628$ &
0.963 & 0 \\
\enddata
\end{deluxetable}

We use a Fisher matrix analysis to estimate the constraints on these 
parameters from the fiducial program defined in \S\ref{sec:fiducial} and
its variations. The Fisher matrix for each experiment consists of 
a model of the covariance matrix for the observable quantities and
derivatives of these quantities with respect to the parameters.
We compute the latter numerically with finite differences and confirm the 
results using analytic expressions when possible.

We model SN data as measurements of the average SN magnitude in each of 
several redshift bins and in a low-redshift calibration sample. 
While our fiducial case assumes that the net 
magnitude error is uncorrelated from one bin to the next, we also 
consider the impact of including a correlated component of the error 
by defining the SN covariance matrix as
\begin{equation}
\label{eqn:covsn}
C_{\alpha \beta}^{\rm SN} = \left\{
\begin{array}{ll}
\sigma_{m,u}^2\, \delta_{\alpha\beta}\,, & \alpha=1~{\rm or}~\beta=1 \,, \\
\sigma_{m,u}^2\, \left({0.2 \over \Delta z}\right) 
\,\delta_{\alpha \beta} + 
\sigma_{m,c}^2 \, \exp\left( -{|z_{\alpha}-z_{\beta}| 
\over \Delta z_c} \right) \,, & \alpha>1~{\rm and}~\beta>1 \,, \\
\end{array}
\right.
%C_{\alpha \beta}^{\rm SN} = \sigma_{m,u}^2\, \left({0.2 \over \Delta z}\right)
%\,\delta_{\alpha \beta} + 
%\sigma_{m,c}^2 \, \exp\left( -{|z_{\alpha}-z_{\beta}| 
%\over \Delta z_c} \right)~,
\end{equation}
where $\Delta z$ is the bin width, $\sigma_{m,u}$ is the uncorrelated error 
in a bin of width $\Delta z=0.2$ (or in the local sample at 
redshift $z_1$), $\sigma_{m,c}$ is the 
correlated error with correlation length $\Delta z_c$, 
and the net error in each bin $z_{\alpha}$ ($\alpha>1$) is 
$\sigma_m = \sqrt{\sigma_{m,u}^2+\sigma_{m,c}^2}$\,.
In general these errors are redshift dependent, but here we assume that 
they are constant for simplicity.
We do not consider possible correlations between the local SN sample
and the high-redshift bins.
For the fiducial forecasts we take $\sigma_{m,c}=0$, so the covariance 
matrix is diagonal. The SN Fisher matrix is then computed as a sum over 
redshift bins
\begin{equation}
\label{eqn:fishersn}
F_{ij}^{\rm SN}=\sum_{\alpha,\beta}{\partial m(z_{\alpha})\over\partial p_i}\,
(C_{\alpha\beta}^{\rm SN})^{-1}\, {\partial m(z_{\beta})\over\partial p_j}~,
\end{equation}
where $m(z_{\alpha}) = 5\log [H_0 \langle D_L(z_{\alpha})\rangle]+\mathcal{M}$ 
is the average magnitude in the bin and the 
derivatives are taken with respect to the parameters of 
equation~(\ref{eqn:forecastparams}).

%\begin{table}[ht]
%\caption{BAO Errors for the Fiducial Program}
%\centering
%\begin{tabular}{r r c c c c}
%\hline\hline
%$z_{\rm min}$ & $z_{\rm max}$ & $V_{\fsky=0.25}~[({\rm Gpc}/h)^3]$ & 
%$\sigma_{\ln(D/r_s)} \,[\%]$ & $\sigma_{\ln(Hr_s)} \,[\%]$  & $r$ \\
%\hline
%  0.000 & 0.072  &  0.010   & 13.386  & 21.881  & 0.409 \\
%  0.072 & 0.149  &  0.075   &  4.895  &  8.002  & 0.409 \\
%  0.149 & 0.231  &  0.217   &  2.873  &  4.697  & 0.409 \\
%  0.231 & 0.320  &  0.449   &  1.997  &  3.265  & 0.409 \\
%  0.320 & 0.414  &  0.781   &  1.515  &  2.476  & 0.409 \\
%  0.414 & 0.516  &  1.218   &  1.213  &  1.983  & 0.409 \\
%  0.516 & 0.625  &  1.761   &  1.009  &  1.649  & 0.409 \\
%  0.625 & 0.741  &  2.407   &  0.863  &  1.410  & 0.409 \\
%  0.741 & 0.866  &  3.148   &  0.754  &  1.233  & 0.409 \\
%  0.866 & 1.000  &  3.970   &  0.672  &  1.098  & 0.409 \\
%  1.000 & 1.144  &  4.860   &  0.607  &  0.992  & 0.409 \\
%  1.144 & 1.297  &  5.799   &  0.556  &  0.909  & 0.409 \\
%  1.297 & 1.462  &  6.770   &  0.514  &  0.841  & 0.409 \\
%  1.462 & 1.639  &  7.758   &  0.481  &  0.785  & 0.409 \\
%  1.639 & 1.828  &  8.745   &  0.453  &  0.740  & 0.409 \\
%  1.828 & 2.031  &  9.718   &  0.429  &  0.702  & 0.409 \\
%  2.031 & 2.249  & 10.664   &  0.410  &  0.670  & 0.409 \\
%  2.249 & 2.482  & 11.576   &  0.393  &  0.643  & 0.409 \\
%  2.482 & 2.732  & 12.443   &  0.379  &  0.620  & 0.409 \\
%  2.732 & 3.000  & 13.261   &  0.368  &  0.601  & 0.409 \\
%\hline
%\end{tabular}
%\label{tbl:baoforecast}
%\end{table}

\begin{deluxetable}{rrccc} 
\tablecolumns{5}
\tablecaption{BAO Errors for the Fiducial Program\label{tbl:baoforecast}}
\tablehead{
$z_{\rm min}$ & $z_{\rm max}$ & $V~[({\rm Gpc}/h)^3]$ & 
$\sigma_{\ln(D/r_s)} \,[\%]$ & $\sigma_{\ln(Hr_s)} \,[\%]$
}
\tablewidth{0pc}
\startdata
  0.000 & 0.072  &  0.010   & 13.386  & 21.881 \\
  0.072 & 0.149  &  0.075   &  4.895  &  8.002 \\
  0.149 & 0.231  &  0.217   &  2.873  &  4.697 \\
  0.231 & 0.320  &  0.449   &  1.997  &  3.265 \\
  0.320 & 0.414  &  0.781   &  1.515  &  2.476 \\
  0.414 & 0.516  &  1.218   &  1.213  &  1.983 \\
  0.516 & 0.625  &  1.761   &  1.009  &  1.649 \\
  0.625 & 0.741  &  2.407   &  0.863  &  1.410 \\
  0.741 & 0.866  &  3.148   &  0.754  &  1.233 \\
  0.866 & 1.000  &  3.970   &  0.672  &  1.098 \\
  1.000 & 1.144  &  4.860   &  0.607  &  0.992 \\
  1.144 & 1.297  &  5.799   &  0.556  &  0.909 \\
  1.297 & 1.462  &  6.770   &  0.514  &  0.841 \\
  1.462 & 1.639  &  7.758   &  0.481  &  0.785 \\
  1.639 & 1.828  &  8.745   &  0.453  &  0.740 \\
  1.828 & 2.031  &  9.718   &  0.429  &  0.702 \\
  2.031 & 2.249  & 10.664   &  0.410  &  0.670 \\
  2.249 & 2.482  & 11.576   &  0.393  &  0.643 \\
  2.482 & 2.732  & 12.443   &  0.379  &  0.620 \\
  2.732 & 3.000  & 13.261   &  0.368  &  0.601 \\
\enddata
\tablecomments{
Column 3 gives the volume of the redshift slice for $\fsky=0.25$. 
In all redshift slices, errors on 
$D/r_s$ and $Hr_s$ are correlated with correlation coefficient $r=0.409$.
}
\end{deluxetable}

For BAO 
we divide the observed volume into bins of equal width in $\ln(1+z)$,
assumed to be uncorrelated, and compute the Fisher matrix
\begin{equation}
\label{eqn:fisherbao}
F_{ij}^{\rm BAO}=\sum_{\mu,\nu,\alpha}{\partial r_{\mu}(z_{\alpha})
  \over \partial p_i}\,
[C_{\mu\nu}^{\rm BAO}(z_{\alpha})]^{-1}\, 
{\partial r_{\nu}(z_{\alpha})\over\partial p_j}~,
\end{equation}
where the measurement vector
${\bf r}(z_{\alpha}) \equiv \{D(z_{\alpha})/r_s, H(z_{\alpha}) r_s\}$,
the sum is over $\mu,\nu = 1,2$ and $\alpha = 1,...N_{\rm bin}$,
and $r_s$ is the sound horizon at recombination (see \S\ref{sec:cmb_lss}), 
for which we use the fitting formula from \cite{hu05},
\begin{equation}
r_s \approx (144.4 \Mpc) \, \left({\om h^2 \over 0.14}\right)^{-0.252}
\, \left({\ob h^2 \over 0.024}\right)^{-0.083}~.
\end{equation}
%Note that the values $\mu,\nu=1,2$ denote the 
%$D(z_\alpha)/r_s$ and $H(z_\alpha)r_s$ measurements, respectively.
We estimate the covariance matrix in each redshift bin using the 
BAO forecast code by \cite{seo07}, which provides estimates of the 
fractional error on distance and the Hubble expansion rate at each 
redshift (relative to $r_s$), $\sigma_{\ln(D/r_s)} =
\sqrt{C_{11}^{\rm BAO}}/(D/r_s)$ and $\sigma_{\ln(Hr_s)} = 
\sqrt{C_{22}^{\rm BAO}}/(Hr_s)$, respectively, as well as the cross correlation
$r = C_{12}^{\rm BAO}/\sqrt{C_{11}^{\rm BAO} C_{22}^{\rm BAO}}$.
For our default forecasts, we start with
the linear theory cosmic variance predictions, corresponding to the 
limit of perfect sampling of the density field within the observed volume
and no degradation of the signal due to nonlinear effects. To approximate 
the effects of finite sampling and nonlinearity, we increase these 
errors by a factor of 1.8 for our fiducial forecasts, which leads to 
parameter constraints comparable to what would be expected with 
sampling $nP=2$ and reconstruction that halves the effects of 
nonlinear evolution. In Table~\ref{tbl:baoforecast} we list the volume
for $\fsky=0.25$ and fiducial BAO covariance matrix elements 
for 20 redshift slices from $0\leq z\leq 3$. The results we obtain are 
only weakly dependent on the number of redshift bins chosen to divide up
the total volume.

The WL Fisher matrix is based on the methodology described by
\citet{albrecht09},
where the explicit formulas are given. It includes both power 
spectrum tomography and cross-correlation cosmography (redshift scaling of 
the galaxy-galaxy lensing signal), but makes no assumption about the 
galaxy bias. The galaxies are sliced into $N_z=14$ redshift bins and we 
consider power spectra in $N_\ell=18$ bins logarithmically spaced over 
$10<\ell<10^4$. We consider all power spectra and cross-spectra of the 
galaxies $g_i$ and the $E$-mode shear $\gamma^E_i$. This leads to $2N_z$ 
scalar fields on the sky, and hence $N_{\rm 2pt} = 2N_z(2N_z+1)/2\times 
N_\ell$ bins in the power spectrum matrix.\footnote{Since we neglect 
magnification bias, some of these spectra, e.g. the correlation of 
high-redshift galaxies with low-redshift shear, are zero for all 
cosmological models.} The length $N_{\rm 2pt}$ vector ${\bf C}$ of power 
spectra incorporates all 2-point information.

Our task is now to construct a model both for ${\bf C}$ and for its 
covariance matrix $\Sigma$, and then to construct the Fisher matrix for 
parameters ${\bf p}$:
\begin{equation}
F_{ij} = \frac{\partial{\bf C}^T}{\partial p^i}\Sigma^{-1}
\frac{\partial{\bf C}}{\partial p^j},
\end{equation}
where $^T$ denotes a matrix transpose. Systematic errors may be 
incorporated as either nuisance parameters ${\bf p}$ (marginalized over 
some prior) or as additional contributions to $\Sigma$:
\begin{equation}
\Sigma_{ij} \rightarrow \Sigma_{ij} + \sigma^2_\varpi \frac{\partial 
C_i}{\partial\varpi} 
\frac{\partial C_j}{\partial\varpi},
\end{equation}
where $\varpi$ is the amplitude of some systematic and $\sigma_\varpi$ is 
the amount over which it is marginalized.

We incorporate in $\Sigma$ the following contributions:
\begin{itemize}
\item
The Gaussian covariance matrix.
\item
The 1-halo contribution to the shear 4-point function, given by Eq.~(A9) 
of \citet{albrecht09}.
\item
Galaxy bias and stochasticity, fully marginalized\footnote{i.e. with 
sufficiently wide prior that no significant information remains.} in each 
bin of $\ell$ and $z$.
\item
The $II$ intrinsic alignment term, obtained by fully marginalizing out the 
shear auto-correlations in each redshift slice.
\item
The $GI$ intrinsic alignment term. It is assumed that Eq.~(\ref{eq:wl:ed}) 
will allow estimation of $P_{e\delta}(k)$ and removal of this term in the 
linear and weakly nonlinear regimes (taken to be $\ell<10^{2.5}$). At 
smaller scales, we impose a weak prior that the $GI$ not exceed present 
upper limits. This is implemented as
\begin{equation}
\frac{\sigma[P_{e\delta}(k)]}{P_\delta(k)} = 0.003\sqrt{N_{\ell,\rm 
nonlin}(N_z-1)},
\end{equation}
where the square root is introduced to prevent many bins from being used 
to ``average down'' this systematic \citep{albrecht09}
\end{itemize}
The photometric redshift errors (one bias parameter for each bin) and 
shear calibration errors (also one bias parameter for each bin) are 
treated as nuisance parameters in the parameter vector ${\bf p}$ and are 
marginalized out before combining with other cosmological probes.

The forecasts for the main SN, BAO, and WL probes are supplemented by 
the expected constraints from upcoming CMB measurements provided by 
the \planck\ satellite. We adopt the Fisher matrix ${\bf F}^{\rm CMB}$ 
constructed by the FoMSWG, 
which includes cosmological constraints from the 70, 100, and 143 GHz 
channels of \planck\ with $\fsky=0.7$, assuming that data collected at 
other frequencies will be used for foreground removal. The noise level
and beam size for each channel comes from the \planck\ Blue Book 
\citep{planck}.
Information from secondary anisotropies of the CMB is not included 
in this Fisher matrix; in particular, constraints from the ISW effect 
(\S\ref{sec:isw})
are removed by requiring the angular diameter distance to the CMB 
to be matched exactly, as described in \cite{albrecht09}.
Additionally, 
the large-scale ($\ell<30$) polarization angular power spectrum 
and temperature-polarization cross power spectrum, which 
mainly contribute to constraints on the optical depth to reionization $\tau$,
are excluded from the forecast and replaced by a Gaussian prior 
with width $\sigma_{\tau}=0.01$. This prior accounts for uncertainty in $\tau$ 
due to limited knowledge of the redshift dependence of reionization, 
which is not included in the simplest models of the CMB anisotropies.
Although $\tau$ does not appear in the parameter set for the Fisher matrices,
marginalization over $\tau$ in the CMB constraints contributes to the 
uncertainty on the primordial power spectrum amplitude $A_s$, 
which in turn affects predictions for the growth of large-scale structure.
%\tbd{Somewhere we need to say how results would change if
%there were no $\tau$ error.}

Combined constraints on cosmological parameters are obtained simply by
adding the Fisher matrices of the individual probes, i.e.\ 
${\bf F} = {\bf F}^{\rm SN}+{\bf F}^{\rm BAO}+{\bf F}^{\rm WL}+
{\bf F}^{\rm CMB}$. Then the forecast for the parameter covariance is
${\bf C} = {\bf F}^{-1}$, and in particular the uncertainty on a given 
parameter $p_i$ after marginalizing over the error on all other parameters
is $\sqrt{[{\bf F}^{-1}]_{ii}}$\,.

Computing the Fisher matrix in the FoMSWG parameter space with a large 
number of independent bins for $w(z)$ gives us the flexibility to 
project these forecasts onto a number of simpler parameterizations, 
including the $w_0$--$w_a$ model for the purposes of computing the FoM.
To change from the original parameter set ${\bf p}$ to some new set 
${\bf q}$, we compute
\begin{equation}
\label{eqn:fisher_reparam}
\tilde{F}_{kl} = \sum_{i,j} {\partial p_i \over \partial q_k}\,
F_{ij}\, {\partial p_j \over \partial q_l}~,
\end{equation}
which gives the Fisher matrix ${\bf \tilde F}$ for the new parameterization.
In particular, projection from bins $w_i$ to $w_0$ and $w_a$ involves 
the derivatives $\partial w_i/\partial w_0 = 1$ and 
$\partial w_i/\partial w_a = z/(1+z)$. We also compute the pivot redshift 
$z_p$ and the uncertainty in the equation of state at that redshift, 
$w_p$. Given the $2\times 2$ covariance matrix $C_{ij}$
for $w_0$ and $w_a$ (marginalized over the other parameters), the pivot
values are computed as \citep{albrecht09}
\begin{eqnarray}
\label{eqn:pivot}
z_p &=& -{C_{12} \over C_{12}+C_{22}} ~, \\
\sigma_{w_p} &=& C_{11} - {C_{12}^2 \over C_{22}} ~, \nonumber
\end{eqnarray}
where the first index corresponds to $w_0$ and the second to $w_a$.

One drawback to the $w_0$--$w_a$ parameterization is that constraints 
on $w(z)$ at high redshift are coupled to those at low redshift by the 
form of the model; for example, if observations determine the value of 
the equation of state perfectly 
at $z=0$ and at $z=0.1$, then it is completely 
determined at high redshift even in the absence of high redshift data.
To specifically address questions related to the ability of dark 
energy probes to constrain dark energy at low redshift vs.\ high redshift,
we define an alternative but equally simple
parameterization in which $w(z)$ takes constant, independent values in 
each of two bins at $z\le 1$ and $z>1$. The projection onto this 
parameterization using equation~(\ref{eqn:fisher_reparam}) requires the 
derivatives $\partial w_i/\partial w(z\le 1) = \Theta(1-z_i)$ and 
$\partial w_i/\partial w(z>1) = 1-\Theta(1-z_i)$, where $\Theta(x)$
is the Heaviside step function equal to 0 for $x<0$ and 1 for $x\ge 0$.

Principal components (PCs) of the dark energy equation of state provide another
way to determine which features of the equation of state evolution 
are best constrained by a given
combination of experiments 
\citep{huterer03,hu02,huterer05,wang05,dick06,simpson06,deputter08,tang08,
crittenden09,mortonson09a,kitching09,maturi09}.
We compute the PCs
for each forecast case by taking the total Fisher matrix for the 
original parameter set (eq.~\ref{eqn:forecastparams}) and 
marginalizing over all parameters other than the 36 binned values of $w_i$.
If we call the Fisher matrix for the $w_i$ parameters ${\bf F}^{w}$, then
the PCs are found by diagonalizing ${\bf F}^{w}$:
\begin{equation}
\label{eqn:fisherdiag}
{\bf F}^{w} = {\bf Q} {\bf \Lambda} {\bf Q}^T \,,
\end{equation}
where ${\bf Q}$ is an orthogonal matrix whose columns are 
eigenvectors of ${\bf F}^{w}$ and ${\bf \Lambda}$ is a diagonal 
matrix containing the corresponding eigenvalues of ${\bf F}^w$.
Up to an arbitary normalization factor, the eigenvectors are equal to 
the PC functions ${\bf e}_i = (e_i(z_1),e_i(z_2),...)$ which 
describe how the binned values of $w(z)$ are weighted with redshift. 
Here we adopt the normalization of \cite{albrecht09},
\begin{equation}
\label{eqn:pcnorm}
\sum_{k=1}^{36} e_i(z_k) e_j(z_k) = \sum_{k=1}^{36} e_k(z_i) e_k(z_j) = 
(\Delta a)^{-1} \delta_{ij} \,,
\end{equation}
where $\Delta a = 0.025$ is the bin width; 
for $i=j$ this condition
approximately corresponds to $\int_{0.1}^1 da [e_i(a)]^2=1$.
With this convention, the columns of ${\bf Q}$ are 
$(\Delta a)^{1/2} \, {\bf e}_i$\,.   The PCs rotate the
original set of parameters to a set of PC amplitudes 
${\bf Q}^T ({\bf 1}+{\bf w})$ with elements
\begin{equation}
\label{eqn:wtopc}
\beta_i = (\Delta a)^{1/2} \sum_{j=1}^{36} e_i(z_j)(1+w_j) \,.
\end{equation}
Combining equations~(\ref{eqn:pcnorm}) and~(\ref{eqn:wtopc}), 
we can construct $w(z)$ in each redshift bin 
from a given set of PC amplitudes as
\begin{equation}
\label{eqn:pctow}
w_i = -1 + \sum_{j=1}^{36} \alpha_j e_j(z_i) \,,
\end{equation}
where $\alpha_i \equiv (\Delta a)^{1/2} \beta_i$\,.
The accuracy with which the $\alpha_i$ can be determined from the data 
is given by the eigenvalues of ${\bf F}^{w}$, 
$\sigma_i \equiv \sigma_{\alpha_i} = (\Delta a/\Lambda_{ii})^{1/2}$, 
and the PCs are numbered in order of increasing variance 
(i.e.\ $\sigma_{i+1} > \sigma_i$).

For constraints that are marginalized over the $w_i$ parameters, 
we impose a weak prior on $w_i$ as suggested by \cite{albrecht09} 
to reduce the dependence of forecasts for $\Delta\gamma$
on the poorly-constrained high redshift $w_i$ values, since arbitrarily 
large fluctuations in $w(z)$ can alter the high redshift growth rate.
We include a weak Gaussian prior with width 
$\sigma_{w_i} = \Delta w/\sqrt{\Delta a}$ by adding to the total Fisher matrix
\begin{equation}
\label{eqn:wprior}
F_{ij}^{\rm prior} = \left\{
\begin{array}{rr}
\sigma_{w_i}^{-2}\, \delta_{ij}\,, & i \le 36 \,, \\
0\,, & i > 36 \,, \\
\end{array}
\right.
\end{equation}
assuming that the parameters are ordered as in equation~(\ref{eqn:forecastparams})
with $p_1=w_1$, $p_2=w_2$, etc. For most forecasts, we use a default 
prior width of $\Delta w = 10$ ($\sigma_{w_i} \approx 63$), which
approximately corresponds to requiring that the average value of $|1+w|$ 
in all bins does not exceed 10. In the next section we also consider how
constraints on certain parameters change with a narrower prior of 
$\Delta w = 1$. For priors wider than the default choice, 
the Fisher matrix computations are subject
to numerical effects arising from the use of a finite number of $w_i$ 
bins to approximate continuous variations in $w(z)$, so we do not 
present results with weaker priors than $\Delta w = 10$.
Note that the construction of PCs of $w(z)$ as 
described above does not include such a prior on $w_i$.

\begin{deluxetable}{ll}
\tablecolumns{2}
\tablecaption{Key to forecast variations\label{tbl:key}}
\tablehead{}
\tablewidth{0pc}
\startdata
$Any\times 4$ & Quadruple fiducial errors (divide Fisher matrix by 16). \\
$Any\times 2$ & Double fiducial errors (divide Fisher matrix by 4). \\
$Any/2$ & Halve fiducial errors (multiply Fisher matrix by 4). \\
\hline
SN-III & Stage III-like SN: total magnitude error of 0.02 per 
$\Delta z=0.2$ bin \\
 & over $0.2\le z \le 0.8$ and in local sample at $z=0.05$. \\
SN$z_{\rm max}$ & Increase max.\ redshift to $\zmax=1.6$
(7 bins with $\Delta z = 0.2$ and 0.01 mag.\ error). \\
SN$-$local & Omit local sample at $z=0.05$. \\
SNc$x$ & Correlated errors: $\sigma_{m,u}=\sigma_{m,c}=0.007$, 
$\Delta z_c=0.2$, with $x$ bins over $0.2\le z\le 0.8$. \\
\hline
BAO-III & Stage III-like BAO, approximating forecasts for BOSS LRGs+HETDEX:\\
  & $(D/r_s,Hr_s)$ errors of $(1.0\%,1.8\%)$ at $z=0.35$, $(1.0\%,1.7\%)$
  at $z=0.6$, \\
  & and $(0.8\%,0.8\%)$ at $z=2.4$. These are ``BAO only'' forecasts for BOSS\\
  & and ``full power spectrum'' forecasts for HETDEX.  \\
BAO$z_{\rm max}$ & Reduce maximum redshift to $\zmax=2$ (20 bins), 
retaining $\fsky=0.25$. \\
\hline
WL-opt & ``Optimistic'' Stage IV case (total error$=2\times\,$statistical).\\
WL-III & Stage III-like WL, approximating forecasts for DES: 5000$\mdeg^2$
  and $n_{\rm eff} = 9\,\arcmin^{-2}$.\\
\hline
CMB-W9 & Fisher matrix forecast for 9-year WMAP data. \\
\enddata
\end{deluxetable}

\begin{deluxetable}{rlllrlllll}
\tablecolumns{10}
\tablecaption{Forecast Uncertainties for Variations of the Fiducial Program\label{tbl:forecasts1}}
\tablehead{
 & Forecast case & $z_p$ & $\sigma_{w_p}$ & FoM & $\sigma_{w(z>1)}$ & $10^3\,\sigma_{\ok}$ & $10^2\,\sigma_h$ & $\sigma_{\Delta\gamma}$ & $\sigma_{\ln G_9}$
}
\tablewidth{0pc}
\startdata
{\it 1} & $[$SN,BAO,WL,CMB$]$ & 0.46 & 0.014 &   664 & 0.051 & 0.55 & 0.51 & 0.034 & 0.015 \\
{\it 2} & $[$SN,BAO,WL-opt,CMB$]$ & 0.39 & 0.013 &   789 & 0.049 & 0.64 & 0.42 & 0.026 & 0.016 \\
\hline
{\it 3} & $[$BAO,WL,CMB$]$ & 0.63 & 0.017 & 321 & 0.054 & 0.56 & 0.99 & 0.034 & 0.015 \\
{\it 4} & $[$SN-III,BAO,WL,CMB$]$ & 0.57 & 0.016 &   433 & 0.053 & 0.56 & 0.75 & 0.034 & 0.015 \\
{\it 5} & $[$SN$\times$4,BAO,WL,CMB$]$ & 0.61 & 0.017 &   353 & 0.054 & 0.56 & 0.91 & 0.034 & 0.015 \\
{\it 6} & $[$SN$\times$2,BAO,WL,CMB$]$ & 0.57 & 0.016 &   433 & 0.053 & 0.56 & 0.75 & 0.034 & 0.015 \\
{\it 7} & $[$SN/2,BAO,WL,CMB$]$ & 0.32 & 0.010 &  1197 & 0.049 & 0.55 & 0.32 & 0.034 & 0.015 \\
{\it 8} & $[$SN$z_{\rm max}$,BAO,WL,CMB$]$ & 0.42 & 0.011 &   841 & 0.050 & 0.55 & 0.40 & 0.034 & 0.015 \\
{\it 9} & $[$SN$-$local,BAO,WL,CMB$]$ & 0.59 & 0.016 & 376 & 0.053 & 0.56 & 0.85 & 0.034 & 0.015 \\
{\it 10} & $[$SNc3,BAO,WL,CMB$]$ & 0.46 & 0.014 &   652 & 0.051 & 0.55 & 0.51 & 0.034 & 0.015 \\
{\it 11} & $[$SNc6,BAO,WL,CMB$]$ & 0.46 & 0.014 &   663 & 0.051 & 0.55 & 0.51 & 0.034 & 0.015 \\
{\it 12} & $[$SNc12,BAO,WL,CMB$]$ & 0.46 & 0.014 &   667 & 0.051 & 0.55 & 0.50 & 0.034 & 0.015 \\
\hline
{\it 13} & $[$SN,WL,CMB$]$ & 0.26 & 0.022 &   152 & 0.321 & 2.13 & 0.72 & 0.038 & 0.022 \\
{\it 14} & $[$SN,BAO-III,WL,CMB$]$ & 0.32 & 0.019 &   299 & 0.120 & 1.19 & 0.57 & 0.035 & 0.017 \\
{\it 15} & $[$SN,BAO$\times$4,WL,CMB$]$ & 0.30 & 0.020 &   245 & 0.145 & 1.16 & 0.65 & 0.036 & 0.018 \\
{\it 16} & $[$SN,BAO$\times$2,WL,CMB$]$ & 0.36 & 0.018 &   380 & 0.087 & 0.76 & 0.58 & 0.035 & 0.016 \\
{\it 17} & $[$SN,BAO/2,WL,CMB$]$ & 0.50 & 0.010 &  1222 & 0.033 & 0.47 & 0.39 & 0.034 & 0.014 \\
{\it 18} & $[$SN,BAO$z_{\rm max}$,WL,CMB$]$ & 0.42 & 0.014 &   547 & 0.071 & 0.66 & 0.52 & 0.034 & 0.015 \\
\hline
{\it 19} & $[$SN,BAO,CMB$]$ & 0.41 & 0.016 &   539 & 0.059 & 0.78 & 0.53 & --- & --- \\
{\it 20} & $[$SN,BAO,WL-III,CMB$]$ & 0.41 & 0.016 &   543 & 0.058 & 0.77 & 0.52 & 0.145 & 0.048 \\
{\it 21} & $[$SN,BAO,WL$\times$4,CMB$]$ & 0.42 & 0.016 &   553 & 0.057 & 0.75 & 0.53 & 0.126 & 0.031 \\
{\it 22} & $[$SN,BAO,WL$\times$2,CMB$]$ & 0.43 & 0.015 &   587 & 0.055 & 0.68 & 0.52 & 0.065 & 0.020 \\
{\it 23} & $[$SN,BAO,WL/2,CMB$]$ & 0.48 & 0.012 &   815 & 0.047 & 0.45 & 0.47 & 0.018 & 0.012 \\
{\it 24} & $[$SN,BAO,WL-opt$\times$4,CMB$]$ & 0.41 & 0.016 &   556 & 0.058 & 0.76 & 0.52 & 0.085 & 0.022 \\
{\it 25} & $[$SN,BAO,WL-opt$\times$2,CMB$]$ & 0.41 & 0.015 &   606 & 0.055 & 0.73 & 0.49 & 0.045 & 0.018 \\
{\it 26} & $[$SN,BAO,WL-opt/2,CMB$]$ & 0.37 & 0.009 &  1397 & 0.040 & 0.52 & 0.30 & 0.017 & 0.013 \\
\hline
{\it 27} & $[$SN,BAO,WL$]$ & 0.31 & 0.020 &   368 & 0.075 & 7.82 & 1.48 & 0.037 & 6.697 \\
{\it 28} & $[$SN,BAO,WL,CMB-W9$]$ & 0.43 & 0.015 &   592 & 0.055 & 1.07 & 0.53 & 0.036 & 0.019 \\
\enddata
\tablecomments{
Forecasts in this table vary the assumptions about a single probe 
at a time from the fiducial program.
With the exception of $w(z>1)$, 
a $w_0$--$w_a$ model for the dark energy equation of state is assumed 
for all parameter uncertainties 
here and in Tables~\ref{tbl:forecasts2} and~\ref{tbl:forecasts3}. 
All forecasts allow for deviations from GR parameterized by
$\Delta \gamma$ and $G_9$.
}
\end{deluxetable}

\begin{deluxetable}{rlllrlllll}
\tablecolumns{10}
\tablecaption{Forecast Uncertainties for Variations of the Fiducial Program (Continued)\label{tbl:forecasts2}}
\tablehead{
 & Forecast case & $z_p$ & $\sigma_{w_p}$ & FoM & $\sigma_{w(z>1)}$ & $10^3\,\sigma_{\ok}$ & $10^2\,\sigma_h$ & $\sigma_{\Delta\gamma}$ & $\sigma_{\ln G_9}$
}
\tablewidth{0pc}
\startdata
{\it 1} & $[$SN,BAO,WL,CMB$]$ & 0.46 & 0.014 &   664 & 0.051 & 0.55 & 0.51 & 0.034 & 0.015 \\
\hline
{\it 2} & $[$SN-III,BAO-III,WL-III,CMB$]$ & 0.42 & 0.032 &   131 & 0.137 & 1.36 & 0.96 & 0.147 & 0.051 \\
{\it 3} & $[$SN-III,BAO-III,WL-III,CMB-W9$]$ & 0.33 & 0.039 &    92 & 0.174 & 2.41 & 1.01 & 0.148 & 0.064 \\
{\it 4} & $[$SN$\times$4,BAO$\times$4,WL$\times$4,CMB$]$ & 0.51 & 0.048 &    52 & 0.179 & 1.32 & 1.98 & 0.128 & 0.033 \\
{\it 5} & $[$SN$\times$2,BAO$\times$2,WL$\times$2,CMB$]$ & 0.49 & 0.026 &   188 & 0.095 & 0.85 & 1.00 & 0.065 & 0.021 \\
{\it 6} & $[$SN/2,BAO/2,WL/2,CMB$]$ & 0.43 & 0.007 &  2439 & 0.027 & 0.34 & 0.26 & 0.018 & 0.011 \\
{\it 7} & $[$SN/2,BAO/2,WL-opt,CMB$]$ & 0.34 & 0.008 &  1832 & 0.035 & 0.55 & 0.26 & 0.023 & 0.014 \\
\hline
{\it 8} & $[$SN-III,BAO-III,WL,CMB$]$ & 0.44 & 0.026 &   169 & 0.126 & 1.20 & 0.89 & 0.035 & 0.017 \\
{\it 9} & $[$SN$\times$4,BAO$\times$4,WL,CMB$]$ & 0.50 & 0.034 &    85 & 0.157 & 1.18 & 1.49 & 0.037 & 0.019 \\
{\it 10} & $[$SN$\times$4,BAO$\times$2,WL,CMB$]$ & 0.57 & 0.026 &   153 & 0.093 & 0.77 & 1.28 & 0.035 & 0.016 \\
{\it 11} & $[$SN$\times$4,BAO/2,WL,CMB$]$ & 0.57 & 0.011 &   891 & 0.033 & 0.47 & 0.53 & 0.034 & 0.014 \\
{\it 12} & $[$SN$\times$2,BAO$\times$4,WL,CMB$]$ & 0.41 & 0.029 &   132 & 0.151 & 1.17 & 1.01 & 0.037 & 0.018 \\
{\it 13} & $[$SN$\times$2,BAO$\times$2,WL,CMB$]$ & 0.49 & 0.023 &   218 & 0.090 & 0.76 & 0.92 & 0.035 & 0.016 \\
{\it 14} & $[$SN$\times$2,BAO/2,WL,CMB$]$ & 0.55 & 0.011 &   966 & 0.033 & 0.47 & 0.49 & 0.034 & 0.014 \\
{\it 15} & $[$SN/2,BAO$\times$4,WL,CMB$]$ & 0.25 & 0.012 &   499 & 0.142 & 1.15 & 0.47 & 0.036 & 0.017 \\
{\it 16} & $[$SN/2,BAO$\times$2,WL,CMB$]$ & 0.27 & 0.011 &   735 & 0.084 & 0.76 & 0.39 & 0.035 & 0.016 \\
{\it 17} & $[$SN/2,BAO/2,WL,CMB$]$ & 0.38 & 0.008 &  1921 & 0.032 & 0.47 & 0.27 & 0.034 & 0.014 \\
{\it 18} & $[$SN$z_{\rm max}$,BAO$z_{\rm max}$,WL,CMB$]$ & 0.40 & 0.012 &   694 & 0.069 & 0.66 & 0.42 & 0.034 & 0.015 \\
\enddata
\tablecomments{
Same as Table~\ref{tbl:forecasts1}, but varying two or three probes
at a time from the fiducial specifications.
}
\end{deluxetable}

\begin{deluxetable}{rlllrlllll}
\tablecolumns{10}
\tablecaption{Forecast Uncertainties for Variations of the Fiducial Program (Continued)\label{tbl:forecasts3}}
\tablehead{
 & Forecast case & $z_p$ & $\sigma_{w_p}$ & FoM & $\sigma_{w(z>1)}$ & $10^3\,\sigma_{\ok}$ & $10^2\,\sigma_h$ & $\sigma_{\Delta\gamma}$ & $\sigma_{\ln G_9}$
}
\tablewidth{0pc}
\startdata
{\it 1} & $[$SN,BAO,WL,CMB$]$ & 0.46 & 0.014 &   664 & 0.051 & 0.55 & 0.51 & 0.034 & 0.015 \\
\hline
{\it 2} & $[$SN,BAO-III,WL-III,CMB$]$ & 0.29 & 0.022 &   239 & 0.129 & 1.35 & 0.59 & 0.147 & 0.051 \\
{\it 3} & $[$SN,BAO$\times$4,WL$\times$4,CMB$]$ & 0.28 & 0.022 &   185 & 0.165 & 1.30 & 0.77 & 0.128 & 0.033 \\
{\it 4} & $[$SN,BAO$\times$4,WL$\times$2,CMB$]$ & 0.28 & 0.021 &   200 & 0.159 & 1.26 & 0.73 & 0.067 & 0.023 \\
{\it 5} & $[$SN,BAO$\times$4,WL/2,CMB$]$ & 0.35 & 0.016 &   373 & 0.115 & 0.98 & 0.54 & 0.020 & 0.014 \\
{\it 6} & $[$SN,BAO$\times$4,WL-opt,CMB$]$ & 0.29 & 0.015 &   361 & 0.102 & 1.21 & 0.57 & 0.042 & 0.020 \\
{\it 7} & $[$SN,BAO$\times$2,WL$\times$4,CMB$]$ & 0.34 & 0.019 &   328 & 0.092 & 0.90 & 0.62 & 0.127 & 0.031 \\
{\it 8} & $[$SN,BAO$\times$2,WL$\times$2,CMB$]$ & 0.35 & 0.019 &   340 & 0.090 & 0.85 & 0.61 & 0.065 & 0.021 \\
{\it 9} & $[$SN,BAO$\times$2,WL/2,CMB$]$ & 0.40 & 0.015 &   502 & 0.078 & 0.67 & 0.51 & 0.019 & 0.013 \\
{\it 10} & $[$SN,BAO$\times$2,WL-opt,CMB$]$ & 0.33 & 0.014 &   506 & 0.072 & 0.83 & 0.49 & 0.033 & 0.017 \\
{\it 11} & $[$SN,BAO/2,WL$\times$4,CMB$]$ & 0.43 & 0.012 &   926 & 0.041 & 0.65 & 0.40 & 0.126 & 0.031 \\
{\it 12} & $[$SN,BAO/2,WL$\times$2,CMB$]$ & 0.45 & 0.011 &  1010 & 0.038 & 0.59 & 0.40 & 0.064 & 0.020 \\
{\it 13} & $[$SN,BAO/2,WL/2,CMB$]$ & 0.54 & 0.008 &  1585 & 0.028 & 0.34 & 0.38 & 0.018 & 0.012 \\
{\it 14} & $[$SN,BAO/2,WL-opt,CMB$]$ & 0.43 & 0.010 &  1251 & 0.035 & 0.55 & 0.35 & 0.023 & 0.015 \\
\hline
{\it 15} & $[$SN-III,BAO,WL-III,CMB$]$ & 0.54 & 0.019 &   346 & 0.060 & 0.77 & 0.79 & 0.146 & 0.048 \\
{\it 16} & $[$SN$\times$4,BAO,WL$\times$4,CMB$]$ & 0.60 & 0.020 &   277 & 0.060 & 0.75 & 0.99 & 0.126 & 0.031 \\
{\it 17} & $[$SN$\times$4,BAO,WL$\times$2,CMB$]$ & 0.60 & 0.019 &   298 & 0.058 & 0.68 & 0.97 & 0.065 & 0.020 \\
{\it 18} & $[$SN$\times$4,BAO,WL/2,CMB$]$ & 0.59 & 0.014 &   486 & 0.049 & 0.45 & 0.75 & 0.018 & 0.012 \\
{\it 19} & $[$SN$\times$4,BAO,WL-opt,CMB$]$ & 0.47 & 0.014 &   568 & 0.049 & 0.64 & 0.56 & 0.026 & 0.016 \\
{\it 20} & $[$SN$\times$2,BAO,WL$\times$4,CMB$]$ & 0.54 & 0.019 &   351 & 0.059 & 0.75 & 0.79 & 0.126 & 0.031 \\
{\it 21} & $[$SN$\times$2,BAO,WL$\times$2,CMB$]$ & 0.55 & 0.018 &   375 & 0.057 & 0.68 & 0.78 & 0.065 & 0.020 \\
{\it 22} & $[$SN$\times$2,BAO,WL/2,CMB$]$ & 0.56 & 0.013 &   567 & 0.048 & 0.45 & 0.65 & 0.018 & 0.012 \\
{\it 23} & $[$SN$\times$2,BAO,WL-opt,CMB$]$ & 0.45 & 0.014 &   619 & 0.049 & 0.64 & 0.52 & 0.026 & 0.016 \\
{\it 24} & $[$SN/2,BAO,WL$\times$4,CMB$]$ & 0.28 & 0.011 &   998 & 0.056 & 0.74 & 0.33 & 0.126 & 0.031 \\
{\it 25} & $[$SN/2,BAO,WL$\times$2,CMB$]$ & 0.30 & 0.011 &  1061 & 0.053 & 0.67 & 0.33 & 0.065 & 0.020 \\
{\it 26} & $[$SN/2,BAO,WL/2,CMB$]$ & 0.35 & 0.009 &  1430 & 0.045 & 0.44 & 0.30 & 0.018 & 0.012 \\
{\it 27} & $[$SN/2,BAO,WL-opt,CMB$]$ & 0.30 & 0.010 &  1242 & 0.049 & 0.64 & 0.30 & 0.026 & 0.015 \\
\enddata
\tablecomments{
Continuation of Table~\ref{tbl:forecasts2}.
}
\end{deluxetable}

\subsection{Results: Forecasts for the Fiducial Program and Variations}
\label{sec:results}

\subsubsection{Constraints in simple $w(z)$ models}
\label{sec:results_simplewz}

We begin with forecasts for which the 36 $w(z)$ bins are projected onto 
the simpler $w_0$--$w_a$ parameter space. 
Tables~\ref{tbl:key}--\ref{tbl:forecasts3}
give the forecast $1\sigma$ uncertainties for
the fiducial program and numerous variations. Each forecast case 
is labeled by a list of the Fisher matrices that are added together, 
and the basic variations we consider are simple rescalings of the 
{\it total} errors for each probe; for example, [SN/2,BAO$\times$4,WL-opt,CMB]
includes the fiducial SN data with the total error halved (i.e.\ the Fisher
matrix multiplied by 4), 4 times the fiducial BAO errors, the 
optimistic version of the WL forecast, and the fiducial \planck\ CMB 
Fisher matrix. 
Note that $/2$ denotes a {\it more} powerful program and $\times 2$
denotes a {\it less} powerful program.
The key in Table~\ref{tbl:key} describes other types of
variations of the fiducial probes.
In some cases we omit a probe entirely, e.g. [SN,BAO,WL]
sums the fiducial Fisher matrices of the three main probes but does not
include the \planck\ CMB priors. Note that even though we assume a
specific systematic error component in computing certain Fisher matrices
(in particular, ${\bf F}^{\rm WL}$), the cases with rescaled errors simply 
multiply each Fisher matrix by a constant factor and thus do not 
distinguish between statistical and systematic contributions to the total
error.

Constraints on the equation of state are given in 
Tables~\ref{tbl:forecasts1}--\ref{tbl:forecasts3} by the DETF FoM and 
the error on $w_p$.
The rule of thumb that $\sigma_{w_a} \equiv ({\rm FoM} \times \sigma_{w_p})^{-1}
\approx 10 \sigma_{w_p}$ 
holds at the $\sim 30\%$ level for 
most of the forecast variations we consider --- i.e., 
at the best-constrained redshift, the value of $w$
is typically determined a factor of ten better than the value of its
derivative.
The forecast tables also list the uncertainty in the high redshift 
equation of state $w(z>1)$ for the alternative
parameterization where $w(z)$ takes 
independent, constant values at $z\le 1$ and $z>1$.
Note that all of these $w(z)$ constraints are marginalized over uncertainties
in $G_9$ and $\Delta\gamma$, so they do {\it not} assume that structure
growth follows the GR prediction.

For the fiducial program outlined in \S\ref{sec:fiducial}, the DETF FoM 
is projected to be around 600--800, depending on whether the WL forecast
uses the default systematic error model or the optimistic model.
This is roughly an order of magnitude larger than the FoM forecast for 
a combination of Stage III experiments 
(e.g.\ see Table~\ref{tbl:forecasts2}, rows 2--3)
and nearly two orders of magnitude larger than current, ``Stage II'' FoM 
values ($\sim 10$).
The equation of state in the $w_0$--$w_a$ parameterization is best measured
by the fiducial set of Stage IV experiments at a redshift $z_p\approx 0.5$
with a $1\sigma$ precision
of $\sigma_{w_p}\approx 0.014$, and the time variation of 
$w(z)$ is determined to within $\sigma_{w_a}\approx 0.11$. 
The fiducial program also yields impressive constraints
of $5.5 \times 10^{-4}$ on $\ok$ and $0.51\hubunits$ on $H_0$.
Forecast $1\sigma$ errors for the modified gravity parameters
are 0.034 on $\Delta\gamma$ and 0.015 on $\ln G_9$.
We caution, however, that the $\ok$, $H_0$, and $G_9$ errors
(but not the $\Delta\gamma$ error) are sensitive to our assumption
of the $w_0$--$w_a$ parameterization 
(see Figures~\ref{fig:contours_sn}--\ref{fig:h_model_sn} below).
CMB constraints make a critical contribution --- the FoM drops
from 664 to 368 if they are omitted entirely
(Table~\ref{tbl:forecasts1}, line 27) --- but the difference
between \planck\ precision and anticipated WMAP9 precision is
modest (line 28) except for $\ok$, where it is a factor of two.

\begin{figure}[p]
\begin{centering}
{\includegraphics[width=3.2in]{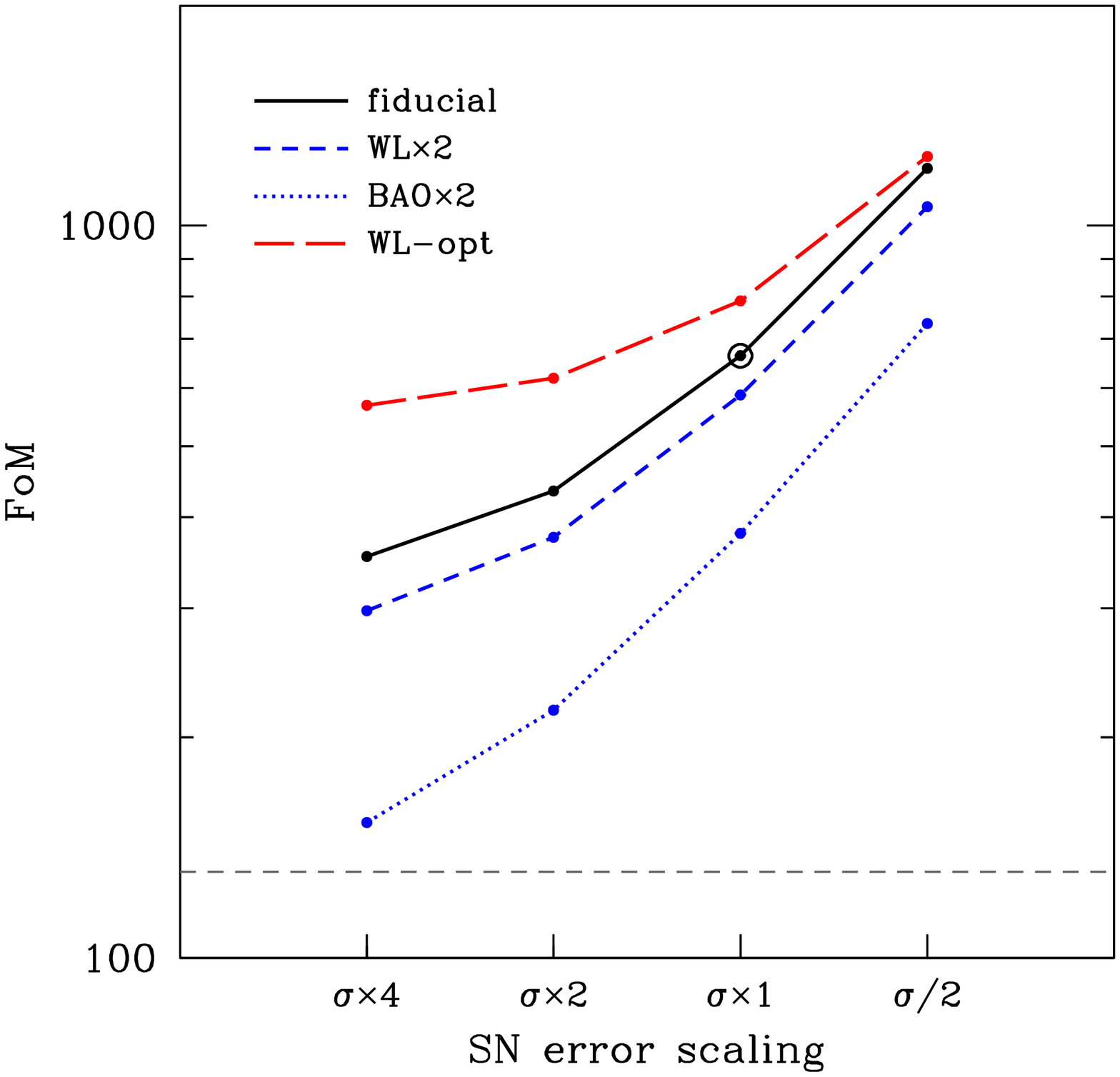}\hfill 
\includegraphics[width=3.2in]{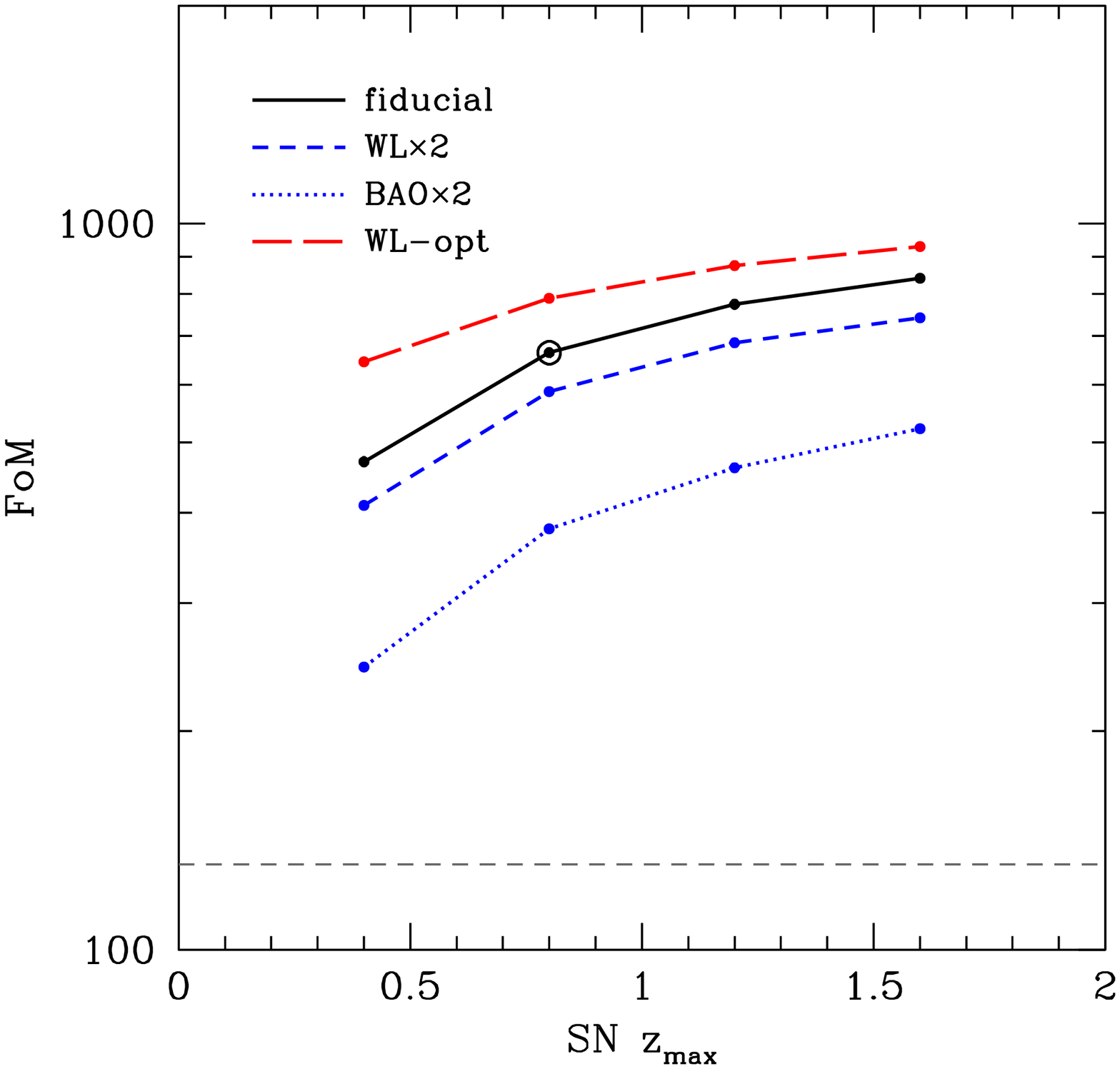}}
{\includegraphics[width=3.2in]{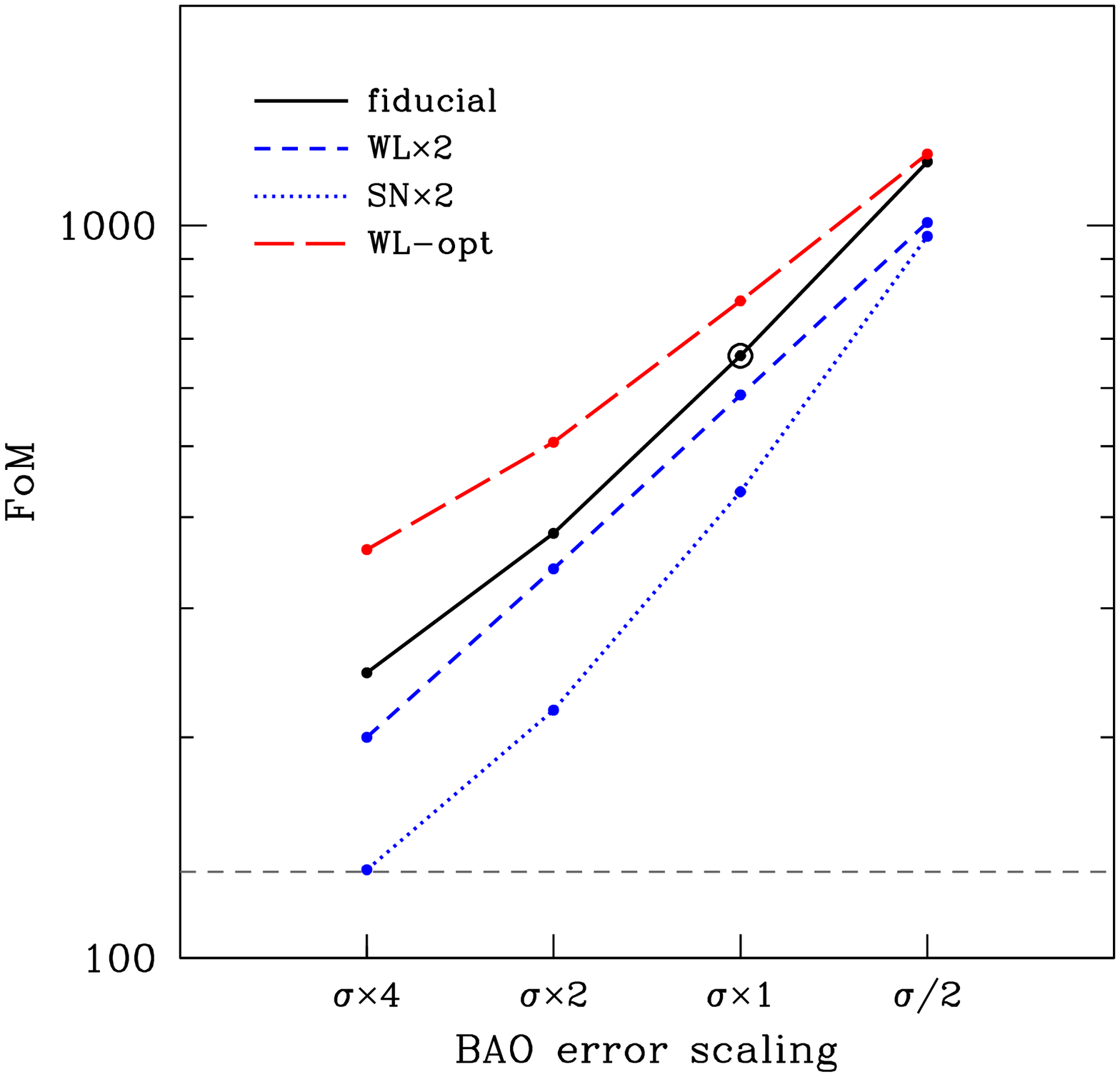}\hfill
\includegraphics[width=3.2in]{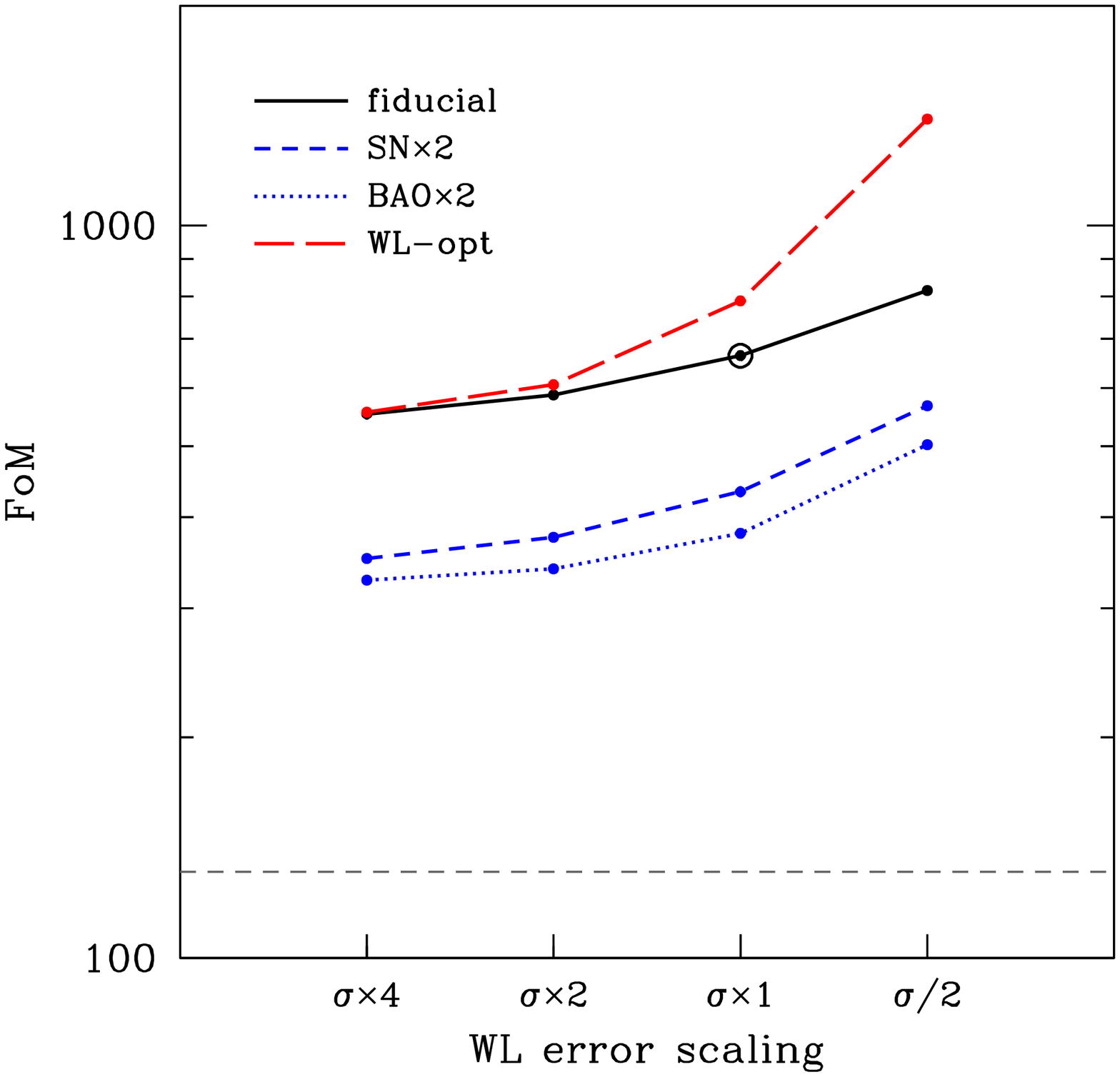}}
\caption{\label{fig:fom}
The DETF FoM, $(\sigma_{w_p}\sigma_{w_a})^{-1}$,
for the fiducial program and simple variants. 
In each panel, the open circle marks the FoM of the fiducial program.
In the upper left panel, the other points along the solid curve
show the effect of scaling the error on the SN measurements by
factors of 2 or 4 while keeping errors for other probes fixed at
their fiducial values.  Dotted, short-dashed, and long-dashed curves show 
the effect of, respectively, doubling the BAO errors, doubling the WL
errors, or adopting the optimistic WL forecasts in which systematic
errors are simply twice the statistical errors.
Other panels show analogous results, but instead of scaling the
total SN error they scale the total BAO error (lower left),
the total WL error (lower right), or the maximum redshift
of the SN constraints (upper right).
In each panel, the dashed gray line marks the forecast performance of 
Stage III probes (including \planck) with FoM=131.
}
\end{centering}
\end{figure}

Figure~\ref{fig:fom} illustrates the key results of
our forecasting investigation, highlighting many aspects
of the interplay among the three observational probes.
In the upper left panel, the solid curve 
shows how the FoM changes as the total SN errors vary from four times fiducial
to half fiducial, keeping the other probes (BAO, WL, and CMB) fixed at 
their fiducial levels.  Other curves show the effect of 
doubling WL or BAO errors or switching to the optimistic WL forecast.
The lower panels show analogous results from varying the BAO or WL errors,
while the upper right panel shows the effect of changing the maximum
redshift of the SN program.
Over the range of variations plotted in Figure~\ref{fig:fom},
the FoM varies from just over 100 to almost 1400.

%\tbd{Rephrased since it seems odd to describe nonuniformity as a ``trend''}
The scaling of the FoM with the forecast errors is not uniform among 
the three main probes. Starting from the fiducial program, the effect of 
doubling or halving errors is greater for BAO than for SN, and greater for 
SN than for WL. This scaling implies that BAO data provide
the greatest leverage in these forecasts.
However, the hierarchy of the three probes is sensitive to the 
assumptions about each experiment; in particular, assuming the 
optimistic version of WL errors promotes WL from 
having the least leverage on the FoM to having the most leverage.
More generally, the 
fact that varying the errors of any individual probe changes the FoM 
noticeably demonstrates the complementarity of the methods.

Unlike many previous FoM forecasts, we marginalize over the 
structure growth parameters $\Delta \gamma$ and $\ln G_9$, which tends
to increase the uncertainties on $w_0$ and $w_a$. In most cases, the
difference between the marginalized constraints and ones obtained under
the assumption of GR ($\Delta \gamma = \ln G_9 = 0$) is small, but 
the difference is greater if WL contributes significantly to expansion 
history constraints; for example, for the fiducial program, the change 
in the FoM due to assuming GR is only $664 \to 771$, whereas with the 
WL-opt forecast the change is $789 \to 1119$.

The local calibrator sample plays an important role in the SN constraints.
Omitting the measurement at $z=0.05$ reduces the FoM from
$664 \to 376$ (Table~\ref{tbl:forecasts1}, line 9).
Even replacing it with a measurement over a broad low-redshift
bin $0 < z < 0.2$, still with an error of 0.01 mag, reduces the
FoM from $664 \to 533$ because it increases degeneracy between
the supernova absolute magnitude scale and dark energy parameters.
Reducing the redshift of the calibrator sample below 0.05 makes
little further difference, and at lower redshifts peculiar velocity
uncertainties may become too large to remove with high precision.
It is also interesting to ask whether it is better to 
go after SNe at high redshifts or to focus on reducing the errors on 
SN data at low redshifts. Comparing the upper panels of Figure~\ref{fig:fom},
we find that the benefit from reducing errors is typically greater than 
that from obtaining SNe beyond $z\sim 1$, at least for the FoM. 
For example, reducing the error per redshift bin from 0.01 mag
(the fiducial value) to 0.005 mag raises the FoM by a factor of 1.80,
but increasing the maximum redshift from 0.8 to 1.6 raises the
FoM by only 1.27 (see Table~\ref{tbl:forecasts1}).
If BAO errors are doubled, the FoM drops substantially, but SN errors
still have much greater leverage than SN maximum redshift.
%We have assumed in these forecasts
%that the error per redshift bin stays constant as the maximum SN redshift
%increases, but in reality higher redshift SNe are likely to have larger 
%systematic errors associated with them, which would diminish the gains 
%from high redshift SNe even more than indicated by the flattening of
%curves in Figure~\ref{fig:fom}.
%Of course, once the systematic errors at $z<0.8$ are saturated,
%pushing to higher redshift may be the only way to continue improving
%SN constraints.

\begin{figure}[th]
\begin{centering}
\includegraphics[width=4.in]{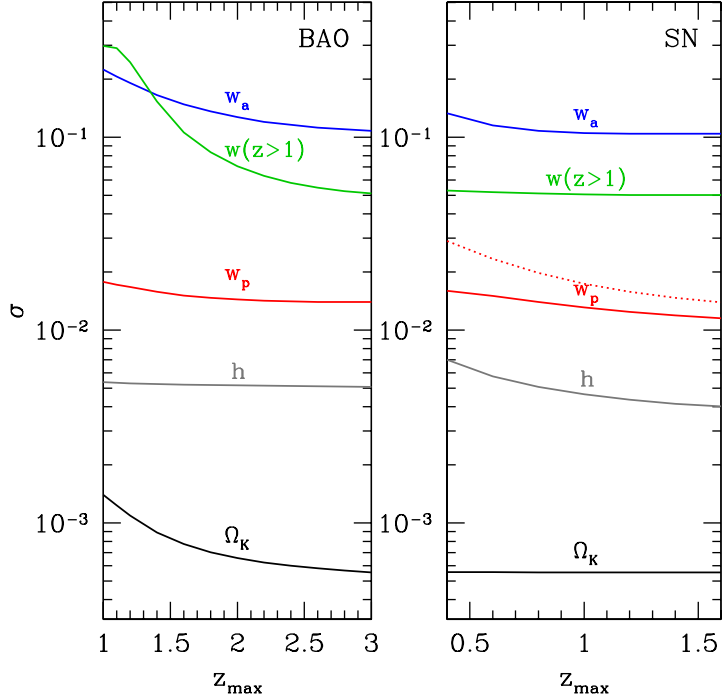}
\caption{\label{fig:zmax}
Variation of $1\sigma$ parameter errors with the maximum redshift 
for BAO at fixed $\fsky$ (left) or for SN with fixed error per
$\Delta z = 0.2$ redshift bin (right). For the solid curves, fiducial 
Stage IV forecasts are assumed for all other probes.  
The dotted curve in the right panel shows the scaling of $\sigma(w_p)$ 
with SN $\zmax$ assuming 4 times larger BAO errors (BAO$\times 4$).
The plotted errors assume a 
$w_0$--$w_a$ parameterization (except for $w(z>1)$).
}
\end{centering}
\end{figure}

The weak dependence of $w(z)$ constraints on the maximum SN redshift extends 
to other parameters as well. Figure~\ref{fig:zmax}
compares the effect on $1\sigma$ errors 
of varying the maximum SN redshift to that of varying
the maximum BAO redshift. For the $w_0$--$w_a$ model, the errors on 
all parameters are relatively insensitive to changes in the maximum SN
redshift at $z\gtrsim 1$, but the errors on $w_a$ and $\ok$ decrease by a 
factor of a few as the maximum BAO redshift increases from $z=1$ to $z=3$.
Likewise, the high redshift equation of state $w(z>1)$ can be determined 
much more precisely as BAO data extend to higher redshifts, but it depends 
little on the maximum SN redshift.  For the fiducial 
Stage IV forecasts, only the Hubble constant error depends significantly on the
depth of SN observations (assuming a $w_0$--$w_a$ model). More pessimistic
assumptions about the achievable BAO errors enhance the importance of 
high redshift SNe for determining $w_p$ (dotted line in Figure~\ref{fig:zmax}),
but the dependence of other parameters on $\zmax$ for the SN data remains weak.

In practice, the impact of the maximum SN redshift on dark energy
constraints will depend crucially on the behavior of systematic errors.
We have assumed in our forecasts here
that the error per redshift bin stays constant as the maximum SN redshift
increases, but in reality higher redshift SNe are likely to have larger 
systematic errors associated with them, which would diminish the gains 
from high redshift SNe even more than indicated by the flattening of
curves in Figure~\ref{fig:fom}.
However, once the systematic errors at $z < 0.8$ are saturated, then
pushing to higher redshift may be the only way to continue improving
the SN constraints.  The gain from the higher redshift SNe then depends
on whether their systematics are {\it uncorrelated} with those at
lower redshift, so that they indeed provide new information.
While there has been considerable recent progress in understanding
and accounting of systematic errors in SN cosmology, there has
been little exploration to date of the correlation of systematics
across redshift bins.  The correlation of systematics may vary
with details of experimental design (e.g., flux calibration), and
it also depends on aspects of the Type Ia supernova population that
are, as yet, poorly understood (e.g., whether there is a mix of
single-degenerate and double-degenerate progenitors that changes
with redshift).
To optimize a specific experiment, one must assess both the
expected behavior of systematics and the observing time required
to discover SNe at different redshifts and to measure them
with adequate photometric and spectroscopic precision.
The SDT report for \wfirst\ \citep{green12} provides a worked
example: with a two-tier strategy (shallow wide fields and
narrow deep fields), the CMB+SN FoM increases steadily as the
maximum redshift is increased from 0.8 to 1.7 at (roughly) fixed
observing time, assuming systematics
that are uncorrelated among redshift bins.
However, reducing the systematics by a factor of two
(from $\approx 0.02$ mag per $\Delta z = 0.1$ bin to
$\approx 0.01$ mag) has a larger impact than raising
$\zmax$ from 0.8 to 1.7.  The contrast is less stark than in 
our Figure~\ref{fig:fom} because the reduction in {\it total}
error is less than a factor of two; with the smaller systematic
errors, the \wfirst\ DRM1 SN survey would be mainly statistics limited.

The behavior in Figure~\ref{fig:fom} can be approximately
understood in terms of the aggregate measurement precision,
a notion we discuss at greater length in \S\ref{sec:forecast_aggregate}
below.  The local $(z=0.05)$ SN bin serves mainly to calibrate
the SN absolute magnitude, so in our fiducial program there
are three $\Delta z = 0.2$ redshift bins with cosmological
information.  
Increasing $\zmax$ to 1.6 changes the number of
non-local bins from three to seven, 
improving aggregate precision by $\sim \sqrt{7/3}$,
and the impact on the FoM is roughly half the impact of
reducing errors by a factor of two while retaining
$\zmax = 0.8$.  If we increase $\zmax$ to 1.6 but 
simultaneously inflate the errors of the non-local
bins by $\sqrt{7/3}$, thus keeping the aggregate precision
of the $z>0.1$ measurements fixed, then the
FoM rises to 749, a 13\% improvement over the fiducial
case, vs.\ a 26\% improvement if we increase $\zmax$ to
1.6 at constant per-bin error.
In this sense, roughly half of the improvement when
extending the redshift limit comes from tightening the
aggregate statistical precision by adding new bins, and
half the improvement comes from the greater leverage
afforded by a wider redshift range.  A similar calculation
for $\zmax=1.2$ (where the corresponding FoM improvements over the 
fiducial case are 9\% and 17\%) leads to the same conclusion.
Ultimately, however, the trade between extending the redshift
range of a SN survey vs. improving the observations at lower
redshift depends on aspects of observational and evolutionary
systematics that are still poorly understood.
This remains an important issue for near-term investigation
with the much more comprehensive data sets that are now
becoming available.

%Dropping $\zmax$ to 0.4 leaves only a single
%non-local redshift bin, and this three-fold reduction is almost but
%not quite as damaging as doubling the errors relative to the
%fiducial case.  

\subsubsection{Constraints on structure growth parameters}
\label{sec:results_growth}

While the DETF FoM is a useful metric for studying the impact of 
variations in each of the dark energy probes, it does not tell the whole story.
Deviations from the standard model might show up in other sectors 
of the parameter space; for example, a detection of non-GR values for
the growth parameters $\Delta \gamma$ and $G_9$ could point to a 
modified gravity explanation for cosmic acceleration that would not 
be evident from measurements of $w(z)$ alone. Thus, even the less
optimistic version of the WL experiment, which adds relatively little to the 
$w(z)$ constraints obtained by the combination of fiducial SN, BAO, and 
CMB forecasts, is a critical component of a program to study cosmic 
acceleration because of its unique role in
determining the growth parameters $\Delta \gamma$ and $G_9$. 

\begin{figure}[ht]
\begin{centering}
{\includegraphics[width=3.2in]{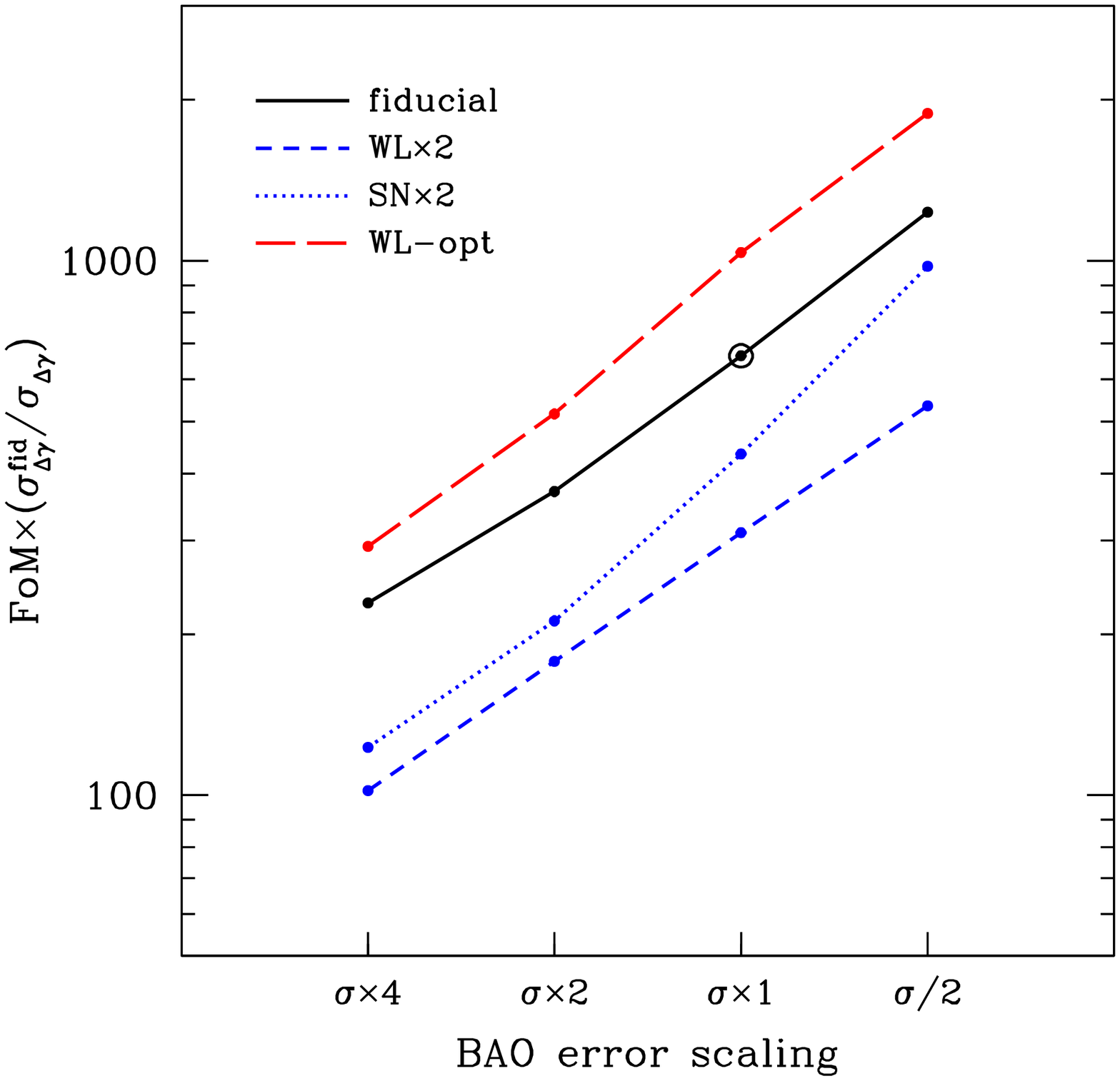}\hfill
\includegraphics[width=3.2in]{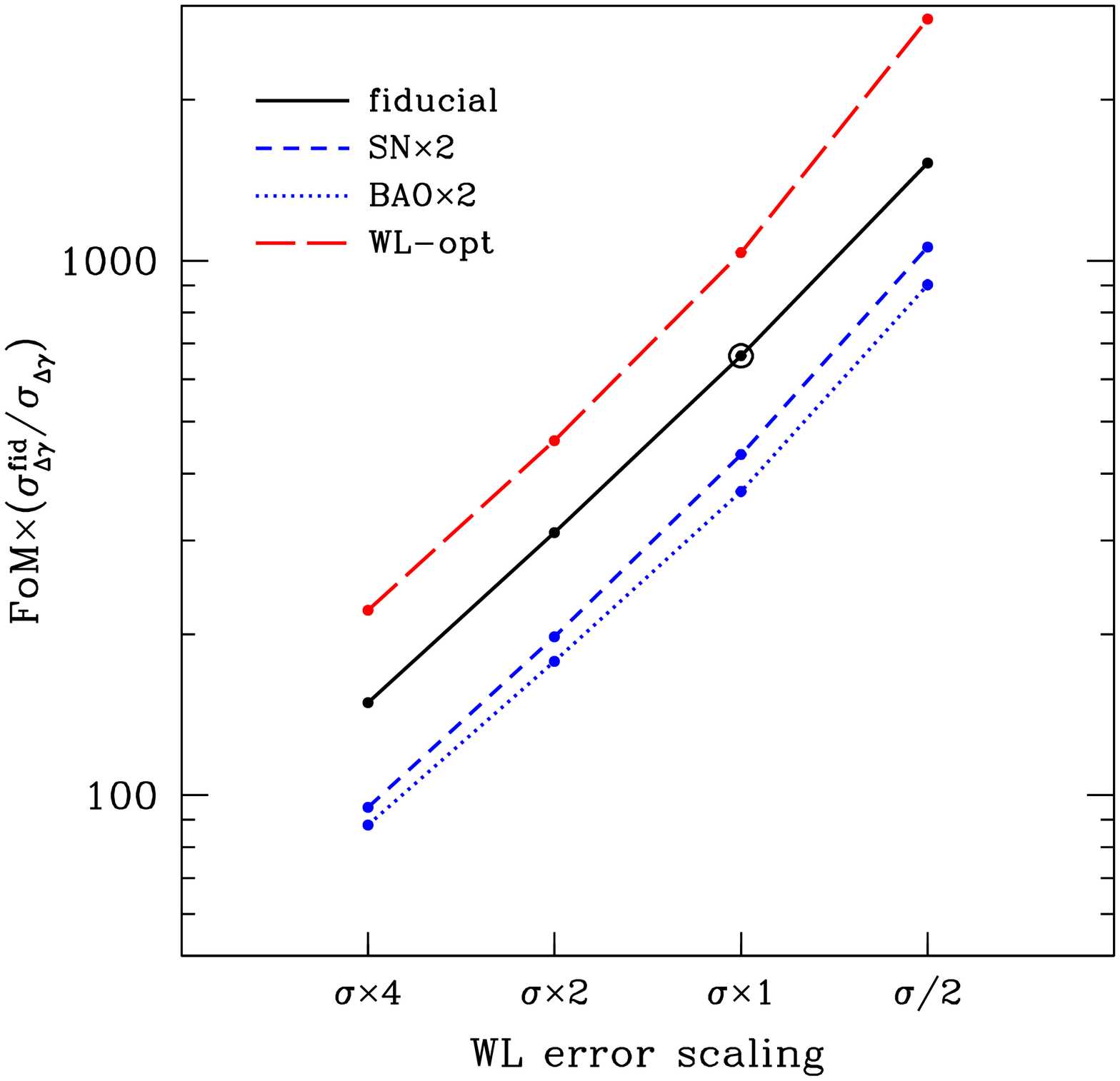}}
\caption{\label{fig:fom_gamma}
FoM scaling with BAO errors (left) and WL errors (right) including
changes in the error on $\Delta \gamma$, normalized to the forecast
uncertainty for the fiducial program, $\sigma_{\Delta \gamma}^{\rm fid}=0.034$.
The fiducial Stage IV forecast is marked by an open circle.
For the Stage III forecast, FoM$\times (\sigma_{\Delta \gamma}^{\rm fid}/
\sigma_{\Delta \gamma}) = 30$.
}
\end{centering}
\end{figure}

The impact of various experiments on the structure growth parameters 
is more evident if we extend the DETF FoM to include $\Delta \gamma$ 
in addition to $w_0$ and $w_a$. As shown in Figure~\ref{fig:fom_gamma}, 
the scaling of this new FoM with respect to WL errors (and, to a lesser 
extent, BAO errors) is much steeper than it is for the usual FoM
(Figure~\ref{fig:fom}). We do not show the scaling with SN errors or 
$\zmax$, since those assumptions do not affect the expected
uncertainties for $\Delta \gamma$ and $G_9$
(see Table~\ref{tbl:forecasts1}, lines 3--12).
One could also consider versions of the FoM that
include uncertainties in $G_9$ and that account for the correlations 
between the structure growth parameters and the dark energy equation of state.

\begin{figure}[th]
\begin{centering}
\includegraphics[width=4.5in]{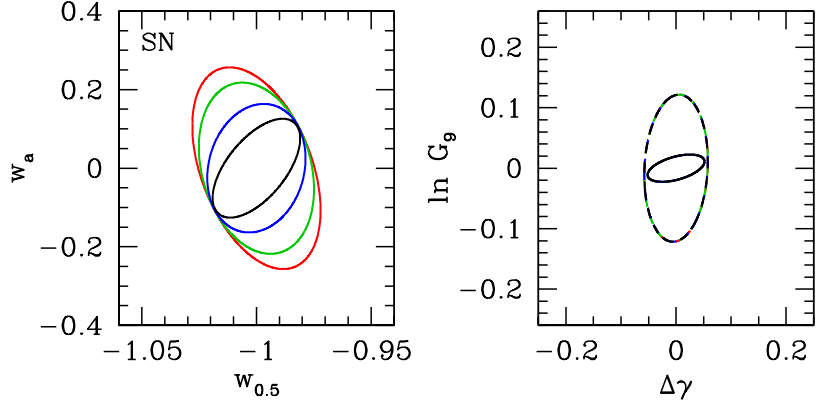}
\caption{\label{fig:contours_sn}
Forecast constraints
(68\% confidence levels) for dark energy and growth parameters, 
varying errors on SN data: fiducial$\times 4$ (red), $\times 2$ (green), 
$\times 1$ (blue), and $/2$ (black). 
In all cases, the fiducial forecasts are used for the other probes 
(BAO, WL, CMB). Contours in the left panel use the value of 
the equation of state at $z=0.5$ (close to the typical pivot redshift), 
$w_{0.5} = w_0+w_a/3$. Dashed contours in the right panel 
show the errors on growth parameters for the binned $w(z)$ 
parameterization, with the default priors corresponding to deviations of 
$\lesssim 10$ in the average value of $w$.
Solid contours assume a $w_0$--$w_a$ parameterization.
In both cases, the $G_9$ and $\Delta\gamma$ constraints are essentially
independent of the SN errors.
}
\end{centering}
\end{figure}

\begin{figure}[th]
\begin{centering}
\includegraphics[width=4.5in]{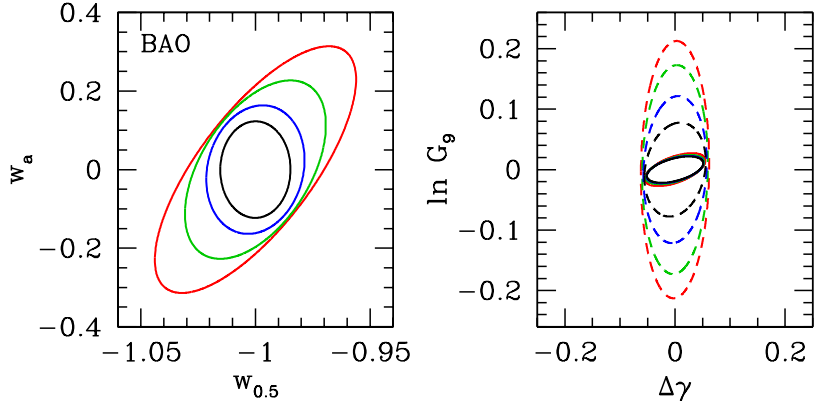}
\caption{\label{fig:contours_bao}
Same as Fig.~\ref{fig:contours_sn}, but varying BAO errors 
from fiducial$\times 4$ (red) to fiducial$/2$ (black).
}
\end{centering}
\end{figure}

\begin{figure}[th]
\begin{centering}
\includegraphics[width=4.5in]{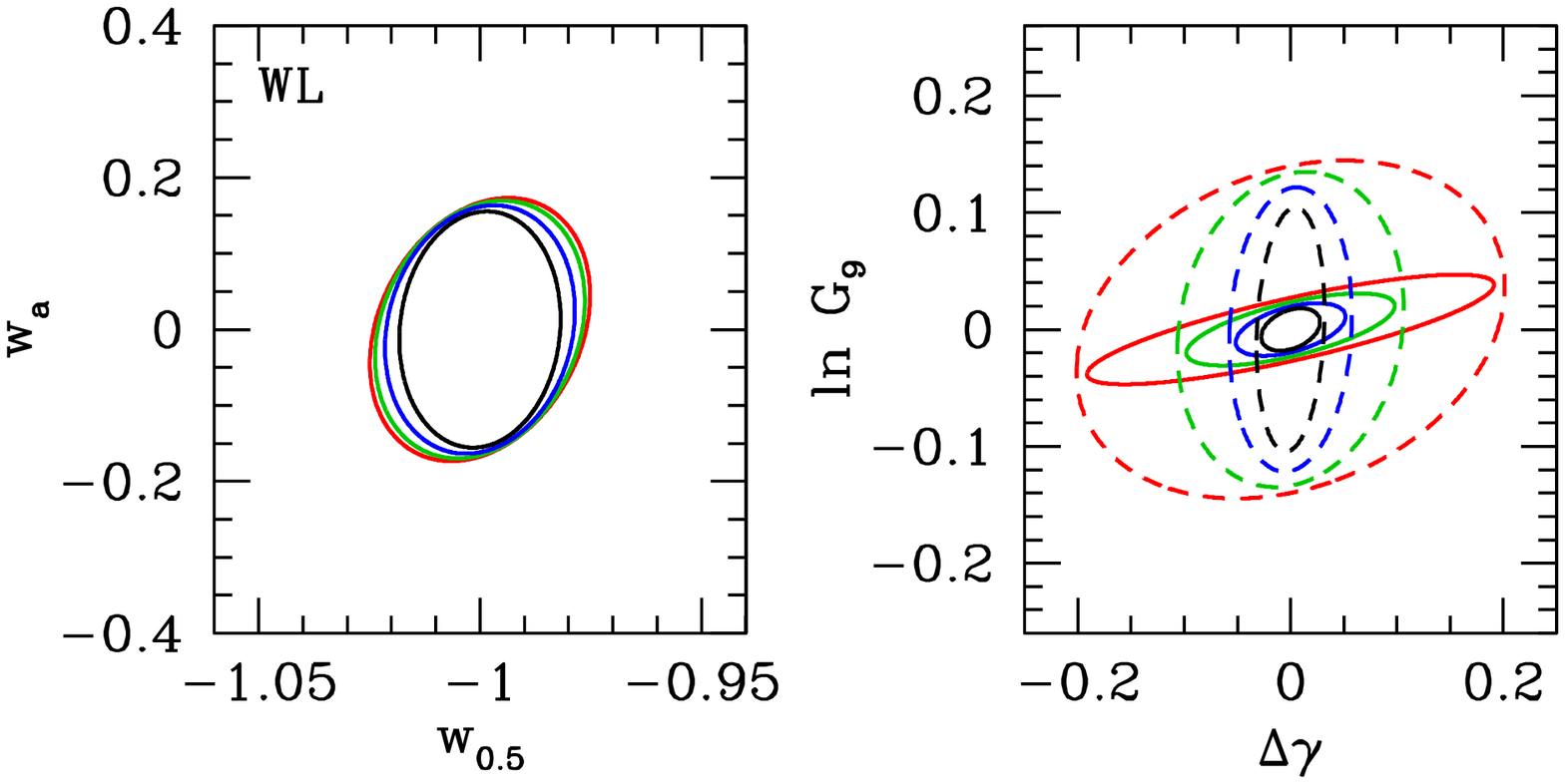}
\includegraphics[width=4.5in]{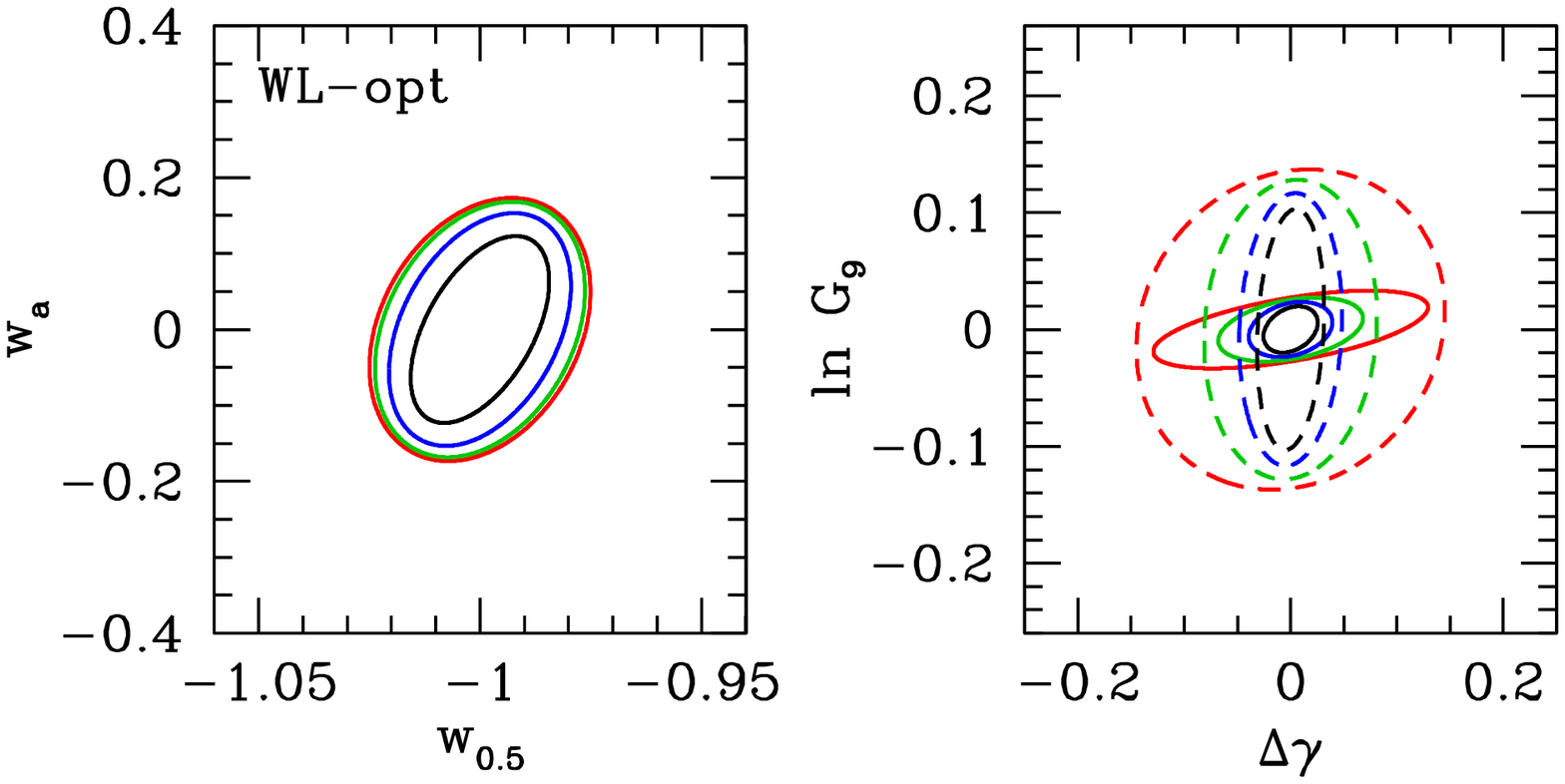}
\caption{\label{fig:contours_wl}
Same as Fig.~\ref{fig:contours_sn}, but varying WL errors 
from fiducial$\times 4$ (red) to fiducial$/2$ (black).
Lower panels assume the optimistic WL forecasts.
}
\end{centering}
\end{figure}

The complementarity between the SN, BAO, and WL techniques is 
further demonstrated 
in Figures~\ref{fig:contours_sn}--\ref{fig:contours_wl}, 
which show the forecast 68\% confidence level contours 
in the $w_{0.5}$--$w_a$ and $\Delta \gamma$--$\ln G_9$
planes after marginalizing over other parameters. 
Instead of $w_0$
we plot $w_{0.5}$, the 
equation-of-state parameter at $z=0.5$, because it is much
less correlated with $w_a$ for most of the forecast scenarios.
In every panel, the blue ellipse shows the error contour of the
fiducial forecast while other ellipses show the effect of
varying the errors of the indicated method.
The opposite orientation of ellipses in 
Figures~\ref{fig:contours_sn} and~\ref{fig:contours_bao} demonstrates
the complementary sensitivity of SN and BAO to $w(z)$: 
the SN data are mainly sensitive to the equation 
of state at low redshift, whereas BAO data measure the equation of state 
at higher redshift. However, the sensitivity to 
the beyond-GR growth parameters comes entirely from WL data,
which provide the only direct measurements of growth, and 
the strength of the $\Delta\gamma$ and $G_9$ constraints
depends directly on the WL errors, as shown in Figure~\ref{fig:contours_wl}.
Conversely, these constraints are very weakly sensitive to the
SN or BAO errors (Figs.~\ref{fig:contours_sn} and~\ref{fig:contours_bao}),
showing that the uncertainties are dominated by the growth measurements
themselves rather than residual uncertainty in the expansion history.
Inspection of Table~\ref{tbl:forecasts1} shows that the $\Delta\gamma$
constraints are essentially linear in the WL errors, while the 
$\ln G_9$ constraints scale more slowly.

Although the $w_0$--$w_a$ parameterization is flexible enough to describe
a wide variety of expansion histories, it is too simple to account for 
all possibilities; in particular, $w(z)$ is restricted to functions that 
are smooth and monotonic over the entire history of the universe.
Because many cosmological parameters are partially degenerate with the 
dark energy evolution, assumptions about the functional form of $w(z)$ 
can strongly affect the precision of constraints on other parameters.
As an example of this model dependence, the 
right panels of Figures~\ref{fig:contours_sn}--\ref{fig:contours_wl}
show how the constraints on the growth parameters weaken (dashed curves)
if one allows the 36 binned $w_i$ values to vary independently instead 
of assuming that they conform to the $w_0$--$w_a$ model. While 
$\Delta \gamma$ forecasts are only mildly affected by the choice of 
dark energy modeling, constraints on the $z=9$ normalization parameter 
$G_9$ depend strongly on the form of $w(z)$.
This dependence follows from the absence of data probing redshifts
$3\lesssim z < 9$ in the fiducial Stage IV program. In the $w_0$--$w_a$
model, dark energy evolution is well determined even at high redshifts,
since the two parameters of the model can be measured from data at $z<3$, 
and thus the growth function at $z=9$ is closely tied to the low redshift
growth of structure measured by WL. 
However, allowing $w(z)$ to vary independently at high redshift 
where it is unconstrained by data decouples the low and high redshift 
growth histories, and therefore $G_9$ can no longer be determined precisely.
In fact, the constraints on $G_9$ in that case depend greatly on the 
chosen prior on $w_i$ (taken to be the default prior of 
$\sigma_{w_i}=10/\sqrt{\Delta a}$
in Figures~\ref{fig:contours_sn}--\ref{fig:contours_wl}).
One important consequence of this dependence on the $w(z)$ model is
that an apparent breakdown of GR via $G_9 \neq 1$ might instead be 
a sign that the chosen dark energy parameterization is too restrictive.

\subsubsection{Dependence on $w(z)$ model and binning of data}
\label{sec:results_modeldep}

\begin{figure}[th]
\begin{centering}
{\includegraphics[width=3.in]{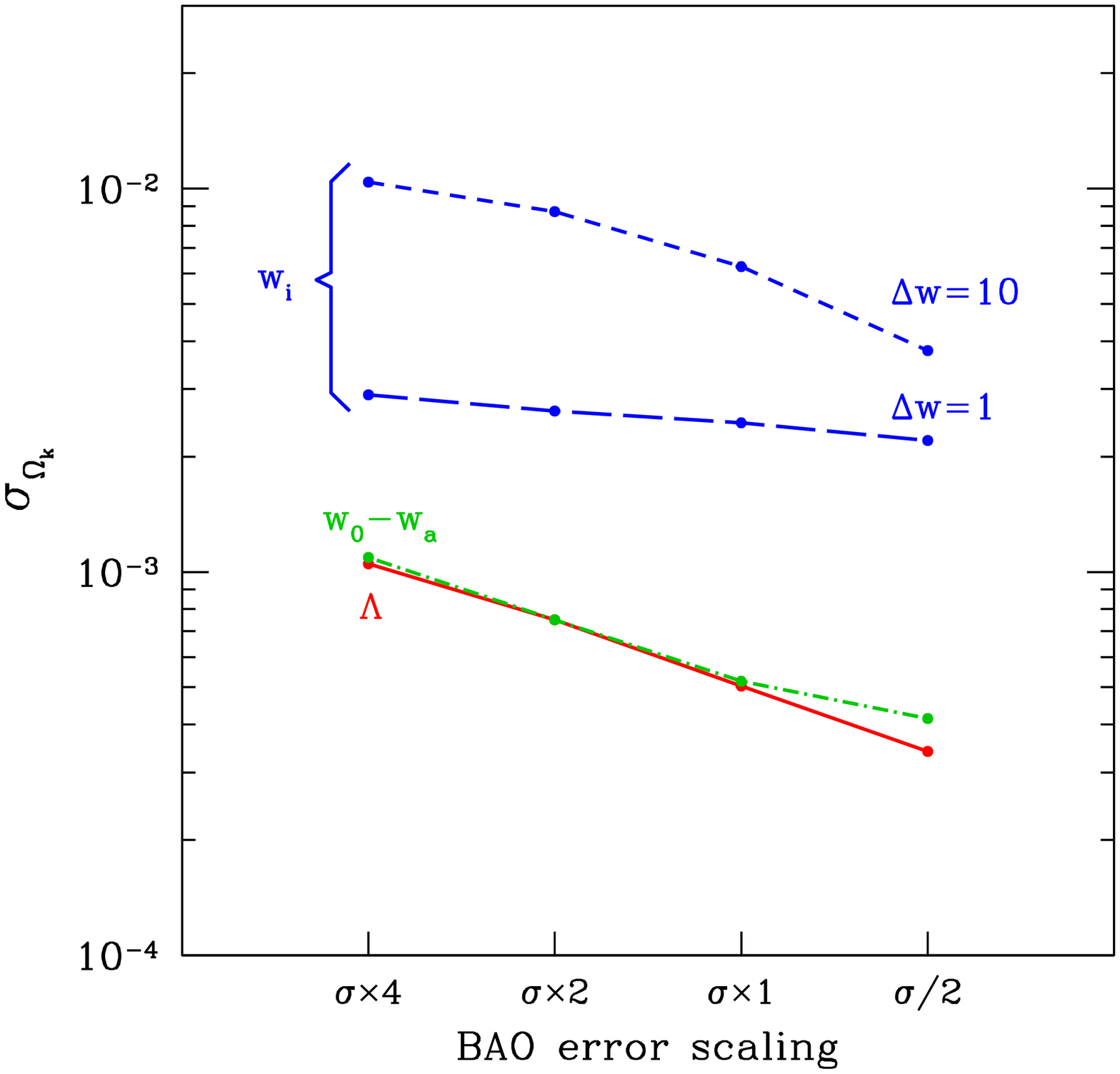}\hskip 0.1in
\includegraphics[width=3.in]{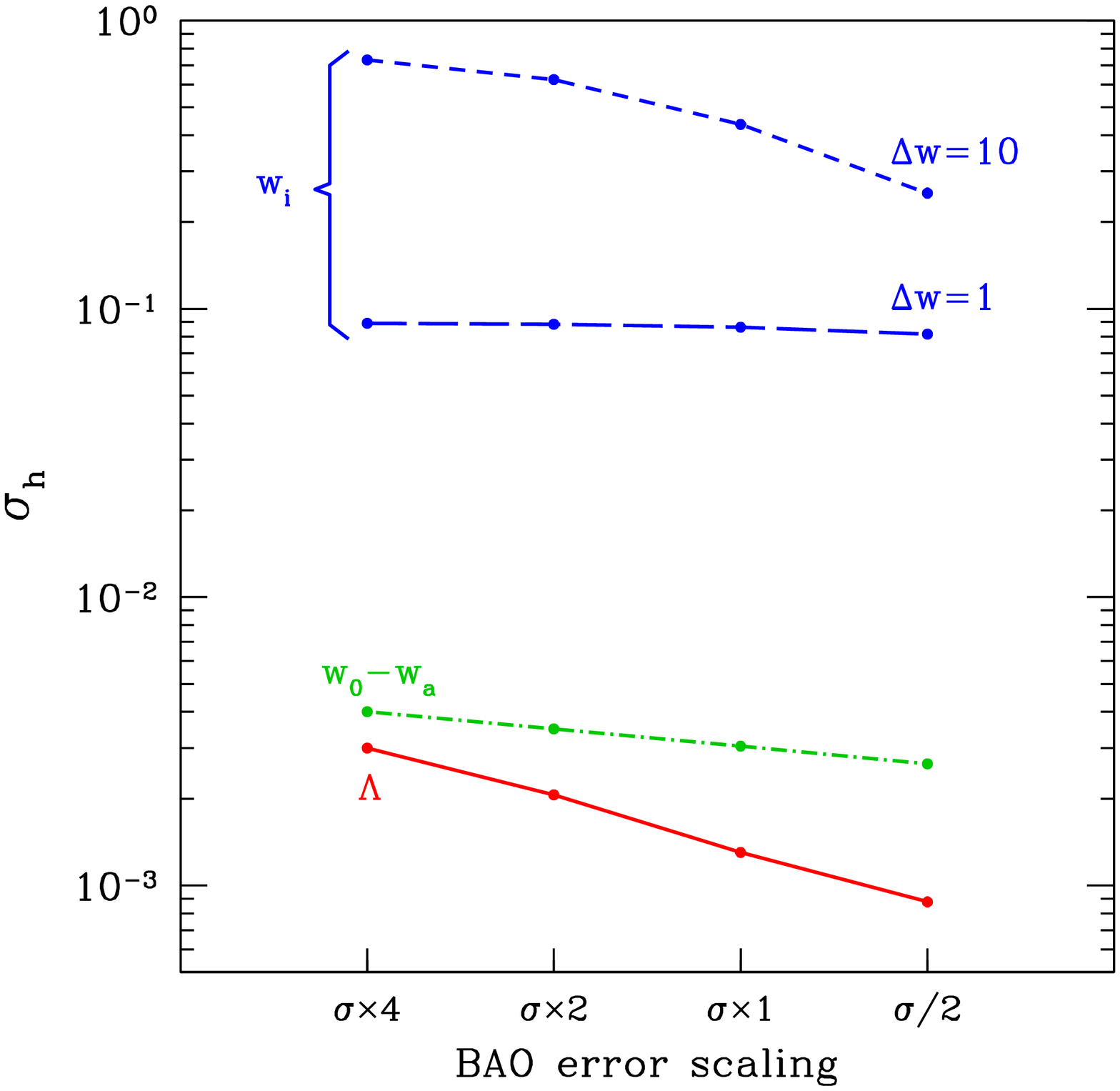}}
\caption{\label{fig:h_ok_model_bao}
Dependence of $\sigma_{\ok}$ (left) and $\sigma_h$ (right)
on BAO errors for various dark energy 
parameterizations and priors. For the $w_i$ curves, the equation of 
state varies independently in 36 bins with Gaussian priors of 
width $\sigma_{w_i}=\Delta w / \sqrt{\Delta a}$.
The fiducial versions of the Stage IV SN, WL, and CMB data are included
in all cases.
}
\end{centering}
\end{figure}

Other parameters are also affected to varying degrees by the choice of 
$w(z)$ model and the priors on the model parameters. 
Figure~\ref{fig:h_ok_model_bao} shows how errors on $\ok$ and $h$
are affected by relaxing assumptions about dark energy evolution.
For the fiducial program and minor 
variants, $\ok$ is very weakly correlated with $w_0$ and $w_a$, resulting
in similar errors on curvature for the $w_0$--$w_a$ and $\Lambda$CDM models.
However, generalizing the dark energy parameterization to include
independent variations in 36 redshift bins can degrade the precision 
of $\ok$ measurements by an order of magnitude or more. In that case, 
the error on $\ok$ is very sensitive to the chosen prior on the value 
of $w_i$ in each bin, and it improves little as the BAO errors decrease. 
This dependence on priors reflects the fact that curvature is 
most correlated with the highest redshift $w_i$ values, which are poorly
constrained by the fiducial combination of data.
Relative to curvature, constraints on the Hubble constant are 
affected more by the choice of dark energy parameterization but less
by priors on $w_i$ in the binned $w(z)$ model.

\begin{figure}[th]
\begin{centering}
\includegraphics[width=3.in]{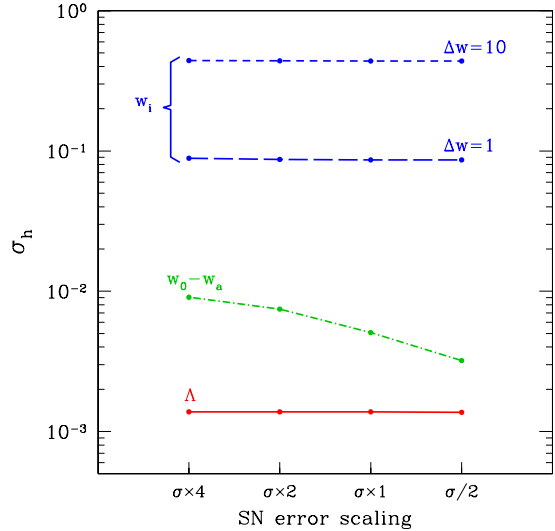}
%{\includegraphics[width=3.in]{h_sn_c4_snlocal.eps}\hskip 0.1in
%\includegraphics[width=3.in]{h_sn_c16_snlocal.eps}}
\caption{\label{fig:h_model_sn}
Dependence of $\sigma_h$ on SN errors for various dark energy 
parameterizations and priors, including the fiducial BAO, WL, and CMB forecasts.
%Constraints in the left panel assume the ``SNc3'' forecast with 
%3 redshift bins at $0.2<z<0.8$ including a correlated systematic error
%between bins with correlation length $\Delta z_c = 0.2$ (see 
%eq.~\ref{eqn:covsn} and Table~\ref{tbl:key} for details). The right panel 
%uses the ``SNc12'' forecast, which uses the same redshift correlation model
%but increases the number of bins at $0.2<z<0.8$ to 12.
}
\end{centering}
\end{figure}

Figure~\ref{fig:h_model_sn} shows the dependence of $\sigma_h$ on 
the precision of SN data for various dark energy parameterizations 
($\sigma_{\ok}$ is nearly independent of the SN errors for this range of
variations around the fiducial forecast; see Table~\ref{tbl:forecasts1}).
If we assume a $w_0$--$w_a$ model for dark energy, Hubble constant errors
strongly depend on the precision of SN data. However, 
Fig.~\ref{fig:h_model_sn} shows that either decreasing or increasing
the number of dark energy parameters can almost completely eliminate the 
dependence of $\sigma_h$ on the SN data. In the case of the simpler 
$\Lambda$CDM model, the combination of the fiducial BAO, WL, and CMB 
forecasts is sufficient to precisely determine all of the model parameters, 
and adding information from SN data has a negligible effect on 
the parameter errors. Adding $w_0$ and $w_a$ to the model
introduces degeneracies between these dark energy parameters and other
parameters, including $h$. Since constraints from SN data help to break 
these degeneracies, reducing SN errors can significantly improve measurement
of the Hubble constant in the $w_0$--$w_a$ model.

As one continues to add
more dark energy parameters to the model, the degeneracies between these
parameters and $h$ increase, but another effect arises that 
diminishes the impact of SN data on $\sigma_h$.
Measurement of the Hubble constant requires relating observed quantities at
$z>0$ (e.g.\ SN distances) to the expansion rate at $z=0$.
In the case of $\Lambda$CDM or the $w_0$--$w_a$ model, the assumed 
dark energy evolution is simple enough that this relation between $z=0$ and
low-redshift observations is largely set by the model. 
However, when we specify $w(z)$ by a large number of independent bins in 
redshift, this relation must instead be determined by the data.
%, which requires
%that the data have sufficient resolution in redshift near $z=0$.
%For the default SN bins of width 0.2 in redshift, nearly 7 of the 36 $w_i$ bins
%fall within the first bin at $0<z<0.2$. Consequently, changes in the 
Since SN data are only sensitive to relative changes in distances,
the lowest-redshift $w_i$ value (centered at $z\approx 0.01$) 
is strongly degenerate with $h$ \citep{mortonson09b}.
%which is strongly degenerate 
%with $h$, have little effect on the SN data \citep{mortonson09b}.
%Dividing SNe into finer redshift bins makes it harder for variations in 
%the lowest-redshift $w_i$ to remain hidden in the SN data, thus restoring
%some of the sensitivity of SN data to the Hubble constant (see the right panel 
%of Figure~\ref{fig:h_model_sn}). 
This degeneracy is partially broken by the local SN sample at $z=0.05$: 
removing it from the forecasts 
increases the error on $h$ from 0.44 to 0.48 in the 
binned $w(z)$ parameterization, and from 0.0051 to 0.0085 in the 
$w_0$--$w_a$ model.
SNe at even lower redshifts are more sensitive to the Hubble constant, 
but they also have larger systematic uncertainties due to peculiar velocities.

%\tbd{Looking at the numbers from the current version of the code,
% I think the statements in this paragraph are wrong.}
%
%Note that because BAO observations are 
%tied to the distance scale of the CMB, they retain their sensitivity
%to $h$ even in the absence of low-redshift constraints. 
%For example, if we drop the 3 lowest-redshift BAO bins ($z<0.231$) entirely 
%from the fiducial forecast with binned $w(z)$ 
%(see Table~\ref{tbl:baoforecast}), 
%then the change in $\sigma_h$ as BAO errors change from half to 
%four times the fiducial errors is $0.22 \to 0.44$, compared to
%$0.14 \to 0.42$ when the three lowest redshift bins are included.
%On the other hand,
%SN data only contribute to measurements of $h$ in the most general
%dark energy parameterizations if they have 
%a low redshift component that can be accurately compared with the 
%sample of SNe at higher redshifts.

%While fine binning in redshift is not {\it necessary} for BAO data to 
%contribute to constraints on general dark energy models, it can help.
For BAO data, the choice of redshift bin width affects
forecasts for models with general equation-of-state variations.
Measurements of $H(z)$ and $D(z)$ in narrower bins are better able to constrain
rapid changes in $w(z)$. They can also reduce uncertainty in the Hubble 
constant by about a factor of two, and in other parameters such as $\Omega_K$,
$\ln G_9$, and $\Delta \gamma$ by a smaller amount, 
relative to measurements in wide bins.  However, in practice one cannot
reduce the bin size indefinitely, since each bin must contain enough 
objects to be able to robustly identify and locate the BAO peak; 
for example, requiring that the bin be at least wide enough to contain 
pairs of objects separated by $\sim 100\hmpc$ along the line of sight 
sets a lower limit of $\Delta z/(1+z)\gtrsim 0.03$. 
We do not attempt to optimize the choice of bins for the simplified 
forecasts in this section, but we note that binning schemes in 
analyses of BAO data aimed at constraining general $w(z)$ variations 
should be chosen with care to avoid losing information about dark 
energy evolution and other parameters. Similar concerns are likely to 
apply for WL data as well.

\subsubsection{Constraints on $w(z)$ in the general model}
\label{sec:results_wzbin}

So far, in the context of general dark energy evolution we have only 
considered the forecast errors on parameters such as $h$ and $\Omega_K$ 
that are partially degenerate with $w(z)$. But how accurately can 
$w(z)$ itself be measured when we do not restrict it to specific 
functional forms? Since the errors on $w_i$ values in different bins 
are typically strongly correlated with each other, it is not very useful 
to simply give the expected $w_i$ errors, marginalized over all other 
parameters. Instead, we can consider combinations of the $w_i$ that 
are independent of one another and ask how well each of these 
combinations can be measured by the fiducial program of observations.

\begin{figure}[th]
\begin{centering}
\includegraphics[width=4.in]{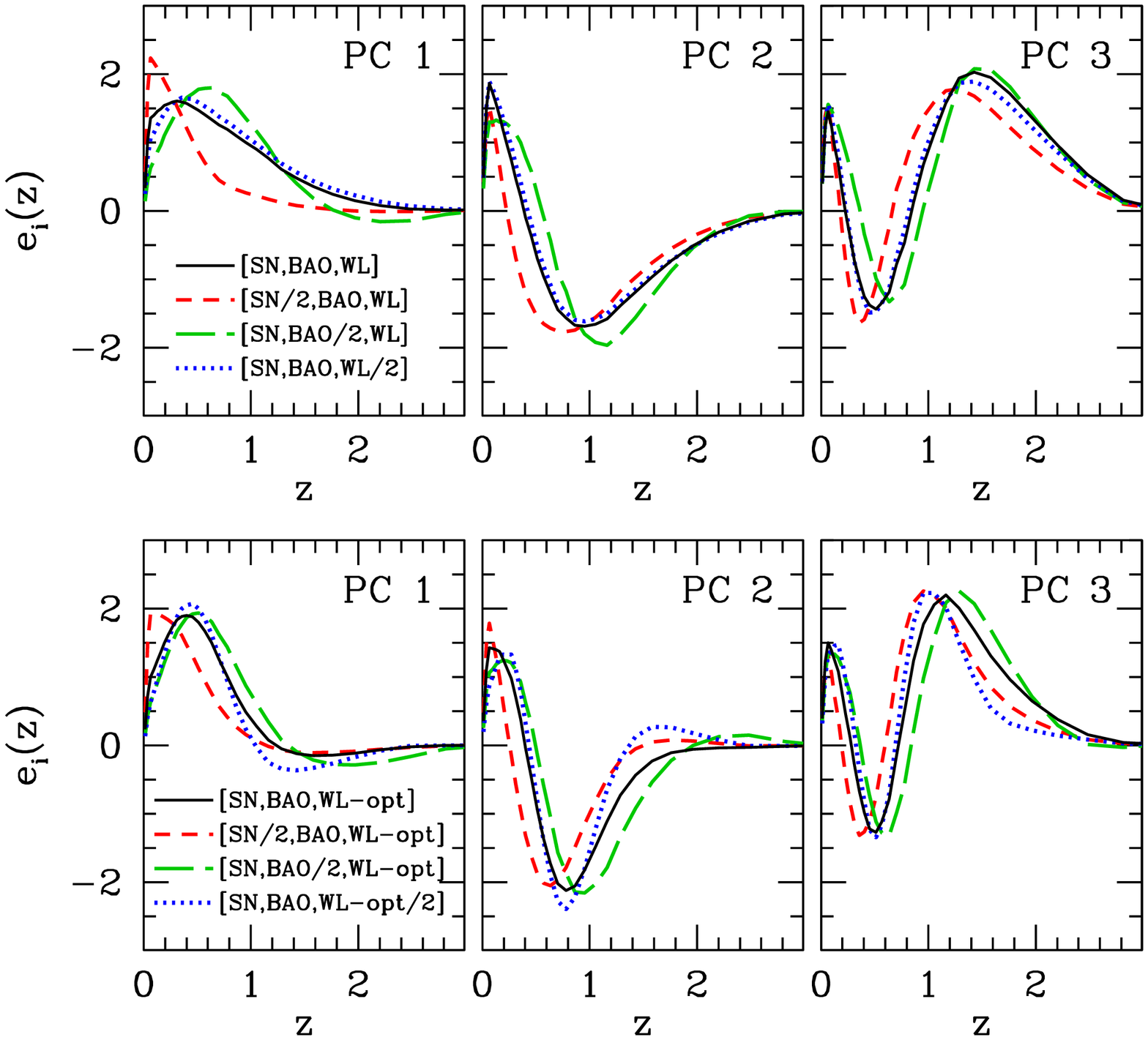}
\caption{\label{fig:pc3}
The three best-measured PCs for the fiducial program (solid curves)
and from programs with SN, BAO, or WL 
errors halved (as labeled). 
The top row uses the fiducial version of the WL forecast, while
the bottom row uses the optimistic WL forecast with reduced systematic
errors. Although not indicated in the plot legends, all forecasts here
include the default \planck\ CMB Fisher matrix. For all PCs shown here, 
$e_i(z)$ is nearly zero for $3<z<9$.
}
\end{centering}
\end{figure}

As mentioned in \S\ref{sec:parameterizations}, many methods for combining
$w(z)$ bins into independent (or nearly independent) components have 
been proposed. Here we adopt the principal component (PC) 
decomposition of the dark energy equation of state.
Starting from the Fisher matrix for the combined acceleration probes, 
the PCs are computed by first marginalizing the Fisher matrix over everything
except for the $w_i$ parameters and then diagonalizing the remaining matrix,
as described above in \S\ref{sec:forecasting}. The shapes of the three
best-measured PCs for the fiducial program (with both fiducial and
optimistic WL assumptions) and some simple variations are plotted in 
Figure~\ref{fig:pc3}. In general, the structure of the PCs is similar
in all cases; for example, the combination of $w_i$ that is most tightly 
constrained is typically a single, broad peak at $z<1$, while the next 
best-determined combination is the difference between $w(z\sim 0.1)$ and 
$w(z\sim 1)$. However, variations in the forecast assumptions slightly 
alter the shape of each PC and, in particular, shift the redshifts at 
which features in the PC shapes appear. Changes in the location of the 
peak in the first PC mirror the dependence of the pivot redshift $z_p$
for the $w_0$--$w_a$ model in 
Tables~\ref{tbl:forecasts1}--\ref{tbl:forecasts3}, with improved SN data
decreasing the peak redshift and improved BAO data increasing it.
The direction and magnitude of these shifts reflects the redshift range
that a particular probe is most sensitive to and the degree to which that
probe contributes to the total constraints on $w(z)$.
Note that so far we have only considered the impact of forecast assumptions
on the functional form of PCs, and not on the precision with which each PC
can be measured.  In general, altering the forecast model changes both 
the PC {\it shapes} and PC {\it errors}, which 
complicates the comparison among
expected PC constraints from different sets of forecasts.

Comparing the top and bottom rows of panels in Figure~\ref{fig:pc3}, 
we see again the contrast between the fiducial WL forecast and the ``WL-opt''
forecast with reduced systematic errors. In the former case, decreasing
WL errors by a factor of two has a negligible effect on the PC shapes
relative to similar reductions in SN or BAO errors. However,
when we take WL-opt as the baseline forecast the PCs depend more on 
the precision of WL measurements and less on that of the SN or BAO data.

\begin{figure}[p]
\begin{centering}
\includegraphics[width=5.in]{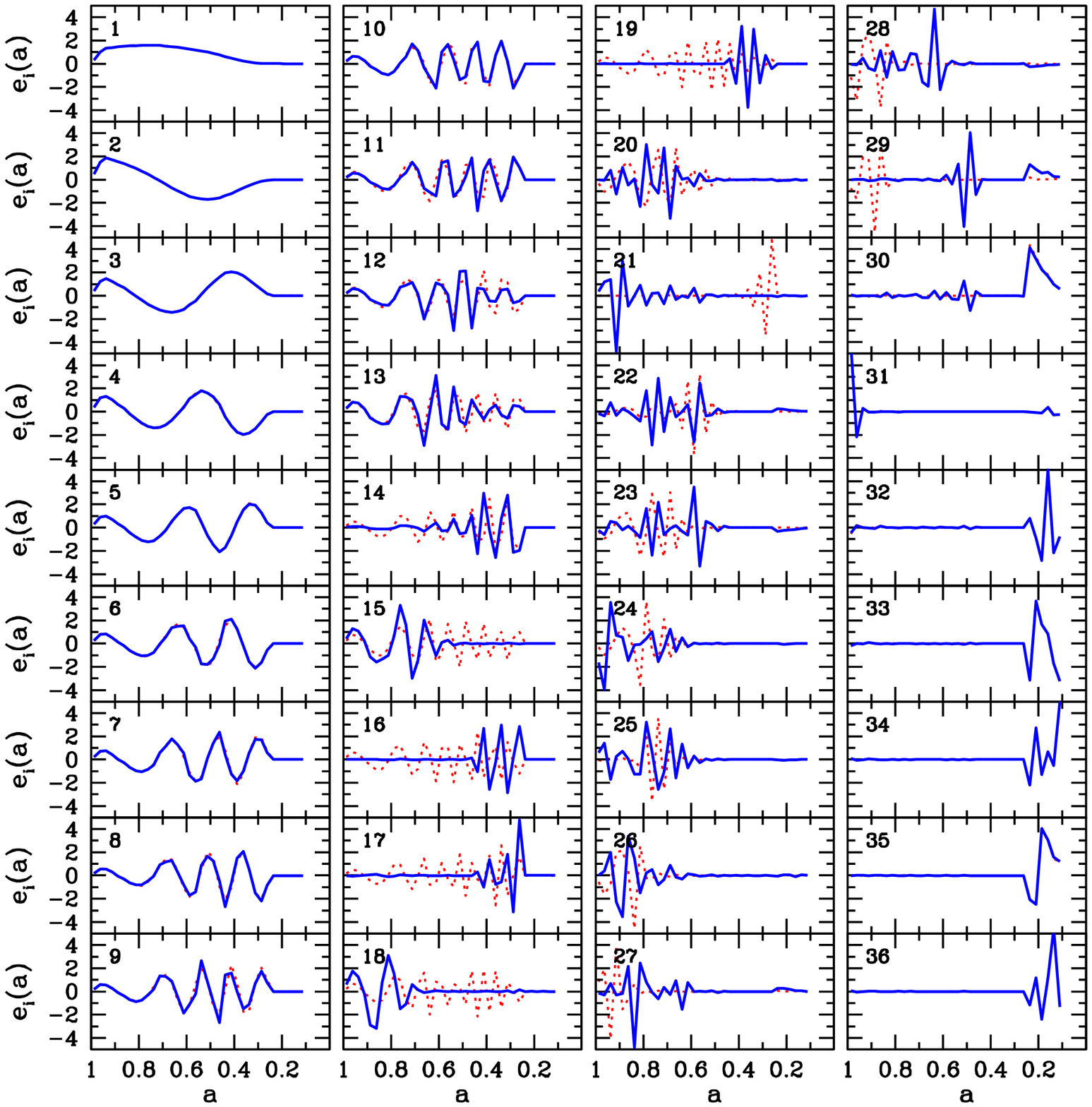}
\caption{\label{fig:pcfid}
PCs for the fiducial program (solid blue curves). Dotted red curves double 
the number of bins used for BAO data from the default choice of 20 to 40.
}
\end{centering}
\end{figure}

\begin{deluxetable}{rrr|rrr|rrr|rrr}
\tablecolumns{12}
\tablecaption{Errors on PC Amplitudes for the Fiducial Program\label{tbl:pcerrors}}
\tablehead{
$i$ & $\sigma_i^{\rm fid}$ & $\sigma_i^{\rm opt}$ & $i$ & 
$\sigma_i^{\rm fid}$ & $\sigma_i^{\rm opt}$ & $i$ & $\sigma_i^{\rm fid}$ & 
$\sigma_i^{\rm opt}$ & $i$ & $\sigma_i^{\rm fid}$ & $\sigma_i^{\rm opt}$
}
\tablewidth{0pc}
\startdata
1 & 0.011 & 0.009 & 10 & 0.135 & 0.102 & 19 & 0.442 & 0.378 & 28 & 1.652 & 1.810 \\
2 & 0.017 & 0.014 & 11 & 0.143 & 0.116 & 20 & 0.779 & 0.413 & 29 & 2.285 & 2.217 \\
3 & 0.026 & 0.019 & 12 & 0.168 & 0.137 & 21 & 0.824 & 0.436 & 30 & 3.243 & 2.973 \\
4 & 0.038 & 0.026 & 13 & 0.180 & 0.150 & 22 & 0.939 & 0.531 & 31 & 6.540 & 6.785 \\
5 & 0.052 & 0.036 & 14 & 0.185 & 0.160 & 23 & 0.978 & 0.609 & 32 & 12.43 & 19.20 \\
6 & 0.067 & 0.047 & 15 & 0.216 & 0.179 & 24 & 1.212 & 0.725 & 33 & 16.59 & 24.78 \\
7 & 0.083 & 0.062 & 16 & 0.252 & 0.240 & 25 & 1.307 & 0.892 & 34 & 25.17 & 46.41 \\
8 & 0.099 & 0.074 & 17 & 0.310 & 0.244 & 26 & 1.457 & 1.036 & 35 & 59.32 & 94.09 \\
9 & 0.115 & 0.089 & 18 & 0.323 & 0.308 & 27 & 1.587 & 1.561 & 36 & 74.12 & 118.0 \\
\enddata
\tablecomments{
$\sigma_i^{\rm fid}$ refers to errors for the fiducial Stage IV program 
(CMB+SN+BAO+WL) and $\sigma_i^{\rm opt}$ to the optimistic WL case 
(CMB+SN+BAO+WL-opt).
}
\end{deluxetable}

\begin{deluxetable}{lccc}
\tablecolumns{4}
\tablecaption{Comparison of Figures of Merit for Selected Forecasts\label{tbl:pcfom}}
\tablehead{
Forecast case & 
$\log_{10} \prod_{i=1}^{36} \left(1+\sigma_i^{-2}\right)^{1/2}$ & 
$(\sum_{i=1}^{36} \sigma_i^{-2})^{1/2}$ & $[\sigma(w_p)\sigma(w_a)]^{-1}$
}
\tablewidth{0pc}
\startdata
$[$SN,BAO,WL,CMB$]$ & 20.2 & 124 & 664 \\
$[$SN/2,BAO,WL,CMB$]$ & 20.8 & 176 & 1197 \\
$[$SN,BAO/2,WL,CMB$]$ & 26.0 & 186 & 1222 \\
$[$SN,BAO,WL/2,CMB$]$ & 21.6 & 140 & 816 \\
$[$SN,BAO,WL-opt,CMB$]$ & 23.0 & 157 & 789 \\
$[$SN/2,BAO,WL-opt,CMB$]$ & 23.4 & 199 & 1242 \\
$[$SN,BAO/2,WL-opt,CMB$]$ & 27.9 & 205 & 1251 \\
$[$SN,BAO,WL-opt/2,CMB$]$ & 26.0 & 240 & 1397 \\
\enddata
\end{deluxetable}

The full set of PCs for the fiducial program is shown in 
Figure~\ref{fig:pcfid}, and the forecast errors on the PC amplitudes 
are listed in Table~\ref{tbl:pcerrors}.
The best-measured, lowest-variance PCs vary smoothly with redshift, 
corresponding to averaging $w(z)$ over fairly broad ranges in $z$. 
There is a clear trend of increasingly high frequency oscillations 
for higher PCs.  Visual inspection of Figure~\ref{fig:pcfid} shows
that the sum of the number of peaks and the number of troughs in
the PC is equal to the index of the PC, a pattern that continues
at least up to PC 13.  Higher PCs often change sign between
adjacent $z$ bins.
High frequency oscillations in $w(z)$ are
poorly measured by any combination of cosmological
data because the evolution of the dark energy {\it density},
which determines $H(z)$, depends on an integral of $w(z)$
(eq.~\ref{eqn:uphi}), and $D(z)$ and $G(z)$ depend
(approximately) on integrals of $H(z)$.  Rapid oscillations
in $w(z)$ tend to cancel out in these integrals.
Many of the most poorly-measured PCs depend on the 
chosen BAO binning scheme, since narrower BAO bins can better sample 
rapid changes in $w(z)$.  As an example, we show how
the PCs of the fiducial program are affected by doubling the number of 
BAO bins in Figure~\ref{fig:pcfid}.

The maximum redshift probed by SN, BAO, and WL data, primarily set by 
the highest-redshift BAO constraint at $z=3$ in our forecasts, imprints a
clear signature in the set of PCs in Figure~\ref{fig:pcfid}.
At high redshift, specifically $z>3$ ($a<0.25$), the first 29 PCs
have almost no weight.  Conversely, PCs 30 and 32-36
only vary significantly at high redshift
and are nearly flat for $z<3$; additionally, the errors on these PCs
are many times larger than those of the first 29 PCs.\footnote{Note that
our $w_i$ parameterization has exactly $(0.25-0.1)/0.025 = 6$ bins at 
$3<z<9$ and 30 bins at $z<3$.  PC 31 parameterizes variations in the
lowest redshift bin $w_1$, which is poorly constrained as discussed in
\S\ref{sec:results_modeldep}.}
Thus, $w(z)$ variations above and below $z=3$ are almost completely decoupled
from each other in the fiducial forecasts, and the high-redshift variations
are effectively unconstrained. CMB data limit the equation of state at 
$z>3$ to some extent, for example, through comparison of the measured distance
to the last scattering surface with the distance to $z=3$ measured in BAO data.
However, such constraints are very weak when split among several independent
$w(z)$ bins at high redshift. Furthermore, since the dark energy density 
typically falls rapidly with increasing redshift, variations in $w(z)$
at high redshift are intrinsically less able to affect observable 
quantities than low-redshift variations, resulting in reduced sensitivity 
to the high-redshift equation of state even in the presence of strong
constraints at earlier epochs. Likewise, variations in $w(z)$ at even 
higher redshifts of $z>9$, where we assume that $w$ is fixed to $-1$, 
are unlikely to significantly affect constraints on $w(z)$ at low 
redshift.\footnote{This partly depends on the choice of 
fiducial model at which the Fisher matrix used to construct the PCs
is computed. Taking a fiducial model with a larger dark energy density 
at high redshift than in $\Lambda$CDM makes the low-redshift PC shapes 
more sensitive to assumptions about the high-redshift equation of state 
(e.g., \citealt{deputter08}).}

\begin{figure}[ht]
\begin{centering}
{\includegraphics[width=3.in]{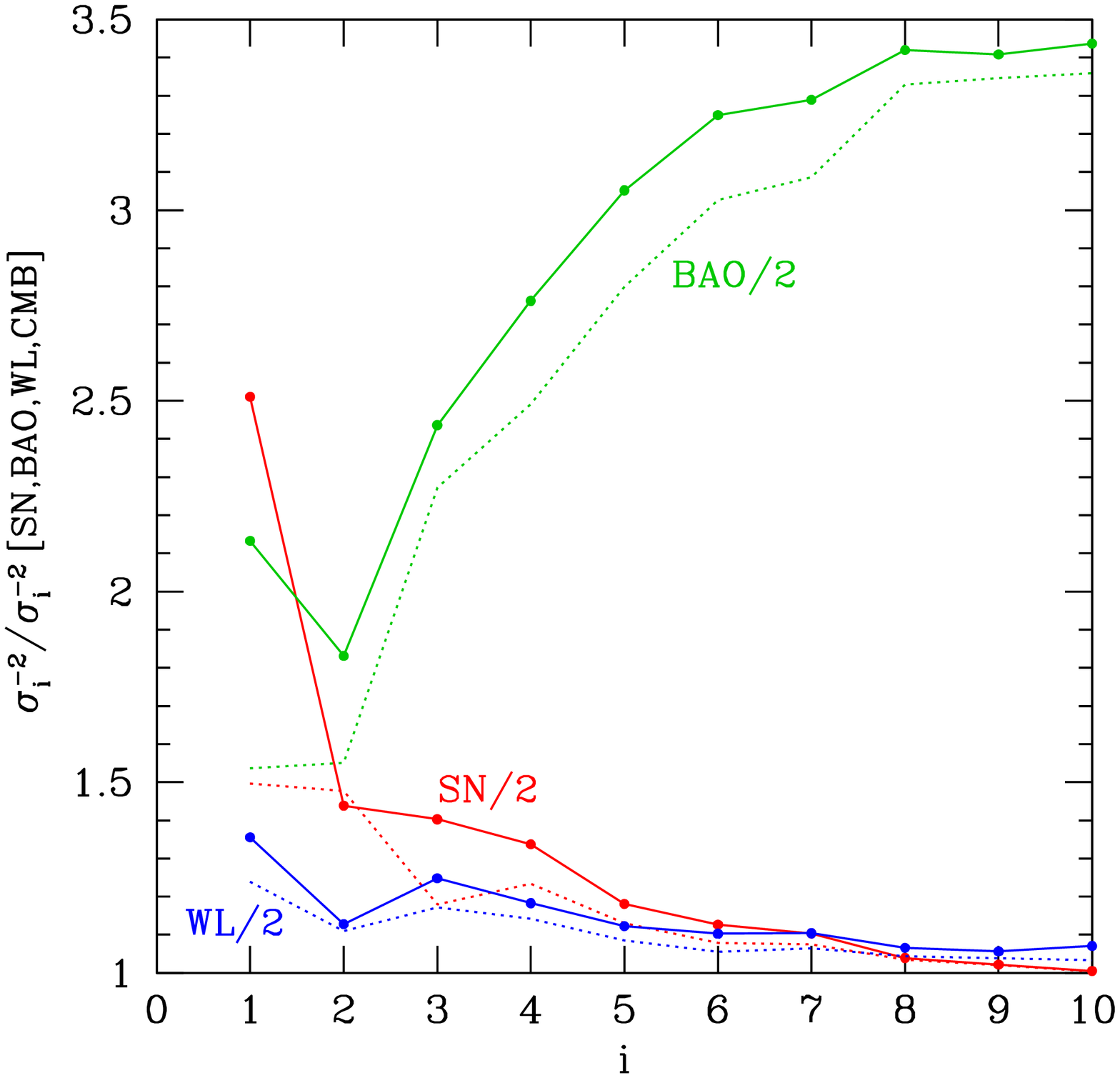}\hskip 0.1in
\includegraphics[width=3.in]{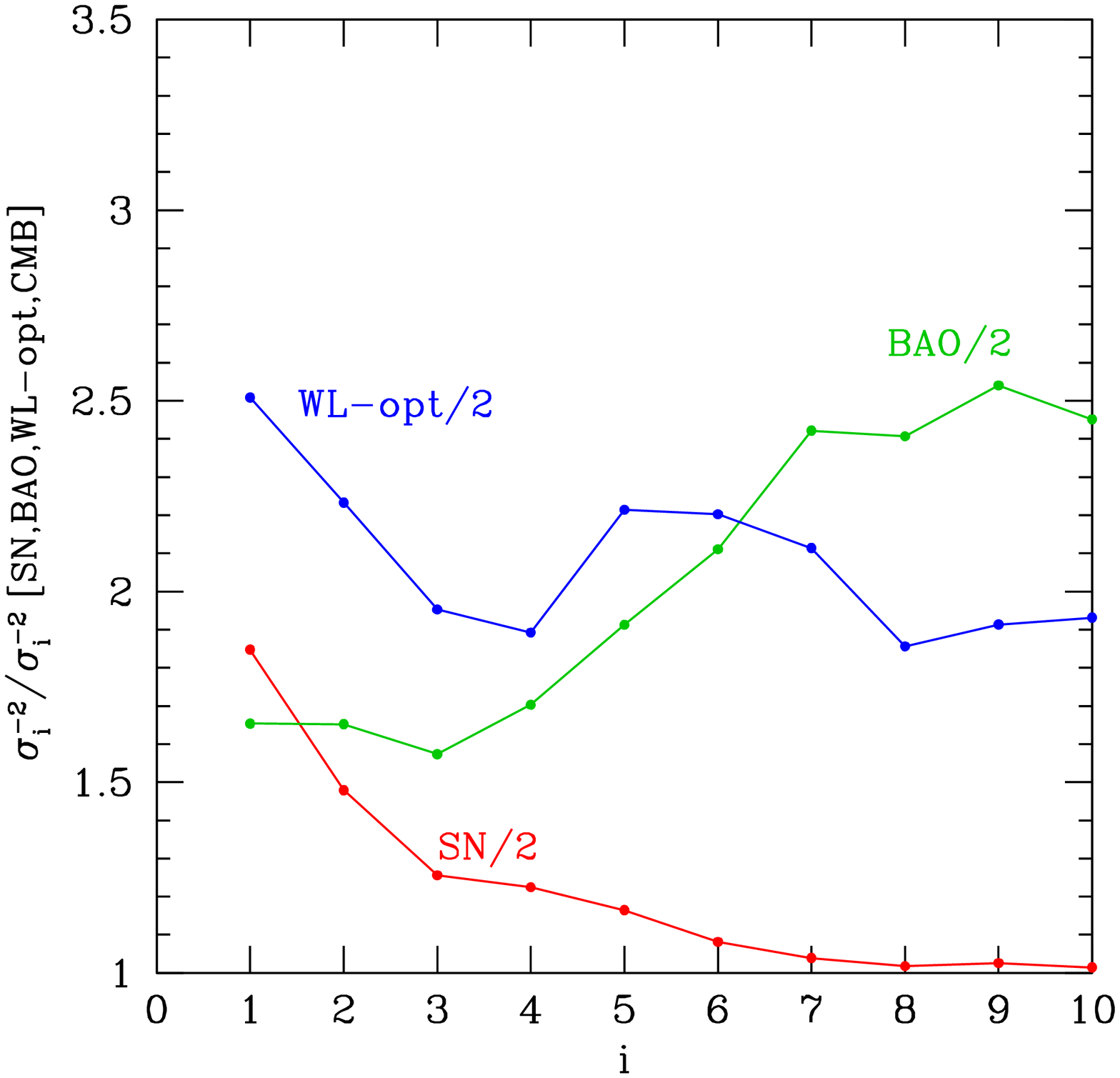}}
\caption{\label{fig:pcsigmaratio}
Ratios of inverse variances of PC amplitudes for variants of the 
fiducial program to the fiducial inverse variances (points and solid curves). 
Each variant divides
SN, BAO, or WL errors by a factor of 2 while keeping other probes 
fixed at the fiducial errors. The left panel assumes the default WL 
forecast and the right panel assumes the optimistic version.
Dotted curves in the left panel use $\hat{\sigma}_i$ instead of $\sigma_i$, 
which describes how well the amplitudes of the {\it fiducial} set of PCs
are expected to be measured by some variant of the fiducial forecast.
}
\end{centering}
\end{figure}

Figure~\ref{fig:pcsigmaratio} shows how the inverse variance $\sigma_i^{-2}$
of the 10 best-measured $w(z)$ PCs increases relative to the fiducial 
program if we halve 
the errors on the SN, BAO, or WL data.  Following \cite{albrecht09},
when computing these ratios $\sigma_{(2)i}^{-2}/\sigma_{(1)i}^{-2}$
(where 1 denotes the fiducial program and 2 the improved program),
we first limit PC variances to unity by making the substitution
$\sigma_i^{-2} \to 1+\sigma_i^{-2}$,
so that uninteresting improvements in the most poorly-measured 
PCs do not count in favor of a particular forecast.
We caution that, as noted earlier, the PC {\it shapes} 
themselves are changing as we change the errors assumed
in the forecast, so $\sigma_{(2)i}^2$ and $\sigma_{(1)i}^2$ are
not variances of identical $w(z)$ components.
However, as shown in Figure~\ref{fig:pc3}, these changes are
not drastic if we consider factor-of-two variations about our
fiducial program.

The differences in $\sigma_i^{-2}$ ratios among improvements in SN, BAO, 
and WL errors is striking.  Relative to the fiducial program, 
reduced SN errors mainly contribute to knowledge of the first few PCs. 
For the fiducial WL systematics, reducing
WL errors helps to better measure several of the 
highest-variance PCs in the plot ($i>10$), 
but it makes little difference to the
well measured PCs.
Reducing BAO errors tightens constraints on 
nearly all of the PCs, with the greatest impact in the intermediate range
between the SN and WL contributions. 
Assuming the optimistic WL errors gives much greater weight to
WL improvements, which now produce the largest improvement in the
first five PCs (right panel of Figure~\ref{fig:pcsigmaratio}).
The trends for reducing SN or BAO errors are similar to before,
but the magnitude of their effect is smaller because they are
competing with tighter WL constraints.
The behavior of the $\sigma_i^{-2}$ ratios of the best-measured
PCs mirrors that shown for the DETF FoM in Figure~\ref{fig:fom}.
With the fiducial WL systematics, BAO measurements have the
greatest leverage, followed by SN, and the impact of reducing
WL errors is small.  With the optimistic WL systematics, on
the other hand, reducing WL errors makes the largest difference,
followed by BAO, followed by SN.

\begin{figure}[ht]
\begin{centering}
{\includegraphics[width=3.in]{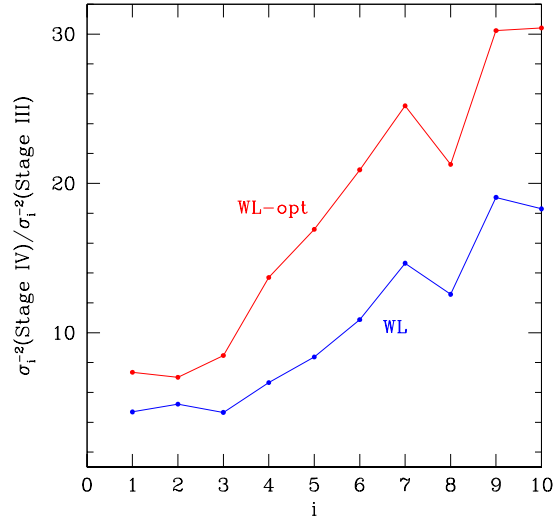}}
\caption{\label{fig:pcsigmaratio_stageiii}
Ratios of inverse variances of PC amplitudes of Stage IV to those of 
Stage III, assuming either the fiducial or optimistic versions of 
the Stage IV WL forecast.
}
\end{centering}
\end{figure}

Dotted curves in the left hand panel show the $\sigma_i^{-2}$ ratios
when we {\it fix} the PCs to be those of the fiducial program.
In this case, the PC errors for the improved programs are no
longer uncorrelated, but the correlation coefficient of errors
among any pair of PCs is less than 0.5 in nearly all cases.
Results are similar to before except for the first component
(first two components for BAO).  These, of course, show less
improvement when they are fixed to be those of the fiducial
program rather than shifting to be the components best
determined by the improved data.
Figure~\ref{fig:pcsigmaratio_stageiii} shows the expected
improvements in $\sigma_i^{-2}$ between our fiducial
Stage III and Stage IV programs.  Consistent with the DETF FoM
plots in Figure~\ref{fig:fom}, the expected improvements are
dramatic, and considerably more so with the optimistic WL assumptions.

The DETF FoM compresses constraints in the $w_0-w_a$ model to a single 
number.  Similar figures of merit for PC constraints have been defined in 
the literature, in various forms, 
each of which may be useful for different purposes.
These include the determinant of ${\bf F}^w$, which characterizes the 
total volume of parameter space allowed by a particular combination of 
experiments in analogy to the DETF FoM for the $w_0$--$w_a$ parameter 
space, 
and the sum of the inverse variances of the PCs, which is typically 
less sensitive than the determinant
to changes in the errors of the most weakly constrained PCs 
\citep{huterer01,bassett05,albrecht06,albrecht07,wang08,barnard08,albrecht09,
crittenden09,amara09,mortonson10,shapiro10,trotta10,march11}.

Examples of these FoMs for the fiducial program and the variants considered
in Figure~\ref{fig:pcsigmaratio} are listed in Table~\ref{tbl:pcfom}.
Here we allow the PC basis to change with the forecast assumptions, 
so ${\bf F}^w$ is diagonal and $\det {\bf F}^w = \prod_{i=1}^{36} 
\sigma_i^{-2}$. As with the ratios of PC variances in 
Figure~\ref{fig:pcsigmaratio}, we restrict the variances to be less than 
unity by replacing $\sigma_i^{-2}\to 1+\sigma_i^{-2}$. 
The other FoM, computed as the sum of inverse variances, requires no such 
prior because PCs with large variances contribute negligibly to the sum.
Note that the choice of PC FoM definition can affect decisions about 
whether one experiment or another is optimal; for example, halving WL 
errors (assuming fiducial systematics)
relative to the fiducial model increases the $\det {\bf F}^w$ 
FoM more than halving SN errors, but the opposite is true for the 
sum of inverse variances, which favors improvements in the best-measured PCs 
and more closely tracks the DETF FoM.
In this case, at least, we regard the latter measure as a better
diagnostic, since the improvements for PCs that are poorly measured
in any case seem unlikely to reveal departures from a cosmological 
constant or other simple dark energy models.
Another virtue of $\sum\sigma_i^{-2}$ (the square of the
quantity tabulated in Table~\ref{tbl:pcfom}) is its sensible
scaling with measurement precision.  If the error of all the
individual cosmological measurements (e.g., $D_L$ values
and WL power spectrum amplitude) is dropped by a factor of
two, as expected if experiments are statistically limited
and data volume is increased by a factor of four, then 
each $\sigma_i$ will drop by a factor of two and
$\sum\sigma_i^{-2}$ will go up by a factor of four, scaling
with data volume just like the DETF FoM.
For $\det {\bf F}^w$, on the other hand, the FoM will go up by
$\approx 2^N$, where $N$ is the number of PCs that have
$\sigma_i$ significantly below one, so there is no obvious
scaling with data volume.

The disagreement between different PC FoMs in Table~\ref{tbl:pcfom} 
highlights one of the difficulties with using PCs or related 
methods for evaluating the potential impact of future experiments.
Forecasts for PCs provide a wealth of information in both the 
redshift-dependent shapes of the PCs and the expected errors on 
their amplitudes, but it is often difficult to interpret what 
this information implies about cosmic acceleration.
Given a set of forecasts for PCs, one can easily compute the expected
constraints on any specific model for $w(z)$ by expressing the 
model in terms of the PC amplitudes (eq.~\ref{eqn:wtopc}); 
this is a potentially useful application, but it makes very limited use
of the available information.
%\tbd{Is the right analogy here really (\ref{eqn:wtopc}) 
%or (\ref{eqn:fisher_reparam})?  Maybe we should give the equation
%here explicitly, or perhaps reference (\ref{eqn:fisher_reparam})
%and say what goes into it?  It seems like a useful point
%to be clear on.}

More generally, we can use the forecast PC shapes and errors to try to 
visualize what types of $w(z)$ variations are allowed by a certain 
combination of experiments. One approach is to generate several random 
$w(z)$ curves that would be consistent with the forecast measurements.
This method is easily implemented with the PCs because the errors on 
different PC amplitudes are uncorrelated. 
One can generate a random realization of $w(z)$ 
by simply drawing an amplitude $\alpha_i$ from a
Gaussian distribution with mean zero and width $\sigma_i$,
then using
equation~(\ref{eqn:pctow}) to compute $w(z)$ corresponding to 
the randomly-drawn $\alpha_i$ values.

\begin{figure}[ht]
\begin{centering}
{\includegraphics[width=3.2in]{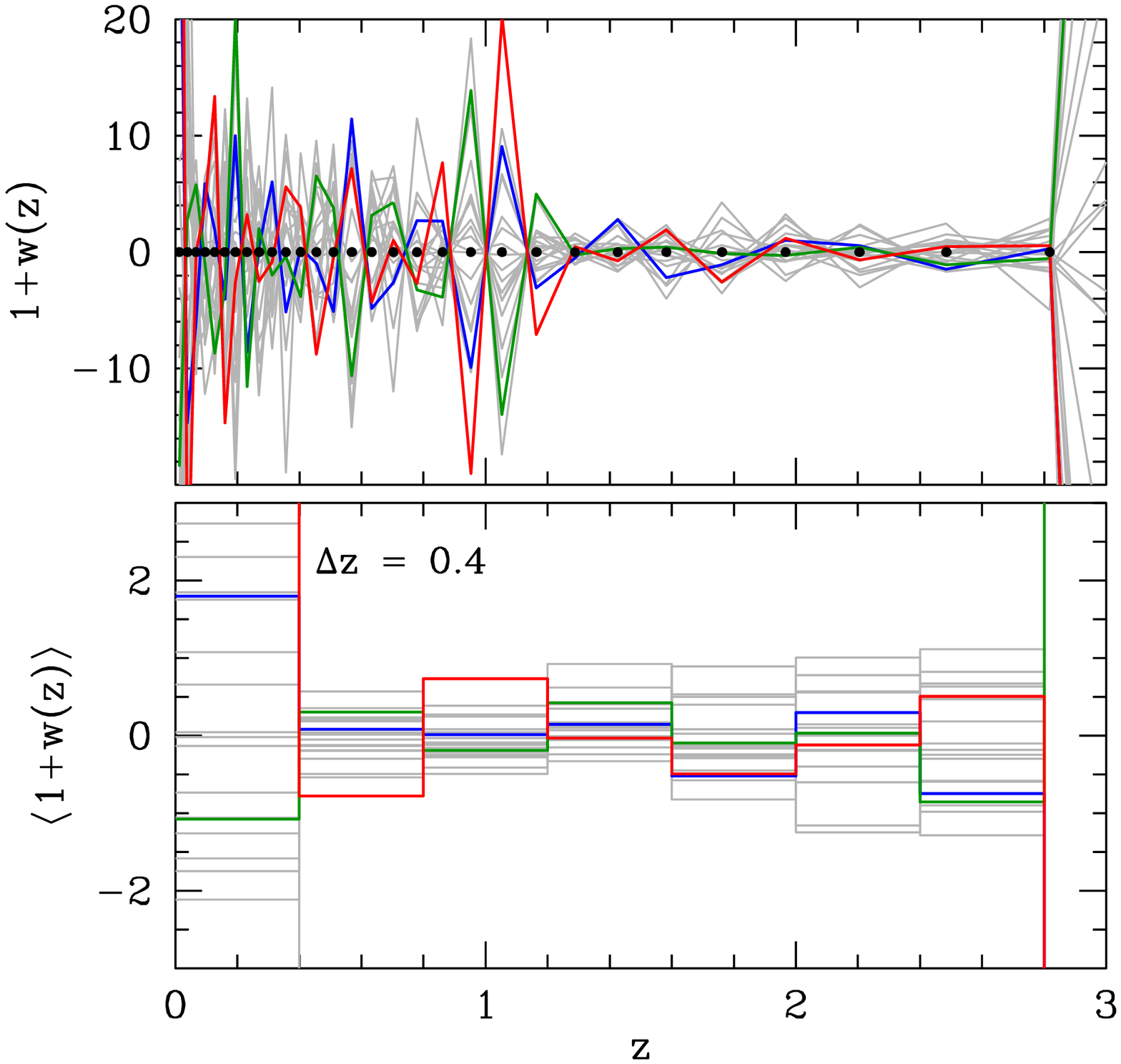}\hfill
\includegraphics[width=3.2in]{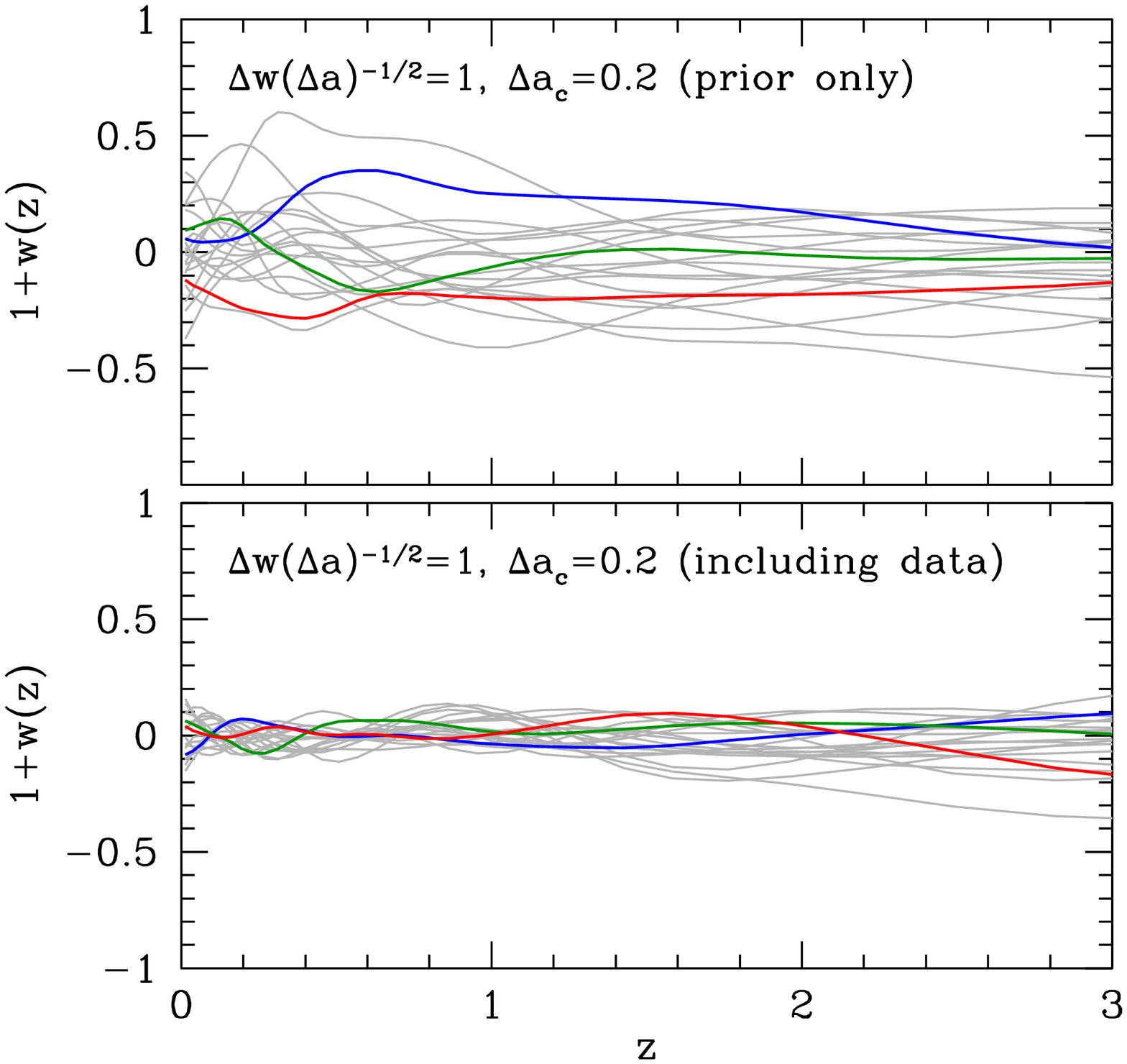}}
\caption{\label{fig:pcw}
Reconstruction of $w(z)$ from PC constraints. {\it Left:} 20 
randomly-generated models that would be indistinguishable from a 
cosmological constant using the fiducial program of experiments.
Three of the 20 models are highlighted (in red, green, and blue) to 
more clearly show examples of the evolution with redshift. 
The lower panel shows the average of $1+w(z)$ in bins of width $\Delta z=0.4$ 
for the same models as in the upper panel.
Points along the $w(z)=-1$ line in the upper panel mark the centers 
of the bins in which $w(z)$ is allowed to vary in our forecasts.
{\it Right:} $w(z)$ reconstruction including a prior of the form in
equation~(\ref{eqn:wcorr2}). The upper panel shows a random selection of 
models consistent with this prior, but without including any data, and 
the lower panel shows examples of models that are allowed by both the 
prior and the data assumed in the fiducial program.
}
\end{centering}
\end{figure}

In the upper left panel of Figure~\ref{fig:pcw}, we use this method to 
plot several $w(z)$ models using the fiducial program PC shapes and 
errors from Figure~\ref{fig:pcfid} and Table~\ref{tbl:pcerrors}, 
respectively. We cut off the plot at $z=3$, since $w(z)$ variations at 
higher redshifts are essentially unconstrained by the fiducial experiments.
Even at lower redshifts, though, the allowed $w(z)$ variations are 
enormous, with $w_i$ values often changing by 10 or more from one bin
to the next. 
(Recall that our prior corresponds to a Gaussian of width
$\sigma_{w_i} \approx 63$ per bin, eq.~\ref{eqn:fisherdiag}.)
Compared to the $\sim 1.5\%$ constraints on $w_p$ in the 
$w_0$--$w_a$ model, this forecast looks rather depressing.
The consequence of allowing the equation of state to be
a free function of redshift is that it is nearly impossible
to say with any certainty what the value of $w$ is at
any specific redshift, because rapid oscillations in $w(z)$ have
tiny effects on observables.
The allowed range of variations would be even larger if
we considered a model with finer $\Delta a$ bins.

The large variations of $w(z)$ in Figure~\ref{fig:pcw} are
driven by the poorly constrained PCs, which have many
oscillations in $w(z)$, peak-to-peak amplitudes
$|\Delta w| \sim 4$, and normalization uncertainties
$\sigma_i \sim 0.1-2.3$ (see Figure~\ref{fig:pcfid}
and Table~\ref{tbl:pcerrors}).
The lower left panel  of Figure~\ref{fig:pcw} shows 
these $w(z)$ realizations averaged over bins of width 
$\Delta z = 0.4$, which  vastly reduces 
the range of variations, 
especially at $z\sim 1$. However, the dispersion of 
$w(z)$ in the bins centered at $z=0.6$ 
and $z=1$ is still about 0.3.
Adding a precise, independent measurement of $H_0$  
reduces the uncertainty in $w(z)$ in the lowest-redshift bin, but it has
little effect at higher redshifts (see \S\ref{sec:forecast_h0}).

Instead of averaging $w(z)$ over wide redshift bins,
one can impose a theoretical prejudice for models with 
smoothly-varying equations of state by adding an
off-diagonal prior to the Fisher matrix, imposing correlations
among the $w_i$ that are closely separated in redshift.
Here we follow \cite{crittenden09}, but we modify their method 
to use scale factor 
rather than redshift as the independent variable
(see also \citealt{crittenden11}), 
adopting a correlation function
\begin{equation}
\label{eqn:wcorr1}
\xi(|a_i-a_j|) = {(\Delta w)^2 \over \pi \Delta a_c} 
\left[1+\left({a_i-a_j \over \Delta a_c}\right)^2 \right]^{-1},
\end{equation}
where $\Delta w$ sets the amplitude of allowed $w(z)$ variations and 
$\Delta a_c$ is the correlation length.
Following the calculation in \cite{crittenden09}, the covariance matrix 
for the $w_i$ bins, which is the inverse of the prior Fisher matrix for those 
parameters, is
\begin{equation}
\label{eqn:wcorr2}
[F_{ij}^{\rm prior}]^{-1}_{(i,j\le 36)} = {(\Delta w)^2 \Delta a_c \over 
\pi \Delta a^2} \left[x_+ \tan^{-1} x_+ + x_- \tan^{-1} x_- - 
2\bar{x}\tan^{-1}\bar{x} + \ln \left( { 1+\bar{x}^2 \over
\sqrt{(1+x_+^2)(1+x_-^2)} } \right) \right],
\end{equation}
where $\bar{x} = |i-j|\Delta a/\Delta a_c$, 
$x_+ = (|i-j|+1)\Delta a/\Delta a_c$, and $x_- = (|i-j|-1)\Delta a/\Delta a_c$.
In the limit $\Delta a_c \to 0$, this reduces to our default diagonal
prior on the $w_i$ parameters with width $\sigma_{w_i} = 
\Delta w / \sqrt{\Delta a}$.

The upper right panel of Figure~\ref{fig:pcw} shows models randomly 
drawn from this prior with $\Delta w/\sqrt{\Delta a}=1$ and $\Delta a_c = 0.2$.
The influence of the correlation function is clearly evident in the 
smoother, lower-amplitude variations of $w(z)$ in these models, 
and yet the range of 
possible models is still much greater than for simpler parameterizations
like $w_0$--$w_a$. Combining this prior with the assumed data set of 
the fiducial Stage IV program, we obtain the $w(z)$ realizations 
plotted in the lower right panel of Figure~\ref{fig:pcw}. Even more so than 
averaging $w(z)$ in wide redshift bins, including this type of prior
significantly narrows the constraints on $w(z)$.
While the particular smoothness prior of~(\ref{eqn:wcorr1}) is 
certainly not unique, this approach of combining PC constraints
dictated by the data sets with theoretically motivated priors on
the behavior of $w(z)$ --- perhaps based on an underlying model
for the potential $V(\phi)$ --- may be the most 
valuable application of the PC approach.

Our constraints on general $w(z)$ models account for the possibility of
modified gravity by marginalizing over the structure growth parameters 
$\Delta \gamma$ and $\ln G_9$. If we instead restrict our analysis to GR
by fixing $\Delta \gamma = \ln G_9 = 0$, the main effect is that the 
dark energy equation of state at high redshifts, $w(3<z<9)$, is better 
constrained because the CMB measurement of the power spectrum amplitude
at $z \sim 1000$ can be more directly related to WL measurements of 
growth at lower redshifts. Because of the additional CMB constraint on 
the distance to the last scattering surface, $w(3<z<9)$ is strongly 
correlated with $\ok$, and therefore assuming GR considerably improves
the determination of spatial curvature in the binned $w(z)$ parameterization. 
For our fiducial forecasts, assuming $\Delta \gamma = \ln G_9 = 0$ lowers 
$\sigma_{\ok}$ by a factor of $\sim 3$ ($0.0075 \to 0.0023$); note that 
this is still several times larger than the error in $\ok$ for the 
simpler $\Lambda$CDM or $w_0$--$w_a$ forecasts.

\subsection{Forecasts for Clusters}
\label{sec:forecast_cl}

We have concentrated so far on the constraints expected for combinations
of CMB, SN, BAO, and WL data, as all of these methods are well studied
and are likely to play a central role in Stage III and Stage IV studies 
of cosmic acceleration.  For other methods we adopt a simplified
approach, first asking how well our fiducial CMB+SN+BAO+WL programs
should predict the basic observables of these methods, then showing how
different levels of precision on these observables would affect
constraints on equation-of-state and growth parameters.  We 
describe our methodology more fully in the next section 
(\S\ref{sec:forecast_alt}), but we begin with a discussion of
clusters, where our analysis of stacked weak lensing calibration
(\S\ref{sec:cl_mass_calibration}) gives a clear quantitative
target for measurement precision.

\begin{figure}[ht]
\begin{centering}
{\includegraphics[width=2.8in]{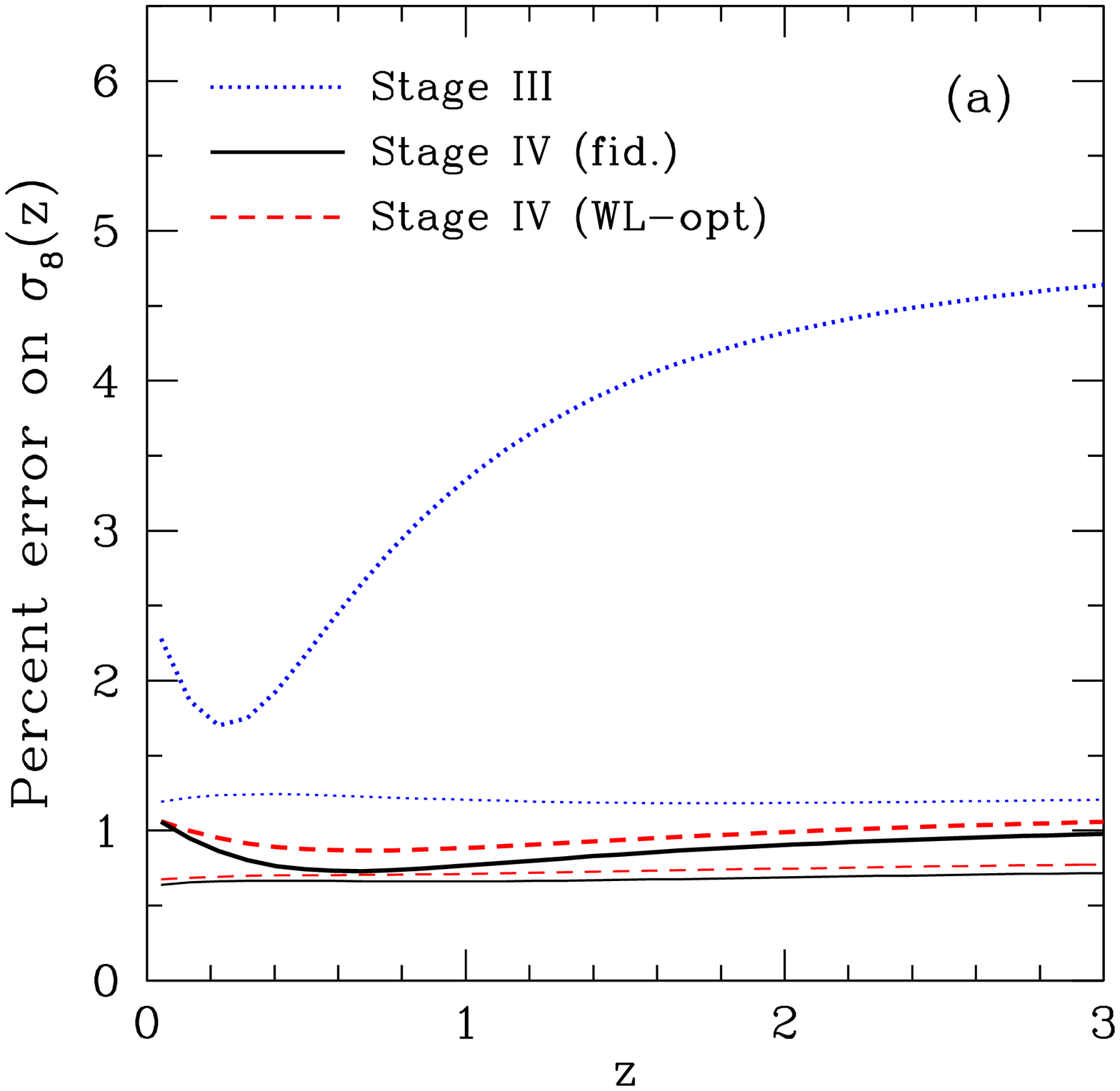}
\includegraphics[width=2.8in]{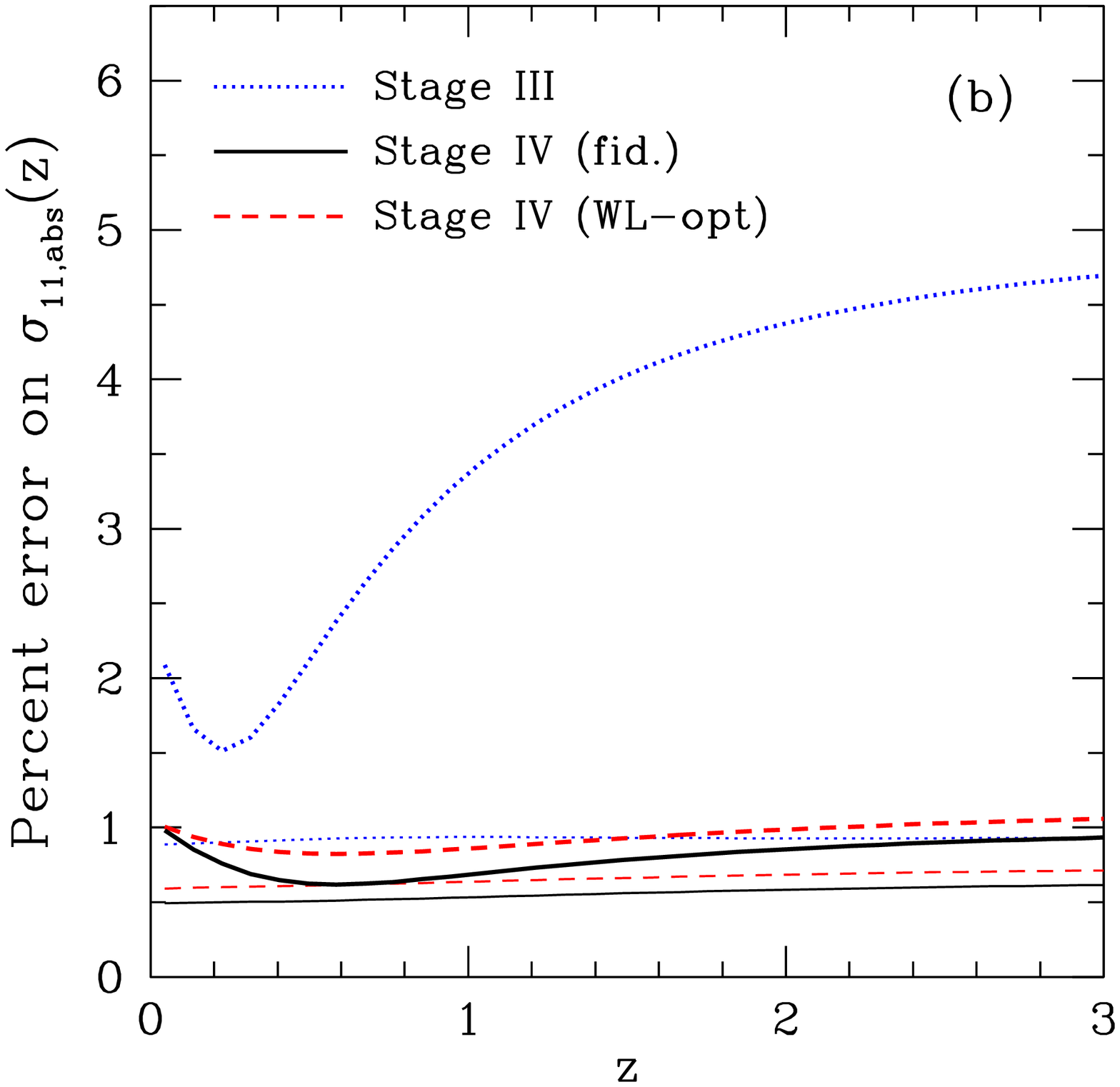}}
{\includegraphics[width=2.8in]{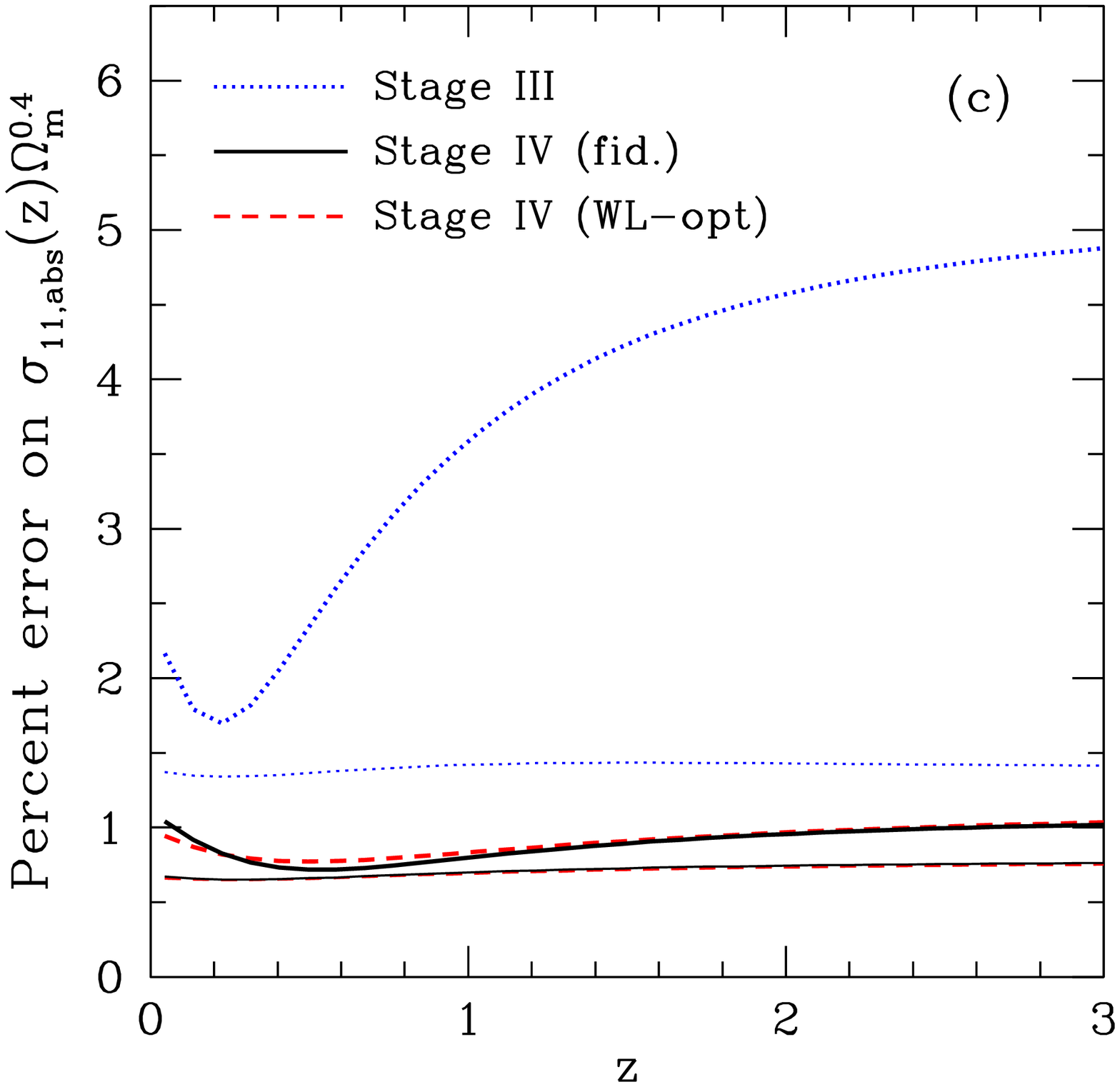}
\includegraphics[width=2.8in]{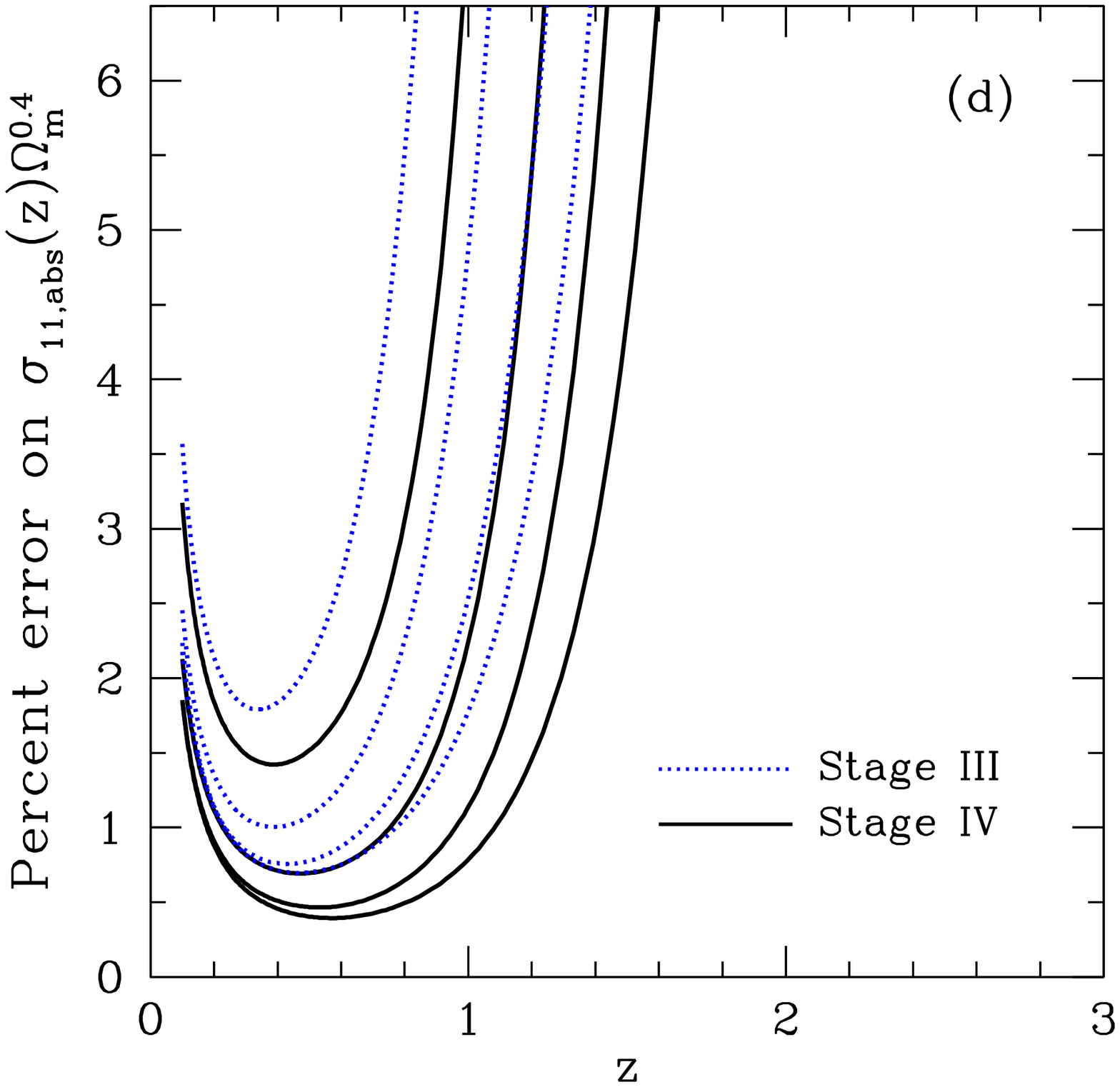}}
\caption{\label{fig:forecast_sigma8}
(a) Predicted fractional errors ($1\sigma$) on $\sigma_8(z)$ from our
fiducial Stage III (dotted) and Stage IV (solid) CMB+SN+BAO+WL
programs, and from the Stage IV program with optimistic WL systematic
assumptions (dashed).  All curves assume a $w_0-w_a$ dark energy
parameterization.  
For each case, the lower, thin curve shows the forecast assuming
GR is correct and the upper, thick curve shows the forecast allowing GR
deviations parameterized by $G_9$ and $\Delta\gamma$.
(b) Like (a), but for $\sigElevz$, the rms matter fluctuation in
spheres of radius $11\Mpc$ (instead of $8\hmpc$).
(c) Like (b), but for the parameter combination $\sigElevz\Omega_m^{0.4}$
that approximates the quantity best constrained by cluster abundances.
(d) Predicted fractional errors in $\sigElevz\Omega_m^{0.4}$ from cluster
abundances in a $10^4\mdeg^2$ survey calibrated by stacked weak 
lensing mass estimates with Stage III ($n_{\rm eff}=10\,\arcmin^{-2}$)
and Stage IV ($n_{\rm eff}=30\,\arcmin^{-2}$) source surface densities
and survey depths (dotted and solid curves, respectively).
From top to bottom, curves correspond to cluster mass thresholds of
8, 4, 2, and $1\times 10^{14} M_\odot$.
}
\end{centering}
\end{figure}

Figure~\ref{fig:forecast_sigma8}a shows the predicted fractional error
($1\sigma$) in $\sigma_8(z)$ for the fiducial Stage III and Stage IV
experimental programs discussed in \S\ref{sec:results}, and
for the Stage IV program with optimistic WL errors.
All curves assume a $w_0-w_a$ dark energy parameterization,
and for each case the lower, thinner curve shows the forecast
assuming GR to be correct while the upper, bolder curve allows
GR deviations parameterized by $G_9$ and $\Delta\gamma$.
Roughly speaking, we would expect a measurement with precision
better than that shown by the upper curve to significantly improve tests
for GR deviations and a measurement with precision better than that 
shown by the lower curve to significantly improve $w_0-w_a$
constraints when assuming GR to be correct.  For Stage IV programs
we predict $\sigma_8(z)$ constraints at the $0.75-1\%$ level
over the full redshift range $0 < z < 3$, with little difference
between the fiducial and optimistic WL assumptions.
In fact, the ``optimistic'' WL assumptions lead to slightly larger
errors in $\sigma_8(z)$ than the fiducial assumptions because for
this quantity doubling the statistical errors has a larger impact
than adding $2\times 10^{-3}$ shear calibration and photo-$z$ errors
(see \S\ref{sec:fiducial}).  For Stage III, the predicted $\sigma_8(z)$
errors are about 1.2\% assuming GR, but they are much larger if
we allow GR deviations, especially at $z>0.8$.  Even for Stage IV,
the good constraints at high $z$ rely on the assumption of a
$w_0-w_a$ equation of state, which allows precise low redshift WL
measurements to be extrapolated to high redshift.
The direct measurements of $z>1$ clustering amplitude are considerably
weaker.  

\begin{figure}[ht]
\begin{centering}
{\includegraphics[width=4.5in]{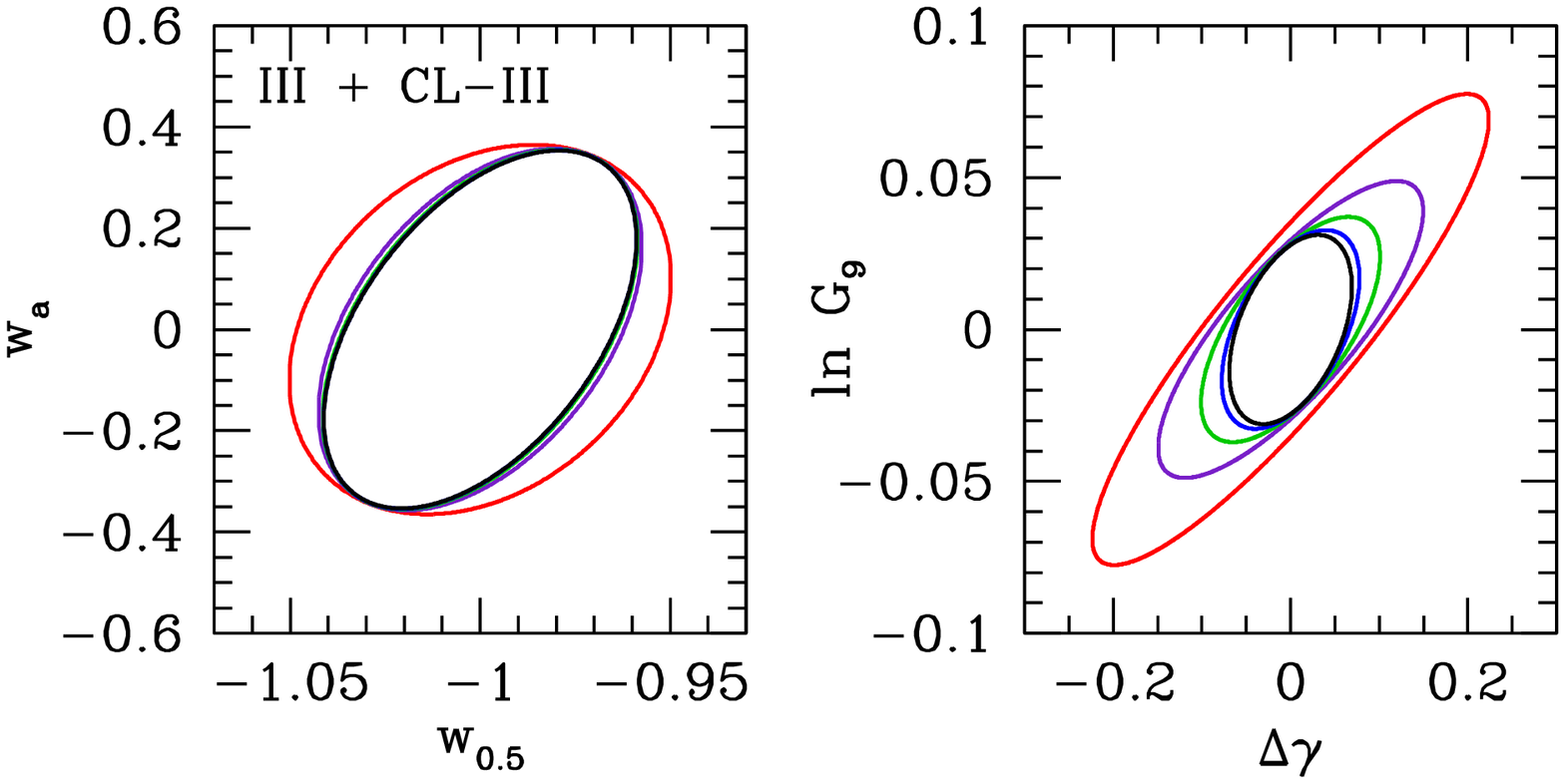}}
{\includegraphics[width=4.5in]{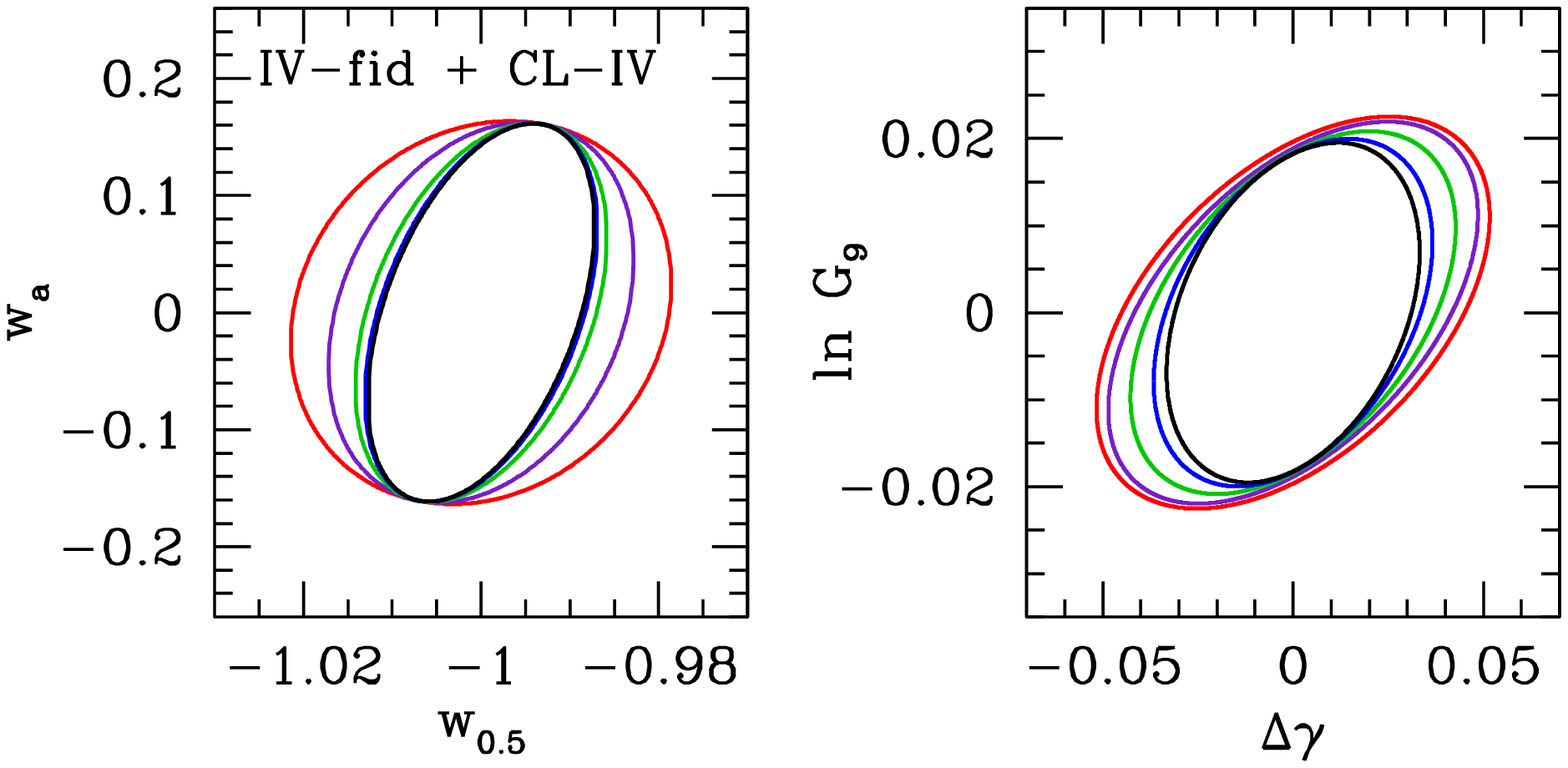}}
{\includegraphics[width=4.5in]{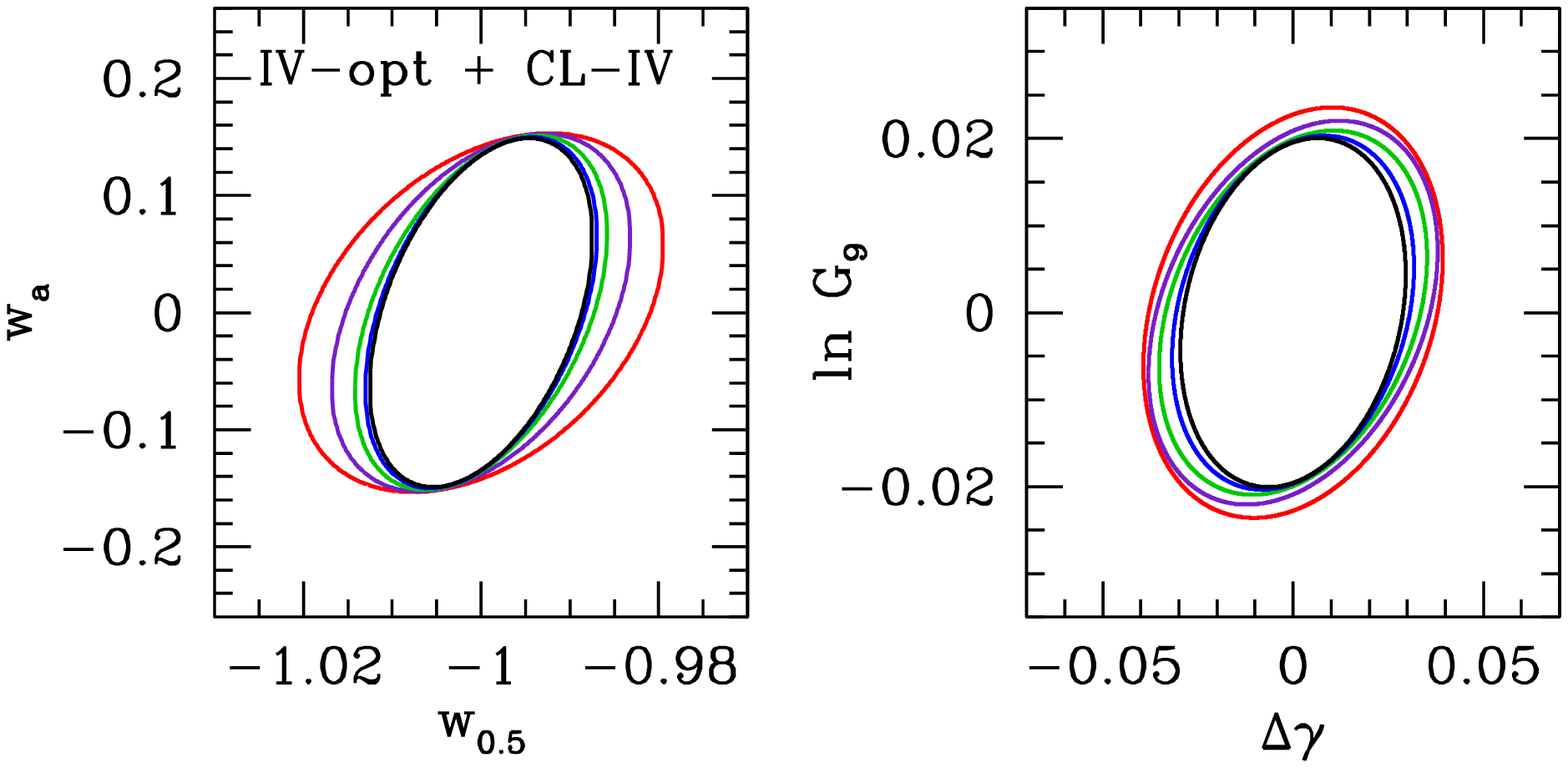}}
\caption{\label{fig:contours_cl}
Predicted constraints ($1\sigma$) on the equation-of-state parameters
$w_{0.5}=w(z=0.5)$ and $w_a$ (left panels) and the growth parameters
$G_9$ and $\Delta\gamma$ (right panels) from our fiducial
CMB+SN+BAO+WL programs combined with cluster abundance measurements
of $\sigElevz\Omega_m^{0.4}$ with the precision shown in 
Fig.~\ref{fig:forecast_sigma8}d.  Top panels show Stage III
clusters with Stage III CMB+SN+BAO+WL, while middle and bottom
panels show Stage IV clusters combined with the fiducial and
WL-opt Stage IV programs, respectively.  Note the change in
axis scale between the top and middle/lower panels.
In each panel, the outermost contour shows the constraints
without clusters, and the remaining contours show the constraints
for cluster mass thresholds of 8, 4, 2, and $1\times 10^{14}M_\odot$
(outer to inner).
}
\end{centering}
\end{figure}

Figure~\ref{fig:forecast_sigma8}b plots $\sigElevz$ errors, which
are tighter than the $\sigma_8(z)$ errors by $\sim 30-50\%$
because uncertainty in $h$ contributes noticeably to the latter.
In \S\ref{sec:cl_prospects} we estimated the errors on $\sigElevz$
achievable with a $10^4\mdeg^2$ cluster survey with weak lensing
mass calibration, assuming Stage III ($10\,\arcmin^{-2}$) or
Stage IV ($30\,\arcmin^{-2}$) effective source densities and survey
depths.  For a mass threshold of $2\times 10^{14} M_\odot$ the
$\sigElevz$ errors at $z\approx 0.5$ are $\sim 1\%$ and $\sim 0.5\%$,
respectively, below the corresponding Stage III and Stage IV
errors shown in Figure~\ref{fig:forecast_sigma8}b.
Furthermore, these cluster errors are per $\Delta z = 0.1$ redshift bin,
so constraints on the clustering amplitude in a smoothly evolving
model can be substantially better if the cluster errors are not
correlated across redshifts.  (The statistical errors should
be uncorrelated, but some forms of weak lensing systematics could
affect many redshift bins in the same direction.)

The cluster errors shown earlier in Figure~\ref{fig:wl_serr}
were derived assuming perfect knowledge of $\om$, with 
$\sigElevz$ as the single parameter controlling the cluster
abundance at each redshift.  In practice, cluster abundances
constrain a parameter combination that is approximately
$\sigElevz\Omega_m^{0.4}$, as discussed in \S\ref{sec:cl}.
The fractional errors in $\om$ from our fiducial CMB+SN+BAO+WL
programs are 2.7\% (Stage III), 1.4\% (Stage IV), and 
1.2\% (Stage IV with WL-opt), making $\Omega_m^{0.4}$ uncertainties
comparable to the fractional errors in $\sigElevz$.  
Figure~\ref{fig:forecast_sigma8}c shows the predicted fractional
errors in $\sigElevz\Omega_m^{0.4}$, which in some ranges are
significantly larger than those for $\sigElevz$.
Finally, Figure~\ref{fig:forecast_sigma8}d
shows our forecast errors on $\sigElevz\Omega_m^{0.4}$ from a 
$10^4\mdeg^2$ cluster survey in which errors are limited by
weak lensing mass calibration statistics.  Here we have
simply set the fractional errors from clusters on 
$\sigElevz\Omega_m^{0.4}$ equal to the ones we derived earlier on
$\sigElevz$, which should be a good but not perfect approximation.
Comparing Figures~\ref{fig:forecast_sigma8}c and ~\ref{fig:forecast_sigma8}d 
shows that cluster errors are competitive with those expected from 
the CMB+SN+BAO+WL combination for cluster mass thresholds
of $\sim 4-8\times 10^{14} M_\odot$ at Stage III or
$1-4 \times 10^{14} M_\odot$ at Stage IV.

Figure~\ref{fig:contours_cl} shows the potential improvement in
equation-of-state and growth parameter determinations from 
including the cluster constraints on $\sigElevz\Omega_m^{0.4}$.
We assume that these constraints have independent errors in
each $\Delta z = 0.1$ bin.  Upper panels show the effect of
adding Stage III cluster constraints 
(dotted curves in Fig.~\ref{fig:forecast_sigma8}d) to the
Stage III CMB+SN+BAO+WL Fisher matrix.
Even adding clusters with an $8\times 10^{14} M_\odot$ mass threshold
substantially improves the
errors on $G_9$ and $\Delta\gamma$, and reducing the
mass threshold to $1-2\times 10^{14} M_\odot$ produces substantial
further gains.  Somewhat surprisingly, the cluster constraints
also lead to significantly smaller errors on the equation-of-state
parameter $w_{0.5}$ and slightly smaller errors on $w_a$.
This improvement largely reflects the additional information about
$\om$, which allows the distance and $H(z)$ constraints from other
probes to translate more directly into $w(z)$ constraints.
We have checked that fixing $\om$ exactly would produce a
still greater improvement in $(w_{0.5},w_a)$ than the gain we
have forecast from clusters, while making little difference to the  
$(G_9,\Delta\gamma)$ errors.

For Stage IV (middle and bottom panels), where we now assume the Stage IV
cluster mass constraints, an $8\times 10^{14} M_\odot$ cluster
sample produces little improvement over CMB+SN+BAO+WL
in $G_9$ and $\Delta\gamma$, but it still leads to noticeable
improvement in $w_{0.5}$.
A $1-2\times 10^{14} M_\odot$ cluster sample produces
substantial gains in both the equation-of-state and growth parameters.
As in the Stage III case, much of the improvement in the equation 
of state comes from the $\om$ information provided by clusters.
However, the cluster constraints
reduce the $w_{0.5}$ error even if $\om$ is held fixed,
so some of this improvement arises from another source,
probably by allowing some WL information to be effectively
transferred from growth to distance.
Adding our Stage IV, $10^{14} M_\odot$ cluster constraint
to the fiducial Stage IV program increases the DETF FoM
from 664 to 1258, and it increases the modified FoM
$[\sigma(w_p)\sigma(w_a)]^{-1}\times[0.034/\sigma(\Delta\gamma)]$
(Figure~\ref{fig:fom_gamma}) from 664 to 1955.
For the WL-opt program, the improvements are
$789 \rightarrow 1363$ and $1037 \rightarrow 2380$, respectively.
For Stage III CMB+SN+BAO+WL, adding Stage III clusters 
leads to improvements of $131\rightarrow 183$ (FoM) 
and $30\rightarrow 137$ (modified FoM).

Our treatment here is simplified because we have ignored
the impact of volume-element changes on the cluster abundance
and have set the scaling index of $\sigElevz\Omega_m^q$ to a constant
value $q=0.4$ instead of including its redshift and mass
dependence.  More importantly, we have assumed that errors
in the cluster abundance will be dominated by the statistical
errors in the weak lensing calibration of the mean mass scale, 
not increased by marginalizing over uncertainties in
mass-observable scatter, incompleteness, contamination,
or theoretical predictions.  The effective mass calibration
uncertanties we are assuming are those in Figure~\ref{fig:wl_merr}.
These are probably pessimistic at $z \ga 1$, where the weak
lensing calibration error exceeds 10\% but one could likely
use other calibration methods (including direct comparison to theory)
to do better; thus, we are underplaying the potential contribution
of high-redshift clusters.
Our approximate calculations confirm the conclusions of
\cite{oguri11b} that clusters calibrated with stacked weak
lensing can make an important contribution to testing
cosmic acceleration models, even in the era of Stage IV
dark energy experiments.
Figure~\ref{fig:forecast_sigma8} also provides a target for
other methods of measuring the matter clustering amplitude,
such as the \lya\ forest (\S\ref{sec:lyaf}).

\subsection{Forecasts for Alternative Methods}
\label{sec:forecast_alt}

We now turn to some of the alternative probes discussed
previously in \S\ref{sec:alternatives}.
For each technique, we first focus on the question of complementarity with 
the primary methods by asking how well the observable quantity measured by a 
particular technique is {\it already} known given the fiducial combination 
of SN, BAO, WL, and CMB data. These predictions provide benchmarks that 
any additional measurement must reach in order to contribute significantly 
to constraints on dark energy or modified gravity parameters. In many 
cases, the precision of the predictions depends strongly on the chosen
parameterization of deviations from the standard paradigm of $\Lambda$CDM 
and GR. We will generally assume a $w_0$--$w_a$ model for the results
in this section, but we note that if one adopts a more general 
parameterization of dark energy then the predictions are normally weaker,
and thus the value of alternative probes is potentially greater.

The covariance matrix for a set of observables ${\bf X}$ measured by a
particular alternative probe can be computed straightforwardly using
the covariance matrix of the cosmological parameters given by the 
inverse of the total Fisher matrix for SN, BAO, WL, and CMB data,
\begin{equation}
\label{eqn:cov_predict}
C_{ij}^{\bf X} = \sum_{k,l} { \partial X_i \over \partial p_k }
\,F_{kl}^{-1}\, { \partial X_j \over \partial p_l }\,,
\end{equation}
where ${\bf p}$ is either the full set of parameters in 
eq.~(\ref{eqn:forecastparams}) or the reduced set with $w_0$ and $w_a$ 
replacing the 36 $w_i$ bins; in the latter case, $\bf F$ is the Fisher 
matrix for the $w_0$--$w_a$ parameterization computed using 
eq.~(\ref{eqn:fisher_reparam}). We compute the full covariance matrices for the
alternative methods, but the plots in the following sections only show the
predicted uncertainties $\sigma_{X_i} = \sqrt{C_{ii}^{\bf X}}$ and do not
reflect the fact that errors on the observables may be correlated.

In addition to computing how well the fiducial SN, BAO, WL, and CMB constraints
predict each observable that would be measured by the alternative techniques, 
we provide several examples to show the improvement in the FoM
and other parameters that would result from a specific 
measurement of that observable.
For these tests, we only consider the impact of measurement of a single 
quantity $X$ at a time, so the total Fisher matrix is modified 
simply by adding the term
\begin{equation}
\label{eqn:fisheralt}
F_{ij}^{\rm alt} = \sigma_X^{-2} \, { \partial X \over \partial p_i }\,
{ \partial X \over \partial p_j }\,,
\end{equation}
where $\sigma_X$ is the assumed uncertainty in the measurement of $X$.

\subsubsection{The Hubble constant}
\label{sec:forecast_h0}

For the Hubble constant, the predicted uncertainty 
from the fiducial probes is simply the value of $\sigma_h$ that comes out
of the Fisher matrix forecasts of the previous section. Assuming a $w_0$--$w_a$
dark energy model, the expected precision on $H_0$ is $0.7\%$ for 
the fiducial Stage IV forecasts and small variations of those forecasts, 
and $1.3\%$ for Stage III (see Tables~\ref{tbl:forecasts1}--\ref{tbl:forecasts3}). 
These are challenging, but probably attainable, targets for future efforts 
to independently measure $H_0$.

\begin{figure}[ht]
\begin{centering}
\includegraphics[width=3.2in]{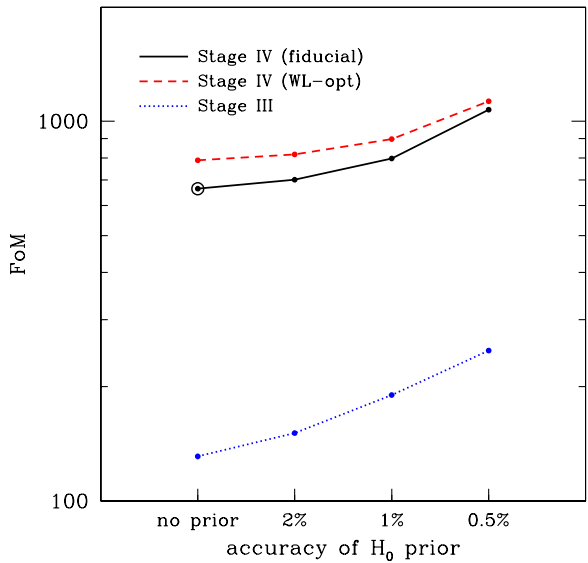}
\caption{\label{fig:forecast_h0}
Dependence of the DETF FoM on the accuracy of additional measurements of the 
Hubble constant for Stage III and IV forecasts from 
\S\ref{sec:results}. The fiducial Stage IV program with FoM$=664$
is marked by an open circle.
}
\end{centering}
\end{figure}

In Figure~\ref{fig:forecast_h0}, we show the effect on the DETF FoM 
of adding a prior on $H_0$ to the fiducial Stage III and IV forecasts.
In all cases, adding a prior with precision that matches the uncertainty 
one would have in the absence of the prior increases the FoM by $\sim 40\%$.
The uncertainties in other cosmological
parameters are affected little by the inclusion
of an independent $H_0$ measurement, as discussed in \S\ref{sec:h0}.

For a more general dark energy parameterization such as the binned 
$w_i$ values, predictions for $\sigma_h$ can be orders of magnitude
weaker than they are for $w_0$--$w_a$ or $\Lambda$CDM 
(see Figs.~\ref{fig:h_ok_model_bao}--\ref{fig:h_model_sn}). 
In this case an independent, local measurement of $H_0$ is vital for accurate
determination of the Hubble constant. However, $H_0$ priors
do not significantly improve dark energy constraints in this case;
an $H_0$ constraint limits the range of $w(z)$ in the lowest-redshift
bin, but since $w(z=0)$ is only weakly 
correlated with the equation of state at higher redshifts by 
SN, BAO, WL, and CMB data, the impact of an additional $H_0$ measurement 
on the equation of state at $z>0$ is small. The improvement in the
DETF FoM in Fig.~\ref{fig:forecast_h0} is largely a consequence of 
the restrictions that the $w_0$--$w_a$ parameterization places on 
the evolution of $w(z)$ between $z=0$ and higher redshifts.
Of course, a discrepancy between directly measured $H_0$ and a
$w_0-w_a$ prediction would already provide the crucial insight
that $w_0-w_a$ is inadequate; it just wouldn't give further direction
about the evolution of $w(z)$.

\subsubsection{The Alcock-Paczynski Test}
\label{sec:forecast_ap}

\begin{figure}[t]
\begin{centering}
{\includegraphics[width=3.2in]{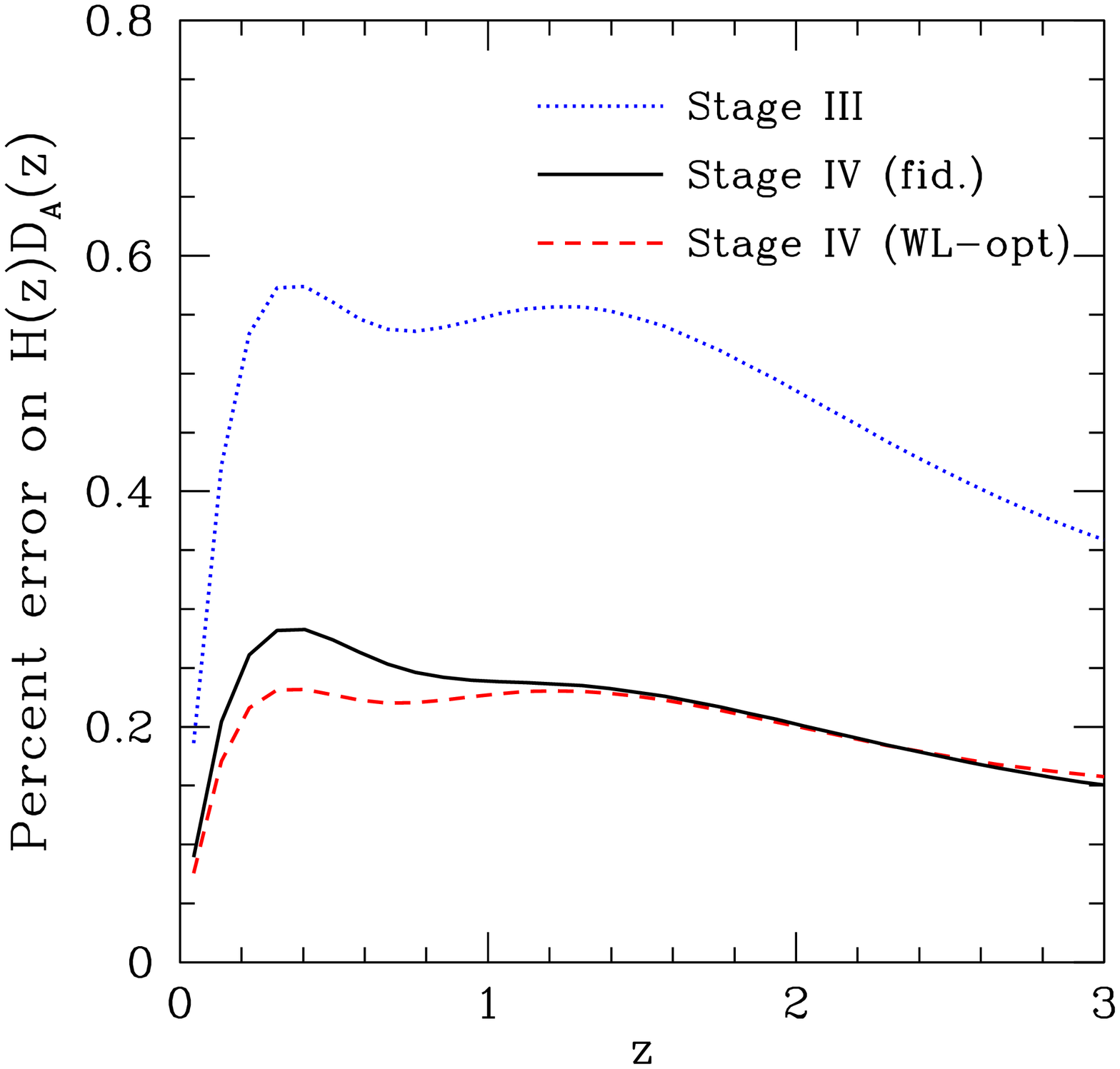}\hfill
\includegraphics[width=3.2in]{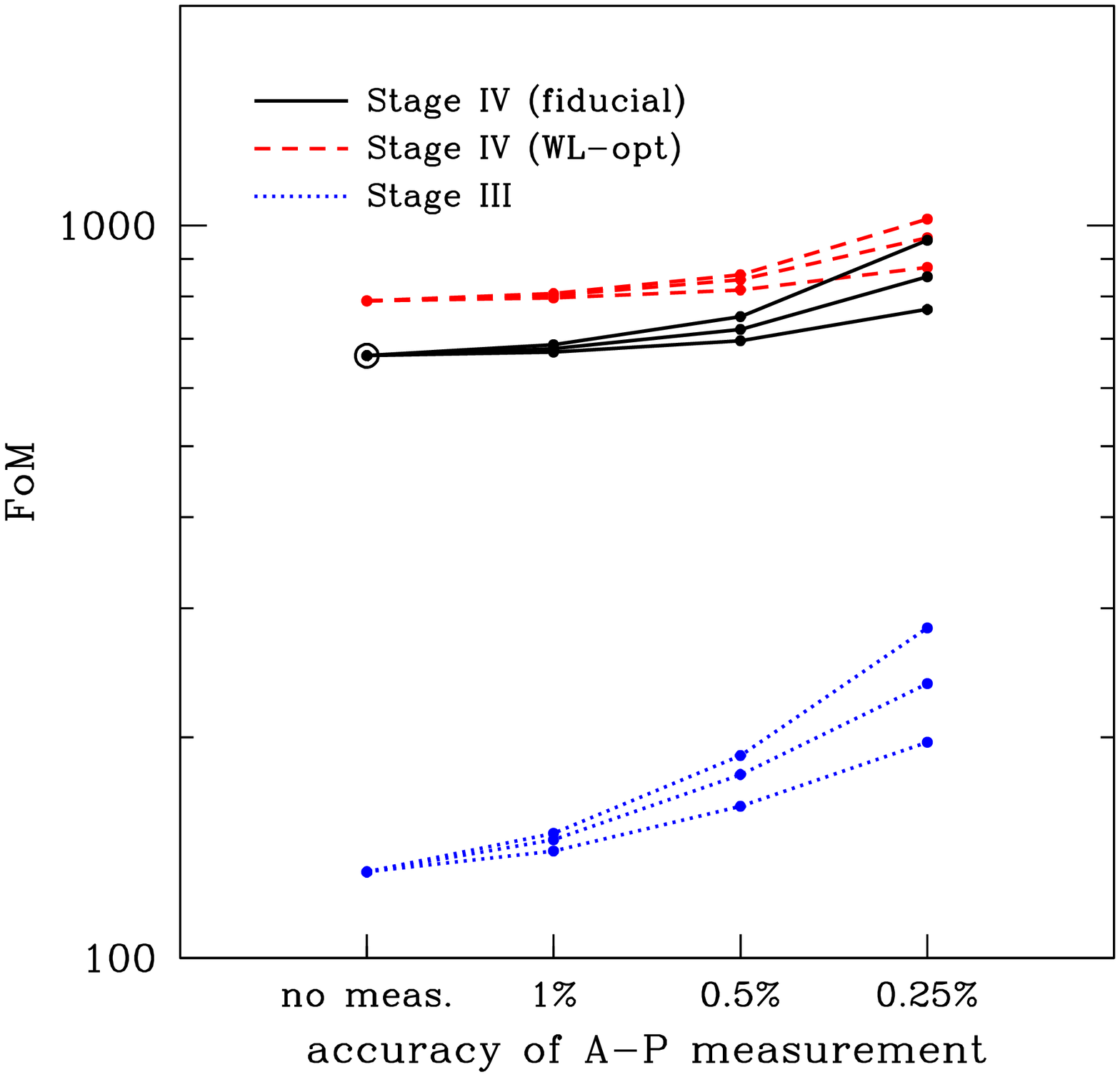}}
\caption{\label{fig:forecast_ap}
{\it Left:} Predicted fractional error ($1\sigma$) of the AP parameter
$H(z)D_A(z)$ from our fiducial Stage III and Stage IV CMB+SN+BAO+WL
programs, assuming a $w_0$--$w_a$ dark energy parameterization.
{\it Right:} Dependence of the DETF FoM on additional measurements of 
$H(z)D_A(z)$ at a single redshift. For each forecast, the three curves from 
top to bottom assume AP measurements at $z=0.5$, $z=1$, 
and $z=2$, respectively.
}
\end{centering}
\end{figure}

For the AP test (\S\ref{sec:ap}), we consider the observable $H(z)D_A(z)$.
Since Stage IV BAO data provide tight constraints on both $H(z)$ and $D_A(z)$,
which are further strengthened by the SN, WL, and CMB measurements, 
it is not surprising that the product $H(z)D_A(z)$ is predicted very
precisely in the combined forecasts. The left panel of 
Figure~\ref{fig:forecast_ap} shows that the uncertainty
in the AP observable is $\sim 0.2\%$ at $0<z<3$ for Stage IV data, and
it is still predicted to sub-percent accuracy with Stage III data. Independent
measurements of the AP observable that are significantly less precise 
than these predictions would contribute little to cosmological constraints.
Note that these results are for a $w_0$--$w_a$ dark energy model. 
If we instead use independently-varying $w(z)$ bins, the uncertainty 
in the AP observable for the Stage IV forecasts increases to $\sim 1\%$ at 
$1<z<3$ and becomes much larger at both lower and higher redshifts, 
although the exact precision of the predictions in this case depends 
strongly on the detailed forecast assumptions such as the prior on 
$w_i$ in each bin or the number of bins used for BAO data.

In the right panel of Fig.~\ref{fig:forecast_ap}, we show the improvement
in the DETF FoM (assuming the $w_0$--$w_a$ parameterization) when 
various measurements of the AP observable are added to the fiducial 
Stage III and IV forecasts. Since the predictions for $H(z)D_A(z)$ are
weakest at $z\lesssim 0.5$, a direct measurement of the AP observable 
at those redshifts has a greater impact on the FoM than measurements at
higher redshifts.\footnote{Note, however, that either decreased SN errors 
or increased BAO errors for any of these forecasts would reduce the 
difference between the predictions at $z<1$ and at $z>1$.}
A $1\%$ measurement of $H(z)D_A(z)$ at $z=0.5$ increases the Stage III 
FoM by about 13\%; a similar improvement in the Stage IV FoM requires
an accuracy of $0.5\%$ at the same redshift.
While the demands suggested by Figure~\ref{fig:forecast_ap} appear stiff,
large redshift surveys in principle have the information to achieve
very high precision on $H(z)D_A(z)$.  The challenge is lowering
systematics to the level needed to achieve this precision.

\subsubsection{Redshift-space Distortions}
\label{sec:forecast_rsd}

\begin{figure}[t]
\begin{centering}
{\includegraphics[width=3.2in]{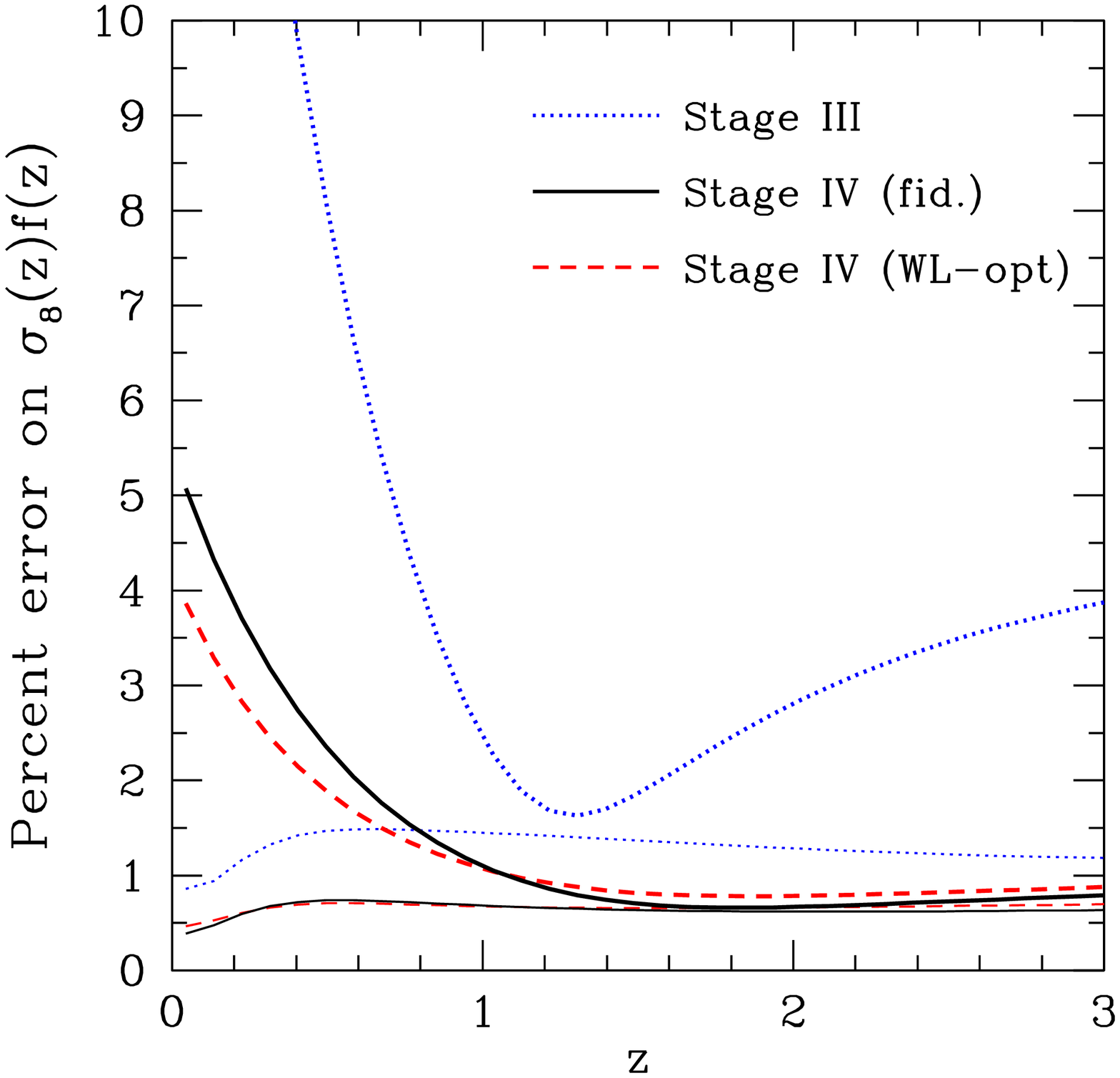}\hfill
\includegraphics[width=3.2in]{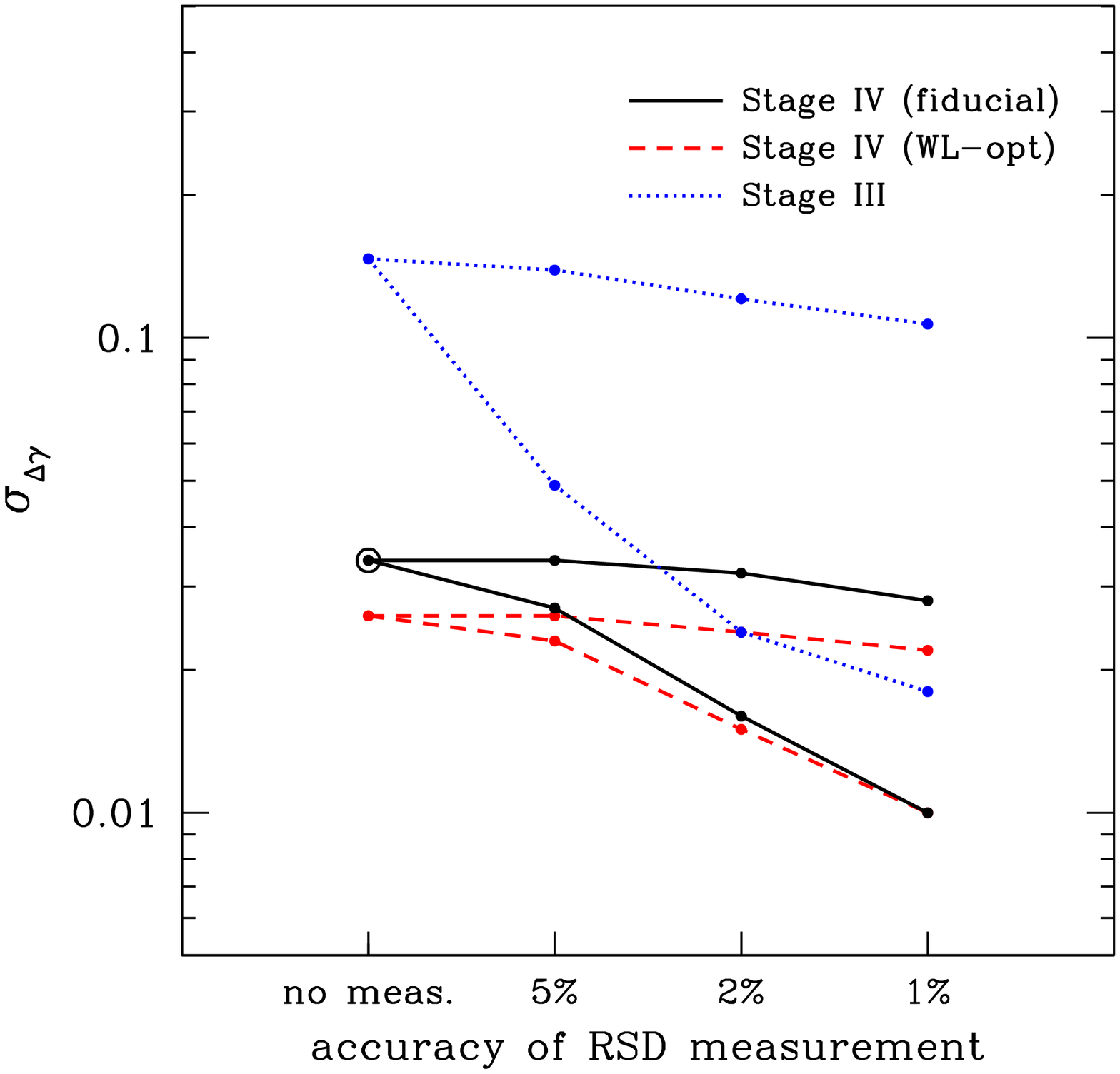}}
\caption{\label{fig:forecast_rsd}
{\it Left:} Predicted fractional error ($1\sigma$) of the RSD observable
$\sigma_8(z)f(z)$ from our fiducial Stage III and Stage IV CMB+SN+BAO+WL
programs, assuming a $w_0$--$w_a$ dark energy parameterization.
The lower, thin curves for each forecast additionally assume GR 
by fixing $\Delta \gamma = \ln G_9 = 0$.
{\it Right:} Improvement in the $1\sigma$ uncertainty on $\Delta \gamma$ 
from an additional RSD measurement at a single redshift.
For each forecast, the lower and upper curves
assume RSD measurements at $z=0.2$ and $z=1$, respectively.
}
\end{centering}
\end{figure}

For redshift-space distortions (RSD; \S\ref{sec:rsd}),
the relevant observable is $\sigma_8(z)f(z)$.
WL data provide some limits on this observable by constraining the structure
growth parameters $\Delta\gamma$ and (in combination with the CMB) $G_9$, 
and through their constraints on the expansion history all of the 
acceleration probes contribute indirectly to the predicted growth 
history. The resulting predictions for Stage III and IV programs are 
plotted in the left panel of Figure~\ref{fig:forecast_rsd}. We show
predictions both for the general case where we marginalize over the 
structure growth parameters and for GR ($\Delta\gamma=\ln G_9=0$).

With the assumption of GR, the RSD observable is predicted to 1--2\%
accuracy for Stage III and 0.5--1\% accuracy for Stage IV.
If we allow modifications to GR through $\Delta \gamma$ and $G_9$, however, 
the uncertainty at $z<1$ increases dramatically. This change is mainly
tied to the freedom to alter the growth rate $f(z)$ at low redshift 
by varying $\Delta \gamma$. Note that the effect of $\Delta \gamma$ 
vanishes at high redshift because $\om(z)$ approaches unity 
and therefore $f(z)\to f_{\rm GR}(z)$ (see equation~\ref{eqn:fullfz}).
At $z\gtrsim 2$, uncertainty in $G_9$ significantly weakens Stage III
predictions of the RSD observable, but the effect on Stage IV predictions
is much smaller.

The DETF FoM can be improved by the addition of precise RSD measurements
if we assume GR; for example, the fiducial Stage IV (Stage III) FoM 
increases by $\sim 10$--$15\%$ with a 1\% (2\%) RSD constraint at $z=1$.
Without assuming GR, the additional information from an RSD
measurement at a single redshift
goes mainly into constraining the structure growth parameters 
(and thus {\it testing} GR).  In this case, the FoM improvement 
from percent-level RSD constraints is $\la 10\%$.
However, percent-level measurements in several redshift bins
can still have an important impact on the FoM.

Low-redshift 
measurements of the RSD observable can contribute significantly to
constraints on $\Delta \gamma$, as shown in the right panel of
Fig.~\ref{fig:forecast_rsd}. For Stage III forecasts, 1--2\% RSD 
measurements at $z=0.2$ reduce the error in $\Delta\gamma$ by nearly an 
order of magnitude, reaching an uncertainty comparable to that expected
from the Stage IV probes. Likewise, the Stage IV constraint on $\Delta\gamma$
can be improved by a factor of a few by the addition of 
percent-level RSD measurements. At higher redshifts, the impact of 
RSD observations on the $\Delta\gamma$ uncertainty is greatly reduced 
due to the diminishing effect of $\Delta\gamma$ on the growth rate 
at high $z$.  This reduced sensitivity at high $z$ is in some sense an
artifact of the $\Delta\gamma$ parameterization; the error on
$\Delta\gamma$ is larger than the error on $\ln f$ by a factor
$|(d\ln f/d\Delta\gamma)^{-1}| = |[\ln\om(z)]^{-1}|$, which for 
the fiducial cosmological model of Table~\ref{tbl:key}
is 1.02 at $z=0.2$ [where $\om(z)=0.373$] and
3.23 at $z=1$ [where $\om(z)=0.734$].

We have computed but not plotted the impact of RSD measurements
on the growth normalization parameter $G_9$.
For Stage IV, the uncertainty in $G_9$ is little
affected by adding RSD measurements at any 
redshift. For Stage III, 1--2\% measurements of $\sigma_8(z)f(z)$ can
reduce the fractional error in $G_9$ by up to a factor of two.
As discussed in \S\ref{sec:gravity}, some modified gravity
theories predict a mismatch between measures of structure
using non-relativistic tracers, which respond to the
Newtonian potential $\Psi$ (eq.~\ref{eqn:metric}), and measures
based on weak lensing, which responds to $\Psi+\Phi$, the sum
of the Newtonian potential and space curvature.
Consistency between RSD and WL, or cluster masses calibrated
by weak lensing, tests for deviations of this sort, in addition 
to the $G_9$ and $\Delta\gamma$ constraints obtained by 
combining the measurements assuming $\Psi = \Phi$.

\begin{figure}[t]
\begin{centering}
{\includegraphics[width=3.2in]{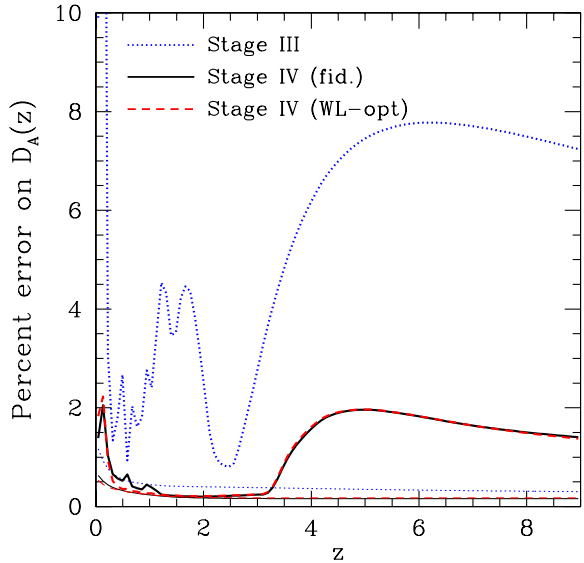}}
\caption{\label{fig:forecast_dz}
Predicted fractional error ($1\sigma$) on the distance $D_A(z)$
from our fiducial Stage III and Stage IV CMB+SN+BAO+WL
programs and the Stage IV program with optimistic weak lensing assumptions.
For each case, the lower, thin curve assumes a
$w_0-w_a$ dark energy parameterization, while the upper,
thick curve represents our binned $w(z)$ model.
}
\end{centering}
\end{figure}

\subsubsection{Distances}
\label{sec:forecast_distance}

As a target for alternative distance indicator methods
(\S\ref{sec:other_di}) and standard sirens (\S\ref{sec:sirens}),
Figure~\ref{fig:forecast_dz} plots the predicted fractional error on
the angular diameter distance from our fiducial Stage III
and Stage IV CMB+SN+BAO+WL programs.
If we assume a $w_0-w_a$ model then the constraints are
tight, better than $\approx 0.25\%$ for Stage IV and $\approx 0.5\%$
for Stage III at all $z>0.5$.
However, the highest redshift distance measurements included in
our forecasts (other than CMB) are BAO measurements at $z=3$, so when we change
to our general $w(z)$ model the distance errors at $z>4$
become dramatically worse, $\approx 2\%$ for Stage IV and 
$\approx 8\%$ for Stage III.  Furthermore, our Stage III 
forecast assumes a 0.8\% distance measurement from HETDEX
at $z=2.4$, which we consider somewhat optimistic because it
assumes that the full power spectrum shape can be used rather
than the BAO scale alone.
At $z<2$ the Stage III curve has a jagged structure that
depends to some degree on the specific choices we have made
in binning and in assigning BAO/WL measurements to particular
redshifts.  

The message to take away is that
Stage III distance errors for the general $w(z)$ model
should be in the $1-2\%$ range at $z<1$, the $3-5\%$ 
range at $1 < z < 2$, and the $6-8\%$ range at $z>4$,
with the errors at $2 < z < 4$ depending on the strength
of BAO measurements from \lya\ emission line galaxies (HETDEX)
or the \lya\ forest (BOSS).  On the Stage III timescale,
alternative distance measurements at $z>1$ with few percent
precision could reveal otherwise hidden departures from 
the $w_0-w_a$ model.  For Stage IV, where we assume powerful
BAO experiments extending to $z=3$, the demands on alternative
distance indicators are much stiffer.  
Even for the general $w(z)$ model, alternative measures at $z>4$
must reach 2\% precision to be competitive.
The Stage IV distance errors in this
model become large at $z < 0.25$, similar to
the several percent errors in $H_0$ seen in
Figure~\ref{fig:h_model_sn}.  As already 
discussed in \S\ref{sec:forecast_h0}, precise low redshift
distance measurements have the potential to reveal late-time
departures from smooth $w(z)$ evolution.

\subsection{Observables and Aggregate Precision}
\label{sec:forecast_aggregate}

We have characterized the performance of the fiducial
program and its variants in terms of their ability to constrain
parameterized models, from the specific
($w_0$-$w_a$ + GR) to the general ($w(z_i)$, $G_9$, $\Delta\gamma$).
An alternative, more model-agnostic approach to characterizing the power of
an experiment is via the aggregate precision with which
it measures its basic observable.
We have already introduced this idea at a few points, most notably
in our discussion of BAO.  By ``aggregate precision'' we mean
the fractional ($1\sigma$) error on an overall factor that
multiplies the observable in all redshift bins (and, if applicable,
angular or mass bins).  For the simple case of an observable $O$ with
independent fractional measurement errors $\Delta\ln O(z_i)$
in $N$ redshift bins, the aggregate
precision follows from the quadrature combination of the individual
errors:
\begin{equation}
\label{eqn:aggregate}
\Delta\ln O_{\rm agg} = 
  \left(\sum_{i=1}^N \left[\Delta\ln O(z_i)\right]^{-2}\right)^{-1/2}.
\end{equation}
One important virtue of forecasting an experiment's aggregate
precision is that it focuses one's attention on the required
control of systematics, especially systematics that are correlated
across redshift bins.

\begin{figure}
\begin{centering}
{\includegraphics[width=2.9in]{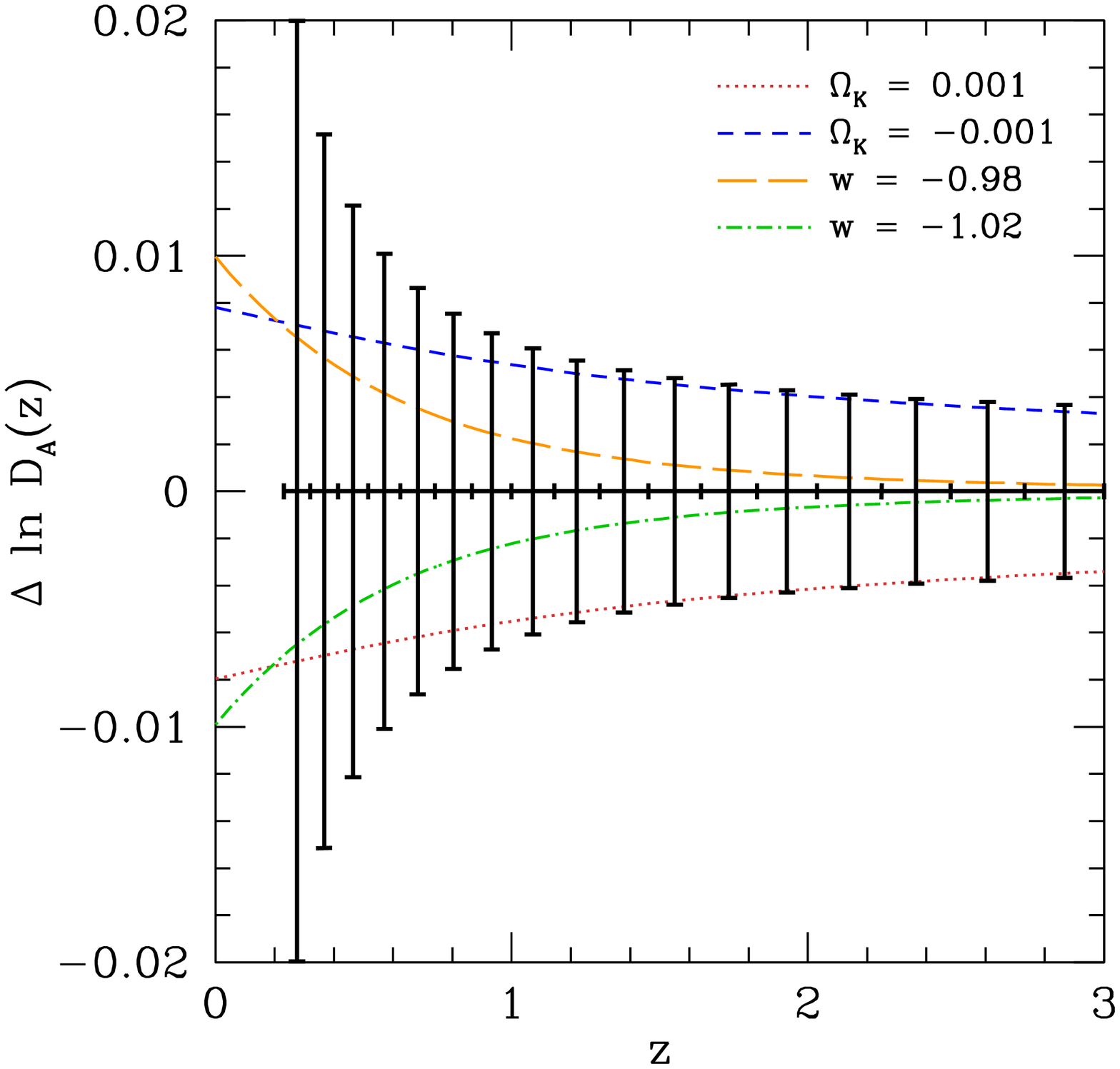}\hskip 0.2truein 
\includegraphics[width=2.9in]{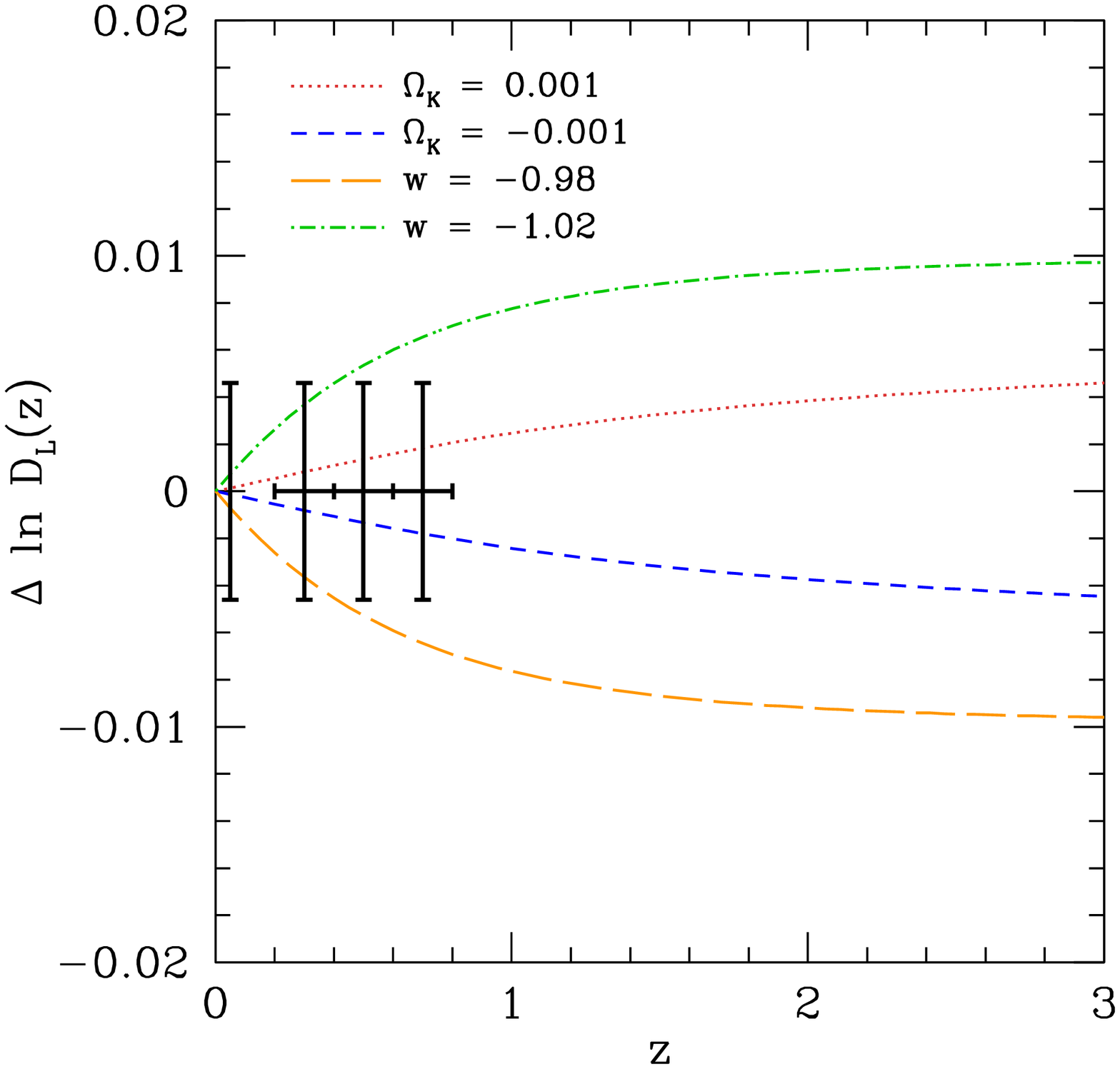}}
{\includegraphics[width=2.9in]{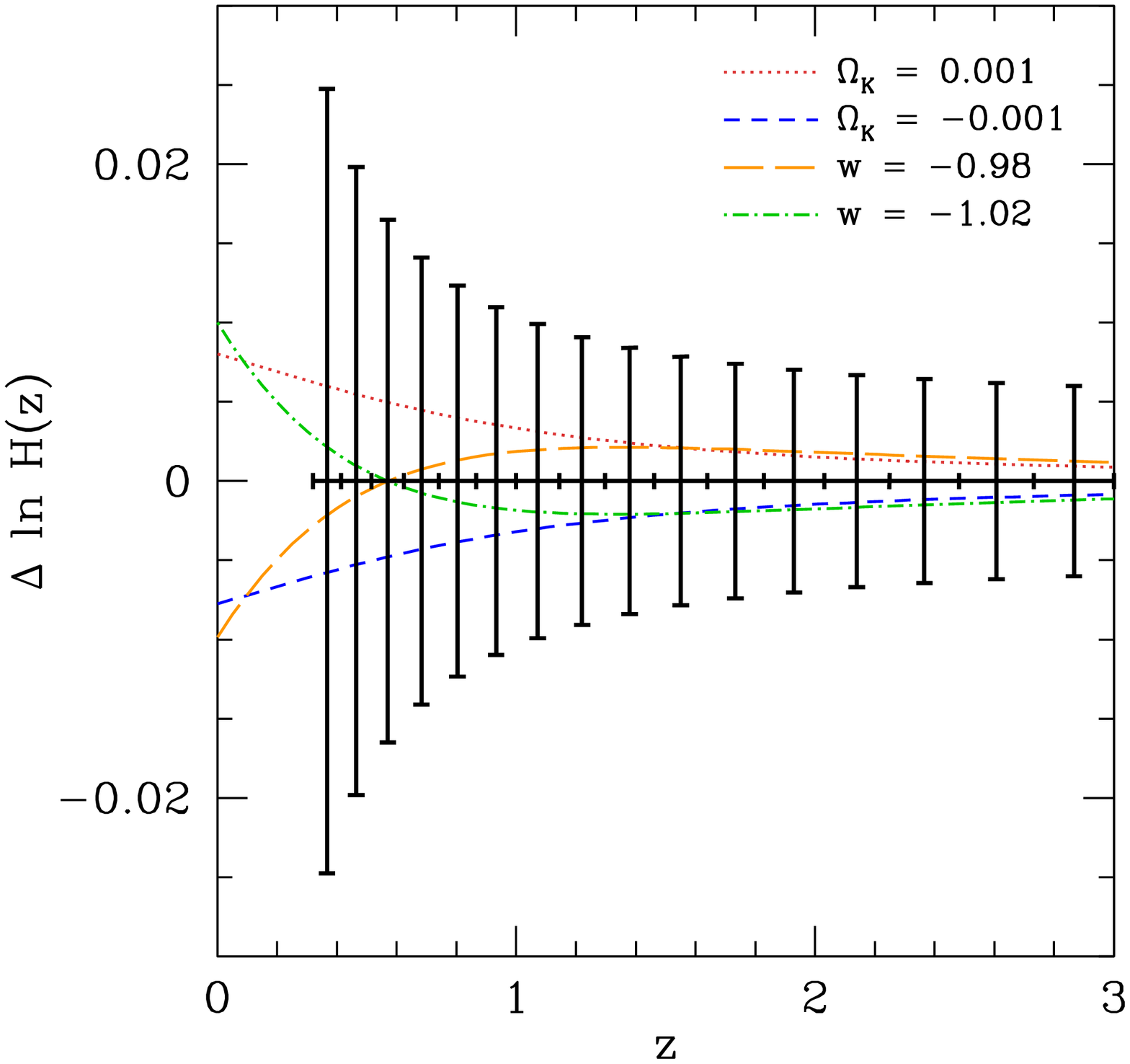}\hskip 0.2truein
{\includegraphics[width=2.95in]{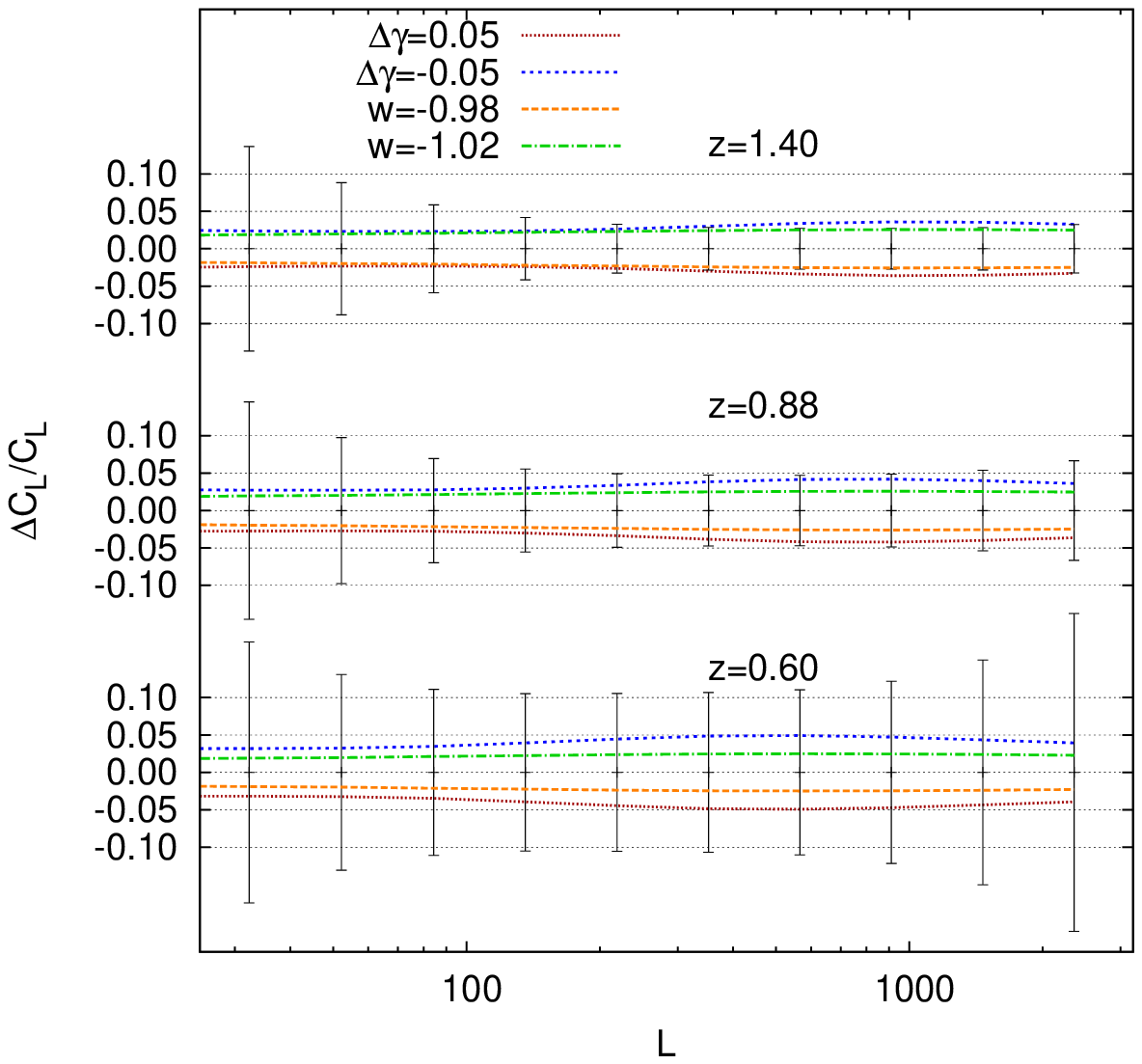}}
}
\end{centering}
\caption{\label{fig:stageiv_errors} 
Forecast errors for fiducial Stage IV SN, BAO, and WL programs, 
compared to model deviations from fiducial \lcdm.
The top panels show the forecast fractional errors in 
angular diameter distance from BAO (left) and luminosity
distance from SNe (right),
in absolute (Mpc) and relative ($\hmpc$) units, respectively.
The lower left panel shows the $H(z)$ error from BAO.
The errors are uncorrelated from bin to bin,
though the $D_A(z)$ and $H(z)$ errors in each bin are correlated 
with each other.
Curves show the fractional changes in the predicted relations relative to 
the fiducial cosmology for CMB-normalized models with
$1+w = \pm 0.02$ or $\ok = \pm 0.001$, as labeled.
Marginalization over $\mathcal{M}$ and $H_0$ allows arbitrary
vertical offsets to the model curves in the SN panel without
changing other cosmological parameters.
The lower right panel shows forecast statistical
errors (no systematics)
in the cosmic shear power spectrum for three of the 14
source photometric redshift bins, in bins of 
$\Delta\log l = 0.2\,$dex.  Statistical errors are approximately
uncorrelated for different $l$-bins at a given $z$, but errors in 
different $z$-bins are correlated.
Curves show the fractional
change in the predicted spectrum in models with $1+w = \pm 0.02$ or
$\Delta\gamma=\pm 0.05$.
}
\end{figure}

Figure~\ref{fig:stageiv_errors} plots the errors on the observables
in our fiducial SN, BAO, and WL programs.  For SN (upper right),
we adopt independent errors of 0.01 mag in each of four redshift
bins, corresponding to fractional errors in luminosity distance
of 0.46\%.  The aggregate measurement precision is therefore
$\Delta\ln D_{L,{\rm agg}} = 0.23\%$.  This is equal to the
aggregate precision forecast for the SN component of the \wfirst\
design reference mission (DRM1) forecast by \cite{green12})
for their ``optimistic'' assumption about 
SN systematics.\footnote{Specifically, that forecast assumes
uncorrelated systematic errors of $0.01(1+z)/1.8$ mag in 16
$\Delta z = 0.1$ redshift bins out to $z = 1.7$.  The total errors
have roughly comparable statistical and systematic contributions.}
However, the measurement in the $z = 0.05$ bin goes mainly
to constraining the nuisance parameter $\mathcal{M}$, the
SN absolute magnitude scale, so it is arguably better to characterize
our fiducial program's aggregate precision as $0.46\%/\sqrt{3} = 0.27\%$,
which is closer to that of the ``conservative'' \wfirst\ 
forecast (0.32\%).
More generally, we note that large local calibrator samples are
likely to achieve high statistical precision, and the systematic
uncertainty in relating this local sample to fainter, redshifted
samples will play a crucial role in determining the cosmological
performance of the SN program.

The left panels show the fractional errors predicted on $D_A(z)$
and $H(z)$ for the fiducial BAO program, as tabulated in 
Table~\ref{tbl:baoforecast}.  These error bars decrease with
increasing redshift because of the greater comoving volume per
$\Delta\ln (1+z)$ bin at high redshift.  Comparison of the
$D_L$ and $D_A$ panels nicely illustrates the complementarity
of SN and BAO as low and high redshift probes, respectively,
though recall that they provide distinct information even at
the same redshift because of relative vs.\ absolute calibration.
The aggregate precision of the BAO measurement is 
$\Delta\ln D_{A,{\rm agg}} = 0.13\%$, tighter
than that from SNe because of the larger number of bins.
Statistical errors in $H(z)$ are larger by a factor of 1.6 in
each redshift bin, so the aggregate precision of the $H(z)$ measurement
is lower by the same factor, 
$\Delta\ln H_{\rm agg} = 0.21\%$.
The $D_A$ and $H(z)$ errors are correlated, with a 
correlation coefficient of $\approx 0.41$ in each redshift bin.

Achieving the goals of our fiducial BAO program --- sampling
the equivalent of $\fsky=0.25$  with $nP \approx 2$ out to
$z=3$ --- will require multiple experiments probing different
redshifts and regions of sky.  While BigBOSS, \euclid, and
\wfirst\ all plan to measure BAO in the range $1 < z < 2$,
it is not clear that they can achieve $\fsky=0.25$ with
$nP\approx 2$ even collectively.  \euclid\ plans to survey
$\approx 14,000\mdeg^2$ over the range $0.7 < z < 2$ in its
6.25-year primary mission, but the forecasts in
\cite{green12}, which are based on the \euclid\ instrument
sensitivity of \cite{laureijs11} and the H$\alpha$ luminosity
function and galaxy bias measurements of
\cite{sobral12} and \cite{geach12}, imply that \euclid\ will
reach $nP < 0.5$ at $z > 1.2$.  BigBOSS plans to survey
$14,000\mdeg^2$ in the northern hemisphere, and a southern hemisphere
equivalent could increase the area to $24,000\mdeg^2$
(limited in the end by Galactic extinction).  Sampling density
forecasts are more uncertain for BigBOSS than for \euclid;
\cite{schlegel11} predict $nP > 2$ out to $z\approx 1.05$
and $nP > 0.5$ out to $z\approx 1.35$, falling to $nP = 0.35$
by $z = 1.65$.\footnote{\cite{schlegel11} use a different
convention, quoting $nP$ for the redshift-space power spectrum
at $k=0.14\invhmpc$ and $\mu=0.6$ instead of the real-space
power spectrum at $k=0.2\invhmpc$.  We have quoted the numbers
from their Table 2.3 as is, with no conversion to our $nP$ 
convention and no independent assessment of the sampling density
the instrument is likely to achieve.}
The \wfirst\ DRM1 of \cite{green12}, with 2.4 years devoted to
high-latitude imaging and spectroscopy, is projected to achieve
$nP \ga 1$ from $1.3 < z < 2.0$, declining to $nP \approx 0.5$
at $z=2.7$.  However, the survey area is only $3,400\mdeg^2$,
so a substantially extended mission would be required to reach
$10^4\mdeg^2$, and the depth is still $nP < 2$.  An implementation
of \wfirst\ using one of the NRO 2.4-m telescopes 
could plausibly
survey $10^4\mdeg^2$ with $nP = 1-2$,
depending on the instrument field of view and the time allocated
to the spectroscopic survey \citep{dressler12}.
In concert with
ground-based surveys covering $z \la 1.2$ and $z>2$, this
offers the best current
prospect of achieving something close to our fiducial BAO
program on the Stage IV timescale.  Breakthroughs in
21cm intensity mapping (see \S\ref{sec:bao_obs_tracers})
could also lead to major progress on this timescale.

As a context for assessing these projected measurement errors, curves 
in these panels show the impact of changing $w$ or $\ok$ in 
``CMB-normalized'' models, as described in \S\ref{sec:dependences}.
These curves are similar to those in Figure~\ref{fig:hzdz}, but
here we have adopted much smaller parameter changes,
$1+w=\pm 0.02$ or $\ok = \pm 0.001$, in line with the
tight constraints expected for Stage IV experiments.\footnote{These
models have $h=0.7030$ ($w=-0.98$), $h=0.7171$ ($w=-1.02$),
$h=0.7157$ ($\ok=0.001$) and $h=0.7045$ ($\ok=-0.001$).
Other parameters can be computed from the conditions
$\Omega_b h^2 = 0.02268$, $\Omega_c h^2 = 0.1119$, 
and $\Omega_\phi = 1-\Omega_c-\Omega_b-\Omega_k$.}
Note that a model that skirts the top of the $1\sigma$
error bars in $N_{\rm bin}$ redshift bins would be ruled out
at the $N^{1/2}_{\rm bin}$-$\sigma$ level.  However, while
one can see the partial tradeoff between curvature and $w$,
these plots do not capture the impact of degeneracy with 
other parameters such as $\om$ and $w_a$.
The behavior of the curves is explained
in \S\ref{sec:dependences} so we will not repeat it here,
but one can see the complementarity of SN and BAO distance
measurements in constraining $w$ and curvature, respectively,
and the roughly constant sensitivity of BAO $H(z)$ measurements
at $1 < z < 3$ to a change in the equation of state.
Some caution is required in interpreting the SN panel because
marginalizing over $H_0$ and $\mathcal{M}$ allows the model
curves to be offset vertically with no change in other parameters,
so the direct information about $w$ and $\ok$ resides in the
{\it slopes} of the curves relative to the data points.

Analogous to the SN and BAO panels of Figure~\ref{fig:stageiv_errors},
the lower right panel shows the projected $1\sigma$ statistical errors
of the WL power spectrum in logarithmic bins $\Delta\log l = 0.2$ dex,
for three of the 14 tomographic bins of source photometric redshift.
Several caveats are in order.  First, while the statistical errors
in different $l$ bins at fixed redshift are independent, errors among
redshift bins are correlated because structure at redshift $z_l$ contributes
to the lensing of all background shells at $z_s > z_l$.  Second, systematics in
shape measurement or photometric redshift calibration will typically
produce errors that are correlated across both redshift and angle;
here we have plotted only statistical errors.  Third, in addition to the 11
auto-correlation power spectra not shown here, our fiducial program
includes $14\times 13$ shear cross-power spectra, and cross-spectra
between shear fields and galaxy density fields.  All of these provide
cosmological information, albeit with correlated errors 
and some loss of constraining power through marginalization over galaxy
bias and intrinsic alignments.  Finally, the shear power spectrum
depends on both geometry and structure growth: for sources at $z_s$
lensed by matter at $z_l$, the expected shear depends on 
$D_A(z_l)$, $D_A(z_s)$, and $\sigma_8(z_l)$.

Correlated errors, the multitude of auto- and cross-correlations, and
the linked parameter dependences mean one cannot characterize the
information content of WL measurements as simply as that of SN or
BAO measurements.  We can nonetheless define an aggregate precision
as the fractional error on the matter fluctuation amplitude $\sigma_8$
with all other parameters --- and thus $D_A(z)$, $H(z)$, and $G(z)$ ---
held fixed.  
From equation~(\ref{eq:wl:ce2}) one can see that the fractional error
on an overall scaling of $D_A(z)$ with fixed matter clustering would 
be similar to this fractional error on $\sigma_8$ with fixed geometry.
%The fractional error on the amplitude of the shear power spectrum,
%which scales as $\sigma_8^2$, would be a factor of two larger.
For our fiducial Stage IV WL program we find an aggregate precision
on $\sigma_8$ of 0.33\%, where our calculation includes marginalization over 
the assumed $2\times 10^{-3}$ systematic uncertainties in shear 
calibration and photometric redshift offsets (and over parameters
describing intrinsic alignments).
The uncertainty in this case is dominated by these systematics,
and the aggregate error is close to the quadrature sum (0.28\%) of these
two fractional contributions.
For the optimistic WL case, with total errors double the statistical
errors (and thus double those plotted in Fig.~\ref{fig:stageiv_errors}), 
the aggregate precision on $\sigma_8$ is 0.14\%.
If we assumed purely statistical errors, as plotted in 
Figure~\ref{fig:stageiv_errors}, then the aggregate precision
would of course be a factor of two higher.
As already discussed in \S\ref{sec:fiducial}, it is likely
that LSST, \euclid, and \wfirst\ will collectively, and perhaps
even individually, exceed the performance of our fiducial
Stage IV program as far as statistical errors are concerned.
The key question is whether they will achieve the tight level
of systematics control that we assume.  In principle these
experiments could collectively achieve an aggregate precision
several times better than that of even our optimistic WL forecast,
with cross-checks between them testing for any experiment-specific
systematics.

For a given photo-$z$ bin, the statistical errors per
$\Delta\log l = 0.2\,$dex bin shown in Figure~\ref{fig:stageiv_errors}
shrink slowly with increasing $l$ because of decreased cosmic variance,
then grow slowly as shape noise errors become dominant
(see \S\ref{sec:wl_stat_errors}).
Errors are smaller for the higher photo-$z$ bins because of the 
larger numbers of source galaxies the larger foreground volume.
Orange and green curves show the impact of $1+w=\pm 0.02$ variations,
which is comparable to the $1\sigma$ error per $\Delta\ln l$ bin for
$z_p = 0.88$ and $z_p = 1.40$.
Red and blue curves show the impact of setting the growth index parameter
to $\Delta\gamma = \pm 0.05$, with all other cosmological parameters
fixed.  Because the logarithmic growth rate is 
$f(z) \approx [\om(z)]^{\gamma+\Delta\gamma}$, a negative
$\Delta\gamma$ corresponds to faster growth and thus higher $C_l$.
These $\Delta\gamma$ and $1+w$ changes have effects of similar
magnitude but with different redshift dependence, so in principle
WL measurements can break the degeneracy between them.
In practice, the strongest degeneracy breaking will likely
come from combining WL data with SN and BAO constraints, which
are independent of $\Delta\gamma$.

\begin{figure}
\begin{centering}
{\includegraphics[width=2.9in]{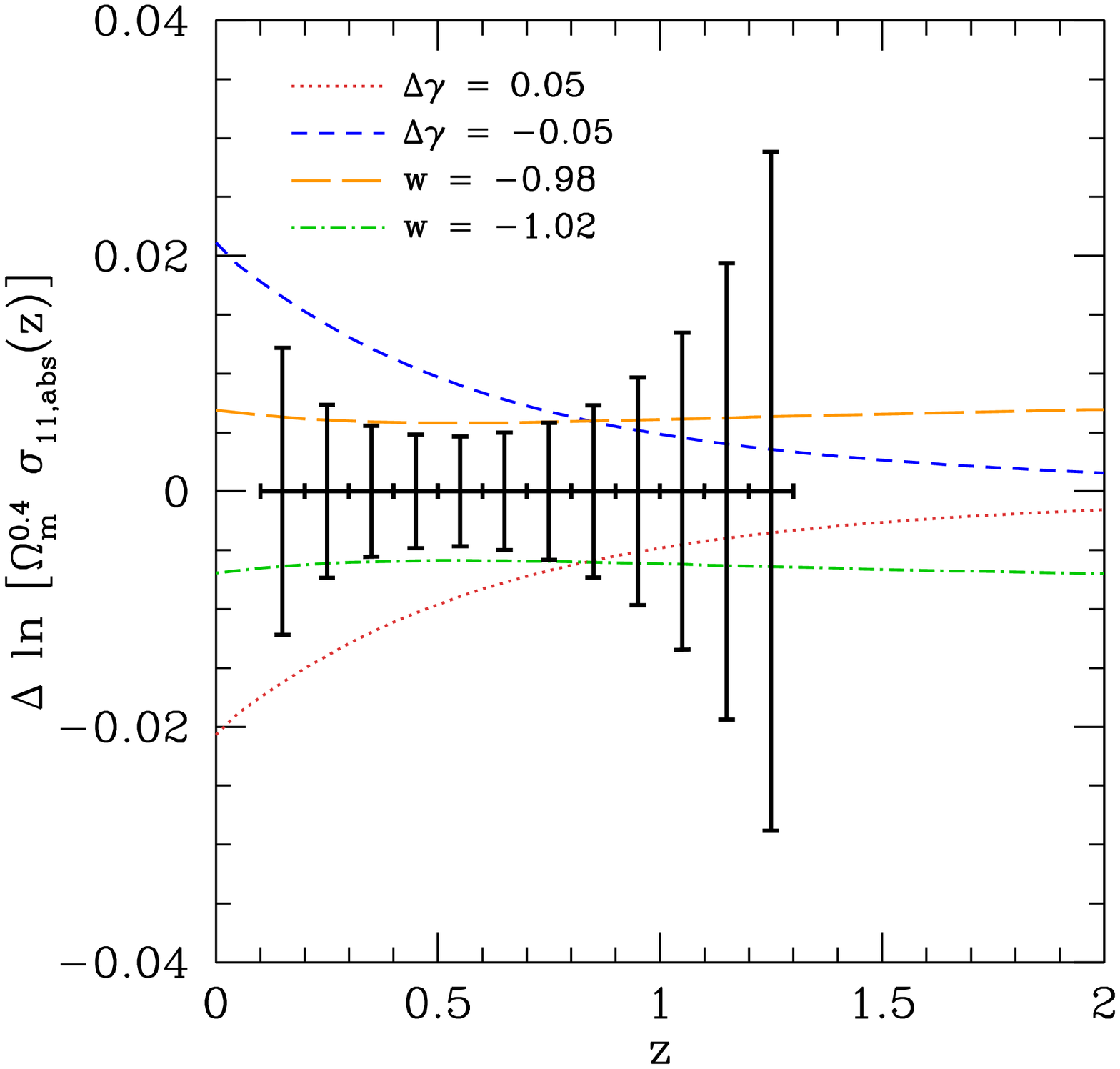}\hskip 0.2truein
\includegraphics[width=2.9in]{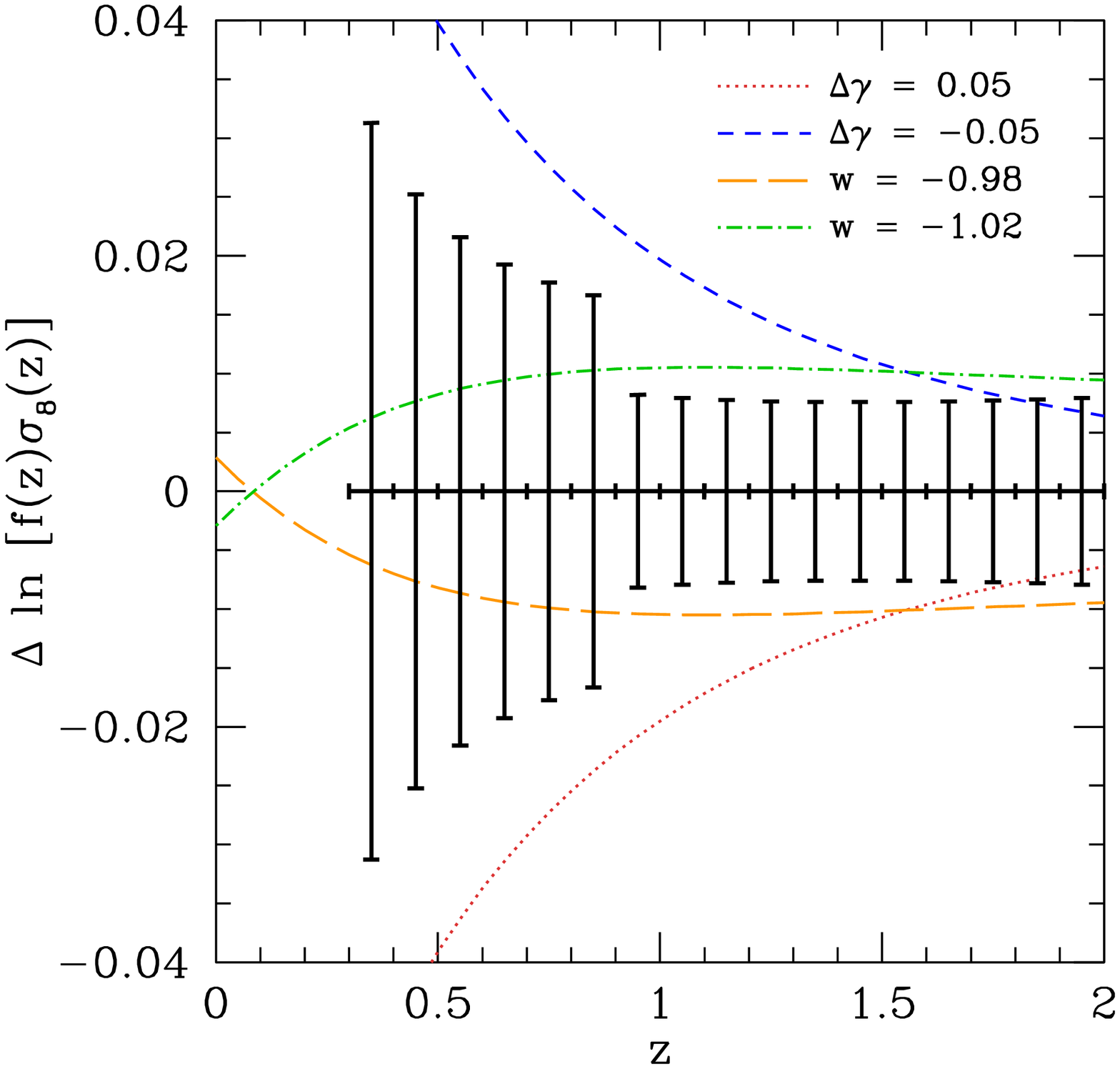}}
\end{centering}
\caption{\label{fig:stageiv_errors2} 
Forecast errors for fiducial Stage IV cluster (left) and RSD (right) programs,
with the assumptions described in the text.
Curves show the predicted deviations from our fiducial \lcdm\ model
for CMB-normalized models with $1+w=\pm 0.02$ and $\Delta\gamma = \pm 0.05$.
}
\end{figure}

Figure~\ref{fig:stageiv_errors2} presents error forecasts for 
two other probes of structure growth, clusters and redshift-space
distortions.  The cluster errors are based on Figure~\ref{fig:wl_serr},
assuming stacked weak lensing mass calibration of 
$M > 2\times 10^{14}M_\odot$ clusters over $10^4\mdeg^2$ with a
WL source density of $30\,\arcmin^{-2}$.  While 
Figure~\ref{fig:wl_serr} is couched in terms of errors on
$\sigElevz$ with other parameters held fixed, here we assume
(as in \S\ref{sec:forecast_cl}) that the constrained quantity
is $\Omega_m^{0.4}\sigElevz$, with the same fractional error;
we caution that this is only an approximate characterization
of the constraints from clusters.
If the errors are dominated by WL shape noise and random cluster
orientations, as assumed here, then they should be essentially
uncorrelated among redshift bins.  The aggregate precision for 
the cluster program is 0.20\%, and since this is comparable to that of 
our other fiducial programs, it is not surprising that clusters
have a significant impact on the expected uncertainties in 
equation-of-state and growth parameters (\S\ref{sec:forecast_cl}).
Achieving this high aggregate precision would demand tight control
of the systematics discussed in \S\ref{sec:cl_systematics},
including the effects of contamination, incompleteness, and
mis-centering, the impact of mass-observable scatter, and the
prediction of the mass function and stacked WL profiles in the
presence of baryonic effects.

The RSD error bars in Figure~\ref{fig:stageiv_errors2} are those
shown previously by the lower solid curve in Figure~\ref{fig:rsderr},
computed with the forecasting code of \cite{white09}.
They assume a galaxy sample like that of our fiducial BAO program
out to $z=2$ and full use of information up to comoving
$k = \kmax=0.2\invhmpc$ at each redshift.
The sharp change in errors at $z=0.9$ is due to the assumed
drop in bias factor as surveys transition from absorption-line
galaxies to emission-line galaxies.  The aggregate precision
on $f(z)\sigma_8(z)$ is 0.22\%, again comparable
to that of our other fiducial programs.  
The ranges $z > 1.4$ and $z < 1.4$ make
equal contributions to this precision.
As discussed in \S\ref{sec:rsd}, we expect the dominant systematic
uncertainty for RSD to lie in theoretical prediction of the
RSD signal in the presence of non-linear gravitational evolution
and galaxy bias, not the measurements themselves.
To realize the full statistical power of Stage IV galaxy redshift
surveys, these theoretical uncertainties must be controlled at 
the 0.2\%-level.
The \cite{white09} forecasts, which assume that $\kmax$ scales
with the non-linear wavelength $k_{\rm nl}(z)$ of the matter
power spectrum, yield smaller errors at high redshift and an
aggregate precision of 0.10\%.

Curves in Figure~\ref{fig:stageiv_errors2} again show the effect of
isolated parameter changes with $1+w=\pm 0.02$ and $\Delta\gamma=\pm 0.05$.
In isolation, either of these changes would
be strongly ruled out by either the fiducial cluster program
or the fiducial RSD program with the assumptions adopted here.
For clusters, the impact of the $w$ changes is comparable to the
$1\sigma$ error bar per $\Delta z = 0.1$ redshift bin over the
range $0.2 < z < 0.8$.  
The impact of $\Delta\gamma$ changes exceeds
the $1\sigma$ error at all $z < 0.8$.  For RSD, the 
$\Delta\gamma = \pm 0.05$ impact exceeds the $1\sigma$ per-bin error
at all $z<1.7$.  In these
CMB-normalized models, where changes to $w$ and $\om$ have counteracting
effects on $f(z)$, the sensitivity of RSD to a constant-$w$ change is
greatest at high redshifts.  The impact of $\Delta w = \pm 0.02$ exceeds
the per-bin $1\sigma$ error for $z \geq 0.9$.

Quantities like the DETF FoM and errors on $\Delta\gamma$ or $\ln G_9$
are useful for optimizing choices in a well defined experimental
program, e.g., area vs. depth or the value of different target
classes in a spectroscopic survey.  However, since we have little
idea where deviations from GR+$\Lambda$ are likely to show up
(if they are there at all), we think that aggregate precision, including
the effects of systematics, is a comparably useful tool for 
providing seat-of-the-pants guidance in a more general situation.
For a given level of aggregate precision, a measurement at low
redshift will typically have more direct sensitivity to dark energy,
a measurement at high redshift will typically have more direct
sensitivity to curvature, and measurements over a range of redshifts
are needed to constrain dark energy evolution.  However, given the
degeneracies among parameters (especially $w$, $\om$, and $\ok$) and
the powerful impact of CMB constraints, it is difficult to identify
a specific redshift range as the optimal one to probe.
For a given method and a given facility, it makes sense to start
where the pickings are easy, in terms of gaining precision relative
to existing knowledge, and move to more difficult terrain when required.
This assessment must also include consideration of where one can most
readily control systematics, which is often but not always at
low redshift.  Extending the redshift range of a method increases
leverage for breaking degeneracies and constraining dark energy 
evolution, but the more important impact is often to improve
the method's aggregate precision by bringing in measurements
with decorrelated errors.  As we have emphasized repeatedly, a full program
should employ multiple methods to take advantage of their complementary
information content and redshift sensitivity and to
cross-check for unrecognized systematics.  Fortunately, 
Figures~\ref{fig:stageiv_errors} and~\ref{fig:stageiv_errors2}
show that several methods have the potential to achieve
$0.1-0.3\%$ aggregate precision in Stage IV experiments, 
a dramatic improvement on the $\sim 1-5\%$ precision that represents
the current state of the art for these methods.

\subsection{Prospects with Many Probes}
\label{sec:forecast_multiprobe}

\begin{figure}[ht]
\begin{centering}
{\includegraphics[width=5.5in]{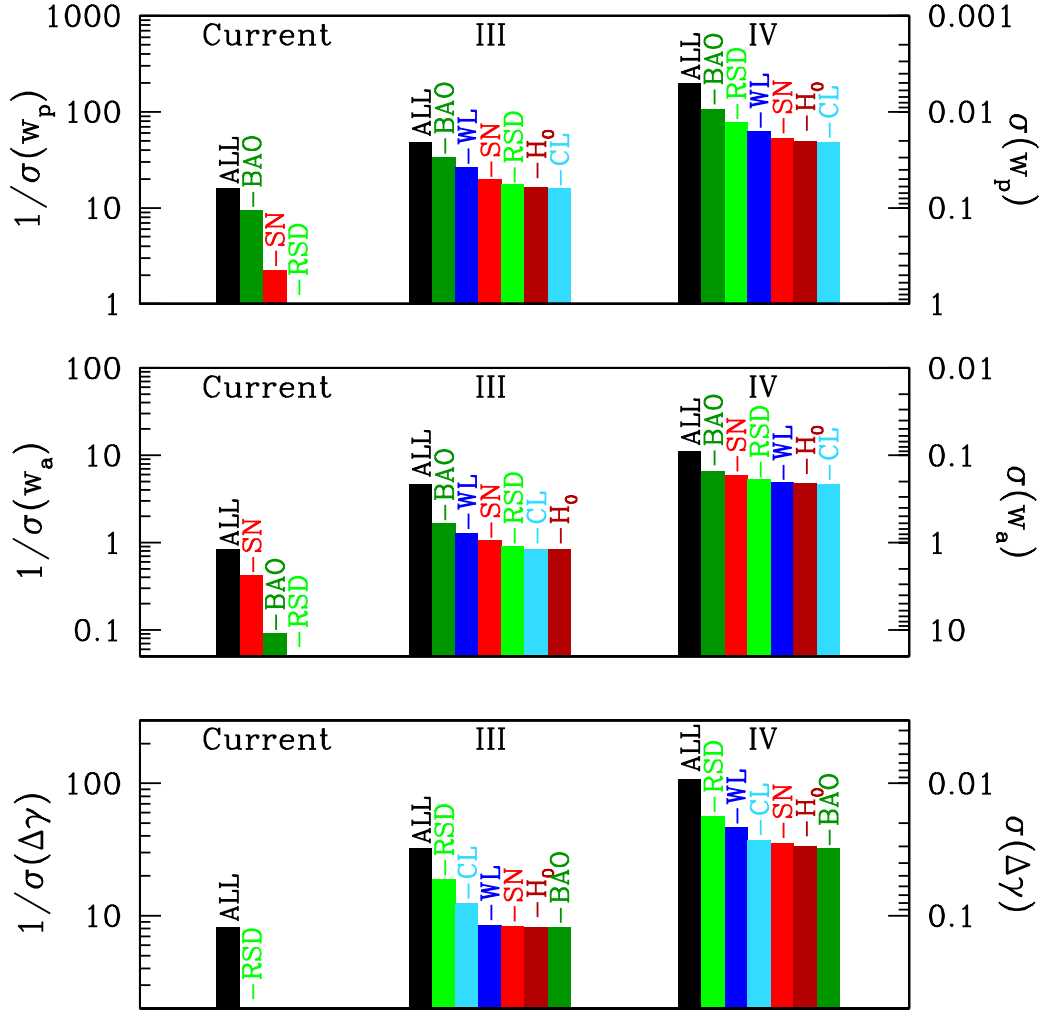}}
\caption{\label{fig:multiprobe}
Forecasts for inverse $1\sigma$ errors (left axis; errors themselves on
the right axis) on $w_p$, $w_a$, and $\Delta\gamma$ from combining
our fiducial CMB+SN+BAO+WL programs with additional constraints 
from redshift-space distortions (RSD), clusters (CL), and direct
$H_0$ measurements.  See text for a description of the errors
assumed for Current (left), Stage III (middle), and Stage IV (right)
forecasts.  Black bars show the results of combining all of these
probes.  Colored bars show the cumulative impact of dropping probes
in succession (``$-$BAO'' should be read as ``minus BAO,'' for example).
When a Stage IV probe is ``dropped'' it is
set to its Stage III precision, and when a Stage III probe is ``dropped''
it is set to its current precision.  
CMB constraints are always retained.
}
\end{centering}
\end{figure}

Section~\ref{sec:results} demonstrates the power of a combined
CMB+SN+BAO+WL experimental program, while
\S\S\ref{sec:forecast_cl}-\ref{sec:forecast_aggregate}
show that other probes could add substantial further sensitivity
to dark energy or modified gravity.  Drawing these results together,
we show in Figure~\ref{fig:multiprobe} the result of combining
our fiducial CMB+SN+BAO+WL programs with representative performance
estimates for clusters, redshift-space distortions, and direct $H_0$
measurement.  (While the AP test could also play an important
role, we consider current understanding of its systematic
uncertainty too limited to allow even representative performance
estimates.)  Top, middle, and bottom panels show inverse errors
on $w_p$, $w_a$, and $\Delta\gamma$, respectively, assuming a
$w_0-w_a$ model with $G_9$ and $\Delta\gamma$ as beyond-GR
growth parameters.  Black bars show the results of combining all
of these probes, while colored bars show the cumulative impact
of successively omitting individual probes (see further 
explanation below).

For Stage IV we assume our fiducial CMB, SN, BAO, and WL constraints,
the cluster and RSD constraints described in 
\S\ref{sec:forecast_aggregate} (Fig.~\ref{fig:stageiv_errors2}),
and an $H_0$ constraint with precision of 1\%.
%and we include cluster (CL) constraints as discussed in 
%\S\ref{sec:forecast_cl}, assuming a cluster sample that reaches
%to $2\times 10^{14}M_\odot$ over $10^4\mdeg^2$, limited by the mass 
%calibration uncertainty from stacked weak lensing with a
%source density of 30 arcmin$^{-2}$ (see \S\ref{sec:cl_mass_calibration}).
%For $H_0$ we assume a precision of 1\%.  For 
%redshift-space distortions (RSD), we take the forecast errors 
%on $f(z)\sigma_8(z)$ that
%\cite{white09} present for a \euclid/\wfirst\ like survey
%(see \S\ref{sec:rsd}).
For Stage III we adopt the CMB+SN+BAO+WL errors summarized
in Table~\ref{tbl:key}.  (Note, in particular, that our assumed
Stage III SN errors are 0.02 mag per $\Delta z=0.2$ bin and 
our Stage IV errors are 0.01 mag per bin, with the same error
for the local calibrator sample at $z=0.05$.
For these plots, though not for others in the paper, we also include
the Union2 SN constraints when computing Stage III and Stage IV.)
For Stage III clusters we assume $5000\mdeg^2$ and a source density of
10 arcmin$^{-2}$ for mass calibration (both appropriate
to DES), while keeping the mass threshold at $2\times 10^{14} M_\odot$.
For $H_0$ we assume 2\% errors, and for RSD we take the
\cite{white09} forecasts for BOSS.
Finally, for current data we take \wmap\ CMB errors,
Union2 SN errors, and the BAO data and errors described in
\S\ref{sec:bao_current}.  
We adopt the RSD errors reported
by \cite{blake11a} from WiggleZ (see \S\ref{sec:rsd}).
We also include a 3\% error on $H_0$, and a 4\% error
on $\sigma_8\Omega_m^{0.4}$ to represent clusters and weak lensing
(see \S\ref{sec:cl_current}).

Beginning with the black bars representing the full combinations,
we see that these projections predict improvements of 
more than an order-of-magnitude for each of the
three parameters --- $w_p$, $w_a$, and $\Delta\gamma$ ---
between current knowledge and Stage IV results.  
These combinations yield $1\sigma$ errors of approximately
0.005 on $w_p$, 0.1 on $w_a$, and 0.01 on $\Delta\gamma$,
testing the $\Lambda$CDM model far more stringently than
it has been tested to date.  Stage III
projections are roughly the geometric mean of current and 
Stage IV constraints in all cases.

It is interesting to ask what the different methods contribute
to this performance, but there is no unique way to decompose
a constraint into a sum of individual contributions, and the
apparent relative importance of different components depends
on how the decomposition is done.  We have attempted one 
form of ``even-handed'' decomposition by dropping individual 
probes in succession, beginning with the probe whose omission
causes the largest increase in the parameter error, then the
probe that causes the largest increase after the first probe
has already been dropped, and so forth.  However, when we
``drop'' a probe we do not omit it entirely; rather, we set 
the error for that probe in the Stage IV forecast equal to
the value we previously assumed for the Stage III forecast,
or we set the error in the Stage III forecast equal
to the value adopted for current data.  Thus, for example, the dark 
green bar in the upper right shows the impact on $\sigma(w_p)$
of replacing the Stage IV BAO constraints with the Stage III
BAO constraints.  The light green bar next to it shows the
impact of {\it also} setting the RSD constraint to the 
Stage III value, the dark blue bar the impact of also setting
the WL constraint to the Stage III value, and so forth.
To give one more example, the light blue bar in the middle of
the bottom panel shows the error on $\Delta\gamma$ using
Stage III WL+SN+BAO+$H_0$ but current constraints for 
RSD and clusters.  If the Stage III WL improvement is also
dropped (dark blue bar) then there is no improvement over
current knowledge of $\Delta\gamma$ because none of the
remaining probes (SN, BAO, $H_0$) directly measures structure
growth.  We {\it always} include CMB constraints, with
WMAP9 errors for current and \planck\ errors for Stage III
and Stage IV.  By construction, the rightmost colored bar
for a given stage matches the black bar of the previous
stage, since we have then set all probes back to their value in the 
previous stage.

We caution against reading too much into the ordering of probes
in Figure~\ref{fig:multiprobe} because it depends in detail on our
assumptions about the expected errors of the individual components;
furthermore, a probe only gains in this plot
based on its {\it differential} improvement between current performance and 
Stage III or between Stage III and Stage IV.
The detailed examination of CMB+SN+BAO+WL in
Tables~\ref{tbl:forecasts1}-\ref{tbl:forecasts3} and the associated
figures provides much more nuanced information.
These caveats notwithstanding, Figure~\ref{fig:multiprobe} demonstrates
several interesting points.  In present data, 
%SNe dominate constraints
%on the equation-of-state parameters, with BAO alone providing
%much weaker constraints.  
BAO make the largest contribution to $w_p$ constraints and
SNe to $w_a$ constraints,\footnote{Interestingly, the roles of BAO
and SNe in the current $w_p$ constraint are reversed relative to the
original {\tt arXiv} posting of our article because of the 
inclusion of the new SDSS DR7 and BOSS measurements, both published in 2012.}
though it is really the combination of the two with CMB data that is
required to achieve interesting constraints in a model space that
allows $w_p$, $w_a$, and $\ok$ to vary simultaneously.
BAO become more powerful in our
fiducial Stage III and Stage IV programs, making the
largest contribution to both the $w_p$ and $w_a$ constraints.  
Current constraints 
on $\Delta\gamma$ rely entirely on RSD, as the cluster constraint
on $\sigma_8 \Omega_m^{0.4}$ is degenerate with $G_9$.
With our adopted error forecast, RSD remains the most
powerful contributor to $\Delta\gamma$ constraints at Stage III 
and Stage IV, outweighing both WL and clusters.  Indeed, with
these errors Stage IV RSD also makes an important contribution
to the $w_p$ measurement.  WL and clusters make significant
contributions to $\Delta\gamma$ constraints but have limited impact on $w_p$
and $w_a$.  

\begin{figure}[ht]
\begin{centering}
{\includegraphics[width=5.5in]{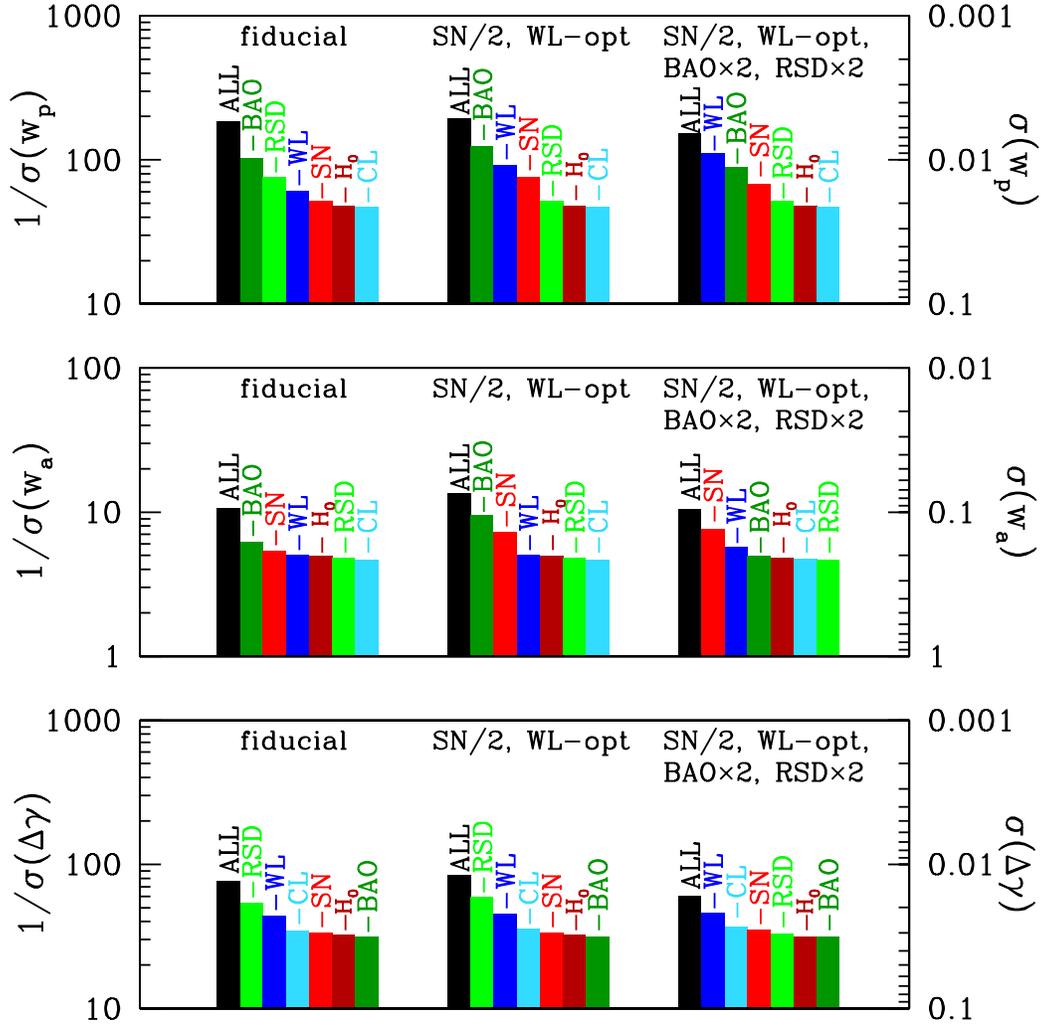}}
\caption{\label{fig:multiprobe2}
Like Figure~\ref{fig:multiprobe}, but showing variations on
the Stage IV fiducial program.
The middle column shows forecasts with the optimistic WL
systematics and SN errors reduced by a factor of two.
The right column shows the effect of, additionally, doubling
the errors on RSD and BAO.  The left column repeats the
fiducial case from Fig.~\ref{fig:multiprobe} for reference.
}
\end{centering}
\end{figure}

Figure~\ref{fig:multiprobe2} shows two variants on the fiducial
Stage IV case.  In the middle column we consider a combined
program with SN errors improved by a factor of two and the optimistic
WL systematics.  The forecast errors on $w_p$, $w_a$, and $\Delta\gamma$
shrink by 6\%, 21\%, and 8\%, respectively.
WL and SN now leapfrog RSD as contributors to the $w_p$
error, though they still contribute less than BAO.
The improvement in $w_a$ is driven by the supernova improvement,
though BAO remains the largest contributor.
The right column shows the effect of,
additionally, doubling the errors on BAO and RSD, since
our fiducial assumptions for these programs are perhaps
overoptimistic compared to the capabilities of planned Stage IV
experiments.  
The forecast errors are larger than they are for the fiducial 
case but by moderate amounts, 20\% ($w_p$), 14\% ($w_a$), and 
28\% ($\Delta\gamma$).  WL leapfrogs BAO to become the strongest 
contributor to $w_p$ precision, while SN and WL both leapfrog BAO
for $w_a$.  With doubled measurement errors, RSD makes only a modest
contribution to the parameter constraints, even for $\Delta\gamma$.

Perhaps the most important message to take from 
Figures~\ref{fig:multiprobe} and~\ref{fig:multiprobe2}
is that these six probes together
with CMB measurements provide a tight web of constraints on
cosmic acceleration models, and that even if one or two methods
prove disappointing, there are others (including ones not shown
in this plot) to take up slack.  We have focused much of our
review on the stiff challenges of controlling systematic errors
at the level demanded by future dark energy experiments.
However, given the ingenuity of the community in devising and
refining analysis methods, we are optimistic that the powerful
data sets provided by these experiments will ultimately lead to 
constraints at the high end of our forecasts.

\vfill\eject
\section{Conclusions}
\label{sec:conclusions}

The first evidence for dark matter emerged from studies
of galaxy clusters in the 1930s \citep{zwicky33}, and
the dark matter problem assumed a central position in cosmology
after technological advances allowed dynamical measurements
in the outer regions of individual galaxies
(\citealt{rubin70,rogstad72}; see review by \citealt{faber79}).
Because it clusters on small scales, dark matter has a
rich phenomenology, and detailed studies of galaxies,
galaxy clusters, large scale structure, the \lya\ forest,
and the CMB have largely pinned down its properties even
though we have yet to identify the dark matter particle
or particles.  The implications of the dark matter problem
have proven even more profound than might have been imagined
in the 1930s, pointing the way to an entirely new form of
matter whose cosmic mean density exceeds that of all baryonic
material by a ratio of 6:1.  There are now several plausible
ideas of what dark matter might be --- ideas that are rooted
in well motivated extensions of the standard model of particle
physics and that (at least in some cases) naturally explain
the observed density of dark matter \citep{bertone05}.
With experimental methods advancing on many fronts, there are
good reasons to hope that dark matter will soon be identified
in particle accelerators, detected directly in underground
experiments, or detected indirectly via its annihilation into
$\gamma$-rays, neutrinos, or cosmic rays.

Evidence for cosmic acceleration began to emerge in the early
1990s, and it rapidly evolved into a near-airtight case following
the supernova discoveries of the late 1990s (see \S\ref{sec:history}).
Whether the cause is a new energy component or a breakdown of
GR, the implications of cosmic acceleration are dramatic,
even more so than those of dark matter.
Cosmic acceleration may ultimately provide clues to the nature
of quantum gravity, or to the structure of the universe on
scales beyond the Hubble volume, or to its history over times
longer than the Hubble time.
There are already many theories
of cosmic acceleration, but none of them offers a convincing
explanation of the observed magnitude of the effect, and
nearly all of them were introduced to explain the observed
acceleration, rather than emerging naturally out of fundamental
physics models.  In contrast to dark matter, most models of
dark energy predict that it is phenomenologically poor,
affecting the overall expansion history of the universe but
little else.  That impression could yet prove incorrect:
other signatures of ``cosmic acceleration physics'' might
appear in small-scale gravitational experiments, in the behavior
of gravity in different large scale environments,
or in non-gravitational interactions.

While the solution to the cosmic acceleration problem could
come from a suprising direction, including theory, there is
a clear experimental path forward through increasingly
precise measurements of expansion history and growth of
structure.  Relative to current knowledge, 
Stage IV experiments can improve the measurement of basic
cosmological observables --- $H(z)$, $D(z)$, and $G(z)$ ---
by one to two orders of magnitude.  Correspondingly, they
can achieve a $1-2$ order-of-magnitude
improvement in constraints on $w$, $2-3$ orders-of-magnitude improvement
in the DETF figure-of-merit, and still greater gains in higher
dimensional parameterizations, including tests of GR violations.
Any robust deviation from a cosmological constant model
would have profound implications, and the greater the precision
and detail with which such a deviation is characterized, 
the greater the direction for understanding its cause.

We have reviewed in considerable detail the four leading
methods --- supernovae, BAO, weak lensing, and clusters ---
and we have briefly discussed some of the emerging new
methods, whose capabilities and limitations are as yet 
less thoroughly explored.  We have also investigated the complementarity
of these methods for constraining theories of cosmic
acceleration.  We have spent little time on the CMB as it
has little {\it direct} constraining power on these theories,
but it does provide crucial constraints on other cosmological
parameters that are essential to precision tests.
We now conclude our article with an editorial recap of our
main takeaway points.

Type Ia supernovae have unbeatable precision for measuring
distances at $z \la 0.5$.  Future surveys can readily achieve
statistical errors of 0.01 mag or less (0.5\% in distance)
averaged over bins of $\Delta z = 0.2$.  The challenge is
getting systematic uncertainties at or below the level of
such statistical errors.  In our view, the key systematics
for SN studies are imperfect photometric calibration, 
evolution in the population of SNe represented at different redshifts,
and the effects of dust extinction.  The first can be addressed
by careful technical design, of the instruments used for
SN surveys and of the observing and calibration procedures.
The second can be addressed by obtaining
high quality observations of the SNe and their
host galaxies that allow one to match the properties
of high and low redshift systems.  The third is best addressed
by working in the rest-frame near-IR, where extinction is low.
Rest-frame IR observations may also mitigate evolution systematics
and improve statistical errors, since current observations indicate
that the scatter in SN luminosities is smaller in the near-IR
than in the optical.

The BAO method complements the SN method in several ways.
SN measure distance ratios relative to local calibrators
(i.e., distances in $\hmpc$), while BAO measure absolute
distances (in Mpc) assuming a calibration of the sound horizon.
SN and BAO measurements at the same redshift therefore provide
complementary information, effectively constraining $H_0$, which
is itself sensitive to acceleration when combined with CMB data.
Spectroscopic BAO measurements that sample a constant fraction
of the sky become {\it more} precise at high redshift because
they cover a greater comoving volume and because they measure
$H(z)$ directly in addition to $D_A(z)$.  (Of course, they also
require a larger number of tracers to probe these larger volumes,
and the tracers themselves are fainter at higher redshifts.)
Cosmic variance limited BAO surveys have roughly constant
sensitivity to dark energy over the range $1 < z < 3$ because
the decreasing dynamical impact of dark energy at higher 
redshifts is balanced by the greater BAO measurement precision.
Furthermore, the BAO method is the only one that we expect to
be statistics-limited even with Stage IV surveys.  
Non-linear matter clustering and non-linear galaxy bias may
shift the BAO peak by more than the statistical errors of
Stage IV experiments, but the shifts can be computed using
theoretical models that are constrained by the smaller scale
clustering data, and moderate fractional accuracy in these
corrections is enough to keep any uncertainty in the corrections
well below the statistical errors.
Thus, we see the main challenge for the BAO method as finding
ways to efficiently map the available structure.
There are several promising ideas, both ground-based and 
space-based, and Stage IV BAO constraints will likely come from
a union of several approaches covering different redshift ranges.

Weak lensing measurements provide sensitivity to both the 
distance-redshift relation and the growth of structure.
The statistical precision achievable with future facilities
is very high, so the challenge is reducing systematic
uncertainties to a level that does not overwhelm these
statistical errors.  The most important problem is reducing
multiplicative shape measurement biases to the level of $\sim 10^{-3}$ 
or below, which requires (among other things) determining
the PSF that affects the galaxy images to very high accuracy.
This is an area of highly active research, and it is not yet
clear what approach will prove most successful; we have
advocated pursuit of a Fourier method that becomes exact
in the limit of high S/N ratio.
Since most shape measurement systematics
depend inversely on the ratio $r_{50}/r_{\rm PSF}$ of galaxy
size to PSF radius, one can mitigate these systematics by
restricting the analysis to larger galaxies, but this gives
up statistical precision by reducing the surface density of
usable sources.  The second major challenge for WL studies
is the measurement and calibration of photometric redshift
distributions, characterizing both their means and their
outlier fractions at the $\sim 10^{-3}$ level or below.
Meeting this challenge requires optical and near-IR imaging
for robust identification of spectral breaks, and large
spectroscopic calibration data sets.  The third systematics
challenge for WL is intrinsic alignment of galaxies.
With continuing theoretical work {\it and} good photometric
redshifts, we believe that this systematic can be kept
subdominant, but it remains a challenging problem.
WL measurements are rich with observables, including higher
order statistics and varied combinations of galaxy-galaxy
lensing, galaxy clustering, and tomography.  Despite the
field's formidable technical obstacles, we think it quite possible
that constraints from WL surveys will eventually {\it exceed}
current forecasts because these additional observables provide
cosmological sensitivity and/or allow systematic uncertainties
to be calibrated away.

Cluster abundance measurements provide an alternative route to
measuring the growth of structure and thus testing the
consistency of GR growth predictions.
In addition, by reducing uncertainty in $\om$ and breaking
other degeneracies, cluster abundance measurements can sharpen
the equation-of-state constraints from SN, BAO, and WL distance measurements.
The key challenge for cluster cosmology is achieving unbiased
and precise calibration of the cluster mass scale.
Realizing the statistical power of future surveys requires absolute
mass calibration accurate at the $0.5-1\%$ level.
In our view, this is only achievable with weak lensing,
because the baryonic physics associated with other observables
is too uncertain to predict them this accurately from first
principles.  We thus see cluster studies as a natural
byproduct of WL surveys and in some sense as a specialized
branch of WL, one that takes advantage of the strong additional
information afforded by knowing the locations of peaks in
the optical galaxy density, X-ray flux, or SZ decrement.
If WL provides the fundamental mass calibration, then the 
shape measurement and photometric redshift uncertainties
that affect WL also affect cluster methods.

While all of these methods can be pursued at ambitious levels
from the ground, all would benefit from the capabilities of
a space mission, especially from the capability of wide-field
near-IR imaging and spectroscopy, which is possible at the
necessary depth only from space.  For SN, a space platform provides
the greater stability and sharp PSF needed for highly accurate
photometric calibration, and it allows observations in the
rest-frame near-IR, which is crucial for minimizing extinction
systematics and may be valuable for reducing evolution systematics.
For BAO, near-IR spectroscopy allows emission-line galaxy surveys
over the huge comoving volume from $1.2 \la z \la 2$, which is 
difficult to probe with ground-based optical or IR observations.
(Intensity-mapping radio methods may be
able to probe this redshift range from the ground, but 
this approach still has significant technological hurdles
to overcome.)
For WL, space observations allow the deep near-IR photometry
that is essential for robust and accurate photometric redshifts,
and they provide stable imaging with a sharp PSF that enables
accurate shape measurements for a high surface-density source
population.  The above considerations motivated both \wfirst\ and
the IR capabilities of {\it Euclid}.  Space-based optical imaging,
the other major element of {\it Euclid}, allows a significantly
sharper PSF and thus potentially more powerful WL measurements,
if the systematic errors are sufficiently well controlled.
More generally, space-based WL measurements can employ a higher
galaxy surface density than ground-based surveys to the
same photometric depth, both because the PSF itself is smaller
and because greater stability and the absence of atmospheric
effects should allow accurate measurements down to a smaller
ratio of $r_{50}/r_{\rm PSF}$.

The current generation of ``Stage III'' experiments such as
BOSS, PS1, DES, HSC,
and HETDEX are collectively pursuing all of
these methods, and they should achieve dark enery constraints
substantially better than those that exist today.
It is crucial that the next generation, Stage IV experiments
maintain, collectively, a balanced program that includes
SN, BAO, and WL, as well as other methods 
(clusters, Alcock-Paczynski, redshift-space distortions) that can be applied
to the same data sets.  There is much more to be gained, and
much lower risk, from doing a good job on all three methods
than from doing a maximal job on one at the expense of the
others.  A balanced program takes advantage of the methods'
complementary information content and areas of sensitivity,
and it allows the best cross-checks for systematic errors.
It is becoming standard practice to trade systematic uncertainties
for statistical errors by parameterizing their impact and
marginalizing --- e.g., over an uncertain shear calibration
multiplier or photometric redshift offset.  While this is a 
powerful strategy for removing biases due to ``known unknowns,''
it does not protect against ``unknown unknowns.''  Any conclusion
about cosmic acceleration will be more compelling if it is demonstrated
by independent methods, and the more interesting the conclusion,
the more crucial this independent confirmation will be.

In \S\ref{sec:forecast} we have provided quantitative forecasts
for a fiducial Stage IV program and for many variants upon it.
Our fiducial SN program assumes 0.01 mag mean errors 
for a local calibrator sample at $z=0.05$ and
in three bins of 
$\Delta z = 0.2$ at $0.2<z<0.8$, uncorrelated from bin to bin.
Our fiducial BAO program assumes mapping 1/4 of the sky to $z=3$,
with errors that are $1.8\times$ the linear theory sample
variance errors over this volume.  Different combinations of
redshift range and sky coverage that have the same comoving
volume yield nearly the same results.  Our fiducial WL program
assumes statistical errors of a $\sim 10^9$-galaxy imaging
survey (more precisely, $10^4\deg^2$ with 23 galaxies/arcmin$^2$),
and systematic errors of $2\times 10^{-3}$ in shear calibration
and photometric redshift calibration.  We also consider an
optimistic case in which the total (systematic + statistical)
errors are simply double the statistical errors, which 
effectively corresponds to total errors $\sim 2-3$ times
smaller than those of the fiducial case.
Our fiducial program corresponds fairly closely to the one
recommended by the Astro2010 Cosmology and Fundamental Physics
panel, and it is a reasonable, probably conservative forecast
of what could be achieved by a combination of LSST, \euclid/\wfirst,
and ground-based BAO and SN surveys.

To quantify the expected performance of this program and its
variants, we considered two dark energy models, one with 
$w_a = w_0+w_a(1-a) = w_p + w_p(a_p-a)$, where 
$a_p = (1+z_p)^{-1}$ is the expansion factor at which $w$
is best constrained, and a second with $w(a)$
allowed to vary freely in each of 36 bins of $\Delta a = 0.025$,
reaching to $z=9$.  In both cases we allowed deviations from 
GR-predicted growth rates characterized by an overall 
multiplicative offset $G_9$ in $G(z)$ and by a
shift $\Delta\gamma$ in the logarithmic growth rate
$d\ln G/d\ln a \propto [\Omega_m(a)]^{\gamma+\Delta\gamma}$.
We focused principally on the expected errors in $w_p$,
$w_a$, $\Delta\gamma$, and $G_9$, including the DETF FoM
defined as $(\sigma_{w_p}\sigma_{w_a})^{-1}$.
While principal components (PCs) of the general $w(a)$ model
allow a much richer characterization of the dark energy history
(and its uncertainties), we regard the combination of the
DETF FoM and the $\Delta\gamma$ error to be as good as any
alternative for characterizing the strength of a combined
program.

The primary results of our forecasting investigation appear in
Tables~\ref{tbl:forecasts1}--\ref{tbl:forecasts3} and, in distilled
form, in Figures~\ref{fig:fom} and~\ref{fig:contours_wl}.
The FoM of our fiducial program is 664, more than five times better
than our Stage III forecast and a roughly 50-fold improvement on
current knowledge.  Within the adopted parameterization, $1\sigma$ 
errors on individual parameters are 0.014 on $w_p$, 0.11 on $w_a$,
0.034 on $\Delta\gamma$, 0.015 on $\ln G_9$, $5.5\times 10^{-4}$
on $\ok$, and $5.1\times 10^{-3}$ on $h$.
All three methods contribute significantly to these constraints.
For our fiducial assumptions, BAO have the greatest leverage on the
DETF FoM, in the sense that halving the BAO errors produces the
greatest increase in the FoM while doubling the BAO errors produces
the greatest decrease.  WL has the least leverage, which implies
that the fiducial BAO and SN measurements constrain the expansion
history well enough that the WL measurements add relatively 
little constraining power.  However, the error on $\Delta\gamma$
scales nearly linearly with the WL errors, since all of the 
information on growth comes from the WL measurements.
(Note that we scale the {\it total} WL errors, equivalent
to multiplying systematic and statistical errors by the same factor.)
Conversely, changing the SN or BAO errors has almost no impact
on the $\Delta\gamma$ constraint.

Changing to our optimistic assumptions about WL systematics
(total errors equal to twice the statistical errors), while
retaining the fiducial SN and BAO assumptions, raises the
FoM from 664 to 789 and lowers the $\Delta\gamma$ error
from 0.034 to 0.026.  
For the optimistic systematics model,
WL measurements have the {\it greatest} leverage on the
DETF FoM instead of the least, and the $\Delta\gamma$
errors continue to scale approximately linearly with the WL errors.
Thus, our conclusions about the power of WL relative to
BAO and SN depend significantly on the assumed importance
of WL systematics, which is difficult to predict at present.

When we move from the $w_0-w_a$ model to the general
$w(z)$ model, the forecast errors on $\Delta\gamma$ barely
change, since it is constrained by differential measurements
of matter clustering over the redshift range of our
fiducial data sets.  The errors on $G_9$, on the other
hand, expand dramatically, because even within GR
the overall amplitude of structure can be shifted 
by the behavior of $w(z)$ outside of our constrained
redshift range (i.e., at $z>3$).  If the amplitude of 
matter clustering proved inconsistent with that of a
$w_0-w_a$, $G_9=1$ model, it would definitely indicate
something interesting, but this measurement alone would
not show whether the unusual behavior arises from a 
violation of GR or from unexpected behavior of $w(z)$
at high redshift.

For variations around our fiducial program, the impact of
reducing the errors of SN measurements is greater than
the impact of increasing the redshift range of these
measurements.  For example, reducing the error per redshift
bin from 0.01 mag to 0.005 mag increases the FoM from
664 to 1197, while increasing the maximum redshift from 
0.8 to 1.6 only raises the FoM to 841.
These scalings imply that the highest priority for SN
studies is to minimize statistical and systematic errors
at $z<1$, and that pushing to higher redshifts is a
lower priority until the reduction in $z<1$ systematics
has been saturated.  At fixed $f_{\rm sky}$, BAO constraints
have a stronger dependence on maximum redshift, because
at higher $z$ the BAO measurements become more precise
and the importance of the direct $H(z)$ measurements
grows.

We have not incorporated cluster abundances into our primary forecasts,
but we have investigated how precisely our fiducial Stage III and
Stage IV programs (CMB+SN+BAO+WL) predict the parameter combination 
$\sigElevz\Omega_m^{0.4}$ that is best constrained by cluster
abundances.  For a $w_0-w_a$ dark energy model, the 
forecast precision is $\sim 1.5\%$ for Stage III and $\sim 0.75\%$
for Stage IV if we assume GR is correct.  If we allow GR
deviations parameterized by $G_9$ and $\Delta\gamma$, then
the forecast precision degrades significantly, especially
for Stage III at $z > 0.5$.  Our analysis in \S\ref{sec:cl_mass_calibration}
indicates that clusters calibrated by stacked weak lensing
should be able to achieve higher precision on $\sigElevz\Omega_m^{0.4}$.
When we add the anticipated cluster constraints
for a $10^4\mdeg^2$ survey with a $10^{14}M_\odot$ mass threshold,
assuming that calibration errors are limited by weak lensing statistics,
we find that the DETF FoM grows by a factor of 1.4 at Stage III 
and 1.9 at Stage IV relative to the fiducial CMB+SN+BAO+WL program.
The error on $\Delta\gamma$ decreases by a factor of 3.2 for Stage III and
by 1.6 for Stage IV.  Cluster studies will be enabled automatically
by large WL surveys, which can be used to identify clusters as
optical galaxy concentrations and to provide mass calibration
for clusters identified by any method (X-ray, SZ, optical).
If they can achieve the limits imposed by weak lensing statistics,
they can add considerable leverage to tests of dark energy
models and deviations from GR.

We have adopted a similar strategy for some of the alternative
probes discussed in \S\ref{sec:alternatives}.  For a 
$w_0-w_a$ dark energy model, the forecast precision on 
$H_0$ is 0.7\% from our fiducial Stage IV program, 1.3\%
for Stage III.  A direct measurement of $H_0$ with 1\%
precision would improve the DETF FoM of the fiducial
Stage IV program by 20\%; a 2\% measurement would improve the 
Stage III FoM by 15\%.   The forecast constraint
on $H_0$ degrades, dramatically, to $\sim 60\%$ in our general $w(z)$ 
model, since large changes in $w$ at low redshift can
affect $H_0$ significantly while having minimal impact
on probes at higher redshift.  Thus, a discrepancy between
direct measurements and $H_0$ constraints from
CMB+SN+BAO+WL data could be a diagnostic for unusual
low-$z$ evolution of dark energy.

The Alcock-Pacyznski parameter $H(z)D_A(z)$ is constrained
to $\sim 0.2-0.3\%$ by our fiducial Stage IV program over
the redshift range $0.2 < z < 3$, setting a demanding
target for AP tests.  The corresponding precision forecast
for Stage III is $\sim 0.5\%$.
Redshift-space distortions and galaxy clustering can
measure the parameter combination $\sigma_8(z)f(z)$, which is 
constrained by our
Stage IV fiducial program to about 5\%
at $z\approx 0.1$, 2.5\% at $z=0.5$, and $\sim 1\%$ beyond
$z=1$, numbers that improve only slightly for the optimistic
weak lensing systematics.  For Stage III, the constraints
at a given redshift are considerably weaker.  In all cases
the constraints are tighter if we assume GR
($\Delta\gamma=0$, $G_9=1$), but the main purpose of
redshift-space distortion analyses would be to test GR growth,
so we regard the looser constraints as the more relevant
targets for such analyses.  This level of precision appears
within reach of large galaxy redshift surveys if theoretical
systematics can be adequately controlled, making redshift-space
distortions a potentially powerful addition to the arsenal
of cosmic acceleration probes.
While WL and redshift-space distortions both probe
structure growth, they have different dependences on the
two distinct potentials that enter the GR spacetime metric
(see \S\ref{sec:gravity}), so a discrepancy between
them could reveal a GR-deviation that might not be captured
by $\Delta\gamma$ alone.  Galaxy redshift surveys designed
for BAO measurements should allow redshift-space distortion
analyses (and AP tests) as an automatic by-product, which may
greatly increase their science return.  Precise measurements
of the shape of the galaxy power spectrum could also reveal
signs of scale-dependent growth, another possible consequence
of modified gravity models, though these may be difficult
to distinguish from other factors that affect the 
power spectrum shape (see \S\ref{sec:gravity}).

The aggregate precision of our fiducial Stage IV program measurements,
in the sense described in \S\ref{sec:forecast_aggregate}, is 
0.23\% in $D_L$ (from SN), 0.13\% in $D_A$ (from BAO),
0.21\% in $H$ (also from BAO), and 0.33\% in $\sigma_8$
for fixed geometry (for fiducial WL; optimistic WL yields 0.14\%).
For the fiducial Stage IV cluster program with weak lensing
mass calibration we forecast 0.20\% aggregate precision on 
$\sigma_8(z)\Omega_m^{0.4}$,
while our fiducial Stage IV RSD forecast yields 
0.22\% aggregate precision on $f(z)\sigma_8(z)$.
The ultimate limits on $H_0$ and Alcock-Paczynski measurements
are still difficult to predict, but sub-percent precision appears
well within reach on the Stage IV timescale.
These forecasts represent a dramatic advance over the current
state of the art, which is roughly 1-5\% for distance measurements
(SN, BAO, $H_0$) and $\sim 5\%$ for structure growth measurements
($\sigma_8$, $f(z)$).  The cosmological measurements of the past two
decades have established a ``standard model'' of cosmology based
on inflation, cold dark matter, a cosmological constant, and a
flat universe.  The measurements of the next two decades will test
that model far more stringently than it has been tested to date.

The future of cosmic acceleration studies depends partly on
the facilities built to enable them, partly on the ingenuity
of experimenters and theorists in controlling systematic errors
and fully exploiting their data sets, and partly on the kindness of nature.
The next generation of 
experiments could merely tighten the noose around $w=-1$,
ruling out many specific theories but leaving us no more
enlightened than we are today about the origin of cosmic acceleration.
However, barely a decade after the first supernova measurements 
of an accelerating universe, it seems unwise to bet that we
have uncovered the last ``surprise'' in cosmology.
Equally important, the powerful data sets required to study cosmic
acceleration support a broad range of astronomical investigations.
These observational efforts are natural next steps in a
long-standing astronomical tradition: mapping the universe
with increasing precision over ever larger scales, from the
solar system to the Galaxy to large scale structure to the CMB.
These ever growing maps have taught us extraordinary things ---
that gravity is a universal phenomenon, that we live in a galaxy populated
by 100 billion stars, that our galaxy is one of 100 billion within 
our Hubble volume, that our entire observable universe has
expanded from a hot big bang 14 billion years in the past,
that the dominant form of matter in the
universe is non-baryonic, and that the early universe was
seeded by Gaussian (or nearly Gaussian) fluctuations that
have grown by gravity into all of the structure that we observe today.
We hope that the continuation of this tradition will lead
to new insights that are equally profound.

\bigskip

We gratefully acknowledge the many mentors, collaborators, and
students with whom we have learned this subject over the years.
For valuable comments and suggestions on the draft manuscript, we thank
Joshua Frieman, Dragan Huterer, Chris Kochanek,
Andrey Kravtsov, Mark Sullivan, and Alexey Vikhlinin.
We also thank the many readers who sent comments in response to
the original arXiv posting of the article, which led to numerous
improvements in the text and referencing.
We gratefully acknowledge support from the National Science Foundation, the
National Aeronautics and Space Administration, the
Department of Energy Office of Science, including NSF grants
AST-0707725, AST-0707985, AST-0807337, and AST-1009505, NASA grant
NNX07AH11G1320, and DOE grants DE-FG03-02-ER40701
and DE-SC0006624.
DW acknowledges the hospitality of the Institute for Advanced
Study and the support of an AMIAS membership during critical
phases of this work.
MM was supported by the Center for Cosmology and Astro-Particle Physics
(CCAPP) at Ohio State University.
CH acknowledges additional support from the Alfred P. Sloan Foundation
and the David \& Lucile Packard Foundation.
ER was supported by the NASA Einstein Fellowship Program,
grant PF9-00068.

\vfill\eject

%\singlespacing

\appendix

\section{Glossary of Acronyms and Facilities}

\noindent
Note that we have not repeated the acronyms of X-ray surveys
listed in Table~\ref{tbl:Xray_catalogs}.
\medskip

{
\parindent=0pt\parskip=0pt

ACS: Advanced Camera for Surveys (on Hubble Space Telescope)

ACT: Atacama Cosmology Telescope

ADEPT: Advanced Dark Energy Physics Telescope

AP: Alcock-Paczynski

BAO: Baryon Acoustic Oscillations

BOSS: Baryon Oscillation Spectroscopic Survey

BigBOSS: Big Baryon Oscillation Spectroscopic Survey

CCD: Charge Coupled Device

CDM: Cold Dark Matter

CFHT: Canada-France-Hawaii Telescope

CFHTLS: Canada-France-Hawaii Telescope Legacy Survey

Chandra: Chandra X-ray Observatory (NASA)

CHIME: Canadian Hydrogen Intensity Mapping Experiment

CMB: Cosmic Microwave Background

COBE: Cosmic Background Explorer

COSMOS: Cosmic Evolution Survey (from Hubble Space Telescope)

CSP: Carnegie Supernova Project

DES: Dark Energy Survey

DEspec: Dark Energy Spectrograph

DESTINY: Dark Energy Space Telescope

DETF: Dark Energy Task Force

DUNE: Dark Universe Explorer

EE50: Encircled Energy 50\%

eROSITA: extended Roentgen Survey with an Imaging Telescope Array

ESA: European Space Agency

ESSENCE: Equation of State: SupErNovae trace Cosmic Expansion

Euclid: Euclid dark energy space mission (ESA)

FKP: Feldman-Kaiser-Peacock (1994) $P(k)$ estimation method

FFT: Fast Fourier Transform

FIRST: Faint Images of the Radio Sky at Twenty-Centimeters (from the VLA)

FWHM: Full Width at Half Maximum

Gaia: Gaia astrometry mission (ESA)

GR: General Relativity

HEAO: High-Energy Astrophysics Observatory (NASA)

HETDEX: Hobby-Eberly Telescope Dark Energy Experiment

HOD: Halo Occupation Distribution

HSC: Hyper-Suprime Camera (for Subaru Telescope)

HST: Hubble Space Telescope

IGM: Intergalactic Medium

IRAC: Infrared Array Camera (on Spitzer Space Telescope)

ISCS: IRAC Shallow Cluster Survey

JDEM: Joint Dark Energy Mission

JEDI: Joint Efficient Dark-energy Investigation

JPAS: Javalambre Physics of the Accelerating Universe Astrophysical Survey
% http://j-pas.org/

JWST: James Webb Space Telescope

KIDS: Kilo-Degree Survey
% http://www.astro-wise.org/projects/KIDS/

LCS: Light Curve Shape

LIGO: Laser Interferometer Gravitational Wave Observatory

LOSS: Lick Observatory Supernova Survey

LRG: Luminous Red Galaxy

LSST: Large Synoptic Survey Telescope

NASA: National Aeronautics and Space Administration

NOAO: National Optical Astronomy Observatories

NVSS: NRAO VLA Sky Survey

Pan-STARRS: Panoramic Survey Telescope and Rapid Response System

PAU: Physics of the Accelerating Universe

PCA: Principal Component Analysis

Planck: Planck CMB satellite (ESA)

PS1: Pan-STARRS 1
% , http://pan-stars.ifa.hawaii.edu/public

PSF: Point Spread Function

PTF: Palomar Transient Factory

RASS: ROSAT All Sky Survey

RCS: Red-Sequence Cluster Survey
% , http://www.astro.utoronto.ca/~gilbank/RCS2

ROSAT: Roentgen Satellite

SDSS: Sloan Digital Sky Survey

SED: Spectral Energy Distribution

SFR: Star Formation Rate

SKA: Square Kilometer Array

SN: Supernovae

SNAP: Supernova Acceleration Probe

SNLS: Supernova Legacy Survey (part of CFHTLS)

SNR: Signal-to-Noise Ratio

SPT: South Pole Telescope

STEP: Satellite Test of Equivalence Principle

SuMIRe: Subaru Measurement of Images and Redshifts

2SLAQ: 2dF and SDSS Large Area Quasar survey

VHS: VISTA Hemisphere Survey
%http://www.ast.cam.ac.uk/$\sim$rgm/vhs/

VLBI: Very Long Baseline Interferometry

VIKING: VISTA Kilo-Degree Infrared Galaxy Survey
% http://www.eso.org/sci/observing/policies/PublicSurveys/sciencePublicSurveys.html

VIRGO: VIRGO gravity wave observatory

VVDS: VIMOS-VLT Deep Survey

UKIDSS: UKIRT Infrared Deep Sky Survey
% http://www.ukidss.org/

WFC3: Wide-Field Camera 3 (on Hubble Space Telescope)

WFPC2: Wide-Field and Planetary Camera 2 (on Hubble Space Telescope)

WFIRST: Wide Field Infrared Survey Telescope

WiggleZ: WiggleZ galaxy redshift survey

WISE: Wide-field Infrared Survey Explorer 
% \cite{wright10}

WL: Weak Lensing

WMAP: Wilkinson Microwave Anisotropy Probe (NASA)

XMM-Newton: X-ray Multi-Mirror Mission 

}

\vfill\eject

\bibliographystyle{elsarticle-harv}
\bibliography{acceleration.bib}

%% Authors are advised to submit their bibtex database files. They are
%% requested to list a bibtex style file in the manuscript if they do
%% not want to use elsarticle-harv.bst.

%% References without bibTeX database:

% \begin{thebibliography}{00}

%% \bibitem must have one of the following forms:
%%   \bibitem[Jones et al.(1990)]{key}...
%%   \bibitem[Jones et al.(1990)Jones, Baker, and Williams]{key}...
%%   \bibitem[Jones et al., 1990]{key}...
%%   \bibitem[\protect\citeauthoryear{Jones, Baker, and Williams}{Jones
%%       et al.}{1990}]{key}...
%%   \bibitem[\protect\citeauthoryear{Jones et al.}{1990}]{key}...
%%   \bibitem[\protect\astroncite{Jones et al.}{1990}]{key}...
%%   \bibitem[\protect\citename{Jones et al., }1990]{key}...
%%   \harvarditem[Jones et al.]{Jones, Baker, and Williams}{1990}{key}...
%%

% \bibitem[ ()]{}

% \end{thebibliography}

\end{document}